%% file: Henderson_Thesis.tex
\documentclass[12pt,twoside]{mitthesis}
\usepackage{lgrind}
\usepackage{graphicx}
\graphicspath{{./figures/}}
\usepackage{amssymb}
\usepackage{amsmath}
\usepackage{url}
\usepackage{rotating}
\usepackage{float}
\usepackage{siunitx}
\usepackage{cite}
\usepackage[nottoc,notlot,notlof]{tocbibind}
\usepackage[pdftex, hyperindex=true, bookmarks]{hyperref}
\usepackage{setspace}
\newcommand{\ratio}{{$\sigma_{e^+p}/\sigma_{e^-p}$} }
\newcommand{\dg}{{$^\circ$} }
\newcommand{\ep}{{$e^- p$} }
\newcommand{\pp}{{$e^+ p$} }
\newcommand{\pmp}{{$e^\pm p$} }
\newcommand{\rtg}{{$R_{2\gamma}$} }
\newcommand{\D}{\mathrm{d}}
\newcommand{\pd}[2]{\frac{\partial #2}{\partial #1}}
\newcommand{\td}[2]{\frac{\mathrm{d} #1}{\mathrm{d} #2}}

\makeatletter
\g@addto@macro\bfseries{\boldmath}
\makeatother

\begin{document}

\setcounter{tocdepth}{3} 
\include{cover}
% Some departments (e.g. 5) require an additional signature page.  See
% signature.tex for more information and uncomment the following line if
% applicable.
%\include{signature}
\pagestyle{plain}
\include{contents}
\include{chap1}
\include{chap2}
\include{chap3}
\include{chap4}
\include{chap5}
\include{chap6}
\include{chap7}
\appendix
\include{appa}
\include{appb}
\include{biblio}
\end{document}

%% file: cover.tex
% -*-latex-*-
% 
% For questions, comments, concerns or complaints:
% thesis@mit.edu
% 

\title{A Precision Measurement of the $e^+p/e^-p$ \\ Elastic Scattering Cross Section Ratio 
at the \\ OLYMPUS Experiment}

\author{Brian Scott Henderson}
% If you wish to list your previous degrees on the cover page, use the 
% previous degrees command:
\prevdegrees{B.S. in Physics, William Marsh Rice University (2010) \\
             B.A. in Mathematics, William Marsh Rice University (2010)}
%\prevdegrees{B.A. in Mathematics, William Marsh Rice University (2010)}
% You can use the \\ command to list multiple previous degrees
%       \prevdegrees{B.S., University of California (1978) \\
%                    S.M., Massachusetts Institute of Technology (1981)}
\department{Department of Physics}

% If the thesis is for two degrees simultaneously, list them both
% separated by \and like this:
% \degree{Doctor of Philosophy \and Master of Science}
\degree{Doctor of Philosophy}

% As of the 2007-08 academic year, valid degree months are September, 
% February, or June.  The default is June.
\degreemonth{September}
\degreeyear{2016}
\thesisdate{August 15, 2016}

%% By default, the thesis will be copyrighted to MIT.  If you need to copyright
%% the thesis to yourself, just specify the `vi' documentclass option.  If for
%% some reason you want to exactly specify the copyright notice text, you can
%% use the \copyrightnoticetext command.  
%\copyrightnoticetext{\copyright IBM, 1990.  Do not open till Xmas.}

% If there is more than one supervisor, use the \supervisor command
% once for each.
\supervisor{Richard Milner}{Professor of Physics}

% This is the department committee chairman, not the thesis committee
% chairman.  You should replace this with your Department's Committee
% Chairman.
\chairman{Nergis Mavalvala}{Curtis and Kathleen Marble Professor of Astrophysics \\Associate Department Head of Physics}

% Make the titlepage based on the above information.  If you need
% something special and can't use the standard form, you can specify
% the exact text of the titlepage yourself.  Put it in a titlepage
% environment and leave blank lines where you want vertical space.
% The spaces will be adjusted to fill the entire page.  The dotted
% lines for the signatures are made with the \signature command.
\maketitle

% The abstractpage environment sets up everything on the page except
% the text itself.  The title and other header material are put at the
% top of the page, and the supervisors are listed at the bottom.  A
% new page is begun both before and after.  Of course, an abstract may
% be more than one page itself.  If you need more control over the
% format of the page, you can use the abstract environment, which puts
% the word "Abstract" at the beginning and single spaces its text.

%% You can either \input (*not* \include) your abstract file, or you can put
%% the text of the abstract directly between the \begin{abstractpage} and
%% \end{abstractpage} commands.

% First copy: start a new page, and save the page number.
\cleardoublepage
% Uncomment the next line if you do NOT want a page number on your
% abstract and acknowledgments pages.
% \pagestyle{empty}
\setcounter{savepage}{\thepage}
\begin{abstractpage}
\input{abstract}
\end{abstractpage}

% Additional copy: start a new page, and reset the page number.  This way,
% the second copy of the abstract is not counted as separate pages.
% Uncomment the next 6 lines if you need two copies of the abstract
% page.
% \setcounter{page}{\thesavepage}
% \begin{abstractpage}
% \input{abstract}
% \end{abstractpage}

\cleardoublepage

\section*{Acknowledgments}

\begin{singlespacing}

Foremost, OLYMPUS was a decade-long project (from conception to the final analyses) involving a collaboration of 60+ people from 6 countries who made
contributions to the various phases of the experiment that are far to numerous to list here.  In particular, of course, I am extremely grateful to my
advisor Professor Richard Milner.  It has been a true privilege to learn from Richard's expertise over the past six years, both in terms of physics and in management of an experiment.
Thank you also to my other committee members Professors Robert Redwine and Robert Jaffe, who both provided excellent guidance in the construction of this thesis.

The staff, technicians, and engineers of the MIT-Bates Research and Engineering Center were not only instrumental to the design and construction of OLYMPUS,
but are also great people and were an absolute pleasure to work with from my first day at MIT.  While I risk
missing somebody by listing names, I learned much of what I know about putting together an experiment from Jim Kelsey, Chris Vidal, Peter Binns, Brian O'Rourke,
and Joe Dodge, who all made several treks to Hamburg to lend their expertise to the construction of the experiment.

Numerous technicians and engineers at DESY worked on the construction of the experiment as well, and while I never worked with any of them closely,
their important contributions must be acknowledged.
Special recognition goes to the DESY accelerator team, led by Frank Brinker, who delivered consistent and well-maintained electron and positron beams to the experiment
throughout our data-taking periods (and would greet the off-going OLYMPUS night shift crew with a hearty ``moin moin'' and handshake at 7 AM as they arrived for the day). 
I would also like to thank the managers of the MIT Engaging computing cluster, who provided a, quite frankly, miraculous resource for the analysis of this experiment.
Although I have never met them in person, they were always quick to resolve any issues we had in using the cluster and patient in putting up with my excessive disk usage.

Looking further into the past, I would like to thank two teachers, in particular, from my junior high and high school years, Christine Moore and Terry Crane,
who encouraged me to think seriously about physics and helped open the door to a much larger academic world.

Among members of the OLYMPUS collaboration, I would like to thank several in particular.  First, thanks to Michael Kohl, Uwe Schneekloth, and Douglas Hasell
for their roles in making OLYMPUS happen and keeping things on track (and especially to the latter for helping take care of us MIT students as we traveled back and forth
from Hamburg and pushed through the analysis).

While not on the author list, Joanne Gregory was an immense help in all aspects of coordinating travel to and from Germany, finding
places to stay while there, and generally navigating LNS and MIT.  Her expertise in managing the hadronic physics group will most certainly be missed.

Several of the OLYMPUS post-docs were monumentally critical in making OLYMPUS happen: Jan Bernauer, Alexander Winnebeck, and J\"{u}rgen Diefenbach.  Each of them
performed apparent miracles in finding last-minute solutions to hardware and software problems, and given another six years in grad school I don't think I could learn everything
they would have had to teach me about experimental physics.

A true place of honor among my acknowledgments belongs to the other OLYMPUS graduate students: Axel Schmidt, Colton O'Connor, Rebecca Russell, and Lauren Ice.
Each of them played a critical role in constructing the OLYMPUS results, taking on responsibilities above a grad student's pay-grade and finding creative, excellent
solutions to many of the challenges we faced.
In addition to their enormous contributions to the experiment, they were great friends who made the whole experience of living in Germany and the grind of building,
operating, and conducting the analysis for the experiment much better.

My parents, Patricia and Scott, are certainly responsible for much of this accomplishment, as they have always been unwaveringly supportive of my education
and my goals, and instilled in me a strong desire to learn and fight for answers.  Thank you, Mom and Dad.

More than anybody else, my wife Hilary has helped me through grad school (even marrying me in the thick of it).  Through the time spent apart in Hamburg, the frustration
when the analysis was stalled, and the crush of work that accompanied the last several months of the project, she has always been loving and supportive, all while going
through law school and the bar exam herself.  Thank you, Hil.

%Richard, committee, rest of the collaboration

% Bates techs (and DESY), Engaging cluster

%Doug, Post-docs

%Penthouse

%Fellow grad students (Axel's plots)

%Parents

%Hilary

\end{singlespacing}

%%%%%%%%%%%%%%%%%%%%%%%%%%%%%%%%%%%%%%%%%%%%%%%%%%%%%%%%%%%%%%%%%%%%%%
% -*-latex-*-

%% file: abstract.tex
%% The text of your abstract and nothing else (other than comments) goes here.
%% It will be single-spaced and the rest of the text that is supposed to go on
%% the abstract page will be generated by the abstractpage environment.  This
%% file should be \input (not \include 'd) from cover.tex.

Measurements of the ratio of the proton elastic form factors ($\mu_pG_E/G_M$)
using Rosenbluth separation and those using polarization-based techniques show a strong discrepancy, which
has persisted both in modern experimental results and in re-analyses of previous data.  The most widely accepted
hypothesis to explain this discrepancy is the treatment of the contributions from hard two-photon
exchange (TPE) to elastic electron-proton scattering in the radiative corrections applied to the Rosenbluth separation measurements.
Calculations of the hard TPE contribution are highly model dependent, but the effect may be measured
experimentally with a precise determination of the ratio of the positron-proton and electron-proton elastic scattering cross sections.

The OLYMPUS experiment collected approximately 4 fb$^{-1}$ of \pp and \ep scattering data at the DORIS storage ring at DESY in 2012, with the
goal of measuring the elastic \ratio ratio over the kinematic range $(0.4 \leq \epsilon \leq 0.9)$, $(0.6 \leq Q^2 \leq 2.2)$ GeV$^2/c^2$ at a
fixed lepton beam energy of 2.01 GeV.  The detector for the OLYMPUS experiment consisted of refurbished elements of the Bates Large Acceptance
Spectrometer Toroid (BLAST) surrounding an internal gaseous hydrogen target, with the addition of multiple systems for the monitoring of the
luminosity collected by the experiment.  A detailed simulation of the experiment was developed to account for both radiative corrections and
various systematic effects.

This work presents preliminary results from the OLYMPUS data, demonstrating that the elastic \ratio ratio rises to several percent at
$\epsilon\approx 0.4$ and indicating a significant contribution from TPE to \pmp scattering.  Additionally, the value of \ratio has been
measured to unprecedented precision at $\epsilon=0.98$, which provides a valuable normalization point for other experimental data.

%% file: contents.tex
  % -*- Mode:TeX -*-
%% This file simply contains the commands that actually generate the table of
%% contents and lists of figures and tables.  You can omit any or all of
%% these files by simply taking out the appropriate command.  For more
%% information on these files, see appendix C.3.3 of the LaTeX manual. 
\tableofcontents
\newpage
\listoffigures
\newpage
\listoftables

%% file: chap1.tex
% Chapter 1
%
% General introduction to problem, sort of extended abstract, description
% of the thesis
%

\chapter{Introduction}
\label{Chap1}

While protons and neutrons comprise nearly all of visible matter by mass, their inner workings remain relatively poorly understood.
Since Rutherford reported the ``anomalous'' discovery that collisions between $\alpha$ particles (helium nuclei)
and nitrogen nuclei could produce hydrogen nuclei (i.e., protons) \cite{doi:10.1080/14786431003659230} in 1919, understanding the role of the proton
in nuclei and the nature of the proton itself has been one of the primary goals of particle physics.  Shortly after his discovery, Rutherford proposed that
the nuclei of atoms consist of protons and related neutral particles of very similar mass \cite{Rutherford374}.  Chadwick discovered Rutherford's
neutral particles in 1932 \cite{Chadwick0,Chadwick1}, and he found that they behaved suspiciously similarly to protons, despite their neutral charge, indicating that
interesting physics hid within the nature of protons and neutrons. The first hint of the proton's internal complexity came with Otto Stern's 1933 discovery
that the magnetic moment of the proton deviated by a factor of 2.79 from the value
expected if it was a point-like particle \cite{stern1,stern2}.  Experiments such as those of Hofstadter at Stanford began to probe the structure of
the proton with elastic electron-proton scattering in the 1950s, demonstrating the proton's finite size \cite{PhysRev.98.217,PhysRev.102.851,RevModPhys.28.214}.  Concurrent with the experiments
of Hofstadter, other experiments were discovering a previously unknown sector of particles related to the proton: a variety of hadronic states varying widely in mass, quantum numbers,
lifetimes, and decay products.  The quark model proposed by Gell-Mann and Zweig in 1964 \cite{GELLMANN1964214,Zweig:570209} gave order to the mess of hadrons, and the MIT-SLAC
experiments in the 1960s and 1970s slowly uncovered  the point-like nature of the proton's, and other hadrons', constituents (i.e., quarks and gluons) \cite{PhysRevLett.23.930,PhysRevLett.23.935}.

While quantum chromodynamics (QCD) comprises a complete quantum field theory of the strong interaction of quarks and gluons \cite{doi:10.1142/9789814525220_0008,Agashe:2014kda}, the non-perturbative
nature of this theory at low energies poses significant difficulties in modeling the interactions that give rise to the bound states of QCD.  Thus,  experimental tests of QCD remain critical to furthering
knowledge of the strong force.  In particular, a complete understanding of the proton, the fundamental bound state of QCD, is crucial.  Despite over a century of both theoretical and experimental study,
basic questions remain about the proton, including whether it is truly stable \cite{pdperk,pdproc}, how its constituents give rise to its total spin ($S=\frac{1}{2}$) \cite{jps,doi:10.1142}, and more seemingly
simple topics such as the physical extent and distribution of charge within protons \cite{Carlson201559,Perdrisat2007694,PhysRevC.69.022201}.  OLYMPUS is chiefly concerned
with a recent puzzle in the latter category.

\section{The Fundamentals of OLYMPUS: Motivation and Goals}

\begin{figure}[thb!]
\centerline{\includegraphics[width=1.0\textwidth]{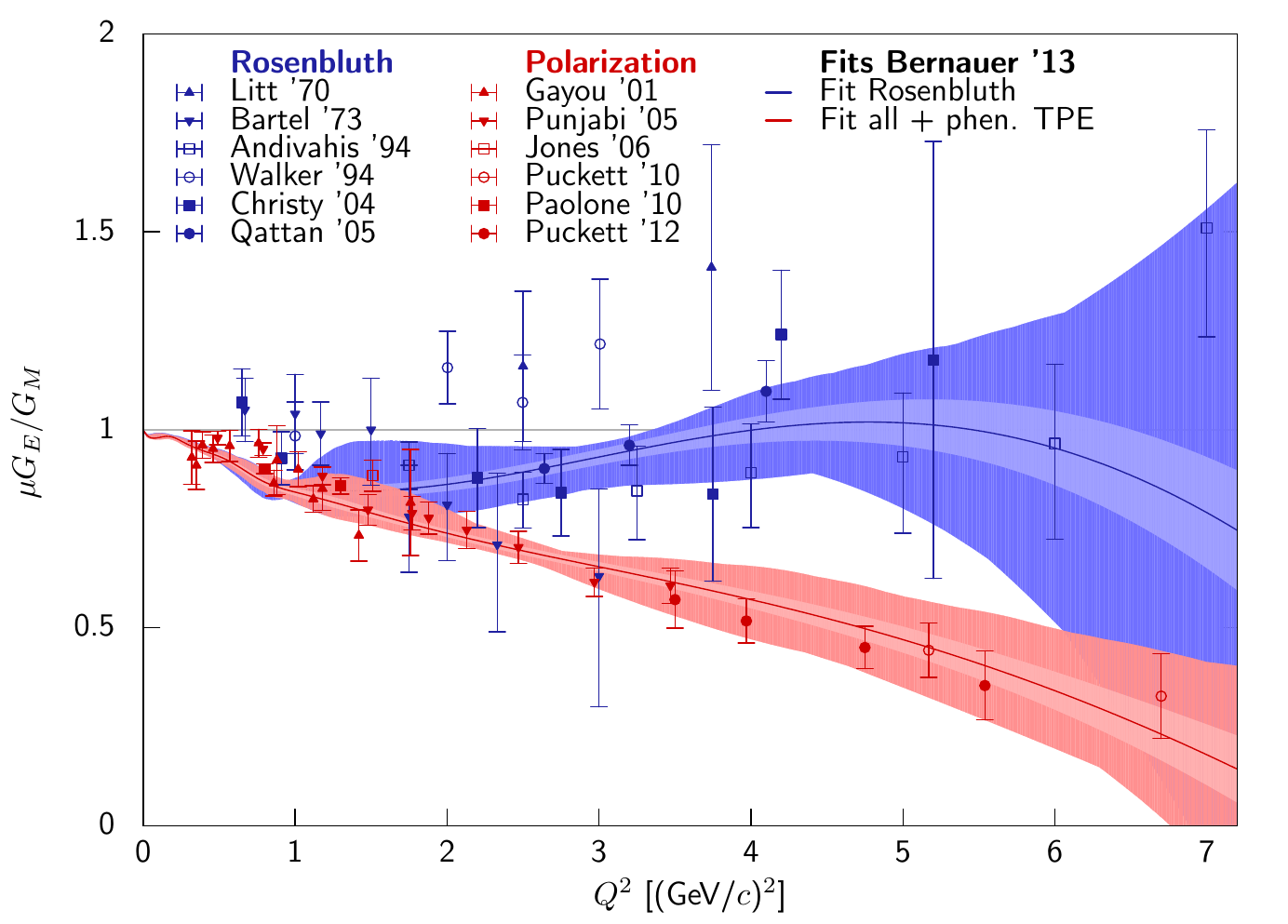}}
\caption[The $G_E/G_M$ discrepancy]{Selection of experimental results on the ratio $\frac{\mu_pG_E}{G_M}$ along with phenomenological fits to the data, illustrating the discrepancy
between experiments using Rosenbluth separation and polarization-based methods.  
(Rosenbluth separation data: \cite{ff3,ff4,ff8,ff9,ff10,ff11}, polarization data: \cite{pol11,pol3,pol12,pol7,pol9,pol13}, phenomenological fits: \cite{BerFFPhysRevC.90.015206}) (Figure reproduced from Reference \cite{Milner:2014}).}
\label{fig:disc}
\end{figure}

The OLYMPUS experiment concerns fundamental properties that were examined by the earliest studies of the proton's finite size and structure, i.e., measurement of the 
form factors $G_E$ and $G_M$ for the elastic scattering of electrons from protons that encode how the structure of the proton affects the electromagnetic interaction between
the two particles.  The form factors are typically represented as functions of the square of the four-momentum transfer that occurs in the \ep interaction $Q^2$, so as to make
them Lorentz invariant representations of the proton's structure.
Measurements of the form factors began in the 1950s with the work of Hofstadter at Stanford \cite{PhysRev.98.217,PhysRev.102.851,RevModPhys.28.214} and continued using the method
originally formulated by Rosenbluth \cite{PhysRev.79.615} (Section \ref{sec:rossep}) into the 1990s, pushing the measurements to higher values of $Q^2$ \cite{ff3,ff4,ff8,ff9}.  In the 1990s,
the development of polarized electron beams and targets provided a new method to study the elastic form factors, offering a means of measuring the ratio $\frac{\mu_pG_E}{G_M}$ as a function of $Q^2$ rather than
a method of measuring them individually \cite{pol11,pol3,pol12,pol7,pol9,pol13}.  When comparing the behavior of $\frac{\mu_pG_E}{G_M}$ as found
by the previous data (shown in blue in Figure \ref{fig:disc}) to the new data based on methods employing polarization (the red data in \ref{fig:disc}), a significant disparity between
the results yielded by each method was discovered.  All data prior to the polarization-based measurements was consistent with a value of $\frac{\mu_pG_E}{G_M}$ that stays relatively constant and near unity as a function of $Q^2$
while the polarization data favors a clear downward sloping trend that is completely inconsistent with the previous results.  Theorists reanalyzed the data of the older experiments
using modern methods \cite{PhysRevC.68.034325} and new experiments using the method of Rosenbluth were conducted \cite{ff10,ff11}, but the discrepancy persisted.

The most widely-accepted hypothesis to explain this discrepancy is that the Rosenbluth method of extracting the form factors from \ep scattering data does not properly account for an effect
that was previously assumed to be negligible: the contribution to the scattering cross section from two photons being exchanged between the particles. Typically, calculations
assume a single photon carrying the full momentum transfer.  Due to the complexity of the proton's state between the exchange of the two photons, this contribution
is not analytically calculable from theory and predictions from various models vary significantly \cite{Blunden:2003sp,Chen:2004tw,Afanasev:2005mp,Blunden:2005ew, Kondratyuk:2005kk, Borisyuk:2006fh,TomasiGustafsson:2009pw}.
Experiments to test this hypothesis were necessary.

\begin{figure}[thb!]
\centerline{\includegraphics[width=1.0\textwidth]{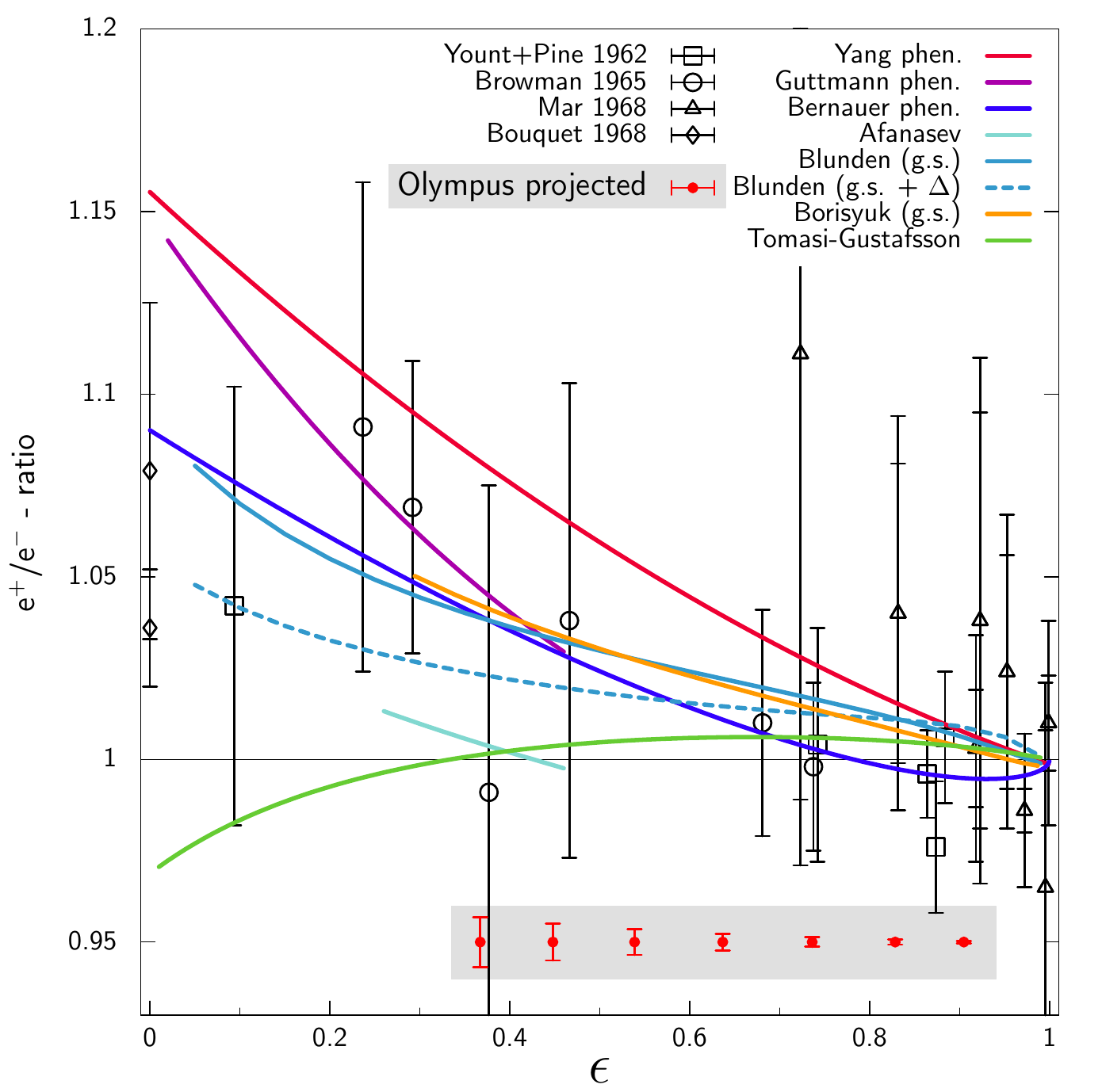}}
\caption[Existing data, model predictions, and projected OLYMPUS uncertainty on \ratio]{The ratio of \pp to \ep scattering at a lepton beam energy
of 2.01 GeV as a function of the kinematic variable $\epsilon$ (see Equation \ref{eq:eps}) as predicted by phenomenological models \cite{BerFFPhysRevC.90.015206,Chen:2007ac, Guttmann:2010au},
theoretical calculations of possible two-photon exchange contributions \cite{Blunden:2003sp,Chen:2004tw,Afanasev:2005mp,Blunden:2005ew, Kondratyuk:2005kk, Borisyuk:2006fh,TomasiGustafsson:2009pw},
and projected uncertainties of the measurement OLYMPUS will provide.  The existing world data on \ratio prior to $\sim$2015 is shown \cite{Yount:1962aa,Browman:1965zz,Bouquet:1968aa,Mar:1968qd}, although these experiments were all at different beam
energies. (Figure reproduced from Reference \cite{Milner:2014}.)}
\label{fig:projections}
\end{figure}

As will be discussed in Section \ref{sec:estpe}, a measurement of the ratio of the cross section for  positron-proton scattering to that of electron-proton scattering provides
an experimental signature of two-photon exchange in the lepton-proton interaction.  To test the hypothesis, and provide a provide a means of discriminating between different theoretical and phenomenological
predictions for the significance of the two-photon exchange contributions, measurements must have very small uncertainties ($\lesssim 1\%$).  The OLYMPUS experiment was designed to achieve this
goal.  The approach taken by OLYMPUS to make this measurement is summarized by the following equation:
\begin{equation}
 R_{2\gamma}\left(\epsilon,Q^2\right) = \frac{\sigma_{e^+p}\left(\epsilon,Q^2\right)}{\sigma_{e^-p}\left(\epsilon,Q^2\right)} = \frac{N_{e^+p,\text{data}  }\left(\epsilon,Q^2\right)}{N_{e^- p,\text{data}}\left(\epsilon,Q^2\right)} \cdot
	  \frac{N_{e^-p,\text{MC}}\left(\epsilon,Q^2,\mathcal{L}_{e^-}\right) }{N_{e^+ p,\text{MC}}\left(\epsilon,Q^2,\mathcal{L}_{e^+}\right)}.
 \label{eq:rat}
\end{equation}
OLYMPUS compared an experimental measurement of the rates of elastic \pmp scattering ($N_{e^\pm p,\text{data}  }\left(\epsilon,Q^2\right)$) with a detailed simulation of the expected
rates in the absence of two-photon exchange ($N_{e^\pm p,\text{MC}  }\left(\epsilon,Q^2,\mathcal{L}_{e^pm}\right)$).  In addition to collecting a high-statistics sample of both \ep and \pp events to control statistical uncertainties, it was also critical
to precisely measure the relative luminosity collected for each of the lepton species ($\mathcal{L}_{e^+}/\mathcal{L}_{e^-}$), and to create a model of the experiment in simulation that
accurately reflected the reality of the experiment to control the systematic uncertainties of the measurement.  This thesis provides a comprehensive discussion of the OLYMPUS
experiment and analysis to produce a measurement of $R_{2\gamma}$ with $\lesssim1\%$ total uncertainty in the kinematic range of $(0.4 \leq \epsilon \leq 0.9)$, $(0.6 \leq Q^2 \leq 2.2)$ GeV$^2/c^2$.
Projections for the precision of the experiment, the world's existing data prior to 2015 on \ratio, and various theoretical and phenomenological predictions for the results of OLYMPUS are shown in
Figure \ref{fig:projections}.

\section{Notes on the Organization and Content of this Work}

The intent of this work is to provide discussion of the essential parts of all elements of the OLYMPUS experiment and complete details on the primary elements which the author
contributed to the experiment.  Topics in the latter category include various elements of the detector construction and calibration (in particular, the hydrogen
target system (Section \ref{sec:target})), the survey of the magnetic field and implementation of the results as a field map for tracking and simulation (Section \ref{sec:magsur}),
the slow control luminosity analysis (Section \ref{sec:sclumi}), all aspects of the 12\dg luminosity analysis from calibration to the event selection analysis (Section \ref{sec:12lumi}),
and a complete independent analysis of the main OLYMPUS \ratio result (Chapter \ref{Chap6}).  Appendix \ref{chap:ed} presents a 3D event display framework created by the author
for the visualization of both data and simulated events for OLYMPUS using ROOT libraries \cite{Brun1997} that is easily adaptable for other experiments.  The other theses written on
OLYMPUS provide detailed descriptions of a number of topics including radiative corrections (References \cite{schmidt} and \cite{russell}), track reconstruction (References \cite{schmidt}
and \cite{russell}), the symmetric M{\o}ller-Bhabha luminosity analyses (Reference \cite{oconnor} and \cite{schmidt}), the analysis and simulation of the time-of-flight detector
system (Reference \cite{russell}), implementation of the surveyed detector geometry in the simulation detector model \cite{oconnor}, and independent \ratio analyses in each.

Rather than reproduce discussion on basic topics of particle physics that are well covered in other works, this thesis assumes basic knowledge of essential particle and nuclear physics concepts,
both in the description of physics processes and the detection of particles, and thus only includes introductions to essential advanced topics that are critical to the motivation and execution of
the OLYMPUS experiment.  Excellent references for any unfamiliar terms and concepts include References \cite{Agashe:2014kda} and \cite{griffiths} for essential
concepts of particle physics, Reference \cite{grupen} for particle detection related topics, and Reference \cite{peskin} for the basics of particle physics theory.

As a guide to the contents of this work, the following summarizes the main topics and goals of each chapter:
\begin{itemize}
 \item Chapter \ref{Chap2} provides an introduction to the fundamentals of proton form factors and their experimental determination and a more detailed discussion
       regarding the proton form factor puzzle, the two-photon exchange hypothesis, and the goals of OLYMPUS.
 \item Chapter \ref{Chap3} describes the essential elements of the design and operation of the OLYMPUS experiment, including conventions for description of the experiment, the beam,
       the hydrogen target, and the detector systems.
 \item Chapter \ref{Chap4} outlines the OLYMPUS analysis strategy, in particular the work done to ensure that the OLYMPUS simulation robustly represented the experimental conditions
       to allow comparison of data and Monte Carlo for the final analyses.
 \item Chapter \ref{Chap5} details the measurement of the relative luminosity of electron and positron data collected in the three independent systems designed for the measurement, with
       particular attention to the slow control and 12\dg luminosities.  For the 12\dg luminosity, all relevant details of the analysis are presented including calibration, hit reconstruction, track 
       reconstruction, simulation implementation, event selection, and a detailed systematic uncertainty analysis.   For the slow control luminosity, a novel Monte Carlo code is presented that
       simulates the density of gas undergoing molecular flow within a geometry.       
 \item Chapter \ref{Chap6} describes the independent analysis of the main cross section result conducted by the author, including aspects of the detector performance and simulation, the analysis
       methods used, and an estimate of systematic uncertainties.
 \item Chapter \ref{Chap7} presents the results of the analysis from the previous chapter and discusses it the context of the other analyses, existing data on \ratio, and the implications
       for the form factor discrepancy and future work.
\end{itemize}

%% file: chap2.tex
% Chapter 2
%
% Motivation/theory
%

\chapter{The Proton, Form Factors, and Theoretical Motivation of the OLYMPUS Experiment}
\label{Chap2}

Fundamentally, OLYMPUS seeks to address the question of how charge and magnetization are distributed within
the proton.  While a complete and consistent theoretical model of the strong force interactions that give rise
to the proton and other hadronic states exists in the form of quantum chromodynamics (QCD)\footnote{See Chapter 9 of Reference \cite{Agashe:2014kda} for a
review of the fundamentals of QCD or Reference \cite{doi:10.1142/9789814525220_0008} for a more detailed discussion.}, calculations of the complex
bound state of three valence quarks, quark-antiquark pairs, and gluons that comprise the proton have proven extremely difficult due to the non-perturbative
nature of QCD at low energies ($\lesssim \Lambda_\text{QCD}\approx 200$ MeV).  While there has been some recent success in calculating properties of the proton and other light bound hadronic
states using lattice QCD \cite{PhysRevD.67.034503,PhysRevLett.92.022001,Durr1224}, full descriptions of the distributions and interactions of particles within bound
states of QCD remain elusive.  Since the proton is the only fundamental, stable bound state of QCD, experimental examination of the proton to determine its
nature is critical to furthering knowledge of the strong force.

Since Rose first suggested in 1948 that the charge distributions within protons and nuclei could
be examined by scattering leptons from protons \cite{PhysRev.73.279}, a variety of experiments have studied the proton in this way over a large
range of energies.  Before proceeding with a description of the OLYMPUS experiment and the analysis of the experiment's data,
it is useful to briefly review the theoretical formalisms regarding the structure of the proton, methods
of probing this structure, and the discrepancy in proton structure measurements that OLYMPUS probed by searching
for two-photon exchange (TPE) in \pmp scattering.  Complete reviews of proton form factors may be found in 
References \cite{Perdrisat2007694} and \cite{0954-3899-34-7-S03}, while References \cite{Arrington:2011dn} and \cite{Carlson:2007sp} specifically
review the physics and experimental landscape of TPE.

\section{Fundamentals and Formalism}

  Since the observation that the proton magnetic moment deviated significantly from the value expected for a point-like spin-$\frac{1}{2}$
  by Otto Stern in 1933 \cite{stern1,stern2}, the nature of the proton's finite structure and the physics that gives rise
  to it has formed a deep and rich field of study in nuclear and particle physics.  While atomic physics-based experiments measuring the Lamb
  shift in hydrogen provide precise measurements of the proton's charge radius \cite{Pohl2010,RevModPhys.80.633,PhysRevLett.84.5496,Sick200362},
  lepton-proton scattering has provided the key means of more comprehensively examining the structure of the proton.  In particular, elastic
  lepton-proton scattering provides insight into the distribution of charge and magnetization within the proton, providing information about
  the size of the proton and the interactions that govern its structure.

  \subsection{Elastic Lepton-Proton Scattering}
  \label{sec:escat}
\begin{figure}[thb!]
  \centerline{\includegraphics[width=0.7\textwidth]{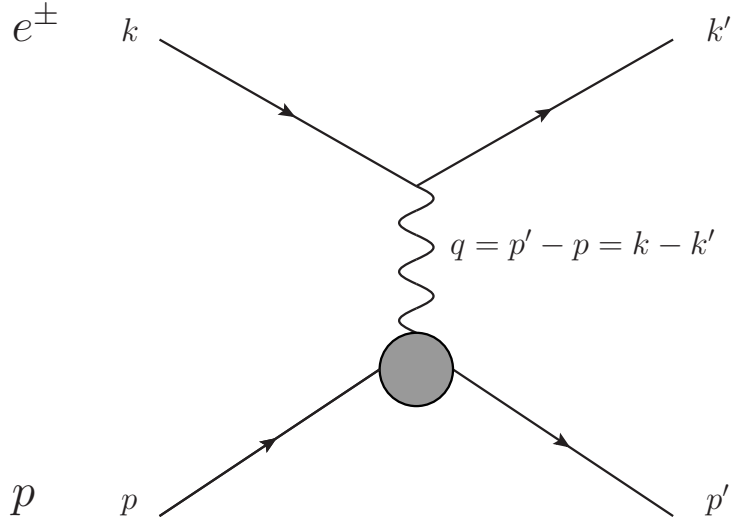}}
  \caption[Feynman diagram of the first order \pmp scattering process]{Feynman diagram representing the first order contribution to the process
  of elastic \pmp scattering, i.e., the Born or single photon exchange approximation \cite{Born1926}.  As in the text, $k$ and $k'$ represent the incoming and
  outgoing four-momenta of the lepton and $p$ and $p'$ those of the proton.  The four-momenta transfered to the proton is $q$.  The circle at the proton
  vertex represents the complex interaction of the proton with the exchanged photon, which is a function of its internal structure and interactions.}
  \label{fig:born}
  \end{figure}

  First, it is useful to establish conventions for the description of elastic scattering events.  For a given \pmp event in which a lepton elastically scatters from
  a proton via the exchange of one or more photons (i.e., via the electromagnetic interaction), the kinematics are
  completely described by the initial and final four-momenta of the lepton (defined to be $k$ and $k'$ respectively), those of the proton
  ($p$ and $p'$ respectively), and the electron and proton masses ($m$ and $M$ respectively) as labeled in Figure \ref{fig:born}.  The four-momentum transfer from the electron to the proton is then:
  \begin{equation}
    q=p'-p=k-k'.
  \end{equation}
  For the kinematics relevant to OLYMPUS, where a beam of leptons strikes protons at rest, and using natural units in which $c=1$:
  \begin{equation}
   k = \left( E_\text{beam},0,0,\sqrt{E_\text{beam}^2-m^2} \right),
  \end{equation}
  \begin{equation}
   p = \left( M,0,0,0 \right),
  \end{equation}
  where $E_\text{beam}$ is the beam energy (2.01 GeV for OLYMPUS) and the initial direction of the lepton is along $z$.  The angle $\theta$ is used to represent
  the polar angle relative to the beam axis ($z$ as described) of the post-scatter three-momentum of the lepton.  Throughout this work, the \textit{lab reference frame} or \textit{lab frame}
  will refer to these convention in which the proton is at rest initially.   For more information on the description of event kinematics as they pertain
  to the conventions used for OLYMPUS, see Section \ref{sec:conv}.
  
  While the variables in the previous paragraph describe the kinematics of the interaction, it is often more convenient to use Lorentz invariant variables
  that allow direct comparison between experiments conducted at different beam energies.  Of particular utility are the square of the four-momentum transfer, $q^2 = q\cdot q$,
  and variable $\epsilon$ that is related to the scattering angle of the electron $\theta$ in the lab frame.  For elastic scattering, note that $q^2<0$ and so it
  is convenient to define:
  \begin{equation}
    Q^2 = -q^2 >0.
  \end{equation}
  The variable $\epsilon$ may be formally defined most easily by first defining two intermediate variables \footnote{The variable $\epsilon$ has a physical
  interpretation as the relative flux of longitudinally polarized virtual photons.  The derivation of this interpretation is well beyond the scope of this work, but is detailed
  in Section 2.4 of Reference \cite{wtf}.}:
  \begin{equation}
   \tau = \frac{Q^2}{4M^2},
  \end{equation}
  \begin{equation}
   \nu = \frac{k\cdot p}{M^2}-\tau.
  \end{equation}
  Then,
  \begin{equation}
   \epsilon = \frac{\nu^2-\tau\left(1+\tau\right)}{\nu^2+\tau\left(1+\tau\right)}.
  \end{equation}
  While this appears to be a cumbersome variable, it has a straightforward interpretation in terms of the lab frame scattering angle of the lepton $\theta$:
  \begin{equation}
   \epsilon = \left[1+2\left(1+\tau\right)\tan^2\left(\frac{\theta}{2}\right) \right]^{-1}.
   \label{eq:eps}
  \end{equation}
  Thus, $\epsilon$ varies from one for very forward scattering in the lab frame ($\theta=0^\circ$) to zero for very backward scattering ($\theta=180^\circ$).
  To interpret $Q^2$ in the lab frame, it is useful to neglect the electron mass ($m\ll M,E_\text{beam}$), which results in:
  \begin{equation}
   Q^2 = 4E_\text{beam}E'\sin^2\left(\frac{\theta}{2} \right),
  \end{equation}
   where $E'$ is the outgoing lepton energy in the lab frame.
   
  Having established conventions for the description of elastic lepton-proton scattering, the question of the cross section for the process and the insight
  it offers regarding the structure of the proton may be addressed.  Since the weak interaction is suppressed by the masses of the $W^\pm$ and $Z$ bosons
  ($M_{W^\pm,Z}\sim \mathcal{O}(100\:\text{GeV})$), it is negligible relative to the electromagnetic interaction (in terms of its contribution to the overall
  interaction cross section).  Additionally, since each photon-fermion interaction vertex in a quantum electrodynamics (QED) Feynman diagram carries a factor of
  $\sqrt{\alpha} \approx \sqrt{\frac{1}{137}}$ to the contribution of that diagram to the matrix element for a given interaction (and thus a factor of $\alpha$ to
  the cross section), often the approximation is made that only lowest-order allowed diagram (that with the fewest number of vertices) is used to compute the matrix element for an interaction.  In the case
  of elastic \pmp scattering, the lowest-order approximation is single photon exchange with no additional initial or final state radiation, as shown in Figure \ref{fig:born}.
  This is known as the Born approximation \cite{Born1926}.
  
  To construct the most general cross section for lepton-proton elastic scattering, first consider the case of a light lepton (spin-$\frac{1}{2}$) scattering from a much more massive, spinless
  target (e.g., a spin-0 nuclei or ``spinless proton''). The only first-order QED diagram that contributes to the process is a diagram akin to that shown in Figure \ref{fig:born}.
  Such a process is known as Mott scattering \cite{Mott425,Mott658}, and the cross section is straightforward to compute from the rules of QED \cite{peskin,griffiths}:
  \begin{equation}
   \sigma_\text{Mott} = \frac{\alpha^2 E' \cos^2\left(\frac{\theta}{2}\right)}{4E^3\sin^4\left(\frac{\theta}{2}\right)}.
  \end{equation}
  The problem of lepton-proton scattering then effectively reduces to properly adjusting the treatment of the proton to account for both its structure and spin.
  
  \subsection{Elastic Form Factors}
  
  To account for the effect of the proton's structure and spin, first consider the general invariant matrix element for the Feynman diagram in
  Figure \ref{fig:born} (using the notation and conventions of Reference \cite{Arrington:2011dn}):
  \begin{equation}
   \mathcal{M}_\gamma = -\frac{e^2}{q^2}j_{\gamma\mu}J_\gamma^\mu,
  \end{equation}
  where $e$ is the electron charge.  The current $j_{\gamma\mu}$ is that of the lepton, and is thus the standard current of a Dirac fermion:
  \begin{equation}
   j_{\gamma\mu} = \overline{u}_e\left(k'\right)\gamma_\mu u_e\left(k\right),
  \end{equation}
  where $u_e$ represents the lepton's Dirac spinor.  The proton's current, $J_\gamma^\mu$, is more complicated.  As first shown by Foldy in 1952 \cite{PhysRev.87.688},
  the most general current for a spin-$\frac{1}{2}$ particle in QED that satisfies current conservation and Lorentz invariance may be written as:
  \begin{equation}
   J_\gamma^\mu =  \overline{u}_p\left(p'\right)\left( \gamma^\mu F_1(Q^2) + \frac{i\sigma^{\mu\nu}q_\nu}{2M} F_2(Q^2) \right) u_p\left(p\right),
  \end{equation}
  where $u_p$ is the proton spinor and $F_1(Q^2)$ and $F_2(Q^2)$ are respectively Dirac and Pauli form factors for elastic scattering from the proton.  In the Born
  approximation, these are functions of $Q^2$ alone (making them Lorentz invariant) and they encode the internal behavior of the proton to all orders in QCD.

  When discussing the structure of the proton it is most useful to recast the form factors as the following linear combinations of the Dirac and
  Pauli form factors $F_1$ and $F_2$:
  \begin{equation}
   G_E(Q^2) = F_1(Q^2)-\tau F_2(Q^2),
  \end{equation}
    \begin{equation}
   G_M(Q^2) = F_1(Q^2)+F_2(Q^2),
  \end{equation}
  which are respectively the electric and magnetic elastic form factors, and were defined by Hand, Miller, and Wilson \cite{ff5}.  These form factors
  are normalized such that $G_E(0) = 1$ and $G_M(0) = \mu_p = 2.793$, i.e., the static values of the proton's charge and magnetic moment in the units of $e=c=\hbar=1$. Parameterizing
  the form factors in this way leads to a simpler form for the final computed differential cross section in the Born approximation, known as the Rosenbluth 
  formula \cite{PhysRev.79.615}\footnote{When first computed by Rosenbluth in 1950, the formula did not take the form shown in Equation \ref{eq:Ros}, but the modern
  representation of the formula is presented here.}:
  \begin{equation}
   \left( \td{\sigma}{\Omega}\right)_\text{Born} = \frac{\sigma_\text{Mott}}{\epsilon(1+\tau)}\left[ \epsilon G_E^2(Q^2) + \tau G_M^2(Q^2)   \right].
   \label{eq:Ros}
  \end{equation}
  
  \subsection{Physical Interpretation of the Elastic Form Factors}

  In addition to simplifying the form of the Rosenbluth formula for the \pmp elastic cross section, the electromagnetic form factors $G_E$ and $G_M$ additionally offer some insight into
  more intuitive properties of the proton's structure, namely the distribution of charge and magnetization within the proton.  As first noted by Sachs in 1962 \cite{PhysRev.126.2256},
  in a reference frame in which both the lepton and proton exchange three-momentum but no energy (and thus the lepton rebounds in the opposite direction from its initial trajectory with the same
  energy)\footnote{This reference frame is known as the Breit frame or ``brick wall'' frame.}
  the form factors $G_E$ and $G_M$ are the Fourier transforms of the charge and magnetization distributions of the proton.  Note, however, that the velocity of this frame relative to the lab
  frame varies as a function of $Q^2$ and that the transformation of spatial distributions between these frames is generally very complex.  While model-dependent methods of conducting such
  transformation have been established \cite{PhysRevC.66.065203}, it is generally quite difficult and caution should be exercised when interpreting the direct physical meaning of $G_E$ and
  $G_M$.

  \section{Determination of the Proton Elastic Form Factors}
  
  While a number of theoretical models exist for computing the proton elastic form factors (see Section 4 of Reference \cite{Perdrisat2007694} for an overview), none have been successful
  in providing a complete and predictive description of the proton's elastic electromagnetic interactions due to complexity of the underlying interactions.  Thus, the determination
  of the values of $G_E$ and $G_M$ has been an active experimental question since they were first formulated.  The first measurements of quantities related to the form factors were
  conducted by Hofstadter and McAllister at Stanford in the mid-1950s, who measured a single form-factor-like quantity (effectively a single factor modifying the Mott cross section, i.e., the combination of
  the effects of two form factors) \cite{PhysRev.98.217,PhysRev.102.851,RevModPhys.28.214}.  As theory progressed, it was realized that experiments were needed that separated the effects
  of the two form factors. The technique of Rosenbluth separation was developed allowing the independent extraction of $G_E^2$ and $G_M^2$. Measurements using this technique
  began in the late 1960s \cite{ff5,ff6,ff14,ff1,ff2,ff3,ff4,ff12,ff7} and continued through the turn of the century \cite{ff8,ff9,ff10,ff11,ff15}.  In the 1990s, the advent of polarized beams
  and targets provided a new method of measuring the ratio $G_E/G_M$, used in experiments conducted at MIT-Bates and Jefferson Lab \cite{pol1,pol2,pol3,pol4,pol5,pol6,pol7,pol8,pol9,pol10,pol11,pol12}.
  The two techniques are briefly
  summarized here, while the results from each method are discussed in Section \ref{sec:discrep}.
  
  \subsection{Rosenbluth Separation}
  \label{sec:rossep}

\begin{figure}[thb!]
  \centerline{\includegraphics[width=0.7\textwidth]{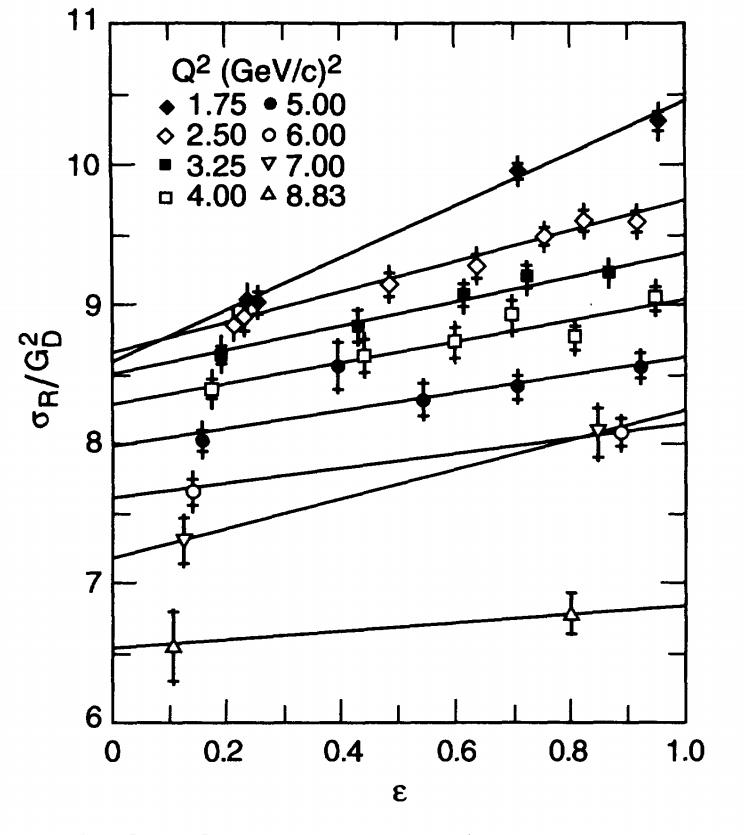}}
  \caption[Example application of Rosenbluth separation]{Application of the Rosenbluth separation technique at eight values of $Q^2$.  (Figure reproduced from
  Reference \cite{ff8}.)}
  \label{fig:rsep}
  \end{figure}

  The Rosenbluth separation technique takes advantage of the form of the Rosenbluth formula (Equation \ref{eq:Ros}) for the \pmp elastic cross section and the dependence of the form factors
  on $Q^2$ alone.  While the exact techniques used by various experiments differ slightly, Rosenbluth separation effectively amounts to rewriting Equation \ref{eq:Ros} in the form:
  \begin{equation}
   \left( \td{\sigma}{\Omega}\right)_\text{reduced} = \frac{\epsilon(1+\tau)}{\tau\sigma_\text{Mott}} \left( \td{\sigma}{\Omega}\right)_\text{exp} = G_M^2(Q^2) + \frac{\epsilon}{\tau} G_E^2(Q^2),
   \label{eq:rsep}
  \end{equation}
  where $\left( \td{\sigma}{\Omega}\right)_\text{exp}$ is the experimentally measured value of the elastic \pmp cross section as a function of $\epsilon$ and $\tau$.
  Since $\tau = \frac{Q^2}{4M}$ and $\epsilon = \left[1+2\left(1+\tau\right)\tan^2\left(\frac{\theta}{2}\right) \right]^{-1}$, measuring the cross section at fixed $Q^2$ over a variety of scattering
  angles $\theta$ and plotting the values of the reduced cross section defined by Equation \ref{eq:rsep} as a function of $\epsilon$ results in a linear function.  Fitting this function yields
  $\frac{1}{\tau}G_E^2(Q^2) = \frac{4M}{Q^2}G_E^2(Q^2)$ as the slope and $G_M^2(Q^2)$ as the intercept.  By conducting these measurements for a variety of $Q^2$ values, the form factors may
  be mapped out.  In most experiments the values of $Q^2$ and $\epsilon$ are selected by adjusting narrow acceptance single- and double-arm spectrometers and collecting
  data in a number of spectrometer configurations \cite{grupen,fernow}.  An example of the application of this technique from Reference \cite{ff8} is shown in Figure \ref{fig:rsep}.  Rosenbluth
  separation is the only technique currently available that allows independent measurement of the individual form factors.
  
  \subsection{Polarization-Based Techniques}

  The Rosenbluth formula (Equation \ref{eq:Ros}) and Rosenbluth separation assume unpolarized lepton beams and protons targets.  As first observed by Akhiezer and Rekalo, the Rosenbluth
  separation technique suffers from the dominance of the $\frac{\tau}{\epsilon}G^2_M$ term at high $Q^2$ (subjecting the extraction of $G_E$ to large uncertainty) \cite{ak1}.  They, and others,
  proposed methods involving experiments with longitudinally polarized electron beams on unpolarized proton targets and measuring the polarization of the recoil proton (i.e., the \textit{polarization 
  transfer} from the lepton to the proton) \cite{ak2,PhysRevC.23.363}.  Additionally, techniques were proposed using unpolarized beams on polarized targets \cite{RevModPhys.41.236}.
  
  Considering the case of a polarized beam on an unpolarized target to illustrate the method, the cross section for elastic $\vec{e}p\rightarrow e\vec{p}$ scattering in the Born
  approximation may be calculated from the rules of QED in a similar fashion to the spin-averaged unpolarized case \cite{ak2,PhysRevC.23.363}.  The differential cross sections
  for the outgoing proton spin to be aligned longitudinally or transversally are respectively:
  \begin{equation}
   \td{\sigma^{(L)}}{\Omega} = h\sigma_{Mott} \frac{E+E'}{M}\sqrt{\frac{\tau}{1+\tau}} \tan^2\left(\frac{\theta}{2}\right) G_M^2,
  \end{equation}
  \begin{equation}
   \td{\sigma^{(T)}}{\Omega} = 2h\sigma_{Mott}\sqrt{\frac{\tau}{1+\tau}} \tan\left(\frac{\theta}{2}\right) G_E G_M,
  \end{equation}
  where $h$ is the electron helicity.  The average longitudinal and transverse polarizations $P_L$ and $P_T$ of the outgoing protons are then proportional to their respective
  outgoing spin cross sections. Taking the ratio of the expressions for the cross sections (and thus polarizations), the unknown electron helicity cancels to yield:
  \begin{equation}
   \frac{G_E}{G_M} = -\frac{E_\text{beam}+E'}{2M}\tan\left( \frac{\theta}{2} \right) \frac{P_T}{P_L}.
  \end{equation}
  This technique provides a robust measurement of the $\frac{\mu_pG_E}{G_M}$ ratio, helped by cancellation of various systematic uncertainties caused by the simultaneous measurement
  of the two polarizations using the same setup, and when paired with Rosenbluth measurements of $G_M$ provides the most sensitive method of extracting $G_E$ at high $Q^2$.  A review
  of the methods used to extract $\frac{\mu_pG_E}{G_M}$ from polarized target experiments may be found in Section 3.2 of Reference \cite{Perdrisat2007694}.

\section{Existing Data and the Proton Form Factor Discrepancy}
\label{sec:discrep}

\begin{figure}[thb!]
  \centerline{\includegraphics[width=0.7\textwidth]{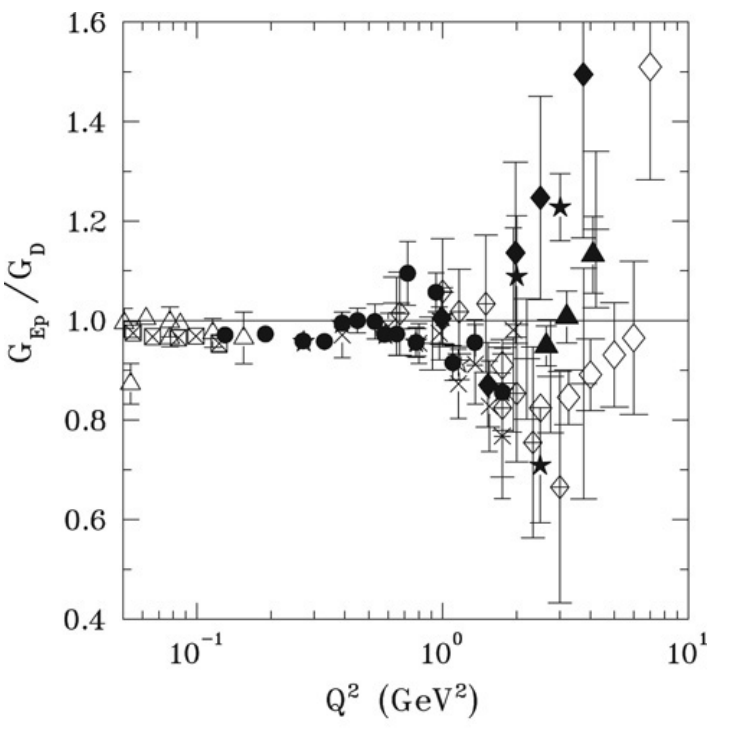}}
  \caption[Selection of Rosenbluth separation measurements of $G_E$]{Selection of experimental measurements of $G_E(Q^2)$ using Rosenbluth separation, normalized to the dipole form factor
  $G_D$ (Equation \ref{eq:dipff}) and proton magnetic moment $\mu_p$.
  (Original data: \cite{ff5,ff3,ff1,ff2,ff4,ff12,ff7,ff13,ff8,ff9,ff10,ff11}) (Figure reproduced from Reference \cite{Perdrisat2007694}).}
  \label{fig:ge}
  \end{figure}

\begin{figure}[thb!]
  \centerline{\includegraphics[width=0.7\textwidth]{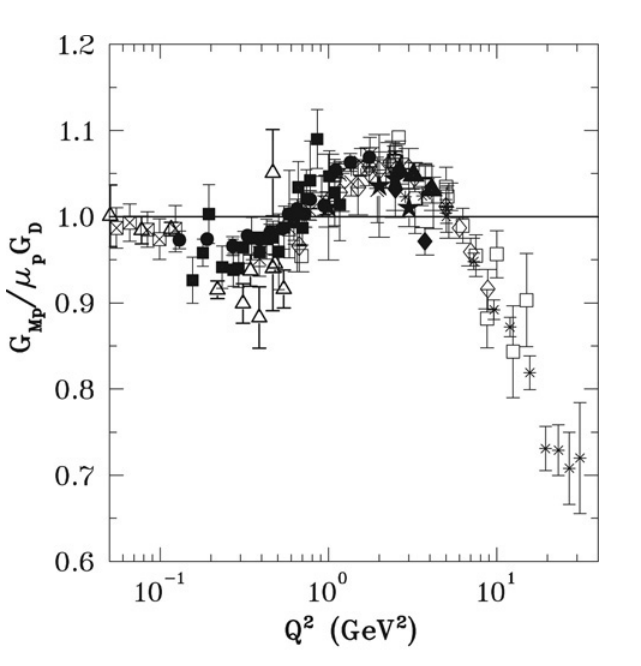}}
  \caption[Selection of Rosenbluth separation measurements of $G_M$]{Selection of experimental measurements of $G_M(Q^2)$ using Rosenbluth separation, normalized to the dipole form factor
  $G_D$ (Equation \ref{eq:dipff}).
  (Original data: \cite{ff5,ff6,ff14,ff3,ff1,ff2,ff4,ff12,ff7,ff15,ff8,ff9,ff10,ff11}) (Figure reproduced from Reference \cite{Perdrisat2007694}).}
  \label{fig:gm}
  \end{figure}

  As noted previously, numerous experiments have measured the proton elastic form factors (or the ratio $\frac{\mu_pG_E}{G_M}$ in the case of the polarization-based measurements) over a range
  of $0.05$ GeV$^2\lesssim Q^2 \lesssim 10$ GeV$^2$.  A selection of Rosenbluth separation data for $G_E$ and $G_M$ is shown in Figures \ref{fig:ge} and \ref{fig:gm} respectively, reproduced
  from Reference \cite{Perdrisat2007694}.  The form factors are typically displayed, as in the figures, normalized to the dipole form factor:  
  \begin{equation}
     G_D = \frac{1}{\left(1+\frac{Q^2}{0.71\:\text{GeV}^2}\right)^2},
     \label{eq:dipff}
  \end{equation}
  which corresponds (in the Breit frame) to an exponentially falling charge or magnetization distribution.  As can be seen in the figures, the dipole model roughly describes the form factor
  data obtained via Rosenbluth scattering below a few GeV$^2$, especially $G_E$, but that the data deviate from the dipole model at higher $Q^2$.  As shown in Figure \ref{fig:disc}, the ratio
  $\frac{\mu_pG_E}{G_M}$ measured using Rosenbluth scattering is $\sim$1 and reasonably flat.
  
  Data from polarization-based measurements, however, provide a starkly different measurement of $\frac{\mu_pG_E}{G_M}$.  As Figure \ref{fig:disc} shows, the data from such experiments
  cluster around a line that decreases as a function of $Q^2$ with a slope of approximately -0.14 GeV$^{-2}$.  At $Q^2\gtrsim 5$ GeV$^2$, global fits to Rosenbluth separation data and
  polarization-based data show more than a factor of two difference in $\frac{\mu_pG_E}{G_M}$.  This is known as the proton form factor ratio discrepancy, and has been an active field
  of research for both theorists and experimentalists.

\section{Possible Causes of the Discrepancy and the Two-Photon Exchange Hypothesis}
\label{sec:posscause}

Given the significant discrepancy between the different methods of measuring the proton form factor ratio (and the consistency of measurements of the same type) discovered
in the 1990s, efforts were made both to verify the discrepancy with modern measurement and to explore possible causes of the discrepancy.  Rosenbluth separation experiments
conducted in 2004-2005 (References \cite{ff10} and \cite{ff11}) and reanalysis of the existing data (References \cite{PhysRevC.68.034325} and \cite{PhysRevC.69.022201}) confirmed the discrepancy at values of $Q^2$ up to $\sim$8 GeV$^2$,
further increasing the call for explanations of the discrepancy
from the standpoint of theory and of understanding the physics implicit in the analysis methods used for each technique.  In particular, the \textit{radiative corrections} applied
for each measurement were examined carefully.

\begin{figure}[thb!]
  \centerline{\includegraphics[width=1.0\textwidth]{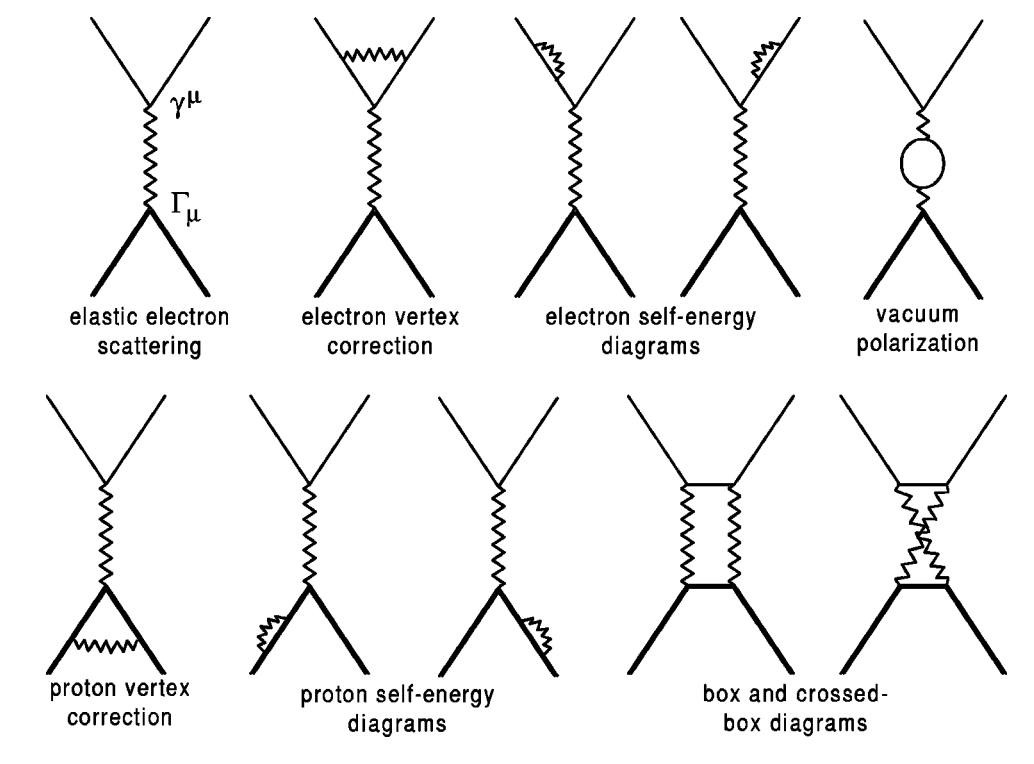}}
  \caption[QED diagrams of up to order $\alpha^2$ that contribute to elastic \pmp scattering]{QED diagrams up to order $\alpha^2$ that contribute to elastic \pmp scattering. 
  (Figure reproduced from Reference \cite{MaximonPhysRevC.62.054320}.)}
  \label{fig:hod}
  \end{figure}

\begin{figure}[thb!]
  \centerline{\includegraphics[width=0.8\textwidth]{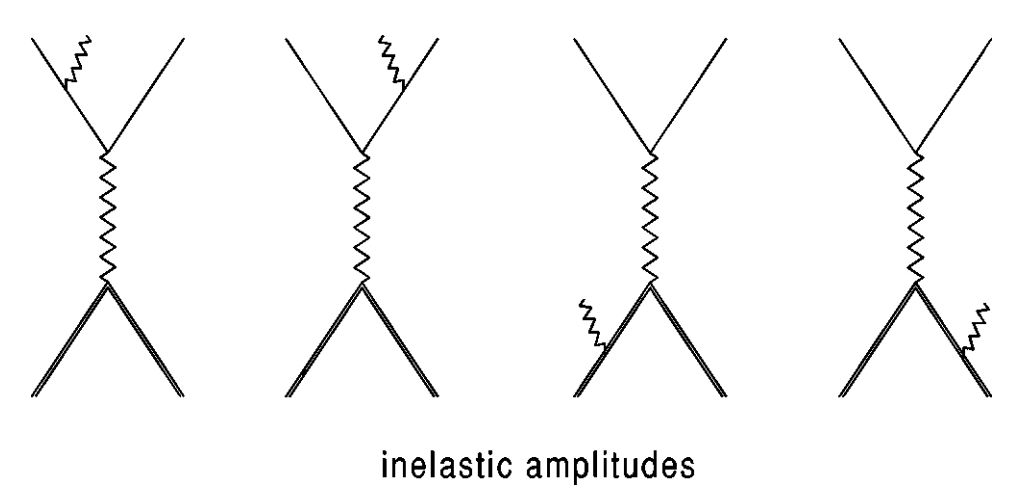}}
  \caption[QED diagrams involving a soft external photon that contribute to elastic \pmp scattering]{QED diagrams involving a soft external photon that contribute to elastic \pmp scattering. 
  (Figure reproduced from Reference \cite{MaximonPhysRevC.62.054320}.)}
  \label{fig:ihod}
  \end{figure}

In general, the term radiative corrections encompasses any shift applied to a raw measured cross section, form factor, etc. to account for the effect of making
the Born approximation.  In practice, for \pmp scattering, this amounts to considering changes to the cross section that occur due to contributions from higher-order
Feynman QED diagrams that are experimentally indistinguishable from elastic scattering by single photon exchange, like those shown in Figures \ref{fig:hod} and \ref{fig:ihod}.
Note that the event types shown in Figure \ref{fig:ihod} may be indistinguishable in an experiment from an event with no externally emitted photon, due to the photon having very small
energy such that the deviation from elastic kinematics is smaller than the resolution of the detector.  Past
experiments (prior to 2000) accounted for the effects of diagrams up to order $\alpha^2$ using the prescription of Mo and Tsai \cite{PhysRev.122.1898,MoRevModPhys.41.205}.
This prescription, however, did not account for the structure of the nucleon and made a number of other approximations.  The method for applying radiative corrections to 
elastic \pmp data was improved upon by Maximon and Tjon \cite{MaximonPhysRevC.62.054320}, who included the effect of the proton form factors in their calculations and
avoided several of the approximations made by Mo and Tsai, which altered the size of the radiative correction (itself typically a correction of $\mathcal{O}(10\%)$ to measured
cross sections) by several percent, an effect not large enough to explain the discrepancy \cite{PhysRevC.68.034325}.  Radiative corrections are also critical to the
OLYMPUS result and the approach used for OLYMPUS is described in Section \ref{sec:radgen}.  A detailed discussion of the OLYMPUS radiative corrections may be found
in References \cite{schmidt} and \cite{russell}.

In these radiative corrections calculations, however, effects due to the two-photon exchange (TPE) diagrams in which both photons carry comparable energy (hard two photon exchange)
were not considered (the diagrams marked ``box'' and ``crossed-box'' in Figure \ref{fig:hod}).   In particular, it is difficult to account for the behavior of the proton
between the proton-photon vertices, since in that leg of the diagram the proton may be off shell or possibly even in an excited state such as a $\Delta^+$.  While early
models of two-photon exchange suggested that the contribution of TPE was considerably smaller than 1\% of the cross section, the validity of these models was limited
to values of $Q^2$ below $\sim$1 GeV$^2$ where the probability of hard TPE is smaller \cite{PhysRev.74.1759,PhysRev.102.537,PhysRev.106.561,PhysRev.113.741,1969190,PhysRev.180.1541,PhysRev.184.1860}.

\subsection{Theoretical Calculations of TPE and their Effect on the $\frac{\mu_pG_E}{G_M}$ Ratio}

\begin{figure}[thb!]
  \centerline{\includegraphics[width=1.0\textwidth]{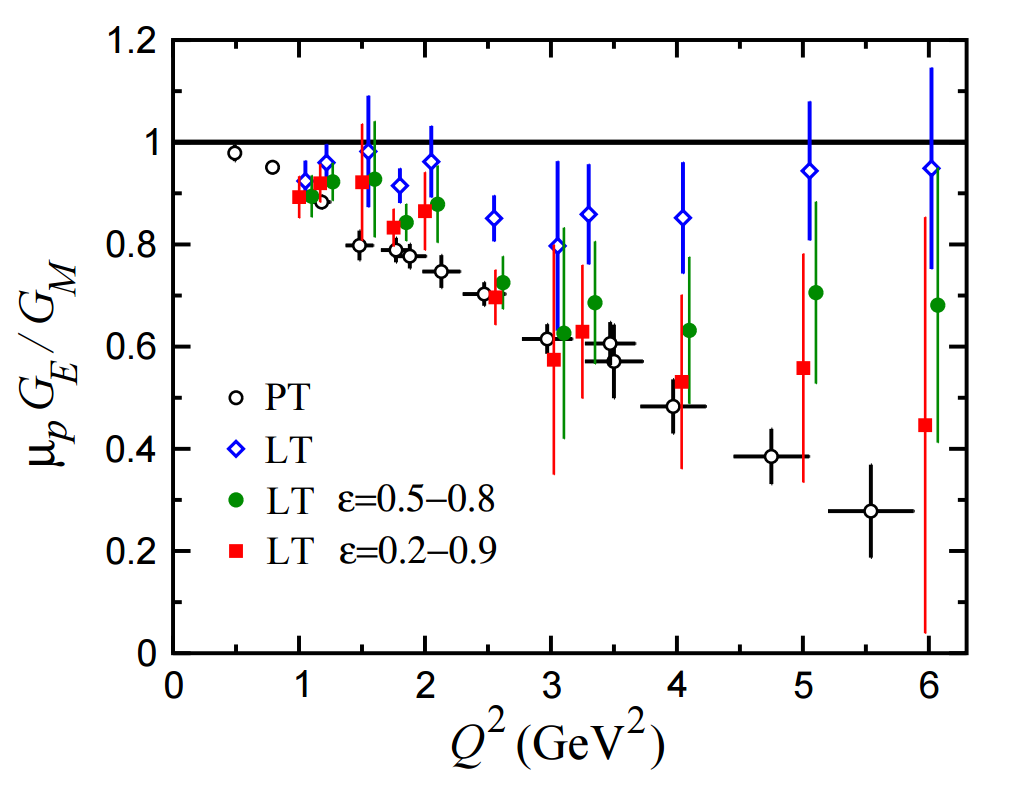}}
  \caption[Example of a theoretical model for the effect of TPE on Rosenbluth separation $\frac{\mu_pG_E}{G_M}$ measurements]{Example of a theoretical model for the
  effect of TPE on Rosenbluth separation $\frac{\mu_pG_E}{G_M}$ measurements, showing the shift of the Rosenbluth separation points
  (``LT'') towards the polarization transfer points (``PT'') for two separate assumptions regarding the amount of TPE that occurs (filled markers).  Note that the TPE model accounts for a significant fraction
  of the form factor discrepancy. (Figure reproduced from Reference \cite{Blunden:2005ew}.)}
  \label{fig:blundpred}
  \end{figure}

  Given that radiative corrections due to the TPE diagrams were the least understood among the corrections of order up to $\alpha^2$, renewed efforts were
  made to better model this effect and examine the effect of such models' predictions on the form factor ratio
  \cite{Blunden:2003sp,Guichon:2003qm,REKALO2004322,Chen:2004tw,Afanasev:2005mp,Blunden:2005ew, Kondratyuk:2005kk,Borisyuk:2006fh,TomasiGustafsson:2009pw}.
  Descriptions of the details of these models are beyond the scope of this work, but an overview may be found in References \cite{Arrington:2011dn} and \cite{Carlson:2007sp}.
  As an example, the effect of the model developed in Reference \cite{Blunden:2005ew} on $\frac{\mu_pG_E}{G_M}$ is shown in Figure \ref{fig:blundpred}, and as can be seen
  such predictions tend to explain at least some of the discrepancy between the Rosenbluth separation and polarization-based measurements.
  
  Since QCD does not provide an unambiguous, calculable
  method for modeling the proton, such predictions vary significantly; models range from accounting for none of the discrepancy to nearly all of it, as can be seen in wide spread
  of model predictions for \ratio due to TPE in Figure \ref{fig:projections}.  Thus, there exists a strong demand for experimental measurements of the contributions of TPE to
  \pmp scattering.

  \subsection{Experimental Signature of TPE}
  \label{sec:estpe}

\begin{figure}[thb!]
  \centerline{\includegraphics[width=1.0\textwidth]{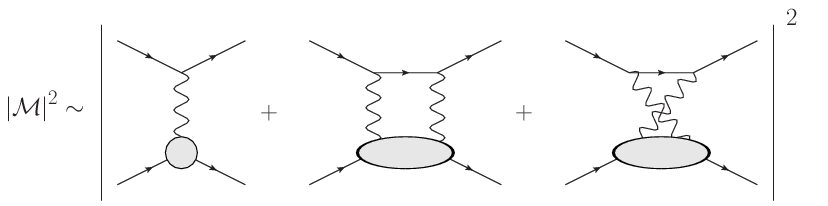}}
  \caption[Contribution of TPE diagrams to the total \pmp matrix element]{A simplified visual representation of the contribution of the TPE Feynman diagrams to the total
  matrix element for \pmp scattering.  Interference terms between the one-photon exchange diagram and two-photon exchange diagrams contribute terms proportional to $(\pm\alpha)^3$ due
  to the lepton vertices, changing sign for \pp and \ep scattering.}
  \label{fig:simpint}
  \end{figure}

  Given the hypothesis of TPE and an established need for experimental measurement of its contribution to \pmp elastic scattering, it was quickly realized that a comparison of \ep and \pp elastic scattering
  cross sections provides direct experimental access to the size of the TPE matrix elements $\mathcal{M_{\gamma\gamma}}$ corresponding to the box and crossed-box diagrams of Figure
  \ref{fig:hod}.  This can be  approximately understood by considering the toy example of the calculation of the total squared matrix element in Figure \ref{fig:simpint}.  Terms of the total
  matrix element $\mathcal{M}$ that represent an interference of the one-photon and two-photon diagrams are proportional $(\pm\alpha)^3$ where the $\pm$ is determined by the charge of the
  lepton.  Thus, the matrix element is shifted downward for electron scattering by these terms and upward for positron scattering, creating an asymmetry that could be measured
  experimentally.  Using $\mathcal{M}_\gamma$ to represent the single photon exchange matrix element (with positive sign, i.e, for \pp scattering) and $\mathcal{M}_{\gamma\gamma}$ for the TPE elements, the ratio
  expressed in terms of an experimental measurement in Equation \ref{eq:rat} may be expressed in terms of the matrix elements as:
  \begin{equation}
   R_{2\gamma}\left(\epsilon,Q^2\right) = \frac{\sigma_{e^+}}{\sigma_{e^-}} \sim 1 + 4\alpha \frac{\mathcal{M}_{\gamma\gamma}}{\mathcal{M}_{\gamma}}.
  \end{equation}
  Thus, an experiment which can measure this cross section ratio at the relevant kinematics provides direct access to the value of the TPE contribution to the cross section,
  which would put valuable constraints on theoretical models and offer insight into what fraction of the form factor ratio discrepancy can be explained by TPE.  An important
  note is that in addition to the Born/TPE interference contribution differing in sign between \ep and \pp scattering, interference between the Born diagram and the externally radiated
  photon (bremsstrahlung) diagrams (Figure \ref{fig:ihod}) and thus careful correction for these effects is required in the analysis of any \ratio experiment attempting to extract
  the TPE contribution.
  
\section{Experiments Measuring \ratio and Physics Goals of the OLYMPUS Experiment}

\begin{figure}[thb!]
  \centerline{\includegraphics[width=1.0\textwidth]{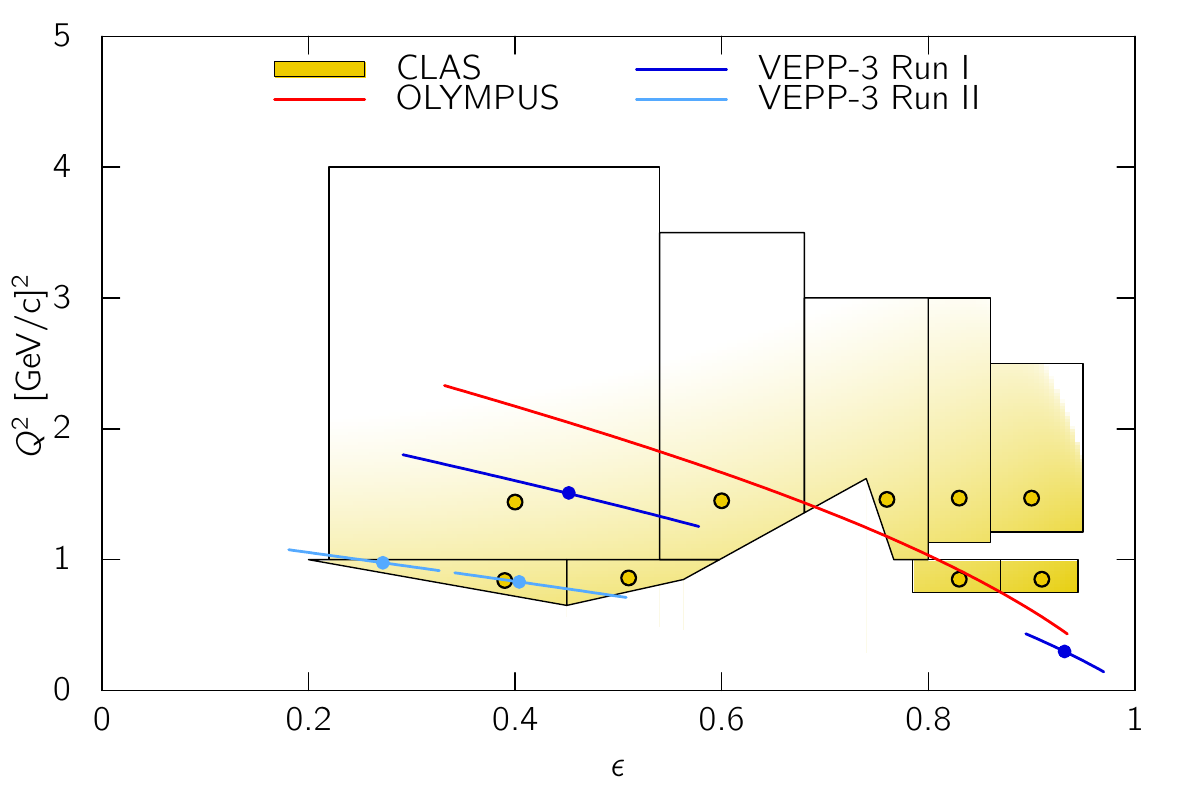}}
  \caption[Kinematic reaches of the three TPE experiments]{Kinematic reaches of the three TPE experiments.  Note that since the CLAS experiment did not have fixed beam energy, the 
  elastic events are spread across the $\left(\epsilon,Q^2\right)$ bins shown, dominated by the higher cross section in the bottom right of the bins. (Figure reproduced from Reference \cite{schmidt}.)}
  \label{fig:reach}
  \end{figure}

Given the spread in theoretical models for TPE and their effect on the value of \ratio (Figure \ref{fig:projections}), experiments measuring \ratio at the accuracy of 1\% or better
at values of $Q^2\gtrsim\mathcal{O}(1\:\text{GeV}^2)$ are required to differentiate the models, with preference towards higher $Q^2$ where the form factor discrepancy is largest.
Existing data on this quantity from the 1960s (shown in Figure \ref{fig:projections}) is much more imprecise than this goal, especially at higher values of $Q^2$
\cite{Yount:1962aa,Browman:1965zz,Anderson:1966zzf,Cassiday:1967aa,Bouquet:1968aa,Mar:1968qd,Bartel:1967aa}.  Thus, three modern experiments have sought to measure \ratio via complimentary
experimental setups:
\begin{enumerate}
 \item OLYMPUS at DESY, Hamburg, Germany \cite{Milner:2014},
 \item CLAS at Jefferson Lab, Newport News, Virginia \cite{PhysRevLett.114.062003,ass}, and 
 \item VEPP-3 at Novosibirsk, Russia \cite{vepp3PhysRevLett.114.062005}.
\end{enumerate}
The relative kinematic reaches of these experiments are shown in Figure \ref{fig:reach}.  Details on these experiments may be found in the cited references, but it is worthwhile
to note the fundamentally different approaches of the experiments. The VEPP-3 experiment operated at lower energies than OLYMPUS with a non-magnetic spectrometer (to avoid differences in \ep and \pp acceptance), while the CLAS experiment
utilized a magnetic spectrometer with a simultaneous $e^+/e^-$ beam produced by pair production from photons.  

\begin{figure}[thb!]
  \centerline{\includegraphics[width=1.0\textwidth]{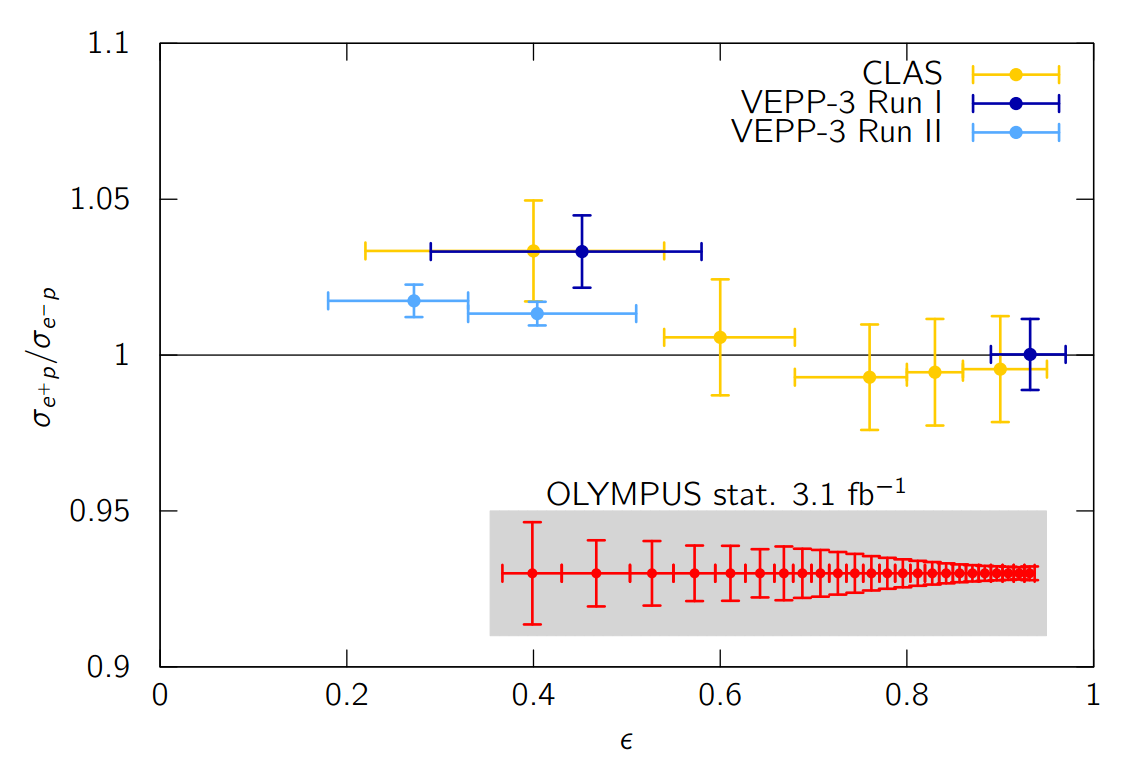}}
  \caption[Results from VEPP-3 and CLAS with the OLYMPUS projections]{Results on \ratio from CLAS \cite{PhysRevLett.114.062003} and VEPP-3 \cite{vepp3PhysRevLett.114.062005}, along with
  projected uncertainties for OLYMPUS (limiting any bin to at most 1\% statistical uncertainty).  Note the different experiments cover different $Q^2$ ranges for the shown $\epsilon$ range, and so
  are not directly comparable on this plot, but that the data are chosen so as to provide a rough comparison. (Figure reproduced from Reference \cite{schmidt}.)}
  \label{fig:crap}
  \end{figure}

Results from the latter two experiments preceded OLYMPUS results, and are shown in Figure \ref{fig:crap}.  As can be seen in the figure, these results are limited in statistics (requiring broad
binning) and, while suggestive of an upward trend in \ratio with decreasing $\epsilon$, do not provide a definitive trend for the TPE contribution.
Given that OLYMPUS collected much higher statistics than either CLAS or VEPP-3 and reaches higher in $Q^2$, OLYMPUS is expected to provide a considerably more definitive ratio result.
It is the goal of OLYMPUS to measure \ratio to a total (statistical+systematic)
uncertainty of better than 1\% across the full kinematic range of the experiment, as shown by the red projected bins of Figure \ref{fig:crap}.  The remainder of this work discusses the OLYMPUS
analysis and results in detail.

%% file: chap3.tex
% Chapter 3
%
% Description of the experiment, spectrometer, etc.
%

\chapter{The OLYMPUS Experiment}
\label{Chap3}

To construct an experimental setup suitable for the physics goals of OLYMPUS, a large
acceptance spectrometer was assembled, capable of exclusively reconstructing $e^\pm p$ 
elastic scattering events while precisely measuring the recorded luminosity for
each lepton species.  To achieve this, the spectrometer consisted of the principal elements
of the Bates Large Acceptance Spectrometer Toroid (BLAST) \cite{Alarcon20001111c,Hasell:2009zza} combined
with several newly constructed luminosity monitors, operated at the DORIS III electron/positron storage
ring at the Deutsches Elektronen-Synchrotron (DESY) in Hamburg, Germany \cite{DORIStab,DORISrep}.  This spectrometer
surrounded a gaseous hydrogen target, internal to the DORIS beamline. During the experiment, the target
was exposed to electron and positron beams from DORIS, with the beam
species changed approximately daily.  Great care was taken to ensure that running conditions for the two different
leptons species were as identical as possible to avoid unwanted systematic differences between the two modes.

Data for the experiment were taken in two periods: January 20, 2012--February 27, 2012 (Run I) and October 24, 2012--January 2, 2013 (Run II).  Approximately
4.5 fb$^{-1}$ of integrated luminosity was acquired, including various calibration runs that are not included
in the sample of data used to construct the \ratio result.  The  vast majority of these data were collected in the second run, after several
improvements to the trigger, target system, and other detector elements were made between runs.

This chapter describes the important details of the experimental setup used to measure the elastic $\sigma_{e^+}/\sigma_{e^-}$ ratio, including both
the detectors used for the reconstruction of \pmp events over a large range of kinematics and the luminosity monitors.  Also discussed
are the essential elements of the data acquisition system (DAQ) and the operation of the experiment, especially as they pertain to
the analysis of the detector data.  Much of this chapter summarizes the complete published descriptions of the experiment in 
References \cite{Milner:2014} and \cite{tdr}.  Additional references that discuss certain elements of the system in greater detail
are cited in the relevant sections.

\section{Conventions for the Description of the Experiment}
\label{sec:conv}

\begin{figure}[thb!]
\centerline{\includegraphics[width=1.1\textwidth]{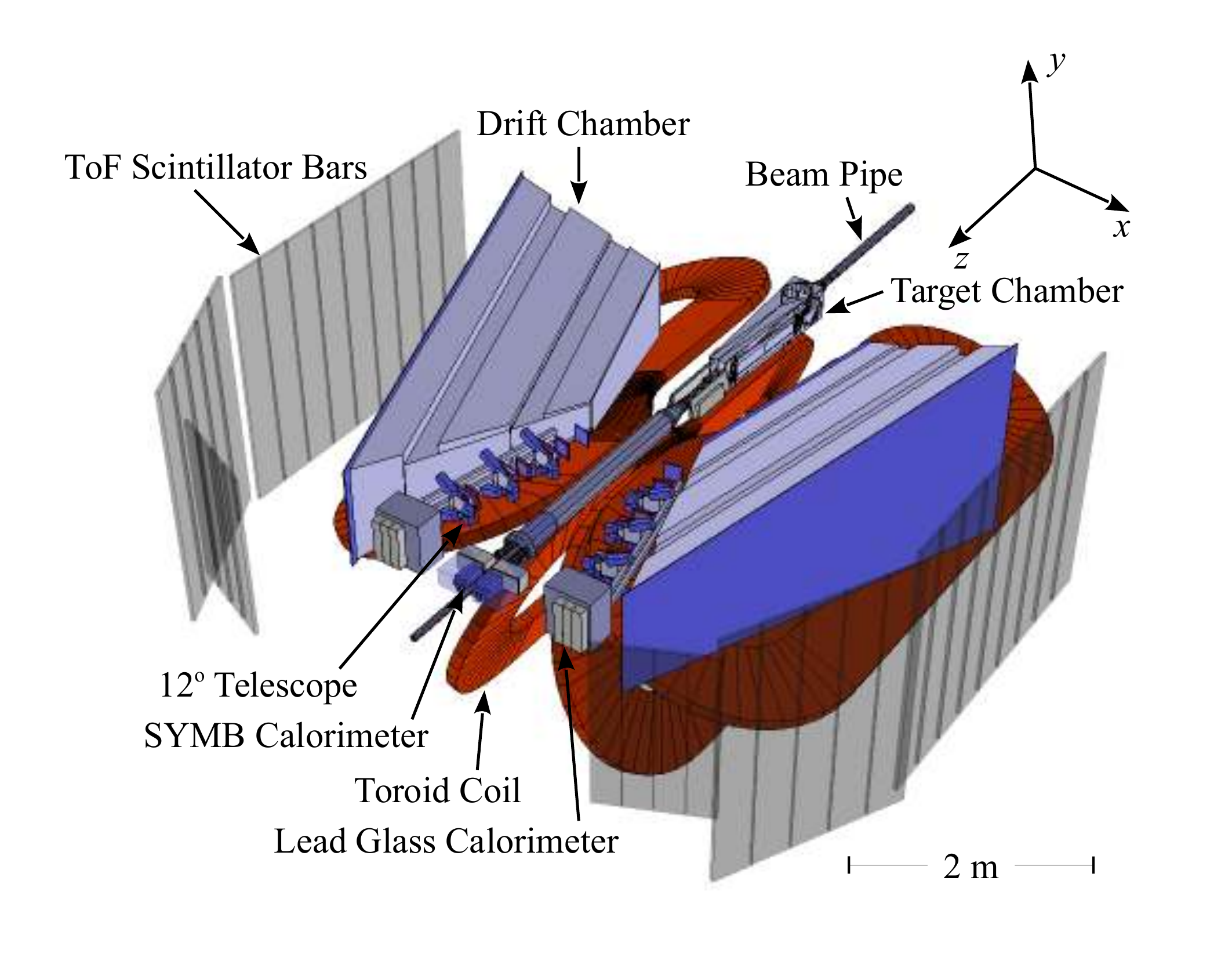}}
\caption[Solid-model of the OLYMPUS detector]{Solid model of the OLYMPUS detector, with essential components, an approximate scale,
and the orientation of the global coordinate
system labeled.  The origin of the global coordinate system is at the center of the target chamber.  The top four ($+y$) toroid coils have been removed to
make other detector systems visible.  Various components that are replicated on each side of the detector are not labeled to avoid redundancy.}
\label{fig:schem}
\end{figure}

Throughout this work, the ``OLYMPUS global'' coordinate system will be used to describe the positions, orientations, and trajectories
of various elements of the experiment, in addition to various ``local'' coordinate systems defined for specific detectors.  Figure
\ref{fig:schem} shows the coordinate axes of the global system relative to the detector setup.  The origin of this system
is defined to be the center of the target cell for the purpose of defining positions throughout the detector.  The beam traversed the experimental setup
approximately along the $+z$-axis through the origin, up to measured beam offsets on the order of a millimeter and angles relative to the
axis on the order of 0.5 mm.  Thus, ``upstream'' refers to the $-z$ direction and ``downstream'' to the $+z$-direction, corresponding to
the movement of beam particles.  In describing the layout of the experiment, the positive $x$-direction is referred to as ``beam left'' and
the negative $x$-direction as ``beam right'', or more succinctly, the left and right sides of the detector.  Unless otherwise noted, references
to coordinate axis in the text and figures refer to this system.

Additionally, standard spherical
coordinate angles (the polar angle $\theta$ and azimuthal angle $\phi$) are used when convenient, especially when describing the kinematics
of particle trajectories as described in Section \ref{sec:escat}.  Note that $\theta$ and $\phi$ for a tracked or simulated particle refer to the angles relative to a coordinate
system aligned with the global system but centered on the scattering vertex for the event in question.

\section{The DORIS III $e^+/e^-$ Storage Ring}

A stable storage ring source of GeV scale positron and electron beams was critical to the OLYMPUS experiment goals.  While
several accelerators were initially assessed as candidates for the location of the experiment \cite{Hillert2006}, the Doppel-Ring-Speicher
\footnote{Doppel-Ring-Speicher translates to ``double-ring storage'' in English.} (DORIS) III
storage ring at the Deutsches Elektronen-Synchrotron, Hamburg, Germany\footnote{From this point forward the accelerator will be referred to as ``DORIS''}
was selected as the facility most capable of delivering both the
desired beam conditions for the experiment and the laboratory support structure for the construction, commissioning, and 
execution of the experiment  \cite{DORIStab,DORISrep}.

Historically DORIS operated as an electron-positron collider, operating at the energy
of 3.5 GeV per beam between 1974 and 1978, and at 5.0 GeV per beam starting in 1978.  Over the course of the 1980s, the ring was converted
to be a synchrotron light source and became a full time light source in 1993.  A key physics discovery, the mixing of the neutral $B$-mesons,
was made at DORIS in 1987 by the ARGUS experiment \cite{ALBRECHT1987245}.  OLYMPUS, which began installation at the former site of the ARGUS
experiment in 2010, took data in 2012 and witnessed the last beams from the facility as the last experiment to run there on January 2, 2013.  Following
OLYMPUS, the accelerator was dismantled \cite{DORIShist}.

\subsection{OLYMPUS at DORIS}

\begin{figure}[thb!]
\centerline{\includegraphics[width=0.6\textwidth]{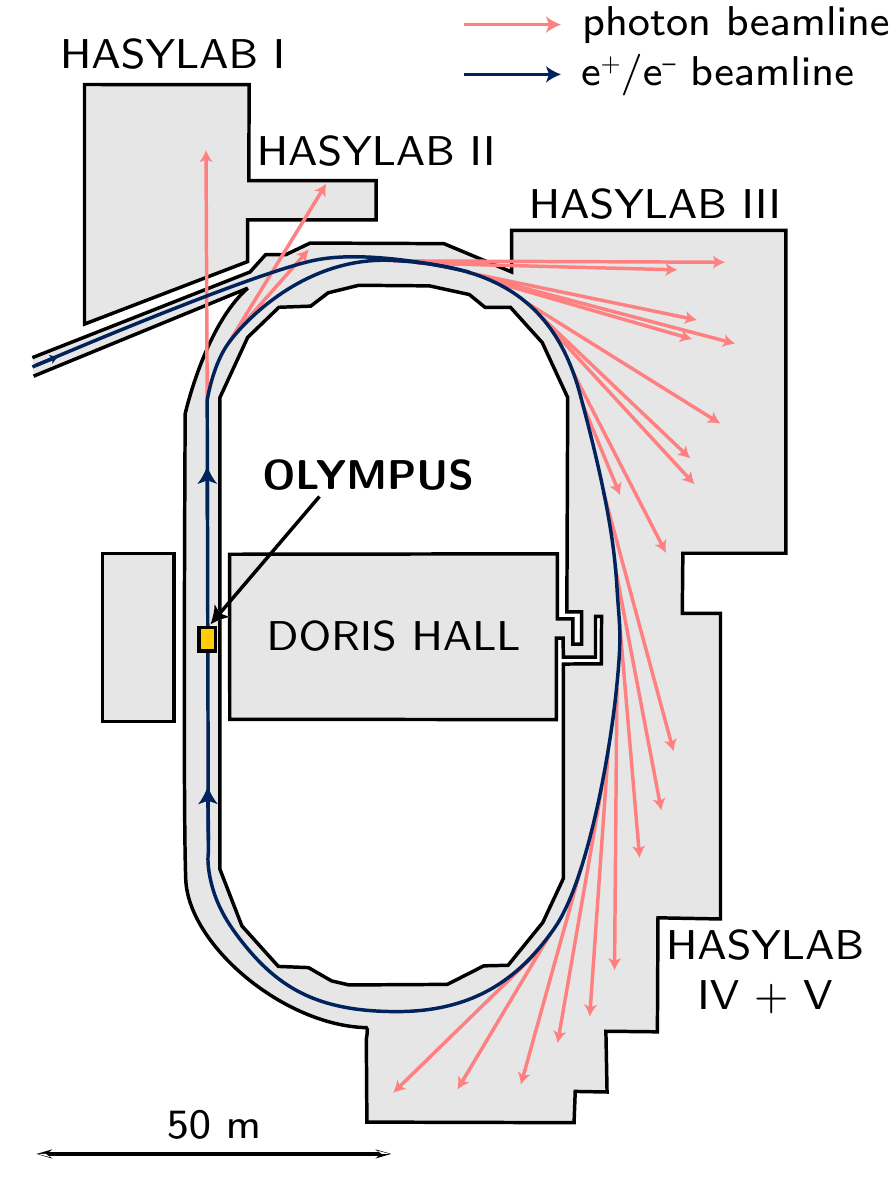}}
\caption[Overhead view of the DORIS III $e^+/e^-$ storage ring]{Overhead view of the DORIS III $e^+/e^-$ storage ring, showing the direction of beam circulation and
the locations of the OLYMPUS
spectrometer in the straight segment of the ring and the synchrotron light experiment
stations around the ring (which were not active during OLYMPUS data taking).  The total length of the beamline was approximately 300 m \cite{Bernauer201420}.}
\label{fig:dorisover}
\end{figure}

Since OLYMPUS was the last experiment to run at DORIS, it was possible to make a number of modifications to the accelerator
to provide for the needs of the experiment.  Additionally, because the accelerator was operated as a synchrotron light source
between the two OLYMPUS data runs, several constraints were placed on the design of the experiment. In particular, the target
system was installed as a permanent part of the beamline, and as such the system was required to handle a variety of conditions.
The experiment was placed in the large straight section of the ring, as seen in Figure \ref{fig:dorisover}.  Several
changes were made to the ring and experiment design to facilitate both the goals of OLYMPUS and synchrotron light production:
\begin{itemize}
 \item a large effort was made to reconfigure the operation of the beam for OLYMPUS running (2.01 GeV, with the beam in ten bunches),
 \item several RF acceleration cavities were relocated away from the detector site,
 \item extra quadrupole magnets were installed on each side of the detector in the beamline to reduce the beam width in the interaction
       region and then return it to its original size for the synchrotron light creation elements of the ring, and
 \item the OLYMPUS target was continuously cooled to allow the operation of the ring in the harsher conditions of synchrotron
       light production (4.5 GeV beam in five bunches, at 150 mA currents).
\end{itemize}
The implementation of the target system to meet these requirements is discussed in Section \ref{sec:target} and in greater
detail in Reference \cite{Bernauer201420}.  Several changes also were required to facilitate the frequent switching between
the lepton species of the beam that OLYMPUS required:
\begin{itemize}
 \item the high voltage pulse power supplies for the pre-accelerator beam extraction line and DORIS injector kicker magnets
       were heavily refurbished,
 \item the septa magnets in the pre-accelerator and injection line were modified to operate in both polarities, and
 \item remote control switches were constructed and installed for the magnet power supplies throughout the accelerator system.
\end{itemize}
The beam position in the interaction region was continuously measured by monitors installed on either side of the target chamber.
Additionally, a dipole reference magnet was installed downstream of the experiment in series with the ring's bending dipoles to continuously
measure the beam current.  This data, along with a number of relevant parameters, was recorded by the accelerator archive systems and was
made available to the OLYMPUS archiving systems.

\subsection{Beam Specifications}
\label{sec:beam}

\begin{figure}[thb!]
\centerline{\includegraphics[width=1.0\textwidth]{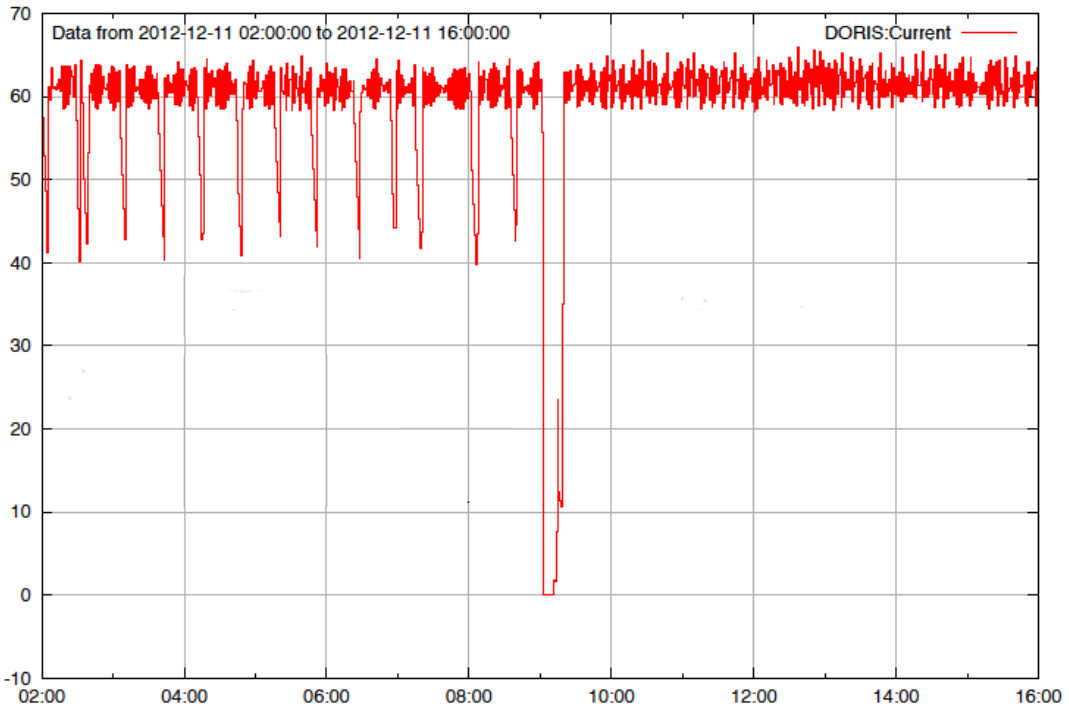}}
\caption[Typical beam current during OLYMPUS operation]{The beam current in mA as a function of time for a typical 14-hour period of OLYMPUS running,
showing the 20 minute pause in operation to 
switch the beam species at 9:00.  To the left of the pause, electron running is shown with the regular larger drops in current caused by the refills
of the PETRA ring.  To the right, positron running is shown, which did not have the top-up interruptions since PETRA also used positrons.  Outside these
interruptions, a stable beam current of around 62 mA was maintained \cite{uwe1}.}
\label{fig:bcur}
\end{figure}

During OLYMPUS operation, the beam was operated at $\sim$2.01 GeV and distributed in ten bunches.  The beam species was
switched approximately daily, but in general a concerted effort was made to keep the running conditions identical for both
species.  Due to the fact that the beam injected into DORIS was at full energy during OLYMPUS running, OLYMPUS ran in ``top-up''
mode throughout the fall run.  In this mode, leptons were added to the beam after the current dropped by only a few percent rather
than waiting until a significant fraction of the beam had decayed to refill.  This allowed OLYMPUS to collect data at a much more
constant beam current, which helped to both increase the collected luminosity and maintain stable data-taking conditions.  Top-up mode
was periodically interrupted during electron running, since approximately every 30 minutes the pre-accelerator system at DESY
was switched to positrons to fill the PETRA ring.  During these times, the current dropped below the top-up level while waiting
for the system to permit a refill.  Typical beam current levels during Run II over the course of a 14 hour period are
presented in Figure \ref{fig:bcur}, showing the stability of the current in top-up mode, the periodic PETRA fill period, and the typical 20 minute daily
pause to switch the beam species.  The beam
parameters for operation of the experiment are summarized in Table \ref{tab:beam}.

\begin{table}[thb!]
\begin{center}
\begin{tabular}{l|l}
Parameter & Value \\
\hline
\hline
Species & $e^+$ or $e^-$ (alternated $\sim$daily) \\
% \hline\hline 
Energy & 2.01 GeV \\
% \hline 
Current & 60--65 mA (top-up mode)  \\
Bunches & 10 \\
Bunch spacing & 96 ns (100 ns every fifth bunch) \\
Lifetime with gas in target & 40--55 minutes \\
Bunch length & 19.5 mm ($\sim$0.1 ns) \\

\end{tabular}

\end{center}
\caption[DORIS beam parameters for OLYMPUS operation]{A summary of the typical parameters of the DORIS beam during OLYMPUS data-taking.}
\label{tab:beam}
\end{table}

\section{The OLYMPUS Internal Hydrogen Target}
\label{sec:target}

Lepton-proton scattering events were generated by placing a windowless, internal gaseous hydrogen
target in the DORIS beamline, which simultaneously provided a high target density of protons while minimizing
adverse effects on the beam due to the presence of the target (increased emittance, uneven bunch charges, etc.).  The target
system consisted of several key components, including the hydrogen gas cell, a beam collimator, a cryogenic cooling system, an aluminum
scattering chamber enclosing the cell, wakefield suppression elements to prevent heating due to the large charge-per-bunch
of the DORIS beams, and a multi-stage vacuum system to remove hydrogen from the beamline to prevent spoiling the ring vacuum.  The primary
goals of the target system were to present an effective thickness of $\sim$3$\cdot10^{15}$ H atoms/cm$^2$ to the beam in view of the detectors
while handling the intense heating conditions of the bunched DORIS beam.

A detailed discussion of these components may be found in \cite{Bernauer201420}, while an essential overview is provided
in this section.  A discussion of the physics of the gas contained within the system and its effect on the experiment's
kinematic acceptance and luminosity is found in Section \ref{sec:sclumi}.

  \subsection{Scattering Chamber}
  
\begin{figure}[thb!]
  \centerline{\includegraphics[width=0.8\textwidth]{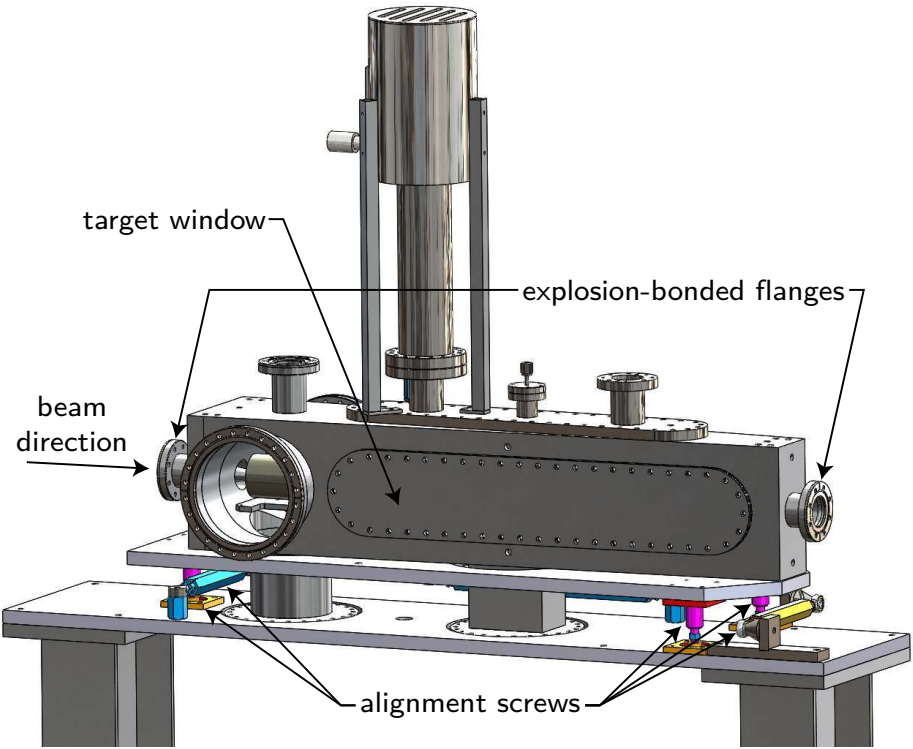}}
  \caption[Schematic of the OLYMPUS scattering chamber]{Schematic of the OLYMPUS scattering chamber, showing the principal elements of the design as viewed from beam
  right \cite{Bernauer201420}.}
  \label{fig:chamber}
  \end{figure}

  The main elements of the target system were contained within a scattering chamber, manufactured from a single solid block of aluminum
  to provide high vacuum integrity.  This chamber was 1200 mm long and 254 mm high.  The chamber tapered from a width
  of 245 mm at its upstream end to 114.3 mm at its downstream end, resulting in a trapezoidal prism shape for the chamber.  This design increased
  the visibility of the target cell within the chamber to the forward detector elements while still providing room for the installation of the collimator 
  and access ports at the upstream end.  The chamber interfaced directly with the beamline
  (and the beamline vacuum) at each end. The chamber was designed at the MIT-Bates Linear Accelerator Center, taking advantage of experience
  gained designing similar targets for BLAST and other experiments \cite{Cheever:2006xt}. Figure \ref{fig:chamber} shows the key elements of the chamber design.
  
  The chamber included several ports to allow access to the inside of the chamber, both for installation and/or repair work and to provide windows for scattered
  particles to escape the target.  The ports were sealed either with Atlas\footnote{Atlas Technologies, Port Townsend, WA, USA} explosion-bonded bimetallic flanges
  or O-rings (in the case of the window ports).  The target windows consisted of 0.25 mm thick 1100 aluminum foil, and subtended a polar angle range of
  8\dg to 100\dg relative to the center of the target to provide a complete view of the cell to the detector systems (with the exception of the 12\dg system
  which could not view the most downstream portion).  The chamber was mounted on an aluminum table, which supported its weight above the vacuum system
  and provided screws for the alignment of the chamber to match the beamline.

  \subsection{Target Cell}
  
\begin{figure}[thb!]
  \centerline{\includegraphics[width=0.8\textwidth]{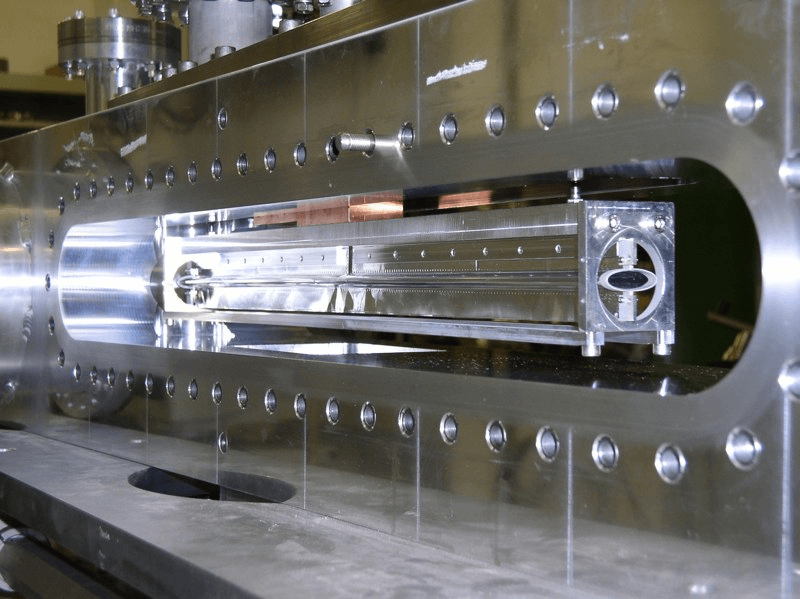}}
  \caption[Photograph of the OLYMPUS target cell]{The OLYMPUS target cell, mounted within the aluminum scattering chamber
           (with the windows and wakefield suppressors removed) \cite{Bernauer201420}.}
  \label{fig:cellphoto}
  \end{figure}

\begin{figure}[thb!]
  \centerline{\includegraphics[width=1.0\textwidth]{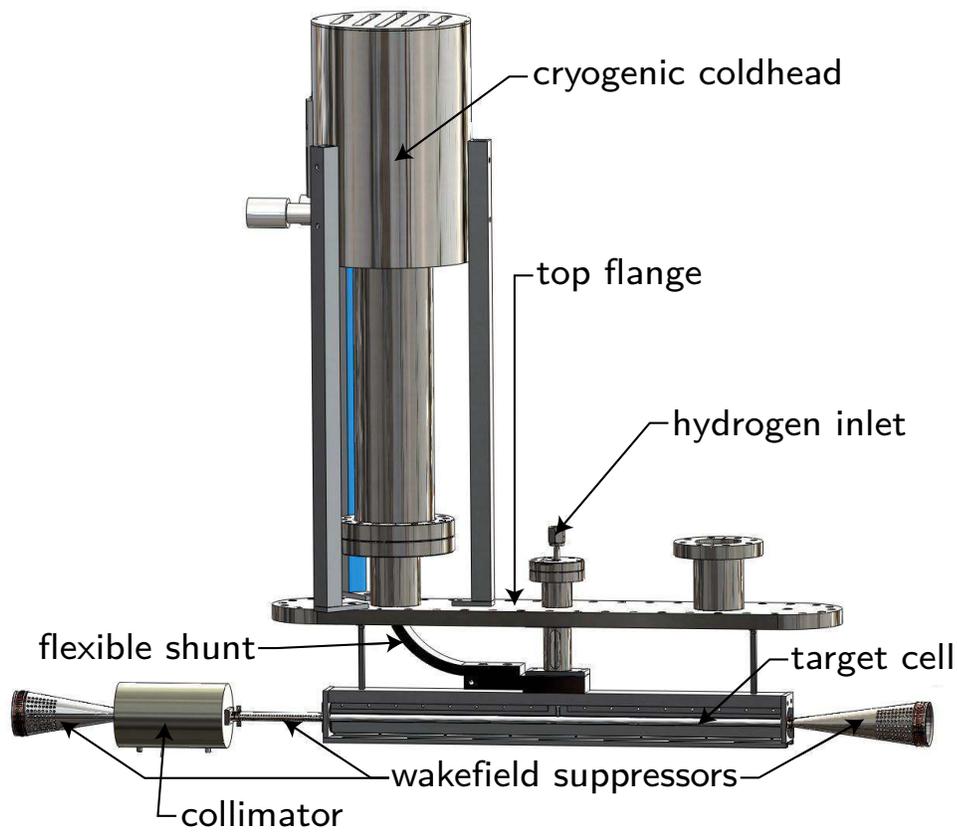}}
  \caption[Components of the OLYMPUS target system within the scattering chamber]{Schematic of the components of the hydrogen target
  system contained within the scattering chamber, including the gas inlet, target cell, connection to the cryogenic system,
  collimator, and wakefield suppressors.  For reference, the origin of the OLYMPUS coordinate system is the center of the target cell
  directly beneath the inlet, and the beam passed from left-to-right (positive OLYMPUS $z$), approximately through the center of the wakefield suppressors,
  collimator, and target cell as shown in the figure. The conal wakefield suppressors at each end connect smoothly to the beamline \cite{Bernauer201420}.}
  \label{fig:tarin}
  \end{figure}

  The target cell, used to contain a considerable concentration of hydrogen in the vicinity of the beam before it escaped to the vacuum system, was a
  600 mm long elliptical aluminum tube through which the beam passed directly.  The elliptical cross section (27 mm wide by 9 mm high) was chosen 
  to mimic the envelope of the DORIS beam to mitigate any unwanted interactions of the beam with solid material in the target system.  The target cells
  used in the experiment were manufactured by molding two sheets of \SI{75}{\micro\meter} thick Goodfellow\footnote{Goodfellow Corporation, Coraopolis, PA, USA}
  aluminum foil and mounting them in an aluminum support frame.  The cells were constructed at Ferrara University/INFN, where similar cells were
  constructed for the HERMES experiment \cite{Airapetian:2004yf}.  A photograph of an OLYMPUS target cell mounted in the scattering chamber (with the windows and other
  internal target system components removed) is shown in Figure \ref{fig:cellphoto}, and its connections to the other internal components
  is shown in Figure \ref{fig:tarin}.  More details on the manufacturing process for these cells are described in Reference \cite{Bernauer201420}.
 
  \subsection{Collimator}
  
  Also among the internal components of the target system shown in Figure \ref{fig:tarin} is a collimator placed on the upstream side of the 
  target cell to block beam halo from entering the OLYMPUS interaction region.  The collimator was manufactured from a solid tungsten cylinder, with an outer
  diameter of 82.55 mm and length of 139.7 mm.  The bore of the collimator was elliptical in cross section, 25 mm by 7 mm at its upstream face and flaring
  to the 27 mm by 9 mm size of the target cell and intermediate wakefield suppressor at its downstream end.  The size of the collimator and its
  bore were chosen based on the results of Monte Carlo simulation of beam halo particles and synchrotron radiation impinging on the collimator,
  and considering the resulting predicted heat loads and forward scattering of unwanted particles into the interaction region.  During the 4.5 GeV
  synchrotron light production operation of DORIS, the collimator was exposed to $\sim$25 W of heating, which was well within the dissipation
  capabilities of the cooling system.  During OLYMPUS operation, the heat load was considerably less ($\sim$1 W).
  
  \subsection{Wakefield Suppressors}
  
  The final internal components of the system shown in Figure \ref{fig:tarin} were the wakefield suppression elements, which served to provide
  a continuous and smooth electrical conductance connection between the beamline interfaces at each end of the target chamber and among the 
  internal target chamber components.  This was required due to the bunched nature of the DORIS beam, which caused strong electromagnetic wakefields
  that could induce considerable heating of the system if not provided a highly conductive means of conveyance in the system.  Three wakefield suppressors
  were present in the system: an elliptical tube element between the cell and the collimator, and two conal elements which flared from elliptical profiles
  at their interfaces with the target system to circular profiles at their interfaces with the beamline.  The elements were coated with a thin layer
  of silver to increase their conductivities and their connections to other system elements included beryllium-copper spring cones to ensure good
  electrical contact.  Holes were drilled through the wakefield suppressors to allow the escape of beam gas, but these holes were placed as far as possible
  from the beam to mitigate their impact on the conductivity of the system.
  
  \subsection{Cryogenic System}
  
  To mitigate the heating caused by the beam and to increase the target density by cooling the H$_2$ gas within the system, the target system
  was actively cooled to temperatures below 75 K whenever beam was in the DORIS ring.  This was achieved using a CryoMech\footnote{CryoMech, Inc., Syracuse, NY, USA}
  AL230 coldhead and CP950 compressor system.  The interface of the coldhead with the components of the target system can be seen in both
  Figures \ref{fig:chamber} and \ref{fig:tarin}.  The cooling system connected to the target cell assembly via a solid copper shunt coated with
  indium at the interface with the cell assembly. The aluminum scattering chamber (exposed to the beam hall atmosphere) was at the temperature of the beam
  hall air and was thus thermally insulated from the cooled elements.  The system was capable of dissipating 36 W at 25 K, sufficient to
  handle the heating caused by both synchrotron light production and OLYMPUS beam conditions.  This was verified by the installation of seven Pt100
  thermocouple temperature sensors along the length of the target that were used to monitor the system temperature whenever beam was in the ring.
  
  \subsection{Vacuum System}
  
\begin{figure}[thb!]
  \centerline{\includegraphics[width=1.0\textwidth]{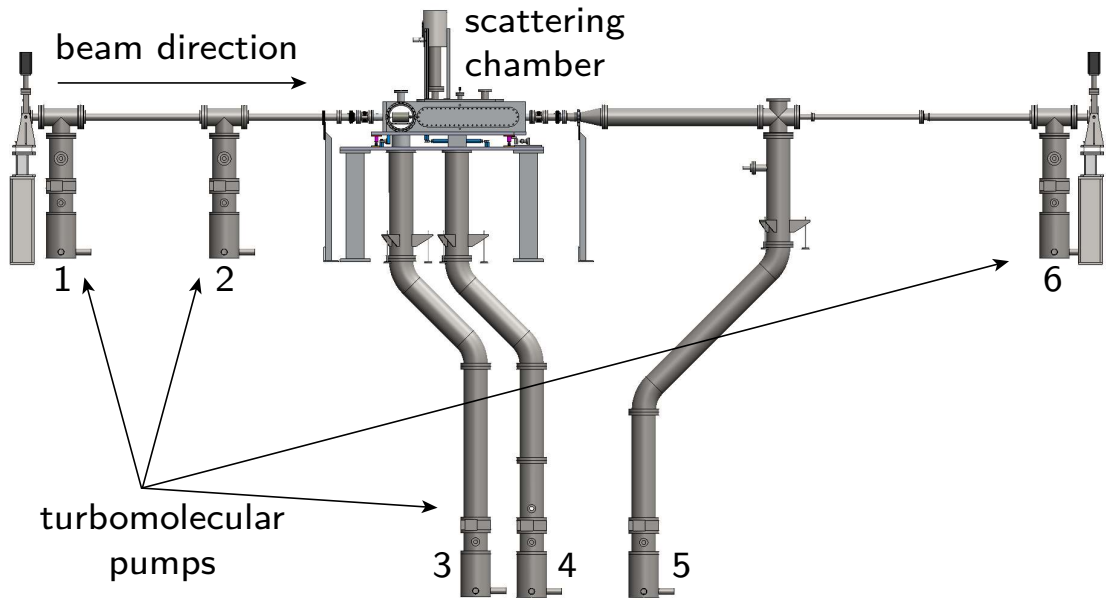}}
  \caption[Diagram of the target system vacuum components]{Diagram of the vacuum components of the OLYMPUS target system, showing the six turbomolecular
  pumps and their high conductance connections to the target chamber and beamline.  The non-evaporable getters were placed at the locations
  of turbomolecular pumps 1, 2, 5, and 7 to assist with the removal of hydrogen at low pressures.  \cite{Bernauer201420}.}
  \label{fig:vac}
  \end{figure}

  Due to the fact that hydrogen gas was flowed directly into the beamline, it was necessary to include a powerful vacuum system to remove the gas from the beamline as it escaped
  the target cell assembly to avoid spoiling the vacuum necessary for the beam.  The system included six turbomolecular pumps (split between
  Osaka\footnote{Osaka Vacuum, Ltd., Osaka, Japan} TG 1100M and Edwards\footnote{Edwards, Crawley, United Kingdom} STP 1003C models).  Since
  the system was required to remove hydrogen from the ring at low pressures, four non-evaporable getters (NEGs) were installed with four of the 
  turbomolecular pumps.  Due to the fact that turbomolecular pumps operate using magnetically levitated rotors, the pumps were required to be placed
  outside of the OLYMPUS magnetic field (either along the beamline from the interaction region or in a pit below the experiment).
  The placement of the pumps is shown in Figure \ref{fig:vac}.  The pumps placed below the experiment were connected to the system with large-diameter
  piping to insure a high conductance connection.  As can be seen in the figure, the pumps were arranged in three stages, each of which reduced
  the pressure by approximately an order of magnitude.
  This reduced the $\sim$10$^{-6}$ Torr pressure inside the chamber to the $10^{-9}$ Torr pressure of the ring.
  
  \subsection{Hydrogen Gas Supply System}
  
  The source of protons for \pmp scattering in the system (and of electrons for the luminosity monitoring processes) was molecular hydrogen gas 
  flowed into the middle of the target cell.  The binding energies of hydrogen atoms and molecules are negligible compared to the GeV
  scale beam energy, and thus the target effectively served as a free proton target.  The H$_2$ gas used for the target was supplied by a
  Parker\footnote{Parker-Hannifin Corporation, Haverill, MA, USA} hydrogen gas generator, which produced ultra-pure ($<$200 ppb impurity) hydrogen
  via the electrolytic dissociation of deionized water.  Within the generator, the dissociation occurred in palladium tubes, which were opaque
  to all atoms and molecules in the system except H$^+$ ions.  The generator produced a pressure of 20 psi on the supply line
  to the target.  This rate of gas production exceeded the target's
  actual supply needs by orders of magnitude, but was done to maintain positive pressure relative to the atmosphere outside the line.
  
  The actual gas flow into the target was controlled by a system of remotely-controlled mass flow controllers (MFCs) and pneumatic solenoid valves
  modeled on the system used for the BLAST target \cite{Cheever:2006xt}, and operated using the slow control system described in Section
  \ref{sec:sc}.  Two buffer vessels of known volume were used to calibrate the MFCs.  The MFCs could provide reliably calibrated flow rates
  between 0.1 and 1.0 standard cubic centimeters per minute (sccm), and the range used during the experiment running was approximately 0.4-0.8 sccm.

  \section{The OLYMPUS \pmp Spectrometer}
  \label{sec:spect}

  The OLYMPUS spectrometer used for the reconstruction of elastic \pmp events was predominantly constructed from various components
  of the BLAST experiment \cite{Hasell:2009zza}.  The detector consisted of gas drift chambers to the left and right of the beam
  for particle trajectory reconstruction, scintillator panels beyond the drift chambers to provide timing and trigger information, and a
  toroidal magnetic field to provide particle bending for momentum reconstruction and background suppression.  The total acceptance of
  the main detector package for \pmp events ranged over approximately $(25^\circ \leq \theta\leq 80^\circ)$ and
  $(-15^\circ \leq \phi \leq 15^\circ) \cup (165^\circ \leq \phi \leq 185^\circ)$ in the particle scattering angles, corresponding to
  the instrumentation of the left and right sides of the detector.  This corresponds to ranges in the kinematic parameters of approximately
  $(0.4 \leq \epsilon \leq 0.9)$ and $(0.6 \leq Q^2 \leq 2.2)$ GeV$^2/c^2$.
  The essential details of these components are described in this section (and in more detail in References \cite{Milner:2014,tdr,Bernauer20169}).

  \subsection{Toroidal Magnet}
  \label{sec:tormag}

\begin{figure}[thb!]
  \centerline{\includegraphics[width=1.0\textwidth]{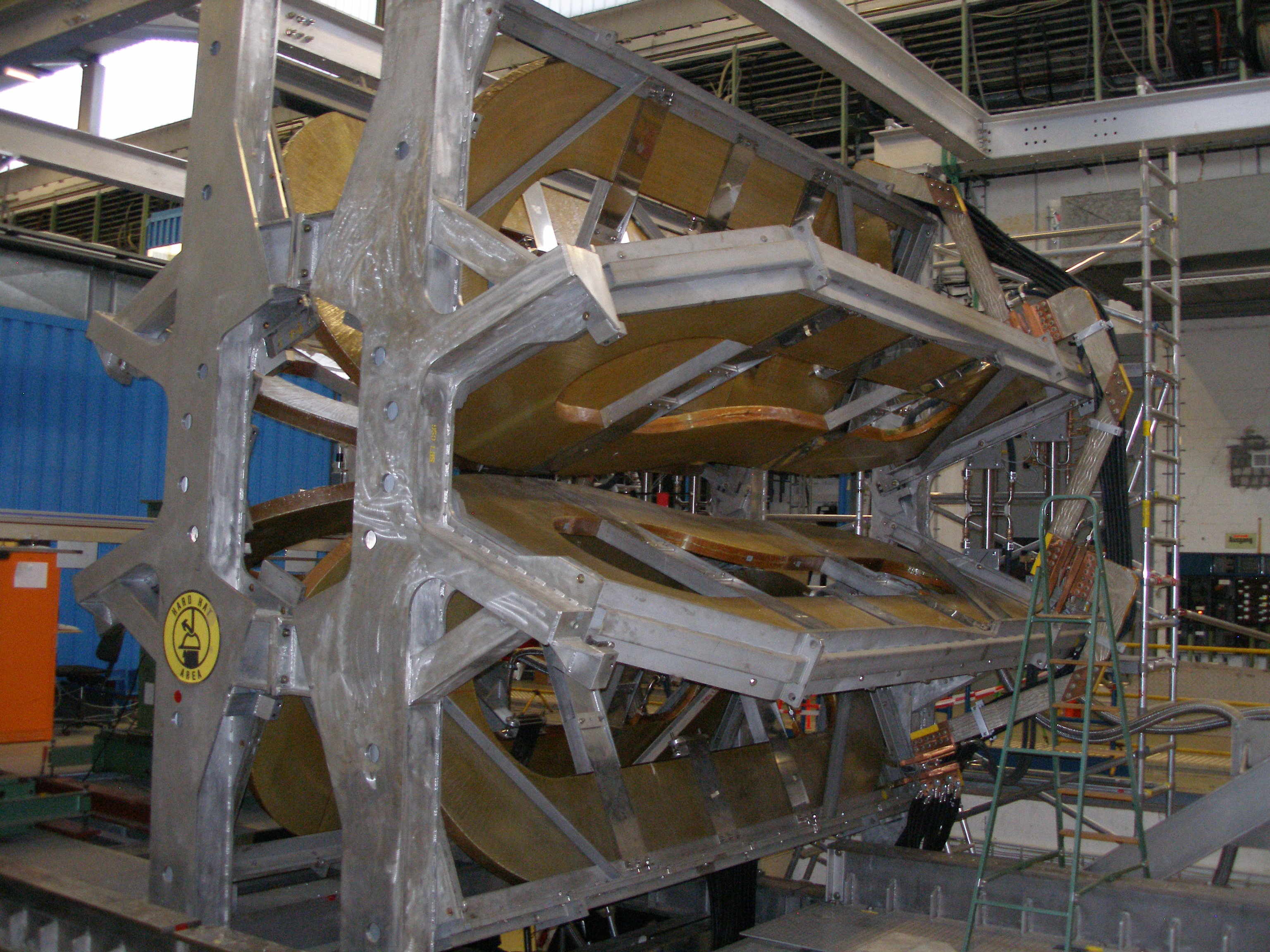}}
  \caption[The OLYMPUS toroid prior to installation]{Photograph of the OLYMPUS toroid prior to installation in the DORIS beamline, showing
  the symmetric arrangement of the eight coils \cite{Milner:2014}.}
  \label{fig:magpho}
  \end{figure}

  Momentum reconstruction for particles detected in the OLYMPUS detector was made possible by the generation of a toroidal magnetic
  field around the target.  Additionally, the field produced
  by the toroid bent low-energy particles (produced by M{\o}ller and Bhabha scattering in the target and other sources of background tracks)
  away from the detector systems to reduce the number of unwanted hits in the detectors.  Eight water-cooled copper coils carrying a nominal
  current of 5000 A arranged symmetrically around the beamline 
  produced this field.  The lower four of these coils are shown in Figure \ref{fig:schem}, while Figure \ref{fig:magpho} shows the complete coil
  configuration prior to installation of the magnet in the DORIS beamline.  The coils pinched towards the beamline in their downstream
  sections, but opened away from each other in the upstream sections to accommodate the target.
  The peak field strength was approximately 0.3 T in the regions of the tracking detectors.    
  
  The coils were originally part of
  the BLAST experiment, and the design of the magnet is described in detail in Reference \cite{Dow2009146}.  A toroidal field was originally
  chosen for the BLAST experiment to minimize the magnetic field near the beam, which was critical to the polarized target experiments
  conducted at BLAST \cite{Hasell:2009zza,doi:10.1142}.  While zeroing the field along the beamline was not critical to OLYMPUS,
  the coils were aligned when installed to minimize the field in this region to avoid perturbing the beam.

\begin{figure}[thb!]
  \centerline{\includegraphics[width=1.0\textwidth]{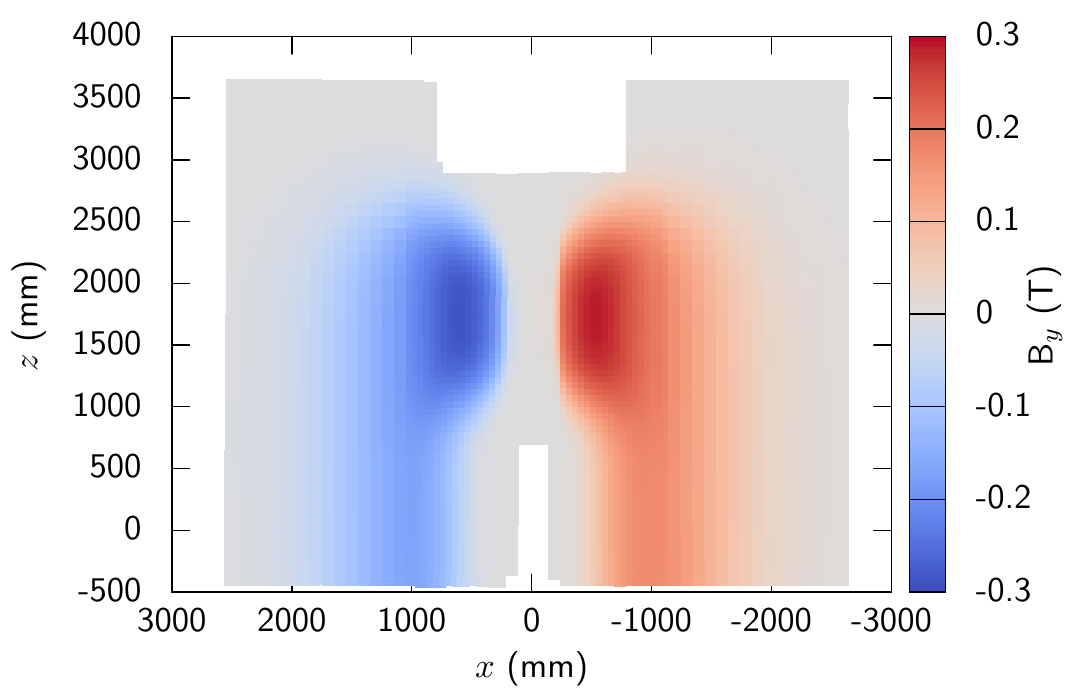}}
  \caption[Map of $B_y$ in the tracking detector region]{Map of the strength of the $B_y$ component of the OLYMPUS magnetic field in the global $y=0$
  plane as measured in the field survey (Section \ref{sec:magsur}).  The regions of strongest field occurred within the drift chamber tracking volumes,
  while the field was much smaller in the vicinity of the target along the $z$-axis \cite{Milner:2014}.}
  \label{fig:bymap}
  \end{figure}

  During data-taking, the toroid was set to produce a field that bent positively charged particles towards positive $\theta$
  trajectories as they moved through the detector system.  The strength of the field in the $y$ direction in the $y=0$ plane bisecting
  the detector systems in shown in Figure \ref{fig:bymap}.  While the original conception of the OLYMPUS experiment included regularly
  switching the toroid polarity to reduce the systematic uncertainties associated with the relative acceptance of positron and electron
  events, this was in practice infeasible.  In the opposite polarity, low energy electrons originating from the target
  were bent into the detectors which caused an intractable background that obscured desired events.  While efforts were made to mitigate this
  effect to permit running with the second polarity (including increasing the field strength and physically shielding the detectors with material
  to stop low energy particles), these changes were not sufficient to create a good environment for elastic event reconstruction.  While some data
  were taken with the opposite polarity for testing purposes, this dataset only represents approximately 13\% of all data collected, since it was
  limited to low luminosity running in which the noise rate was sufficiently low to permit event reconstruction.
  
  To ensure that the magnetic field was well understood for the purposes of track reconstruction and event simulation, a large effort was undertaken
  after the conclusion of data-taking to survey the field throughout the detector volumes.  This effort is described in Section \ref{sec:magsur} and
  Reference \cite{Bernauer20169}.
  
  \subsection{Time-of-Flight (ToF) Scintillator System}
  
  The time-of-flight (ToF) scintillator system, also inherited from the BLAST experiment \cite{Hasell:2009zza}, consisted of 36 scintillator
  bars (18 per side) arranged in walls on each side of the detector, as shown in Figure \ref{fig:schem}.  These bars played the critical role of providing timing
  signals for the trigger and the readout of the main detector elements, as described in Section \ref{sec:trig}.  The bars were arranged in panels
  of four, five, and nine bars on each side with the panels arranged to point the normal vector of the panel plane approximately towards the target.
  The forward panel bars measured $119.4\:\text{cm}\times17.8\:\text{cm}\times2.54\:\text{cm}$, while the bars in the rear two panels were larger:
  $180.0\:\text{cm}\times26.2\:\text{cm}\times2.54\:\text{cm}$.  This arrangement guaranteed that the acceptance of the ToF bars for tracks
  originating from the target completely included that of the drift chambers and 12\dg luminosity telescopes.
  
  Each bar consisted of a solid block of Bicron\footnote{Bicron, Solon, OH, USA}
  BC-408 scintillator, a plastic scintillator designed for applications requiring fast response times (0.9 ns) over large areas such as in the OLYMPUS
  experiment \cite{bicron}.  While new BC-408 exhibits attenuation lengths on the order of two meters, due to the age of the scintillator bars
  and their exposure to radiation throughout the BLAST and OLYMPUS experiments, the scintillators' attenuation in the bars during OLYMPUS
  running was significantly worse.  While efforts were made to replace the most damaged bars, a large analysis effort was made to properly
  account for the state of the scintillator, which is described in detail in Reference \cite{russell}.
  
\begin{figure}[thb!]
  \centerline{\includegraphics[width=1.0\textwidth]{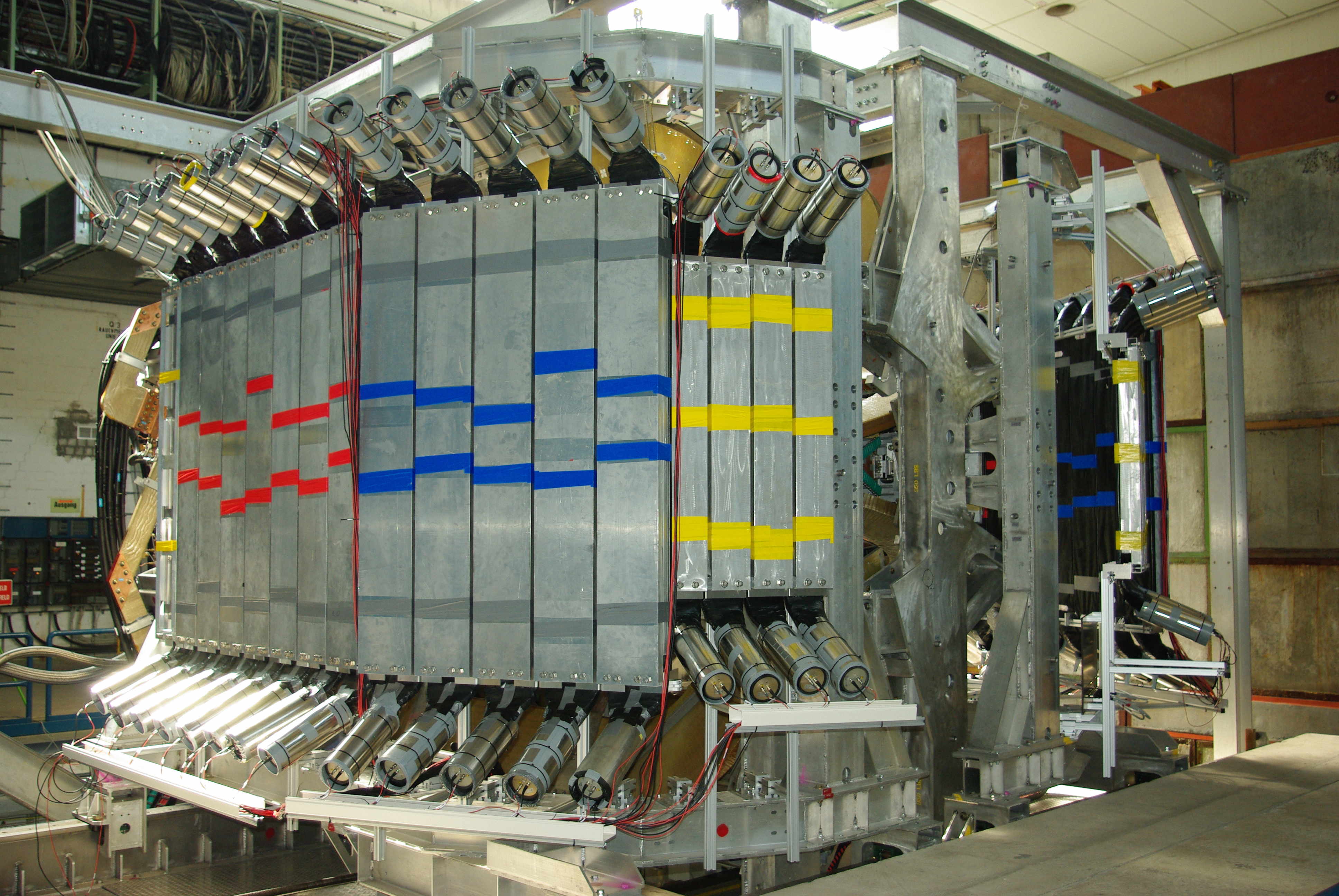}}
  \caption[Arrangement of the right side ToF detectors]{Photograph of the right side of the ToF detector system prior to installation of the detector
  package within the DORIS beamline, showing the orientation of the PMTs
  perpendicular to the direction of the toroidal magnetic field \cite{Milner:2014}.}
  \label{fig:tofpho}
  \end{figure}

  Each bar was instrumented with two 3-inch diameter Electron Tubes\footnote{Electron Tubes, Ltd., Ruislip, Middlesex, United Kingdom}
  9822B02 photomultiplier tubes, connected to the top and bottom of each bar via Lucite\footnote{Lucite International, Southampton, Hampshire, United Kingdom}
  light guides.  The light guides where arranged to orient their connected PMTs approximately perpendicularly to the toroidal magnetic field
  to minimize the effect of the field on the PMT gain, as shown in Figure \ref{fig:tofpho}.  The PMTs were also wrapped in mu-metal shielding to further reduce any such
  effects.  Signals from the detectors were passed to dedicated ADC and TDC channels for each PMT, processed using constant fraction discriminators
  for the downstream 16 bars on each side and leading-edge discriminators for the upstream 2 bars on each side.  This distinction was due to the fact
  that the upstream bars were not included in the original design, and were added only after it was realized that a larger margin of safety was
  desired to cover the full acceptance of the 12\dg telescopes.
  
  The signals from the PMTs were rapidly digitized for processing by the trigger.  When triggered, these signals produced a common-start signal
  for the ToF TDCs and a common-stop signal for the drift chamber TDCs.  The total ADC counts were recorded as a measure of the energy
  deposition in the ToF bars, and the ToF TDC information for each bar could be used to estimate both the vertical position at which the bar was hit 
  and the flight time of the particle from the vertex to the bar.  This analysis is described in Section \ref{sec:recon}.

  \subsection{Drift Chamber Tracking System}
  \label{sec:thegdwcs}
  
  The main detectors that provided hit position information for reconstruction of \pmp events were two large drift chambers mounted between the toroid
  coils on the left and right sides of the detectors.  Functioning as a typical drift chamber, charged particles passing through the gas contained within these 
  chambers induced ionization, and the electric field created by the wires held at potential throughout the chamber caused the resulting electrons to drift
  towards dedicated detection wires.  Better position resolution was achieved by recording the signal time using TDCs, with a common-stop signal provided by the fast signal of the time-of-flight
  system.  Since drift speeds of the electrons in the gas were on the order of 5-6 cm/\SI{}{\micro\second}, the arrival of the drifting electrons across the centimeter-scale cells occurred
  well after the ToF signal.  This extended drift time was converted to a distance from the wire at which the particle passed (Section \ref{sec:ttd}), which provided positional
  information for track reconstruction.
  
\begin{figure}[thb!]
  \centerline{\includegraphics[width=0.8\textwidth]{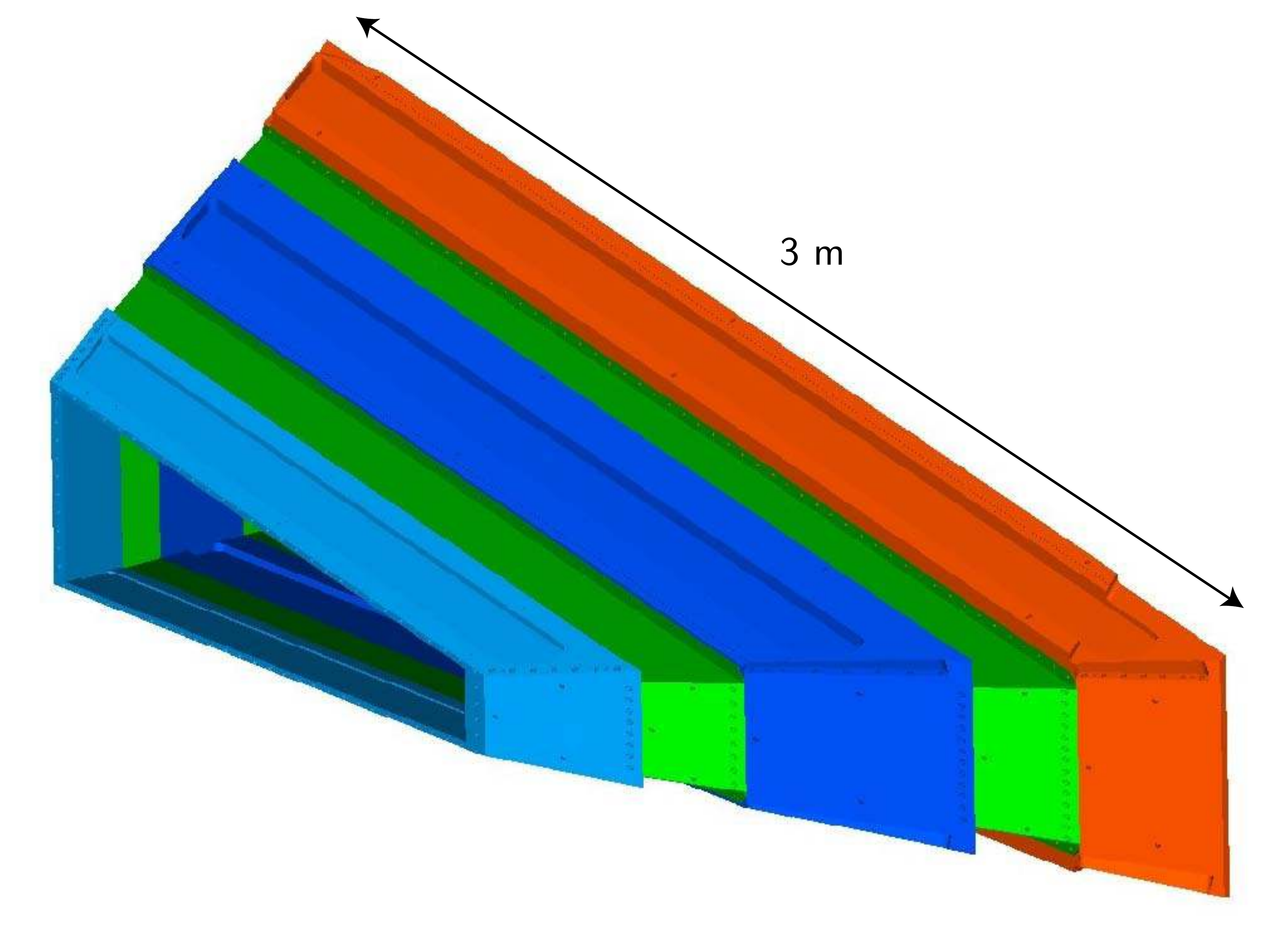}}
  \caption[Schematic of an assembled OLYMPUS drift chamber]{An isometric schematic of one of the drift chambers, illustrating how the three frames were assembled to create
  a single gas volume \cite{Milner:2014}.}
  \label{fig:dframe}
  \end{figure}

\begin{figure}[thb!]
  \centerline{\includegraphics[width=0.95\textwidth]{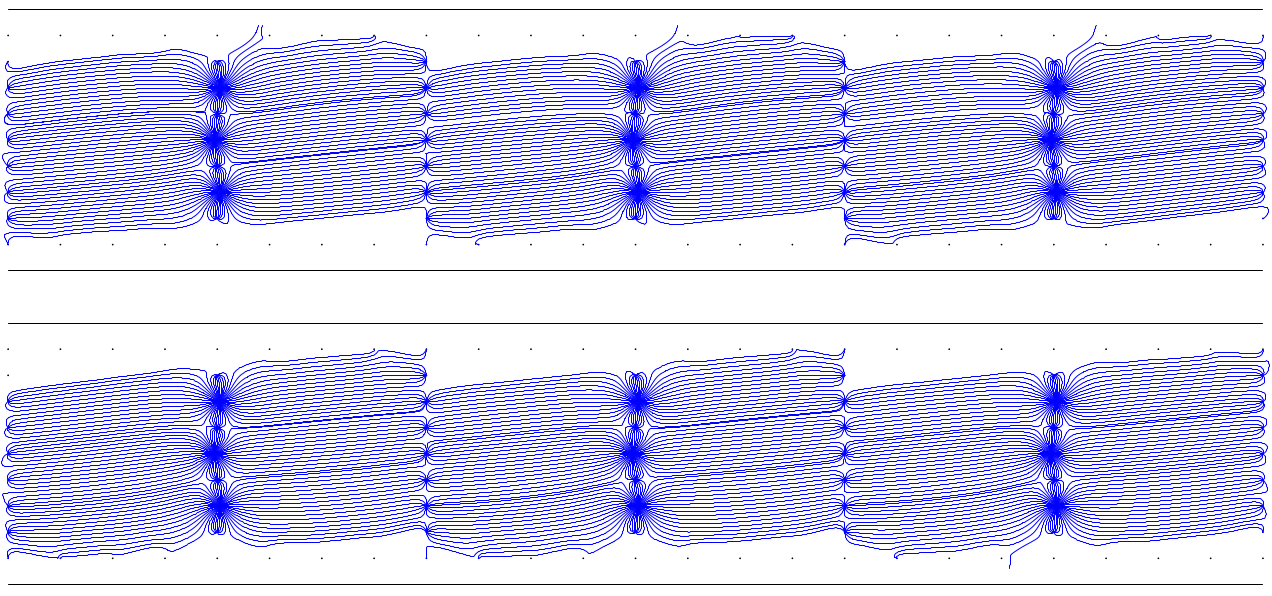}}
  \caption[Lines of drift within the OLYMPUS drift chamber cells]{A view along the direction of the wires forming the rectangular drift cells, with the lines of electron drift induced in the cells
  by the voltages on the wires.  Note that the lines predominantly tend towards the three sense wires on which signals were recorded.  The drift lines are tilted relative to the axes of the cells
  due to the interaction of the OLYMPUS magnetic field with the drifting electrons, creating a drift component in the direction of $\mathbf{E}\times\mathbf{B}$ \cite{Milner:2014}.}
  \label{fig:cells}
  \end{figure}

  Once again, these detectors were inherited from the BLAST experiment \cite{Hasell:2009zza}, but
  underwent a complete refurbishment for OLYMPUS.  The refurbishment included a complete rewiring of the chambers as well as new voltage distribution electronics and connections
  for all wires.  Each chamber consisted of three trapezoidal frustum-shaped aluminum frames joined to create a single
  gas volume, a schematic of which is shown in Figure \ref{fig:dframe}.  Thin plastic windows closed off the gas volume on the inner and outer faces of the connected frame
  assemblies. Each of the three frames contained two layers of wire ``cells''.  A cell
  was formed by an array of wires that were arranged to form a rectangular box region used to create an electric drift region, as shown in Figure \ref{fig:cells}.  The wires were held at stepped voltages
  from ground at the boundary between two cells to 2800 V at the center of a cell.  In the center of the cell, an additional column of wires contained
  three sense wires held at 3900 V to attract drifting electrons to generate the detector's signals.  To help resolve the ambiguity of a single time recorded on a sense wire,
  the three sense wires were offset from the center plane of the cell
  by $\pm0.5$ mm to create a small difference in drift times from each side of the cell. 
  Each cell was 78 mm by 40 mm, and extended from the top inner face of the frame to the bottom
  inner face.  The two layers of cells in each frame extended parallel to each other, 20 mm apart, arranged such that wires in layer were at $\pm5^\circ$ from vertical to
  create a stereo angle for 2D position reconstruction.  A total of 318 cells were present in the two drift chambers, containing 954 sense wires and approximately 10,000 total wires.
  
  The chambers were filled with an Ar:CO$_2$:C$_2$H$_6$O gas mixture at a ratio of approximately 87.4:9.7:2.9. The argon and carbon dioxide were mixed at a 9:1 ratio by a dedicated
  mixing system, while the ethanol was introduced by bubbling the mixed gas through liquid ethanol at 5 $^\circ$C.  Unfortunately, this method of introducing ethanol to the system
  resulted in fluctuation in the concentration of ethanol during data-taking, which altered the drift properties of the gas.  This required careful calibration of the drift chamber
  time-to-distance calibration, as described in Section \ref{sec:ttd}\footnote{The introduction of ethanol to the drift chamber gas was part of an attempt to reduce the number of noise
  hits in the drift chambers, particularly in the inner layers.  This, however, was not a sound decision from the standpoint of achieving that goal.  While it did confer the benefit
  of slowing the drift speed of electrons in the gas (thus increasing the resolution somewhat), ethanol is typically introduced to drift chambers so as to reduce the amount of carbon
  ``whisker'' build-up on the wires due to other organic compounds in the drift gas, which can cause dark currents on the signal lines \cite{blum}.
  For an Ar:CO$_2$ mixture, however, whisker growth is not expected and has not been reported to be observed \cite{dont}.  Thus, the introduction of ethanol was very unlikely to reduce the level
  of noise in the OLYMPUS drift chambers, and the instability of the ethanol level certainly was more of a detriment than was balanced by the minor benefit of decreased drift speed.}.
  The gas pressure in the chambers was maintained at about 1 inch
  of water above local atmosphere to create positive pressure within the chambers, corresponding to a flow rate of approximately 5 L/min.
  
  Pulse signals from arriving drift electrons on sense wires were first decoupled from the high voltage on the voltage distribution electronics, and then discriminated and amplified
  on dedicated front-end electronics attached directly to distribution boards on the chambers.  The discriminated signals were passed to LeCroy\footnote{Teledyne LeCroy, Chestnut Ridge, NY, USA}
  1877 multi-hit TDC modules.  Conversion of recorded drift times to distances from the wires and the method of reconstructing those distances
  into trajectories are discussed in Section \ref{sec:recon}.

\section{$12^\circ$ Tracking Telescope Luminosity Monitors}
  
    To provide one method of luminosity measurement for the experiment, a dedicated telescope tracking system was developed for the reconstruction
    of elastic \pmp scattering events in which the lepton is scattered with $\theta\approx 12^\circ$.  The system tracked leptons independently of the
    drift chambers, while the drift chamber data were used to reconstruct the corresponding proton near the upper end of the chambers' $\theta$ acceptance.
    Hit positions of passing charged particles were reconstructed in the detector plane elements of the telescope, and these hits were reconstructed
    to find the particles' trajectories, as described in Sections \ref{sec:12hit} and \ref{sec:12track}.
    The system took advantage of the rapid increase of the elastic \pmp cross section at small $\theta$ to produce a high statistics count of events.
    Using the system as a luminosity measurement requires the assumption that TPE is small at forward angles, and thus absent this
    assumption and given another method of normalizing the luminosity between species, the 12\dg telescopes provide an additional tracking point
    for the measurement of \ratio forward of the drift chambers.
    
\begin{figure}[thb!]
    \centerline{\includegraphics[width=0.95\textwidth]{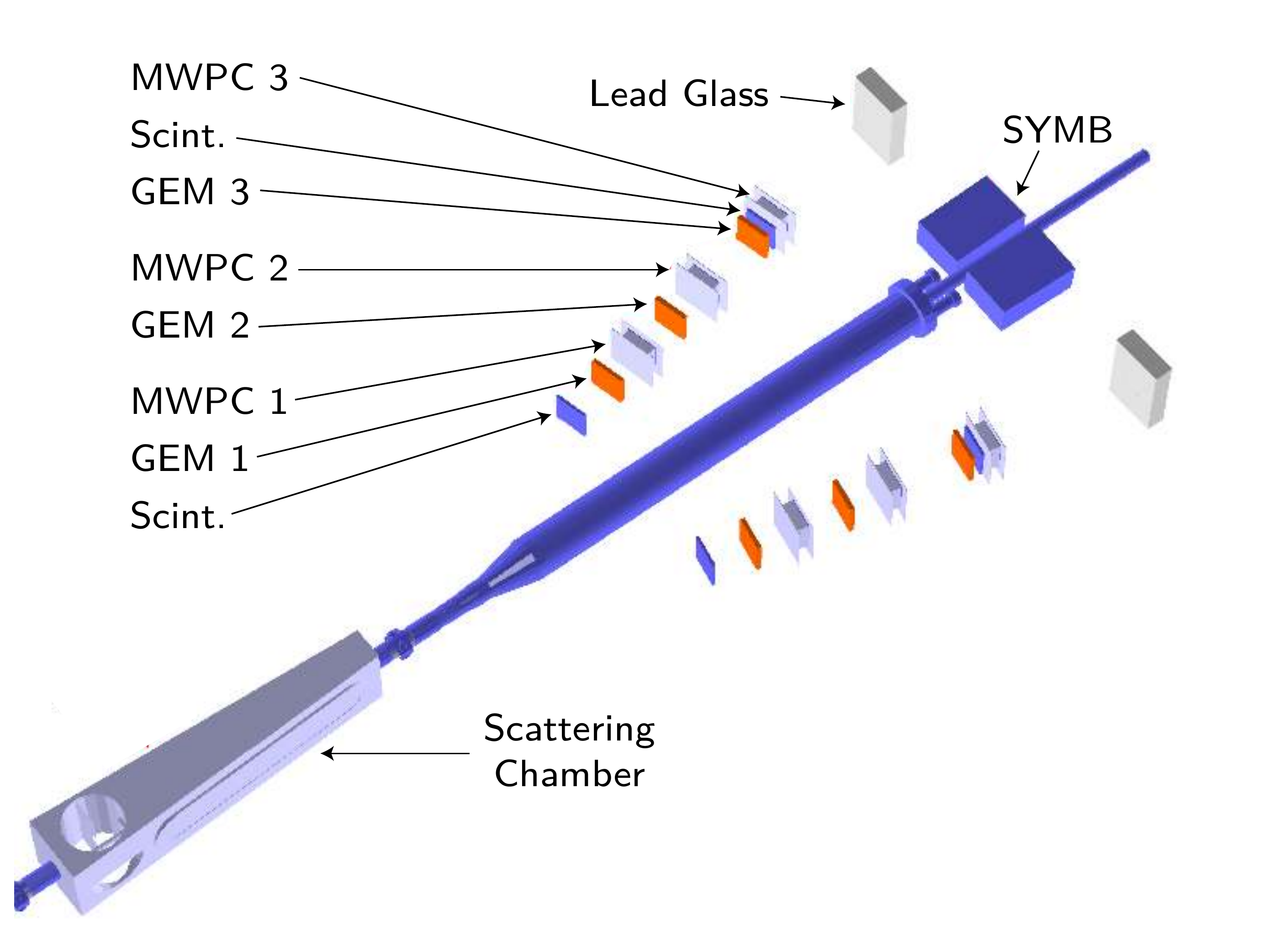}}
    \caption[Diagram of the forward detector systems]{Schematic view from above of the forward detector systems, including the 12\dg luminosity monitor telescopes
    and Symmetric M{\o}ller/Bhabha calorimeter, relative to the target chamber and beamline \cite{Milner:2014}.}
    \label{fig:fordet}
    \end{figure}

\begin{figure}[thb!]
    \centerline{\includegraphics[width=1.0\textwidth]{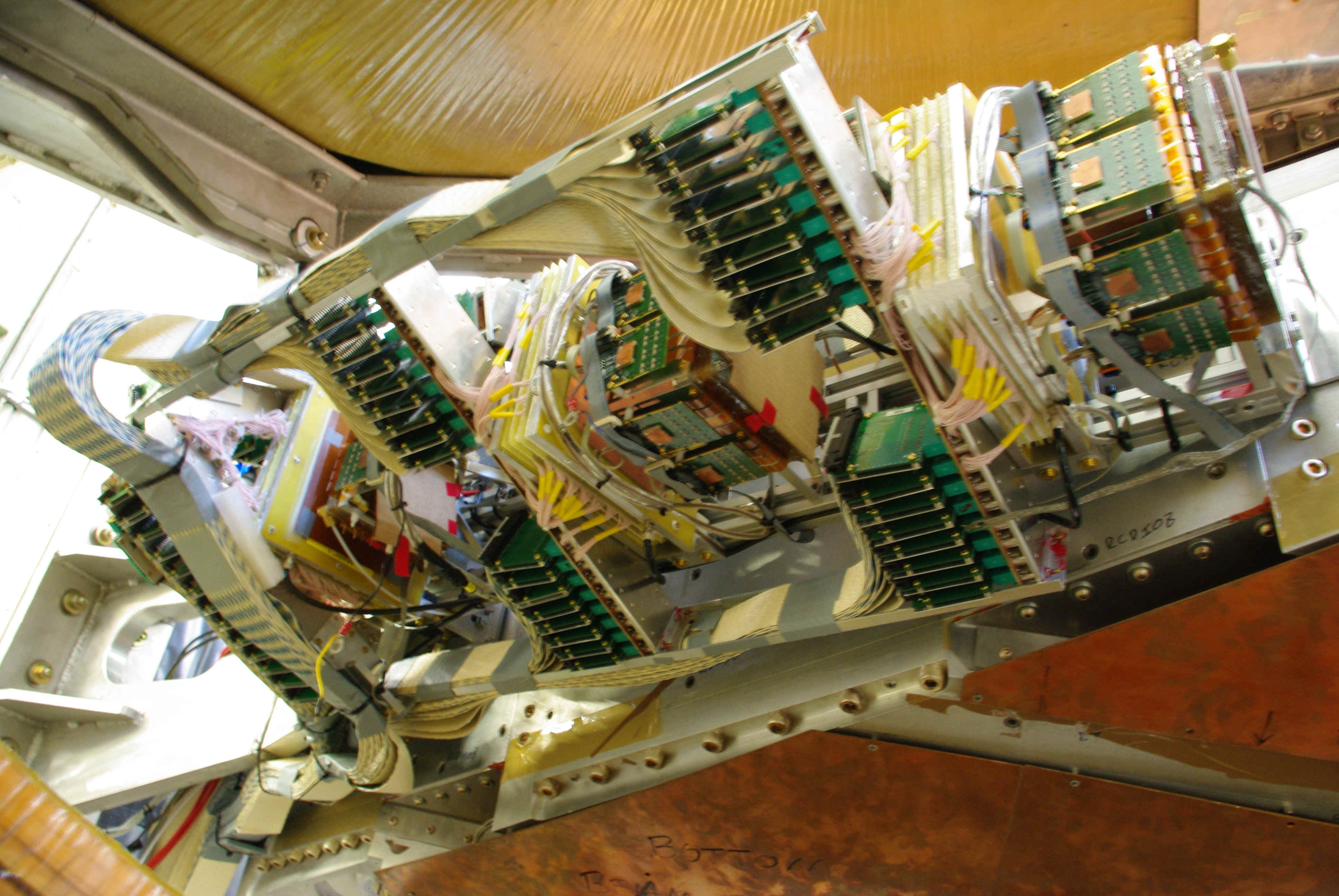}}
    \caption[Photograph of the right-side 12\dg tracking telescope]{Photograph of the right-side 12\dg tracking telescope, showing the detector elements
    and their readout electronics \cite{Milner:2014}.}
    \label{fig:12photo}
    \end{figure}

    Two telescopes were constructed as units and then mounted to the forward side of each of the drift chambers.  Each system consisted of three gas electron
    multiplier (GEM) detectors interleaved with three multi-wire proportional chambers (MWPCs) for tracking, with two scintillator tiles instrumented with
    silicon photomultipliers (SiPMs), and a Pb glass calorimeter array for triggering.  The essential layout of these systems is shown in Figure
    \ref{fig:fordet}, while Figure \ref{fig:12photo} shows a photograph of one of the telescopes with its mounting brackets and readout electronics.
    Six planes and two detector technologies were utilized to provide high resolution
    tracking of the high momentum forward events, and to provide a mechanism of cross-checks and calibration within the detector system.  Ultimately,
    the GEM detectors were not used in the final luminosity analysis due to problems with the stability of their efficiencies (see Section
    \ref{sec:hahahaha}), but were useful in the calibration of the system and are being considered for use in future experiments \cite{refId0,Balewski:2014pxa}.
    
    This section describes the essential details of these detector systems, while the luminosity analysis of the elastic \pmp scattering events
    recorded by the 12\dg telescopes in conjunction with back-angle portions
    of the drift chamber tracking system is presented in Section \ref{sec:12lumi}.  The value of \ratio at $\sim$12\dg using alternate
    luminosity measurements is considered in Section \ref{sec:12TPE}.
    
    \subsection{Silicon Photomultiplier (SiPM) Scintillator Tiles}
    \label{sec:sipm}
    
    The main trigger signals for the 12\dg telescopes were provided by two $120\:\text{mm}\times120\:\text{mm}\times4\:\text{mm}$
    scintillator tiles in each telescope, consisting of solid blocks of Eljen\footnote{Eljen Technology, Sweetwater, TX, USA} EJ-204 plastic
    scintillator.  Charged particles passing through the planes induced scintillation light, which was detected by the instrumentation of the planes to produce
    a recorded signal.  As shown in Figure \ref{fig:fordet}, the tiles were placed at the target ends of the telescopes
    and between the last two tracking planes in each telescope so that they would not be the limiting elements in the acceptance of
    the detectors.  Each tile was instrumented with two Hamamatsu\footnote{Hamamatsu Photonics K.K., Hamamatsu, Japan} silicon photomultiplier (SiPM)
    multi-pixel photon counters (MPPCs) mounted on opposite sides of the square tiles to homogenize the total recorded light yield across the horizontal
    acceptance of the tiles.  Additionally, each tile was wrapped in Millipore\footnote{EMD Millipore, Billerica, MA, USA} Immobilon-P diffuse reflector
    to further boost the light yield.  Analog signals from the MPPCs in each tile were summed and passed to a constant fraction discriminator
    to produce the effective fast trigger signal from each tile.  Basic parameters of the SiPM tiles and their operation are summarized in Table
    \ref{tab:sipm}.
    
        The operation of the 12\dg system trigger is described in Section \ref{ss:12dtrig} and the performance of the scintillator tiles
    is analyzed in Section \ref{sec:12perf}, but in general the tiles provided a very high efficiency ($>$99\%), uniform trigger for the telescopes
    throughout data-taking.
    
  \begin{table}[thb!]
  \begin{center}
  \begin{tabular}{l|l}
  Parameter & Value \\
  \hline
  \hline 
  Scintillator type & Eljen EJ-204 \\
  Size & $120\:\text{mm}\times120\:\text{mm}\times4\:\text{mm}$ \\
  %\hline
  Reflective coating & Millipore Immobilon-P \\
  %\hline
  SiPM type & Hamamatsu S10931-050P (3600 pixels) \\
  %\hline
  Typical gain & $7.5\cdot 10^5$ \\
  %\hline
  Typical bias voltage & 72 V \\
  %\hline
  Typical dark count rate & $\sim$0.8 MHz \\
  %\hline
  Preamplification & 25x \\
  %\hline

  \end{tabular}
  
  \end{center}
  \caption[Parameters of the SiPM tiles and their operation]{A summary of essential parameters of the SiPM tiles used in the 12\dg telescope
  trigger and their operation \cite{dief1}.}
  \label{tab:sipm}
  \end{table}
    
    \subsection{Lead Glass Calorimeters}
    \label{sec:lg}
    
\begin{figure}[thb!]
    \centerline{\includegraphics[width=1.0\textwidth]{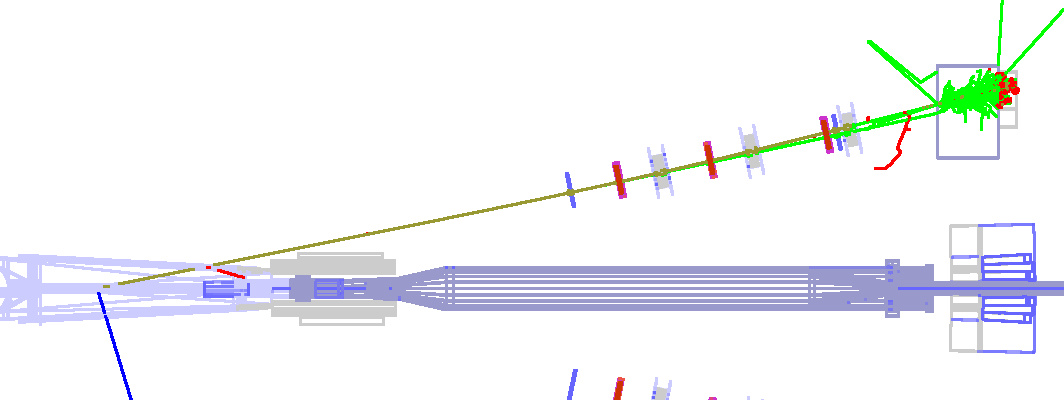}}
    \caption[Simulated lead glass trigger event]{Overhead view of a simulated $e^+p$ event in the left 12\dg telescope,
	      showing the showering in the steel beam behind the telescope
             (mass of photons in green) and resulting energy deposits from electrons in the penetrating shower (red).}
    \label{fig:lgevent}
    \end{figure}

    Lead glass calorimeters were mounted to the steel beams of the detector frame behind each 12\dg telescope to
    provide an independent trigger for the system for calibration and performance evaluation of the SiPM scintillator
    tile trigger system.
    Each calorimeter consisted of three rectangular prisms of lead glass instrumented with photomultiplier tubes at one
    end.  Particles passing through the 12\dg telescopes would shower in the steel beam beyond the telescope, and the calorimeters
    recorded the remnants of this shower exiting the beam as shown in Figure \ref{fig:lgevent}.  The signal from
    this system was relatively noisy, with a number of events that did not correspond to trackable 12\dg events, due to the
    conditions in the detector hall, but the sample was sufficient to evaluate the performance of the SiPM tile trigger, as
    described in Section \ref{sec:12perf}.
    
    \subsection{Gas Electron Multiplier (GEM) Detectors}
    \label{sec:gemdet}
    
\begin{figure}[thb!]
    \centerline{\includegraphics[width=0.75\textwidth]{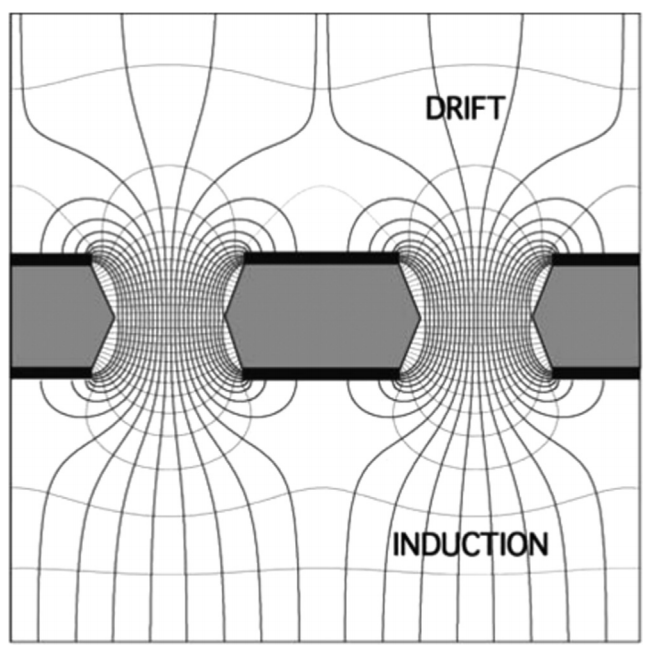}}
    \caption[Electric field in the vicinity of a GEM foil]{Electric field in the vicinity of holes of a GEM foil.
    Note the high field gradients in the immediate vicinity of the small holes.  (Figure reproduced from \cite{Sauli20162}.)}
    \label{fig:gemfield}
    \end{figure}

    Three of the tracking elements in each telescope were triple gas electron multiplier (GEM) detectors \cite{Sauli20162}.  A relatively
    new detector technology, a GEM detector amplifies the initial ionization of gas by a charged particle by drifting the ionization electrons
    towards small holes ($\sim$\SI{100}{\micro\meter} diameter) in a copper-coated Kapton\footnote{E. I. du Pont de Nemours and Company, Wilmington, DE, USA} foil,
    which is held at a fixed potential to attract the drifting electrons \cite{grupen}.  The high field gradients in the vicinity of the holes,
    shown in Figure \ref{fig:gemfield}, induce an avalanche of secondary electrons to multiply the initial ionization signal by factors on the order of 1000.
    Several such foils may be stacked together with order millimeter spacing and stepped potentials to further multiply the signal while avoiding high potential
    gradients across the whole detector which could cause unwanted discharge.  The final multiplied electron signal is typically collected on a readout
    plane beyond the final multiplication foil, which is arranged in a pixel or strip layout to provide precise hit position information by
    leveraging the high statistics of the position distribution of the arriving electrons.
    
\begin{figure}[thb!]
    \centerline{\includegraphics[width=0.95\textwidth]{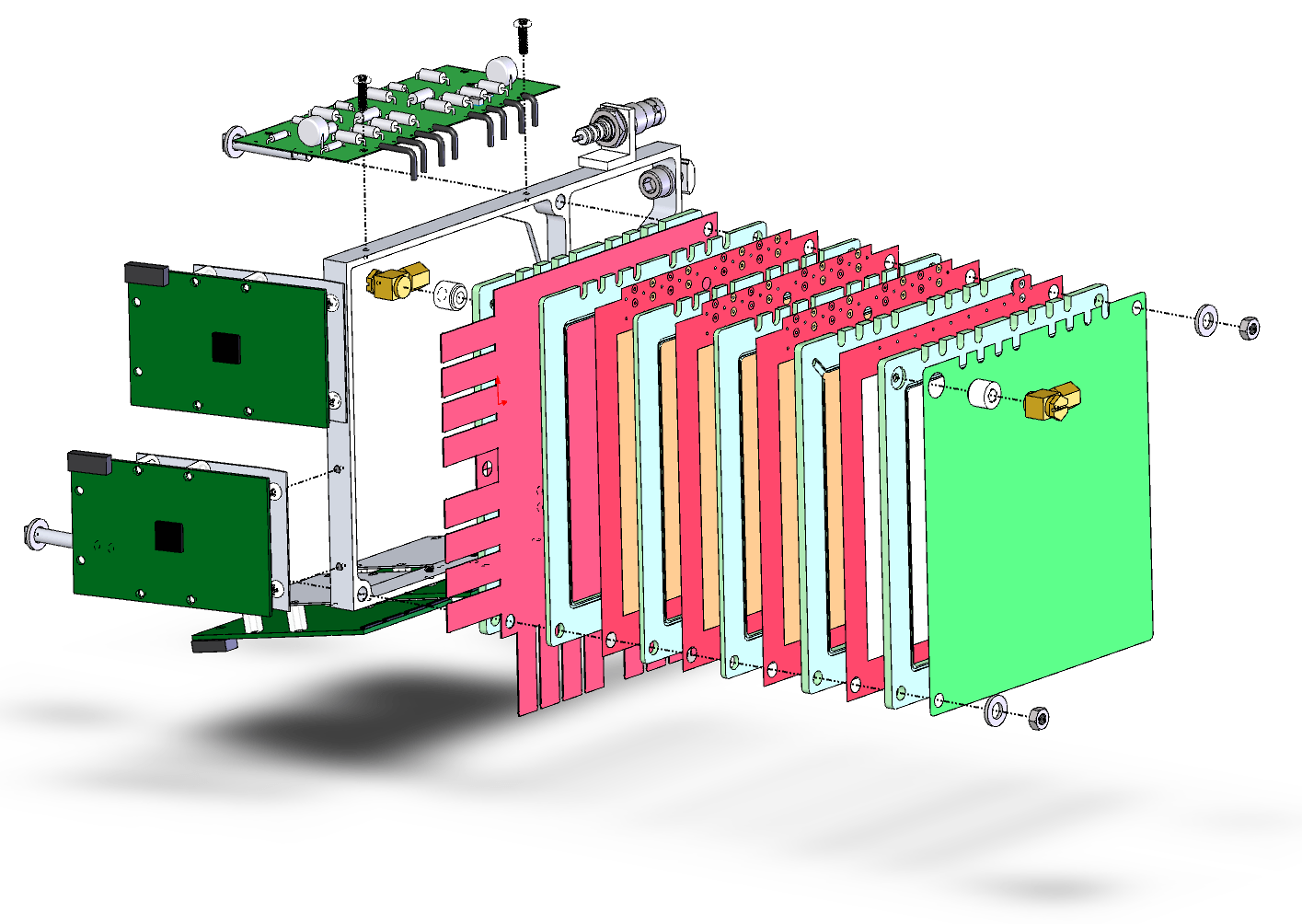}}
    \caption[Exploded view showing the components of an OLYMPUS GEM detector]{ Exploded view schematic of one of the OLYMPUS triple-GEM detectors,
    showing the component layers from the readout electronics and foil on the left, the three GEM foils with their support frames in the middle, and the
    cathode foil and pressure containment layer on the right \cite{Milner:2014}.}
    \label{fig:gemex}
    \end{figure}

\begin{figure}[thb!]
    \centerline{\includegraphics[width=0.8\textwidth]{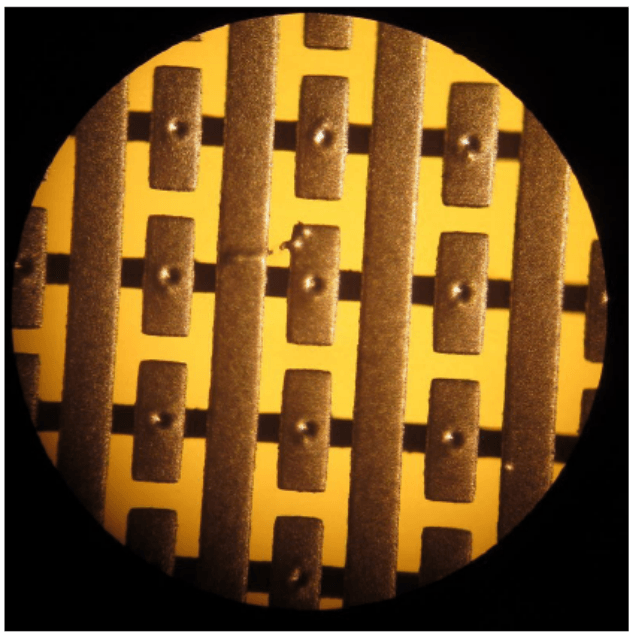}}
    \caption[Micrograph of the GEM detector readout plane pattern]{ Micrograph of the GEM readout plane layout, showing the perpendicular strip/pad
    pattern with the pads connected by vias beneath the strips.  Note the small defect near the center, a repaired short between a pad and
    strip that was repaired by breaking the connection \cite{kohl1}.}
    \label{fig:gemro}
    \end{figure}

    The OLYMPUS GEM detectors were designed at the MIT-Bates Linear Accelerator Center and assembled at Hampton University.  Each detector unit
    consisted of three GEM foils (with the amplification holes) in between a cathode and readout foils.  The total active area of each detector
    was $100\:\text{mm}\times100\:\text{mm}$.  The readout board and GEM foils were spaced 2 mm from one another, with a 3 mm gap between
    the cathode and the first multiplication foil, the arrangement of which is shown in Figure \ref{fig:gemex}.  The five foils were enclosed in a pressurized volume filled with a premixed Ar:CO$_2$ 70:30
    gas mixture such that none of the foils were subjected to a pressure gradient that could cause deformation.  The GEM foils, manufactured by
    TechEtch\footnote{TechEtch, Inc., Plymouth, MA, USA}, consisted of \SI{50}{\micro\meter}-thick Kapton coated with \SI{5}{\micro\meter} of
    copper on each side.  The holes in the foil were \SI{70}{\micro\meter} in diameter, arranged in a triangular pattern over the entirety of the 
    foil with \SI{140}{\micro\meter} spacing.  The readout boards (also from TechEtch) consisted of strips and pads of copper arranged to provide
    two dimensional information from charge collection on a Kapton substrate.  The pattern of strips and pads on a readout foil is shown in Figure
    \ref{fig:gemro}.  The pitch of pads/strips along each direction was \SI{400}{\micro\meter}
    and the relative sizes of the strips and pads was optimized to balance the charge sharing between the strips and pads.  Each readout board
    had 250 strip/pad channels along each of the two plane dimensions.
    
    The signals from the readout pads and strips of a single plane were routed to four readout cards (two per dimension), each with 128 channels (leaving 12 total
    unused channels on each GEM detector).  Each card included an APV25-S1 analog pipeline chip \cite{French:2001xb}, which sampled each of the 128 channels
    at a 40 MHz rate.  When triggered, the pipeline was read out in a single multiplexed data line to the data acquisition system managed by a multi-purpose
    digitizer (MPD) \cite{Musico:2011lia}.  The digitizer included VME-based fast ADCs and a field-programmable gate array (FPGA) for the creation
    and readout of the GEM's data signal.
    
    The analysis of the GEM detector data, including the development of a new dedicated hit-finding algorithm for the detectors, is discussed in
    Section \ref{sec:12lumi}.  While the GEM detectors were not used in the final analyses of this work (as discussed in Section \ref{sec:hahahaha}),
    an effort was undertaken to understand their output for the purposes of using GEM data for calibration when possible and to facilitate their
    possible use in future experiments.
    
    \subsection{Multi-Wire Proportional Chambers (MWPCs)}
    \label{ss:mwpcdet}
    
    The remaining three tracking planes in each telescope consisted of multi-wire proportional chambers (MWPCs), constructed at the Petersburg Nuclear Physics Institute (PNPI).  Each detector
    consisted of three planes of anode sense wires, each with a cathode wire plane on each side at 2.5 mm spacing.  Similar to the drift chambers in principle,
    particles passing through the MWPC planes
    induced ionization in the gas filling the detectors.  This created electrical signals on the sense wires as the electrons and ions drifted towards the wires, which
    were held at high voltage.  For the MWPCs, however, no timing or other signal information was maintained and thus only hit/no hit information
    was available from each wire.    The sense wires in each of the three
    anode planes were \SI{25}{\micro\meter} in diameter, consisted of gold-plated tungsten, and were spaced by 1 mm along the plane. The cathode wires 
    were \SI{90}{\micro\meter}-diameter beryllium bronze wires and were spaced by 0.5 mm.  The middle anode wire plane was oriented so that the wires
    were vertical, while the wires in the other two planes were oriented at $\pm30^\circ$ from vertical to provide two-dimensional reconstruction information,
    while prioritizing the resolution in the horizontal position as it was most important to momentum reconstruction.  The wires were connected to a CROS-3
    readout system, detailed in Reference \cite{Uvarov:cros3}, that interfaced with the main OLYMPUS DAQ.

    The anode planes were contained within a pressure volume filled with an Ar:CO$_2$:CF$_4$ 65:30:5 mixture to provide the medium for ionization.
    This gas mixture and the voltage anode plane operation voltage (3200 V) were chosen by consideration of simulations of the detectors
    using the GARFIELD gas ionization and drift software package \cite{Veenhof:1998tt} and experience operating similar detectors at the HERMES
    experiment \cite{Andreev:2001kr}.
    
    Hit information from the MWPCs consisted of ``yes/no''-type decisions from the wires in each plane, which limited their resolution but greatly
    simplified their operation and readout.  Combined information from all three planes provide better hit resolution in the horizontal direction and strong
    noise rejection.  The performance of the MWPCs and analysis of their data, including their critical role in the determination of the 12\dg luminosity estimate,
    are discussed in detail in Section \ref{sec:12lumi}.
    
\section{Symmetric M{\o}ller/Bhabha Scattering Luminosity Monitor}

As a second source of luminosity monitoring, OLYMPUS included a forward ($\theta\approx 1.3^\circ$) calorimeter system designed for the detection of
symmetric M{\o}ller/Bhabha (SYMB) scattering events. The detector was designed to register scattering events involving
the atomic electrons of the hydrogen gas in the target and beam leptons in which
the outgoing particles scatter at symmetric angles from the direction of the beam.
Three principal processes contribute to this measurement between the two beam species:
\begin{enumerate}
 \item elastic $e^-e^-\rightarrow e^-e^-$ scattering (M{\o}ller scattering) \cite{moller},
 \item elastic $e^+e^-\rightarrow e^+e^-$ scattering (Bhabha scattering) \cite{bhabha}, and
 \item $e^+e^-\rightarrow \gamma\gamma$ annihilation.
\end{enumerate}
Additionally, and critically to the final method of luminosity determination from this system as described in Section \ref{sec:symblumi},
the detector also registered the very far forward leptons from \pmp events.  As particles entered the SYMB detector, they induced \v{C}erenkov radiation,
which was detected by the photomultipliers that instrumented the system.  At such small $\theta$, the cross sections for this processes
are very large (on the order of a barn/sr), and thus the SYMB calorimeter was subject to very high rates.  As such, the detector operated at
a much higher rate than the rest of the detector, internally counting events using a digital histogramming system to count events in which
energy was registered in one or both sides of the calorimeter.  The nature of these histograms and the conditions under which they are filled
is described in Section \ref{sec:symbro}.  This section covers the essential details of the SYMB and its operation,
while Reference \cite{PerezBenito20166} provides a full description of the system. The analysis of the SYMB data and the resulting
luminosity estimate is summarized, although not detailed, in Section \ref{sec:symblumi}.

\begin{figure}[thb!]
  \centerline{\includegraphics[width=1.0\textwidth]{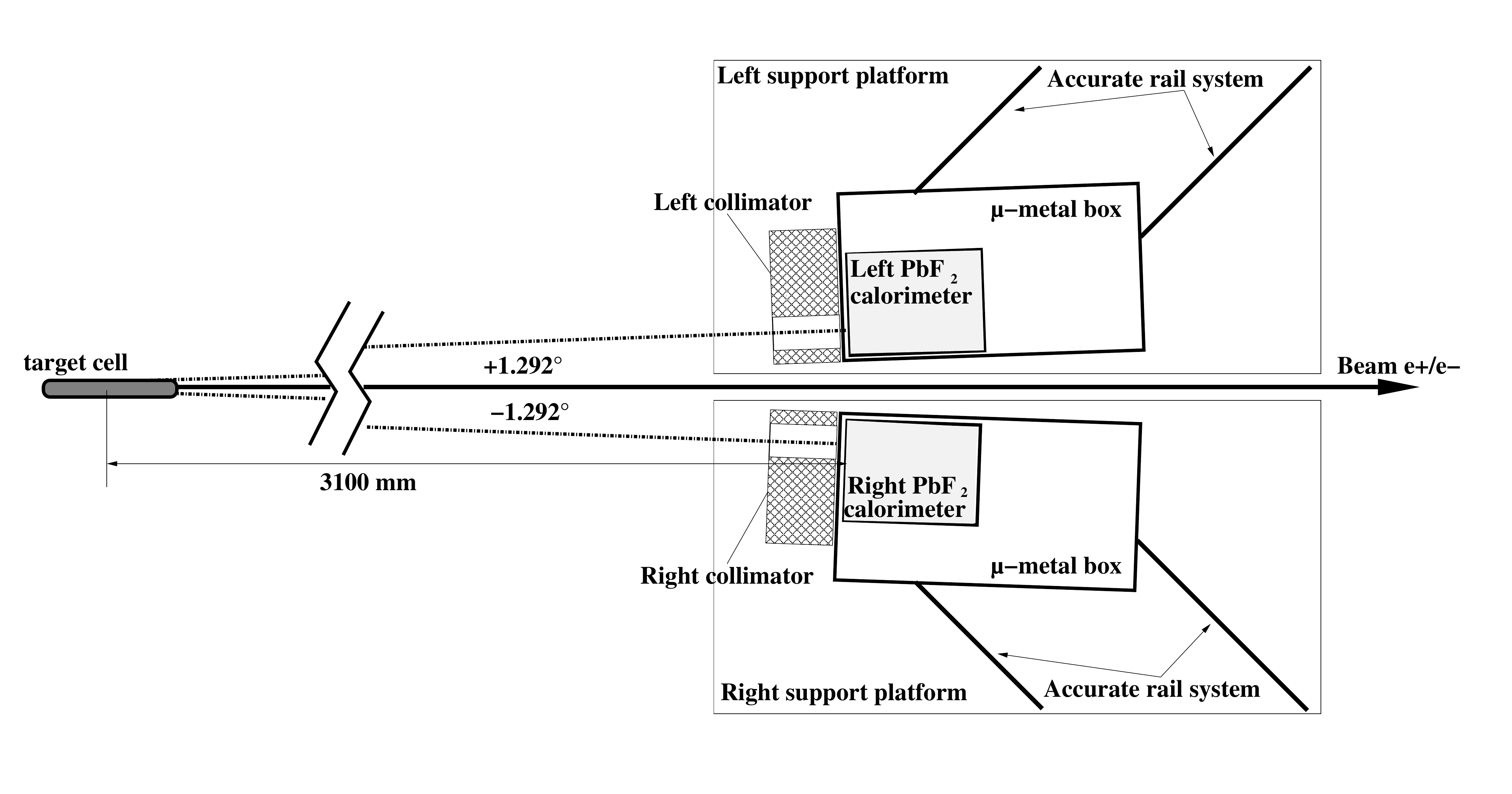}}
  \caption[Schematic of the SYMB calorimeter system]{Schematic overhead view of the SYMB detector system, showing the approximate placement
  of the collimators, crystal arrays, and shielding boxes of the detector.  Note that the horizontal scale of the figure is broken in the 
  beamline region to illustrate the placement of the detector relative to the target \cite{Milner:2014}.}
  \label{fig:symbsc}
  \end{figure}

\begin{figure}[thb!]
\includegraphics[width=0.49\textwidth]{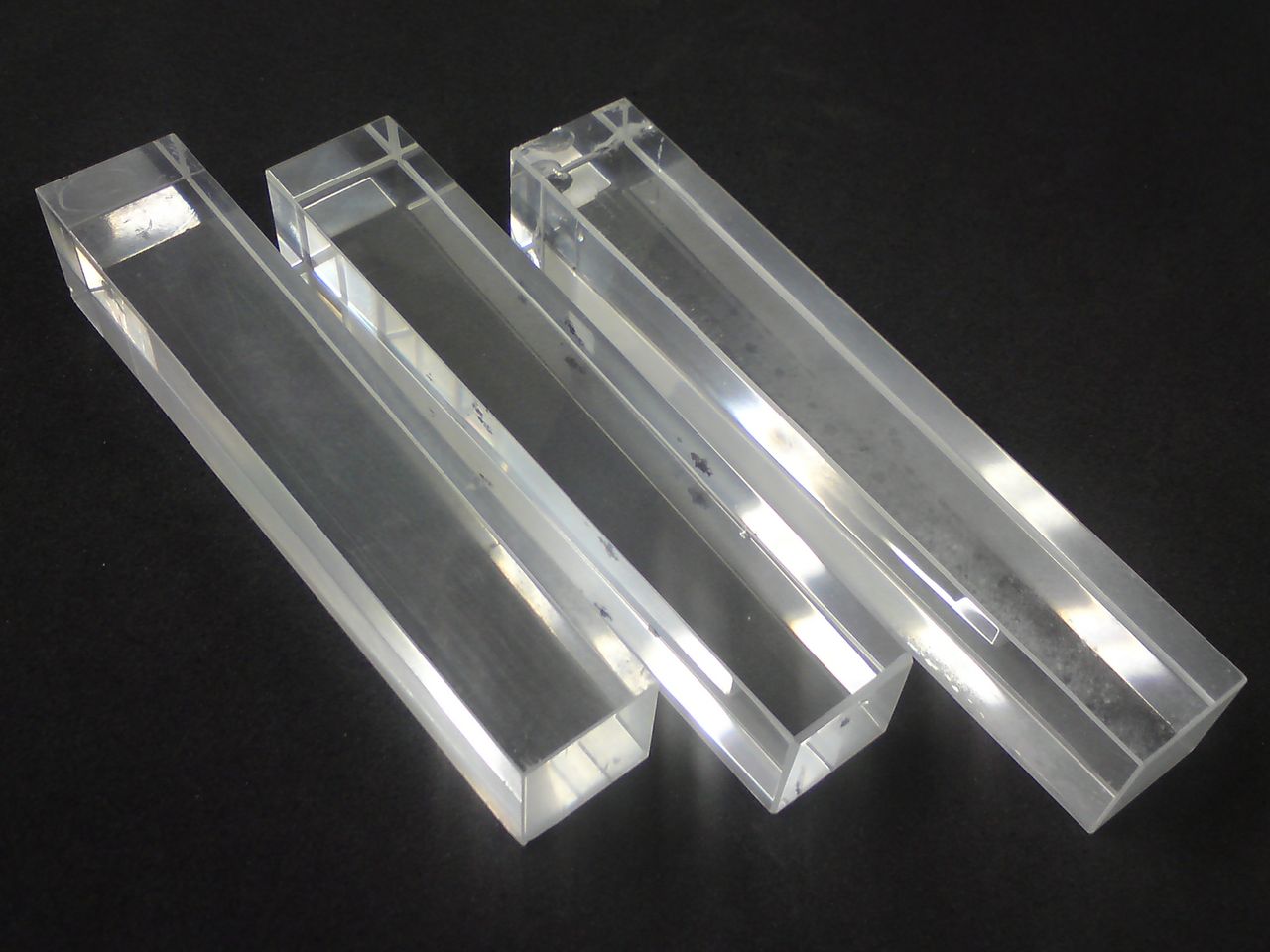}
\includegraphics[width=0.49\textwidth]{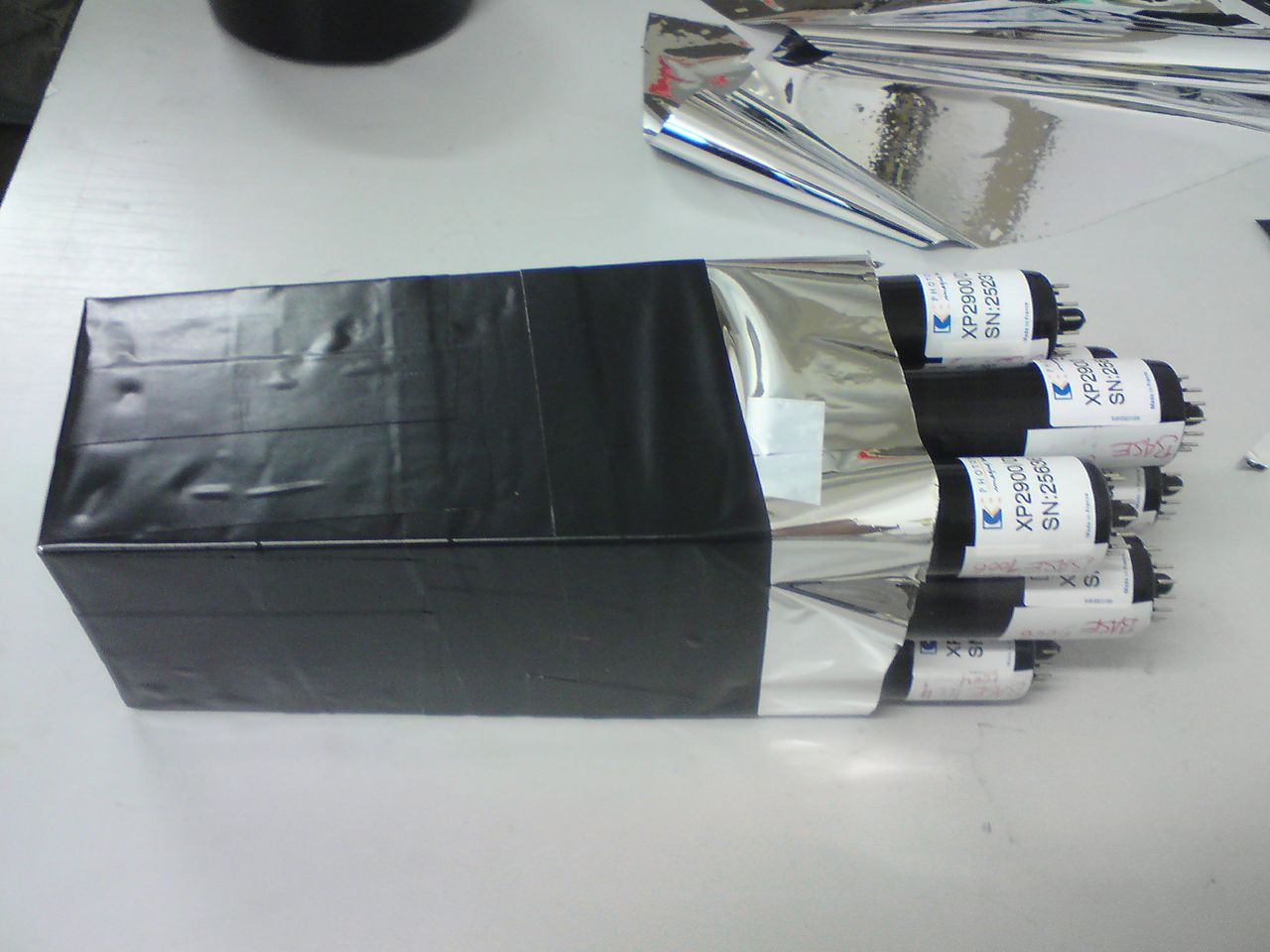}
  \caption[Photographs of the SYMB calorimeter crystals and assembly]{Three of the PbF$_2$ crystals of the SYMB detector (left) and one of the nine-crystal calorimeter
  assemblies instrumented with PMTs (right)  \cite{Milner:2014}.}
  \label{fig:crys}
  \end{figure}

Figure \ref{fig:fordet} shows the placement of the SYMB system near the beamline and approximately three meters from the target chamber,
while Figure \ref{fig:symbsc} shows a more detailed schematic of the components of the system.  The detectors were constructed at
Johannes Gutenberg-Universit\"{a}t using expertise and materials from the A4 experiment \cite{KOBIS1998625,A4M}.  Two calorimeters were
constructed, each consisting of a $3\times 3$ array of PbF$_2$ crystals.  Each crystal was approximately $26\:\text{mm}\times26\:\text{mm}\times160\:\text{mm}$
in size.  In the $3\times 3$ array, the long dimension of the crystals was placed parallel to the direction of incoming particles and corresponded to
approximately 15 radiation lengths.  The 78 mm square cross section of the array contained approximately two Moli\`{e}re radii \cite{Baunack:2011pb}.
Examples of the crystals and one of the SYMB nine-crystal assemblies are shown in Figure \ref{fig:crys}.

Since PbF$_2$ is a pure \v{C}erenkov radiator, the response time
of the crystals is extremely fast (as there is no scintillation delay), making it an ideal material for a high-rate electromagnetic calorimeter
\cite{ANDERSON1990385}.   Millipore reflective paper was wrapped around each individual crystal to increase the contained light yield.
Each of the 18 crystals was instrumented with a
Philips\footnote{Koninklije Philips N.V., Amsterdam, the Netherlands} XP 29000/01 PMT to collect light produced in the crystals, chosen for their
fast (20 ns) response time.  This response time was the limiting factor in the SYMB operation rate, and thus the detector could operate at up
to 50 MHz.  Each PMT was connected to a dedicated ADC channel to measure the effective energy deposited for an event.
Lead collimators, 100 mm thick with 20.5 mm diameter holes, were placed in front of each array to shield the crystals and
define the acceptance for symmetric events.  Since the collimator holes defined the acceptance of the calorimeters, they were carefully
surveyed.

\section{Slow Control System}
\label{sec:sc}
    
  Elements of the experiment slower than the event trigger rate were controlled by a slow control system that monitored, set, and recorded several
  parameters at all times during the operation of the experiment.  These parameters included voltages set via communication with high voltage supplies,
  temperature readings, the target gas supply system settings, DORIS beam parameters, and a variety of other settings.  The software linking the slow
  control system to the hardware elements of the experiment was implemented using the Experimental Physics and Industrial Control System (EPICS) \cite{epics}.
  In addition to the EPICS system, a server was implemented for the system in Python using the FLASK micro-framework\footnote{\url{http://flask.pocoo.org/}}.
  Finally a JavaScript- and HTML-based graphical interface for this server provided the ability to control the system from any computer with secure access
  to the internal network at DESY (and appropriate logins for the OLYMPUS system).  The software was run on three dedicated Linux VME computers, with appropriate interface
  cards for the various hardware elements of the experiment.  The user interface included the ability to change settings such as voltages,
  gas flow, vacuum controls, etc., real-time plots and warnings regarding the operational conditions of the detector and beam,
  as well as a variety of presets for all parameters appropriate for different running conditions.  This made the experiment
  easy to monitor and operate, not only from the control room at DORIS (which was staffed at all times during the experiment), but from anywhere with Internet connectivity.
  Additionally, the slow control continuously sampled readout information from various systems and recorded these in a PostgreSQL database, which was properly synced
  with the data acquisition system to associate the history of the readout with the recorded detector data.
    
\section{The OLYMPUS Trigger and Readout}
\label{sec:trig}

  The relatively high luminosity associated with the fixed-target nature of the OLYMPUS experiment, in conjunction
  with the fact that a complete detector readout required much more time ($\sim$0.1 ms) than the beam bunch spacing, required
  the experiment to employ a trigger system to select events of interest for readout from the fast detector system information of the experiment
  and reject events that would not correspond to elastic \pmp events.
  The primary trigger signals were the discriminated signals from the ToF scintillators and the 12\dg telescope scintillators, although information
  from the drift chambers, lead glass calorimeters, and DORIS accelerator information also contributed to the system.  Notably, the DORIS
  bunch clock provided the reference time signal for the ToF and drift chamber TDCs.  When a trigger condition was met, information from all
  relevant detector channels was read out and the system was gated (i.e., prevented from collecting further events) until the readout was complete.
  The time the trigger was closed for readout amounted to ``dead time'', which was typically $\sim$30\% during normal OLYMPUS
  operation.  Thus, in order to maximize the usefulness of the dataset, the trigger was implemented to increase
  the fraction of recorded events that included reconstructible elastic \pmp events, while simultaneously providing a sufficient number
  of events for various testing and calibration purposes.
  
  In addition to the main trigger, which focused on selecting elastic \pmp events, a number of other triggers provided more open configurations at a prescaled rate.  The trigger was implemented
  using a VME FPGA, which permitted the combination of up to 16 fast input signals to produce up to 16 output trigger conditions
  in parallel that could be passed to the data acquisition system.  Additionally, each of the output conditions could be prescaled to reduce to the rate at which
  an individual condition produced a detector readout below the natural rate of occurrence for that condition.  This was useful for certain test trigger conditions, which
  would have dominated the data sample without prescaling.  The critical elements
  of the trigger system and conditions are described in the following subsections.    

  \subsection{Main Kinematic Trigger and Second-Level Drift Chamber Trigger}
    
\begin{figure}[thb!]
    \centerline{\includegraphics[width=1.5\textwidth]{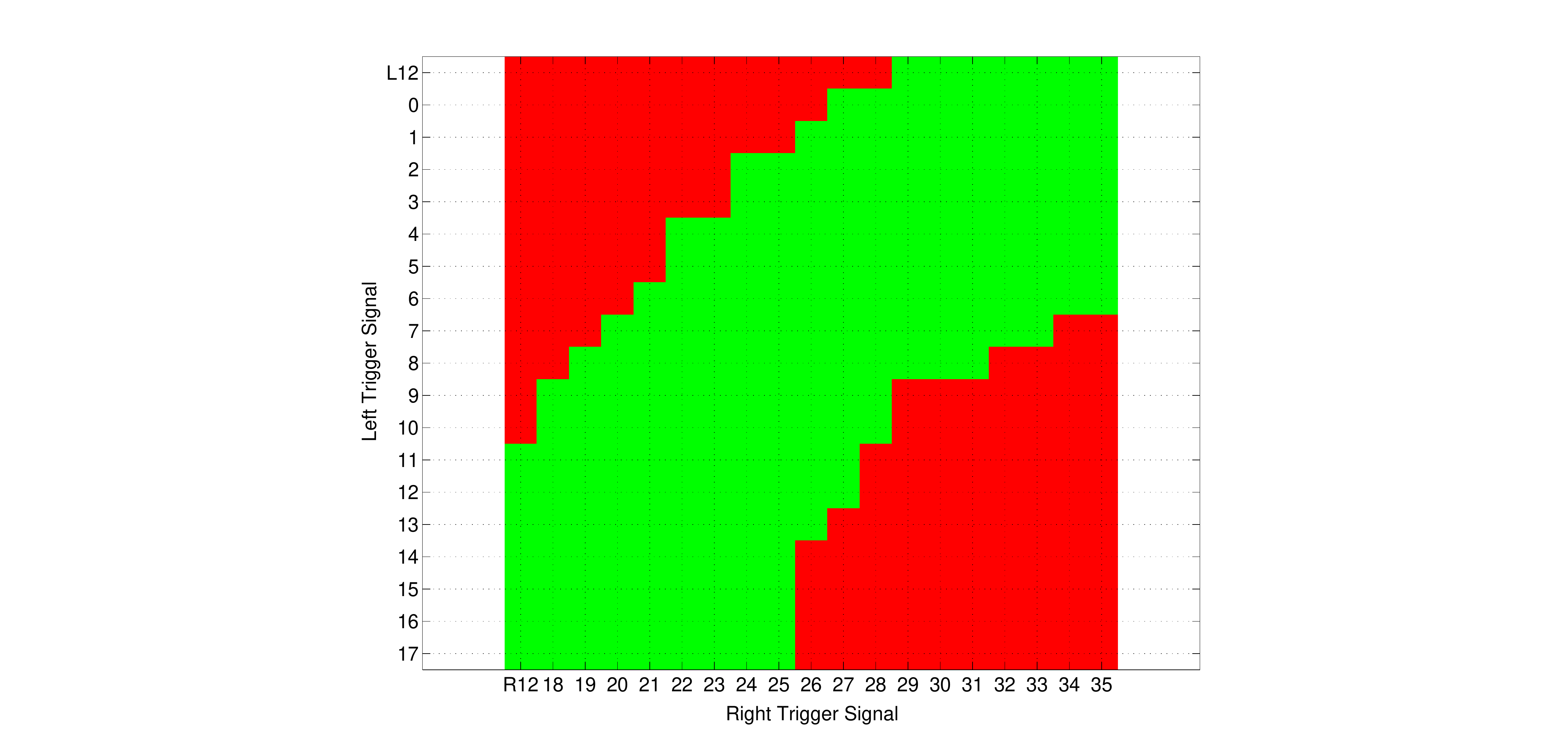}}
    \caption[Map of allowed OLYMPUS trigger conditions]{Map of the corresponding left and right fast trigger signal pairs from the OLYMPUS
    ToF bars (labeled by index) and 12\dg telescope scintillators (labeled L12 and R12, for left and right) that would result in a detector
    readout (green) versus event rejection (red) for the main kinematic and 12\dg luminosity triggers.  This configuration was used for the
    majority of production data-taking runs.}
    \label{fig:trigcon}
    \end{figure}

    The main OLYMPUS trigger combined information from the ToF scintillators and the drift chambers to produce events with an
    enhanced purity of elastic \pmp events.  The basic trigger condition required that a ToF bar on each side of the detector
    recorded a hit (defined as coincidence between the top and bottom PMTs of each bar), such that the left and right bar pairing
    corresponded to a conceivable left/right pairing for the kinematics of elastic scattering from the gas target.  The map of such
    allowed pairings is shown in Figure \ref{fig:trigcon}.  The allowed left/right pairings were determined by Monte Carlo simulation of both
    \ep and \pp events. All pairings that occurred for elastic events in the simulation of either species were allowed for all runs,
    and an additional buffer of one bar on each side of
    the allowed block of bars on the opposite side corresponding to a specific bar was added to avoid rejecting good elastic \pmp events.
    The primary advantage of this system was that it disallowed events that paired very forward bars on each side, since these bars
    had the highest hit rates but kinematically could not correspond to elastic events.  This trigger was not prescaled (i.e., it was allowed to
    induce data readout at its natural rate of occurrence).
  
    Additionally, during Run I it was determined that the restrictions of the kinematic left/right ToF pairings in the trigger still
    resulted in a high number of events with insufficient data in the drift chambers to permit reconstruction.  Thus, for the primary
    production run, a ``second-level'' trigger element was added to the main trigger which required that at least one wire in each of the middle
    and outer drift chamber regions on each side recorded a good time for an event to be accepted.  Since events without such information could
    not be reconstructed anyway, this condition did not reject any events that could be considered part of the final data sample. The condition
    reduced the false trigger rate due to ToF noise, beam hall radiation conditions, etc. by approximately a factor of ten, which greatly
    improved the dead time of the experiment to allow a much higher efficiency of luminosity collection.

  \subsection{$12^\circ$ Telescope Trigger}
  \label{ss:12dtrig}
  
    The second primary element of the trigger system was the dedicated trigger for the 12\dg telescopes, which required the coincidence
    of both SiPM scintillator tiles (Section \ref{sec:sipm}) in one of the 12\dg arms and a ToF hit (top/bottom PMT coincidence) in one of the back seven bars
    on the opposite side.  These conditions are shown as ``L12'' and ``R12'' in Figure \ref{fig:trigcon}.  This trigger was also not prescaled
    so as to produce the highest possible statistics for the 12\dg elastic \pmp luminosity measurement.  The system could also be triggered
    using the lead glass calorimeters mounted behind the telescope, although this trigger was prescaled due to its high rate of undesired
    events and was used predominantly for measurement of the SiPM tile efficiencies (Section \ref{sec:12eff}).
  
  \subsection{SYMB Readout}
  \label{sec:symbro}
  
    The readout of the SYMB calorimeter was handled independently of the main trigger system due to the way in which the SYMB detector produced
    digital histograms of events internally.  More detail on this system may be found in Reference \cite{PerezBenito20166}.  The detector was
    designed to rapidly sum the light yield in the nine crystals on each side of the calorimeter, apply basic conditions to the recorded
    yield in each side, and then record the event in a digital histogram if all conditions were met.  Each digital histogram was
    essentially a 2D histogram of the energy recorded by each side of the detector system (represented by the total ADC count).  
    The readout system was based on the similar
    system used for the A4 experiment, where the crystals used in the SYMB detector were originally designed and used \cite{KOBIS1998625,A4M}.
    The detector could operate at a rate up to 50 MHz, which was predominantly limited by the 20 ns response time of the PMTs used
    to instrument the PbF$_2$ crystals.  Two basic conditions were applied to the SYMB PMT signals to generate a readout in the digital histograms:
    \begin{enumerate}
     \item that the total energy deposited in the nine crystals of one of the calorimeters exceeded a given threshold, and
     \item that the central crystal of the array meeting the threshold requirement have the largest recorded energy deposition
           of the nine (the ``local max'' requirement).
    \end{enumerate}
    If either side of the detector met these requirements, both sides were read out to fill either the left- or right-master histogram
    depending on which side met the conditions.  If both sides simultaneously met the conditions, the coincidence histogram was filled.
    These histograms were filled at a rate much higher than the main trigger rate, and thus the data acquisition system did not record
    the histograms on an event-by-event basis.  Instead, the integrated histograms were read out approximately every 70,000 events (or
    at the end of a run) to limit the readout time spent on SYMB data acquisition.
  
  \subsection{Additional Triggers}
  \label{ss:addtrig}
  
    Other triggers used during the experiment, typically designed to provide information for tests, calibration, etc. for the detectors,
    were prescaled to reduce their rate of occurrence in the dataset or were used for occasional dedicated tests runs.  Such triggers included:
    \begin{enumerate}
     \item conditions involving looser requirements on combinations of ToF hits, including allowing hits with only one PMT firing in a bar (either
           with or without a coincident hit in the opposite side of the detector), useful for calibration of the ToFs,
     \item triggers allowing the readout of the 12\dg telescopes with only registered SiPM or lead glass hits (i.e., with no requirement of
           a ToF bar),
     \item a ``clock'' trigger for random readout of the detector, and
     \item dedicated triggers for recording of cosmic ray events or other special event topologies.
    \end{enumerate}
    Such triggers will be described as necessary in the description of any calibrations for which they were used.  In general, these
    special triggers were employed with varying prescale factors throughout the experiment, but the main triggers (the kinematic and
    SiPM tile 12\dg trigger) were run without prescaling for all data production runs.
    
  \subsection{Readout}
  
  The readout system was implemented by the Universit\"{a}t Bonn group, based on experience developing similar systems for operation
  of the ELSA accelerator and its experiments \cite{Hillert2006,bonndaq}.  The system ran on VME CPU modules, and operated as a ``synchronous system'';
  when a trigger occurred all detectors were read out simultaneously and no further triggers were accepted until the readout was complete.
  While such a system increased the dead-time of the experiment (the fraction of the running time in which the trigger was gated and events were
  not accepted), the advantage of avoiding possible synchronization of data errors was judged to be of sufficient benefit to the experiment to justify
  the increased dead-time.  The readout systems for the detectors were arranged in a master-slave architecture, in which a master module managed a number
  of slave modules responsible for interaction with particular systems and handled the gating of the trigger until all slave modules completed their readout.
  The data acquisition system collected all relevant detector readout data for a given event (ADC counts, TDC counts, slow control parameters, etc.), and produced
  an output ZEBRA format file \cite{zebra}.  This ZEBRA file was then converted to a ROOT tree format to facilitate easier access to the data \cite{Brun1997}.
  More information on this system can be found in References \cite{Milner:2014} and \cite{bonndaq}.
    
  \section{Experiment Operation}

\begin{figure}[thb!]
    \centerline{\includegraphics[width=1.0\textwidth]{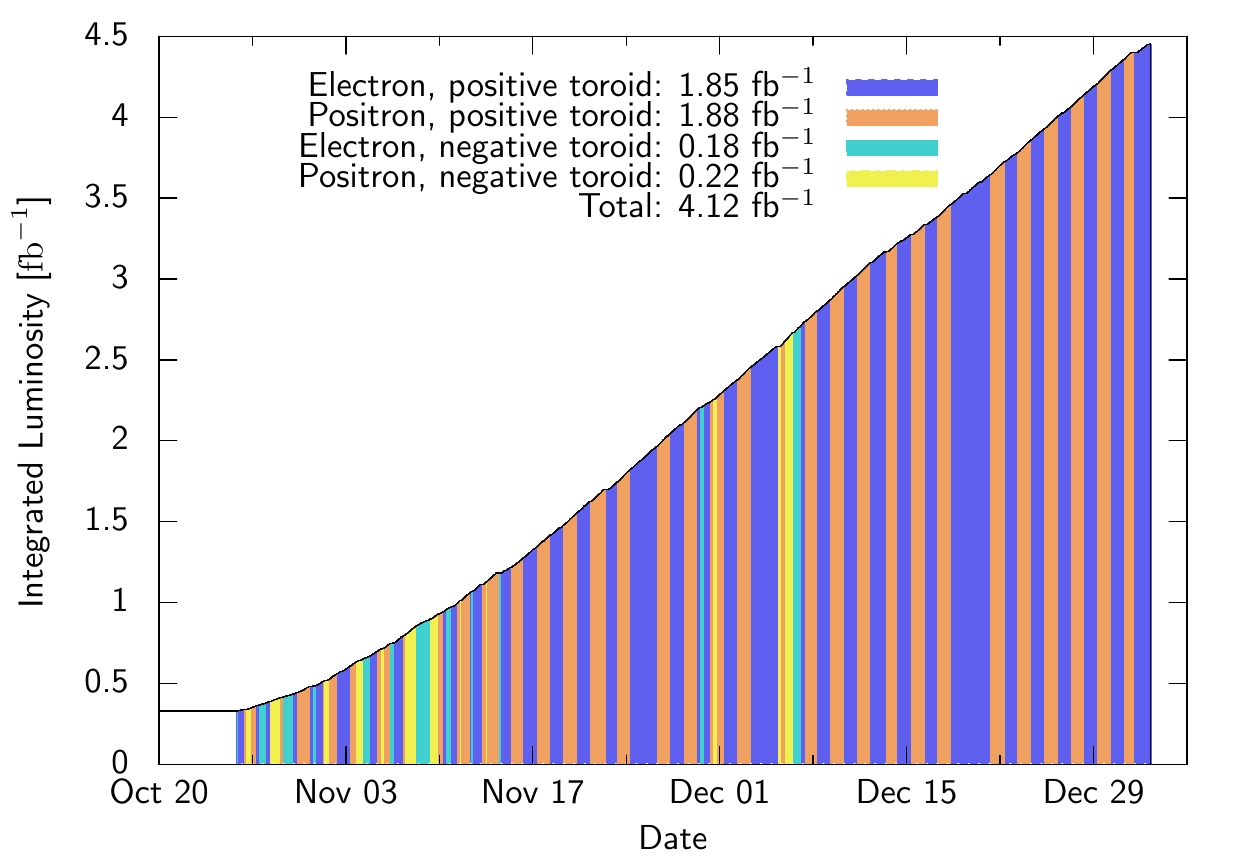}}
    \caption[Approximate integrated luminosity collected by OLYMPUS during Run II]{The integrated luminosity collected by the experiment, as measured by the slow control
    (Section \ref{sec:sclumi}), during the second OLYMPUS run, separated by the various beam/toroid configurations \cite{Milner:2014}.}
    \label{fig:lot}
    \end{figure}

  During data-taking, a two or three person team operated the experiment 24 hours a day from the control room in the DORIS ring building, which was shared
  with the DESY accelerator operators to facilitate communication between the OLYMPUS and accelerator operators.  The OLYMPUS shift crew was responsible for operating
  the slow control system and monitoring its various readouts, as well as continuously monitoring the data using low-level analysis that was conducted as data were taken.
  A dedicated run plan dictated the configuration of the beam, detector, etc., to maintain stable running during the main production periods.  The experiment
  collected approximately 4.5 fb$^{-1}$ of data during Run II (which comprises the data set for this work), of which approximately 3.1 fb$^{-1}$ is considered
  excellent data in which the configuration and condition of the detector were optimal.  It is this latter dataset that is used in this work for the analyses presented, although
  at a later date an effort may be made to include portions of the data not used.  The collection of luminosity over time for Run II is shown in
  Figure \ref{fig:lot}.
  
%% file: chap4.tex
% Chapter 4
%
% Section discussing the overall analysis strategy, since it so uniquely
% depends on MC and a description of that MC
%

\chapter{Analysis Strategy, Detector Calibration, and Monte Carlo Simulation}
\label{Chap4}

In the original design of the OLYMPUS experiment \cite{tdr}, the plan for the operation and analysis of the experiment involved regularly switching the
polarity of the toroidal magnet and computing $R_{2\gamma}$ as a super-ratio of the experimentally measured elastic event counts $N_{e^\pm p,\pm B}$ in the four combinations of
species and toroid polarity, normalized to the luminosity collected for each orientation:
\begin{equation}
 R_{2\gamma} = \frac{\sigma_{e^+p}}{\sigma_{e^-p}} \approx \sqrt{ \frac{N_{e^+p,+B}N_{e^+p,-B}}{N_{e^-p,+B}N_{e^-p,-B}}\cdot \frac{\mathcal{L}_{e^-p,+B}\mathcal{L}_{e^-p,-B}}{\mathcal{L}_{e^+p,+B}\mathcal{L}_{e^+p,-B}}}.
\label{eq:idrat}
\end{equation}
Measuring $R_{2\gamma}$ using the method of Equation \ref{eq:idrat} confers several advantages, primarily relating to the cancellation of systematic effects due to the differences
in the relative acceptance of \pp and \ep events events due to physical detector bounds and detector efficiency.  This cancellation occurs due to the fact that an \pp event of given kinematics traversing the
detector in positive field polarity take the same path through the detector as an \ep event with the same kinematics in negative polarity (and vice versa), while two such events
in traversing the detector in the same field polarity take slightly different paths through the system due to their opposite bending, and thus are subject to different acceptances,
detection efficiencies, etc.  Since the acceptance of the OLYMPUS detector for exclusive \pmp events is primarily dominated by the acceptance of the lepton, the slight difference
in proton acceptances for \ep and \pp events of the same vertex kinematics that does not cancel between the field polarity configurations was expected to be negligible.

This approach, while ideal in principle, suffers from a number of complications that make it impractical:
\begin{enumerate}
 \item In the negative field polarity M{\o}ller scattering electrons are swept into the first layers of the drift chamber tracking system, causing a saturation of
       certain regions of the detector and spoiling the cancellation of tracking efficiency between the two polarities, as described in Section \ref{sec:tormag}.
 \item The reversal of the magnetic field reverses the direction of the $\mathbf{E}\times\mathbf{B}$ ionization electron drift in the drift chambers, significantly changing the calibration of recorded drift
	time-to-distance (TTD) from the wire at which a particle passed.  This required change in detector calibration between polarities again spoils the ideal cancellation of efficiencies.
	See Figure \ref{fig:cells} for an illustration of the effect on the drift lines	and Section \ref{sec:ttd} for a description of the TTD calibration.
 \item The negligibility of the different acceptance for protons from \pp and \ep events of the same kinematics in opposite toroid polarities depends on having uniform efficiency for
       track reconstruction throughout the acceptance.  As will be discussed in Section \ref{sec:specperf}, this was not the case for OLYMPUS (and would likely be an impractical requirement due to the small likelihood
       of individual channels, voltage supplies, etc. in a 954-channel system performing identically).
\end{enumerate}
For these reasons, it was determined that OLYMPUS could achieve better systematic uncertainties by taking data with a single toroid polarity setting, and making a detailed effort
to properly account for the acceptance differences of \ep and \pp events in simulation.

Conducting a measurement of $R_{2\gamma}$ based on Equation \ref{eq:rat} demands careful consideration of all aspects of the detector system and the analysis that affect the relative
acceptance of \ep and \pp events and proper implementation of such effects in the Monte Carlo simulation.  This chapter describes the essential strategy used in the OLYMPUS
analysis to achieve this goal, including essential information on detector survey and calibration, a brief
introduction to event reconstruction in the main spectrometer, and a description of the advanced Monte Carlo simulation employed in the analysis.  Chapters \ref{Chap5} and \ref{Chap6}
provide more detailed information on the specifics of the luminosity and main cross section ratio analyses, including the measurement and simulation implementation of effects such as detector
efficiencies, resolutions, etc.\ for individual detector systems.

\section{Calibration of the Spectrometer Position and Magnetic Field}
\label{sec:speccal}

  Due to the fact that OLYMPUS operated using a single field polarity, knowledge of the positions of the detectors, their solid angle coverage, and the magnetic field
  throughout the detector volumes was crucial to properly determining the acceptance of the OLYMPUS detector for \pp and \ep events.  The acceptance of the detector was
  determined by careful survey of both the detector positions and the magnetic field, and great care was taken to properly represent the results of these measurements in
  the representations of the detector system used for the Monte Carlo simulation and particle track reconstruction.  This section describes the essential methods of the position
  and field surveys and their implementations in the OLYMPUS analysis.

  \subsection{Detector Position Survey and Modeling}
  \label{sec:detmod}
  
  The primary optical surveys of the detectors, target chamber, and beamline elements were conducted by the DESY survey and alignment group (MEA2) \cite{mea2}.  The surveys were conducted
  using a laser tracker system, which provided precise measurement of a polar and azimuthal angle relative to the mounted position of the tracker and a laser-ranged distance from the tracker
  to reflective prism targets placed on all elements of the detector, beamline, support frames, detector hall walls and floors, etc.  The position of the laser tracker relative
  to the center of the OLYMPUS coordinate system (Section \ref{sec:conv}) could be reconstructed via measurements of known points throughout the detector hall, which were conducted regularly
  throughout the survey and anytime the laser tracker was moved.  This allowed reconstruction of any measured point in the OLYMPUS coordinate system from the information provided by 
  the laser tracker.  All detector elements, the target chamber, and beamline elements were surveyed at least twice: once in late 2011 prior to Run I and again after Run II in Spring 2013.
  Certain individual detector elements were surveyed more frequently in the interim to check for shifting positions.
  
  Physical objects that were part of the detector, target system, beamline, etc. were modeled for the analysis using Geometry Description Markup Language (GDML), an XML-based language
  for the description of geometries that is compatible with both the ROOT and GEANT4 frameworks \cite{gdml,Brun1997,Agostinelli:2002hh}.  This allowed the use of a single geometry model
  in all OLYMPUS analysis applications (simulation, track reconstruction, visualization, etc.), greatly reducing the possibility of geometric errors in the analysis. The solid models of experiment elements
  in the geometry model were constructed on the basis of both survey data and original design specifications for the elements.  The placement and orientations of the elements in the
  geometry were determined from the survey data.  The survey data were converted to coordinates in the OLYMPUS global coordinate system, and a global fit was performed for the corresponding
  points on the modeled detector elements to determine their positions and rotations.  Redundancy and overdetermined placements in the survey dataset allowed for the identification and removal
  of inconsistencies and erroneous survey data.  The final survey fit provided accurate placement of objects in the solid model to better than \SI{100}{\micro\meter} for most elements, although
  some elements with less-determined survey data (such as the GEM planes) had slightly higher uncertainties \cite{bernauer3}.  Complete details on the implementation of the OLYMPUS solid
  model may be found in Reference \cite{oconnor}.
  
  \subsection{Beam Position}
  
  At all times during OLYMPUS data-taking, the position of the DORIS beam was monitored by two beam position monitors (BPMs), placed slightly upstream and downstream of
  the target chamber along the beamline.  This provided a measurement of the central position of the DORIS beam at two points near the OLYMPUS target with an uncertainty
  of $\sim$\SI{100}{\micro\meter}.  The high degree of precision was obtained by conducting a detailed survey of the positions of the BPMs, as described in the
  previous section, and then performing a series of calibration measurements after they had been removed from the beamline \cite{bpm}.  The calibration was conducted by 
  mounting the BPMs with a current carrying wire passed through them which simulated the beam current.  The position of this wire was varied and the BPM readout was matched
  to the surveyed position of the wire.  The wire position was varied well beyond the range of beam positions that occurred during OLYMPUS data-taking, but measurements were
  focused on the regions most relevant to the experimental conditions. The surveyed wire positions were fitted in a similar fashion as the detector survey data so as to produce a mapping between
  the BPM readout data and beam positions in the OLYMPUS global coordinate system.  Additional tests, such as reversing the direction of current in the wire, were conducted so as to simulate the
  oppositely-charged beam species to provide estimates of systematic effects.
  
  \subsection{Magnetic Field Survey and Modeling}
  \label{sec:magsur}
  
  Since electrons and positrons bend in opposite directions in a magnetic field, precise knowledge of the OLYMPUS magnetic field was critical to determining the detector acceptance
  for implementation in the simulation and tracking.  To achieve this, a detailed survey of the OLYMPUS magnetic field was conducted in situ before moving the toroid coils after
  the experiment.  The detector elements (chosen to be non-magnetic and thus neither influence or be influenced by the field) were removed, permitting access to probe all areas
  within the volumes corresponding to the detector acceptance.  This effort is described in detail in References \cite{Bernauer20169} and \cite{schmidt}, but is briefly summarized here.
  
  The measurement was conducted by constructing a system in which a three-dimensional Hall probe was mounted to a system of translation tables and support brackets that allowed movement
  of the probe throughout the volume relevant to OLYMPUS track trajectories, from the target chamber to the locations of the ToF scintillators (the outermost detectors).  Positions were
  systematically scanned in 50 mm steps in the inner tracking region (which affects trajectories most strongly) and 100 mm steps in the outer region in all three spatial dimensions.  Approximately
  36,000 points were surveyed. During this survey, the position of the Hall probe was constantly monitored using a system of theodolites so that its position at each measurement point could be reconstructed.  After the measurements
  were performed, a fitting procedure similar to that used for the detector elements was used to reconstruct the probe positions corresponding to the field measurements.
  
  Since the implementation of the field for tracking and reconstruction requires knowledge of the field at all locations (rather than a grid of points), a model was developed
  to compute the field at any location in the detector system.  This model consisted of approximating the OLYMPUS toroid coils as filaments and computing the field due to the current using
  the Biot-Savart law.  This allowed the computation of the magnetic field at arbitrary location and additionally provided the capability of numerically computing the spatial
  derivatives of the field components.  The placement of the toroid coils in the model was determined by initially allowing their positions to float and fitting their
  positions and rotations so as to best fit the survey data position/field points.  This fit achieved an average residual of 18.7 G, and was in general much better throughout the critical
  tracking volumes.  The coil model, however, was not a good approximation near the coils (where the approximation of modeling the toroid elements as thin filaments broke down) and thus
  the field was interpolated directly from the data in such regions.  
  
  The model calculation, however, was too slow to be used directly for 
  simulating or reconstructing particle trajectories for the OLYMPUS analysis, and so a fast and precise
  interpolation scheme was developed\footnote{The remainder of this section is predominantly reproduced from Reference \cite{Bernauer20169}, which was written by and describes
  the work of the author.}. The coil model was used to pre-calculate the magnetic 
  field vector and its spatial derivatives on a regular $50$~mm~$\times$~$50$~mm~$\times$~$50$~mm grid covering the entire 
  spectrometer volume, so that the field could be interpolated between surrounding grid points.

  The interpolation scheme had to balance several competing goals:
  \begin{itemize}
  \item minimizing the memory needed to store the field grid,
  \item minimizing computation time for field queries, and
  \item faithfully reproducing the coil model in both the field and its derivatives.
  \end{itemize}
  To achieve this, an optimized tricubic spline interpolation scheme was developed based
  on the routine of Lekien and Marsden \cite{NME:NME1296}.  For each point $P$ in the grid,
  24 coefficients were calculated using the coil model (8 per component of the vector magnetic field):
  \begin{multline}
  C_{i,P} = \left\{ B_i, \pd{x}{B_i}, \pd{y}{B_i},\pd{x\partial y}{B_i},\pd{z}{B_i},\pd{x\partial z}{B_i},\pd{y\partial z}{B_i},\pd{x\partial y \partial z}{B_i}\right\}
  \\ \text{for} \:\: i\in \left\{ x,y,z \right\}.
  \end{multline}
  
\begin{figure}[hptb]
  \centerline{\includegraphics[width=0.8\textwidth]{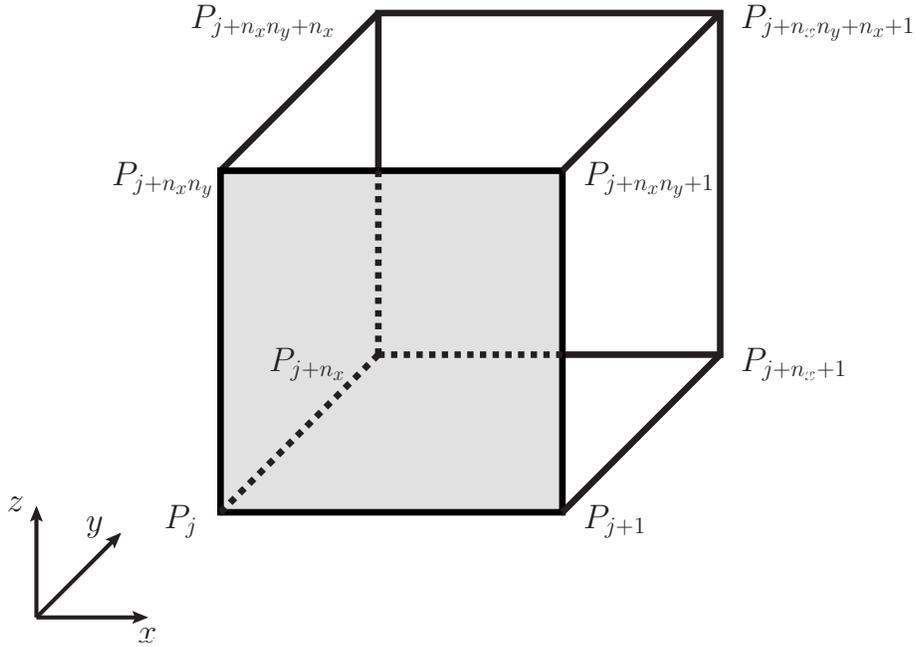}}
  \caption[Grid box indexing scheme for the magnetic field interpolation]{A generalized box in the interpolation grid identified by its lowest-indexed grid point $P_j$, where $n_x$ and 
  $n_y$ are the $x$ and $y$ dimensions of the grid in units of grid points. \label{fig:grid}}
  \end{figure}

  For the interpolation, it is convenient to consider the grid in terms of boxes defined
  by eight grid points, as shown in Figure \ref{fig:grid}, and define box-fractional 
  coordinates $x_f,y_f,z_f \in [0,1] $ parallel to the global axes spanning each box.  
  Each point in the grid is labeled with an integer index $j$, such that stepping from
  point $P_j$ one unit in $x$ reaches point $P_{j+1}$. Stepping one unit in $y$ from
  point $P_j$ reaches $P_{j+n_x}$, where $n_x$ is the size of the grid in the $x$ direction.
  Stepping from point $P_j$ one unit in $z$ reaches point $P_{j+n_xn_y}$, where $n_y$ is the 
  size of the grid in $y$ direction. Then, a local tricubic spline can be defined for each 
  field component in the box:
  \begin{equation}
  B_i(x,y,z) = \sum_{l,m,n=0}^3 a_{i,lmn} x_f^ly_f^mz_f^n \:\:\:\: i\in \left\{ x,y,z \right\},
  \end{equation}
  where the coefficients $\left\{a_{i,lmn}\right\}$ are functions of
  the set of the 64 parameters $\left\{C_{i,P}\right\}$, where $P$ is any of the eight grid points at 
  the vertices of the box.  
  This function is a 64-term polynomial for each box and is $C^1$ at the box boundaries. 
  The coefficients $\left\{a\right\}$ can be computed from the parameters $\left\{C_{i,P}\right\}$ following the 
  prescription in Reference \cite{NME:NME1296}. This prescription, however, requires three $64\times64$ matrix 
  multiplications per box. Once completed for a given grid box, these multiplications can be stored 
  for future use, but this adds to the size of the grid in memory, approaching a factor of 8 for
  large grids.

  To avoid these costs, the spline was refactored so that the parameters $C_{i,P}$ can be used 
  directly as coefficients. The resulting basis functions take the form:
  \begin{gather}
  f_0(x_i) = \left(x_i-1\right)^2 \left(2x_i+1\right) \\
  f_1(x_i) = x_i\left(x_i-1\right)^2 \\
  f_2(x_i) = x_i^2\left(3-2x_i\right) \\
  f_3(x_i) = x_i\left(x_i-1\right)
  \end{gather}
  where $ i\in \left\{ x_f,y_f,z_f \right\}$.  The interpolation then takes the form:
  \begin{equation}
  \label{eq:spline}
  B_i(x,y,z) = \sum_{l,m,n=0}^3 b_{i,lmn} f_l(x_f) f_m(y_f) f_n(z_f) \:\:\:\: i\in \left\{ x,y,z \right\},
  \end{equation}
  where each coefficient $\left\{b_{i,lmn}\right\}$ is one of the parameters $C_{i,P}$. The correspondence
  between $\left\{b_{i,lmn}\right\}$ and $C_{i,P}$ is shown in Table \ref{tab:coef}.

  \begin{table}[hptb]
  \tabcolsep=0.15cm
  \begin{center}
  \begin{tabular}{l | c c c c c c c c}
    &  $B_i$  &  $ \pd{x}{B_i}$  &  $ \pd{y}{B_i}$  &  $\pd{x\partial y}{B_i}$  &  $\pd{z}{B_i}$  &  $\pd{x\partial z}{B_i}$  &  $\pd{y\partial z}{B_i}$  &  $\pd{x\partial y \partial z}{B_i}$ \\
  \hline
  $P_{j}$  &  000  &  100  &  010  &  110  &  001  &  101  &  011  & 111 \\
  $P_{j+1}$  &  200  &  300  &  210  &  310  &  201  &  301  &  211  & 311 \\
  $P_{j+n_x}$  &  020  &  120  &  030  &  130  &  021  &  121  &  031  & 131 \\ 
  $P_{j+n_x+1}$  &  220  &  320  &  230  &  330  &  221  &  321  &  231  & 331 \\
  $P_{j+n_xn_y}$  &  002  &  102  &  012  &  112  &  003  &  103  &  013  & 113 \\ 
  $P_{j+n_xn_y+1}$  &  202  &  302  &  212  &  312  &  203  &  303  &  213  & 313 \\
  $P_{j+n_xn_y+n_x}$  &  022  &  122  &  032  &  132  &  023  &  123  &  033  & 133 \\
  $P_{j+n_xn_y+n_x+1}$  &  222  &  322  &  232  &  332  &  223  &  323  &  233  & 333 \\
  \end{tabular}
  \end{center}
  \caption[Mapping of coefficients for the tricubic spline defined by Equation \ref{eq:spline}]{Mapping of the coefficients $\left\{b_{i,lmn}\right\}$ (defined in Equation
  \ref{eq:spline}) to the field values and
  derivatives at the grid points contained in the box with lowest-indexed point $P_j$.  Entries
  in the table are the values of $lmn$ corresponding to each combination of point and
  coefficient on the interpolation box. \label{tab:coef}}
  \end{table}

  With this interpolation scheme, the procedure for querying the field map consisted
  of determining the grid box containing the queried point, computing the box fractional
  coordinates of the queried point in that box, and then applying the tricubic spline
  interpolation for each of the three field components independently.  Special care was
  taken to optimize the speed and number of arithmetic operations in the routine (e.g., by pre-computing factors
  such as the basis functions that are used repeatedly and by converting division operations to
  multiplications whenever possible). Additionally, the coefficients for each grid
  point were arranged in a single array so that, for any box in the grid, the 192 coefficients
  associated with the 8 points of the box occurred in 16 continuous blocks of 12 coefficients,
  permitting rapid reading of the entire array and facilitating single instruction, multiple data (SIMD)
  computing, further increasing the speed of field queries. This scheme provided a precise and fast field
  implementation for the detector model for both simulation and track reconstruction.

\section{Event Reconstruction}
\label{sec:recon}

  For exclusive elastic \pmp event reconstruction, data were combined from the drift chambers and ToF scintillators, in conjunction with the trajectory bending caused
  by the magnetic field, to provide complete reconstruction of particle trajectories (the event vertex and momentum vector at the vertex).  While track reconstruction
  algorithms for these detectors existed from the BLAST experiment \cite{crawford}, OLYMPUS operated with higher particle energies, different background conditions, different drift gas properties, and
  more demanding precision goals than BLAST. Thus, required new tracking algorithms were required.
  
  In essence, track reconstruction in the OLYMPUS detector amounted to finding the best solution for a particle trajectory given a set of loci that is derived from data
  from the drift chambers and time-of-flight scintillators given the knowledge of the magnetic field throughout the volume of the detector. The locus of points corresponding to a ToF
  hit consisted of a horizontal band of points across a bar with a good top-bottom PMT hit combination, reconstructed from the time difference of the hits in the top and bottom PMTs of the bar.
  These ToF hits were relatively low resolution ($\sigma\approx 10$ cm in the vertical direction and limited to width of the bar in the horizontal direction), but provided valuable information when
  properly weighted in the track reconstruction.  The locus of points from a wire chamber hit consisted of two lines parallel to the wire with a valid time recorded, approximately equidistant from the wire in the
  plane parallel to the wire chamber faces.  The locus consisted of two lines due to ambiguity of a single time in representing a hit that passed either upstream or downstream of the wire.  This
  ambiguity was resolved for each wire hit through a combination of the 0.5 mm stagger between successive wire planes, use of hits in adjacent cells to limit a track location, and global fit information.
  The conversion of the drift time recorded on a wire to the distance of the locus from the wire is in general a complicated problem, and is discussed in the next section.
  
\begin{figure}[thb!]
  \centerline{\includegraphics[width=1.0\textwidth]{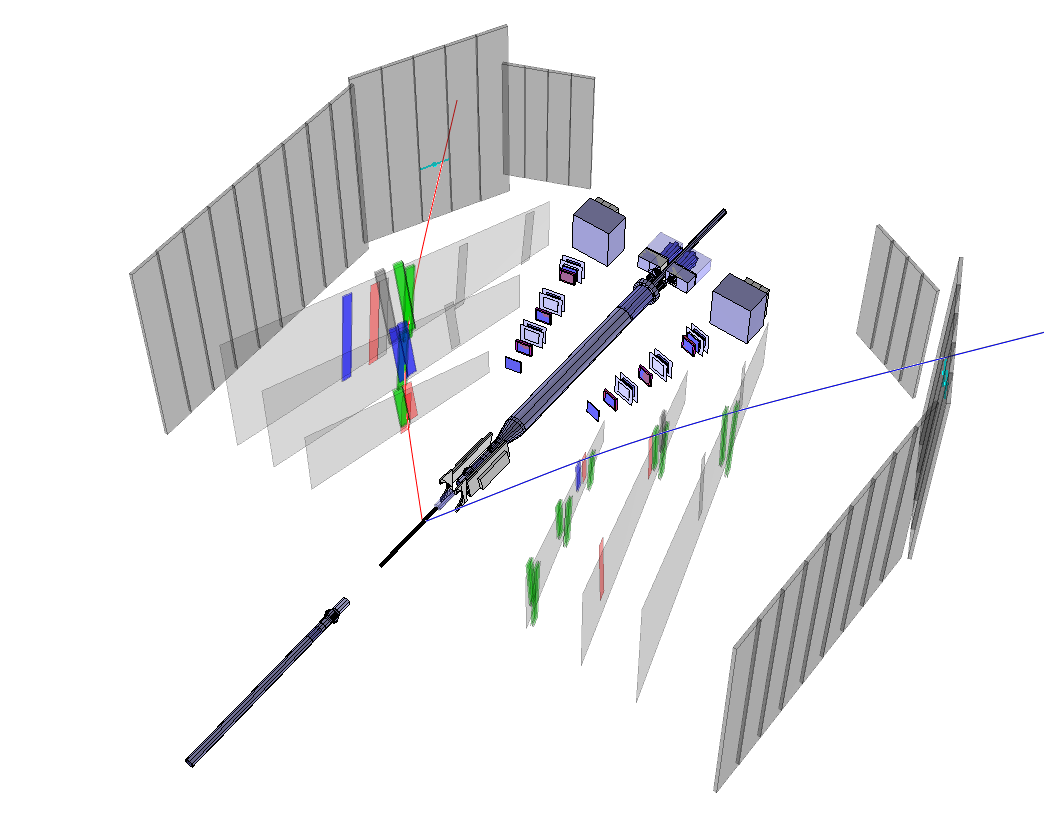}}
  \caption[Event display of a reconstructed \ep event]{A reconstructed elastic \ep event where the electron (red track) was detected in the left drift chamber
  and the proton (blue track) in the right drift chamber.  The cyan lines in the ToF bars represent the horizontal lines corresponding to the loci of points corresponding to the 
  PMT time difference used to estimate the vertical position of the ToF hit.  The color-filled sections of the drift chamber planes indicate wires with valid times for the event, color-coded
  by the number of wires with valid times in a given cell (green indicates all three wires in a cell fired, blue two wires, and red one).  The scattering chamber and toroid coils have been
  removed from the display for clarity.}
  \label{fig:elase}
  \end{figure}

  Ideally, a reconstructed track in the OLYMPUS detector had valid wire times in all 18 wire planes and a good ToF hit fitting the trajectory suggested by those wire hits.  Often a track had
  multiple valid hits in a single wire plane due to the track crossing the boundary between two wire chamber cells and ionizing gas in the active regions of both, which provided a valuable constraint
  on the location of the track by automatically resolving the wire-side decision for that plane.  In practice, inefficiencies in the chambers caused a track to have fewer than 18 hits in the drift
  chambers, but typically a track could be reconstructed with fewer hits as in the example event reconstruction shown in Figure \ref{fig:elase}.
  
  \subsection{Time-to-Distance (TTD) Conversion for the Drift\\Chambers}
  \label{sec:ttd}
  
  The conversion of recorded drift times (effectively the elapsed time between the ionization of the gas in the drift chamber to the signal time on the wire after drift) to the 
  corresponding distance from the wire at which the track passed in the drift chambers (the time-to-distance (TTD))
  was a complicated function of the electric fields generated by the wires, the magnetic
  field in the vicinity of the individual wires, the angle at which the track passed relative to the normal of the relevant wire plane, and individual irregularities of the wires.
  Due to the complexity of the problem, several models were tested for the TTD conversion.  These models varied from models of very few parameters fit only to simulations of the drift
  gas using the Monte Carlo frameworks GARFIELD and MAGBOLTZ \cite{Veenhof:1998tt,Biagi1999234} to spline models of hundreds of parameters fit iteratively to the experiment data.
  Ultimately, it was found that a model between these two extremes yielded the best results, both in terms of final tracking resolutions and ability to properly reconstruct the maximum number
  of tracks.
  
\begin{figure}[thb!]
  \centerline{\includegraphics[width=0.9\textwidth]{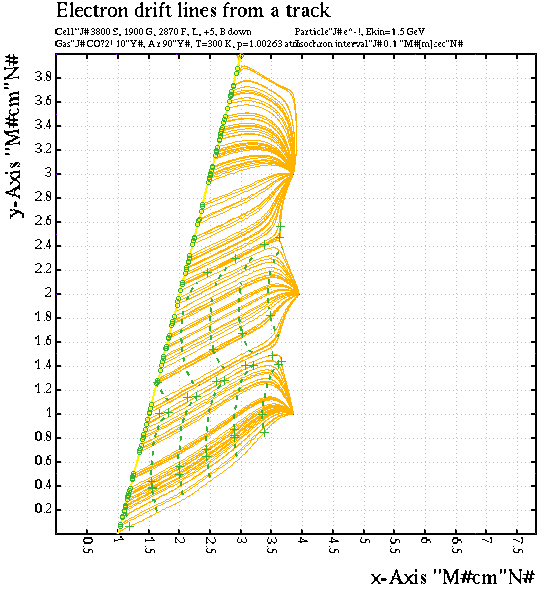}}
  \caption[Simulated lines of electron drift in an OLYMPUS drift chamber cell]{Simulated lines of electron drift (yellow) from points of ionization along a particle trajectory (open green circles)
  to the wires in a simulated OLYMPUS drift chamber cell using GARFIELD for the drift simulation and MAGBOLTZ for the determination of the drift gas properties \cite{Veenhof:1998tt,Biagi1999234}.  The drift
  lines are angled relative to the symmetric axes of the drift cell due to the $\mathbf{E}\times\mathbf{B}$ drift induced by the local magnetic field. The 
  three points at which drift lines converge correspond to the positions of the sense wires.  The green dashed lines represent isochrons, lines of equal drift time to a given wire.}
  \label{fig:drift}
  \end{figure}

  The goal of the TTD function is to convert the time recorded by a wire to the position in the plane of the wire at which the track passed.  Figure \ref{fig:drift} shows an
  example of a simulated track passing through an OLYMPUS drift chamber cell as viewed looking along the wires.  Using the coordinate system specified in the figure, the goal of the TTD
  is to reconstruct the position of the particle trajectory in the $y=1$ cm, $y=2$ cm, and $y=3$ cm planes.  Assuming the ideal condition in which there are no noise times for a wire, the
  earliest time on a wire will correspond to an ionization from the passing track.  Even ignoring the stochastic nature of the points along the trajectory that have ionizations, the earliest
  ionization point with the shortest drift time to the wire is in general not in the wire plane and is a function of the local magnetic field (which gives rise to the $\mathbf{E}\times\mathbf{B}$
  drift in the local $y$ direction) and the angle relative to the local $y$ axis at which the track passes.
  
\begin{figure}[thb!]
  \centerline{\includegraphics[width=1.00\textwidth]{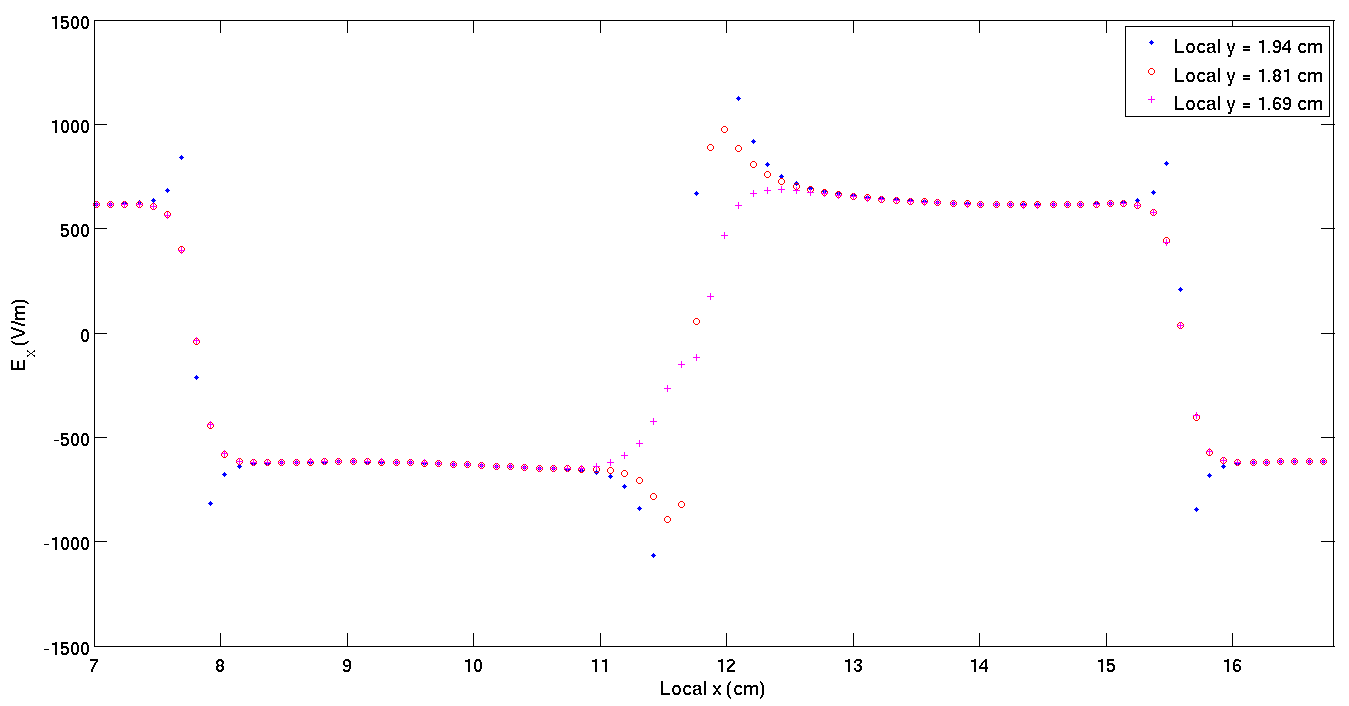}}
  \caption[Calculated electric field strength for a simulated drift chamber cell]{Calculated electric field strength in the local $x$ direction for a middle wire in a simulated OLYMPUS drift cell.  The wire
  in question is at $x\approx12$ cm and the correspond ground wires are are $x\approx8$ and $x\approx16$ cm.  For the bulk of the cell the electric field is roughly constant, but changes rapidly in the 
  vicinity of the sense wires and ground planes.}
  \label{fig:cellEx}
  \end{figure}

  The basic behavior of drifting electrons in an OLYMPUS drift cell may be understood by considering the electric field the electrons experience as they traverse the cell, as shown in Figure
  \ref{fig:cellEx}.  For the bulk of the cell, the arrangement of wires provides an approximately uniform drift field of $\sim$600 V/cm, which in conjunction with the resistance to drift of the gas, gives rise
  to an approximately constant drift velocity for the electrons.  In this region, the TTD function is approximately linear for all other conditions such as magnetic field strength and track incidence
  angle held fixed.  The TTD function must model different behaviors near the sense and ground wires.  Near the sense wire, the drift field rapidly increases, causing the drift electrons to accelerate
  towards the wire.  This effectively compacts the drift distances from the region near the wire into a shorter range of times than in the linear region.
  Near the ground planes, the relatively weak field causes electrons in this region to slowly accelerate, spreading many similar distances from the wire to a wide
  range of drift times.
  
  The final TTD model used in the OLYMPUS analysis accounted for these effects by modeling distance from the wire as a function of recorded drift time as a cubic function near
  the wire, a linear function in the bulk of the cell, and steeper linear plateau near the ground planes.  Additionally, this function was adjusted by the trigonometric factors
  introduced by the angle of the Lorentz drift and the angle of the track relative to the wire plane.  The function was given additional freedom along the length of each wire
  by making all parameters of the function polynomials in the global $\phi$ of the track.  More details on the derivation and exact forms of these functions may be found in
  Reference \cite{schmidt}.
  
\begin{figure}[thb!]
  \centerline{\includegraphics[width=1.00\textwidth]{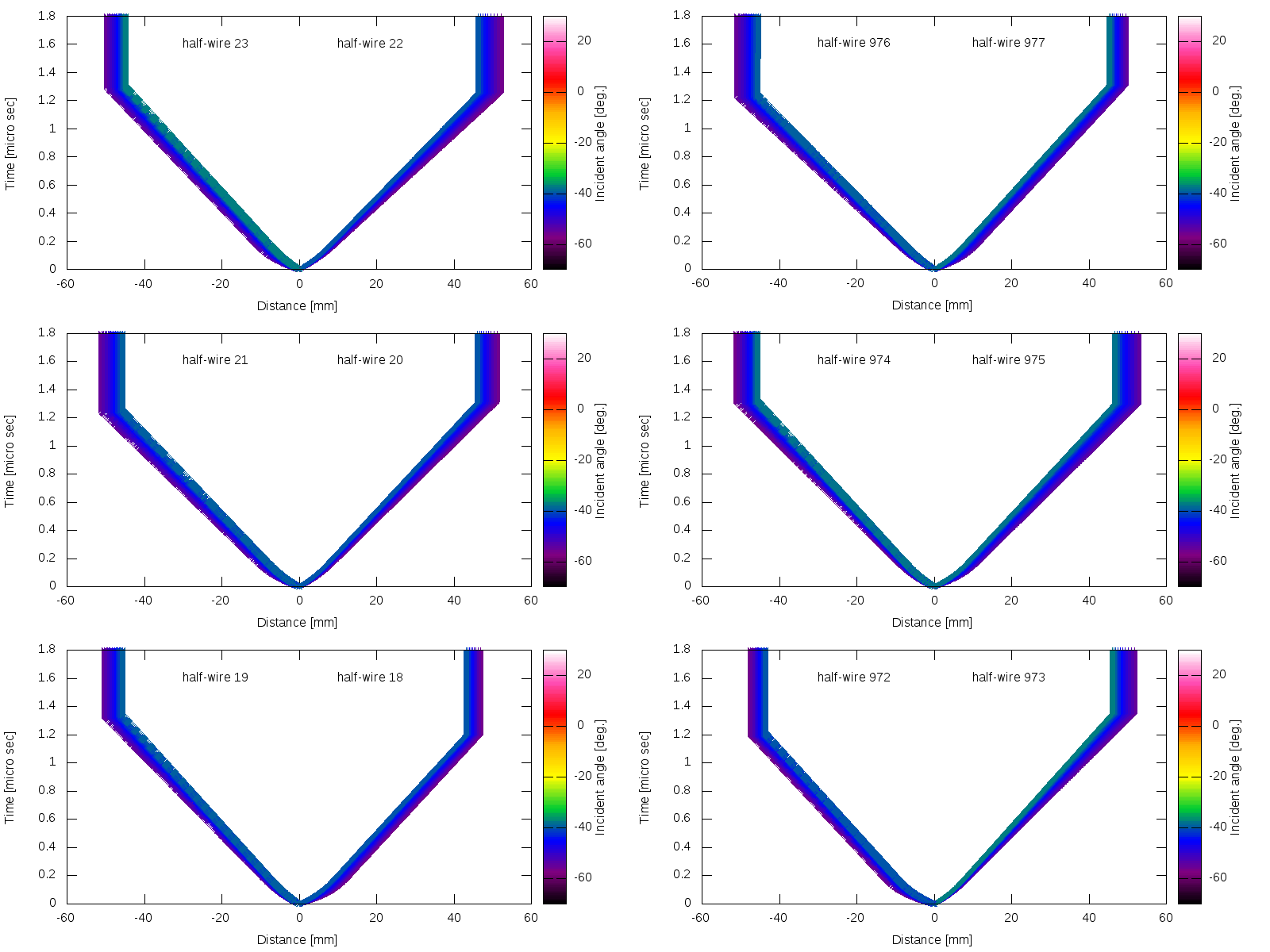}}
  \caption[Example fitted TTD functions for two drift chamber cells]{Example TTD fit results for two cells that were near the downstream ends of the drift chambers in the same positions
  left and right of the beamline for fixed track $\phi$.  Note the spread in the TTD over several millimeters caused by the angle of incidence of the track.  For this case, the functions in the
  corresponding left and right cells are extremely similar, as would be expected in the absence of any imperfections in wire voltages, placement, etc.}
  \label{fig:ttdex}
  \end{figure}

  Rather than attempt to determine the parameters of the functions in each region based on the ideal physics of the drift cell (i.e., the Lorentz angle, the drift velocities,
  the radius of the cubic acceleration region, the position of the start of the plateau region, etc.), the parameters were iteratively fit to data tracks.
  This allowed considerably more freedom in accounting for imperfections in cell voltages, wire time offsets, and the effects of the magnetic field in three dimensions.  This also
  allowed for better handling of the fact that, as discussed in Section \ref{sec:thegdwcs}, the fraction of ethanol in the drift gas fluctuated and thus caused changes in the TTD relation.
  To account for this, the data runs were examined by hand and grouped according to the width of their drift time distributions.  A fit of the TTD model to a GARFIELD simulation was used as a seed for the
  first iteration of tracking the data.  Each individual wire was fit to two functions for the TTD on the upstream and downstream sides of the wire for each group to account for imperfections
  in individual wires and cells.  This seed was constructed for the TTD group with the distribution width closest to average width among the groups and then the iterated solution
  for that group was used as the seed for the next nearest groups in widths and those groups were iterated to produce a solution, which was then used to seed the next group, and so on.  Each group was
  iterated (tracked using the results of the previous TTD fit and then refitted to the TTD model) at least twice, after which it was found that the
  fit residuals did not typically continue to improve.  Typically, average residuals between the TTD function and the reconstructed track distances were on the order of a few tenths of a millimeter.
  Examples of the resulting TTD functions are shown in Figure \ref{fig:ttdex}.
  
  \subsection{Track Reconstruction}
  \label{sec:track}
  
  With the TTD functions established, wire hit loci could be passed to a tracking algorithm for the full reconstruction of particle trajectories.  This process is covered
  in considerably more detail in References \cite{schmidt} and \cite{russell}, but is briefly described here for completeness.  Multiple tracking algorithms were developed
  for OLYMPUS, in part as a control on systematic uncertainties due to tracking efficiency and in part due to the challenging nature of OLYMPUS tracking (caused by the high
  rates of noise hits in the inner portions of the drift chambers, the ambiguity of the upstream/downstream decision for hits on an individual wire, etc.).
  
\begin{figure}[thb!]
  \centerline{\includegraphics[width=0.7\textwidth]{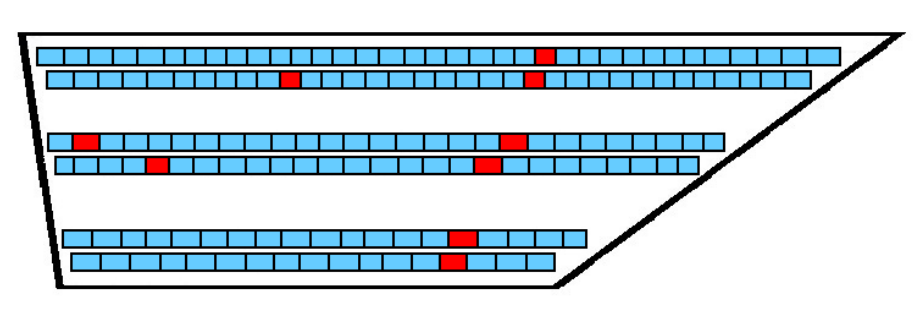}}
  \centerline{\includegraphics[width=0.7\textwidth]{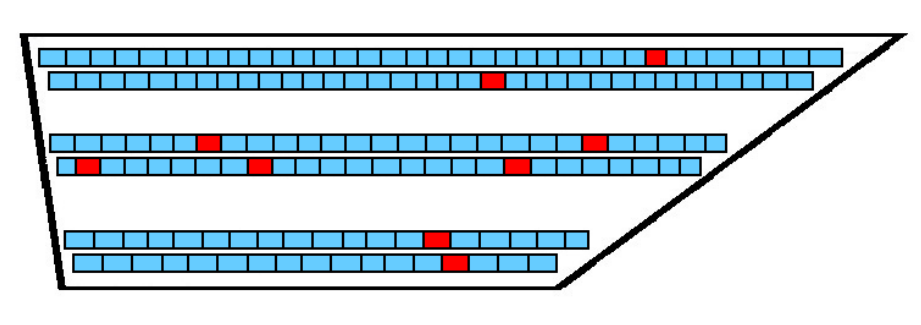}}
  \caption[Toy example of the tracking pattern library function]{Toy example of tracking pattern matching for the OLYMPUS track reconstruction algorithms.  In the top figure, the six cells
  obvious to the eye as corresponding to a track form a pattern match while the noise hits from the from the three cells to the left of the pattern were not passed to the track fitter.  In
  the bottom figure, no pattern was matched and thus no attempt was made to track the event \cite{oconnor1}.}
  \label{fig:pattern}
  \end{figure}

  The tracking algorithms used two common components, which were designed to reduce noise and increase the speed of track reconstruction.  The first of these was
  a pattern library of the combinations of wire hits that could reasonably correspond to a track of given kinematics so as to eliminate wires from consideration that
  were not included in a possible track combination and to avoid attempting to track events in which no pattern was present  \cite{DELLORSO1990436}.  This pattern library was generated from a very large set of
  simulated events extending throughout and beyond the kinematic ranges and particle species possible for the OLYMPUS running conditions.  This additionally prevented tracks from events with combinations
  of noise hits from mimicking elastic tracks that could contaminate the sample and were difficult to simulate.  A toy example of a matched pattern and a rejected event are shown
  in Figure \ref{fig:pattern}.  To account for inefficiencies in the wire chambers, patterns were allowed a tolerance of missing one complete cell layer (i.e, one of the six) in matching
  library patterns.
  
  The second common component was a model fit to simulation data that provided positions in the wire planes corresponding to initial track parameters known as \textit{Fasttrack}.  Other
  tracking routines (including that used in the 12\dg system (Section \ref{sec:12track})) iteratively simulate particle trajectories to minimize the residuals between the trajectory and
  hit positions.  Such an approach, however, would be much too slow for the OLYMPUS reconstruction.  To avoid the need of iteratively propagating simulated trajectories, a dense library of simulated
  events was generated and a generalized spline of the initial simulated track parameters was fit to the positions of the trajectories in each wire plane.  Then, when reconstructing tracks, the 
  spline function was queried for a given set of kinematic parameters to interpolate the resultant wire plane locations of the trajectory.  In addition to providing much higher speed than
  simulated propagation, the spline function also allowed the computation of derivatives with respect to track parameters which is useful for many tracking algorithms.
  
  With these tools in place, multiple algorithms were considered for the final fit of trajectories to data.  The method used for the main final analysis was based on the
  Elastic Arms Algorithm (EAA) \cite{OHLSSON1,OHLSSON2}.  The essential principle of the algorithm is to begin with template tracks (i.e., the ``arms'') and deform them through
  an iterative effective temperature cooling procedure to produce a global best fit the data.  Numerous additions to this procedure were made in order to optimize the algorithm for the purposes
  of OLYMPUS, which are discussed in References \cite{schmidt} and \cite{russell}.  In general, this algorithm performed well for OLYMPUS, which is discussed quantitatively in
  Section \ref{sec:reconrev}.
  
  The other algorithms created for OLYMPUS included early iterations in which iterative simulated propagation was used (the predecessor of the 12\dg system tracking algorithm
  discussed in Section \ref{sec:12track}) and approaches involving a local fitting of track elements to collections of hits to then build complete trajectories.  While these algorithms
  were used as checks of the main EAA algorithm, they were not used for the final analyses presented in this work and thus are not discussed further here.

\section{The OLYMPUS Monte Carlo Simulation}
\label{sec:sim}

  As noted previously, the analysis strategy of OLYMPUS required that the simulation be an accurate and detailed representation of the experiment and the experimental
  conditions in all conceivable ways.  In particular, the results of simulation were converted to the format of the raw data (i.e., TDC counts, ADC counts, etc.) after application
  of the relevant resolutions on such quantities and then reconstructed using exactly the same methods as used on raw data.  In this way, biases from simulation approximations
  were minimized, providing a robust framework for the comparison of data and simulation.  This section describes the procedure by which ``simulated raw data'' were produced from
  the Monte Carlo framework; the reconstruction and analysis proceeded as described for the raw data for all elements of the experiment.
  
  One of the notable advantages of this approach is that it permits a complete treatment of the effects of radiative corrections on the final data analysis, since the size of the radiative
  corrections for any experiment depends on the acceptance and resolution of the detector systems used and the way in which events are selected from the data sample.
  This arises from the fact that radiative events are only distinguishable from
  purely elastic events if they result in a change to the momentum of the particle from its elastic momentum by an amount that can be distinguished by the detector according to its
  resolution.  Additionally, since radiative events may change trajectories they may be pushed in and out of the acceptance or into regions of different detection efficiency
  relative to the pure elastic trajectory.  The OLYMPUS analysis accounts for all such effects by full representing the detector in simulation including its physical acceptance, efficiencies,
  and resolutions and by applying identical analyses to the data and simulation so that the effects of the choices made in the selection of elastic events are equally represented in data and
  simulation.  Although full Monte Carlo treatment of radiative corrections is common in modern higher energy experiments, most existing treatments of radiative corrections for \pmp scattering
  are designed for single-arm (inclusive), high momentum resolution experiments and only allow for an adjustment of the size of the correction based on a single event selection parameter (typically
  effectively amounting to the deviation of the lepton energy from the purely elastic energy) \cite{PhysRev.122.1898,MeisterPhysRev.130.1210,MaximonPhysRevC.62.054320,MoRevModPhys.41.205,PhysRevC.64.054610}.
  The full Monte Carlo method used for OLYMPUS provides a higher level of confidence in the proper handling of radiative effects,
  especially those that are opposite in sign for electrons and positrons, than classical methods by allowing radiative corrections to be properly treated in an exclusive event selection
  and reducing the uncertainties associated with the single-arm momentum resolution of the experiment. The VEPP-3 TPE experiment utilized a similar approach to radiative corrections
  as OLYMPUS and developed an \pmp radiative event generator \cite{vepp3PhysRevLett.114.062005,esepp}, while the CLAS TPE experiment applied a more classical approach
  to radiative corrections to their data \cite{ass,PhysRevC.64.054610}.
  
  \subsection{Procedure}
  
  The procedure of simulating events and producing simulated detector data proceeded in three steps:
  \begin{enumerate}
   \item generation of event vertices (initial positions and momenta of particles in the event),
   \item propagation of all generated particles through the detector geometry using GEANT4 with all relevant physics processes activated \cite{Agostinelli:2002hh}, and
   \item digitization of the energy depositions, positions, etc. recorded by GEANT4 into ``raw data'' quantities.
  \end{enumerate}
  The propagation step of the procedure utilized the well-verified physics processes simulated by GEANT4, and depended critically on the detailed representation of the detector in the simulation
  as discussed in Section \ref{sec:speccal}.  The event generation and digitization procedures were developed particularly for the purpose of the OLYMPUS simulation and are
  discussed in the following sections.
  
  \subsection{Event Generation}
  \label{sec:gen}

  The first step in the OLYMPUS simulation was to generate event vertices by a Monte Carlo procedure.  In general, this amounted to specifying a physical process to occur in the 
  detector system, drawing an event vertex from a specified target distribution, and drawing the momenta of all particles involved in the process from the distributions relevant
  to the physics process.  Options for the target distribution included realistic distributions generated from simulation of the target system (Section \ref{sec:tarsim}), fixed-vertex,
  events generated isotropically in the target region, and approximations to the simulated target distribution.  Options for the particle/momentum distributions included purely elastic
  \pmp (Born approximation) kinematics, various test generators with isotropic momentum/angle distributions, an implementation of the ESEPP generator produced by the VEPP-3
  TPE experiment \cite{esepp}, a new radiative \pmp generator developed for OLYMPUS (see next section), as well as generators for $e^\pm e^-$ events for the SYMB system that
  included a new radiative generator for such processes \cite{spuds}.
  
  Notably, most of the OLYMPUS generators were \textit{weighted} in that each event in simulation carries a scalar weight, which represents its contribution to the integral over
  all Monte Carlo events that is compared with data.  This approach provides two significant advantages:
  \begin{enumerate}
   \item events may be drawn more isotropically in phase space and then weighted appropriately so as to reduce the statistical error of the simulation in areas
	 of phase space where the true cross section is small (such as high $\theta$ in \pmp scattering) without simulating far more events than necessary in high cross section regions, and
   \item an event can carry multiple weights corresponding to different models of radiative corrections, protons form factors, physical approximations, etc., allowing a single sample of
         simulation events to test multiple physics models simultaneously without propagating, digitizing, and reconstructing the simulation events multiple times.
  \end{enumerate}
  The first advantage allowed OLYMPUS to achieve a very high statistical precision on the simulation throughout the entirety of the accepted phase space so as to make the statistical uncertainty due
  to the Monte Carlo effectively negligible next to the data statistical and systematic uncertainties.  The second provided a rich platform for the use of various radiative corrections models,
  form factor models, etc. with the OLYMPUS data, which simplifies the comparison of OLYMPUS results with previous experiments and provides a strong indicator of the systematic uncertainties
  on the OLYMPUS results associated with such effects.
  
  For production simulation, i.e, the simulation results designed for the direct data/Monte Carlo comparison, the target distribution was specified as the resulting distribution
  from the molecular flow Monte Carlo simulation described in Section \ref{sec:tarsim} and the physics processes were determined by the OLYMPUS radiative event generator.

  \subsubsection{The OLYMPUS Treatment of Radiative Corrections and Radiative Event Generator}
  \label{sec:radgen}
     
    As previously noted, an new radiative \pmp event generator was developed for use with the OLYMPUS simulation framework. The development, testing, and application of this
    generator are discussed in great detail in References \cite{schmidt} and \cite{russell}, while only the essential details are provided here.  This generator was designed
    to implement a variety of prescriptions for radiative corrections and the proton form factor, while providing a direct interface with the OLYMPUS analysis framework.  In general, previous prescriptions
    for radiative corrections to elastic \pmp scattering took the basic form:
    \begin{equation}
     \text{d}\sigma_\text{exp.} = \text{d}\sigma_\text{Born}\cdot\left( 1 + \delta(\theta,\Delta E)\right),
    \end{equation}
    where $\text{d}\sigma_\text{exp.}$ is the experimentally measured elastic cross section, $\text{d}\sigma_\text{Born}$ is the Born (single-photon exchange) cross section, and $\delta$ is the
    radiative correction factor, which is typically computed as a function of the scattering angle $\theta$ and the effective energy resolution for the distinction of
    radiative events for the experiment $\Delta E$.  Typically, results for elastic \pmp scattering were reported as the extracted Born cross section after the subtraction of the correction
    $\delta$ for comparison between experiments.  The first of the prescriptions for computing the correction $\delta$ was developed by Mo and Tsai in the 1960s \cite{PhysRev.122.1898,MoRevModPhys.41.205}.
    Additionally, Meister and Yennie developed an approach based on the early work of Tsai that made additional approximations for the purposes of facilitating computation \cite{MeisterPhysRev.130.1210}.
    These correction prescriptions were the standard for nearly four decades until Maximon and Tjon published a new prescription in 2000 that reduced the number
    of approximations made relative to Mo and Tsai, accounts approximately for the structure of the proton via the introduction of the dipole form factor (Equation \ref{eq:dipff}), 
    and reformulates the contributions due to soft two-photon exchange \cite{MaximonPhysRevC.62.054320}.  Each of these prescriptions was designed for single-arm (inclusive) experiments,
    and are thus formulated to rely on a cut in the lepton $\Delta E$.  A prescription developed by Ent \textit{et al.}\ in 2001 reformulates the procedure of Mo and Tsai to produce
    a method that allows the calculation of $\delta$ as a function of the missing energy reconstructed from exclusive detection of the lepton and proton, extending the applicability
    of such models to coincidence experiments \cite{PhysRevC.64.054610}.
    
    An extension to models for elastic \pmp radiative corrections, first proposed by Yennie \textit{et al.}\ in 1961 \cite{YENNIE1961379}, replaces the $(1+\delta)$ correction with a factor
    of $\exp(\delta)$, a procedure known as \textit{exponentiation}.  As demonstrated in Reference \cite{YENNIE1961379}, this allows for the correction to account for the emission
    of multiple soft photons (i.e., that do not distinguishably change the event kinematics) and prevent the infrared divergence that occurs for the $(1+\delta)$ prescription, which
    effectively only treats single photon emission.  The soft approximation breaks down, however, as $\Delta E$ increases, necessitating a transition to consideration of such higher
    $\Delta E$ events as hard bremsstrahlung processes.
    
    The OLYMPUS radiative generator incorporates a wide variety of these prescriptions via the calculation of multiple weights.  The details of the implementations of these
    methodologies is discussed in References \cite{schmidt} and \cite{russell}, but the essential capabilities provided by the generator include the following:
    \begin{enumerate}
     \item implementation of the prescriptions of Mo and Tsai, Meister and Yennie, and Maximon and Tjon,
     \item exact tree-level calculation of the bremsstrahlung matrix element (avoiding the soft photon approximation made by previous approaches),
     \item separate weights for exponentiated and non-exponentiated corrections,
     \item treatment of the vacuum polarization diagrams either from calculations including all leptons in the loop or using a data-driven approach (which, in principle,
           includes all possible particles in the loop) \cite{vacpolweb,vacpolpres},
     \item weights representing the Born and soft-photon approximations, and
     \item and proper application of radiative corrections for different proton elastic form factor models.
    \end{enumerate}
    These various assumptions are represented across multiple weights computed for each simulated event, allowing the OLYMPUS data to be analyzed under a wide variety of
    radiative corrections models and physical assumptions, facilitating both comparison to previous elastic \pmp data and providing extensive insight into the systematic effects
    of radiative corrections for the final OLYMPUS results.
    
    The generator was tested extensively prior to use for the final OLYMPUS analysis, including comparisons to the ESEPP generator \cite{esepp} and comparisons to the model
    of Maximon and Tjon \cite{MaximonPhysRevC.62.054320} in the appropriate regions of phase space, as described in References \cite{schmidt} and \cite{russell}.  In general,
    the OLYMPUS radiative generator performed extremely well under all tests and provided a robust platform for the analysis of OLYMPUS results.
  
  \subsection{Digitization of the Detector System}
  
  In general, every effort was made to represent the elements of the OLYMPUS experiment in simulation to accurately represent the conditions under which the experiment operated.
  This involved accounting for a number of factors via the generation of simulation events for each individual data run (each of which included $\sim$1$\cdot 10^{6}$ triggers) that
  properly accounted for any effects such as beam position, target gas temperature, etc. that were subject to time variation.  Such parameters were provided to the simulation from
  the slow control data for each data run, and they were used to adjust the conditions such as the target gas density, the position of the beam with respect to the locations
  of generated events, the magnetic field strength, etc.\ on an event-by-event basis to properly negate the possibility of such effects altering the final data/simulation comparison.
  
  Regarding the digitization of detector elements, individual systems were treated so as to properly account for their resolutions and efficiencies (as measured from experimental data)
  in the generation of simulated hits.  The exact procedures used for different systems are discussed in the relevant sections  describing the analyses using each detector, but in general
  this was achieved by applying appropriate uncertainty to simulated TDC and ADC values, testing simulated hits against data-driven maps of simulated efficiency, and the elimination of
  data from the final analysis corresponding to times when detectors were behaving unpredictably and could not be properly modeled in this fashion.  This approach created a detailed model
  of the OLYMPUS experiment in simulation, accounting for detector imperfections, a wide variety of time-varying effects that could otherwise introduce systematic uncertainties, and
  a method of completely accounting for effects of detector acceptance and elastic event selections in the treatment of the radiative corrections applied to the experiment results.

%% file: chap5.tex
% Chapter 5
%
% Luminosity determination, focusing on 12 degree system since
% that was my focus
%

\chapter{Determination of the Luminosity}
\label{Chap5}

As noted in Chapter \ref{Chap1}, the OLYMPUS result on \ratio depends equally on two key elements of the data: the measurement of relative \pmp
rates as a function of angle and the measurement of the relative luminosity collected between the two species modes.
Ideally, each of these quantities should be known individually to better than $\pm1\%$ total (statistical and systematic) uncertainty
 so as to provide an overall uncertainty of less than $\pm1\%$ on the final measurement.  To this end, OLYMPUS
employed three functionally independent methods of luminosity determination:
\begin{enumerate}
 \item calculation of the luminosity from the effective target density and beam current (the ``slow control luminosity''),
 \item dedicated (separate from the main tracking system) forward ($\epsilon \approx 0.98$, $\theta \approx 12^\circ$) elastic \pmp event
       reconstruction (the ``12\dg luminosity''),
 \item and very forward ($\theta\approx 1.3^\circ$) integrating calorimetric measurements of elastic $e^\pm e^-$ (M{\o}ller and Bhabha
       scattering \cite{moller,bhabha}), $e^+e^-\rightarrow \gamma\gamma$ annihilation, and elastic \pmp events (the ``Symmetric M{\o}ller/Bhabha (SYMB) luminosity'').
\end{enumerate}
Each of these methods was sensitive to different physics processes and systematic uncertainty contributions, making 
them a comprehensive and complementary set of measurements for the luminosity determination.  In particular, the slow control 
determination provided a real-time estimate of the luminosity during data-taking independent of event reconstruction.  The 12\dg system provided
a direct normalization of elastic \pmp with detectors independent from the drift chambers in a region where two-photon exchange (TPE) is expected to
be small. The SYMB calorimeter used extremely forward \pmp scattering (where TPE is expected to be extremely small) in conjunction
with the independent $e^\pm e^-$ process to provide an overall normalization decoupled from the TPE measurement of interest.

Throughout this chapter, reference will be made to the ``absolute luminosity'' and the ``species-relative'' luminosity ratio measurements,
with important distinctions drawn between them.  For clarity, the absolute luminosity $\mathcal{L}_{e^\pm}$ for a given period of data-taking refers to the value
of the integrated luminosity that would be used in extracting the absolute cross section of elastic events:
\begin{equation}
 \sigma_{e^\pm p}\left(\epsilon,Q^2\right) = \frac{N_{e^\pm p}\left(\epsilon,Q^2\right)}{\mathcal{L}_{e^\pm}},
 \label{eq:abslumi}
\end{equation}
where $N_{e^\pm p}\left(\epsilon,Q^2\right)$ is the number of elastic \pmp events reconstructed in a given data
bin characterized by the phase space point $\left(\epsilon,Q^2\right)$ (assuming perfect acceptance) and $\sigma_{e^\pm p}\left(\epsilon,Q^2\right)$
is the total integrated cross section for elastic \pmp scattering into that phase space bin.  The species-relative luminosity
ratio is then defined as the dimensionless ratio of the absolute luminosities for each lepton species:
\begin{equation}
 R_\text{lumi} = \frac{\mathcal{L}_{e^+}}{\mathcal{L}_{e^-}},
 \label{eq:rellumi}
\end{equation}
and ultimately it is this quantity that appears in Equation \ref{eq:rat} and is critical to the determination of
$R_{2\gamma}$.  Notably, many systematic uncertainties cancel in the ratio of the luminosities to allow much more
precise determination of $ R_\text{lumi}$ than of the individual luminosities.  This is discussed in detail in Section
\ref{ss:12sys}.  Since the OLYMPUS data are in principle valuable for certain measurements involving the absolute
cross section (such as elastic form factor measurements), the absolute luminosity is considered in addition to
$ R_\text{lumi}$ to provide it for possible future use.

The author's primary effort regarding the luminosity analysis consisted of the determination of the slow control
and 12\dg luminosity estimates, and thus this chapter will cover each of those in detail in Sections
\ref{sec:sclumi} and \ref{sec:12lumi}.  Additionally, Section \ref{sec:tarsim} addresses the efforts that were made
to properly determine and simulate the shape of the target gas distribution generated by the OLYMPUS target system. This
had important implications for the relative acceptance of the two lepton species in the various detector systems.
Section \ref{sec:symblumi} provides a brief discussion of the SYMB system
luminosity analysis and the results arising from it. Section \ref{sec:alllumi} summarizes the results of the different
luminosity analyses and sets the stage for the primary OLYMPUS results on \ratio.

\section{Slow Control Luminosity and Target Gas Distribution Determination}
\label{sec:sclumi}

The slow control luminosity determination served to provide a real-time estimate of the luminosity determination during
OLYMPUS data-taking independent of any of the event triggers and reconstruction.  While not as precise, either for the
absolute or species-relative luminosity, the slow control estimate provided an important baseline for run simulations
and a cross check for more precise measurements.  Additionally, as part of the slow control luminosity determination, concerted
efforts were made to understand the shape of the target gas distribution used in the experiment, which affected both the luminosity
and main \ratio results via the varying acceptance for \ep and \pp events as a function of position in the target.  This was achieved
by developing a gas molecular flow Monte Carlo simulation with accurate representations of the both the target system geometry and the
physics of the gas within the system.  This section will cover the various elements of the slow control luminosity determination in
detail, paying particular attention to the target gas simulation.

\subsection{Principle of the Measurement}

The slow control luminosity determination makes use of the basic definition of instantaneous luminosity
for a fixed target experiment \cite{bettini}:
\begin{equation}
 \td{\mathcal{L}_\text{SC}}{t} = \frac{I_\text{beam}(t)}{e}\int_\text{target}\rho_p(z,t)\:\D z = \frac{2I_\text{beam}(t)}{e}\int_\text{target}\rho_{\text{H}_2}(z,t)\:\D z,
\label{eq:lumidef}
\end{equation}
where $I_\text{beam}(t)$ is the beam current, $e$ the electron charge, $\rho_p(z,t) = 2\rho_{\text{H}_2}(z,t)$ represent the density of protons/hydrogen molecules
in the target along the beam, and the integral runs over the length of the beam in the target $z$.  Thus, a good estimate of the luminosity using
this method includes measurement of the beam current as a function of time and a model for the number and spatial distribution of molecules present in the target as a
function of the target input gas flow, temperature, and geometry.

During data-taking, the target and beam conditions were monitored using the slow control system described in Section \ref{sec:sc}.  On an event-by-event
basis, the increment to the integrated slow control luminosity was calculated as the following variation of the integral of Equation \ref{eq:lumidef}:
\begin{equation}
 \Delta\mathcal{L}_\text{SC} = Q_{\text{H}_2} \cdot \frac{I_\text{beam}}{e} \cdot \Delta t_\text{DTC} \cdot n_T \cdot\sqrt{\frac{75\:\text{K}}{T}},
 \label{eq:sclcalc}
\end{equation}
where $Q_{\text{H}_2}$ is the input flow rate of H$_2$ molecules into the target cell in standard cubic centimeters
per minute (sccm)\footnote{The unit standard cubic centimeter per minute is defined as the flow rate of 1 cubic centimeter of gas at temperature 0 $^\circ$C
and pressure 1.01 bar passing a given point per minute.  For reference, this defines $1\:\text{sccm}=4.477962\cdot 10^{17}\:\text{particles/s}$.   For 1 sccm of
H$_2$ flow, note that the rate of proton flow is twice the gas particle flow.}, $I_\text{beam}$ is the beam current in A, $e$ the electron charge in C,
$\Delta t_\text{DTC}$ the trigger livetime (i.e., the elapsed time after being ``dead-time-corrected'' (DTC) for the time during detector readout when
the event triggers were not open), $n_T$ the effective total thickness of the gas target in cm$^{-2}$ presented to the beam as calculated by the target
simulation described in Section \ref{sec:tarsim} for 75 K and 1 sccm H$_2$ flow rate, and $T$ is the target temperature in K.  Note that the integrated
luminosity scales linearly with the beam current (number of beam particles on target), trigger live time, and gas flow rate into the target (number of particles
in the target), as would be intuitively expected.  The $1/\sqrt{T}$ dependence of the target thickness arises from the fact that the H$_2$ molecules
rapidly thermalize with the target walls, and thus their average velocity goes as the $\sqrt{T}$ behavior of the mean of the Maxwell-Boltzmann distribution \cite{maxwell1,maxwell2},
effectively changing the average amount of time that a single hydrogen molecule spends in the target by the inverse factor (further discussed in
Section \ref{sec:tarsim}).  The total integrated slow control luminosity for a running period is simply the sum over the values of $\Delta\mathcal{L}_\text{SC}$ for
each event in the period.  Also, note that
$\Delta t_\text{DTC}$ may be replaced by $\Delta t$, the simple elapsed time, to convert to a measure of the ``delivered'' integrated luminosity,
but that the ``collected'' luminosity is more relevant since this corresponds to the time when the detector was actually active and thus corresponds
to the luminosity that the detector system measures.

\subsection{Simulation of the Target Gas Distribution}
\label{sec:tarsim}

The target gas distribution used in the OLYMPUS simulation was determined using a newly-developed, standalone Monte Carlo
simulation of the molecular flow of hydrogen molecules within the target system.  This section reviews the physics
of gas flow relevant to the system, explains the limitations of standard gas density calculations that make them insufficient
for the requirements of the experiment, and describes the new simulation.

\subsubsection{Relevant Details Regarding the Target System}

The internal components of the target system are shown in Figure \ref{fig:tarin}, including the various system components
in which hydrogen gas was contained and subject to exposure to the beam.  Qualitatively, gas density was highest in the region
of the target cell directly below the inlet and tapered towards the edges of the system where gas could escape into the beamline
via holes in the wakefield suppressors or their connections to the beamline at the ends.  The flow rate was controlled by
mass flow controllers (as detailed in Section \ref{sec:target}) and experimental data were only taken when the gas
flow was in a steady state (rate of input from the inlet matching the rate of gas collection by the pumps), as indicated by
the achievement of steady pressure in the target system.  The temperature of the target cell assembly was measured as part of the slow control
system by seven equally-spaced thermocouples along the system parallel to the beam.  Unfortunately, not all of these thermocouples were calibrated
for absolute temperature measurements.

\subsubsection{Physics of the Target System Gas}

Examining the components of Equation \ref{eq:sclcalc}, the $\Delta t_\text{DTC} $ was provided by the OLYMPUS trigger system (Section \ref{sec:trig}),
$I_\text{beam}$ was provided by the DESY accelerator system, and $Q_{\text{H}_2}$ and $T$ were monitored by the target and gas flow system (Section
\ref{sec:target} and Reference \cite{Bernauer201420}).  The effective target thickness, however, is a much more complicated factor, involving
the dynamics of gas flow in conjunction with the specifics of the OLYMPUS target geometry and vacuum system.  The physics of gas flow varies
considerably as a function of the gas pressure, nature of the conduit, and nature of the gas.  A complete discussion of these phenomena may
be found in Reference \cite{rothvac}, but a brief discussion of the topic is provided here to establish the physics important to the flow regimes
present in the OLYMPUS target.  The flow regimes of sparse gases are typically classified by the \textit{Knudsen number} \cite{Knudsen,0034-4885-49-10-001}:
\begin{equation}
 \text{Kn} = \frac{\lambda}{D},
\end{equation}
the dimensionless ratio of the mean free path $\lambda$ between gas molecule collisions and the typical diameter $D$ of the conduit. For
$\text{Kn}\gtrsim0.5$, the behavior of the gas is dominated by its interaction with the conduit and intermolecular interactions are
considered negligible, a region known as \textit{molecular flow} where the dynamics of the gas flow simplify considerably.  Ideally,
the OLYMPUS target system would be in this regime since this would make an accurate representation of the target gas much more
feasible.  To determine if this was the case, the hydrogen may be approximated as an ideal gas with velocities distributed according
to the Maxwell-Boltzmann distribution.  In this case, the mean free path may be analytically calculated:
\begin{equation}
 \lambda = \frac{k_BT}{\sqrt{2}\pi\xi^2P} = 2.33\cdot 10^{-20}\:\left[ \frac{\text{Torr}\cdot\text{cm}^3}{\text{K}} \right]\:\frac{T}{\xi^2 P},
\end{equation}
where $k_B$ is the Boltzmann constant, $P$ the gas pressure, and $\xi$ the effective molecule diameter.  This approximation was verified experimentally
by Sutherland and others in thermodynamic conditions similar to those in the OLYMPUS target \cite{sutherland}.

The region of highest pressure in the target gas system, and thus lowest mean free path, is the narrowest aperture  (examining only the parts
near the cell that immediately affect the distribution of gas in the cell).  This region corresponded to the inlet tube,
as shown in Figure \ref{fig:tarin}.  The inlet was several centimeters long and of diameter $D=0.5$ cm, and was cooled by the cryogenic system
to temperatures as low as 35 K (in the absence of beam heat load on the system).  While the pressure inside the inlet is difficult to estimate
and was not directly measured, note that for the hydrogen molecular diameter experimentally determined to be  $\xi=4.04$ \AA~\cite{weast}
and the conservative (resulting in the lowest mean free path) case of $T=35$ K:
\begin{equation}
  \lambda P = 5 \cdot 10^{-4}\:\text{cm}\cdot\text{Torr}.
\end{equation}
Thus, to consider the gas to be in the molecular flow regime ($\text{Kn} = \frac{\lambda}{D} \gtrsim 0.5$) requires:
\begin{equation}
  P \lesssim 2 \cdot 10^{-3}\:\text{Torr}.
\end{equation}
Since this is three orders of magnitude larger than the pressure measured in the scattering
chamber during normal running conditions (a region of only approximately 1--2 orders of magnitude greater in volume than the inlet) \cite{Bernauer201420},
it is reasonable to assume that the gas was characterized by molecular flow throughout the target system.

\subsubsection{Molecular Flow}

As noted, in the molecular flow regime it is assumed that interaction between mole\-cules in the gas are negligible
and interactions with the walls of the conduit dominate the behavior of the flow.  In particular, for the case of
the OLYMPUS target system where the gas is relatively low in density and the conduit walls (the target cell, inlet,
wakefield suppressors, etc.) are being actively cooled, the gas rapidly thermalizes to the temperature of the conduit.
The gas can be understood as a collection of individual molecules that propagate along straight lines between
collisions with the walls of the conduit with speeds distributed according to the Maxwell-Boltzmann distribution
of temperature $T$ (hence the $1/\sqrt{T}$ dependence of Equation \ref{eq:sclcalc}).  Note that the collisions
at the walls of the conduit are not reflection-like (i.e, angle of incidence equal to the angle of reflection),
since the walls are fundamentally ``rough'' on the scale of molecular collisions.  The rebounds from the wall
are typically modeled to be distributed as the cosine of the angle from the surface normal vector independent
of the incident angle, with no preference for the azimuthal angle relative to the normal vector (known as ``Knudsen's cosine
law'' \cite{kcos}).  Note that the preference is to rebound normal to the surface, with vanishing probability to rebound
along the surface.  While it is not excluded that there may be a finite time between collision with the wall and the reemission
of the molecule, this effect will equilibrate and become negligible after the gas flow is well established and many
molecules are in the system.

\subsubsection{Conductance Modeling}

Having established the molecular flow nature of the flow of the gas through the target system, the traditional approach to
determining the resultant gas density inside the target cell is to define the \textit{conductance} $C$ of the target cell
(with dimensionality of volume per unit time), analogous to
an electrical conductance.  By construction, this conductance obeys a corresponding molecular flow analog of Ohm's law:
\begin{equation}
 C = \frac{Q_T}{\Delta P},
\end{equation}
where $Q_T$ is the \textit{throughput} (with dimensionality pressure times volume per unit time) 
and $\Delta P$ the pressure difference across the system.  A more useful quantity than the
throughput is the \textit{pumping speed} $S$, defined as the throughput divided by the pressure
at the entrance to the system (in this case the inlet of the target cell)\footnote{The name ``pumping
speed'' arises from the fact that the system of interest often terminates at a vacuum pump, but this need
not be the case and the term can describe the rate of gas flow through any conduit.}.
Defining $P_\text{in}$ as the inlet pressure to the target cell and $P_\text{out}$ as the pressure
at the exit to the cell, then $S=Q_T/P_\text{in}$ and:
\begin{equation}
 C = \frac{S P_\text{in}}{P_\text{in}-P_\text{out}} = \frac{S}{1-\frac{P_\text{out}}{P_\text{in}}}.
\end{equation}
This result is important in that it implies that for $P_\text{out}\ll P_\text{in}$, the pumping speed
goes to the conductance, independent of the precise value of the conductance.  Thus, while the pressure
in the target inlet tube and at the exit to the target cell are not well known, it is satisfied that 
$P_\text{out}\ll P_\text{in}$ and so the conductance of the system may be calculated to determine the
rate at which molecules pass through the cell (and thus the density of molecules inside the cell).

The computation of the conductance of an arbitrary conduit, however, is quite difficult.  Typically,
such calculations require computation of integrals over all possible straight-line particle trajectories between collision points,
rely on assumptions that the geometry has uniform cross section, and ignore boundary conditions (i.e., assume
the conduit is very long).  Analytical calculations for several simple geometries may be found in Reference
\cite{rothvac}.  For the original OLYMPUS simulation, the target density was modeled using the Steckelmacher analytical computation
of the conductance for a long tube of elliptical cross section \cite{0022-3727-11-4-011} to model the OLYMPUS
target cell.  Since the model had constant cross section, the resulting predicted gas density distribution was triangular,
peaking at the inlet and linearly declining to zero at each end of the target cell (similar to a voltage dropping
across a long, uniform resistor), shown as the ``Old Slow Control MC'' distribution in Figure \ref{fig:tardists}.
While initially expected to provide a sufficiently good model of the true
target density, comparison of analyses of simulation and data via tracking in multiple detectors indicated that
the simulated triangular target model did not predict the shape of the measured distribution well near the ends of the cell
and underestimated the density of the target by approximately 20\%.

In retrospect, this model failed due to the fact that it ignored the end conditions of the target cell, which directly connect to wakefield suppressors.
While the wakefield suppressors were manufactured with holes to allow the escape of gas (see Figures 7--9 in Reference
\cite{Bernauer201420}), the model severely underestimated the containment of gas by these elements and could not
make any prediction regarding the shape of the gas distribution in the regions near the ends of the cell.  Additionally, the effective
conductance for gas escape was not the same at each end of the cell due to the connection of the collimator, and thus
the distribution of gas was additionally not symmetric about the inlet port.  It was,
however, completely impractical to use analytical methods to compute the conductance of the complete target system
geometry, due to both the elliptical cone shapes of the collimator bore and wakefield suppressors and the lack
of gas escape handling in such calculations.

\subsubsection{Simulation of Molecular Flow and Conduit Geometry}

Due to the fact that the acceptance of \ep and \pp events in the detectors can differ as a function of
target vertex position and the desire to have the slow control luminosity calculation more accurately
represent the overall target thickness, a new method was developed for computing the target density to
ensure an accurate representation in the simulation.  As discussed, the complexity of the complete geometry
made any attempt at an analytical approach infeasible, which naturally suggests Monte Carlo simulation methods.
While there exists commercial software for molecular flow \cite{comsol}, this software is quite expensive
and designed for engineering applications that do not necessarily have similar conditions to internal
gas targets in nuclear and particle physics experiments.  For this reason, a new, standalone molecular
flow and geometry simulation known as ``TargetSim'' written in C++ was developed for the purpose of the OLYMPUS experiment\footnote{The 
source code for TargetSim, along with basic documentation, is available from the author for academic applications.}.

\begin{figure}[thb!]
\centerline{\includegraphics[width=1.0\textwidth]{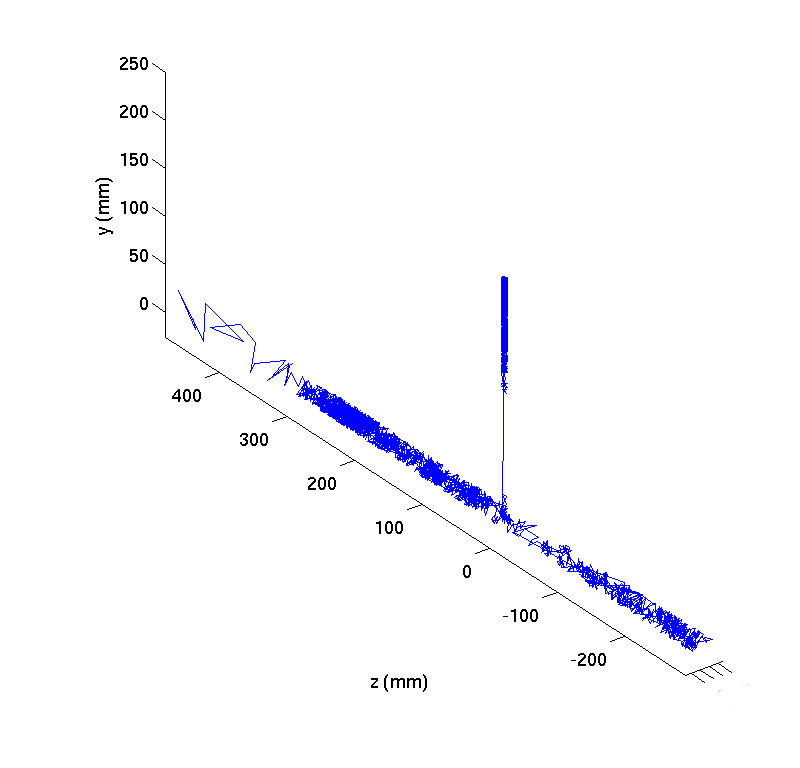}}
\caption[Simulated path of a hydrogen gas molecule in the target system]{Simulated path of a hydrogen molecule inside the OLYMPUS
target system, beginning at the top of the inlet, passing into the upstream portion of the target cell, and eventually leaving
the system through the downstream wakefield suppressor.  Coordinates are approximately the OLYMPUS global coordinates.}
\label{fig:typpath}
\end{figure}

In general, TargetSim follows the approach of propagating particles (i.e., molecules or atoms of gas) of given mass through a modular geometry.
Molecules traverse the geometry starting from a source distribution defined by the user (i.e., a position and velocity
distribution for particles entering the system).  Each input particle is individually tracked to its next collision with a
wall of the geometry or until it reaches a designated exit of its current geometry unit.  At the exit to a geometry unit,
a particle may pass into another geometry unit, in which case its next intersection or exit point is calculated for that unit,
or out of the system (i.e., to a location where it will reach a vacuum pump).  At each wall collision point, the gas molecule
is assumed to thermalize with the local temperature of the wall, and is then re-emitted with speed distributed according
to the Maxwell-Boltzmann distribution and direction according to Knudsen's cosine law.  This process is iterated for each
particle until it reaches a system-exit condition.  Such a propagated path in the OLYMPUS target system is shown in Figure
\ref{fig:typpath}.

Along the particle's path, the time occupation of the particle
in three-dimensional spatial bins is recorded.  Then, at the end of the trajectory, the time occupation histogram
may be divided by the total time the particle spent in the system to create a position density distribution histogram for
a single particle. For a collection of many particles, the average of such distributions will converge to the position distribution
of the target system under steady-state conditions.  Furthermore, the average time a particle spends in the system may
be used in conjunction with the steady-state input flow rate to determine the average number of particles inside
the system at a given time, and thus to place an overall normalization on the position distribution as a function of
flow rate for direct implementation in the simulation and slow control luminosity calculation.

The definition of the geometry in TargetSim is a user input to the program, which must include the following information for each
unit in the geometry:
\begin{enumerate}
 \item equations or inequalities defining the surfaces of the geometry that a particle may strike,
 \item a method of calculating the surface normal vector at any point on the strikable surfaces of the geometry,
 \item a defined temperature function for each strikable point on the geometry,
 \item an analytical or numerical method for computing the intersection of a line (i.e., a particle trajectory) with the strikable surfaces, and
 \item conditions defining which points on the strikable surfaces are exit points and into which geometry unit or system exits those exit points pass
       the particle.  
\end{enumerate}
Geometry units are defined as instances of a general C++ geometry unit class with member variables and functions that provide
the above information to a propagator function which handles the Monte Carlo drawing of new particle directions, generates the
time occupation histograms, and records other requested information about the particle trajectory.  The version of TargetSim used
for OLYMPUS includes implementations of cylindrical tubes, elliptical tubes, and elliptical cones as geometry units (including
some elements with holes that correspond to system exit locations).  In principle, however, any geometry description that can be
implemented to satisfy the requirements above can be used as a geometry unit, thus making TargetSim extremely flexible to handle 
a variety of molecular flow systems.

\subsubsection{Simulation Implementation and Results}

For the OLYMPUS implementation of TargetSim, gas particles were generated at the top of the cylindrical hydrogen inlet tube, 250 mm above the top of
the elliptical target cell, at room temperature with trajectories distributed as $\cos\theta$ relative to the downward pointing
vector along the inlet tube, which has been shown to be a good approximation for molecular flow in a long cylindrical tube \cite{ZHANG2012513}.  The geometrical
elements included in addition to the inlet were the elliptical target cell, the elliptical cone internal bore of the collimator, and the three 
wakefield suppressors (all shown in Figure \ref{fig:tarin}).  The system exit conditions were at the holes of the wakefield suppressors and the ends of the
wakefield suppressors that attach to the beamline.  Simulations in which the cylindrical beamline was expanded beyond the wakefield suppressors and in
which the target chamber was implemented as a box containing the system were conducted so as to assess the probability of a particle
reentering the system after one of the aforementioned exit conditions. This effect was found to be negligible, and so to increase the speed of the simulation,
additional beamline and target chamber elements were not included in the geometry.  The surfaces of the system were assumed to be
at constant temperature, as suggested by the limited calibrated temperature information available from the thermocouples.

\begin{figure}[thb!]
  \centerline{\includegraphics[width=1.2\textwidth]{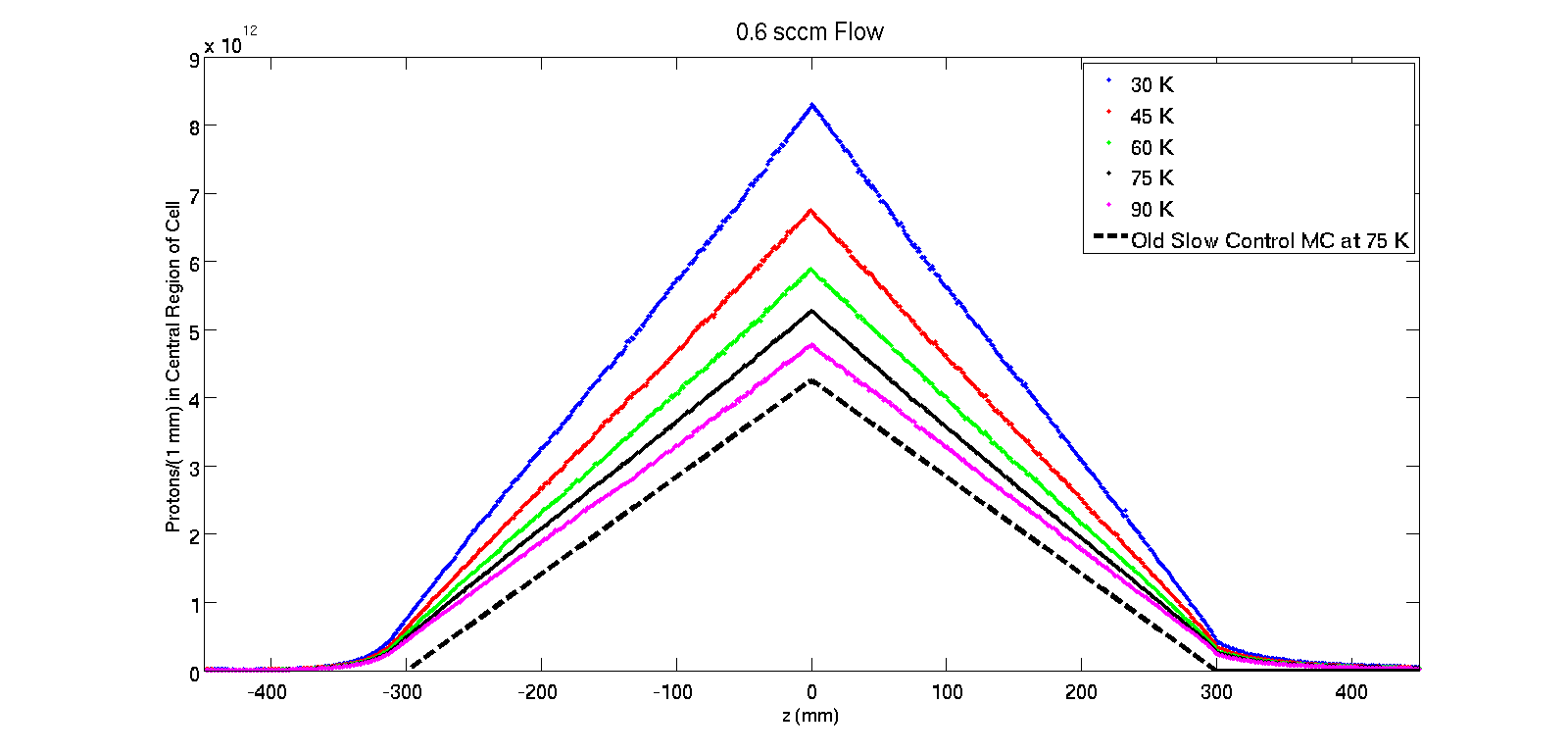}}
  \caption[Predicted target gas distributions from TargetSim]{Predicted target gas density distributions for the region of the system near
  the beam from TargetSim, for 0.6 sccm input flow and a range of typical constant temperatures for running conditions, compared to the prediction
  from the elliptical tube conductance calculation of Reference \cite{0022-3727-11-4-011} (``Old Slow Control MC at 75 K'').  Note that while the central regions of the simulation
  predications are approximately triangular as predicted by the conductance, the distributions are greater in magnitude and asymmetric
  relative to the triangular conductance prediction due to the increased and asymmetric resistance to flow of the additional target system
  components on each end of the cell.}
  \label{fig:tardists}
  \end{figure}

The results of the simulation for typical OLYMPUS running conditions are shown in Figure \ref{fig:tardists}, including a comparison
to the results of the Steckelmacher elliptical tube conductance calculation.  As can be seen, the target simulation predicts
the $\sim$20\% increase in density relative to the conductance calculation that was suggested by other detector systems, and predicts
a more complicated shape for the target gas distribution extending beyond the $\pm300$ mm of the target cell.  Note that the upstream
($-z$) end of the distribution is especially important to the 12\dg luminosity telescope measurement since events from this range
may be in the telescope tracking acceptance.  The simulation reproduces the $1/\sqrt{T}$ dependence of the gas occupation of the target
expected from the Maxwell-Boltzmann distribution of velocities.  Thus, this factor may be computed analytically, as previously claimed
in the discussion of Equation \ref{eq:sclcalc} so as to avoid requiring simulations at each individual measured temperature from the dataset.

\begin{figure}[thb!]
  \centerline{\includegraphics[width=1.2\textwidth]{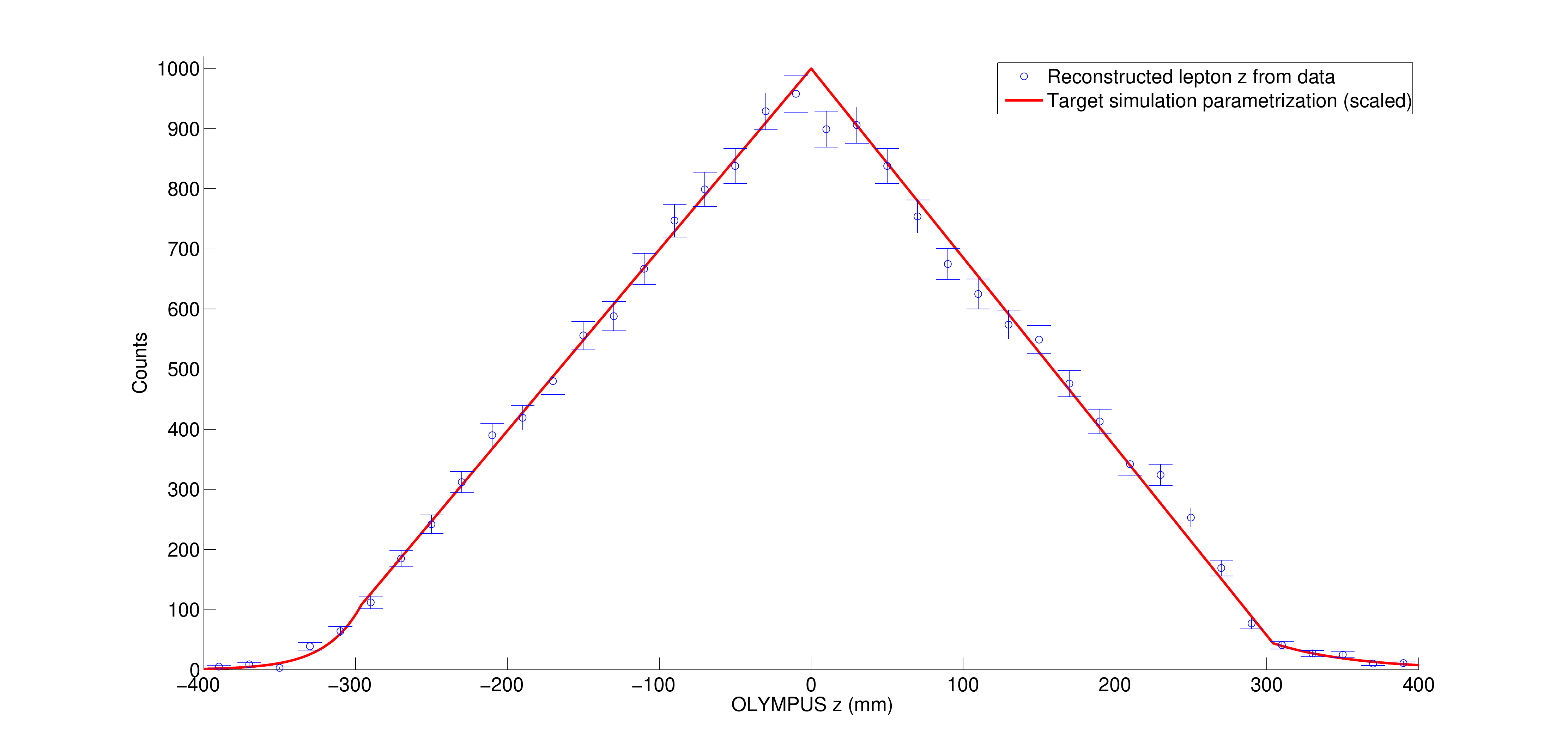}}
  \caption[Comparison of target distribution simulation to data]{Track vertex $z$ distribution for elastic events with lepton $\theta$ within 0.5\dg of 42$^\circ$,
           compared to the predicted distribution from the raw target simulation distribution.  Note that the simulation distribution is normalized to have
           equal integral to the data distribution for comparison of the distribution shapes in this plot.}
  \label{fig:tarcomp}
  \end{figure}

To implement the results of the target simulation in the main OLYMPUS simulation to properly represent the vertex distribution of tracks,
the shape of the distribution was parametrized using piecewise polynomial fits to its shape. The distribution was normalized
by using the effective total target thickness of $n_T = 7.9095\cdot 10^{15}$ protons/cm$^{-2}$ predicted by the simulation at $T=75$ K and  $Q_{\text{H}_2}=1.0$
sccm and then correcting for the measured flow and temperature as a function of time using Equation \ref{eq:sclcalc}.  Comparison of the parametrization of
the simulation to data is complicated by the fact that the extended angular acceptance of the OLYMPUS tracker distorts the reconstructed $z$ vertex
position.  By narrowing the event selection to a small $\theta$ range, however, data and the raw target distribution prediction may be approximately
compared.  Such a comparison is shown in Figure \ref{fig:tarcomp}, for reconstructed elastic events selected for lepton $\theta$ within 0.5\dg of 32\dg.
In general, the simulation predicts the shape of the distribution very well, including the slopes on each side of the triangle and the behavior near
the ends of the cell.

Due to the strong indications that TargetSim produces a better target distribution prediction (both in terms of shape and normalization) than
previous methods, the results of TargetSim were incorporated into all generated simulation datasets for the OLYMPUS experiment.

\subsection{Systematic Uncertainty and Discussion of the Slow\\Control Luminosity}
\label{sec:scsum}

Note that, despite the apparent success of TargetSim in predicting the shape and normalization of the gas density of the target,
several uncertainties make it unreasonable to use the slow control luminosity on its own as a precise luminosity determination.  As
previously noted, the temperature measurements along the target cell were not well calibrated, and thus it is impossible
to determine the exact temperature of the target cell as a function of time and whether or not the temperature was uniform across the system at any given
time.  Even a very plausible 5 K shift in temperature alters the absolute slow control luminosity by $\sim$3.2\%.  Given that the target
temperatures during electron running were $\sim$10 K higher than during positron running due to different beam conditions, any non-linearity
or other misunderstood aspects of the temperature measurement could induce a false asymmetry in the slow control luminosity measurement
on the order of a percent.  While beam energy was well constrained, the input flow was calibrated by filling buffer volumes in the gas
supply system, a process with only percent level precision and that doesn't account for any losses in the several-meter-long input line
from the supply system to the gas inlet.  Adding these to the fact that the simulation is not yet definitively experimentally verified,
it is reasonable to ascribe a systematic uncertainty to the slow control luminosity of $\delta_{\text{SC,abs}}=\pm 5\%$ absolute
and $\delta_{\text{SC,rel}}=\pm 2\%$ relative.

While not viable as a standalone precision measurement, the slow control luminosity provided a valuable approximate
 benchmark for the other luminosity monitors, and more
importantly the results of TargetSim provided a more accurate representation of the target gas distribution for use in the simulation.
The versatility and generality of the TargetSim code make it a good candidate for use in future experiments with molecular
flow regime gas targets and other applications.

\section{Luminosity Determined Using the 12$^\circ$ System}
\label{sec:12lumi}

To take advantage of the rapidly increasing elastic lepton-proton scattering cross section
at small lepton scattering angles ($\theta$) a dedicated tracking system consisting of two six-plane
telescopes was constructed as a means of providing a luminosity normalization point for
the measurement of \ratio in the main tracking volume.  While subject to possible differences
in the \ep and \pp elastic cross sections due to TPE, these effects are universally expected to be
small (i.e., less than the experimental precision) at the kinematics accepted by the telescopes.
This uncertainty is addressed in Section \ref{ss:12sys} and an extraction of the value of
\ratio from the system using an independent luminosity determination from the SYMB is presented
in Section \ref{sec:12TPE}.

The detectors and trigger of the 12\dg telescopes are described in Chapter \ref{Chap3}, while
this section addresses the analysis of the data from the system.  The analysis included full representation of the system in
the OLYMPUS Monte Carlo simulation, hit reconstruction,
track reconstruction, and event selection to produce final yield of elastic \pmp events for the luminosity
determination.  Additionally, this section discusses the performance of the system, the comparison of reconstructed
data to Monte Carlo simulation, and the resulting luminosity extraction.

Please note that throughout this section, ``12$^\circ$'' will be used as a shorthand to refer to events in which the lepton
is reconstructed in the aforementioned tracking telescopes and to the detector system as a whole, even though the actual lepton scattering angles
accepted by the telescope varied over a range of several degrees around $\theta=12^\circ$.

  \subsection{Principle of the Measurement}
  
\begin{figure}[thb!]
  \centerline{\includegraphics[width=1.0\textwidth]{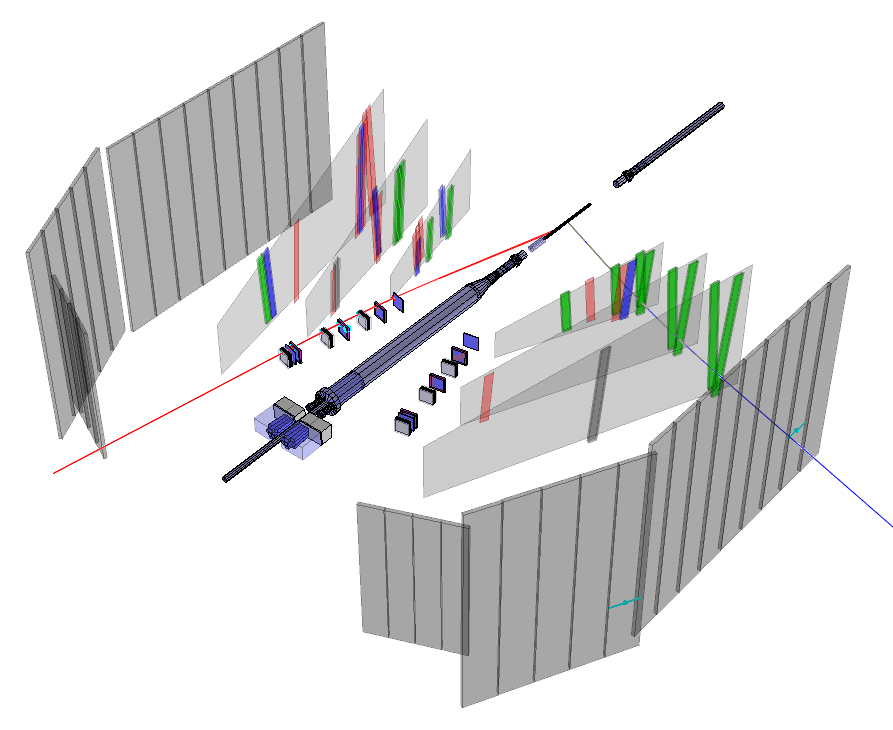}}
  \caption[Event display of a typical 12\dg telescope event]{A reconstructed elastic \ep event where the electron (red track) was detected in the right 12\dg telescope
  and the proton (blue track) in the left drift chamber.  Note the proper rejection of uncorrelated hits in the drift chamber to find the good selection of hits that
  corresponded to the proton track.  The scattering chamber and toroid coils have been removed from the display for clarity.}
  \label{fig:typ12}
  \end{figure}

  Since the cross section for elastic \pmp scattering increases rapidly as $\theta$ decreases, scattering events at forward angles are a 
  natural choice for statistically precise luminosity measurements due to the high rate of events that can be sampled.  With this principle
  in mind, OLYMPUS included dedicated forward tracking elements to expand the acceptance for elastic \pmp events to scattering angles as small as 9\dg for positrons
  and 11\dg for electrons.  The most backward-going recoiling protons from elastic \pmp events where the lepton is accepted by the 12\dg telescope had $\theta\approx76^\circ$,
  meaning they were within the drift chamber acceptance described in Section \ref{sec:spect}.  This allowed the exclusive reconstruction of elastic events, providing
  a strong lever against background contamination while maintaining a high statistics data sample.  Figure \ref{fig:typ12} shows an event display of such an event
  from the data set.
  
  In essence, the determination of the luminosity from the 12\dg system involved the reconstruction of possible particle hits and tracks in the telescopes and drift chambers for events
  passing the 12\dg trigger (Section \ref{ss:12dtrig}), and then testing the resulting \pmp pairs (selecting on the beam species for elastic kinematics via a series
  of cuts (Section \ref{sec:12ana}).  Similar to the main \pmp analysis method (Section \ref{sec:mainana}), this procedure was followed for both experimental data
  and digitized Monte Carlo data.  Then, for a given simulated integrated luminosity $\mathcal{L}_\text{MC}$ corresponding to a set of data (correctly simulated beam parameters,
  detector efficiencies, etc.), the measured integrated luminosity in the 12\dg system for the set of data is simply a function of the number of elastic events $N$ accepted in data and simulation
  and the simulated luminosity:
  \begin{equation}
   \mathcal{L}_{\text{12}^\circ} = \frac{N_\text{data}}{N_\text{MC}\left(\mathcal{L}_\text{MC}\right)} \cdot \mathcal{L}_\text{MC}.
   \label{eq:l12}
  \end{equation}
  While simple in principle, this method requires a deep understanding of the conditions under which data were taken so that the simulation
  properly replicates any conditions that could have affected the elastic event yield.  While the simulation makes every attempt
  to faithfully reproduce the data, as described in Section \ref{sec:sim}, simulation parameters are in general considered as possible
  sources of systematic uncertainty and are exhaustively analyzed in Section \ref{ss:12sys}.
  
  In a single data run file (typically with $\sim$1.0$\cdot 10^6$ triggers, lasting $\sim$20 minutes), approximately 
  10,000 total accepted \ep events or 19,000 accepted \pp events (due to the differences in the trigger noise conditions between beam species)
  were recorded.  Since simulation could be run to arbitrary statistical precision, the data rate determined the overall statistical uncertainty
  on the 12\dg luminosity estimate: approximately 1\%/run or 0.5\%/hour.  Combining the entire data run, the statistical precision
  is on the order of 0.01\% and is thus negligible compared to various systematic uncertainties.
  
  \subsubsection{Constraints of a Single Arm Measurement}
  
  For the purpose of the analysis presented in this work, the luminosity was determined via the use of exclusively reconstructed \pmp events
  rather than with inclusive events in which the lepton is reconstructed in one of the 12\dg  telescopes, but no requirement is placed on the proton.  While
  the latter has the advantage that it would completely separate the 12\dg measurement from dependence on drift chamber data and make its independence
  as a monitor more robust, the differing beam environments between $e^-$ and $e^+$ running made a single-arm inclusive measurement extremely difficult.
  In particular, during electron running the rate of hits in the ToF bars that were part of the 12\dg system trigger were considerably higher
  than during positron running, which led to an increased rate of triggers from non-elastic events for $e^-$ running.
  
  When attempting a single-arm
  analysis, it was found that the ratio of the positron rate to the electron rate was approximately 2\% lower than the ratio found by an exclusive
  analysis due to this increased background in the electron sample.  While using information such as ToF meantime and energy deposition could
  recover some of this difference, it could not discriminate against events with multiple ToF hits in the 12\dg trigger window and the simplicity of the 
  ToF-only data did not provide a strong separation of good elastically scattered protons.  Additionally, as will be discussed in Section \ref{sec:hahahaha},
  the final analysis made use only of the MWPCs as tracking elements.  Due to this, the resolution on the kinematic parameters of the reconstructed
  lepton was quite limited.  This made a background subtraction scheme like the one used for the main analysis (Section \ref{sec:backsub}) a dubious
  approach.  Due to this, it was determined that the systematic uncertainty inherent in an inclusive measurement was greater than the systematic
  uncertainty introduced by requiring proton track information from the drift chambers.  Thus, this work predominantly considers the exclusive measurement.
  
  \subsubsection{Possible Contribution of TPE}
  \label{sec:12posstpe}
  
  Fundamentally, the measurement of the luminosity using elastic \pmp scattering in the 12\dg system as a normalization point for the measurement of
  \ratio is limited by the fact that it involves the same physics process that is being examined for TPE contributions.  While the vast majority of theoretical
  and phenomenological models predict that TPE should be small (or at least negligible compared to systematic effects) in the kinematic region accepted
  by the 12\dg system \cite{BerFFPhysRevC.90.015206,Chen:2007ac, Guttmann:2010au,Blunden:2003sp,Afanasev:2005mp,Chen:2004tw,
  Blunden:2005ew, Kondratyuk:2005kk, Borisyuk:2006fh,
  TomasiGustafsson:2009pw}, there exists very little experimental evidence to validate these predictions.  The effect of this assumption is considered as a systematic uncertainty
  for the luminosity determination, and is discussed in detail in Section \ref{ss:tpe12sys}.  Also considered in this work, however, is the measurement of the value of \ratio in
  this kinematic region, using the SYMB system to provide the luminosity normalization, and this is discussed in Section \ref{sec:12TPE}.
  
  \subsection{Discussion of the Exclusion of the GEM Detectors}
  \label{sec:hahahaha}
  
\begin{figure}[thb!]
  \centerline{\includegraphics[width=1.15\textwidth]{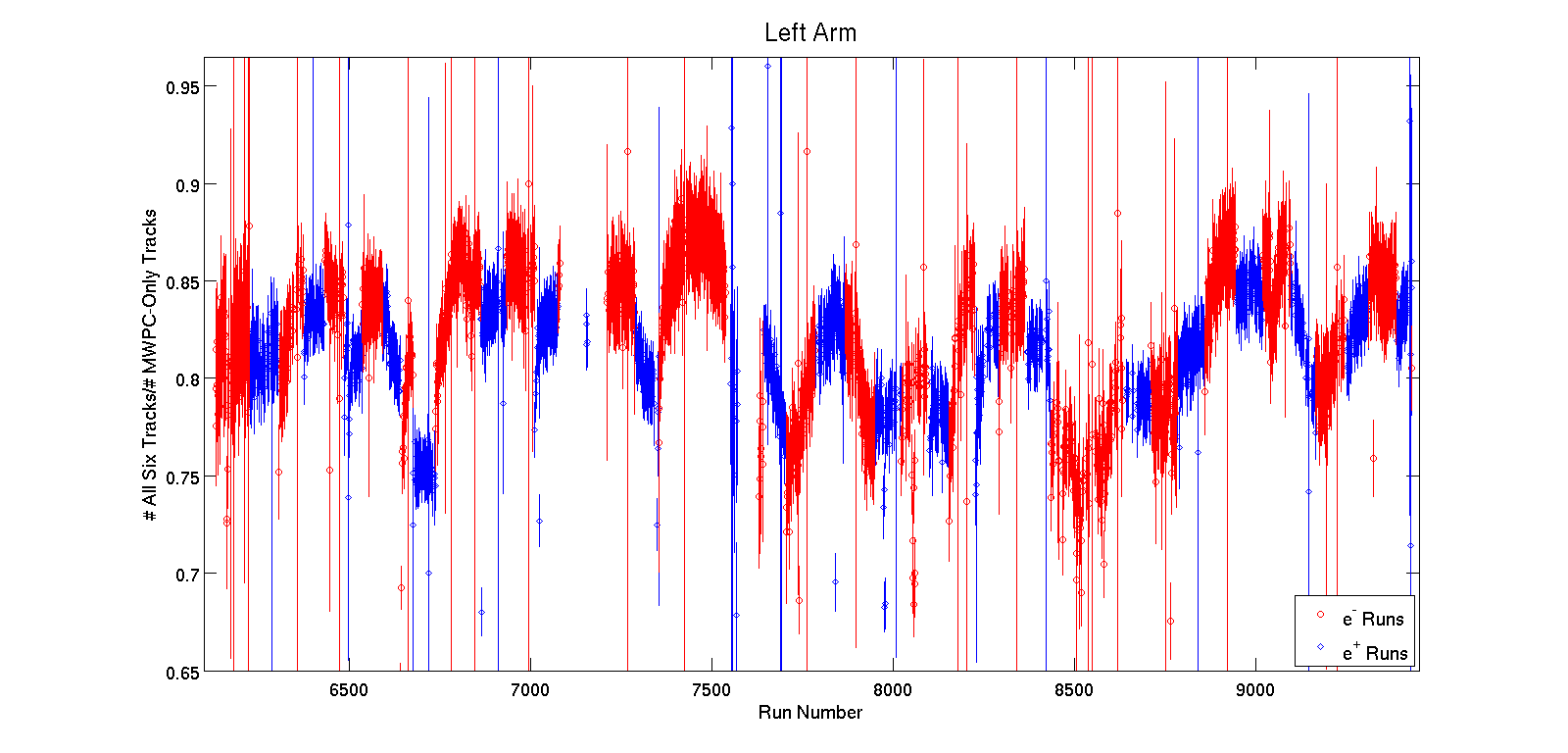}}
  \caption[Time-varying efficiency of the GEM detectors]{Ratio of accepted six-plane 12\dg system tracks in the left arm to the number of accepted MWPC-only three plane tracks
  as a function of OLYMPUS run index.  The strong structures that vary with run index indicated a severe problem with the GEMs, which prevented their use for the final luminosity
  analysis.}
  \label{fig:gemblow}
  \end{figure}

  As previously noted, the GEM detectors that were part of the 12\dg telescopes were not utilized in the luminosity analysis presented in this work.  When closely examined
  in the course of studying their performance during data-taking, it was found that they exhibited a strong, time-dependent variation in their efficiency for detection
  of particles on the order of 10\%.  The essence of this issue is illustrated in Figure \ref{fig:gemblow}, which shows the variation in the number of six-plane accepted tracks (i.e., events in which
  each GEM contributed a hit) relative to the number of accepted tracks using only the MWPC hits.  While it would be expected that few six-plane tracks would be found (both due
  to the smaller acceptance covered by all six planes relative to the MWPCs only and the influence of single plane inefficiencies), it would be expected that the value of the ratio
  would be constant to within statistical variation over the course of data-taking. Furthermore, this variation in efficiency was
  found to be almost entirely correlated between the GEM planes within a telescope (i.e., all three GEMs dropped in efficiency
  together, possibly on an event-by-event basis).  This was discovered by noting that the structures present in Figure \ref{fig:gemblow} persist even when relaxing the tracking conditions
  to demand only four planes.  To require four planes, however, at least one GEM must supply a hit for the track.  Requiring even a single GEM induced the structure in the measured
  yield of elastic events, indicating that the efficiency of the GEM planes varied in a strongly correlated way.
  
  When this was discovered, a large effort was undertaken to attempt to determine its cause and rectify the issue to permit the use of the GEMs in the 12\dg
  analysis.  The algorithms for hit-finding in the GEMs (Section \ref{sec:12hit}) and track reconstruction in the 12\dg telescopes (Section \ref{sec:12track}) were completely redone
  in an attempt to solve the problem.  A great deal of improvements were made through this process, including improving the overall efficiency of the GEM planes and tracking, but
  no changes that were made significantly affected the observed time dependence in the GEM hit yields.  While the various changes in the efficiency over time could be roughly
  correlated with changes in beam conditions, these effects were not sufficiently quantifiable to produce a solution.  Based on this, it is theorized that the root cause of the issue
  was a saturation effect in the GEM readout electronics that caused a varying, unknown deadtime for GEM hits, which gave rise to the observed inefficiency effects.
  Ultimately, the exact cause of this time-varying efficiency was not definitively identified, but seemed to be associated with the readout system of the GEMs and its behavior
  as a function of hit rate/beam conditions.  This should be diligently kept in mind when considering the use of the OLYMPUS GEMs in future experiments  \cite{refId0,Balewski:2014pxa}.
  
  Due the correlation in this efficiency variation, the GEMs were fundamentally prevented from measuring event rates (absolute or
  relatively).  Any attempt to correct for the time-variation (i.e., by varying the simulated efficiency of the telescope and/or planes over time)
  would amount to a manual scaling of the simulated yield, effectively nullifying the measurement of the elastic rate as an indicator of the luminosity.
  While it is conceivable that GEM hits could be used if present without requiring them, this still would require implementing the time variance in the efficiency in
  simulation in order to replicate the tracking resolution of the data in the simulation. Due to the uncertainty this would introduce if not properly accounted for 
  (and the fact that it would still introduce a manually-inserted variation in detector response into the simulation), it was decided that the GEMs could not be
  part of the actual yield determination in the 12\dg system.  Hit information from the GEMs was still useful in tracking for applications in which absolute rate information
  was not required, such as measuring the efficiencies of the MWPCs and SiPM scintillators (after checking to ensure no cross-system correlation was present) and measuring detector
  misalignment using tracks.
  
  \subsection{Hit Reconstruction}
  \label{sec:12hit}
  
    The first step in reconstructing particle trajectories for particles in the 12\dg system was to generate hit positions in the detector planes from the system's raw data.
    A new hit-finder for the GEMs was written ``from-scratch'' for the GEMs, partially in an attempt to remedy the issues described in Section \ref{sec:hahahaha}.  While the new algorithm
    did not remedy the time-dependent issue, it did significantly improve the performance of the GEMs in many regards.  Due to this, and the fact that the GEMs may be used in future
    experiments, this new hit-finder is described here.  For the MWPCs, a functional hit-finder was available based on the code used for the HERMES experiment MWPCs \cite{Ackerstaff1998230},
    and so only the basic elements of this algorithm and the improvements made are discussed.
  
    \subsubsection{GEM Detectors}
    \label{sec:gemhit}
    
\begin{figure}[thb!]
    \centerline{\includegraphics[width=0.95\textwidth]{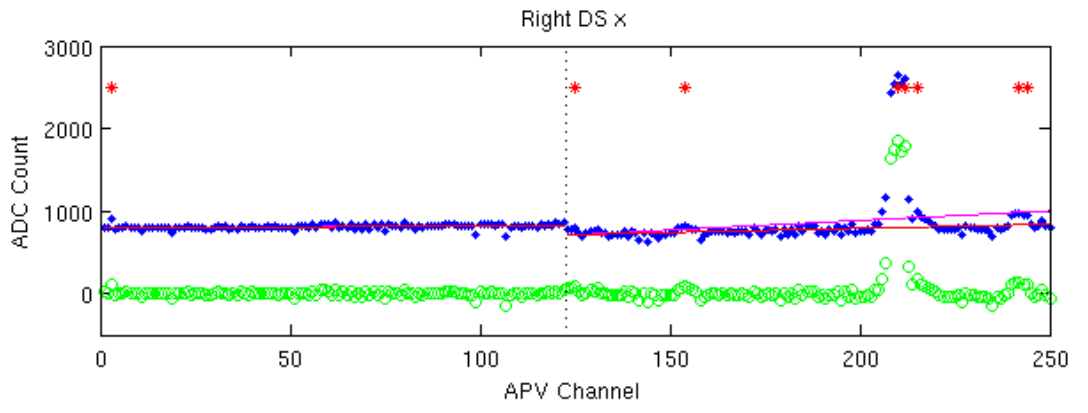}}
    \caption[Example of a 1D GEM hit in the raw ADC data]{An example of how a charged particle passing through one of the OLYMPUS GEM planes creates a signal visible in the 1D data from one of
    the strip/pad patterns on the readout plane.  The blue points correspond to the raw ADC counts registered on each ADC change (each channel corresponds to a single readout strip or connected
    strip of pads), and the green represents the ADC data after a baseline/pedestal subtraction.  Red points indicate a first pass attempt at identifying local maxima in the data as a first step in
    hit-finding.  In this event, a clearly identifiable hit occurs around APV Channel 210 with a possible smaller hit around Channel 240.  The vertical dotted line represents the break between the two
    separate APV cards used to readout a single dimension of the plane.}
    \label{fig:g1d}
    \end{figure}

    While the GEMs were not used in the final analysis, a large effort was undertaken to improve their hit-finding routines relative to the original algorithms used, not only in the process
    of attempting to solve the issues discussed in Section \ref{sec:hahahaha}, but also to increase their resolution and efficiency for usage in calibration analyses.  The two crossing
    patterns of the readout planes of the GEMs (described in Section \ref{sec:gemdet}) essentially provided independent 1D hit information.  An example of such a 1D hit is shown in 
    Figure \ref{fig:g1d}.  The basic strategy used in hit-finding for GEM detectors like the ones used in OLYMPUS is to find hit candidates (peaks/maxima) in the 1D data samples,
    and then combine 1D hits in the same planes to form 2D hit candidates.  Note, however, that the number of candidates scales multiplicatively with the number of 1D hits from each
    axis.  To combat this, most GEM systems (including the OLYMPUS detectors) are designed to share the charge signal as equally as possible between the dimensions of the readout so
    that the magnitude of the 1D hits can be compared so as to indicate better which pairs of of 1D hits go together to properly form a 2D hit \cite{1748-0221-7-03-C03042,kgem}.  
    
\begin{figure}[thb!]
    \centerline{\includegraphics[width=1.4\textwidth]{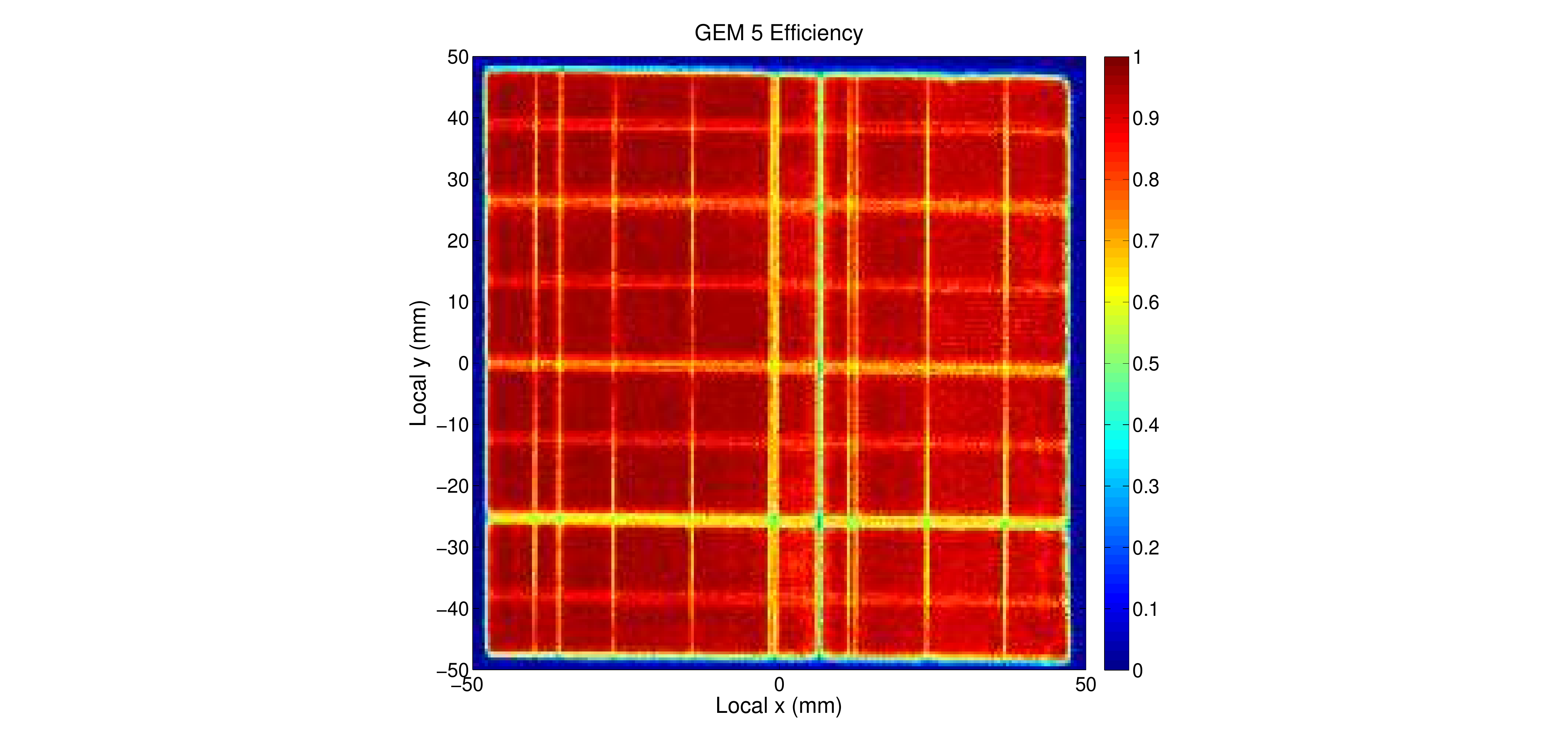}}
    \caption[Example GEM efficiency map using a previous hit-finding algorithm]{Efficiency for hit reconstruction of GEM 5 (the furthest downstream GEM in the right telescope), using
    the previous hit-finding algorithm.  Note the periodic occurrence of inefficient regions, as well as other random strips with low efficiencies caused by the issues discussed in the text.
    The improved efficiency of this GEM, as well as of the others, is shown in Figure \ref{fig:gemeff}.}
    \label{fig:badeff}
    \end{figure}

    The OLYMPUS GEMs presented several specific challenges with regard to hit-finding that were addressed in the improved hit-finder (relative to the original software used on the experiment)
    that is described here:
    \begin{enumerate}
     \item strips with weak amplification and/or bad data transfer, which occurred both randomly among the ADC channels and in a periodic pattern at the edges of the four connectors
           used used to read out each APV (see as a grid pattern in Figure \ref{fig:badeff}),
     \item different ADC baseline/pedestals between the two APVs covering a single plane dimension, which made it difficult to reconstruct hits at the boundary between two APVs (seen
           in the blue points in Figure \ref{fig:g1d}),
     \item channel-to-channel ADC pedestal variation on top of the common-mode variation, and 
     \item other large-scale problems such as failed APV cards on the middle GEM in the left telescope.     
    \end{enumerate}
    The new hit-finding algorithm addressed these issues to significantly increase the overall efficiencies of the detectors and to reduce the structural inefficiencies shown
    in Figure \ref{fig:badeff}, but did not affect the time-dependent behavior of the GEM hit yields discussed in Section \ref{sec:hahahaha}.
    
\begin{figure}[thb!]
    \centerline{\includegraphics[width=1.15\textwidth]{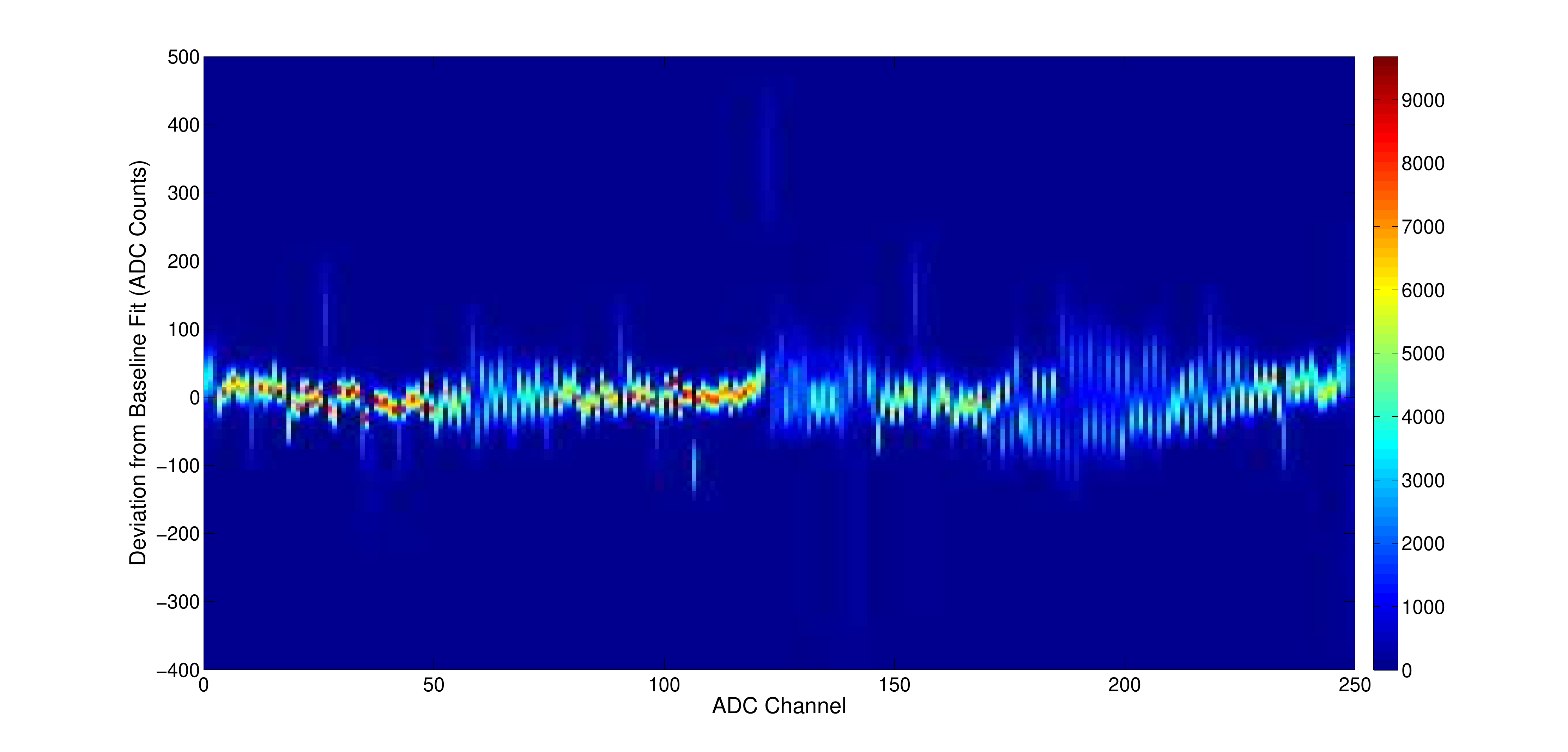}}
    \caption[Channel-to-channel noise in the GEM ADC counts]{Histogram of the average deviation of the ADC count in each channel along one dimension of one of the GEMs.  As can be seen
    channels typically exhibited a clear mean deviation that could be used to correct the pedestal subtraction for that channel and a definable width that could be used to set the noise
    threshold for hit candidates for each channel.}
    \label{fig:c2c}
    \end{figure}

\begin{figure}[thb!]
    \centerline{\includegraphics[width=1.00\textwidth]{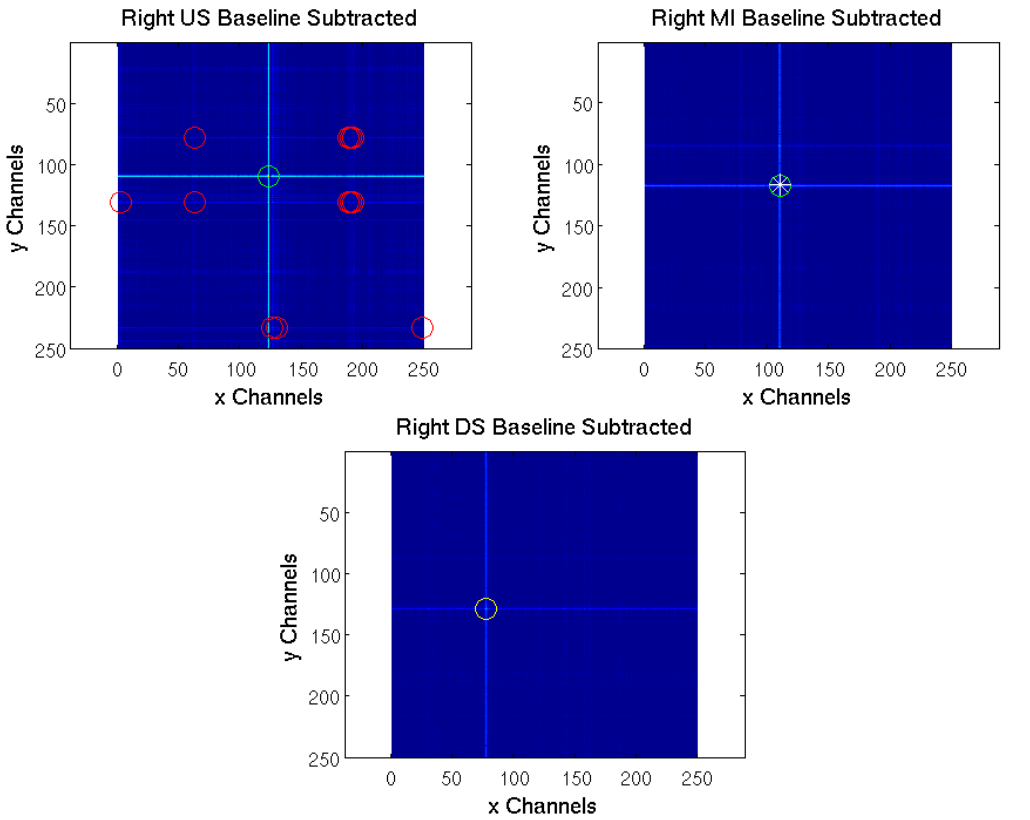}}
    \caption[Hits found by the new GEM hit-finder for a single event]{Data from the right GEM telescope for a single event (2D histograms of ADC counts) showing the hits identified
    by the original GEM hit-finding algorithm (white stars) and the new algorithm (circles, colored red, yellow, or green by increasing hit score).  The three strongest hits from each
    plane reconstruct well to an elastically scattered lepton that would have been missed by the old algorithm due to the missed hits in the upstream (US) and downstream (DS) planes.  The upstream
    hit was missed in the old algorithm due to it $x$ location near the APV card boundary, while the downstream hit was missed due to its relatively weak signal and being near the $y$ APV card
    boundary.  For the latter, the new baseline subtraction methods allowed the hit to be separated from the noise.}
    \label{fig:ghits}
    \end{figure}

    The new GEM hit-finder proceeded as follows\footnote{The source code (C++) for this hit-finding algorithm, along with basic documentation, is available from the author for academic
    applications.}:
    \begin{enumerate}
     \item Raw channel-by-channel ADC counts were mapped to their corresponding local coordinates in the GEM planes ($x$ and $y$).
     \item A first pass was made over all events in a given data run file in which a 12\dg trigger fired in which a line was fit to the ADC count as a function of channel number (properly ordered)
           for each APV card (two cards per dimension per plane) to calculate the common-mode pedestal for the channels.  A rudimentary 1D peak-finding algorithm was used to remove points
           near a possible hit so as to avoid biasing the baseline fit.  This procedure is illustrated by the magenta (without peak removal) and red (with peak removal) lines shown in Figure
           \ref{fig:g1d}.  The previous hit finding algorithm did not correct for possible hits in the baseline removal, which made the procedure prone to rejecting hits, especially
           near the edge of an APV card.  This created regions of strong inefficiency along the central axes of the planes.
     \item After the first pass over the data, the mean and variance over all baseline fits of the separation of a single ADC channel count from the fit baseline on each event were calculated.
           While the baseline fluctuated event-by-event, it was found that single channels exhibited predictably low or high counts relative to neighboring channels with a width that varied by
           channel as well, as can be seen in Figure \ref{fig:c2c}.  The mean deviations and width of the distributions for each channel saved to adjust the pedestal subtraction from the value
           suggested by the baseline subtraction alone.  No such channel-to-channel correction was made in the previous algorithm, which allowed the new algorithm to recover hits from channels
           with lower average counts and reject spurious highs from high-count channels.
     \item The data were then passed-over a second time, again event-by-event.  Each ADC channel was adjusted by the baseline+deviation pedestal computed in the first pass.  Local maxima in the ADC
           channels were identified (using thresholds based on the widths from Step 3), and all pairings of $x$ and $y$ minima were considered.  Pairings were scored based on the relative match of
           the $x$ and $y$ signal strengths as well as the absolute total strength of the pairing.  The user then chose how to select which hits to pass to the tracking algorithm from among the scored
           hits (e.g., with minimum accepted scores, passing the highest-scoring hits, etc.).  A visualization of identified hits with comparison to the hits identified by the old algorithm is
           shown in Figure \ref{fig:ghits}.
    \end{enumerate}
    This new hit-finder succeeded in increasing the overall efficiencies of the GEMs by 5-10\% and greatly reduced the structure in the efficiencies across the planes. The resulting efficiencies
    are shown in Figure \ref{fig:gemeff} in Section \ref{sec:12eff}. While the time-dependence of the GEM efficiencies did not ultimately arise from a hit-finding issue, this improved hit-finder
    was useful in allowing the GEMs to provide better data for calibration of the 12\dg system and should be useful for future experiments that use the OLYMPUS GEMs.

    \subsubsection{MWPC Detectors}
    
\begin{figure}[thb!]
      \centerline{\includegraphics[width=0.75\textwidth]{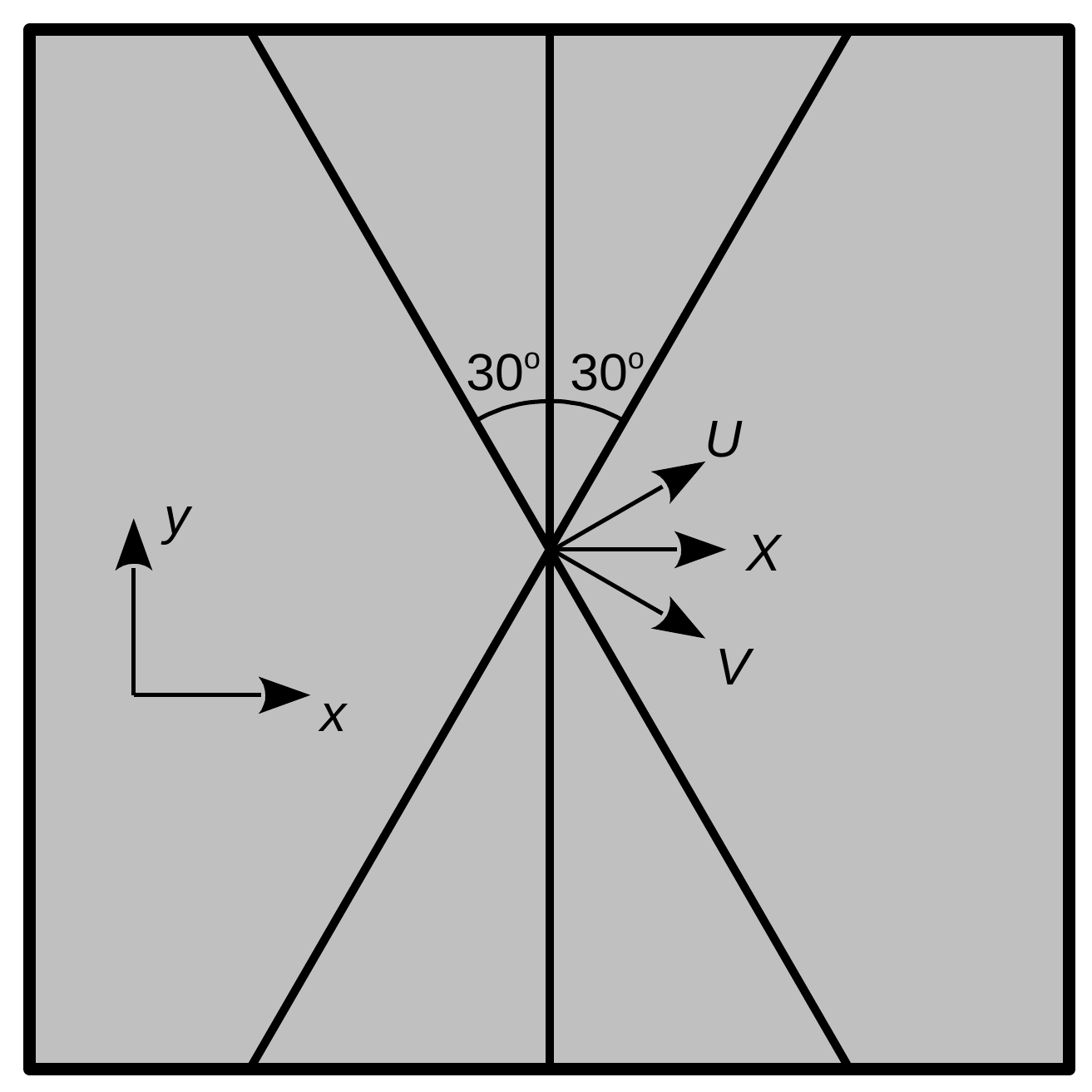}}
      \caption[Local coordinate system for the MWPC planes]{Local coordinate system used for the MWPC detector hit finding.  The $U$, $X$, and $V$ coordinates measure distances
      along the perpendiculars to the different wire orientations, which may be converted via linear combinations to the local coordinates $x$ and $y$ that are used for the final
      hit positions.  The local $x$ coordinate points away from the beamline in both telescopes, while $y$ points up in the left MWPCs and down in the right MWPCs.}
      \label{fig:uxv}
      \end{figure}

    Due to the relative simplicity of the MWPC hit information (see Section \ref{ss:mwpcdet}), hit-finding for the MWPCs did not require
    the same complexity of analysis as the GEM hit-finder.  Initial hit decisions for single wires were produced by software adapted from the software used for the HERMES experiment
    MWPCs \cite{Andreev:2001kr,Ackerstaff1998230}
    adapted by the PNPI group.  While, in principle, single wire 1D hits could be passed to the track reconstruction algorithm used for the 12\dg system, the combination of 
    information from the three wires in a chamber provided an important means of rejecting noise hits.  Recalling that the MWPC wires in chamber were arranged in three planes
    such that the wires were oriented at $-30^\circ$, $0^\circ$, and $+30^\circ$ relative to vertical in the three planes, local coordinates $U$, $X$, and $V$ may be defined
    perpendicular to the wire orientations in each plane, in additional to local $x$ and $y$ coordinates perpendicular to the sides of the chamber face.  This system is illustrated
    in Figure \ref{fig:uxv}.  The single wire hit decisions were given corresponding $U/X/V$ coordinates, which can then be converted to the plane coordinates via the following
    linear combinations:
    \begin{equation}
     x_X = X,
    \end{equation}
    \begin{equation}
     x_{UV} = \frac{U+V}{\sqrt{3}},
    \end{equation}
    \begin{equation}
     y_{UV} = U - V.
    \end{equation}
     Note that there are only three linearly independent combinations that form $x$ and $y$ since there are three inputs $U$, $X$, and $V$
     (e.g.; $y_{XV} = \sqrt{3}X-2V = \sqrt{3}\left(x_X-x_{UV}\right)+y_{UV}$).  In constructing MWPC hits, all combinations of hits on the $U$, $X$, and $V$
     planes within a single chamber on a given event were considered.  Any combinations with $\left|x_X-x_{UV}\right|>4$ mm were rejected as bad combinations.  The 4 mm
     cutoff was chosen by studying tracks that were reconstructed with the other five planes of the telescope, and identifying hits in the unused sixth plane that appeared
     correlated with the track (i.e., that were good hits).  The boundary was chosen to ensure that such good hits were not cut.  While this cut was wide, it was chosen
     to exclude as few good hits as possible since hits from all three MWPCs in a telescope were required for tracks in the final analysis.  The resulting higher hit multiplicity was handled
     by the tracking algorithm described in the next section.
     
     The final local hit coordinates for a given plane that were passed to the tracking algorithm were a weighted average of the independent $x$ constructions (since $x$ corresponded to
     the bending direction of the field and thus was the direction in which maximum resolution was desired), while $y_{UV}$ was used for the $y$ coordinate.  That is:
     \begin{equation}
      x = \frac{2x_X+3x_{UV}}{5},
      \label{eq:mwpcx}
     \end{equation}
     \begin{equation}
      y = y_{UV}.
      \label{eq:mwpcy}
     \end{equation}

  \subsection{Tracking in the $12^\circ$ Telescope}
  \label{sec:12track}
  
  While the reconstruction of the scattered proton in 12\dg events was conducted using the main drift chamber tracker
  described in Section \ref{sec:recon}, the 12\dg telescopes utilized a separate tracking system that allowed
  greater flexibility than the main tracker.  This was possible due to the relative simplicity of reconstruction in
  the 12\dg telescopes compared to the drift chambers in which the reconstruction is complicated by the uncertainty
  in the drift time-to-distance calibration and the duplicity of each recorded hit.  While the drift chamber tracking was
  constrained to generate event vertices along the line of the beam due to its kinematic look-up library (``fasttrack''),
  the 12\dg tracker had full freedom to determine all five kinematic parameters of tracks passing through the telescopes.  This allowed
  the surveyed positions and orientations of the 12\dg system to be well-verified, since the tracking was able to reproduce measured
  beam positions and expected target distributions with no input of such information.
  
\begin{figure}[thb!]
    \centerline{\includegraphics[width=1.15\textwidth]{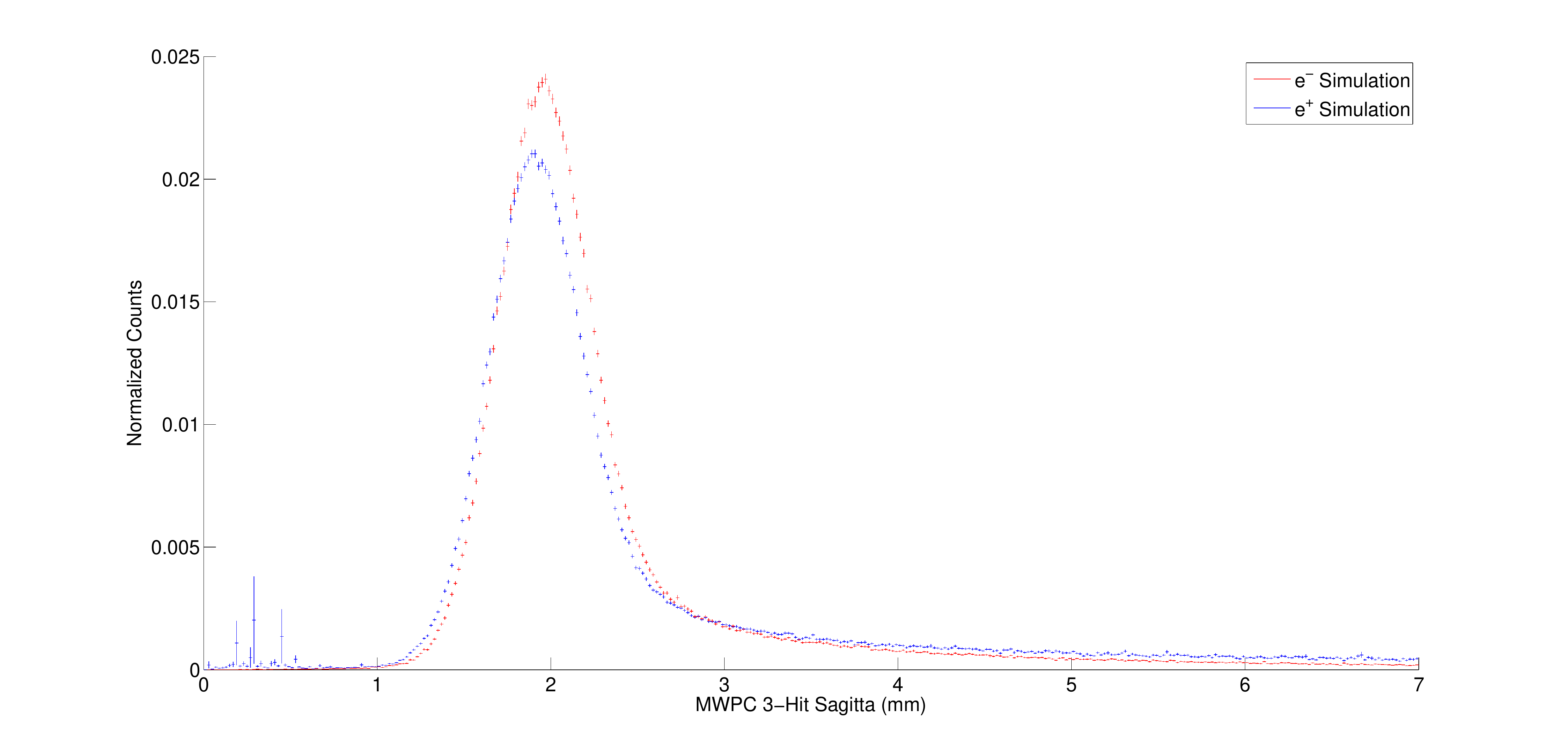}}
    \caption[Distribution of track sagittas in the MWPC telescopes (simulated)]{Distribution of the sagittas for track candidates consisting of three MWPC hits in simulation for each
    species in simulation.  The clear peak value, similarity of the distributions to those seen in data, and verification that events in the long tail of the distribution corresponded
    to unwanted events permitted the sagitta to serve as a useful cut against unwanted hit combinations to save time when tracking events.}
    \label{fig:mwpcsag}
  \end{figure}

  The first stage of the 12\dg tracker was a candidate forming algorithm, which was necessary due to the very high rate of 
  particles scattered in the regions forward of the drift chambers (either from the target region or upstream of experiment) that
  caused a high rate of noise hits in the 12\dg detector planes.  For the tracking used in the final analysis, only the MWPC hits
  were used and so track candidates were first formed of all possible combinations of a hit in each of the three planes in the telescope
  for a given event.  For each of these three hit combinations, the sagitta (i.e., the distance of the middle hit from the line connecting
  the hits in the inner and outer planes) was computed.  Using a large library of simulated events created using the radiative elastic event generator
  for both lepton species, the expected distribution of the value of the sagitta of MWPC tracks for good events was constructed.  The distribution of
  sagittas in simulation is shown in Figure \ref{fig:mwpcsag}.  While the peak value of the sagitta for elastic events was slightly offset between the two
  species due to the magnetic field, both peaked near 2 mm and had similar widths.  Visual inspection of events with sagittas greater than 5 mm indicated
  that the vast majority of these events involved hard scattering from a metallic element of the detector, and thus these events were not good events
  for the purpose of the analysis.  Thus, for both species a cut was placed at 5 mm for the maximum value of the sagitta for a candidate to be passed to the track
  fitting algorithm.  In data, this cut 25\%--30\% of three-hit candidates (predominantly with sagittas greater than 10 mm, a region where simulation indicates there
  were no good events).  When tracking with the GEMs additional information regarding the correlation of hits positions in adjacent GEM/MWPC pairs could be used to
  further clean the track candidate sample; this is not discussed in detail here as it was not a component of the final analysis.
  
  The hits belonging to a selected candidate were passed as local coordinates in the detector planes to the tracking algorithm.  The tracking algorithm utilized
  the GEANT4E extension to GEANT4, that propagated tracks through the simulated OLYMPUS geometry and magnetic field according to the most probable energy loss rather
  than determining energy losses via Monte Carlo \cite{geant4e,Agostinelli:2002hh}.  In this scheme, a track propagated through a given geometry and field with given
  initial conditions behaves deterministically as it traverses the detector system.  This propagation was iterated with the initial conditions ($\theta$, $\phi$, $\left|\mathbf{p}\right|$,
  $y$, and $z$ at the event vertex) free to vary so as to minimize the residuals between the recorded hit positions in data and the propagated hit positions generated by
  the GEANT4E simulation.  The minimization of the residuals was conducted using a Levenberg-Marquardt minimization routine, as implemented by the C/C++ Minpack libraries \cite{levmar,minpack,cminpack}.
  While a relatively slow method of tracking compared to other algorithms, the relatively small number of 12\dg events (compared to those handled by the drift chamber tracking) in the OLYMPUS
  trigger sample allowed the use of the tracker to take advantage of its complete kinematic freedom in performing alignment calibrations, assessing resolutions, etc.
   
  \subsection{System Performance}
  \label{sec:12perf}
  
    This section discusses various relevant aspects of the performance of the 12\dg system during the OLYMPUS data runs and the methods used
    to make the performance assessments.  In general, the system performed consistently and well throughout data-taking (with the exception of
    the previously mentioned GEM issues).  The redundancy of the system with the inclusion of GEM data allowed precise assessment of the efficiencies
    and resolutions of the trigger and MWPC planes, providing high confidence in the simulation implementation of those detectors.
  
    \subsubsection{Detector Efficiencies}
    \label{sec:12eff}
    
      To provide an accurate measure of both the absolute and species-relative luminosities, it was critical to
      determine and properly simulate the efficiencies of the various components of the 12\dg system.  In general, these
      efficiencies were measured to very high precision using fully reconstructed events with the detector of interest
      removed, which was made possible by the high redundancy of the 12\dg telescopes. The methods and results of the
      efficiency determinations for each subsystem are described in this section.
    
      \paragraph{SiPM Scintillator Planes}
      
      The performance of the SiPM trigger planes was evaluated using the lead glass calorimeters and the associated
      trigger described in Sections \ref{sec:lg} and \ref{ss:12dtrig}.  Since electron and positron events were
      distributed differently across the trigger planes, it was important to measure any inconsistencies in efficiency
      across the planes for implementation in the simulation.
      
\begin{figure}[thb!]
      \centerline{\includegraphics[width=1.15\textwidth]{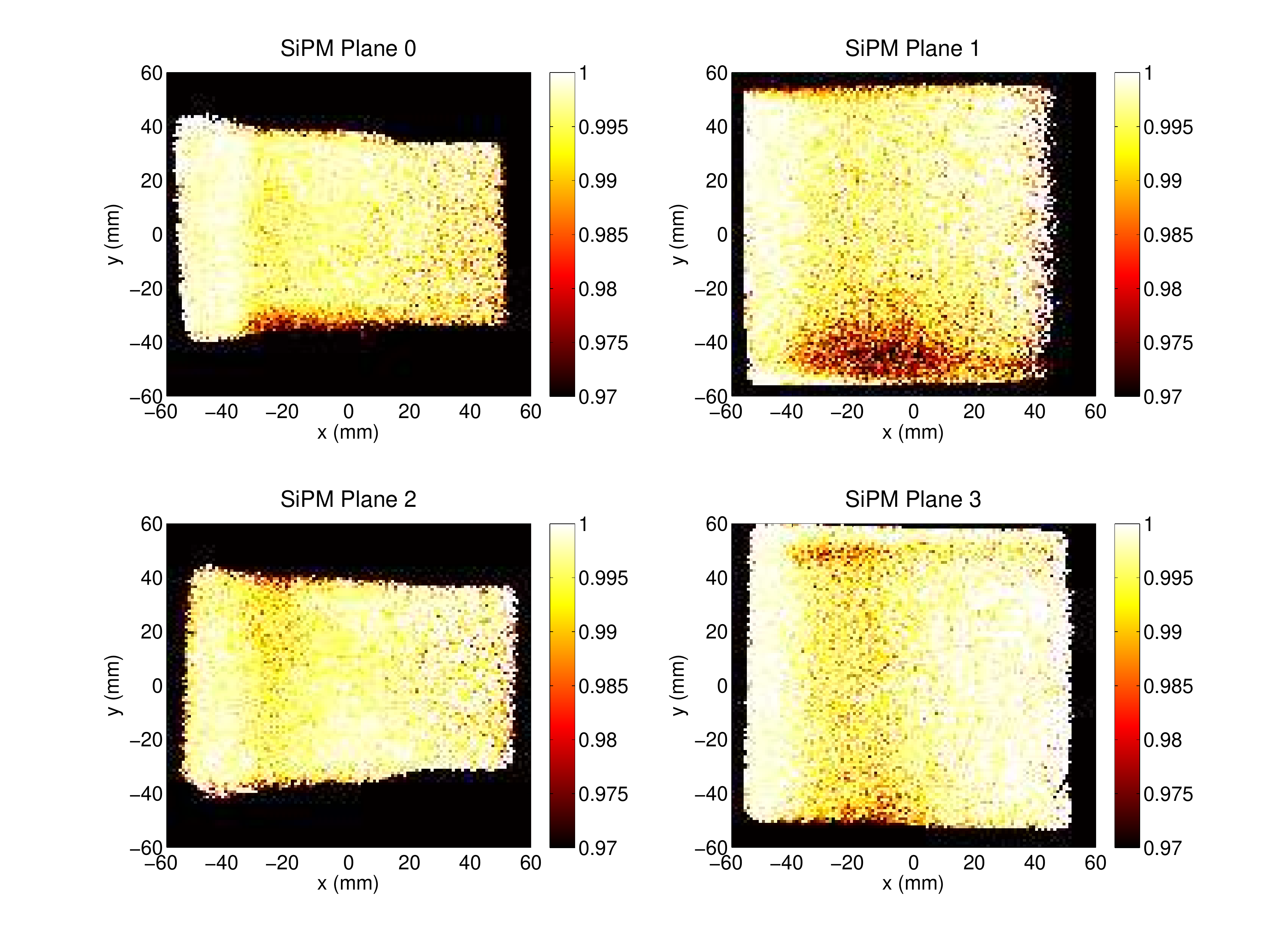}}
      \caption[Efficiencies of the SiPM scintillator tiles for the 12\dg trigger]{Efficiency maps for the four SiPM-instrumented
      scintillator tiles used for the 12\dg trigger.  The $x$ and $y$ coordinates in each plot represent the local detector coordinates
      of each plane, which follow the convention in which $x$ points away from the beamline and $y$ points up in the left sector
      and down in the right.  Note that due to acceptance constraints (determined by the outermost detectors in the
      telescopes (MWPCs 2 and 5), the inner planes (0 and 2) in each telescope are not completely illuminated by tracks.}
      \label{fig:sipmeff}
      \end{figure}

      A large sample of lead glass trigger events was compiled, and all possible six-plane tracks from the 12\dg telescopes
      in this sample were constructed.  Six-plane tracks were used so as to achieve the best projected position resolution
      for the tracks in the scintillator planes.  The standard elastic sample cuts for the 12\dg analysis (described
      in Section \ref{sec:12ana}) were used to ensure that the tracks represented good, relevant events for the purpose
      of the study.  For each track, the trajectory of the track was propagated using the GEANT4E tracker to the planes
      of the scintillator tiles.  For each plane, if the plane recorded a hit for that event the corresponding position
      bin was marked efficient for the event. Approximately 3 million tracked events per telescope were used to generate the efficiency
      maps shown in Figure \ref{fig:sipmeff}, which were implemented in the digitization of the 12\dg trigger for the
      simulation.  In the regions of the planes corresponding to the kinematic acceptance of the telescopes
      for tracks, the efficiency of the planes was in excess of 97\% and typically higher ($\gtrsim$99\%).  By comparing subsets
      of the data sample, it was determined that no significant time dependence or beam species/running condition dependence
      affected the SiPM plane efficiencies. 
      
      \paragraph{MWPC Detectors}
      
\begin{sidewaysfigure}
      \centerline{\includegraphics[width=1.05\textwidth]{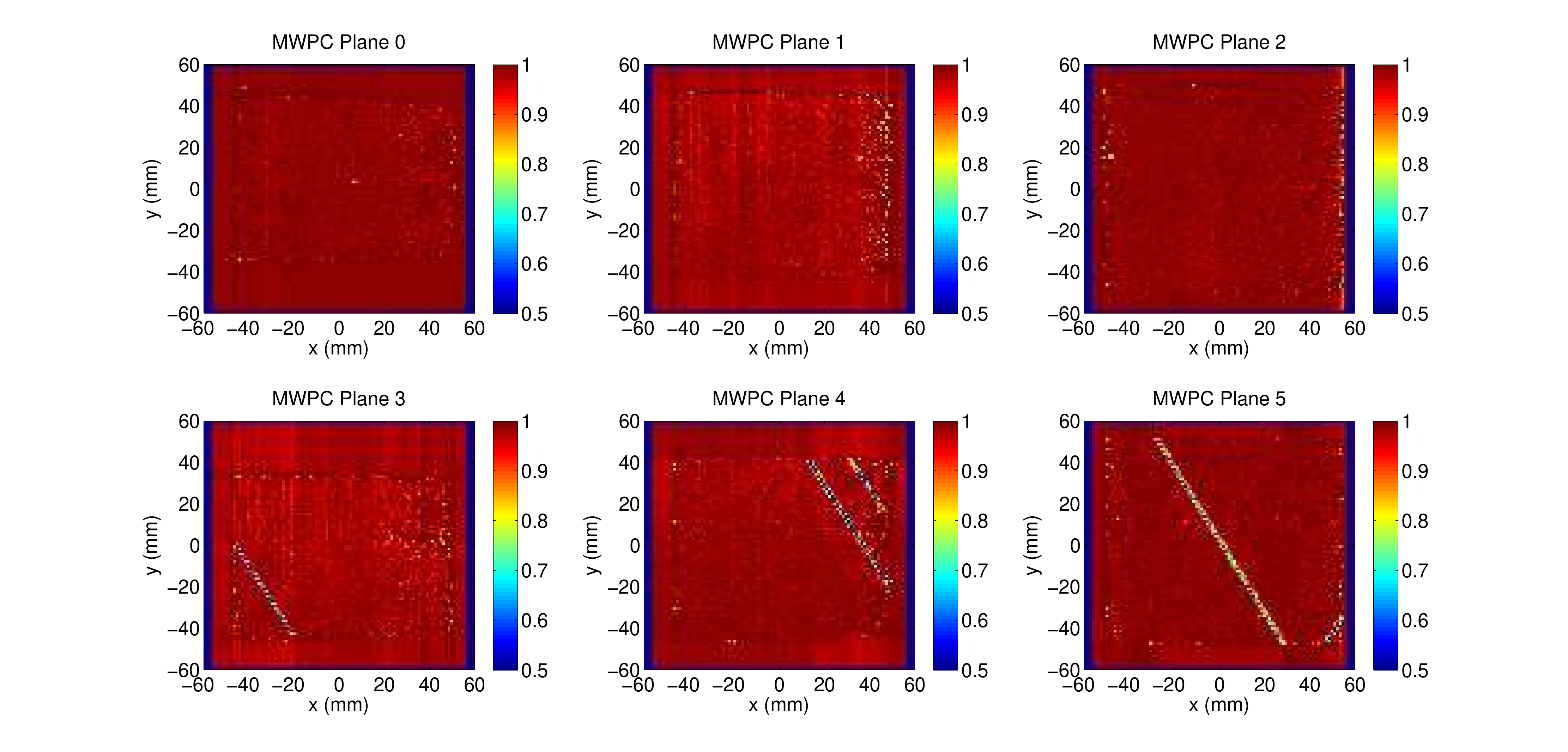}}
      \caption[Efficiencies of the MWPC detectors]{Efficiency maps for the six multi-wire proportional chambers used in the 12\dg
      luminosity telescopes.  The $x$ and $y$ coordinates in each plot represent the local detector coordinates
      of each plane, which follow the convention in which $x$ points away from the beamline and $y$ points up in the left sector
      and down in the right.  It is notable that all six chambers exhibit remarkably consistent and high ($>$98\%) away from the five known
      inactive wires in the right side detectors (Planes 3-5).}
      \label{fig:mwpceff}
      \end{sidewaysfigure}

      As the MWPC planes were critical to the ultimate measurement of the luminosity in the 12\dg system, proper implementation of their
      efficiencies in the simulation was also critical for the same reasons as for the SiPM plane efficiencies.
      Due to the redundancy of the 12\dg tracking system, it was straightforward to measure the MWPC plane efficiencies using the main
      trigger and tracking with hits from the other five planes in the telescope aside from the plane being assessed.  For a large 
      fraction of the data set (including data from all time periods used in the final analysis), tracks were reconstructed in the
      12\dg system using all five-plane combinations to check the efficiency of the sixth unused plane in each set using the same
      methodology as described for the SiPM planes.  The MWPC planes were found to have extremely high and consistent efficiency, both
      across the face of each plane and as a function of time.  Notably, the three planes in the right sector telescope had five known
      inactive wires (due to bad connections, malfunctioning ADC channels, etc.) that were easily identified by this analysis, serving as a proof
      of the validity of the analysis.  Figure \ref{fig:mwpceff} presents the efficiency maps for each plane.
      
      \paragraph{GEM Detectors}
      
\begin{sidewaysfigure}
      \centerline{\includegraphics[width=1.00\textwidth]{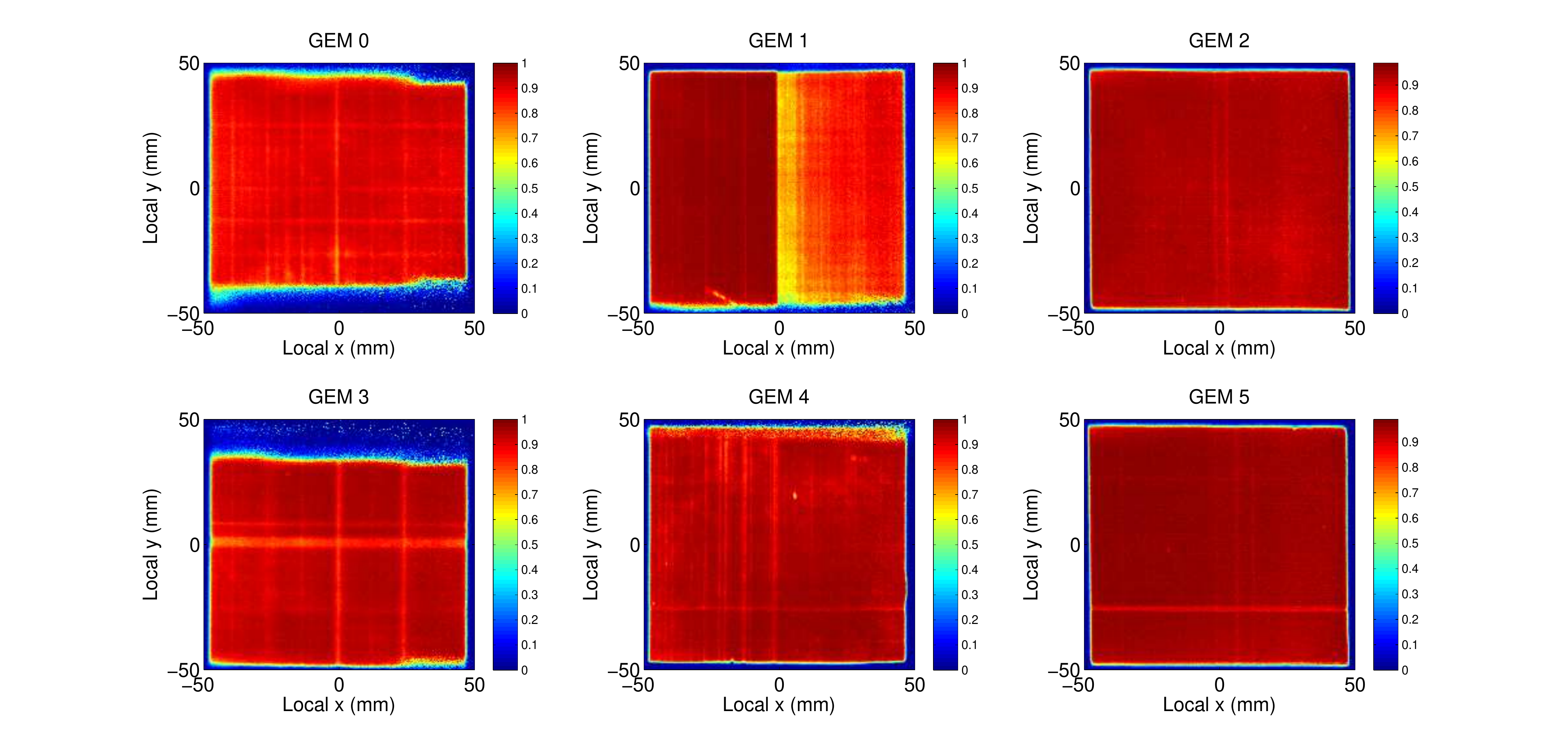}}
      \caption[Average efficiencies of the GEM detectors]{ Efficiency maps for the six GEMs used in the 12\dg
      luminosity telescopes, using the new GEM hit-finder described in Section \ref{sec:gemhit}.  Note that GEM 1 operated with a known bad APV for its $+x$ data (and no instrumentation for its
      $y$ data), and the inner GEMs (0 and 3) were not completely illuminated due to constraints in the acceptance from the other detector and trigger planes. The efficiency with the new algorithm was significantly better than
      that with the previous hit-finding method (Figure \ref{fig:badeff}).  The efficiency data shown here was sampled for specific runs, but was found to vary approximately uniformly across each
      plane as a function of time as discussed in Section \ref{sec:hahahaha}.
      The $x$ and $y$ coordinates in each plot represent the local detector coordinates
      of each plane, which follow the convention in which $x$ points away from the beamline and $y$ points up in the left sector
      and down in the right.  }
      \label{fig:gemeff}
      \end{sidewaysfigure}

      The same method used for the MWPC efficiencies was applied to measure the efficiencies of the individual GEM planes, the results
      of which are shown in Figure \ref{fig:gemeff}.  While the efficiency with the new hit-finding algorithm was improved significantly
      over that with the old algorithm (Figure \ref{fig:badeff}, the time-dependence problem of the overall efficiency was not solved, as previously
      discussed.  Some ``striping'' from weak channels remained in the new efficiency maps (as well as small regions of lower efficiency due to defects
      in the readout planes in GEMs 1 and 4, as well as the larger region in GEM 1 due to a known bad APV), but in general the new hit-finder recovered
      a significant number of hits in the regions in which the GEM data was challenging. 
      
    \subsubsection{Hit Resolution}
    
    The hit resolution in the MWPCs was fundamentally limited by the discrete nature of the detector's readout, which provided hits at specific wire locations spaced by
    approximately 1 mm. While the placing of the wires was not exact (due to the wires being soldered by hand to the detector frames), the uncertainty on this placement ($\mathcal{O}(0.1\:\text{mm})$)
    was considerably less than the wire spacing and randomly distributed about the nominal placement, and thus effectively negligible to the
    overall hit position resolution.  Propagating an uncertainty of 0.5 mm on each individual MWPC plane coordinates ($U$, $X$, and $V$) through Equations \ref{eq:mwpcx} and \ref{eq:mwpcy},
    and approximately accounting for the correlation of hits in the three planes results in the approximate reconstructed hit uncertainties of:
    \begin{equation}
     \Delta x \approx 0.25\:\text{mm},
    \end{equation}
    \begin{equation}
     \Delta y \approx 0.50\:\text{mm},
    \end{equation}
    in the local MWPC coordinates.  In general, this calculation was found to be consistent with both data and simulation.  Regarding the GEM hit resolution, a careful
    study was not completed for this work since the GEM hit-finder was not fully optimized after it became clear that the use of the GEMs in final luminosity analyses
    was precluded.  In general, however, resolutions of order \SI{50}{\micro\meter} are likely achievable.
    
    \subsubsection{Lepton Tracking Efficiency}
    
    A lower bound on the lepton tracking efficiency was found by comparing the number of simulation events for which the generated kinematics were within
    the cuts described in Section \ref{sec:12ana} and a candidate set of hits was produced according to the scheme discussed in Section \ref{sec:12track} to
    the number of successfully tracked leptons from such candidates.  This study found that the tracking algorithm produced a valid lepton track for 96.67\% of such candidates,
    with a very small lepton species difference (discussed in Section \ref{ss:12sys}).  Note, however, that this is a conservative lower bound on the tracking efficiency
    since many of the events missed correspond to events in which the lepton underwent a hard scatter from an element of the target or detector system.  This process occurs
    in both simulation and data, and such events cannot be reasonably reconstructed and used in an elastic \pmp event sample.  Visual inspection of such events indicated that
    tracks missed in the inefficient sample very frequently corresponded to such hard-scatter anomalies.
    
    \subsubsection{Lepton Tracking Resolutions}
    
    The resolution of the tracking in the 12\dg system (as it pertains to the analysis in which the GEM detectors were not included) was fundamentally limited
    by the resolution of the MWPC hit reconstruction rather than limitations of the tracking scheme.  Table \ref{tab:12res} summarizes some of the main tracking resolution
    parameters estimated from the combined dataset used in the analysis.  The resolutions were not found to vary between beam species, and the simulation digitization was
    developed so as to properly replicate the data resolutions to the extent possible.  Examples of the distributions of parameters presented in the table may be found in Section \ref{sec:12comp}.
    Since the MWPCs were designed to sacrifice resolution in the vertical direction for extra sensitivity to the horizontal direction (the direction of the field bending), the resolution
    in reconstructed $\theta$ was considerably better than that in $\phi$.  In general, the momentum resolution is quite wide.  While this complicates event selection, it also reduces
    uncertainties due to radiative corrections that could otherwise be large due to the relatively narrow acceptance of the telescopes.  Naturally, reconstructed parameters involving
    \pmp pairs convolve the resolutions of the 12\dg and drift chamber tracking, which is briefly discussed in Section \ref{sec:recon} but covered in more detail in References \cite{schmidt}
    and \cite{russell}. Inclusion of the GEM detectors improves the resolution by a significant
    factor (nearly an order of magnitude in some parameters), but no complete quantitative study of this has been completed.
    
      \begin{table}[thb!]
      \begin{center}
      \begin{tabular}{|l|c|}
      \hline 
      Parameter & Estimated Resolution \\
      \hline\hline 
      Vertex $y$ position & 3.5 mm \\
      \hline
      Vertex $z$ position & 50 mm \\
      \hline
      Lepton momentum & 310 MeV \\
      \hline
      $\theta$ & 0.1\dg \\
      \hline
      $\phi$ & 0.4\dg \\
      \hline
      \end{tabular}
        
      \end{center}
      \caption[Approximate tracking resolutions in the 12\dg telescopes]{A summary of approximate estimates of the resolution achieved for several kinematic parameters
      for the tracking of \pmp leptons in the 12\dg telescopes.}
      \label{tab:12res}
      \end{table}
    
  \subsection{Method of Analysis}
  \label{sec:12ana}
  
  While single-arm inclusive measurements of the luminosity using the 12\dg system were investigated, it was ultimately determined that an exclusive measurement (requiring
  the detection of the protons in the \pmp scattering events) was the method that could achieve the smallest overall systematic uncertainty in the final absolute and relative
  luminosity determinations.  This is primarily due to the fact that in a single-arm measurement, contamination from background (i.e., non-elastic \pmp events that mimic
  the signal of such an event (a $\sim$1.9 GeV forward lepton with a rearward ToF hit)) would be an appreciable fraction of the sample.  While a simple cut on the meantime of the ToF hit
  rejected a large fraction of this background, the relatively poor resolution associated with tracking only with the three planes of the MWPCs in each telescope made it difficult to properly
  model and subtract the remaining background due to random ToF hits in the appropriate timing window.  Since it was known that rates of random ToF hits during electron beam operation were
  notably higher than during positron running, it was determined that an inclusive measurement would entail a larger systematic uncertainty than an exclusive measurement (which introduces
  uncertainty from the tracking of the proton).  An exclusive measurement does, however, heavily suppress the background leaving a very clean \pmp sample.  Tests following the background
  subtraction scheme described in Section \ref{sec:backsub} determined that for the exclusive selection of 12\dg events the background fraction was $\mathcal{O}(0.1\%)$ and that the
  species-relative difference in the background was on the order of the statistics of the entire dataset ($\mathcal{O}(0.01\%)$) and thus negligible.
  
  The final analysis combined leptons tracked using the methods described in Section \ref{sec:12track} and protons tracked in the drift chambers using
  the elastic arms algorithm tracking scheme discussed in Section \ref{sec:track}.  The methodology and application to both simulation and data are described
  in the following sections.  This scheme produced the values of the event counts $N_\text{data}$ and $N_\text{MC}$ that were used in Equation \ref{eq:l12}
  to determine the integrated luminosity.
  
  \subsubsection{Exclusive Event Selection Scheme}
  
  The exclusive event selection proceeded as follows, looping over all data and simulated event triggers:
  \begin{enumerate}
   \item Determine if the trigger condition met the 12\dg trigger requirements (Section \ref{ss:12dtrig}), and reject the event if not.
   \item Using the information from the trigger (which determined which 12\dg telescope was active for the event), search for hit in
         one of the valid trigger ToF bars (Figure \ref{fig:trigcon}) with a large enough meantime ($\gtrsim12$ ns) to allow the possibility that
         the hit corresponded to an elastically scattered proton.  This cut was chosen so as to be only a very wide selection against fast lepton backgrounds
         from the target and to err on the side of inclusiveness rather than rejecting possible protons.  If no such ToF hit was present, the event was rejected.
   \item Create all possible pairing of properly charged leptons according to the track bending (i.e., $e^+$ or $e^-$ determined by the beam species) in the specified
	 telescope and identified proton tracks in the opposite side drift chambers.  If no such pair exists, reject the event.
   \item Select the pair from among the created pairs (multiple-pair events accounted for $\ll$1\% of events at this step), which minimizes the resolution-weighted sum of the cut 
         parameters listed in the next step.
   \item Check if the event passes the following cuts (the chosen values of these cuts based on a comprehensive systematic uncertainty analysis are described in the fiducial and 
         elastic cut portions of Section \ref{ss:12sys}):
   \begin{enumerate}
    \item Fiducial cut on the reconstructed position of the event vertex in global $y$ as determined by the lepton track (the proton tracks did not have flexibility in this dimension)
    \item Fiducial cut on the reconstructed position of the event vertex in global $z$ as determined by the average of the two tracks
    \item Elastic kinematics cut on the coplanarity ($\Delta\phi\approx180^\circ$) of the two tracks
    \item Elastic kinematics cut on the correlation of the placement of the global $z$ event vertex of the two tracks ($\Delta z$)
    \item Elastic kinematics cut on the beam energy reconstructed from the angles of the two tracks, assuming elastic kinematics ($E_{\text{beam},\theta}\approx 2010$ MeV)
    \item Elastic kinematics cut single-arm missing energy of the lepton over the square of the energy as computed by the expected
          elastic energy from the reconstructed $\theta$ ($\Delta E'_\theta/E'^2\approx 0$) 
   \end{enumerate}
   \item Add the event to the count of elastic events $N$ if all steps are passed.      
  \end{enumerate}
  Note that only one cut that considers reconstructed momentum was used due to the relatively poor momentum resolution of the MWPCs and the difficulty of precisely tracking
  particles in the back-angle portions of the drift chamber where the tracks traverse the wire layers nearly perpendicularly (resulting in few crossings between cells, which
  help to provide more precise track positions for tracking).  In the course of determining the final event selection method, a number of cut schemes were tested and the boundaries
  of the cuts were systematically tested as described in Section \ref{ss:12sys}.  This methodology was chosen for its robustness to the weaknesses of the 12\dg measurement (poor
  momentum resolution) and its stability under variance of the cuts.

  \subsubsection{Simulation of 12\dg Events}
  
  Simulated 12\dg events (used to produce $N_\text{MC}$) were treated identically to data events, following the analysis strategy discussed in Chapter \ref{Chap4}. The simulation
  was conducted in the same framework as described for the main analysis in Section \ref{sec:sim}, and
  the creation of digitized hits in the ToFs and drift chambers (i.e., values from simulation converted to mimic the structure of the experimental data) was conducted using
  the same methods as for the main \ratio analysis.  The digitization of the 12\dg system elements was straightforward, due to the discretized nature
  of the MWPC $U$, $X$, and $V$ plane hits discussed in Section \ref{sec:12hit}.  Recorded energy depositions from the primary particles were associated with discrete wire positions
  and then simulated MWPC hits were reconstructed in an identical fashion as data hits.  Based on the distribution of hits from data that registered on multiple adjacent wires, simulated
  MWPC hits were allowed to also register on multiple wires and rates randomly drawn from the data distribution.  SiPM plane digitization for the trigger was also straightforward, in that it only
  involved matching the energy deposition threshold in simulation to a value similar to experimental conditions.  Due to the fact that good 12\dg events produced SiPM hits well above the 
  experimental threshold, the choice of this threshold in simulation was effectively negligible.  After hits were constructed in the SiPM planes and MWPCs, they were tested by position
  against the efficiency maps presented in Section \ref{sec:12eff} using a random number drawn against the efficiency.  Hits passing the efficiency test were packaged in the same format
  as the hits produced from data, and from this point simulation data were treated identically to experimental data, ensuring a robust comparison.
  
  \subsection{Comparison of Data and Simulation}
  \label{sec:12comp}

\begin{figure}[thb!]
  \centerline{\includegraphics[width=1.15\textwidth]{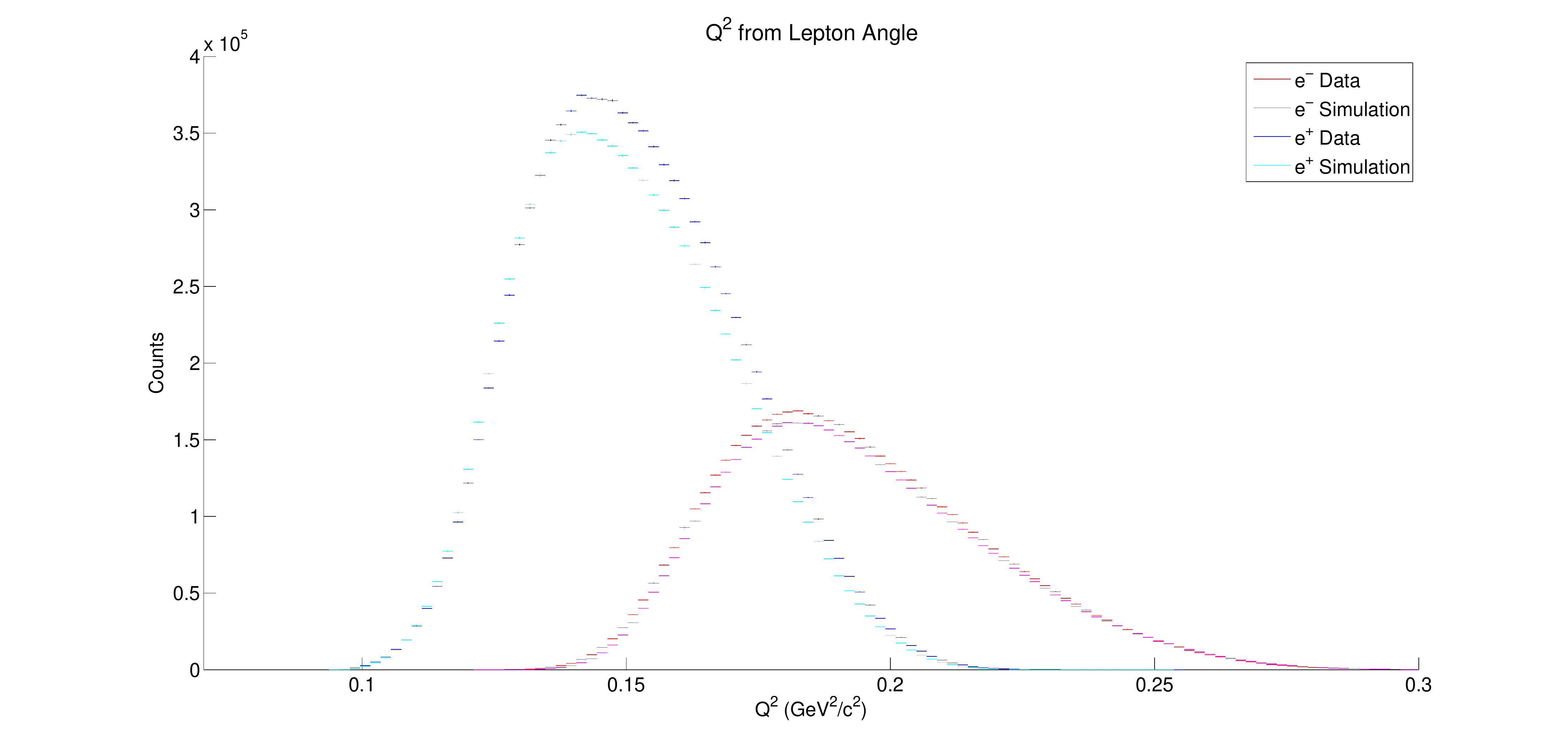}}
  \caption[Distributions of $Q^2$ for 12\dg leptons in data and simulation]{Distributions of reconstructed $Q^2$ for 12\dg leptons of each species, in data and simulation
  for the entirety of the data sample.}
  \label{fig:12q2}
  \end{figure}

\begin{figure}[thb!]
  \centerline{\includegraphics[width=1.15\textwidth]{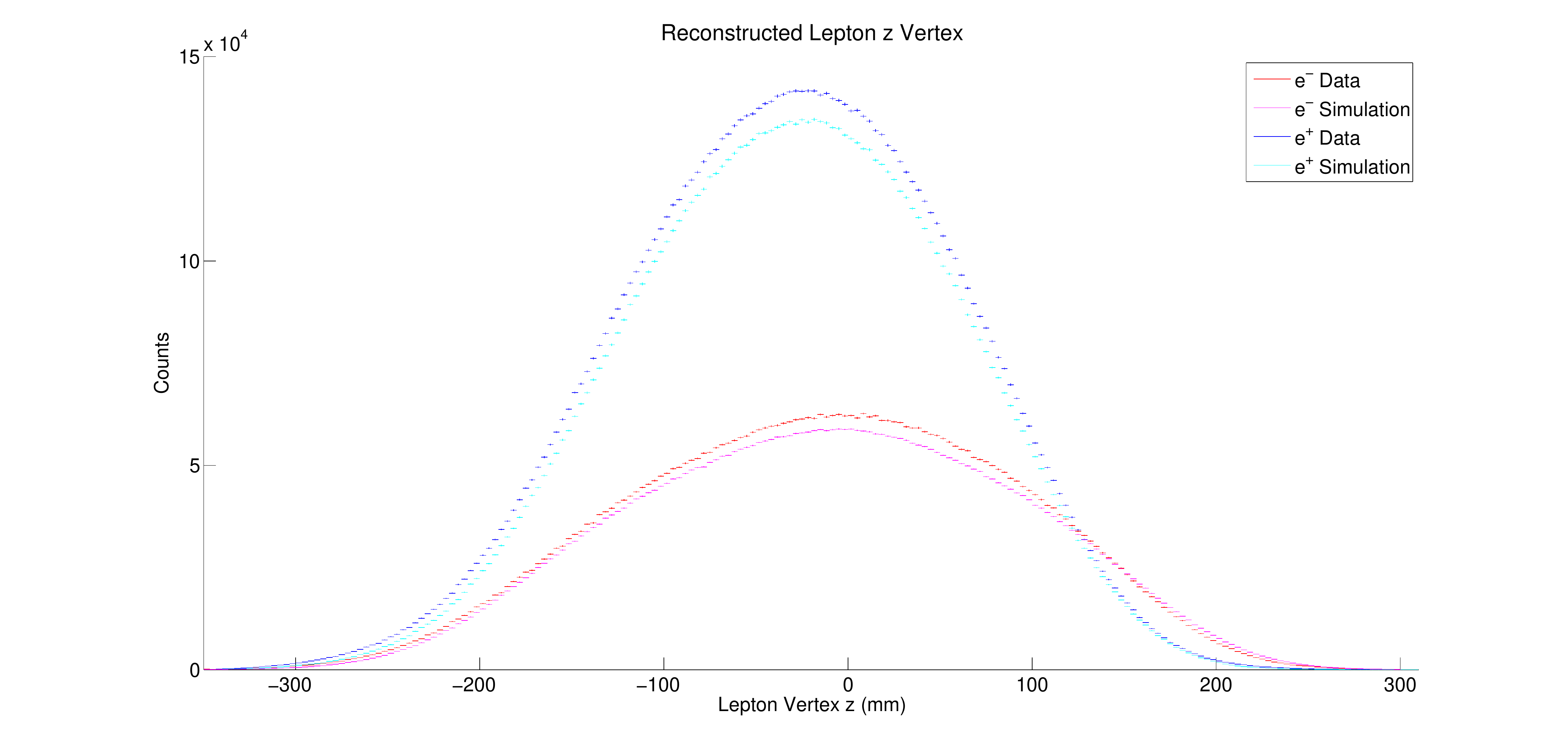}}
  \caption[Distributions of vertex $z$ for 12\dg leptons in data and simulation]{Distributions of reconstructed vertex $z$ for 12\dg leptons of each species, in data and simulation
  for the entirety of the data sample.}
  \label{fig:12z}
  \end{figure}

\begin{figure}[thb!]
  \centerline{\includegraphics[width=1.15\textwidth]{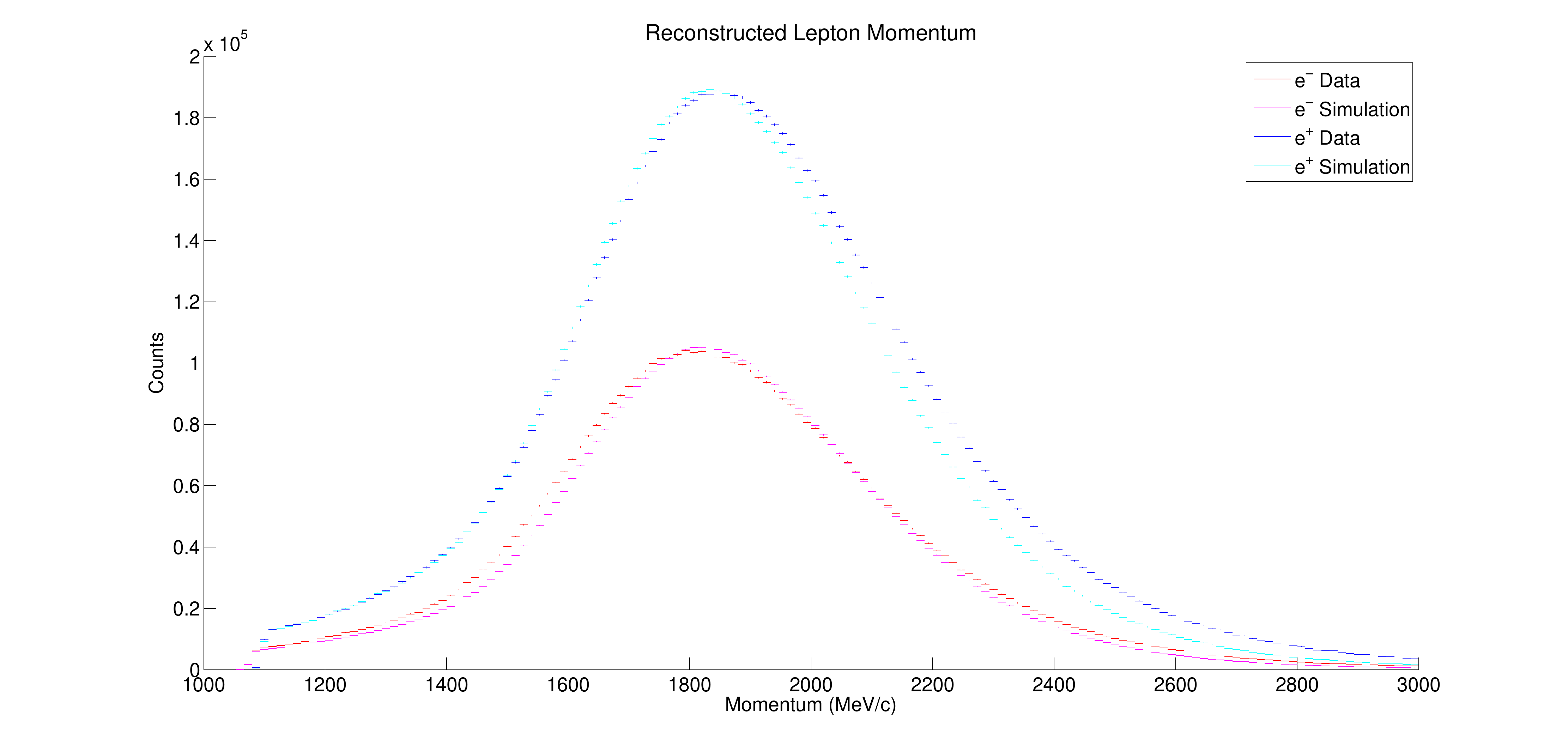}}
  \caption[Distributions of reconstructed momentum for 12\dg leptons in data and simulation]{Distributions of reconstructed momentum for 12\dg leptons of each species, in data and simulation
  for the entirety of the data sample.}
  \label{fig:12mom}
  \end{figure}

\begin{figure}[thb!]
  \centerline{\includegraphics[width=1.15\textwidth]{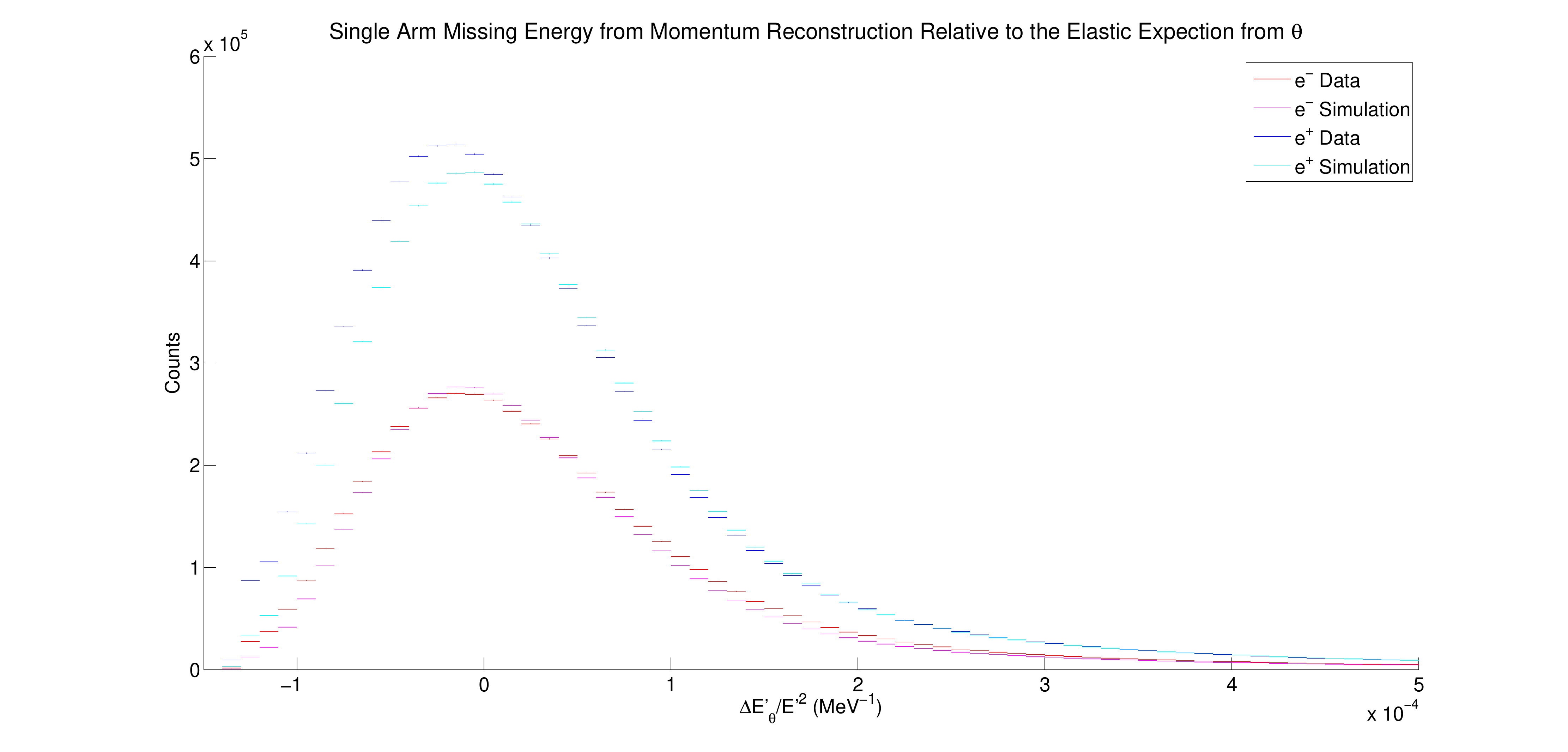}}
  \caption[Distribution of single-arm missing energy for 12\dg leptons in data and simulation]{Distributions of the single-arm missing energy $\Delta E'_\theta/E'^2$
  for 12\dg leptons of each species, in data and simulation for the entirety of the data sample.}
  \label{fig:12de}
  \end{figure}

  This section provides several figures showing distributions of various kinematic parameters reconstructed by the 12\dg telescope analysis for both lepton species
  in data and simulation so as to provide a basic overview of the nature of the data and establish the validity of the data/simulation comparison.  While plots of
  all kinematic parameters would fill many pages, several representative plots that are particular relevant to the analysis are shown.  In general, the data/simulation
  agreement is very good to within resolutions and the conservative cuts applied to the data avoid regions of disagreement due to slight differences in resolution.  Note
  that since the final 12\dg luminosity measurement (Section \ref{sec:12res}) found a result several percent above the slow control estimate for the absolute luminosity (which was used to generate
  the simulation), the integrals of the distributions shown differ by that factor.  Uncertainties caused by these residual data/simulation differences were accounted for when
  considering the systematic uncertainties due to the various applied cuts. The distributions presented are:
  \begin{itemize}
   \item $Q^2$ reconstructed from the lepton angle (Figure \ref{fig:12q2}),
   \item vertex $z$ reconstructed by the lepton track (Figure \ref{fig:12z}),
   \item reconstructed lepton momentum (Figure \ref{fig:12mom}), and 
   \item $\Delta E'_\theta/E'^2$, the single arm lepton missing energy relative to the expected energy at the reconstructed $\theta$ (Figure \ref{fig:12de}).
  \end{itemize}
  
  \subsection{Systematic Uncertainties}
  \label{ss:12sys}
  
  Since any uncertainty in the luminosity determination directly affects the uncertainty on
  the final \ratio ratio result, it was critical to consider all possible causes of systematic
  uncertainty in the extraction of the 12\dg luminosity, especially any such factors that affect
  the relative luminosity of the lepton species.  Naturally, the uncertainty of the species-relative
  luminosity determination is considerably smaller than the uncertainties of an absolute luminosity determination
  due to the numerous systematic effects that are the same for both species, making the 12\dg an effective system
  for the task at hand (although not necessarily a precise absolute luminosity monitor).  Note that no systematic effects that
  explicitly arise from the GEM detectors are formally discussed or computed since the GEM data did not directly contribute to
  the luminosity measurements in this work. The systematic uncertainties for the
  12\dg system are discussed in the following sections, and the absolute and relative luminosity
  uncertainties are summarized in Table \ref{tab:12ds}.  A discussion of the interpretation and implications of these
  estimates may be found at the end of this section following a discussion of the details of each contribution estimate.
  
  Regarding the methods used to compute the systematic uncertainties and the way in which they should be
  interpreted, the values presented in this section are estimates of the maximum range
  that the values of the extracted absolute and relative luminosities could have due to conservative 
  estimates of the plausible range over which each effect may vary.  Due to this, these are not to be interpreted
  as Gaussian uncertainties in most cases (as many of the effects are non-Gaussian), and, in general, the
  interpreted Gaussian uncertainties corresponding to the values in Table \ref{tab:12ds} are less than or equal to the
  presented values (as is discussed at the end of this section).  Only effects found to amount to a contribution of 0.01\%
  or greater are discussed here, since this is
  the order of the overall statistical uncertainty of the dataset of 12\dg events.
  
\begin{table}[thb!]
  \begin{center}
  \begin{tabular}{|l|c|c|}
  \hline 
  Uncertainty Source & Relative (\%) & Absolute (\%) \\
  \hline\hline 
  ToF trigger efficiency ($\delta_{\epsilon_\text{ToF}}) $ & $\pm0.19$ & $\pm0.25$ \\
  \hline
  SiPM trigger efficiency ($\delta_{\epsilon_\text{SiPM}}$) & $\pm0.01$ & $\pm0.10$ \\
  \hline
  MWPC plane efficiency ($\delta_{\epsilon_\text{MWPC}})$ & $\pm0.01$ & $\pm0.05$ \\
  \hline
  Magnetic field ($\delta_{B})$ & $\pm0.15$ & $\pm0.35$ \\
  \hline
  Lepton tracking efficiency ($\delta_{\epsilon_{e,\text{track}}})$ & $\pm0.18$ & $\pm0.86$ \\
  \hline
  Proton tracking efficiency ($\delta_{\epsilon_{p,\text{track}}})$ & $\pm0.10$ & $\pm0.80$ \\
  \hline
  Beam position/slope ($\delta_\text{BPM}$) & $\pm0.01$ & $\pm0.01$ \\
  \hline
  Beam energy ($\delta_{E_\text{beam}}$)& $\pm 0.02$ & $\pm 0.02$ \\ 
  \hline
  Detector position ($\delta_\text{det}$) & $\pm 0.02$ & $\pm 0.20$ \\
  \hline
  Fiducial cuts ($\delta_\text{fid}$) & $\pm 0.12$ & $\pm 0.22$ \\
  \hline
  Elastic cuts ($\delta_\text{elas}$) & $\pm 0.27$ & $\pm 1.63$ \\
  \hline
  Radiative corrections ($\delta_\text{rad}$) & $\pm0.08$ & $\pm0.45$ \\
  \hline
  Elastic form factors ($\delta_\text{ff}$) & $\pm0.14$ & $\pm1.20$ \\
  \hline
  TPE at $\theta = 12^\circ$ ($\delta_\text{TPE}$)* & $\pm0.10 $ &  $\pm0.10 $\\
  \hline\hline
  Total including TPE uncertainty ($\delta_{12^\circ,\text{TPE}}$) & $\pm0.47\%$ & $\pm2.44\%$ \\
  \hline
  Total without TPE uncertainty ($\delta_{12^\circ}$) & $\pm0.46\%$ & $\pm2.44\%$ \\
  \hline
  \end{tabular}
  
  \end{center}
  \caption[Systematic uncertainties of the 12\dg luminosity determination]{A summary of the contributions to the systematic uncertainty
  in the determination of \ratio and of the absolute single-species luminosity
  in the 12\dg monitors in percent, as discussed in detail in
  Section \ref{ss:12sys}.  Absolute uncertainties are averaged between the species for the purpose of quoting a single number.  These uncertainties may be
  considered to be independent, in general, and thus are added in quadrature to produce the total
  uncertainty estimate. Note that the TPE uncertainty (marked by *) contributes when the 12\dg result is used as a determination
  of the relative luminosity for the \ratio result, but is not included for a measurement of TPE at $\theta \approx 12^\circ$ using an
  independent luminosity extraction.}
  \label{tab:12ds}
  \end{table}
  
  \subsubsection{ToF Trigger Efficiency}
\begin{figure}[thb!]
  \centerline{\includegraphics[width=1.15\textwidth]{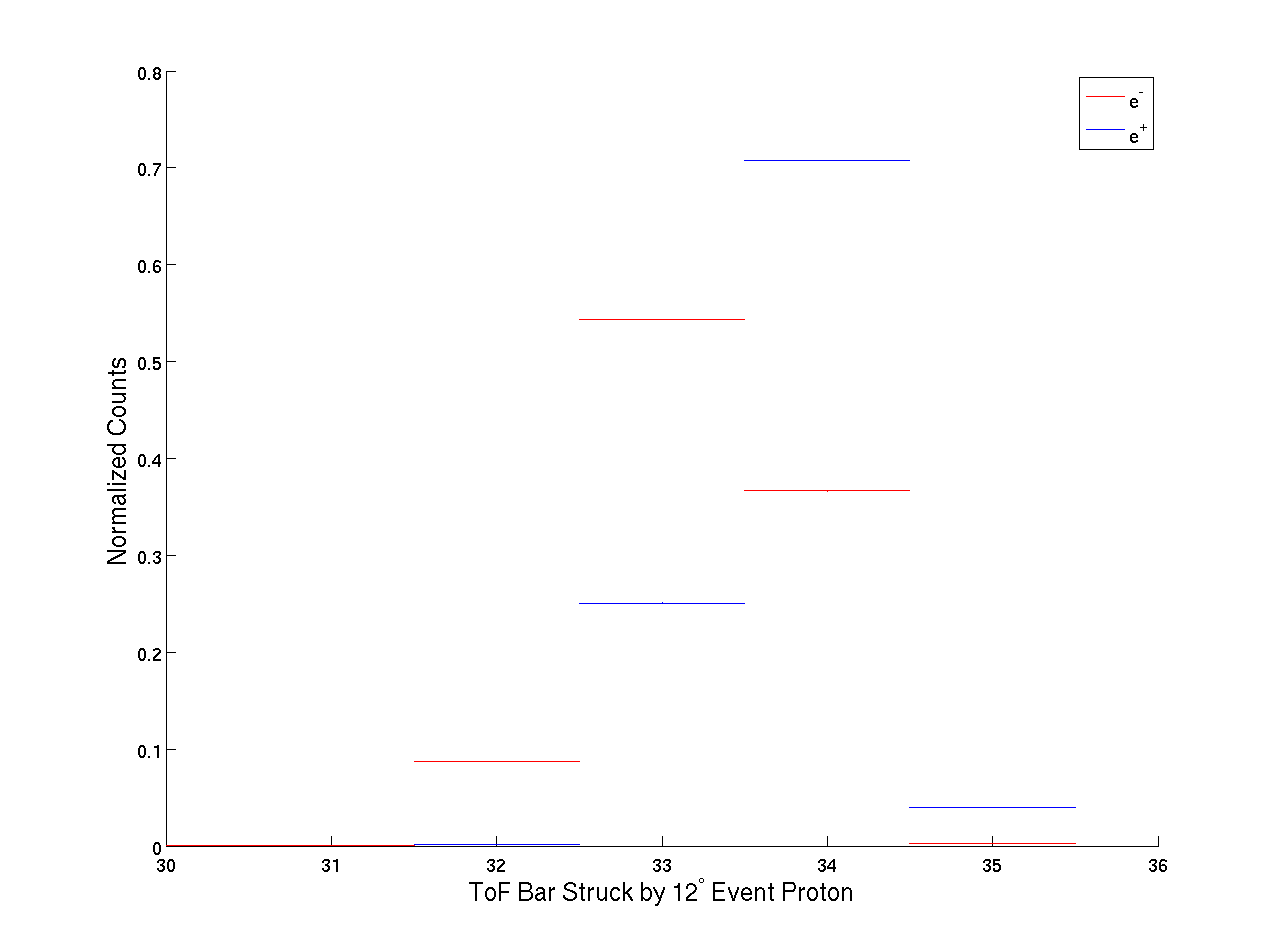}}
  \caption[Distributions of proton ToF hits for $e^+p$ and $e^-p$ 12\dg events]{Distributions of the ToF bars struck
  by the protons in elastic $e^+p$ and $e^-p$ events in which the lepton was recorded by the left 12\dg telescope, normalized for comparison
  between events of the different lepton species.  Note that 12\dg $e^+p$ events have protons at more backward
  angles (higher ToF index) than $e^-p$ events due to the toroidal field which caused positively charged particles
  to be out-bending. Due to the differences in these distributions, any anisotropy in ToF bar efficiencies could
  introduce a systematic shift in the \ratio determination in the 12\dg system.}
  \label{fig:12tdist}
  \end{figure}

  Due to the fact that protons from $e^+p$ and $e^-p$ scattering events are distributed differently
  among the ToF scintillator bars at backwards angles (Figure \ref{fig:12tdist}) and the fact that a recorded ToF hit is
  required for the 12\dg trigger (Section \ref{ss:12dtrig}), any unaccounted for anisotropy among the efficiencies
  of the ToF bars would introduce a shift in the relative \ratio measurement.  Since the 12\dg trigger is only
  concerned with the efficiency of generating hits for protons of relatively low momentum ($\sim$10$^2$ MeV/$c$) in the ToFs
  (which deposit a large amount of energy in the 
  scintillator) and since those protons strike the central region of the bar (the most efficient region for generating
  a signal in both photomultipliers), the expected efficiency for the ToF
  element of the 12\dg trigger is essentially unity from physics considerations alone.  Due to issues with the PMT
  couplings and electronics used to readout the ToF bars, however, some inefficiencies were likely present in the
  experiment that are difficult to account for in the simulation, which assumes a nearly perfect efficiency for
  12\dg event protons.
  
\begin{figure}[thb!]
  \centerline{\includegraphics[width=1.15\textwidth]{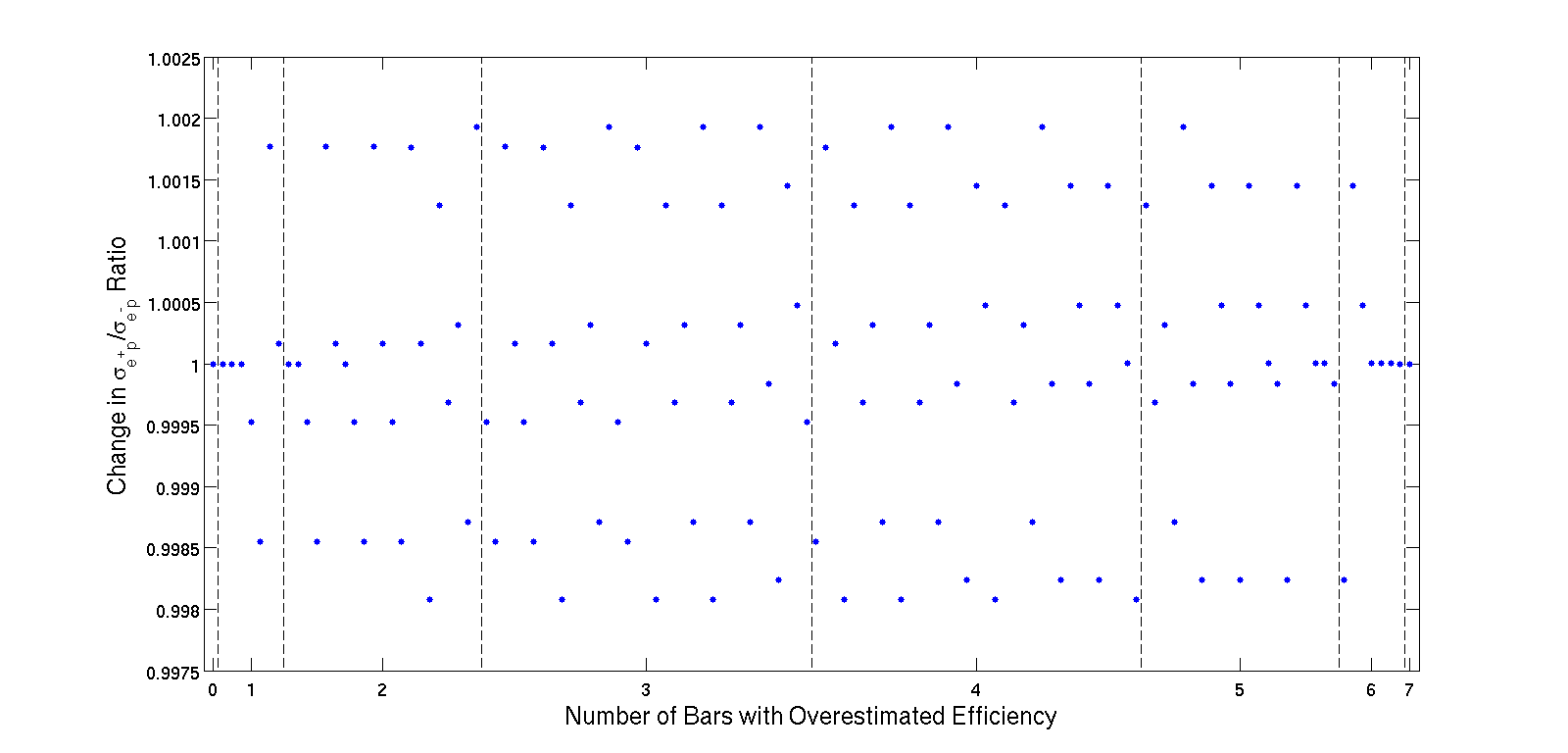}}
  \caption[Effect of ToF efficiency uncertainty on the 12\dg luminosity ratio]{Estimated possible effect on the extracted 12\dg luminosity ratio due to
  uncertainty in the efficiency of the ToF trigger. The uncertainty of 0.5\% in the efficiency of each bar was applied for each
  subset of bars involved in the trigger, as described in the text.  The maximal deviation was $\delta_{\epsilon_\text{ToF,rel}} = 0.19\%$.}
  \label{fig:tofsys}
  \end{figure}

  Via estimates of the ToF efficiencies from both dedicated triggers (Section \ref{ss:addtrig}) and ``scintillator sandwich'' measurements,
  it was determined that the efficiency for proton detection can be conservatively estimated to be in excess of 99.5\% for the protons
  of interest \cite{russell1}.  Due to the fact that the Monte Carlo assumed near perfect efficiency for the events
  of interest, the effect of unaccounted for inefficiencies in the simulation could be estimated by rejecting events
  in specific bars with an artificial efficiency of 99.5\%, and examining the effect on the electron-positron ratio
  and the absolute luminosity estimates.  This rejection was computed for all subsets of the seven bars that are a
  part of the 12\dg trigger for each side (i.e., reducing the efficiency of all single bars, pairs of bars, triplets of bars, etc.).
  The resulting effect on the \ratio ratio is shown in Figure \ref{fig:tofsys}.  Since the maximal error on the ratio can
  occur for multiple combinations of small numbers of bars, the systematic uncertainty of \ratio in the 12\dg system
  was conservatively assigned to be the maximum value of the deviation in the ratio under this method ($\delta_{\epsilon_\text{ToF,rel}} = 0.19\%$).
  The uncertainty of the absolute luminosity for each species from this effect was taken to be the mean effect from all 0.5\% efficiency
  drop combinations, i.e., $\delta_{\epsilon_\text{ToF,abs}} = 0.25\%$.
  
  \subsubsection{SiPM Trigger Efficiency}
  
  While electron and positron tracks in the 12\dg telescopes illuminated approximately the same total regions of the SiPM planes due to the telescope
  acceptance being constrained by other elements, the distributions of the illumination of the planes by the two different species may
  be different due to the change in the cross sections for elastic scattering across the ranges of $\theta$ accepted for each species.  The SiPM
  plane efficiencies computed in Section \ref{sec:12eff} and shown in Figure \ref{fig:sipmeff} are in general quite good ($>$99\%) throughout
  their active regions. The small anisotropies in the efficiency could lead to an artificial difference in the extracted electron and positron
  luminosities.  The anisotropies in the SiPM plane efficiencies are on the percent level, and taking the behavior of SiPM Planes 1 and 3 as
  the worst case scenarios, a model of a 1\% efficiency drop with an uncertainty of the drop in efficiency of 0.5\%
  in a band of local $x$ of width $\sim$30 mm provides a conservative test case for determining the effect on the 12\dg trigger.  The relative
  normalized illumination of the SiPMs between $e^+$ and $e^-$ hits differs by at most 0.1\%, as determined by simulation.  So, assuming 
  the maximally asymmetric case between species, in which the lepton hit distributions differ by 0.1\% in the aforementioned model
  uncertain region (corresponding to 30\% of the detector plane), the resulting change in the ratio of the number of accepted hits between lepton
  species is shifted by less than 0.01\%.  Thus, even for a very conservative estimate in the uncertainty in the knowledge of the
  SiPM plane efficiencies, the high efficiency, uniformity of the efficiency, and small acceptance of the telescopes (making the hit distributions
  for each species very insensitive to differing slopes in the cross sections for each species as a function of $\theta$), the effect
  on the relative extracted luminosity is small and may be conservatively quoted as $\delta_{\epsilon_\text{SiPM,rel}}=0.01\%$.
  
  Assigning an overall conservative  uncertainty of $0.1\%$ to the overall efficiencies of the individual SiPM planes from the statistics of the
  estimate in Section \ref{sec:12eff} and track reconstruction, this directly contributes an overall trigger efficiency uncertainty of
  $\delta_{\epsilon_\text{SiPM,abs}}=0.1\%$ to the absolute luminosity measurement.
  
  \subsubsection{MWPC Plane Efficiency}
  
  The uncertainty introduced due to MWPC plane inefficiency is similar in nature to that introduced by the SiPM plane inefficiency, but the
  MWPC planes, especially in the left telescope, exhibited even more consistent efficiency than the SiPM planes, as discussed in 
  Section \ref{sec:12eff} and shown in Figure \ref{fig:mwpceff}.   In the left telescope
  (MWPC Planes 0--2) the anisotropies are on the scale of a few tenths of a percent and millimeter width, but are relatively randomly distributed across
  the planes rather than the larger scale bands in the SiPMs.  Given this difference, but still accounting for the fact that all three
  MWPCs in a telescope must have hits to generate a track, the uncertainty of the relative luminosity determination is estimated
  to be no more than half the value more rigorously computed for the SiPMs: $\delta_{\epsilon_\text{MWPC,rel,left}}=0.005\%$.
  
  For the right side MWPCs (Planes 3--5), the situation is somewhat different due to the presence of the inactive wires, which amount
  to hard acceptance edges for 12\dg events.  While the possibility of using two-wire hits to mitigate the effect of inactive wires
  was explored, it was found that the spurious hits due to noise introduced by accepting such hits would amount to a more detrimental
  effect than implementing a model of the inactive wires in the simulation to account for their effect.  To estimate the possible
  errors introduced by uncertainty in the simulation model, first note that the inactive wires correspond to about 0.8\% of the total
  active area of the three planes.  Then, if the illumination of these regions differs between species on the
  level of 0.1\% and the uncertainty in the placement and inefficiency of these regions (hits may be constructed in these regions
  due to hits on the surrounding wires in the plane containing the inactive wire) is assumed to be $\pm$5\% for the sake of providing
  a wide estimate, the induced asymmetry remains smaller than the quoted left side relative uncertainty.
  
  As an additional test, the simulation of the MWPC inactive wires was tested in two separate ways.  In the first method (the method
  ultimately used for the analysis), simulation hits were treated on the wire-plane level (as data hits are treated in the main analysis)
  and thus the inactive wires provided no hit information and lowered the efficiency in the region of the hit reconstruction planes
  surrounding them.  In the second method, the inefficiency due to the wires was implemented via the maps shown in Figure \ref{fig:mwpceff}.
  The difference in the overall accepted track rates in the simulation between these methods was found to be on the order of 0.01\%.  While
  the wire-plane hit reconstruction digitization method demonstrably better mimics the data (both in terms of methodology and comparison
  between the resultant of hit and track distributions), the difference between the methods is taken as the uncertainty in this case.  Since the
  estimates for the left and right side differ but are each small effects in the overall uncertainty, for the simplicity of assigning a single number to
  the MWPC efficiency effect, the uncertainty is estimated as: $\delta_{\epsilon_\text{MWPC,rel}}=0.01\%$ for both sides.  Since this is a considerably
  smaller than other uncertainties, this choice does not significantly impact the final systematic uncertainty estimate.
 
  The MWPC efficiencies were computed with extremely high statistics and with very precise five-plane tracking, as discussed in
  Section \ref{sec:12eff}, and thus the overall uncertainty of the MWPC efficiencies is at most 0.02\%.  Since all three MWPC
  planes in a telescope are required to reconstruct an accepted 12\dg event, the absolute luminosity uncertainty from this source
  is conservatively estimated as $\delta_{\epsilon_\text{MWPC,abs}}=0.05\%$.
  
  \subsubsection{Magnetic Field}

  As described in Section \ref{sec:magsur} and Reference \cite{Bernauer20169}, a large effort was undertaken to properly model the
  magnetic field of the OLYMPUS spectrometer for simulation and track reconstruction since any uncertainty in the field directly
  corresponds to uncertainty in the acceptance of electron and positron scattering events.  Uncertainty in the field in the 12\dg
  telescope region can occur due to uncertainty in the measurement of the vector components of the field, uncertainty in the
  reconstructed position of the Hall probe used for the measurements, and errors/residuals in the field model used to fit and
  interpolate the field for the simulation and reconstruction.
  
  The region surrounding the 12\dg telescopes is among the hardest to model due to the telescopes occupying the region near the
  ``pinch'' of the toroid (i.e., the region where the coils most-closely approach each other).  As noted in the description of
  the field model, regions near the coils are sensitive to the thin filament model used to approximate the toroid coils resulting
  in residuals between the field model and the survey measurements.  Field components in the OLYMPUS $y$ direction are the strongest experienced
  by 12\dg tracks and affect the in/out-bending of tracks. Field components in the OLYMPUS $x$ result in approximately azimuthal bending. Components
  of the field along $z$ are nearly parallel to 12\dg tracks and are less than 50 G throughout the region experienced by 12\dg tracks, and
  thus any uncertainties in this field are negligible.  Additionally, the $x$ component of the field is also very small in this
  region, especially in the area near the target cell where deflections would most strongly affect the telescope acceptance. Simulations
  indicate that shifts of even $\sim$100\% in the magnitude of $B_x$ are lesser effects than realistic uncertainties in $B_y$,
  and thus are sub-dominant effects.
  
\begin{figure}[thb!]
  \centerline{\includegraphics[width=1.15\textwidth]{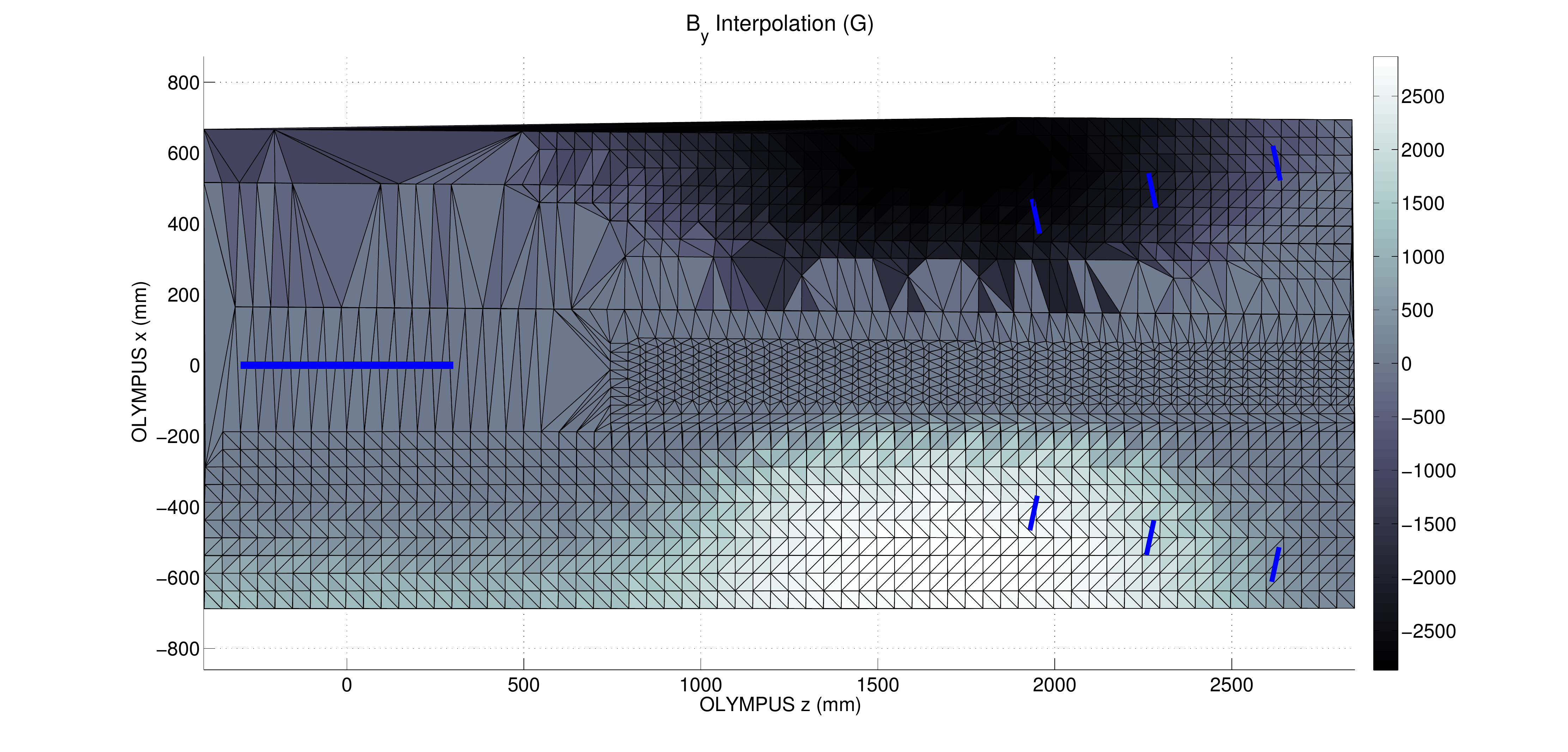}}
  \caption[Magnetic field model for $B_y$ in the 12\dg region]{Field model interpolation calculation for $B_y$ in the region
  important for the 12\dg telescopes, on a triangulated grid of the survey measurement points.  Approximate positions of the 
  target cell and MWPC tracking planes are marked in blue for reference.}
  \label{fig:fy}
  \end{figure}

\begin{figure}[thb!]
  \centerline{\includegraphics[width=1.15\textwidth]{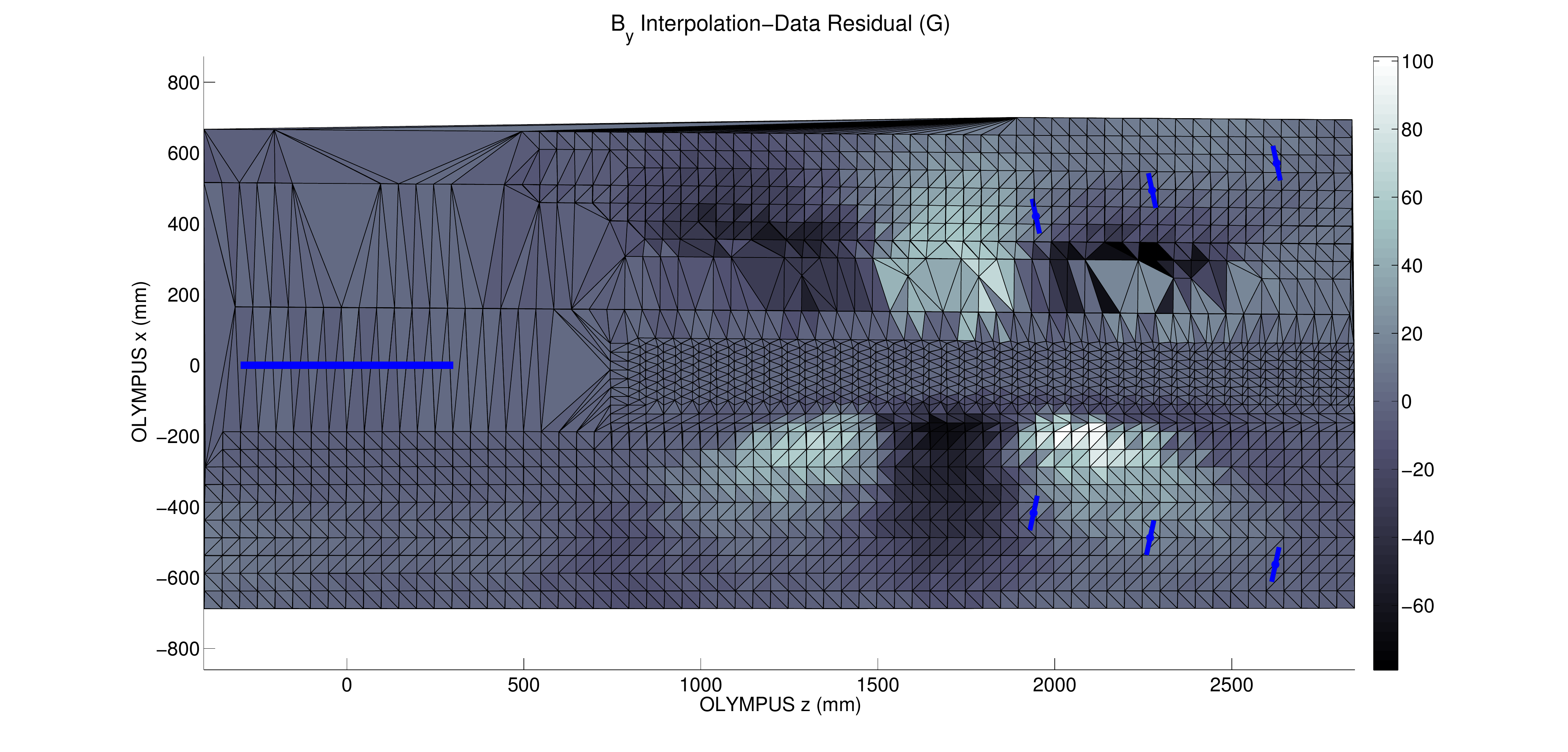}}
  \caption[Survey data/field model residuals for $B_y$ in the 12\dg region]{Residual between the field model interpolation calculation for $B_y$ and
  the measurements at surveyed points in the region
  important for the 12\dg telescopes, on a triangulated grid of the survey measurement points.  Approximate positions of the 
  target cell and MWPC tracking planes are marked in blue for reference.}
  \label{fig:fyres}
  \end{figure}

  Having established $B_y$ as the most critical element of the field for the 12\dg system, the magnitudes of uncertainties in this
  region were examined.  Figure \ref{fig:fy} shows the field model calculation for the magnitude of $B_y$ in the region, while
  Figure \ref{fig:fyres} shows the residual in $B_y$ between the field model and the survey measurements.  Comparing the gradients
  in $B_y$ experienced by 12\dg tracks between survey points (spaced by either 5 or 10 cm), and noting that the uncertainty of
  the position of the survey points is at least an order of magnitude smaller than the spacing, any shift in the field grid relative
  to the true position may be neglected compared to the residuals between the survey and the model shown, which are on the order of tens of gauss
  (or a few percent of the magnitude of $B_y$).  It should be noted, however, that the residuals are not uniformly distributed, and
  exhibit two key elements that lessen their effect:
  \begin{enumerate}
   \item residuals along the first meter of the track, where deflections most strongly affect the acceptance, are extremely small compared
         to the pinch region, and
   \item tracks experience regions of approximately equal in magnitude positive and negative residuals as they approach the telescopes
         leading to at least some cancellation of effects on the acceptance due to these regions.
  \end{enumerate}
  Thus, a model for determining the uncertainty in acceptance due to field uncertainties that systematically scales
  the field on the order of a percent represents a conservative estimate of the acceptance uncertainty since a full systematic
  shift in all field components separates the acceptances of the two lepton species to a significantly higher degree than
  the observed deviations in the field model.
  
\begin{figure}[thb!]
  \centerline{\includegraphics[width=1.15\textwidth]{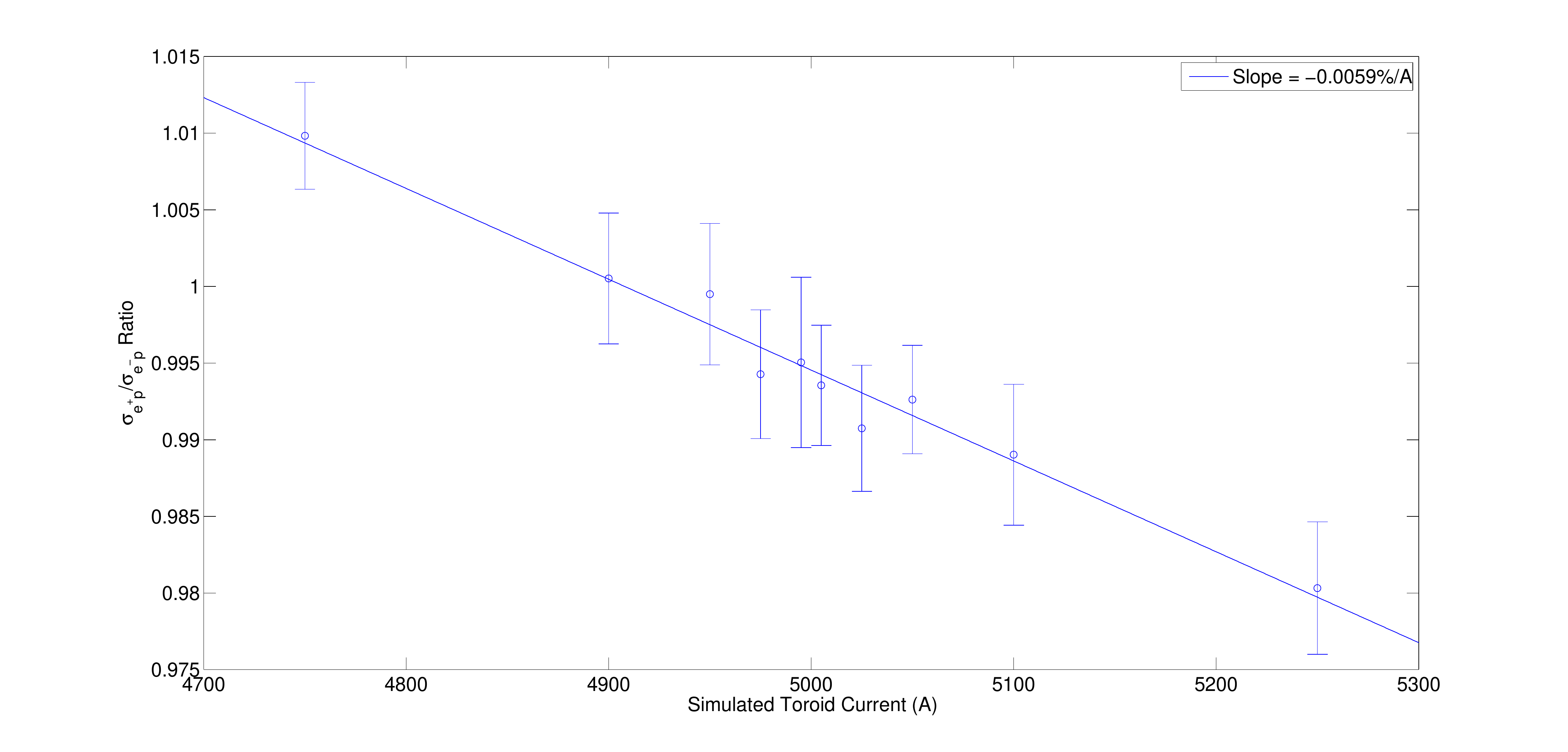}}
  \caption[Effect of magnetic field uncertainty on the relative 12\dg luminosity]{Effect of scaling the magnetic field (from the nominal setting of
  5000 A for the toroid current) on the extraction of \ratio in the 12\dg system.  Note that the error bars on the points are statistical, but that
  the statistical uncertainty between points is highly correlated due to the fact that each point arises from the same generated events.}
  \label{fig:magsys}
  \end{figure}

\begin{figure}[thb!]
  \centerline{\includegraphics[width=1.15\textwidth]{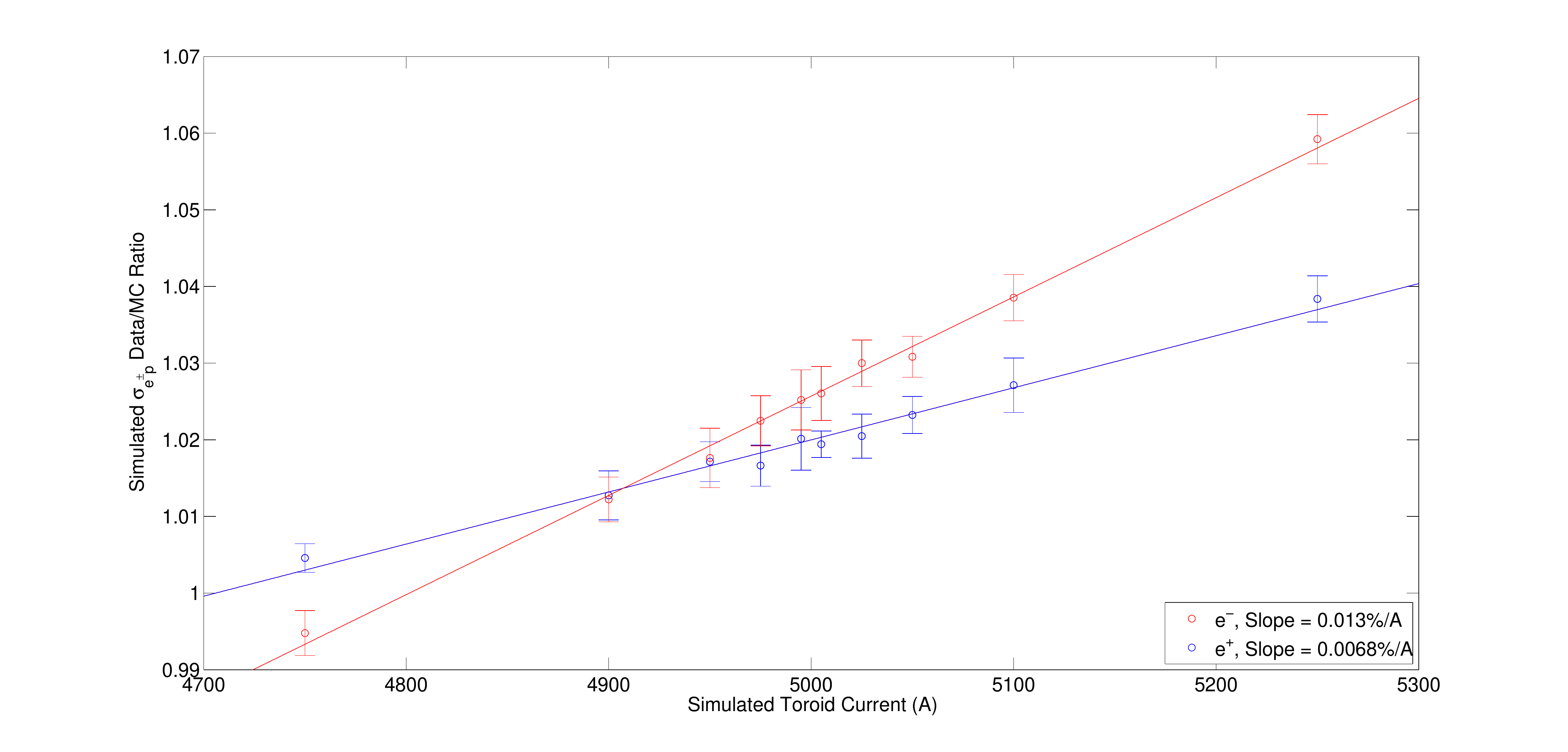}}
  \caption[Effect of magnetic field uncertainty on the absolute 12\dg luminosity]{Effect of scaling the magnetic field (from the nominal setting of
  5000 A for the toroid current) on the extraction of the absolute luminosities in the 12\dg system.  Note that the error bars on the points are statistical, but that
  the statistical uncertainty between points is highly correlated due to the fact that each point arises from the same generated events.}
  \label{fig:magsysabs}
  \end{figure}

  To examine the effects of field scaling uncertainty, simulations were conducted in which a set of generated events was propagated
  using several scaled fields (implemented by varying the toroid current over $\pm$5\% from its nominal value) but reconstructed
  using the nominal field model.  The effects on the species-relative and absolute luminosity extractions
  are shown in Figures \ref{fig:magsys} and \ref{fig:magsysabs}, respectively.  Linear models were found to be good descriptions for the
  variation in the extracted relative luminosity and the individual species absolute luminosities as a function of toroid current, and
  the resulting fits are shown in the figures.  The integrated average residual seen by 12\dg track is considerably less than the 18 G average
  RMS residual quoted in Reference \cite{Bernauer20169}, indicating that a shift of the field model of 0.5\% in the tracking/bending regions 
  (i.e., a shift in the toroid current of 25 A) represents a conservative estimate of the field error.  Thus, using the results of the simulations
  shown in Figures \ref{fig:magsys} and \ref{fig:magsysabs}, the relative and absolute systematic uncertainties due to the field may be estimated
  as $\delta_{B,\text{rel}} = \pm0.15\%$ and $\delta_{B,\text{abs}} = \pm0.35\%$, respectively.
  
  \subsubsection{Lepton Tracking Efficiency}
  
  Given the way in which the leptons in the 12\dg arm were tracked with the GEANT4E tracker and the minimal number of hits
  available for each track in the final 12\dg analysis (three total hits in a telescope from the three MWPC planes), a track
  could be fit to the vast majority of 12\dg events having a hit in each of the three planes that passed the track candidate
  selection criteria (Section \ref{sec:12track}).  This was due to the kinematic flexibility
  of tracker to fit to the three points aside from rare failure modes of the either GEANT4E or the Levenberg-Marquardt minimization
  routine.  Anytime a lepton track was produced for an event, it was considered individually and with all possible proton track pairings,
  and thus the systematics of the quality of the reconstruction are captured in the consideration of the effects of the elastic
  and fiducial cuts applied to the tracks.
  
  To estimate the possible effect on the relative luminosity extraction, the rate of successful tracking for track candidates
  was examined for each species in the simulation.  Note that not all hits produced in simulation are components of a desired
  lepton track due to hits produced by secondaries, unusual hard scattering that causes tracks to deviate heavily from standard trajectories,
  etc., and thus some rejection of events is expected.  For positron runs the simulation candidate-to-track efficiency was found
  to be 96.76\%, while for electrons the efficiency was slightly lower at 96.58\%.  While it is extremely unlikely that this difference
  is entirely due to an asymmetry in rejection of good events, as indicated by visual inspect of such events,
  the contribution to the systematic uncertainty due to this effect is conservatively estimated as the difference in these efficiencies:
  $\delta_{\epsilon_{e,\text{track,rel}}} = 0.18\%$
  
  An exact determination of the effect on the absolute luminosity extraction is difficult since the tracking inefficiency is not
  necessarily representative of the fraction of ``good events'' lost due to lepton tracking failures.  To assess the fraction of
  events in the sample of missed tracks that likely would have been counted as a good event if reconstructed, approximately
  100 such events were inspected by eye using the visualization routine for the simulation propagation (see Appendix \ref{chap:ed}).
  Making a generous assessment, perhaps a quarter of such tracks were within the acceptance of the telescope but ultimately not 
  reconstructed.  Thus, the uncertainty of the absolute luminosity from this effect is taken to be a quarter of the overall
  inefficiency: $\delta_{\epsilon_{e,\text{track,abs}}} = 0.86\%$
  
  \subsubsection{Proton Tracking Efficiency}
  
  While the effect of proton tracking has been minimized to the extent possible in the analysis due to the difficulty of tracking
  the high $\theta$ protons in 12\dg events (as discussed in Section \ref{sec:12ana}), the uncertainty in the reconstruction efficiency for the exclusive events may
  provide some species-relative effects due to the difference in the relative distributions of the protons in $e^+p$ and $e^-p$ scattering (i.e.,
  the protons from each event type are distributed differently in the drift chamber and thus sample different cells).  As discussed
  in Section \ref{sec:recon}, two methods of track reconstruction were utilized for the OLYMPUS drift chambers and thus available
  for reconstruction of the protons in 12\dg events: EAA and SA. The EAA
  tracker used for the reconstruction of the leptons in the main 12\dg analysis is quite efficient in the region relevant to 12\dg
  events, but the SA tracker suffers from notable inefficiencies in the region.  Due to this, the SA tracker is not useful for a final
  analysis, but may be used to estimate the magnitude of the effects of drift chamber and tracking inefficiency on the 12\dg result.
  
  To make such an estimate, the 12\dg analysis was re-run using protons reconstructed using the SA tracker and the resulting difference between
  the SA and EAA analyses was used to estimate the effect of proton tracking efficiency.  Note that at the time this analysis was performed
  the SA tracker was not in its final state and improved somewhat in its performance at back angles for subsequent analyses.
  While the absolute values of the measured
  $e^\pm p$ cross sections drop on the order of a percent in the SA analysis due to the missing protons, the change in the relative
  measurement is not as drastic and can be used as an estimate in the uncertainty caused by a large inefficiency in the proton reconstruction
  on the relative measurement.  The induced differences in the value of \ratio at $\theta \approx 12^\circ$ due to the two different
  tracking methods are shown in Table \ref{tab:12saeaa}.  The size of the effect was found to be on the order of 0.1\%.  Since the
  SA tracking used for this analysis represents a significantly worse proton reconstruction capability than the EAA tracking used
  for the main 12\dg analysis, this was taken as a conservative estimate of the effect of proton tracking on the luminosity
  ratio extraction: $\delta_{\epsilon_{p,\text{track,rel}}} = 0.2\%$.  For the estimate of the effect on the absolute extraction, approximately
  half of the decrease in the absolute extraction between SA and EAA reconstruction was taken as a rough estimate of this uncertainty:
  $\delta_{\epsilon_{p,\text{track,abs}}} = 0.8\%$.
  
  \begin{table}[thb!]
  \begin{center}
  \begin{tabular}{|l|c|c|c|}
  \hline
  Event side & \ratio with EAA & \ratio with SA & Difference\\
  \hline\hline 
  Lepton left & 1.000 & 1.002 & 0.2\% \\
  \hline 
  Lepton right & 0.994 & 0.993 & 0.1\% \\
  \hline
  \end{tabular}
  
  \end{center}
  \caption[Difference in the determination of \ratio in the 12\dg system between EAA and SA proton tracking]{Difference in the determination of \ratio in the 12\dg system between EAA and SA proton tracking.}
  \label{tab:12saeaa}
  \end{table}
  
  \subsubsection{Beam Position and Slope}
  
\begin{figure}[thb!]
  \centerline{\includegraphics[width=1.15\textwidth]{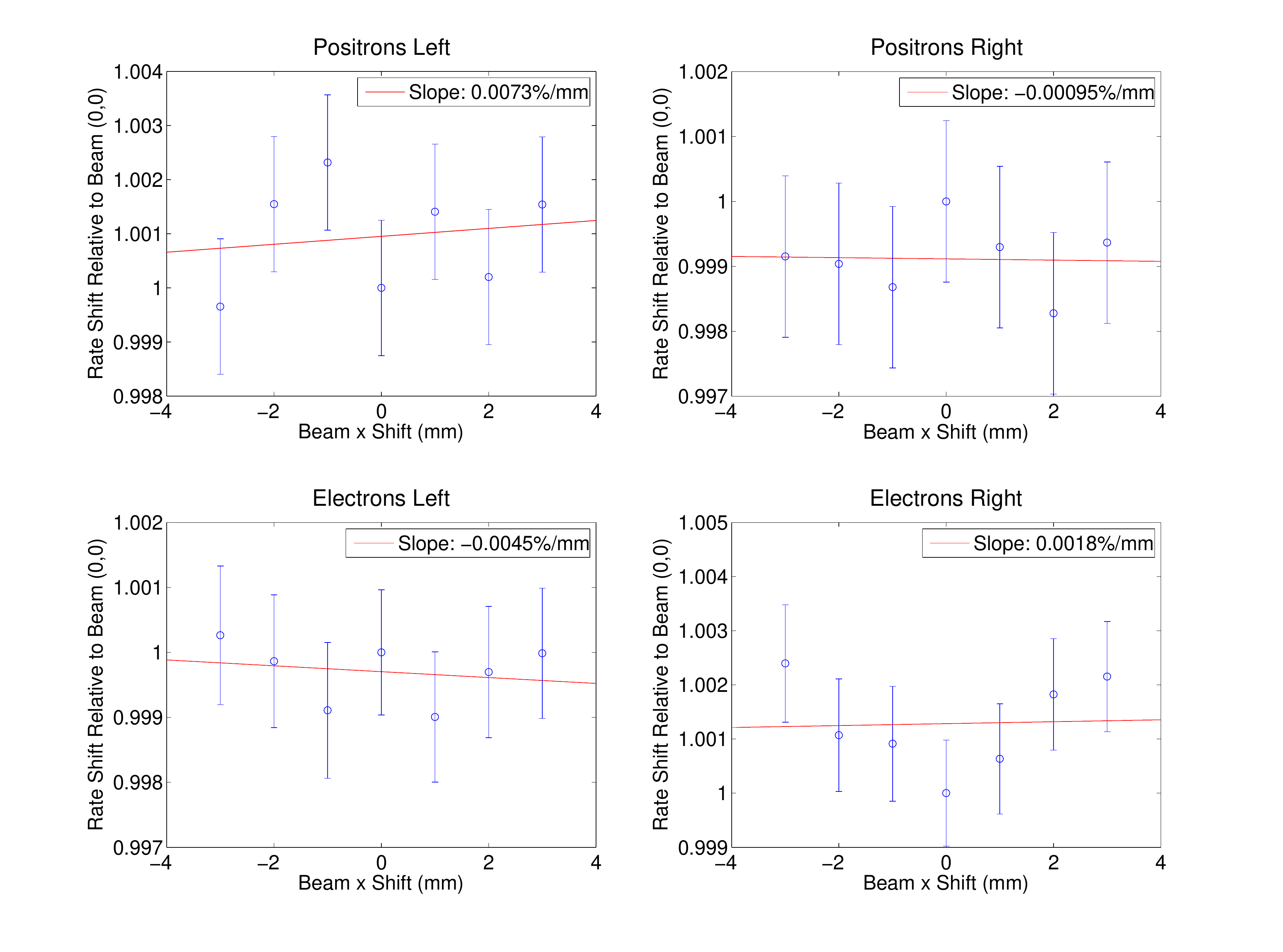}}
  \caption[Effect of beam $x$ shifts on the 12\dg luminosity]{Effects of varying the simulated beam $x$ position on the rates for each species
  in each 12\dg telescope for fixed $y_\text{beam} =0$.  Note that the error bars are slightly overestimated due to correlation between the
  Monte Carlo data sets.  For each combination, the shifts in rates are at most on the order of 0.1\%/mm in both the absolute and species-relative
  measurements.}
  \label{fig:bxsys}
  \end{figure}

\begin{figure}[thb!]
  \centerline{\includegraphics[width=1.15\textwidth]{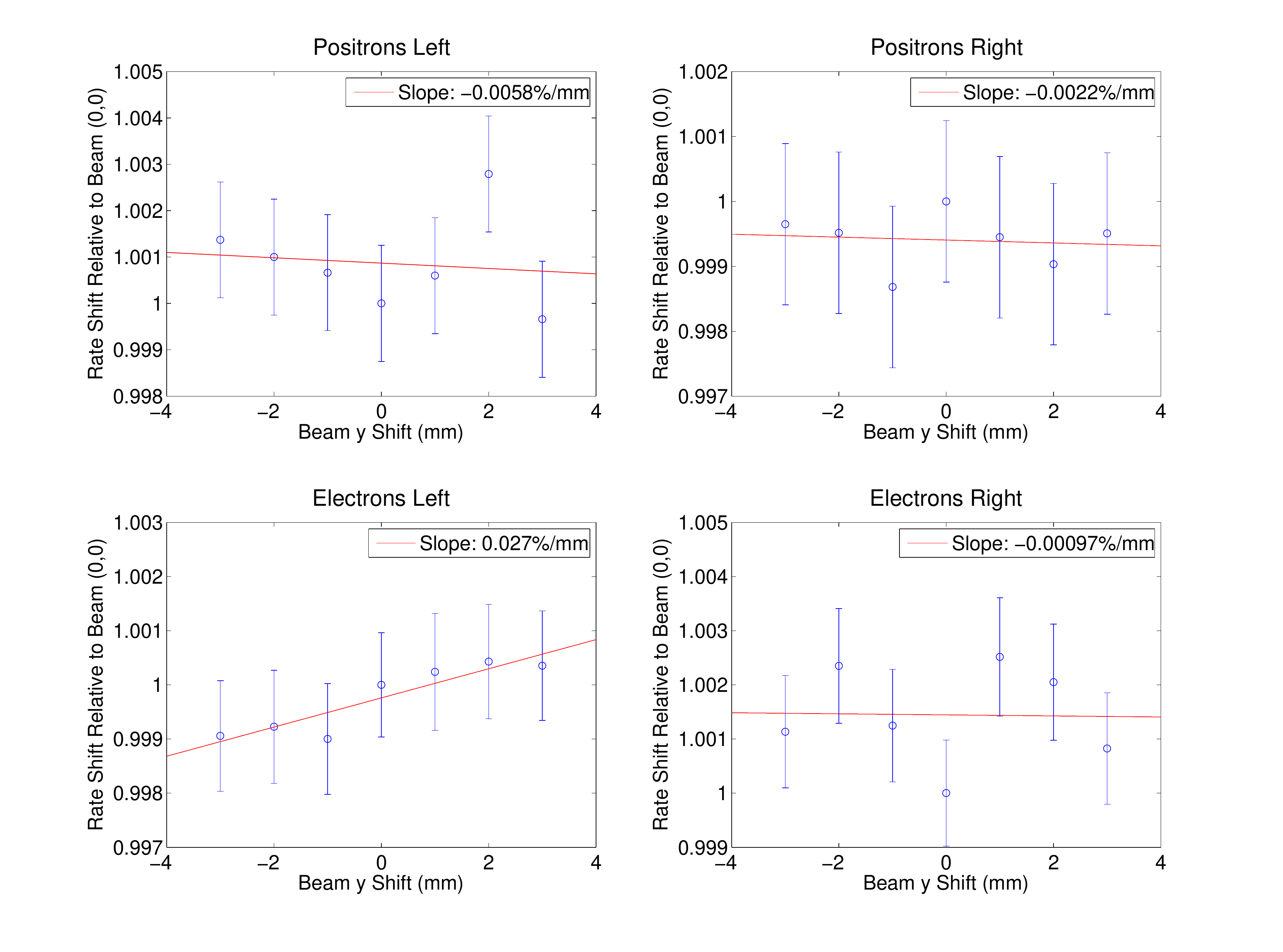}}
  \caption[Effect of beam $y$ shifts on the 12\dg luminosity]{Effects of varying the simulated beam $y$ position on the rates for each species
  in each 12\dg telescope for fixed $x_\text{beam} =0$.  Note that the error bars are slightly overestimated due to correlation between the
  Monte Carlo data sets.  For each combination, the shifts in rates are at most on the order of 0.1\%/mm in both the absolute and species-relative
  measurements.}
  \label{fig:bysys}
  \end{figure}

  The quoted uncertainty of the OLYMPUS beam position reconstruction from the beam position monitor surveys and model fits was
  \SI{100}{\micro\meter} on the absolute beam positions and \SI{20}{\micro\meter} on the species-relative shift error \cite{bernauer1}.
  To assess the sensitivity of the 12\dg system to errors in the knowledge of the beam position, simulated data sets were generated at a
  variety of beam shifts that exceeded the aforementioned expected position uncertainty by an order of magnitude (i.e., shifts of several
  mm).  Figures \ref{fig:bxsys} and \ref{fig:bysys} show the shifts in the absolute rates for each lepton species in each 12\dg telescope
  for shifts in the beam $x$ and $y$ positions, respectively, while the other position dimension is held fixed.  Taking the largest
  slopes as a conservative estimate of the effects of beam position results in an uncertainty in the absolute rate due to beam position
  uncertainty of approximately $0.027\%$/mm and an uncertainty in the species-relative luminosity of $0.033\%$/mm.  In addition to the
  data shown in the figures, several other beam positions and slopes on the order of several mm at each BPM were tested as well and observed
  to exhibit no effects larger than those shown in the figures.  It is also notable that the 12\dg system is insensitive to any uncertainties
  in the shape of the beam profile (i.e., something deviating from the Gaussian envelope described in Section \ref{sec:beam}) since such effects
  are on much smaller length scales than the beam shifts described in this section.
  
  Since the uncertainty in the beam position reconstruction (both absolute and relative) is approximately an order of magnitude less than
  \SI{1}{\mm}, the systematic uncertainties of the relative and absolute luminosity measurements in the 12\dg system
  due to the beam position and slope is very conservatively estimated to be $\delta_\text{BPM,rel} = \pm0.01\%$ and $\delta_\text{BPM,abs} = \pm0.01$.
  Thus, the 12\dg monitors were extremely robust to changes in beam position and slope providing a good complement to the SYMB system, which exhibits
  a much more notable dependence on beam shifts.
  
  \subsubsection{Beam Energy}
  
\begin{figure}[thb!]
  \centerline{\includegraphics[width=1.15\textwidth]{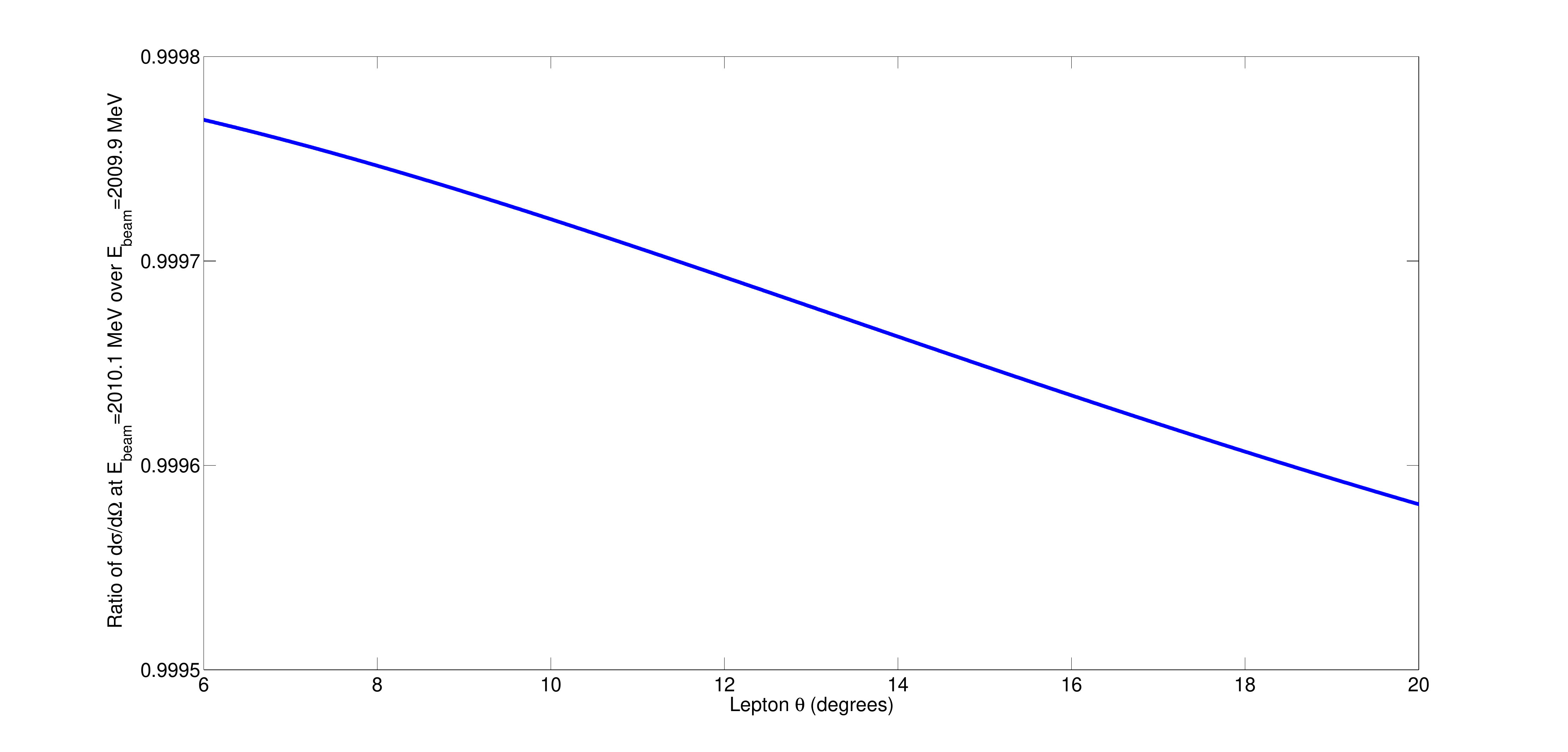}}
  \caption[Effect of $E_\text{beam}$ uncertainty on the Rosenbluth cross section near $12^\circ$ ]{Ratio of the Rosenbluth $e^\pm p$ cross section (using the Bernauer
  form factor parametrizations \cite{BerFFPhysRevC.90.015206}) in the vicinity of the 12\dg acceptance for a beam of $E_\text{beam} = 2010+0.1$ MeV to that at $E_\text{beam} = 2010-0.1$ MeV, i.e. $\pm 1\sigma$
  around the nominal beam energy.}
  \label{fig:ebeamsys}
  \end{figure}

  The $1\sigma$ uncertainty of the DORIS beam energy was determined by the DESY accelerator group to be $\pm 0.1$ MeV for each
  species ($E_\text{beam}\approx 2010$ MeV) \cite{brinker1}.  While  the two species beams had slightly different measured energies ($\Delta E\approx0.5$ MeV), this was accounted
  for in the generation of simulated events using the measured values of the DORIS dipole magnet current saved in the slow control database throughout
  data-taking.  Thus, it is the uncertainty on the knowledge of the relative absolute energies (rather than the absolute
  energy) difference that affects the uncertainty of the final results.  The 0.1 MeV uncertainty on the individual beam energies was estimated by
  testing the DORIS beam with various perturbations and was continuously stabilized with a system of correction magnets.  Due to the importance
  of precision for the OLYMPUS results, this control of the beam energy was significantly better than the $\sim$5 MeV ($\sim$0.1\%) beam energy
  uncertainty that was present when DORIS was operated as in $e^+e^-$ collider mode for the ARGUS experiment \cite{Albrecht:1996gr,DORIStab}.  To estimate the magnitude of the
  effect of this uncertainty on the extracted luminosity in the 12\dg system, the Rosenbluth cross section (Equation \ref{eq:Ros}) was computed as a 
  function of $\theta$ for the nominal beam energy and beam energies $\pm 1\sigma$ from it.  Note that shifts in radiative corrections due to the
  beam energy are considerably smaller effects, and thus the variation in the Rosenbluth cross section is a sufficient approximation for this
  estimate.  Figure \ref{fig:ebeamsys} shows the ratio of this
  cross section at $+1\sigma$ beam energy to that at $-1\sigma$, i.e. the effective shift that would occur in the ratio if one lepton species
  has a beam energy lower than expected from the energy measurement by 0.1 MeV and the other higher by 0.1 MeV.  The maximum deviation of this ratio in the 12\dg acceptance
  is approximately 0.04\%, and the uncertainty of \ratio at 12\dg may thus be reasonably estimated as $\delta_{E_\text{beam,rel}} = 0.02\%$ with
  a similar effect on the absolute cross section: $\delta_{E_\text{beam,abs}} = 0.02\%$
  
  \subsubsection{Detector Position}
  
  Due to the large number of degrees of freedom available for uncertainties in the positions of detectors relevant to 12\dg
  event reconstruction (positions and rotations of each of the MWPC and SiPM planes, and additional degrees of freedom
  form the ToFs and drift chambers), it was not practical nor instructive to approach this uncertainty via a simulation-based method.
  Notably, the acceptance of the 12\dg system is almost entirely dependent on the position of the outermost plane used in each telescope
  (i.e., MWPC planes 2 and 5) since they cover the smallest solid angle ranges relative to the other detector elements (both in the
  12\dg and the proton tracking detectors).  Thus, the dominant uncertainties from acceptance issues related to poorly reconstructed
  detector positions arise from the outermost planes.  While there could be additional uncertainties from poor track reconstruction due
  to uncertain knowledge of the offsets between planes, these were minimized by aligning the detectors relatively using both events at
  nominal magnetic field and straight tracks from zero field runs tracked in the rest of the planes excluding the one being studied.
  For this reason, uncertainties in the placement of the outermost planes may be considered the dominant source of detector
  position systematic error.  Since for even only one plane there are six degrees of freedom in the placement of the detector, a simulation-based
  approach still would require a relatively impractical computing investment and thus a method involving considerations of the
  effects of geometry and the $e^\pm p$ cross section is used instead.
  
  First, considering the rotational degrees of freedom of the detectors, basic geometrical arguments may be used to estimate the size of the effect.  Leptons
  striking the 12\dg detector planes do so approximately perpendicularly to the plane, and so any rotation of the plane relative to this standard
  incidence angle would reduce the effective acceptance of the plane.  While, of course, not all 12\dg tracks strike the planes truly
  perpendicularly, this approximate case is sufficient for estimating the size of the effect.  For a small rotation by angle $\psi$ about an axis
  in the plane of the MWPC detector, the acceptance area presented to an incoming track varies as $\cos\psi$.  Given that uncertainties of the
  rotations are on the order of 0.2$^\circ$ for the MWPC survey \cite{bernauer2}, this would correspond to a change on the order
  of less than a part in $10^5$ in the acceptance.  Thus, the uncertainty due to errors in the rotational placements is nearly certain
  to be minimal for both the relative and absolute luminosities.
  
\begin{figure}[thb!]
  \centerline{\includegraphics[width=1.15\textwidth]{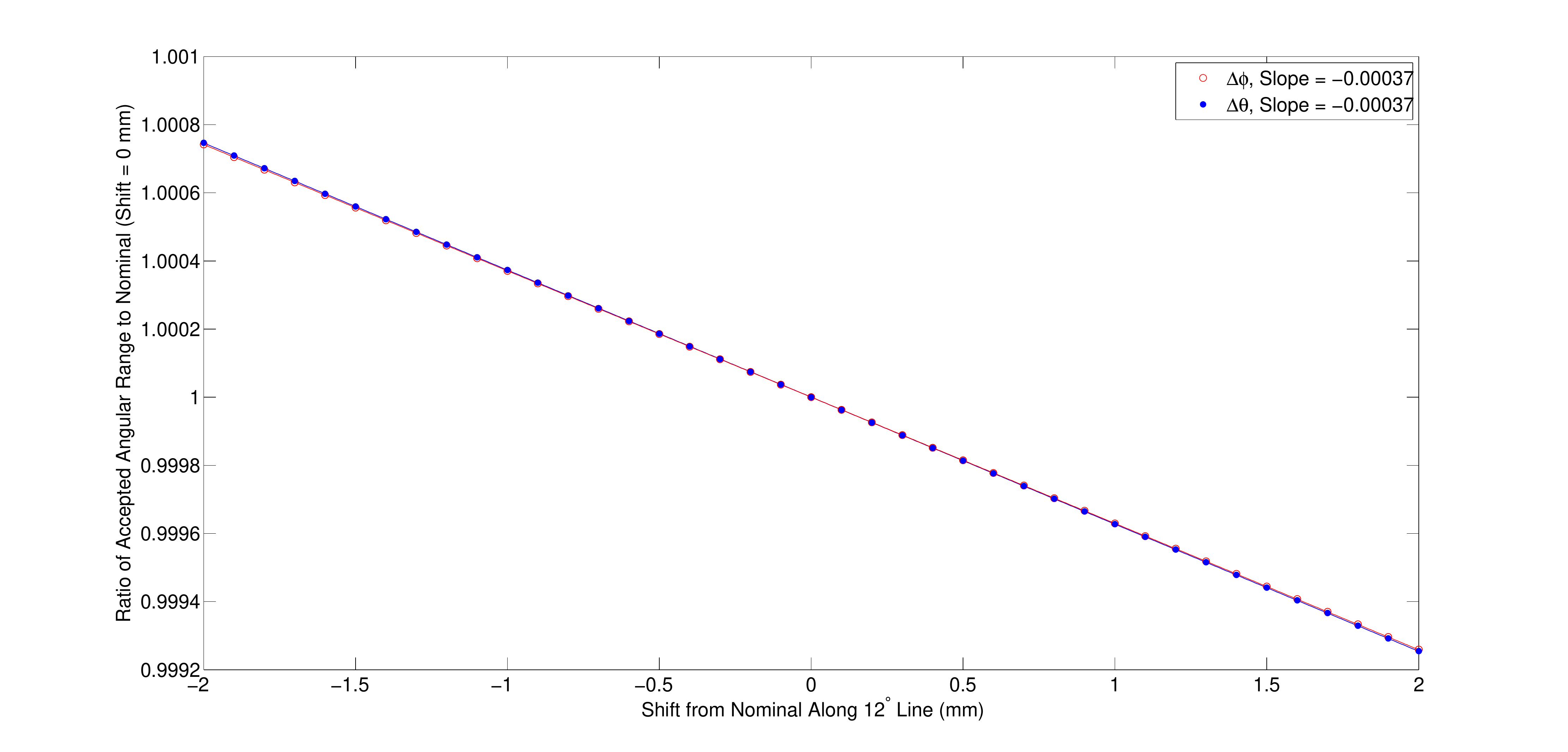}}
  \caption[Effect on angular acceptance of the 12\dg system for shifts toward/away from the target center]{Effect on the solid angle encompassed
  by the outermost MWPC planes when shifted toward or away from the target cell center along the 12\dg line.  The effect in both
  $\theta$ and $\phi$ is approximately 0.04\%/mm.}
  \label{fig:zs}
  \end{figure}

\begin{figure}[thb!]
  \centerline{\includegraphics[width=1.15\textwidth]{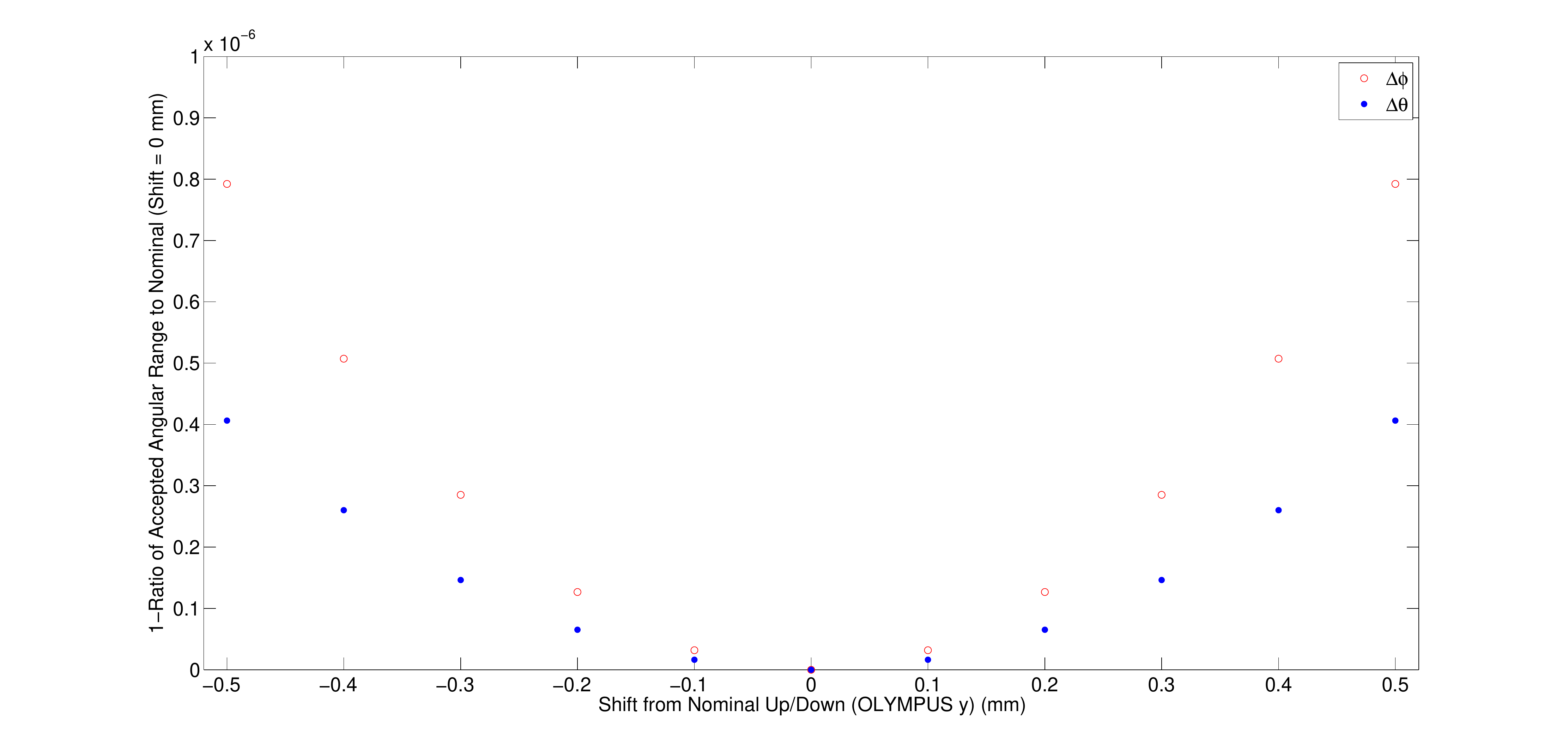}}
  \caption[Effect on angular acceptance of the 12\dg system for shifts up/down]{Effect on the solid angle encompassed
  by the outermost MWPC planes when shifted up or down.  The effect in both
  $\theta$ and $\phi$ is $\sim$1$\cdot 10^{-7}$ for shifts on the order of \SI{200}{\micro\meter}.}
  \label{fig:ys}
  \end{figure}

  To consider the positional degrees of freedom the detectors, geometric considerations in conjunction with a
  Rosenbluth cross section calculation (similar to that conducted
  for the beam energy uncertainty estimate) were used.  The residuals of the MWPC position reconstruction from the survey were approximately
  \SI{200}{\micro\meter} \cite{bernauer2}, and thus shifts of that order were considered for each of the three available degrees of
  freedom.  The outermost MWPCs were nominally located \SI{2.68}{\meter} from the center of the target along the $\theta=12^\circ$
  lines in the OLYMPUS $x$-$z$ plane, and the survey positions differed from these values on the order of several mm.  At this
  position, the MWPC planes' extent (110 mm$\times$110 mm) covers a geometrical solid angle (not necessarily corresponding to a track
  phase space solid angle) range of $\phi = \pm5.6^\circ$ and $\theta = 12\pm1.2^\circ$. Shifting the detector towards or away from
  the target along the 12\dg line changes these ranges on the order 0.04\%/mm, as can be seen in Figure \ref{fig:zs}, and so is on the order
  of hundredths of a percent for shifts on the order of \SI{200}{\micro\meter}.  Similarly, the results of shifting the detector up or
  down are shown in Figure \ref{fig:ys}.  In this case, the results are even smaller, amounting to changes of order $10^{-7}$ for shifts
  of hundreds of micrometers.  Since shifts in neither of these directions significantly change the cross section of $e^\pm p$ events sampled
  (the first does not change the mean or central value of $\theta$, while the second is predominantly a shift in $\phi$ under which
  the cross section is invariant), these geometrical factors are a good estimator of the cross section change due to such shifts.
  
\begin{figure}[thb!]
  \centerline{\includegraphics[width=1.15\textwidth]{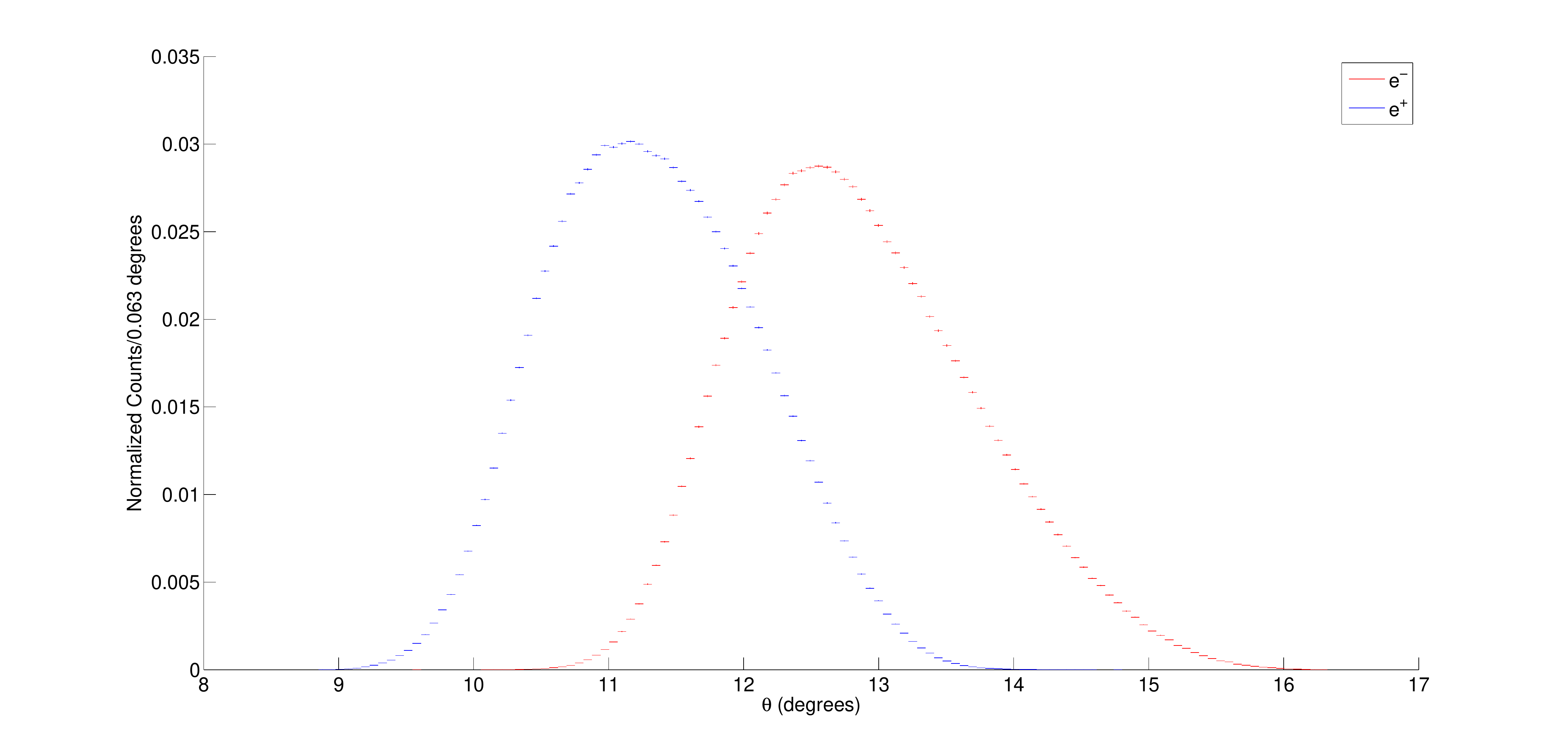}}
  \caption[Normalized lepton $\theta$ distributions of accepted $e^+p$ and $e^-p$ events in the 12\dg luminosity system]{Normalized distributions of
  the lepton $\theta$ for electron- and positron-proton elastic scattering events accepted by the
  12\dg system.}
  \label{fig:tddist}
  \end{figure}

  For the final direction of position shift for an MWPC plane, perpendicular to the $\theta = 12^\circ$ line in the $y=0$ plane (i.e., the local $x$ of the
  detectors), more care must be taken since this involves a shift in the $\theta$ acceptance of the detector and thus a change in the sampled cross
  section, which rapidly changes as a function of $\theta$.  Additionally, different $\theta$ ranges are sampled for each species, further
  adding possible sources of uncertainty.  For reference, the normalized $\theta$ distributions accepted for each species are shown in Figure
  \ref{fig:tddist}.
  
\begin{figure}[thb!]
  \centerline{\includegraphics[width=1.15\textwidth]{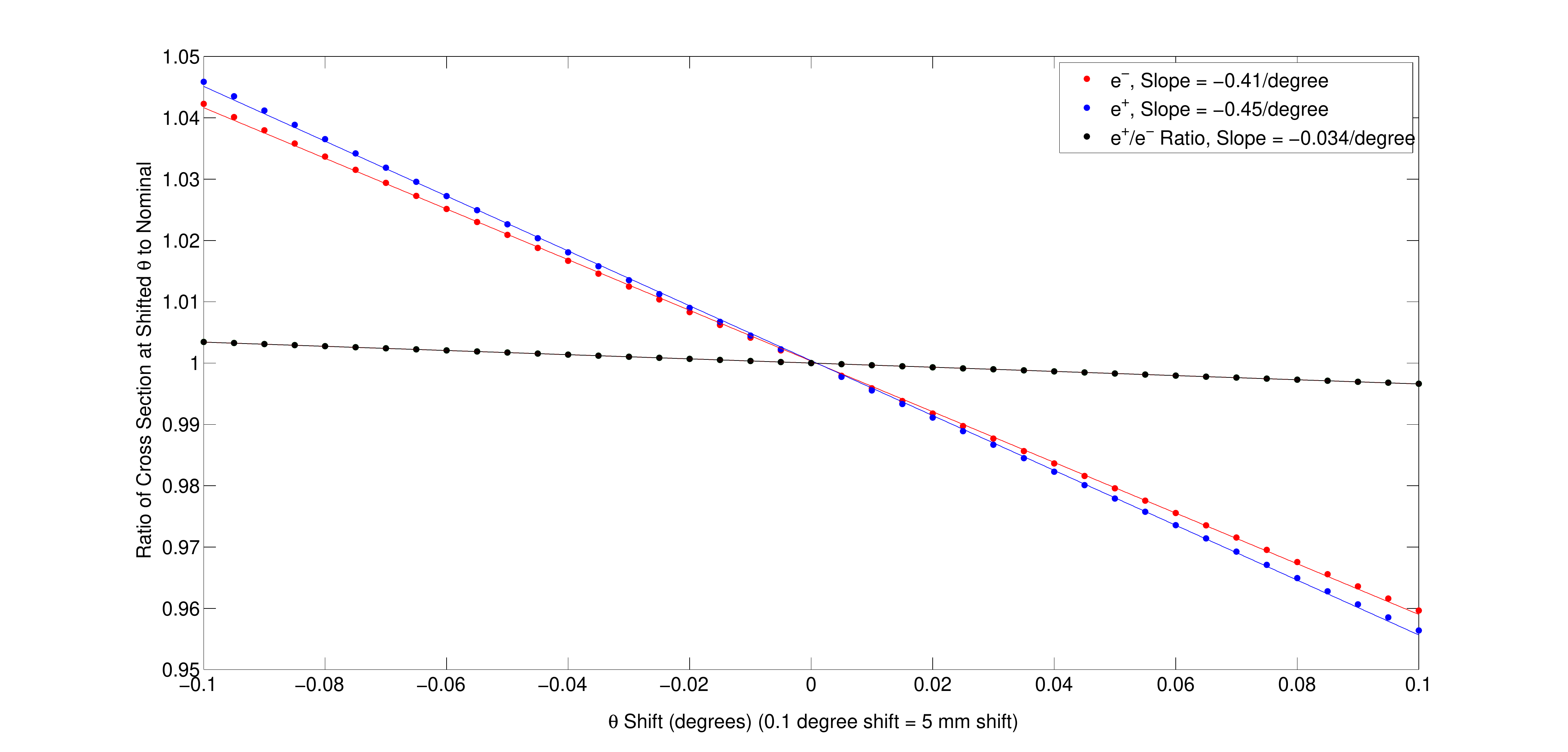}}
  \caption[Effect of shifts in the $\theta$ acceptance of the 12\dg telescopes]{Effect on the simulated cross section
  in the 12\dg acceptance, and thus on the luminosity extraction, due to shifts in the $\theta$ acceptance of the detectors
  away from nominal for each species individually and the ratio.  Note that expected shifts from survey uncertainties are
  considerably smaller than the range of shifts presented.}
  \label{fig:ts}
  \end{figure}

  A shift in this direction corresponds to a shift of a $0.02^\circ$/mm in $\theta$ for small shifts with
  a negligible change in the total $\theta$ coverage (and, of course, no change in $\phi$).  Since the distributions
  shown in Figure \ref{fig:tddist} are of approximately the same width, the cross-section at the mode angle for each species acceptance
  is taken as an approximate stand-in for the cross section in the acceptance and the resultant shifts in the $e^+p$ and $e^-p$
  cross sections for $\theta$ shifts on the order of several hundredths of a degree were computed using the Rosenbluth
  formula.  The effects on the error that would be expected due to such shifts on the simulated cross section, and thus
  on the luminosity, are shown for the individual lepton species and the ratio of the species in Figure \ref{fig:ts}.  Since a
  \SI{200}{\micro\meter} shift corresponds to a shift of approximately $0.004^\circ$, the resultant relative effects due to this uncertainty
  on the luminosity extractions are 0.014\%.
  
  Combining the various effects discussed in this section, and noting that for both the relative and absolute luminosities shifts
  in effective $\theta$ are the dominant contributers, the estimate for the systematic uncertainties due to detector position
  are $\delta_\text{det,rel} = 0.02\%$ and $\delta_\text{det,abs} = 0.20\%$.
  
  \subsubsection{Fiducial Cuts}
  
  Since the 12\dg telescopes had limited acceptance and poor reconstruction resolution due to the use of only the MWPC planes
  in the final analysis, fiducial cuts were only made on the reconstructed $y$ and $z$ positions of the lepton vertex in
  the OLYMPUS global coordinate system as described in Section \ref{sec:12ana}.  Note that any attempt to make a fiducial cut in $\phi$ is extremely dangerous
  since deviations of the OLYMPUS magnetic field from a perfect toroidal field experienced by 12\dg tracks focus electrons
  and defocus positrons in $\phi$ and thus any fiducial cut within the reconstructed distributions explicitly introduces
  a false shift between the species.  The possible systematic effects of this focusing, however, are accounted for in the
  previous considerations of the effect of the magnetic field and shifts in the detector position/acceptance.
  
\begin{figure}[thb!]
  \centerline{\includegraphics[width=1.2\textwidth]{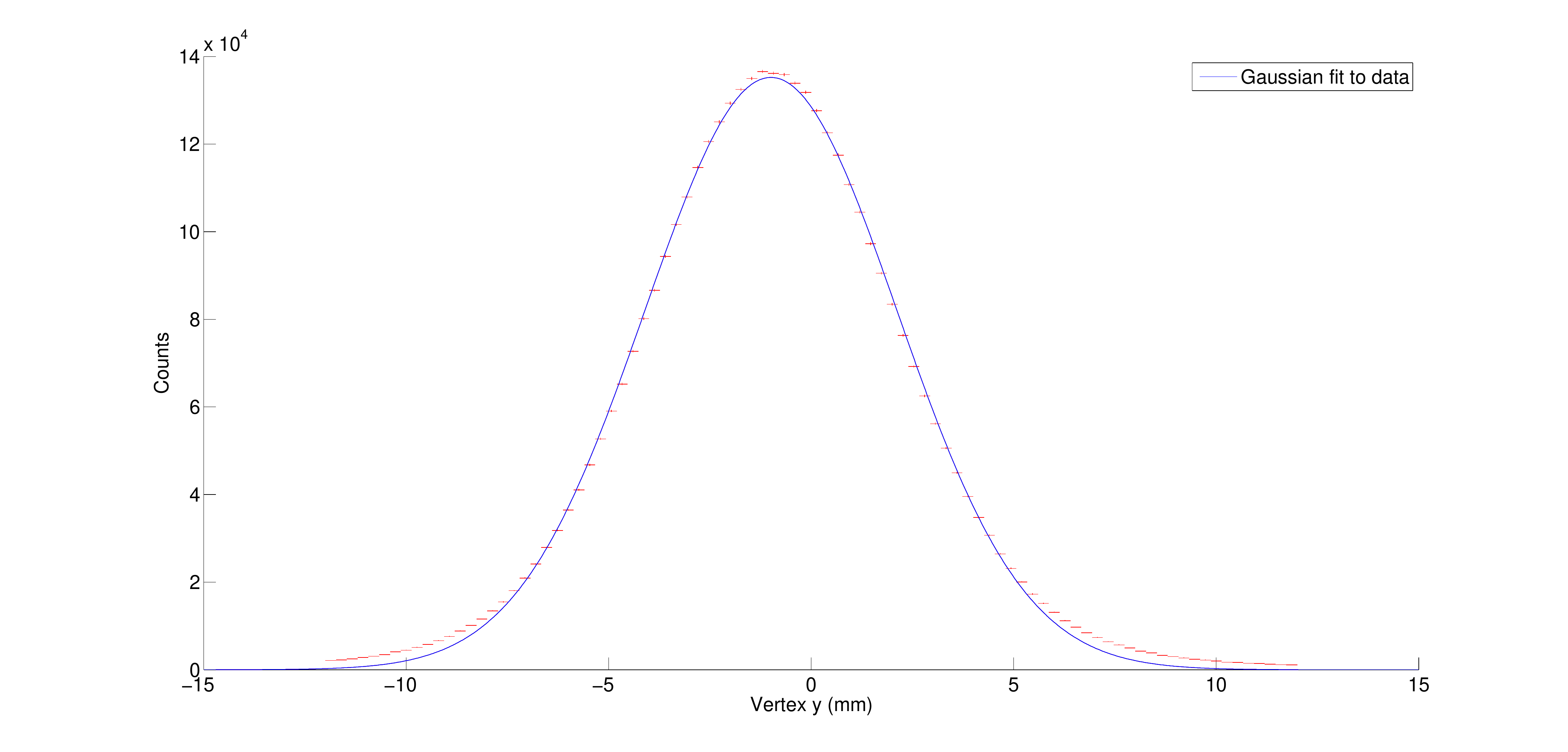}}
  \caption[Distribution of $y$ vertex positions for left-going 12\dg electrons]{Distribution of
  reconstructed $y$ vertex positions for left-going 12\dg electrons with a Gaussian fit applied
  for reference.}
  \label{fig:sampy}
  \end{figure}

\begin{figure}[thb!]
  \centerline{\includegraphics[width=1.15\textwidth]{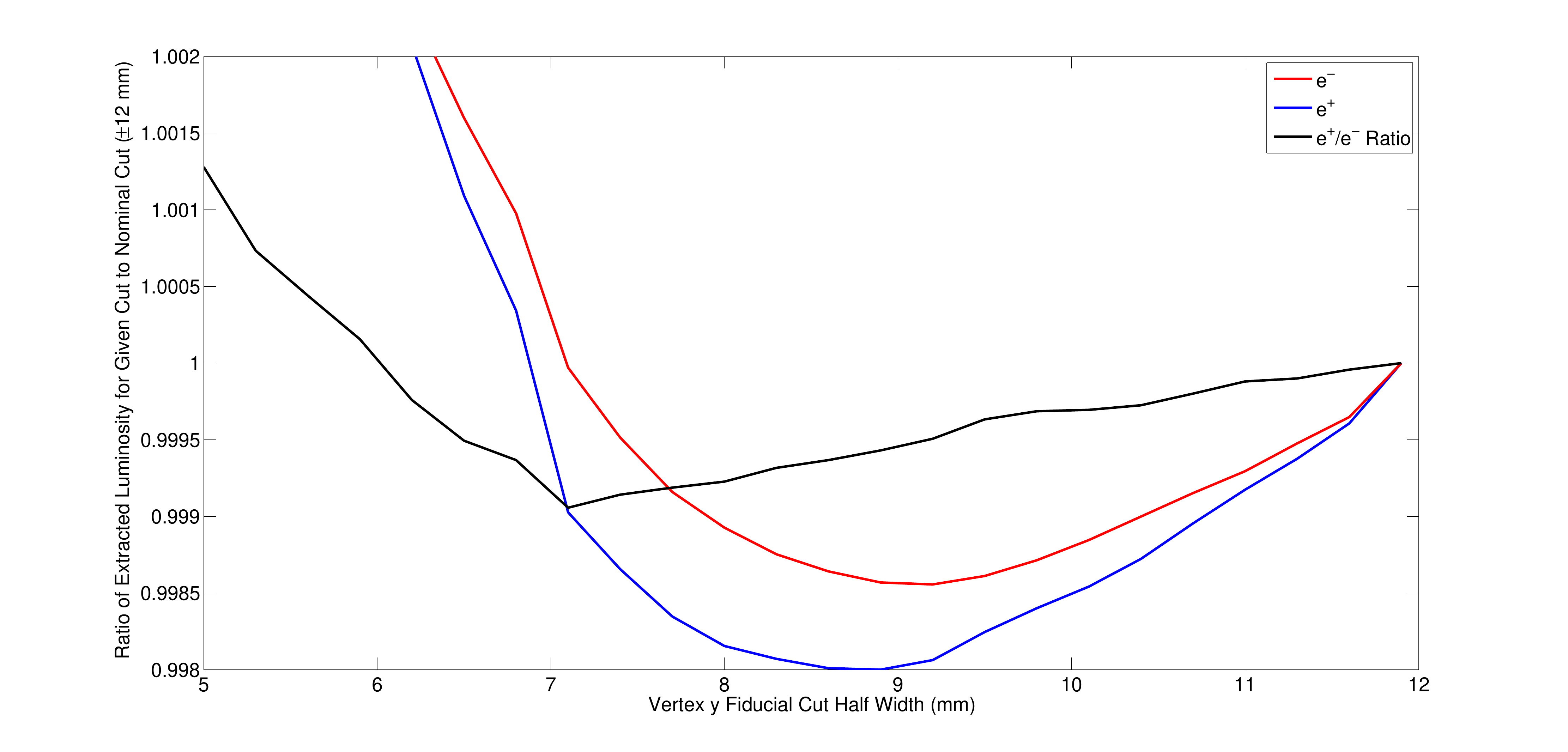}}
  \caption[Effect of vertex $y$ fiducial cut on the extracted 12\dg luminosity]{Ratio of the absolute and species-relative
  luminosities extracted in the 12\dg for varying $y$ fiducial cuts to the value at the nominal cut of $\pm12$ mm.  Cuts smaller
  than 8 mm begin to cut into the Gaussian region shown in Figure \ref{fig:sampy} and are thus not considered reasonable cuts.}
  \label{fig:yfid}
  \end{figure}

  For both the $y$ and $z$ fiducial cuts, the cuts were made at the widest reasonable values due to the system's lack
  of position reconstruction resolution and based on the simulation distributions that were generally free of background
  and mis-reconstructed tracks.  Thus, to test the effects of these cuts each value was further constrained to
  reasonably tighter cuts in each case.  A sample histogram of reconstructed $y$ vertex positions (for left going electrons)
  with a Gaussian fit is presented in Figure \ref{fig:sampy}, which is representative of these distributions in the
  different configurations.  Note that the nominal fiducial cut was symmetric and placed well into the tails
  of the distributions at $\pm12$ mm to avoid cutting good events in the tails.  The effect of varying the $y$ fiducial cut is shown in Figure \ref{fig:yfid}.
  Taking the smallest reasonable cut as $\pm8$ mm, since at cuts smaller than that the good Gaussian region of the distribution
  is cut into, the maximal effect of reasonable cuts on the absolute luminosity is 0.2\% while the relative luminosity is
  changed by at most 0.07\%.
  
  The question of the $z$ fiducial cut is somewhat more complicated than the $y$ cut due to the fact that the target
  distribution has an irregular shape (described in Section \ref{sec:tarsim}) and the poor 12\dg resolution
  caused by the limitation to three-plane tracks in combination with the small-angle lever arm for this reconstruction.
  Additionally, the acceptance of the 12\dg events is constrained by the target chamber on the downstream side, leading the magnetic
  field to create a difference in the reconstructed $z$ distributions for each species.  The reconstructed distributions
  for each species are shown in Figure \ref{fig:12z}.  The nominal fiducial cut was placed as $-380$ mm $<z<250$ mm.  Due to the fact
  that the 12\dg system has an unrestricted view of the upstream end of the target, the cut on that end has considerably less effect
  than the upstream cut as can be seen in the distributions.  To estimate the effect of the downstream cut, the distributions were
  examined and an upper bound of $z=225$ mm was determined to be a reasonable estimate of the tightest reasonable cut.
  The results are summarized in Table \ref{tab:12zcut}.  The effect on both the absolute and relative luminosities was found to be
  on the order of 0.1\%.  
  
  \begin{table}[thb!]
  \begin{center}
  \begin{tabular}{|l|c|}
  \hline
  Luminosity measurement & Nominal/Tight Cut  Ratio \\
  \hline\hline 
  $e^+$ Absolute & 1.0003 \\
  \hline 
  $e^-$ Absolute &  0.9991 \\
  \hline
  $e^+/e^-$ Relative &  1.0012 \\
  \hline
  \end{tabular}
  
  \end{center}
  \caption[Effect of varying the upstream $z$ fiducial cut for 12\dg events]{Change to the absolute and relative 12\dg luminosity extractions
  that occurs when making an upstream $z$ fiducial cut at 225 mm relative to the nominal cut at 250 mm.}
  \label{tab:12zcut}
  \end{table}
  
  Taking the two fiducial cut effects together by combining them in quadrature, the contributions to the absolute and relative 
  uncertainties are estimated as $\delta_\text{fid,abs} = 0.22\%$ and $\delta_\text{fid,rel} = 0.12\%$.
  
  \subsubsection{Elastic Cuts}
  
  Due to the relatively poor resolution of the 12\dg telescope for most kinematic variables, the philosophy for the
  final analysis regarding cuts on such variables was to keep such cuts wide so as to avoid cutting into regions
  of good data.  The fact that background contributions were very small in exclusively reconstructed 12\dg events permits
  such a philosophy, but it is important to assess the possible effects of such cuts by examining the effects of tightening
  cut boundaries.  As described in Section \ref{sec:12ana}, the four kinematic elastic cuts made in the 12\dg analysis were:
  \begin{enumerate}
   \item coplanarity ($\Delta\phi$),
   \item vertex $z$ correlation ($\Delta z$),
   \item beam energy reconstructed from the two track $\theta$ values assuming elastic kinematics ($E_{\text{beam},\theta}$), and
   \item and the single-arm missing energy of the lepton over the square of the energy as computed by the expected
         elastic energy from the reconstructed $\theta$ ($\Delta E'_\theta/E'^2$).
  \end{enumerate}
  Following a similar procedure as was used for the assessment of the fiducial cuts, the boundaries for each of these
  cuts was tightened over a reasonable range and the effect on the resultant luminosity extraction was computed.  Since each of these
  cuts was quite broad, tightening the cuts has a larger systematic effect on the luminosity extraction than broadening the cuts
  (since relatively few events are outside the cuts), and thus the following studies primarily concern examination
  of the effects of tighter cuts.
  
\begin{figure}[thb!]
  \centerline{\includegraphics[width=1.15\textwidth]{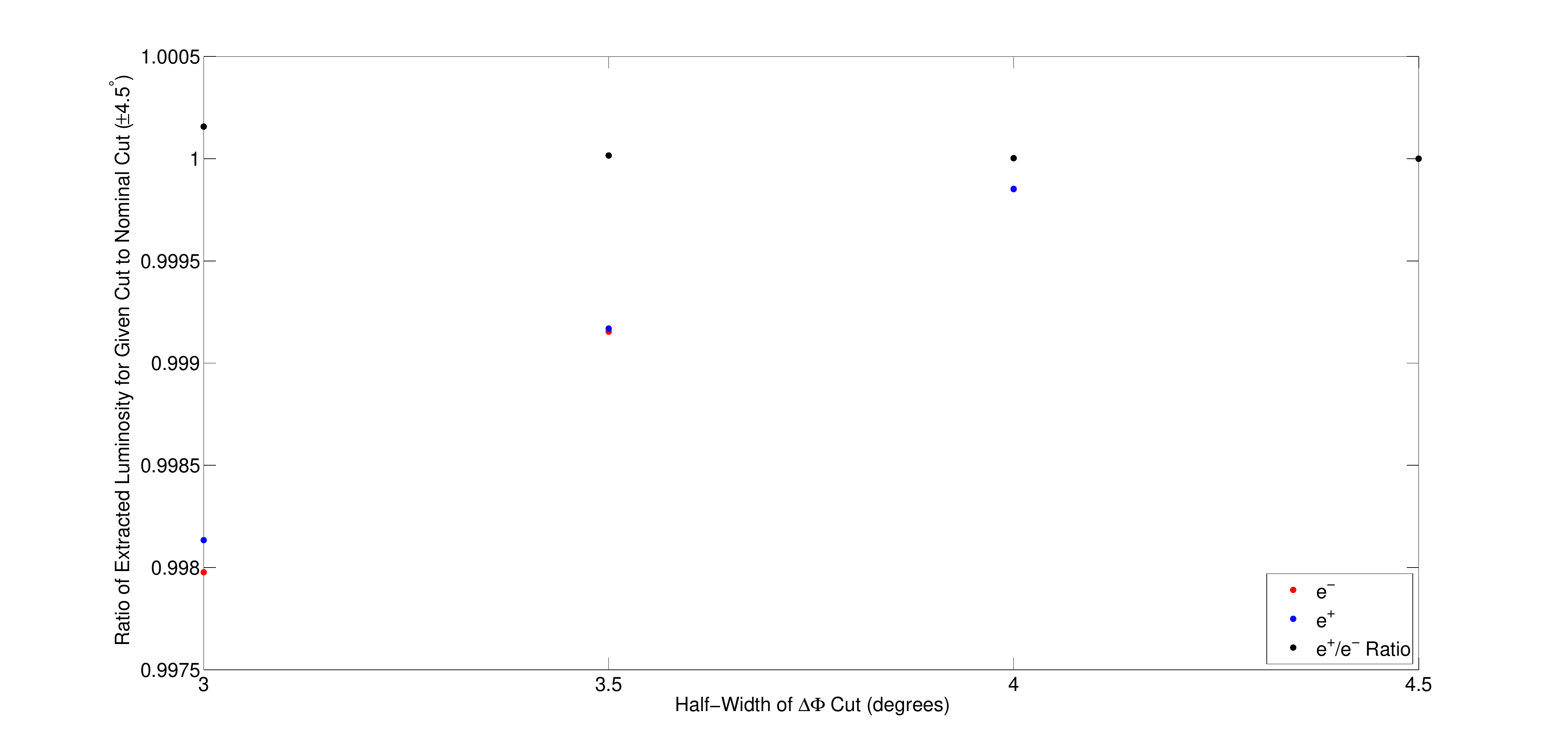}}
  \caption[Effect of coplanarity cut on the extracted 12\dg luminosity]{Ratio of the absolute and species-relative
  luminosities extracted in the 12\dg system for varying $e^\pm p$ track coplanarity ($\Delta\phi$) cuts to the value at the nominal cut of $\pm4.5^\circ$.}
  \label{fig:phicut}
  \end{figure}

  The nominal coplanarity ($\Delta\phi$) cut was placed conservatively at $\pm4.5^\circ$ around $180^\circ$ due to the lack
  of resolution in the lepton $\phi$ reconstruction and general uncertainty in the quality of proton reconstruction.
  This cut was tightened to $\pm3.0^\circ$ in four steps to examine the uncertainty associated with this cut, where
  $\pm3.25^\circ$ was judged to be the narrowest reasonable cut.  The results of this study are shown in Figure \ref{fig:phicut}.
  Tightening the cut lowers to the absolute
  luminosity estimate (as the data resolution in $\phi$ is slightly broader than in simulation), but leaves the relative
  luminosity stable.  The maximum effect on the luminosity from this cut is determined to be 0.01\% relative and 0.15\% absolute.
  
\begin{figure}[thb!]
  \centerline{\includegraphics[width=1.15\textwidth]{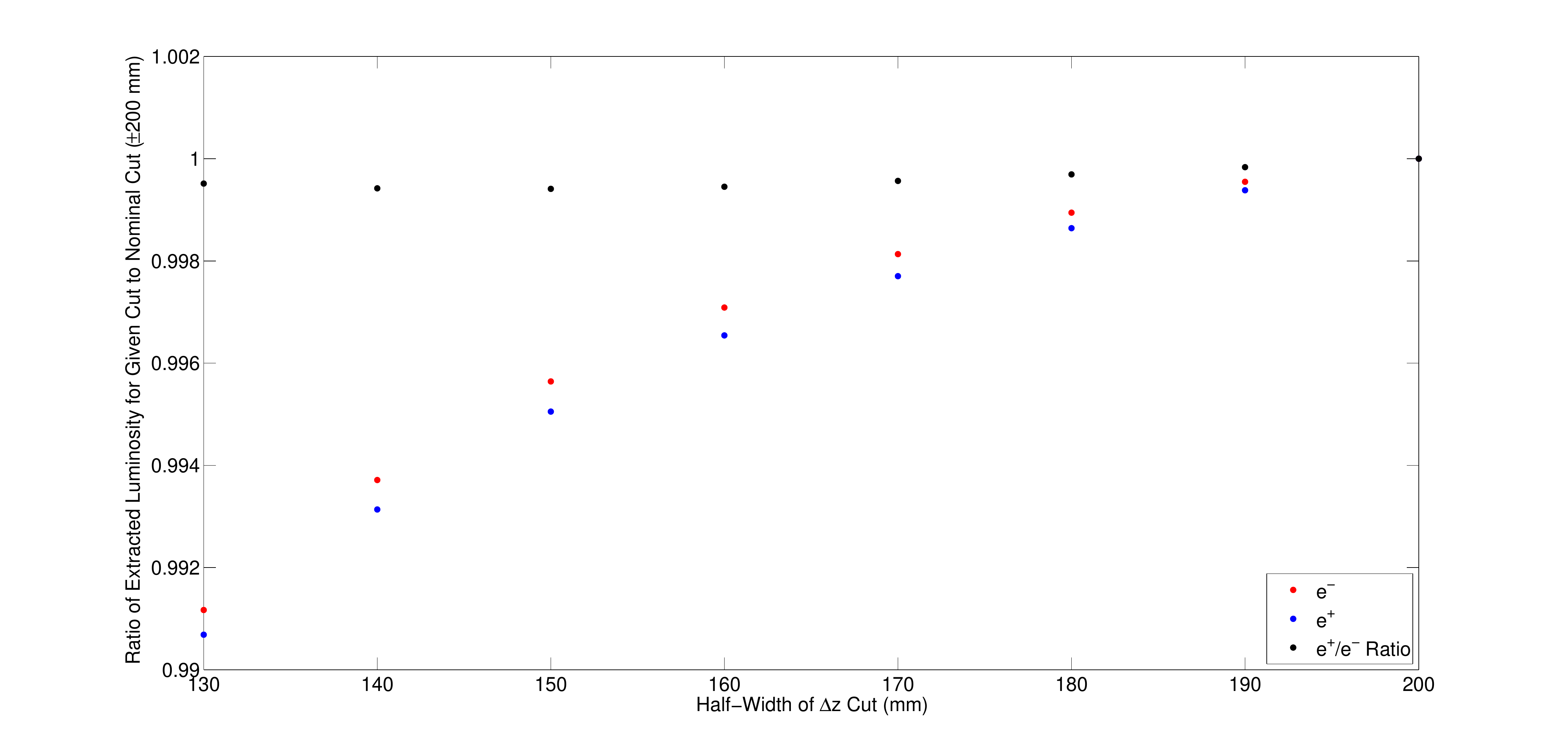}}
  \caption[Effect of vertex $z$ correlation cut on the extracted 12\dg luminosity]{Ratio of the absolute and species-relative
  luminosities extracted in the 12\dg for varying vertex $z$ correlation cuts ($\Delta z$) to the value at the nominal cut of $\pm200$ mm.}
  \label{fig:zcut}
  \end{figure}

  Varying the cut on the vertex $z$ correlation between the lepton and proton track from the nominal value of $\pm 200$ mm produced
  effects similar to those observed for the coplanarity cut.  Figure \ref{fig:zcut} shows the results of the $z$ cut study, in which the
  cut was varied towards the minimum reasonable value of $\pm130$ mm. From this information, the maximum effect on the luminosity
  extraction is estimated as 0.05\% relative and 0.8\% absolute.
  
\begin{figure}[thb!]
  \centerline{\includegraphics[width=1.15\textwidth]{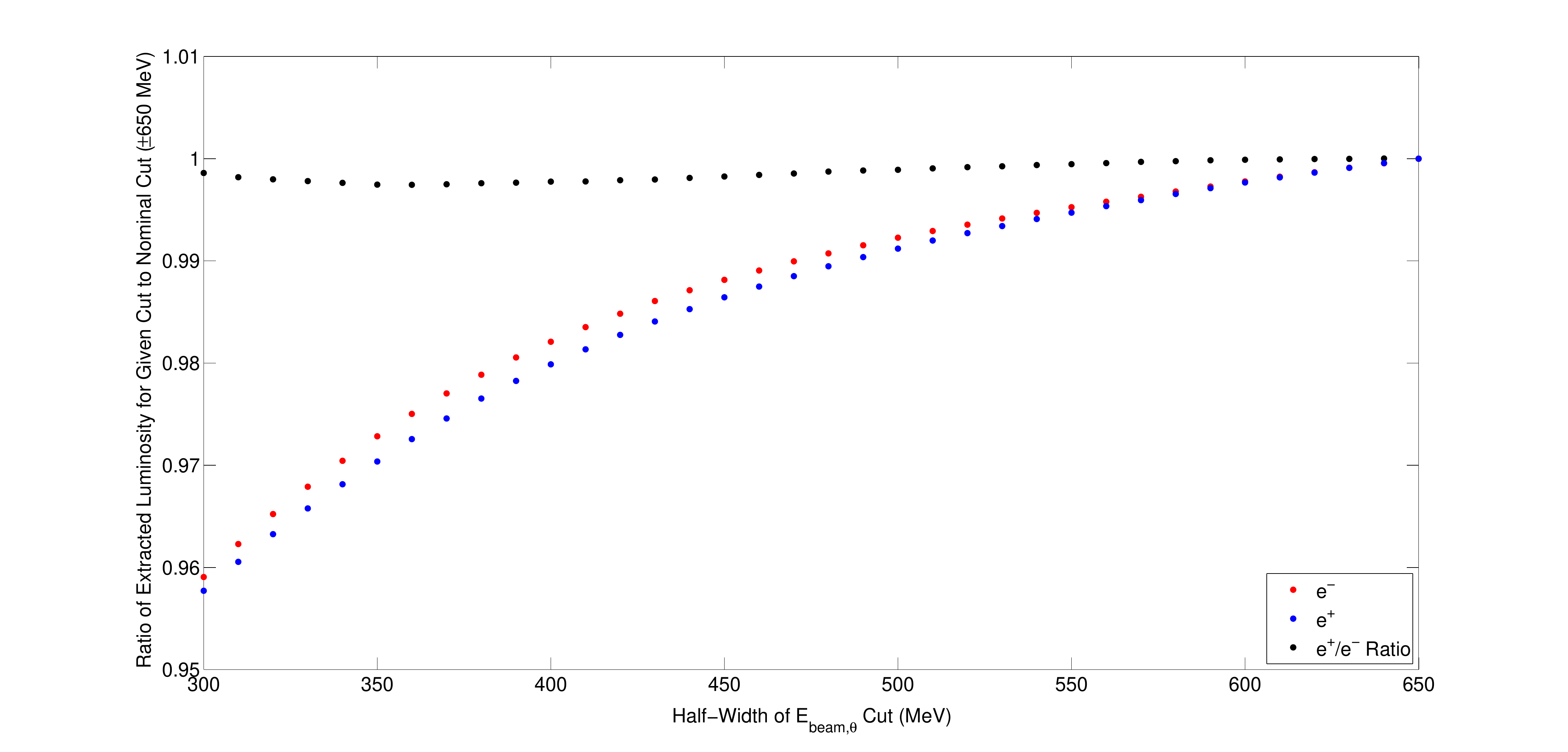}}
  \caption[Effect of reconstructed beam energy from angles cut on the extracted 12\dg luminosity]{Ratio of the absolute and species-relative
  luminosities extracted in the 12\dg system for varying cuts on the beam energy reconstructed from track angles assuming elastic kinematics ($E_{\text{beam},\theta}$)
  to the value at the nominal cut of $1950\pm650$ MeV.}
  \label{fig:ebangsys}
  \end{figure}

\begin{figure}[thb!]
  \centerline{\includegraphics[width=1.15\textwidth]{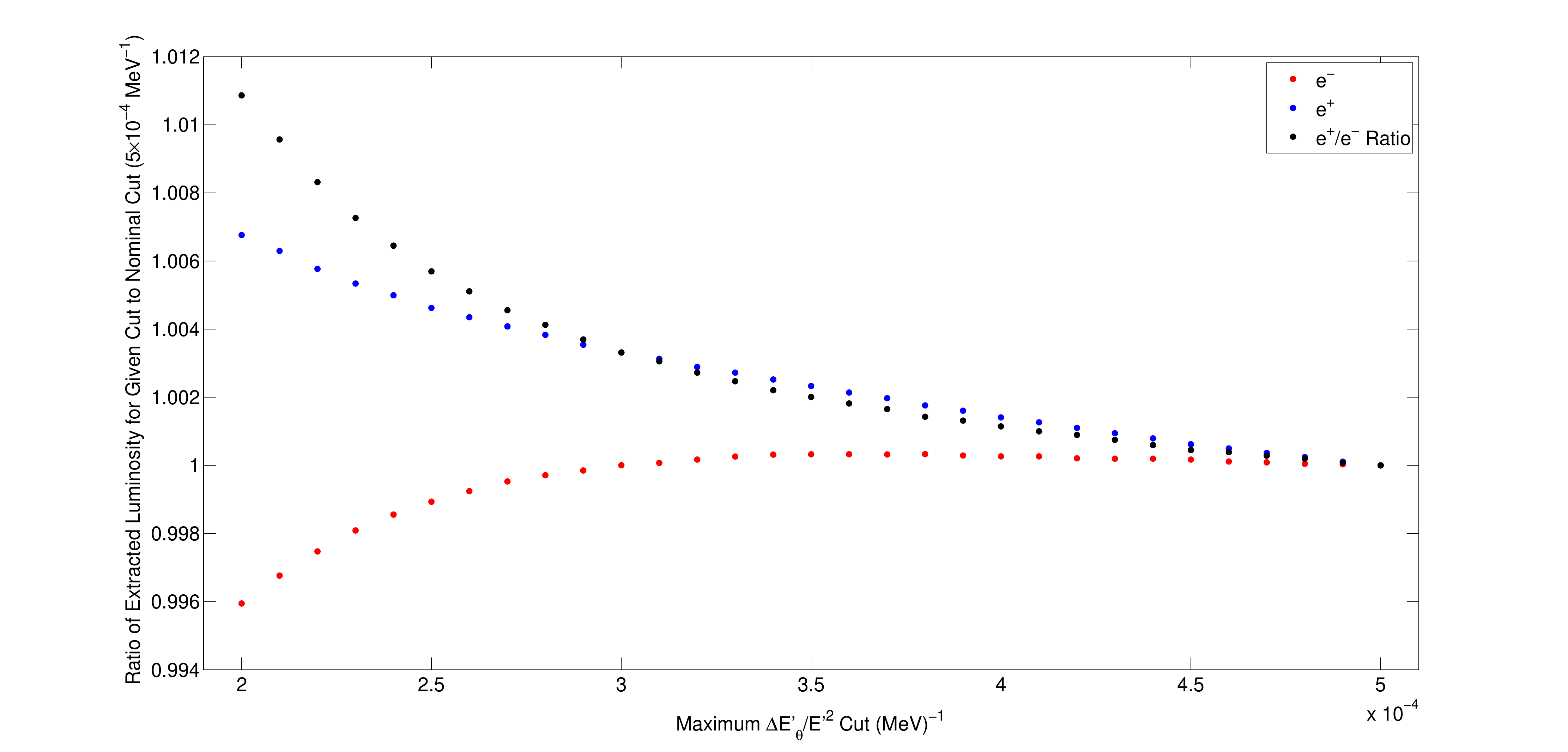}}
  \caption[Effect of lepton missing energy cut on the extracted 12\dg luminosity]{Ratio of the absolute and species-relative
  luminosities extracted in the 12\dg system for varying cuts on missing energy of the lepton relative to the expected
  elastic energy from the lepton $\theta$ over the square of the energy ($\Delta E'_\theta/E'^2$)
  to the value at the nominal cut of $<$5$\cdot 10^{-4}$ MeV$^{-1}$.}
  \label{fig:deesys}
  \end{figure}

  The final two cuts ($E_{\text{beam},\theta}$ and $\Delta E'_\theta/E'^2$) are heavily correlated as they both involve the expected
  elastic energy of the lepton from the lepton $\theta$.  First they will be considered separately in a fashion similar to the previous
  two cut studies, and then their correlation will be discussed to determine the overall contribution from these cuts to the systematic
  uncertainty.  For the $E_{\text{beam},\theta}$ cut, the range of the cut was varied from the nominal value of $1950\pm650$ MeV down
  to $1950\pm300$, while the nominal cut of $\Delta E'_\theta/E'^2 < 5\cdot 10^{-4}$ MeV$^{-1}$ was varied down to $\Delta E'_\theta/E'^2 < 2\cdot10^{-4}$ MeV$^{-1}$
  Since these cuts are very sensitive to multiple aspects of the detector resolution, the studied cut range was extended well beyond
  values considered reasonable.  Figures \ref{fig:ebangsys} and \ref{fig:deesys} show the results of the studies for the 
  $E_{\text{beam},\theta}$ and $\Delta E'_\theta/E'^2$ cuts, respectively.
  
  For the $E_{\text{beam},\theta}$ cut (Figure \ref{fig:ebangsys}), the effect on the relative luminosity is quite flat even for the tightest cuts, and thus
  the contribution of error to the relative luminosity from this effect alone may be estimated to be 0.18\% (the maximum
  deviation from unity in the study).  For the absolute luminosity, the deviation from the nominal value
  quickens for cuts below a half-width of 450 MeV, indicating that this region is not a valid cut region due to its sensitivity
  to distribution rapidly changing in this area.  The deviation from unity at $\pm450$ MeV is thus taken as the estimate
  of the absolute uncertainty due to this effect: 1.40\%.
  
  Variance of the $\Delta E'_\theta/E'^2$ cut (Figure \ref{fig:deesys}) causes a smaller absolute effect than the $E_{\text{beam},\theta}$ angles cut, but
  notably shows different behaviors between species.  Due to the poor momentum resolution for reconstructing the leptons, a cut
  at $\Delta E'_\theta/E'^2 < 3.5\cdot10^{-4}$ MeV$^{-1}$ was considered the tightest reasonable cut, resulting in an uncertainty
  of 0.20\% for the relative luminosity and 0.24\% for the absolute value for this effect alone.
  
  Comparing the values from these two effects, they individually indicate a similar magnitude of uncertainty for the relative
  luminosity while differing in the absolute luminosity contribution.  The larger effect of $E_{\text{beam},\theta}$ on the absolute
  luminosity is likely due to uncertainty in the proton tracking in data, which broadens this distribution relative to simulation in
  which the recoil protons are more accurately reconstructed.  While it is known that these effects are correlated, it is difficult to
  fully decouple them due to the complexity of the variables.  Thus, as a very conservative estimate, they are considered independent
  and are added in quadrature for the final estimate for elastic cut uncertainty contribution to the luminosity.
    
  Combining the four cut variation effects in quadrature the total uncertainties due to the elastic cuts
  amount to $\delta_\text{elas,rel} = 0.27\%$ and $\delta_\text{elas,abs} = 1.63\%$ for the relative and absolute
  extractions, respectively.  The fact that these  uncertainties are large relative to other sources demonstrates
  that the GEMs would have been a valuable contribution to the 12\dg measurement had it been feasible, due to the
  large improvement they provide to tracking resolution.  It is notable that the MWPCs, originally designed to provide
  alignment/efficiency measurement capability for the GEMs and extra low-resolution track points, performed extremely
  well to provide a luminosity measurement of sufficient precision to satisfy the OLYMPUS goals.
 
  \subsubsection{Radiative Corrections}
  
\begin{figure}[thb!]
  \centerline{\includegraphics[width=1.0\textwidth]{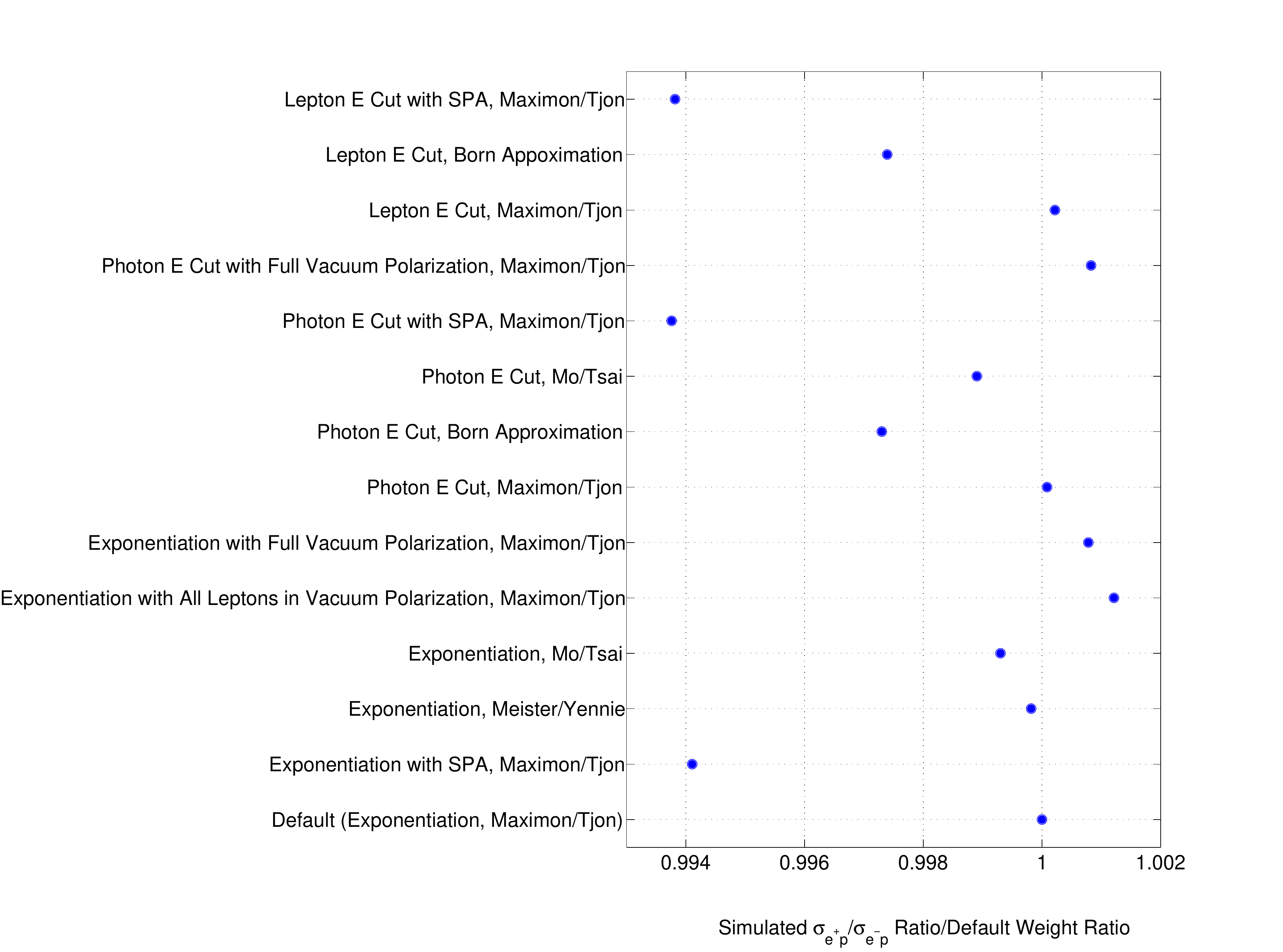}}
  \caption[Effect of radiative corrections uncertainty on the 12\dg luminosity ratio]{Effects of applying different 
  radiative corrections schemes to the simulation on the resulting extracted relative 12\dg luminosity.  The maximal deviation
  between realistic models was $\delta_\text{rad,rel} = \pm0.08\%$.  Details on the three main
  correction schemes included above (Maximon \& Tjon, Meister \& Yennie, and Mo \& Tsai) may be found in
  \cite{MaximonPhysRevC.62.054320}, \cite{MeisterPhysRev.130.1210},
  and \cite{MoRevModPhys.41.205}, respectively, and a more detailed discussion of these methods may be found in 
  Section \ref{sec:radgen}.}
  \label{fig:rcsys}
  \end{figure}

\begin{figure}[thb!]
  \centerline{\includegraphics[width=1.0\textwidth]{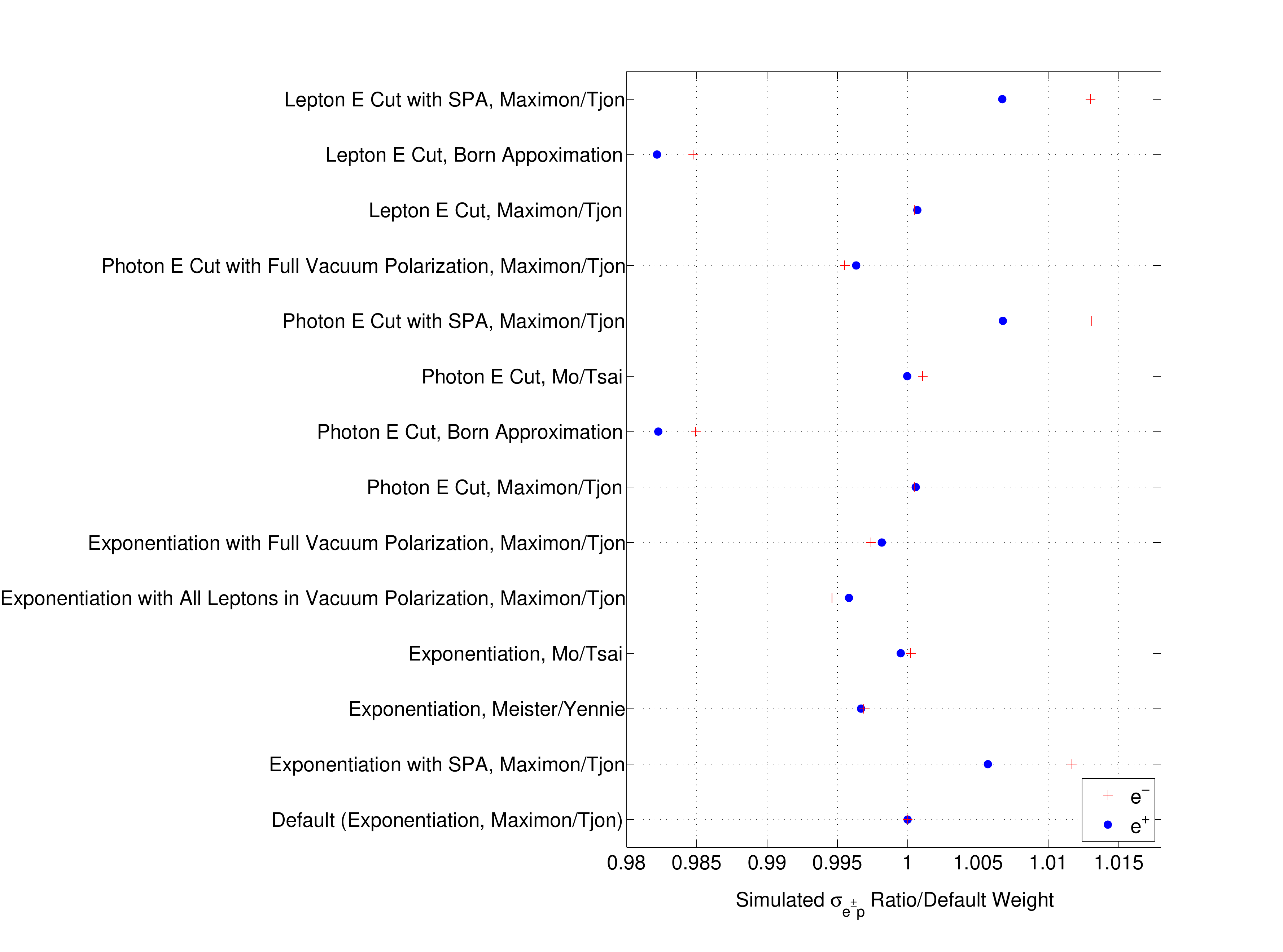}}
  \caption[Effect of radiative corrections uncertainty on the absolute 12\dg luminosity]{Effects of applying different 
  radiative corrections schemes to the simulation on the resulting extracted absolute 12\dg luminosity for each species.  The maximal deviation
  between realistic models was $\delta_\text{rad,abs} = \pm0.45\%$.  Details on the three main
  correction schemes included above (Maximon \& Tjon, Meister \& Yennie, and Mo \& Tsai) may be found in References
  \cite{MaximonPhysRevC.62.054320}, \cite{MeisterPhysRev.130.1210},
  and \cite{MoRevModPhys.41.205}, respectively, and a more detailed discussion of these methods may be found in 
  Section \ref{sec:radgen}.}
  \label{fig:rcsysabs}
  \end{figure}

  To assess the effect of radiative corrections on the extracted relative and absolute luminosities, the experiment's
  simulation included a number of radiative corrections schemes, implemented via multiple stored event weights as
  described in Section \ref{sec:radgen}.  Figure \ref{fig:rcsys} shows the effect on the extracted species
  relative luminosity for the available radiative corrections schemes, while Figure \ref{fig:rcsysabs} shows
  the effect on the absolute luminosity extractions for each species.  The simulated data shown in the figures
  includes most of the Run II dataset (several thousand runs) and therefore the statistical uncertainty, which is
  also highly correlated between the points, is extremely small relative to the variation in the points due to the weights.
  Several models included in the radiative corrections (in particular the Born approximation and soft photon approximation (SPA))
  are included only for comparison with historical experimental data and are not considered realistic models.  Additionally,
  the inclusion of full vacuum polarization effects derived from $e^+e^-\rightarrow\text{hadrons}$ cross sections is
  preferred to the inclusion of only lepton vacuum polarization effects \cite{Actis2010,vacpolweb,vacpolpres}.  Further details on the different radiative
  corrections schemes included in the figures may be found in Section \ref{sec:radgen}.
  
  After discounting the previously aforementioned unrealistic models, the spread in the remaining results was
  used to estimate the systematic uncertainty introduced by radiative corrections.  The maximally different results
  occurred between the Mo \& Tsai (Reference \cite{MoRevModPhys.41.205}) correction under the photon $\Delta E$ method and the methods with full vacuum polarization
  included for both the absolute and relative measurements.  The radiative correction uncertainty was thus conservatively
  taken to be the maximal spread: $\delta_\text{rad,rel} = \pm0.08\%$ and $\delta_\text{rad,abs} = \pm0.45$.

  \subsubsection{Elastic Form Factors}
  
\begin{figure}[thb!]
  \centerline{\includegraphics[width=1.15\textwidth]{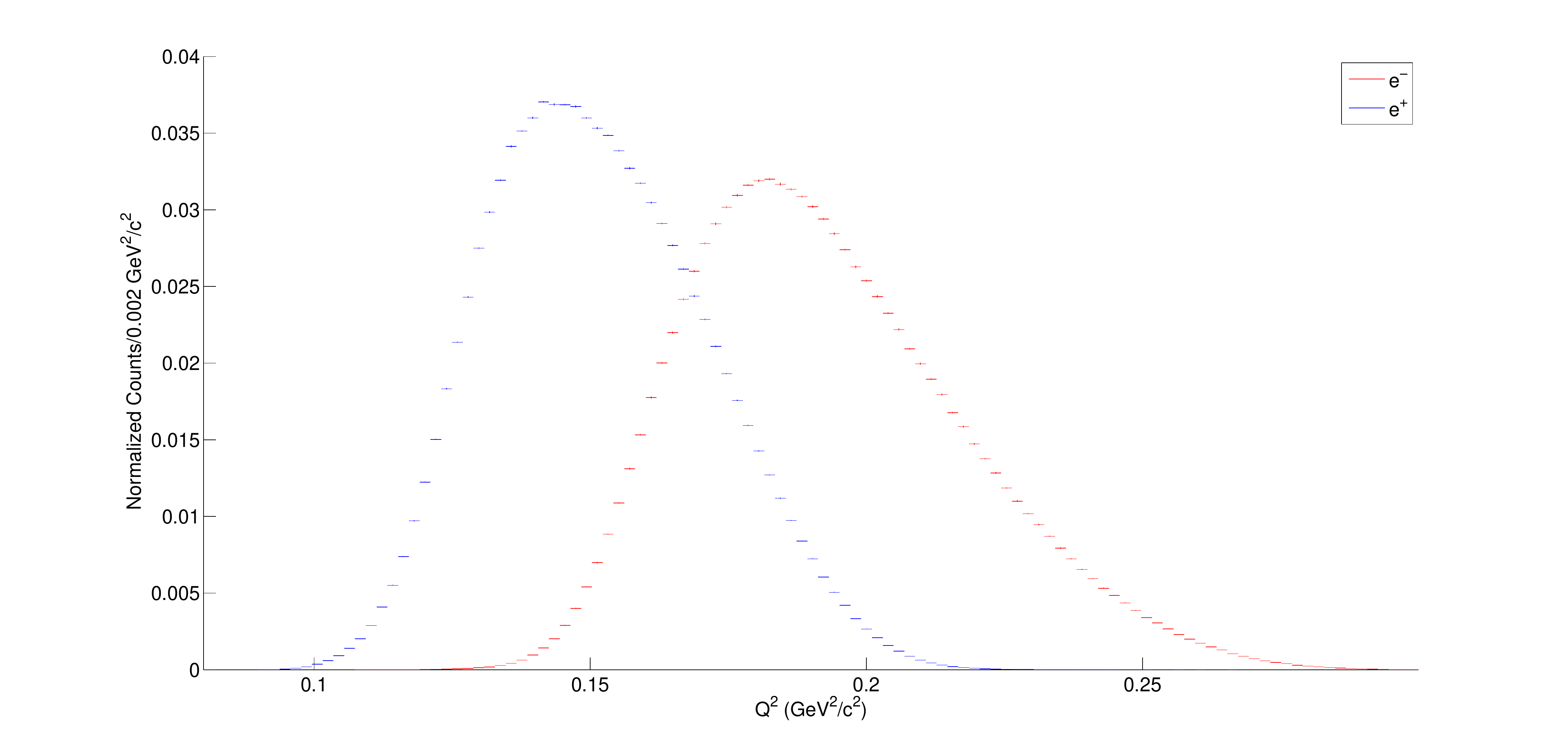}}
  \caption[Normalized $Q^2$ distributions of accepted $e^+p$ and $e^-p$ events in the 12\dg luminosity system]{Normalized distributions of
  the $Q^2$ computed from the reconstructed lepton $\theta$ for electron- and positron-proton elastic scattering events accepted by the
  12\dg system.  Note the differences of the systematic $\sim$0.3 GeV$^2/c^2$ shift between the acceptance ranges of the two species and the
  longer tail at higher $Q^2$ values for electrons caused by the in-bending of the negatively-charged electrons from the upstream end of the target cell.}
  \label{fig:12degq2}
  \end{figure}

  Due to the fact that the $Q^2$ range accepted by the 12\dg telescopes for each of the lepton species is slightly different
  in the single toroid polarity, there is an uncertainty introduced from the uncertainty in the magnitude of the proton
  elastic form factors and the variation in the form factors at small $Q^2$.  Figure \ref{fig:12degq2} shows the distribution in $Q^2$ of
  events of each lepton species in the 12\dg telescopes, which notably exhibits a shift of approximately 0.3 GeV$^2/c^2$ between
  the two event types. Thus, the form factor (a function of $Q^2$) will not cancel completely in the ratio of the integrated acceptances
  for each species.  Additionally, any uncertainty in the magnitude of the form factors directly contributes systematic uncertainty
  to the absolute luminosity extraction.
  
\begin{figure}[thb!]
  \centerline{\includegraphics[width=1.0\textwidth]{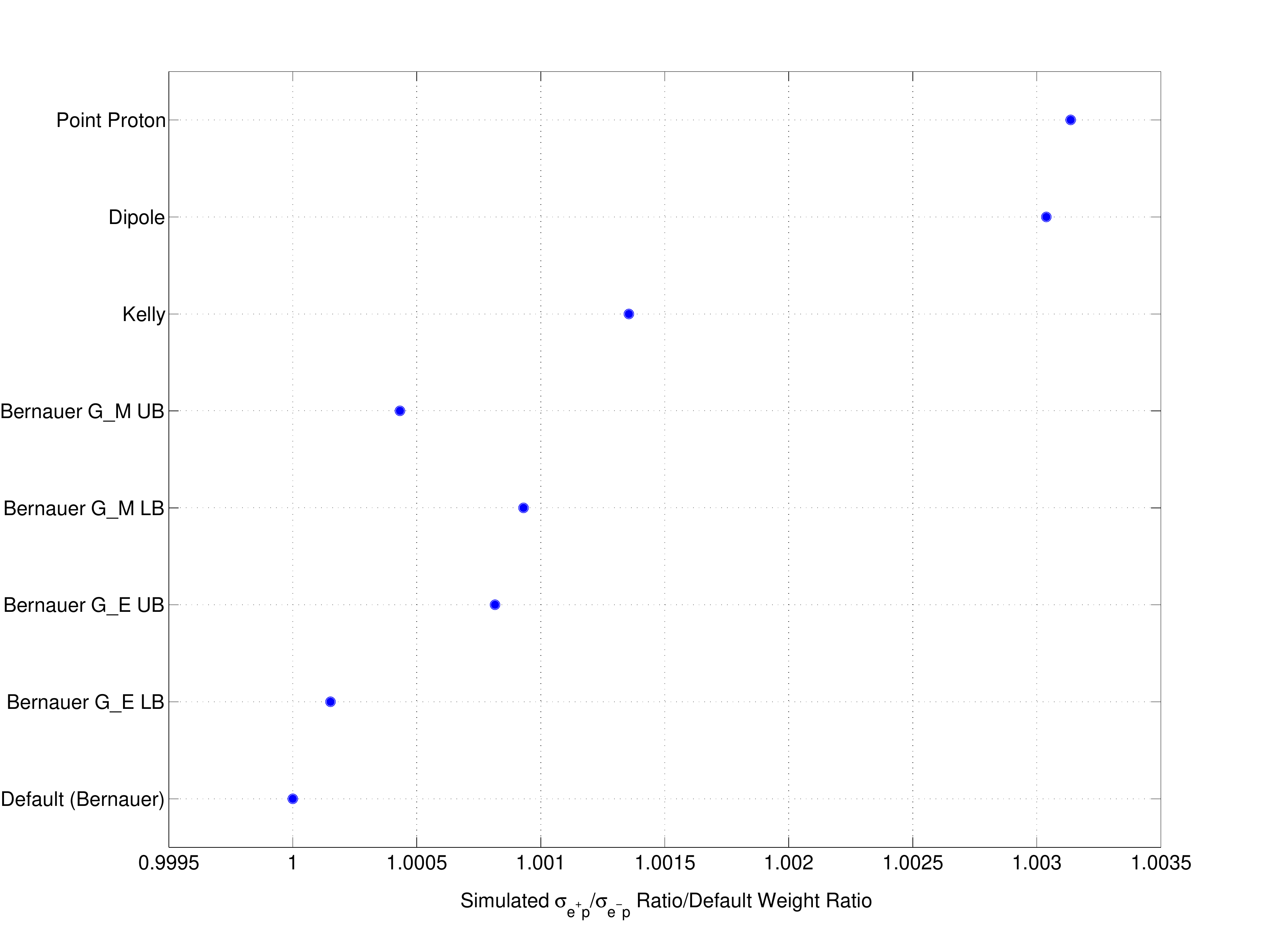}}
  \caption[Effect of form factor uncertainty on the relative 12\dg luminosity]{Effects of applying different
  form factor models to the simulation on the resulting extraction of the species-relative 12\dg luminosity
  ratio, including the unphysical dipole and point proton models.  The systematic uncertainty was estimated as the spread in the Bernauer and Kelly parametrizations
  (References \cite{BerFFPhysRevC.90.015206} and \cite{KellyPhysRevC.70.068202}, respectively): $\delta_\text{ff,rel} = \pm0.14\%$.
  Note that radiative corrections were applied to these results using the default method (exponentiation,
  Maximon \& Tjon \cite{MaximonPhysRevC.62.054320}) in each case, but that the effects of the changing form factor were accounted for in calculating
  the radiative corrections.}
  \label{fig:ffsys}
  \end{figure}

\begin{figure}[thb!]
  \centerline{\includegraphics[width=1.0\textwidth]{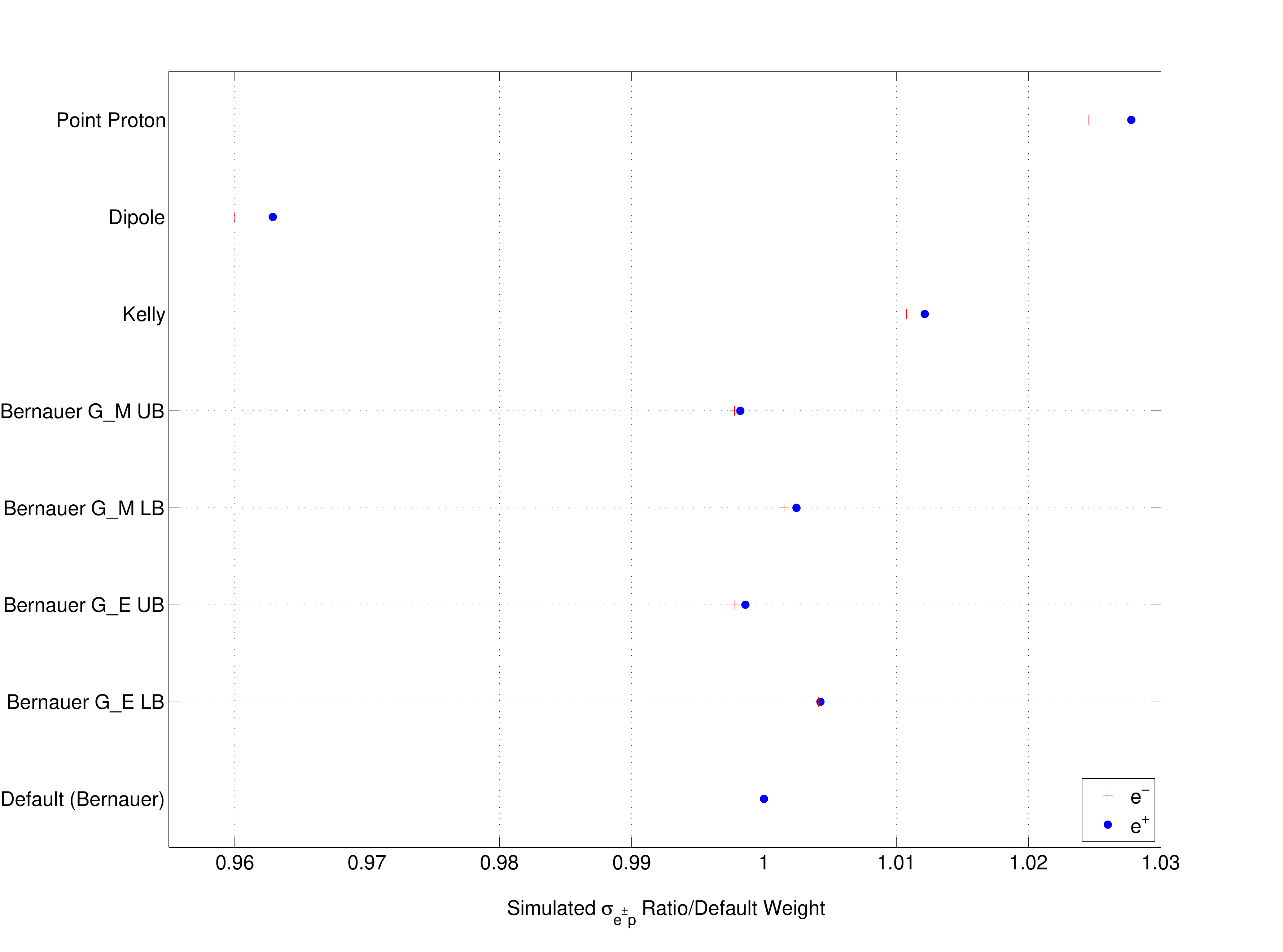}}
  \caption[Effect of form factor uncertainty on the absolute 12\dg luminosity]{Effects of applying different
  form factor models to the simulation on the resulting extraction of the absolute 12\dg luminosity
  for each species, including the unphysical dipole and point proton models.  The systematic uncertainty was estimated as the spread in the Bernauer and Kelly parametrizations
  (References \cite{BerFFPhysRevC.90.015206} and \cite{KellyPhysRevC.70.068202}, respectively): $\delta_\text{ff,rel} = \pm1.2\%$.
  Note that radiative corrections were applied to these results using the default method (exponentiation,
  Maximon \& Tjon \cite{MaximonPhysRevC.62.054320}) in each case, but that the effects of the changing form factor were accounted for in calculating
  the radiative corrections.}
  \label{fig:ffsysabs}
  \end{figure}

  To estimate the uncertainty in the 12\dg luminosity estimate due to uncertainties in the proton form factors, a similar
  method was used as in the previous section to estimate the effects of radiative corrections uncertainties.  The OLYMPUS
  simulation includes additional event weights for several form factor models including, the Bernauer, et al.\ fits
  (Reference \cite{BerFFPhysRevC.90.015206}) and the Kelly model (Reference \cite{KellyPhysRevC.70.068202}), as well as dipole and
  point-like proton models for reference.  Most notably, the Bernauer model exhibits a structure in $G_M$ in the vicinity 
  of $Q^2 \approx 0.2$ GeV$^2/c^2$ (Figure 20.b. in Reference \cite{BerFFPhysRevC.90.015206}) that is not present in the majority of form factor fits that creates disagreement in the
  prediction of different form factor models within the 12\dg system acceptance.  The spread in the resulting luminosities when using the Bernauer and Kelly
  models, which does not exhibit the aforementioned structure in $G_M$, was taken as an estimate of the systematic uncertainty due to the form factors, discounting the spread due to the
  unrealistic  proton and dipole form factors.  Figures \ref{fig:ffsys} and \ref{fig:ffsysabs} show the results of this study
  for the species-relative and absolute luminosity extractions, respectively, in which it was found that the uncertainties due
  to the form factors could be estimated as $\delta_\text{ff,rel} = \pm0.14\%$ and $\delta_\text{ff,rel} = \pm1.2\%$.
  
  \subsubsection{TPE at 12\dg}
  \label{ss:tpe12sys}
  
  As noted in Section \ref{sec:12posstpe}, the underlying physics process measured by the 12\dg system is the same process ($e^\pm p$ elastic scattering)
  that is under examination in the main detector for TPE contributions and thus it is not excluded that a difference in the $e^+p$ and
  $e^-p$ cross sections at $\theta \approx 12^\circ$ due to TPE could exist and systematically shift any attempt to determine the
  relative luminosity from this measurement.  To estimate the possible size of this shift, several methods may be considered:
  \begin{enumerate}
   \item comparison with the other OLYMPUS luminosity measurements (amounting to a measurement of \ratio at $Q^2 \approx 0.165$ GeV$^2/c^2$, $\epsilon \approx 0.98$
         as is discussed in Section \ref{sec:12TPE}),
   \item comparison with existing data, and
   \item examination of the spread of theoretical predictions for the value of \ratio in 12\dg region.
  \end{enumerate}
  The question of the first point is addressed in Section \ref{sec:12TPE} and effectively does not treat the 12\dg result as a
  luminosity measurement, so here it is attempted to estimate this uncertainty from outside sources.  From the standpoint of data
  constraints, the VEPP-3 experiment (Reference \cite{vepp3PhysRevLett.114.062005}) used forward elastic $e^\pm p$ scattering for
  luminosity normalization in a similar fashion as OLYMPUS would use the 12\dg data as a standalone luminosity normalization and
  thus does not provide a useful constraint on the value of \ratio at high $\epsilon$/low $Q^2$.  The CLAS experiment, which
  used photon-induced $e^+e^-$ pair production to balance luminosities, provides a measurement of \ratio in the small angle bin with
  average $Q^2 = 0.232$ GeV$^2/c^2$, $\epsilon = 0.915$ of $R_{2\gamma} = 0.991\pm 0.009$ (including systematic and statistical uncertainty) and 
  a variety of measurements at $\epsilon > 0.88$ and $Q^2<0.9$ GeV$^2/c^2$ that all scatter around $R_{2\gamma}=1$ within uncertainties on the
  order of 1\% \cite{rimal,ass}. Most of the data prior to the modern experiments, predominantly from the 1960s, consists of data at low $Q^2$ but
  higher $\epsilon$ (References \cite{Yount:1962aa,Browman:1965zz,Anderson:1966zzf}) or are at much higher energies/$Q^2$ (References \cite{Cassiday:1967aa,
  Bouquet:1968aa,Mar:1968qd}).  One older experiment (Reference \cite{Bartel:1967aa}), provides a measurement at a comparable kinematic
  point to the 12\dg system of $R_{2\gamma}=1.012$, but quotes uncertainties on the order 3.0\%.  Thus, at best, existing data constrains this uncertainty to order 0.5\%, and
  no precise measurement of \ratio at very high $\epsilon$ exists in previous data.
  
  Examining various theoretical and phenomenological models for TPE in this region (References \cite{BerFFPhysRevC.90.015206,Chen:2007ac, Guttmann:2010au,Blunden:2003sp,Chen:2004tw,Afanasev:2005mp,
    Blunden:2005ew, Kondratyuk:2005kk, Borisyuk:2006fh,TomasiGustafsson:2009pw}), most predict $R_{2\gamma}$ to be near unity but vary on the order of
  a tenth of a percent, as exemplified in Figure \ref{fig:projections}, and thus are all roughly consistent with the previously aforementioned experimental
  data.  Taking into account these available sources of experimental and theoretical estimates, it is reasonable to apply an additional 0.1\% uncertainty
  on the relative and absolute luminosities when the 12\dg is taken as a standalone normalization point for the data set: 
  $\delta_\text{TPE,abs} = \delta_\text{TPE,rel} = \pm 0.1\%$.  Any contribution OLYMPUS can make in reducing this uncertainty via
  the combination of multiple luminosity measurements would be valuable for the constraint of the VEPP-3 normalization point
  and the various models.
  
  \subsubsection{Additional Discussion on Systematic Uncertainties}
  
  As can be seen in Table \ref{tab:12ds}, the 12\dg system total systematic uncertainty for the relative luminosity is  well below 1\% and thus
  provides a sufficiently good measurement for the \ratio analysis to meet the OLYMPUS goals.  Additionally, the system performed admirably
  in providing an approximate absolute luminosity measurement (or cross section measurement if combined with an alternate luminosity measurement).
  Furthermore, note that the dominant systematics (ToF trigger efficiency, lepton tracking, efficiency, magnetic field, and elastic cuts)
  are all estimated in a non-Gaussian fashion and thus represent ``box-like'' limits.  Those wishing to interpret the systematic uncertainty
  as a Gaussian $1\sigma$ error could approximate this by multiplying the standard deviation of the uniform distribution of unity width
  ($1/\sqrt{12}$), although this is not necessarily a correct interpretation or good estimate of the systematic uncertainty.  Certain contributions 
  to the uncertainty are correlated between the left and right telescopes, reducing somewhat
  the estimate of the relative uncertainty between the telescopes.  The details of this will be discussed in the comparison of the left
  and right side results in Section \ref{sec:12res}.
  
  \subsection{Results}
  \label{sec:12res}

  With the complete analysis method and systematic uncertainty determination, the final results for the 12\dg luminosity may be considered.
  The results in this section are quoted as ratios of the measured 12\dg luminosity (calculated separately for detection of the lepton in
  the left and right telescopes and for the combined statistics of both sides) to the luminosity used to generate the simulation event sets
  (i.e., the slow control luminosity for a given run).  This amounts to reformulating Equation \ref{eq:l12} as:
  \begin{equation}
   \frac{ \mathcal{L}_{\text{12}^\circ} } {\mathcal{L}_\text{MC} } = \frac{\mathcal{L}_{\text{12}^\circ}}{\mathcal{L}_\text{SC}} = \frac{N_\text{data}}{N_\text{MC}\left(\mathcal{L}_\text{SC}\right)}.
  \end{equation}
  Note that the value of ${ \mathcal{L}_{\text{12}^\circ} }/{\mathcal{L}_\text{SC} }$ is expected to be unity only within the uncertainty of the slow control luminosity measurement, which, as discussed
  in Section \ref{sec:scsum}, is on the order of several percent in both the relative and absolute luminosities.
  
  To provide a scale in units of integrated luminosity, a typical data run consisted of $\sim$1.5$\cdot 10^{36}$ cm$^{-2}$ of recorded integrated luminosity. Thus, the entire data
  set used for the analysis in this work ($\sim$2200 runs) corresponds to $\sim$3.1$\cdot 10^{39}$ cm$^{-2}$ ($\sim$3.1 fb$^{-1}$) of data, approximately equally split between the two lepton species.
  
\begin{sidewaysfigure}
  \centerline{\includegraphics[width=1.05\textwidth]{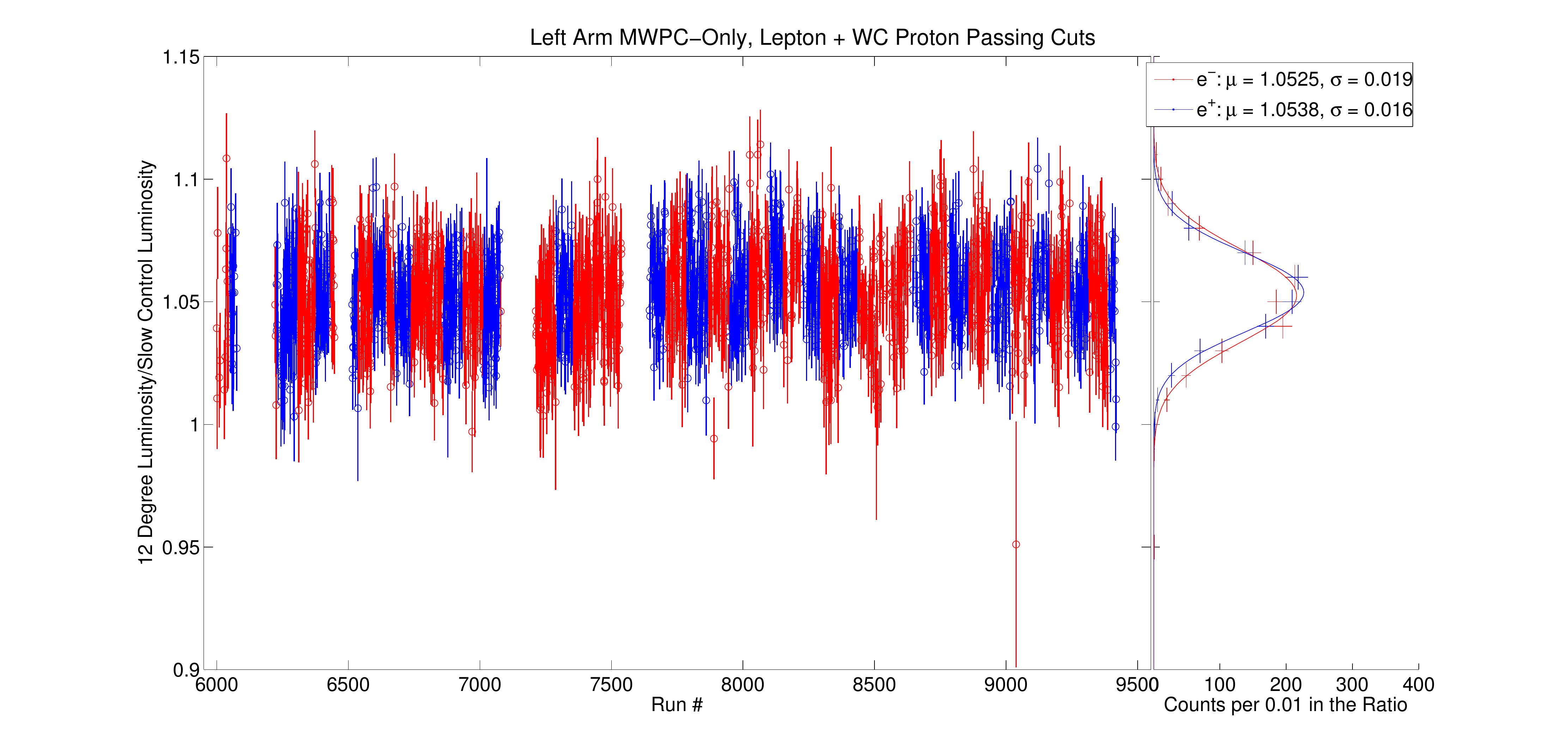}}
  \caption[Luminosity determined by the left 12\dg arm by run]{Run-by-run estimate of the integrated luminosity collected by the OLYMPUS experiment relative to the slow control estimate,
  as measured by examining \pmp events in which the lepton was detected in the left 12\dg telescope in coincidence with a proton in the right drift chamber, subject to the analysis described
  in Section \ref{sec:12ana}.  The run-by-run values are histogrammed in the right-hand plot and fitted to Gaussian
  distributions for each species to produce the estimates of the luminosities for each species over the full data set.}
  \label{fig:12left}
  \end{sidewaysfigure}

\begin{sidewaysfigure}
  \centerline{\includegraphics[width=1.05\textwidth]{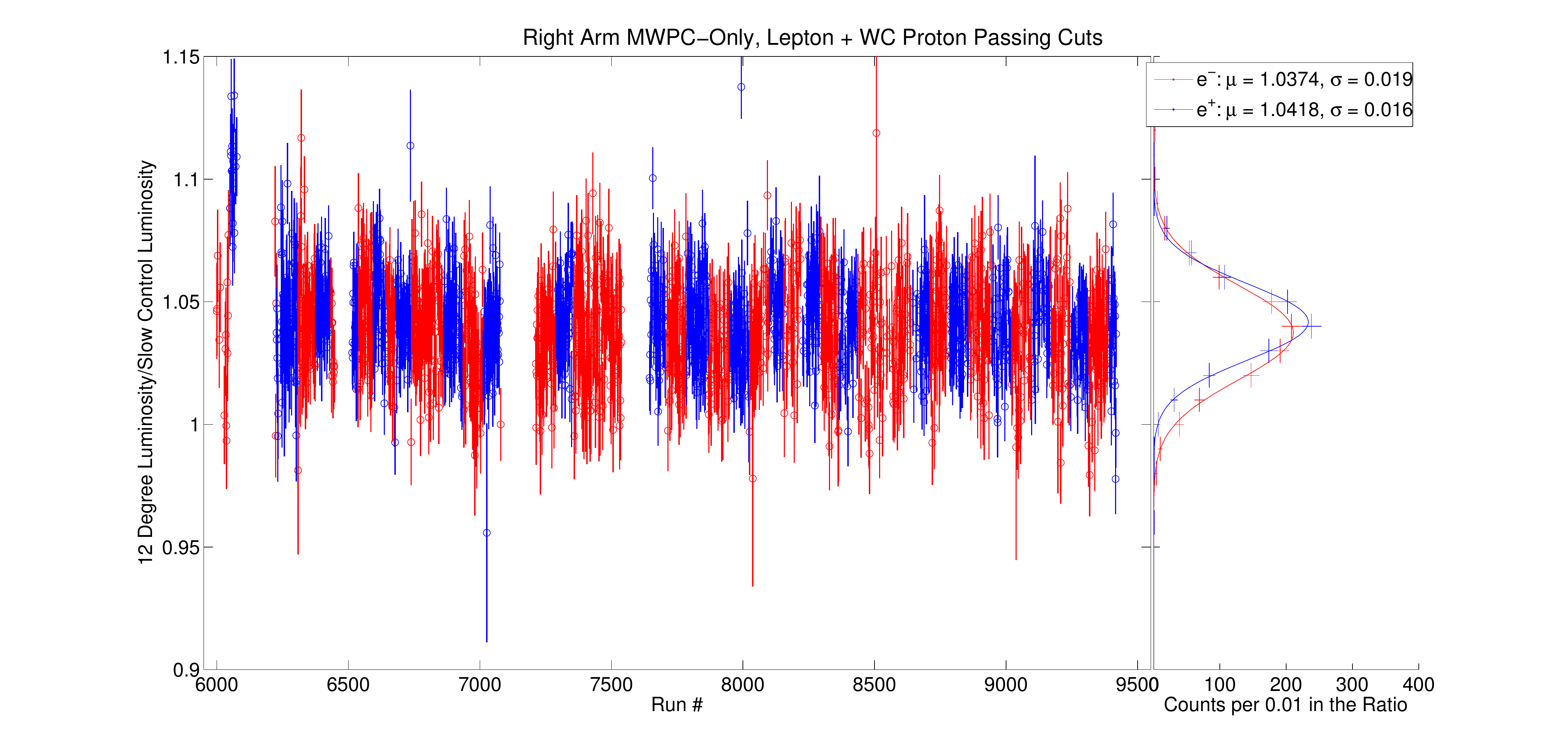}}
  \caption[Luminosity determined by the right 12\dg arm by run]{Run-by-run estimate of the integrated luminosity collected by the OLYMPUS experiment relative to the slow control estimate,
  as measured by examining \pmp events in which the lepton was detected in the right 12\dg telescope in coincidence with a proton in the left drift chamber, subject to the analysis described
  in Section \ref{sec:12ana}.  The run-by-run values are histogrammed in the right-hand plot and fitted to Gaussian
  distributions for each species to produce the estimates of the luminosities for each species over the full data set.  The outlier positron runs in the vicinity of Run 6050 were taken while
  the MWPC data acquisition was malfunctioning, and thus they are excluded from the estimate.}
  \label{fig:12right}
  \end{sidewaysfigure}

\begin{sidewaysfigure}
  \centerline{\includegraphics[width=1.05\textwidth]{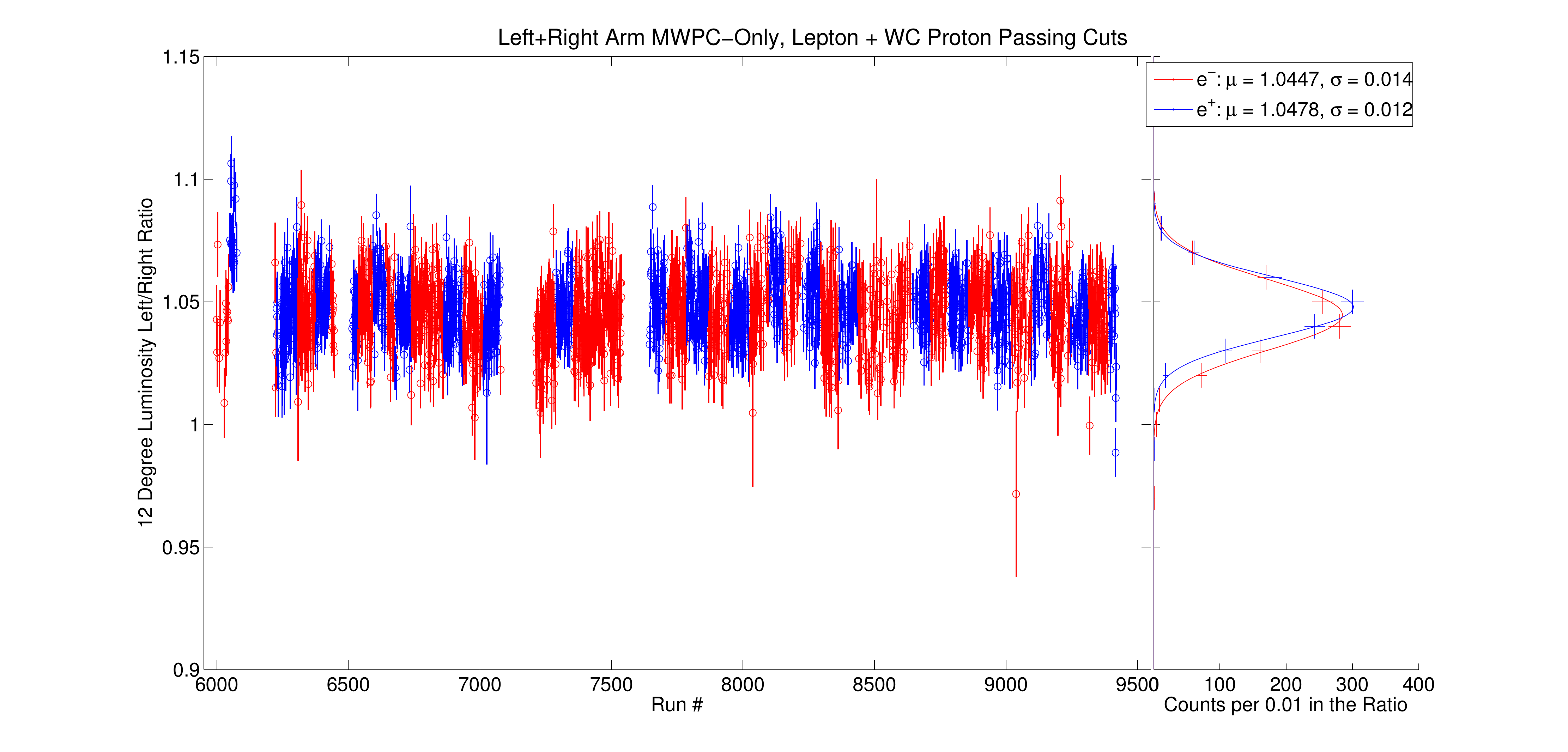}}
  \caption[Luminosity determined by the combined left and right 12\dg arm results by run]{Run-by-run estimate of the integrated luminosity collected by the OLYMPUS experiment relative
  to the slow control estimate,
  as measured by examining \pmp events in which the lepton was detected in either 12\dg telescope in coincidence with a proton in the opposite side drift chamber, subject to the analysis described
  in Section \ref{sec:12ana}.  The run-by-run values are histogrammed in the right-hand plot and fitted to Gaussian
  distributions for each species to produce the estimates of the luminosities for each species over the full data set. The outlier positron runs in the vicinity of Run 6050 were taken while
  the MWPC data acquisition was malfunctioning, and thus they are excluded from the estimate.}
  \label{fig:12comb}
  \end{sidewaysfigure}

  The results of the 12\dg luminosity estimates are summarized in Table \ref{tab:12results}, showing the measurements and associated uncertainties (as determined in Section \ref{ss:12sys})
  for detection of the lepton in the left and right arms as well as the measurement combining the statistics of the two samples.  Figures \ref{fig:12left}, \ref{fig:12right}, and \ref{fig:12comb}
  show the run-by-run estimates for the left arm, right arm, and combined measurements respectively.  For each run-by-run sample, the values are histogrammed (weighted by their statistical significance)
  to provide a means of determining the effective values of ${ \mathcal{L}_{\text{12}^\circ} }/{\mathcal{L}_\text{SC} }$ for the full dataset.
  Statistical errors on the datasets were computed at the 95\% confidence bound on the fit to the mean of this histogram.  For all estimates, the statistical uncertainty is effectively negligible
  relative to the systematic uncertainties. In general, the widths of the histogrammed distributions of run-by-run 
  luminosities were found to be only slightly larger ($\sim$0.1\%) than the mean statistical error associated with the estimate from individual runs (1.5\% and 1.7\% for single-telescope measurements of 
  \pp and \ep runs respectively) and the distributions are well represented by Gaussian distributions,
  indicating that only minor time-varying systematic effects were present in the slow control luminosity.  As will be discussed in Section \ref{sec:alllumi},
  the SYMB run-by-run estimates show similar features as the 12\dg, indicating that the run-by-run variance was indeed due to effects in the slow control estimate rather than the 12\dg estimate.
  
    \begin{table}[htb!]
  \begin{center}
  \begin{tabular}{|l|c|}
  \hline
  Measurement & Value \\
  \hline\hline 
  $\mathcal{L}_{\text{12}^\circ,e^+,\text{L}}/\mathcal{L}_{\text{SC},e^+}$ ($e^+$ Left + $p$ Right)                         & $1.0538 \pm 0.0003\:(\text{stat.}) \pm 0.0244\:(\text{syst.})$ \\
  \hline
  $\mathcal{L}_{\text{12}^\circ,e^-,\text{L}}/\mathcal{L}_{\text{SC},e^-}$ ($e^-$ Left + $p$ Right )                        & $1.0525 \pm 0.0003\:(\text{stat.}) \pm 0.0244\:(\text{syst.})$ \\
  \hline
  $(\mathcal{L}_{\text{12}^\circ,e^+,\text{L}}/\mathcal{L}_{\text{12}^\circ,e^-,\text{L}})/(\mathcal{L}_{\text{SC},e^+}/\mathcal{L}_{\text{SC},e^-})$ & $1.0012 \pm 0.0004\:(\text{stat.}) \pm 0.0046\:(\text{syst.})$ \\
  \hline\hline
  $\mathcal{L}_{\text{12}^\circ,e^+,\text{R}}/\mathcal{L}_{\text{SC},e^+}$ ($e^+$ Right + $p$ Left)                         & $1.0418 \pm 0.0003\:(\text{stat.}) \pm 0.0244\:(\text{syst.})$ \\
 \hline 
  $\mathcal{L}_{\text{12}^\circ,e^-,\text{R}}/\mathcal{L}_{\text{SC},e^-}$ ($e^-$ Right + $p$ Left)                         & $1.0374 \pm 0.0003\:(\text{stat.}) \pm 0.0244\:(\text{syst.})$ \\
  \hline
  $(\mathcal{L}_{\text{12}^\circ,e^+,\text{R}}/\mathcal{L}_{\text{12}^\circ,e^-,\text{R}})/(\mathcal{L}_{\text{SC},e^+}/\mathcal{L}_{\text{SC},e^-})$ & $1.0042 \pm 0.0004\:(\text{stat.}) \pm 0.0046\:(\text{syst.})$ \\
  \hline\hline
  $\mathcal{L}_{\text{12}^\circ,e^+}/\mathcal{L}_{\text{SC},e^+}$                                                  & $1.0478 \pm 0.0002\:(\text{stat.}) \pm 0.0244\:(\text{syst.})$ \\
  \hline
  $\mathcal{L}_{\text{12}^\circ,e^-}/\mathcal{L}_{\text{SC},e^-}$                                                  & $1.0447 \pm 0.0002\:(\text{stat.}) \pm 0.0244\:(\text{syst.})$ \\
  \hline
  $(\mathcal{L}_{\text{12}^\circ,e^+}/\mathcal{L}_{\text{12}^\circ,e^-})(\mathcal{L}_{\text{SC},e^+}/\mathcal{L}_{\text{SC},e^-})$         & $1.0030 \pm 0.0003\:(\text{stat.}) \pm 0.0046\:(\text{syst.})$ \\
  \hline
  \end{tabular}
  
  \end{center}
  \caption[Summary of the 12\dg luminosity results]{Summary of the results of the measurement of the luminosity in the 12\dg system for the measurements using the left and right telescopes for the detection
  of the scattered lepton separately, as well the combined result.  Determination of the quoted systematic uncertainties is detailed in Section \ref{ss:12sys}.  The statistical errors represent the 95\%
  confidence interval for the average value of the ratio computed over the entire data set.  The measurements are presented as the average adjustment to apply to the integrated slow control luminosity (SCL)
  for the entire data set so as to compute the estimate of the integrated luminosity from the 12\dg measurements for each lepton species.  The $e^+/e^-$ ratios presented here would be measurements
  of \ratio assuming perfect relative  slow control luminosity ($\mathcal{L}_{\text{SC},e^+}/\mathcal{L}_{\text{SC},e^-}=1$), but are better interpreted as normalization points assuming no TPE at 12\dg
  for the relative luminosity or as measurements of
  ${N_{e^+p,\text{data}  }\left(\epsilon,Q^2\right)}/{N_{e^- p,\text{data}}\left(\epsilon,Q^2\right)}$ at $\epsilon \approx 0.98$, $Q^2 \approx 0.165$ GeV$^2$ that may be normalized by
  an independent luminosity measurement (e.g., that of the SYMB system (Section \ref{sec:symblumi})) to compute $R_{2\gamma}$ as in Equation \ref{eq:rat}.  See Section \ref{sec:12TPE} for latter analysis.}
  \label{tab:12results}
  \end{table}
  
  In general, the absolute luminosities determined by the 12\dg systems are several percent above the slow control estimate, while the species-relative luminosity determination is within a few tenths
  of a percent of unity.  The left and right estimates agree well to within uncertainties, providing further evidence for the validity of the measurements.
  These estimates provide an effective normalization for the main OLYMPUS results and, with the addition of the SYMB luminosity estimate, an additional measurement
  of \ratio in the vicinity of $\theta\approx 12^\circ$, as will be discussed in Chapters \ref{Chap6} and \ref{Chap7}.
  
\section{Luminosity Determined Using the SYMB System}
\label{sec:symblumi}

  As originally designed, the symmetric M{\o}ller/Bhabha calorimeter system was to provide a luminosity measurement completely independent from \pmp scattering, examining only
  the processes $e^\pm e^-\rightarrow e^\pm e^-$ and $e^+ e^-\rightarrow \gamma\gamma$ involving beam leptons and atomic electrons from the hydrogen gas within the target.  This method
  of measuring the luminosity would avoid any assumptions regarding the magnitude of TPE effects in \pmp scattering at forward angles that are necessary when using the 12\dg measurement
  as a luminosity normalization.  The original measurement principle was to count such lepton-lepton scattering events in which the outgoing particles scatter symmetrically about the
  beam axis ($\theta\approx 1.29^\circ$ at $E_\text{beam}=2.01$ GeV) and then detected in coincidence in the left and right calorimeters via the deposit of approximately 1 GeV of energy
  in each and, as in the other OLYMPUS analyses, compare the measured rates to those expected from Monte Carlo simulation.  The collimators in front of the calorimeters limited the acceptance
  of the detector to such events near the symmetric scattering angle.  This method was to provide a very high
  statistics measurement of the integrated luminosity using the very high rate of forward lepton-lepton scattering, with statistical uncertainties far smaller than any other element of the analysis.
  Unfortunately, for a number of reasons briefly described in the next section and detailed in Reference \cite{oconnor}, this method of luminosity determination was untenable for the OLYMPUS analysis
  in that it would have introduced an unacceptably large systematic uncertainty to the final \ratio analysis.
  
  To provide a luminosity measurement from the SYMB system that avoided the problems associated with the original method, a new analysis was developed in which the ratio of rates of multiple event
  types was detected in the SYMB calorimeters.  In particular, two types of events were considered for the analysis:
  \begin{itemize}
   \item coincidence symmetric lepton-lepton scattering events (as would have been used in the original analysis), and 
   \item the detection of \textit{simultaneous} (i.e., from the same beam bunch) coincidence symmetric leptons and additionally a single lepton with energy $\sim$2 GeV corresponding
         to a very forward elastic \pmp scatter resulting in the deposition of $\sim$1 GeV in one calorimeter and $\sim$3 GeV in the other.
  \end{itemize}
  While this method greatly reduces the statistical power of the measurement, taking the ratio of multiple event types reduces systematic uncertainties associated with detector efficiency
  and eliminates the need to simulate the three separate lepton-lepton event types. This analysis, referred to as the \textit{multi-interaction event} (MIE) method,
  and its results are described in Section \ref{sec:mielumi} and Reference \cite{schmidt}
  provides complete detail on the method.
  
  In general, this section provides a brief summary of the analyses and results from the SYMB system for the purpose of establishing the necessary results for the main analysis
  described in the remaining chapters.   Complete discussions of these analyses may be found in the theses of O'Connor and Schmidt (References \cite{oconnor} and \cite{schmidt}).
  
  \subsection{Discussion of the Untenability of the Original Coincident Event Analysis}
  
\begin{figure}[thb!]
  \centerline{\includegraphics[width=0.9\textwidth]{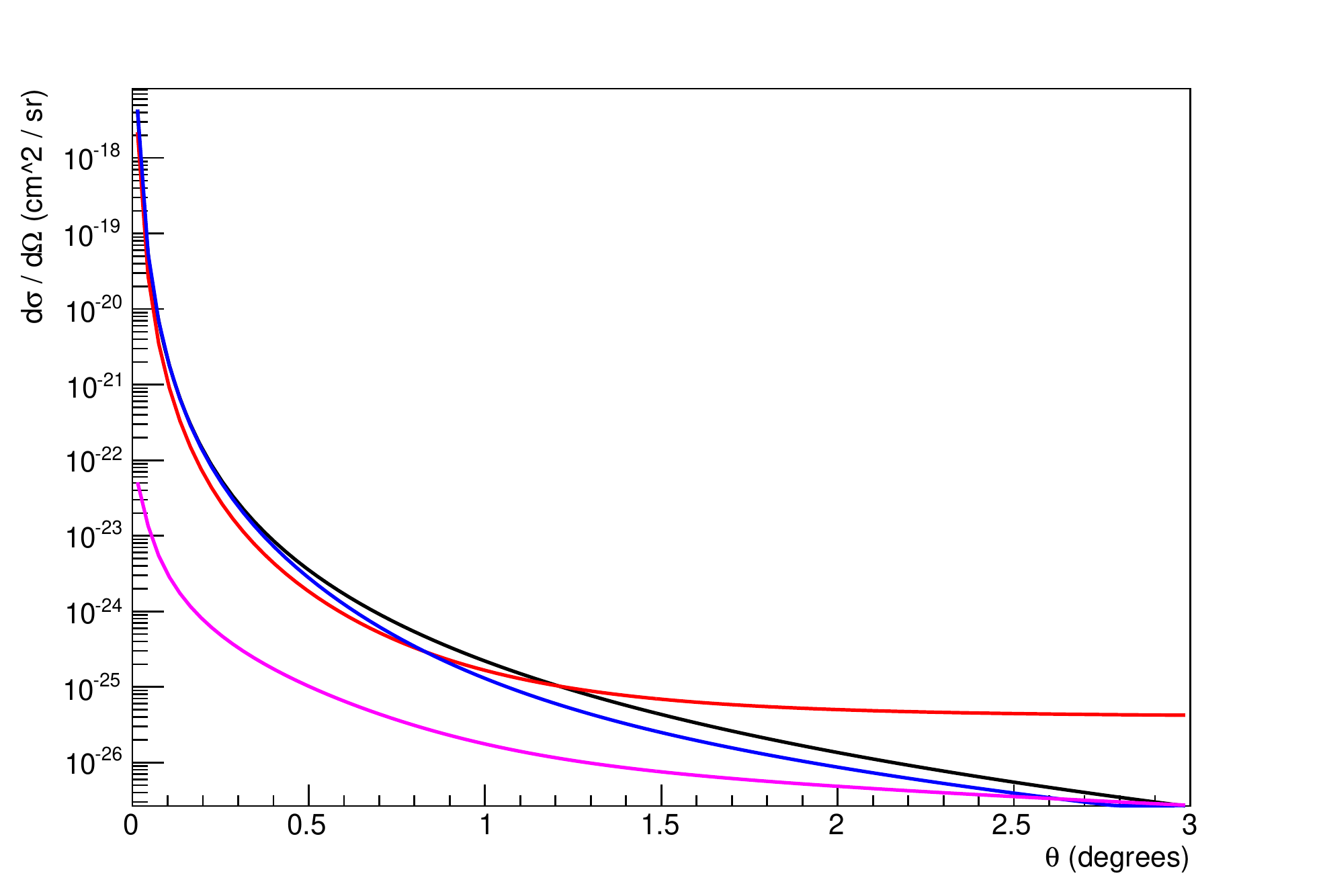}}
  \caption[Tree-level differential cross sections for SYMB processes]{Calculated tree-level differential crosses sections for M{\o}ller scattering (red), Bhabha scattering (blue),
  pair annihilation (magenta), \ep elastic scattering (black) in the vicinity of the SYMB calorimeter acceptance.
  (Figure reproduced from \cite{PerezBenito20166}.)}
  \label{fig:cs3}
  \end{figure}

  As originally conceived, the SYMB analysis would have involved comparing the rates of symmetric $e^-e^-\rightarrow e^-e^-$ scattering for electron beam running
  to the rates of symmetric $e^+e^-\rightarrow e^+e^-$ and $e^+e^-\rightarrow \gamma\gamma$ for positron beam running, normalized to simulation of all three processes
  in a similar fashion as the other OLYMPUS analyses.  This concept was based on a forward calorimeter luminosity monitoring scheme used by the HERMES experiment to 
  measure the relative luminosity of positron beams of different polarizations incident on a hydrogen target via $e^+e^-\rightarrow e^+e^-$ and $e^+e^-\rightarrow \gamma\gamma$
  events \cite{Benisch2001314}.  In converting this concept to a measurement of the relative luminosities of different beam species, however, several important
  differences were not properly considered.  In particular, the cross sections for the aforementioned processes change rapidly as a function of $\theta$ in the region
  of the SYMB detectors, and additionally the cross sections for the different processes vary in $\theta$ with different slopes in the region of interest ($\theta\approx1.3^\circ$),
  as shown in Figure \ref{fig:cs3}.  This difference makes the original method extremely sensitive to small shifts in beam position, beam slope, and the placement of the detectors
  (both in absolute space and relative between the two collimators).  Additionally, the comparison of cross sections for multiple processes involves the simulation of all such processes
  and the application of proper radiative corrections methods for each, introducing an additional large systematic uncertainty to any luminosity result using such a method.  Reference \cite{oconnor}
  provides a full analysis of these effects.  It was determined that given the survey uncertainty, beam position uncertainty, available radiative corrections schemes at the time
  of the analysis, and other systematic effects there was at least a 2.8\% systematic uncertainty in any species-relative luminosity measurement made using this method.  Use of such
  an imprecise luminosity would spoil the resulting uncertainty on the $R_{2\gamma}$ determination well beyond the 1\% uncertainty goals of the experiment.  Analyses that attempted to
  use this method found values of $\mathcal{L}_{e^+}/\mathcal{L}_{e^-}$ that deviated by several percent or more from the measurements of all other systems, and additionally found
  that the measured rate varied inexplicably by several percent over time.
  
  In addition to the inherent sources of uncertainty in this method, it is also believed that the electronics used in the SYMB detector to assess the local-max condition (requiring
  that the central crystal in the calorimeter have the highest energy deposited) and the left/right coincidence module may have had failure modes that affected the detector during
  data-taking.  Regarding the latter, the coincidence histogram showed behavior away from the main (1 GeV, 1 GeV) peak that could not be reproduced in simulation, and was possibly
  due to a small timing error in the window of coincidence used by the coincidence module between the two calorimeters.  Regarding the former, it was hypothesized that for events with
  large energy deposition in a single calorimeter, the comparator module which determines if the central crystal has the highest deposition could be saturated at its maximum value leading
  to an erroneous local maximum event rejection \cite{compsat}.  These two possible issues pose additional concerns for the coincidence measurement, and must be avoided or suppressed
  in any alternative methods of extracting a luminosity measurement from the SYMB.
  
  \subsection{The Multi-Interaction Event Luminosity Determination}
  \label{sec:mielumi}
  
\begin{figure}[thb!]
  \centerline{\includegraphics[width=0.9\textwidth]{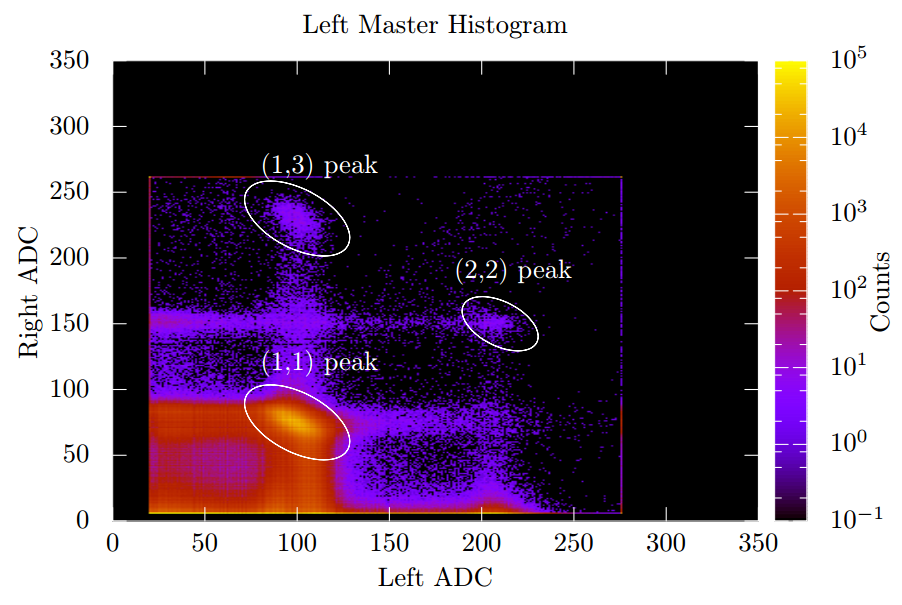}}
  \caption[Example of the SYMB left-master histogram in data]{The left-master histogram from SYMB data, which was filled for any event in which the left SYMB calorimeter
  met its energy deposition (ADC count) threshold with the center crystal having the highest deposition regardless of the ADC count in the right calorimeter.  The visible
  peaks correspond to the various combinations of scattering events that can occur in coincidence (i.e., from the same beam bunch), as described in the text.
  (Figure reproduced from \cite{schmidt}.)}
  \label{fig:lmmie}
  \end{figure}

  With the originally proposed method for producing a luminosity analysis from the SYMB system invalidated, other methods were sought to provide a means
  of estimating the luminosity using the SYMB that was immune to the problems that were believed to be responsible for the failure of the coincidence
  method.  A method was developed that took advantage of the fact that, in addition to the (1 GeV, 1 GeV) peak corresponding to coincident symmetric
  leptons, the SYMB single-side-master histograms showed additional peaks corresponding to other event types (e.g., occurrence of two lepton-lepton
  events in coincidence creating a peak at (2 GeV, 2 GeV), a lepton-lepton event in coincidence with an \pmp event in which the \pmp lepton deposits
  $\sim$2 GeV in one of the calorimeters creating a peak at (1 GeV, 3 GeV), etc.).  An example of the left-master histogram showing these
  peaks is shown in Figure \ref{fig:lmmie}.  The MIE analysis compares the relative rates of symmetric lepton-lepton events and events in which
  a symmetric lepton-lepton event occurs in coincidence with an elastically scattered \pmp lepton in the right side calorimeter as recorded in the left-master
  histogram (the (1,1) and (1,3) peaks in Figure \ref{fig:lmmie}) and normalizing to the simulated rate for the \pmp leptons (noting that the coincidence lepton-lepton
  rate cancels in the ratio, eliminating the need to simulate it).  This offers several important advantages over the original lepton-lepton coincidence method:
  \begin{enumerate}
   \item By taking a ratio of rates, any species-dependent detector efficiency variations are canceled to first order.
   \item By using one of the side-master histograms, any unknown problems with the electronics for the coincidence histogram are irrelevant.
   \item Since the high energy deposition (3 GeV) occurs in the right calorimeter while the left side receives only 1 GeV for the events of interest,
         the left side is not near the comparator saturation region in which its local-max condition may erroneously fail.  Since the method examines
         the left-master plot, no requirements are placed on the right-side deposition making it irrelevant if the right side local max comparison fails
         due to the high energy deposition.
   \item The need to simulate the three processes of M{\o}ller scattering, Bhabha scattering, and pair annihilation is eliminated, leaving only simulation
         of the \pmp lepton rate.
   \item The MIE method is considerably less sensitive to effects that introduce very large systematics to the original coincidence method such as beam position and detector position
         since the rate of coincident leptons cancels in the ratio.
  \end{enumerate}
  The disadvantages of the MIE method include a significant reduction in statistical precision relative to the coincidence method (since it depends on two events being recorded
  from a single beam bunch) and the fact that only the left-master histogram provides the necessary (1 GeV, 3 GeV) peak needed for the analysis since the right-master histogram
  ADC range was not set to include 3 GeV events from the left side.  These drawbacks, however, are greatly outweighed by the robustness of the method.
  
  This analysis is detailed in full in Reference \cite{schmidt}, but the essential details are provided here for the purpose of establishing the method
  for use as one of the luminosity estimates for the final $R_{2\gamma}$ result.
  
  \subsubsection{Principle of the Measurement}
  
  As noted, the MIE method compares the number $N_{(1,1)}$ of symmetric lepton-lepton events (i.e., M{\o}ller scattering, Bhabha scattering, and pair annihilation) to
  the number $N_{(1,3)}$ of events where a symmetric lepton-lepton event occurs in the same beam bunch in which an elastically \pmp lepton reaches the right side calorimeter 
  recorded by the left master histogram, normalized to the expected cross section of \pmp leptons in the right calorimeter from simulation $\sigma^\text{MC}_{e^\pm p\rightarrow \text{R}}$
  and the number of beam bunches $N_b$ in the data sample.  As derived in Reference \cite{schmidt}, this estimate must also account for the variance in integrated
  luminosity delivered by single beam bunches $v_b$, and the average cubed integrated luminosity of single beam bunches $\left< \mathcal{L}_b^3 \right>$ that enter the formula in
  higher order terms due to effects caused by the fluctuation of the likelihood of multiple simultaneous events occurring with fluctuations in the beam bunch charge and the possibility of more
  than two events occurring simultaneously (which ``moves'' events out of the peaks of interest).  In principle, these higher order terms also depend on the total luminosity (the quantity
  being measured) and the cross-section for all processes that may enter the measurement $\sigma_\text{tot}$.  These may be safely estimated as the slow control luminosity
  and the dominant (1,1) cross section respectively since they enter only in the higher order terms making any error in their determination a small effect. The result of the derivation
  in Reference \cite{schmidt} provides the following formula for the integrated luminosity of a given data sample using the MIE method:
  \begin{equation}
   \mathcal{L}_\text{MIE} = \frac{N_{(1,3)}N_b}{N_{(1,1)}\sigma^\text{MC}_{e^\pm p\rightarrow \text{R}}} - \frac{v_bN_b^2}{\mathcal{L}_\text{SC}} -
      N_b\sigma_\text{tot}\left[\left(\frac{v_bN_b}{\mathcal{L}_\text{SC}} + \frac{\mathcal{L}_\text{SC}}{N_b}   \right)^2 - \frac{N_b\left< \mathcal{L}_b^3 \right>}{\mathcal{L}_\text{SC}} \right].
  \end{equation}
  For reference, the second term is approximately 1\% of the leading term, while the third term is approximately 0.1\%.  Since the uncertainty of the measurement is on the order of
  the third term, consideration of higher orders was unnecessary.
  
  The beam parameters $N_b$ and $v_b$ were provided by the DORIS beam parameter archives associated with the slow control system.
  The procedure for determining the count rates in data ($N_{(1,3)}$ and $N_{(1,3)}$) is detailed in Reference \cite{schmidt},
  but essentially amounted fitting the centroids of the relevant peaks in the left master histogram, placing rectangular box cuts around those centroids, and then
  integrating the boxed regions.  It was determined that the choice of box size significantly affect the absolute value of $\mathcal{L}_\text{MIE}$, but did not as significantly
  affected the species-relative measurement $\mathcal{L}_{\text{MIE},e^+}/\mathcal{L}_{\text{MIE},e^-}$.  Given that no clear method existed to determine the exact proper size
  of the box cut, no absolute luminosity determination is quoted for the MIE method and the size of the box cut was tested as a systematic uncertainty of the relative measurement.

  \subsubsection{Systematic Uncertainties}
  \label{sec:miesys}
  
  As previously noted, the uncertainty on the absolute measurement of the luminosity from the MIE method is quite large (at least several percent)
  due to the large variation in absolute count rates that occurs when the sizes of the cut boxes are varied. Additionally, there is no a priori methodology
  for determining the ``correct'' box size, and thus it is extremely difficult to place a specific estimate of the absolute luminosities.
  The analysis is capable, however, of producing an estimate of the species-relative luminosity with
  uncertainty small enough to meet the goals of OLYMPUS.
  
  The main sources of uncertainty for the species-relative MIE luminosity estimate are summarized in Table \ref{tab:miesys}, as analyzed in Reference \cite{schmidt}.
  The total systematic of the MIE analysis for the species-relative luminosity was found to be $\delta_\text{MIE}=\pm 0.27\%$, dominated by the contributions from
  uncertainty in the beam position and slope, uncertainty in the modeled positions and orientations of the calorimeters and collimators, and the choice of the size
  of the cut boxes used in the analysis.  Estimates of the effects were conducted in a manner similar to those described in Section \ref{ss:12sys} for the 12\dg
  system (the exact methods are described in \cite{schmidt}).  Due to the relatively small magnetic field in the region traversed by particles from the target to the SYMB ($\lesssim50$ G),
  magnetic field plays a small role in the MIE uncertainty, while beam energy and radiative corrections are comparably-sized effects for the MIE and 12\dg analyses.  Since the \pmp events
  in the SYMB calorimeters are much further forward than those in the 12\dg system ($\epsilon = 0.99975$, $Q^2 = 0.002$ GeV$^2$), any uncertainty due to TPE effects is expected to be
  much smaller than other effects that were tested (since the TPE uncertainty was 0.1\% and the TPE contribution must go to zero at $\epsilon = 1$).
 
  \begin{table}[thb!]
  \begin{center}
  \begin{tabular}{|l|c|c|}
  \hline 
  Uncertainty Source & Relative (\%) \\
  \hline\hline 
  Beam position/slope ($\delta_\text{BPM}$) & $\pm0.21$ \\
  \hline
  Detector/collimator position  ($\delta_\text{geo}$)& $\pm0.13$ \\
  \hline
  Cut box sizes ($\delta_\text{cuts}$) & $\pm0.10$ \\
  \hline
  Magnetic field ($\delta_{B})$ & $\pm0.05$ \\
  \hline
  Radiative corrections ($\delta_\text{rad}$) & $\pm0.03$ \\
  \hline
  Beam energy ($\delta_{E_\text{beam}}$) & $\pm0.01$  \\
    \hline\hline
  Total ($\delta_\text{MIE}$) & $\pm0.27\%$ \\
  \hline
  \end{tabular}
  
  \end{center}
  \caption[Systematic uncertainties of the SYMB MIE luminosity determination]{A summary of the contributions to the systematic uncertainty
  in the determination of \ratio from the SYMB MIE luminosity estimate. These uncertainties may be
  considered to be independent, in general, and thus are added in quadrature to produce the total
  uncertainty estimate.}
  \label{tab:miesys}
  \end{table}
  
  Due to the relative simplicity of the data output of the SYMB system, there are relatively few identifiable possible causes of systematic uncertainties for the
  MIE analysis.  The system, however, is very sensitive to effects such as beam position and detector geometry (effects to which the 12\dg is quite insensitive).
  In general, the sensitivities to systematic effects are very complimentary between the 12\dg and MIE analyses, with large effects such as tracking efficiency and magnetic
  field in the 12\dg system being either irrelevant or much smaller effects in the MIE analysis.  The statistical precision of the MIE analysis is comparable to
  the 12\dg analysis, and thus effectively negligible.  Due to this, the 12\dg and MIE luminosity estimates provide an important cross check for the luminosity used in the
  final \ratio analysis and present the opportunity to present a measurement of \ratio in vicinity of 12\dg as well (Section \ref{sec:12TPE}).
  
  \subsubsection{Results}
  \label{sec:mieres}
  
\begin{sidewaysfigure}
  \centerline{\includegraphics[width=1.05\textwidth]{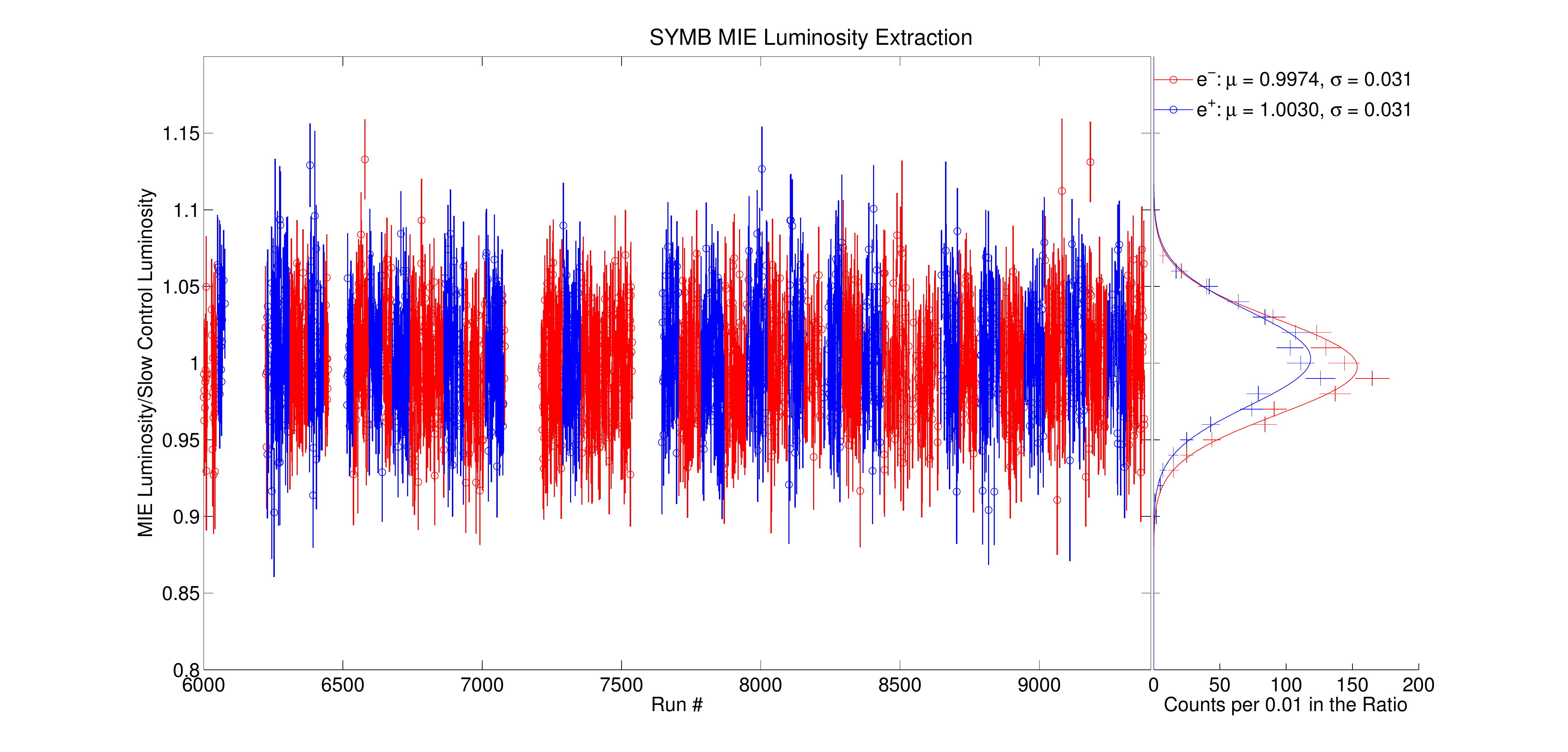}}
  \caption[Luminosity determined by the multi-interaction event SYMB analysis by run]{Run-by-run estimate of the integrated luminosity collected by the OLYMPUS experiment relative to the slow control estimate,
  as measured by the SYMB multi-interaction event analysis.  The data and analysis methods are from Reference \cite{schmidt}. The run-by-run values are histogrammed in the right-hand plot and fitted to Gaussian
  distributions for each species to produce the estimates of the luminosities for each species over the full data set.}
  \label{fig:mie}
  \end{sidewaysfigure}

  Since the MIE analysis is only able to use the left-master histogram (since the right-master histogram range was not set so as include the necessary (3 GeV, 1 GeV) peak
  for the MIE calculation), the MIE analysis produces a single estimate for the relative luminosity.  Similarly to the 12\dg results (Section \ref{sec:12res}), this estimate is quoted as a ratio
  relative to the slow control luminosities for each species. As computed in Reference \cite{schmidt}, the estimate of the species relative luminosities relative to slow control
  for the dataset of interest was:
  \begin{equation}
   \frac{\mathcal{L}_{\text{MIE},e^+}}{\mathcal{L}_{\text{MIE},e^-}} \cdot\frac{\mathcal{L}_{\text{SC},e^-}}{\mathcal{L}_{\text{SC},e^+}} = 1.0055 \pm 0.0010\:(\text{stat.}) \pm 0.0027\:(\text{syst.}).
   \label{eq:mie}
  \end{equation}
  This result is very consistent with the 12\dg estimate and the expected uncertainties associated with the slow control determination.  This result may be used either as
  a luminosity normalization point at very forward angles, or combined with the 12\dg estimate to provide a high-confidence estimate for the main result.  The run-by-run luminosity
  estimate from the MIE method and the projected fit distributions are shown in Figure \ref{fig:mie}.
  
\section{Discussion of the Luminosity Analyses}
\label{sec:alllumi}

\begin{sidewaysfigure}[htb!]
  \centerline{\includegraphics[width=1.05\textwidth]{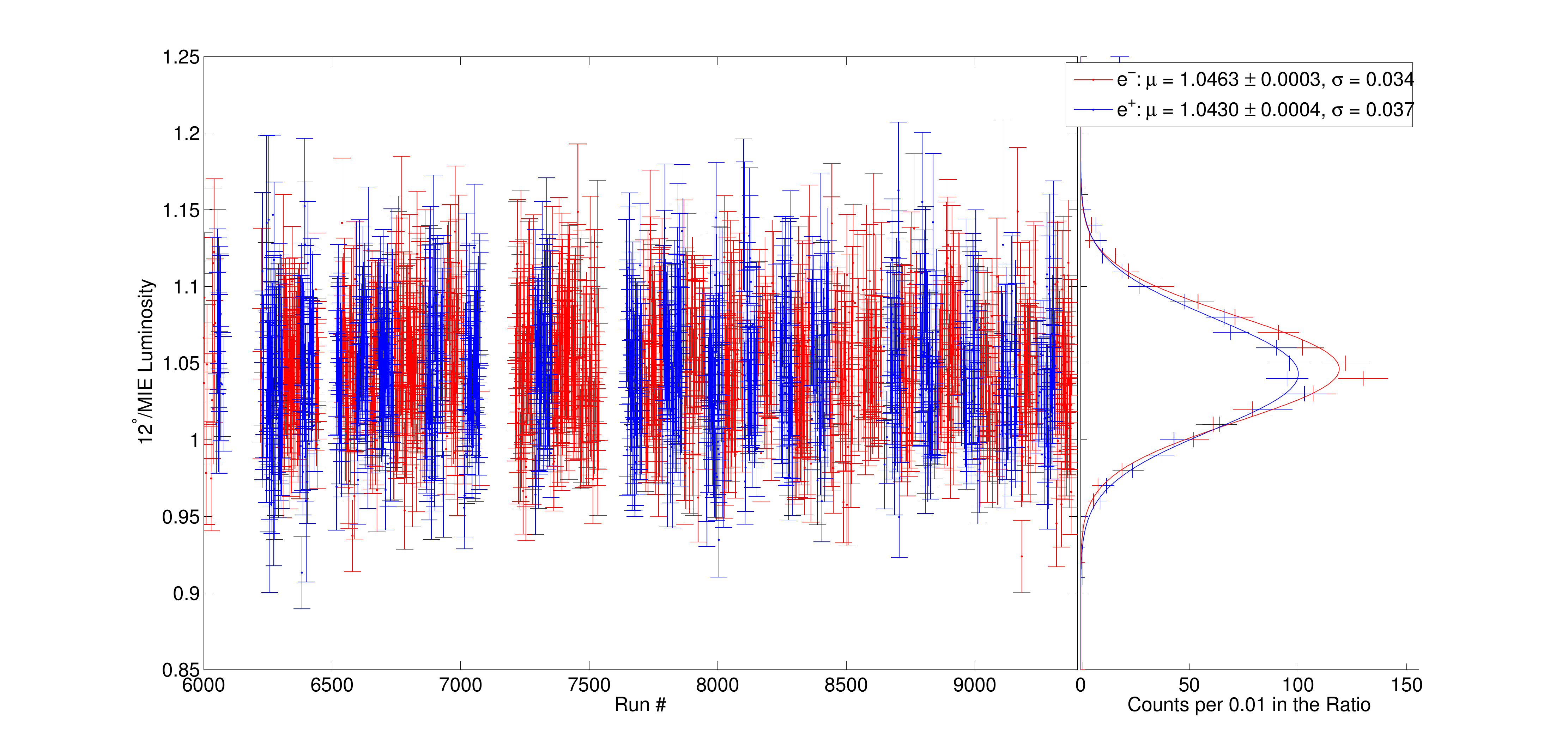}}
  \caption[Ratio of the 12\dg and MIE luminosities run]{Run-by-run ratio of the combined left+right 12\dg luminosity (Figure \ref{fig:12comb}) and the MIE luminosity (Figure \ref{fig:mie}).
  The run-by-run values are histogrammed in the right-hand plot and fitted to Gaussian
  distributions for each species to produce the estimates of the luminosities for each species over the full data set.}
  \label{fig:12mie}
  \end{sidewaysfigure}

  In general, despite the issues with the performance of the GEMs in the 12\dg telescopes and the SYMB coincidence event analysis, the species-relative luminosity measurements for OLYMPUS
  achieved the necessary level of uncertainty to permit an overall uncertainty on the measurement of $R_{2\gamma}$ of less than 1\%.  The excellent agreement of the MWPC-only 12\dg
  \pmp analysis and the MIE analysis in the SYMB system provides a high degree of confidence in the luminosity measurements, since each system is subject to very different
  systematic uncertainties.  Figure \ref{fig:12mie} shows the run-by-run ratio of the combined left/right 12\dg and MIE estimates.  Notably, this run-by-run ratio shows less time variance
  in the individual luminosity estimates relative to the slow control luminosity, indicating that the two methods captured systematic effects that were not accounted for in the slow control
  estimate.  While the MIE method sacrifices the complete exclusion of \pmp scattering that would have been part of the coincidence symmetric lepton-lepton scattering
  method, the additional reduction in uncertainty provided by the MIE analysis via the cancellation of efficiency uncertainties in the ratio of counts and the reduction of the number
  of required simulated physics processes and the low uncertainty on the estimate of possible TPE for such forward scattering ($\epsilon = 0.99975$) make the estimate very robust.  Additionally,
  the careful analysis performed for the 12\dg system provided not only a high precision relative luminosity measurement but also an absolute luminosity estimate for each species with uncertainty
  of only a few percent, which may be useful for future physics analyses with the OLYMPUS data.
  
  For the \ratio analysis that follows in the remaining chapters, the luminosity measurements may be combined into a single average normalization point or taken individually to provide
  either an estimate of the systematic effects of the variation of the relative luminosity scale or to provide an measurement of $R_{2\gamma}$ in the vicinity of $\theta=12^\circ$.  These different
  cases are discussed in Section \ref{sec:thegoddamnresults}.  For the case of the averaged single normalization point, the two analyses (weighted by their uncertainties) provide an estimate of
  the species-relative integrated luminosity relative to the slow control luminosity over the course of the full dataset under consideration of:
  \begin{equation} 
   \frac{\mathcal{L}_{e^+}}{\mathcal{L}_{e^-}} \cdot\frac{\mathcal{L}_{\text{SC},e^-}}{\mathcal{L}_{\text{SC},e^+}} = 1.0048 \pm 0.0024\:(\text{combined stat. + syst.}).
   \label{eq:avrellumi}
  \end{equation}

%% file: chap6.tex
% Chapter 6
%
% Main TPE Analysis
%

\chapter{The $e^+p/e^-p$ Cross Section Ratio Analysis}
\label{Chap6}

The final piece of the OLYMPUS result for $R_{2\gamma}$ is the analysis of the elastic \ratio ratio in the main tracking volumes
of the detector.  As has been discussed previously, this involves not only a robust method of selection elastic \pp and \ep events
from the data sample, but also proper representation of the detector system in the simulation so that the elastic selection method
may be applied equally to the data and the simulation.  This chapter discusses the performance of the main tracking detectors (the
drift chambers and time-of-flight (ToF) scintillators), the implementation of the tracking detectors in the simulation, and a method
of selecting elastic events and conducting background subtraction to produce a result for \ratio over the full acceptance of the detector.
Additionally, a preliminary estimate of the systematic uncertainty in the \ratio analysis is discussed in Section \ref{sec:mainsys}.
The results of this analysis are presented in Chapter \ref{Chap7}.

Note that the analysis described here is only one of several conducted using the OLYMPUS data to produce a \ratio result.  Multiple
analyses using unique methods were conducted by different members of the OLYMPUS collaboration so as to provide an estimate of the 
systematic uncertainty in the result due to choices made in the elastic event selection and background selection.  These analyses utilized
different methods for particle-type identification, combinations of kinematic cuts to produce the elastic event sample, models for the remaining
background after cuts, and orderings of the various steps in the analysis to provide providing a robust examination of the effects of analysis
decisions on the final result.  Information on several of these analyses may be found in References \cite{schmidt}, \cite{russell}, and \cite{oconnor},
and Section \ref{sec:indana} discusses the comparison of the analyses.

\section{Spectrometer Performance and Modeling in Simulation}
\label{sec:specperf}

  Characterization of the detectors involved in the reconstruction of elastic \pmp events was critical to the analysis, as it allowed
  a detailed implementation of the detector response in simulation so as to ensure that the simulated detector accurately represented
  the acceptance of the detector during the experiment.  In particular, it was critical to model the efficiency and resolutions of the
  drift chambers and ToFs in detail so that reconstruction of simulated tracks could occur on equal footing as tracks in experimental data.
  This section describes the measurements and implementation of these parameters in simulation, particularly focusing on the drift chamber
  efficiencies.

  \subsection{Drift Chambers}
  \label{sec:wcperf}
  
    Since the drift chambers were the main reconstruction detectors for OLYMPUS, it was critical to properly model them in the Monte Carlo,
    especially with regard to efficiencies (which affected the effective acceptance of the detector) and resolutions (which affected the validity
    of applying the same elastic event selections to data and simulation).  Each of these quantities was measured throughout the chamber using the 
    experimental data and globally-fit track information.
  
    \subsubsection{Efficiency}

\begin{figure}[thb!]
    \centerline{\includegraphics[width=1.15\textwidth]{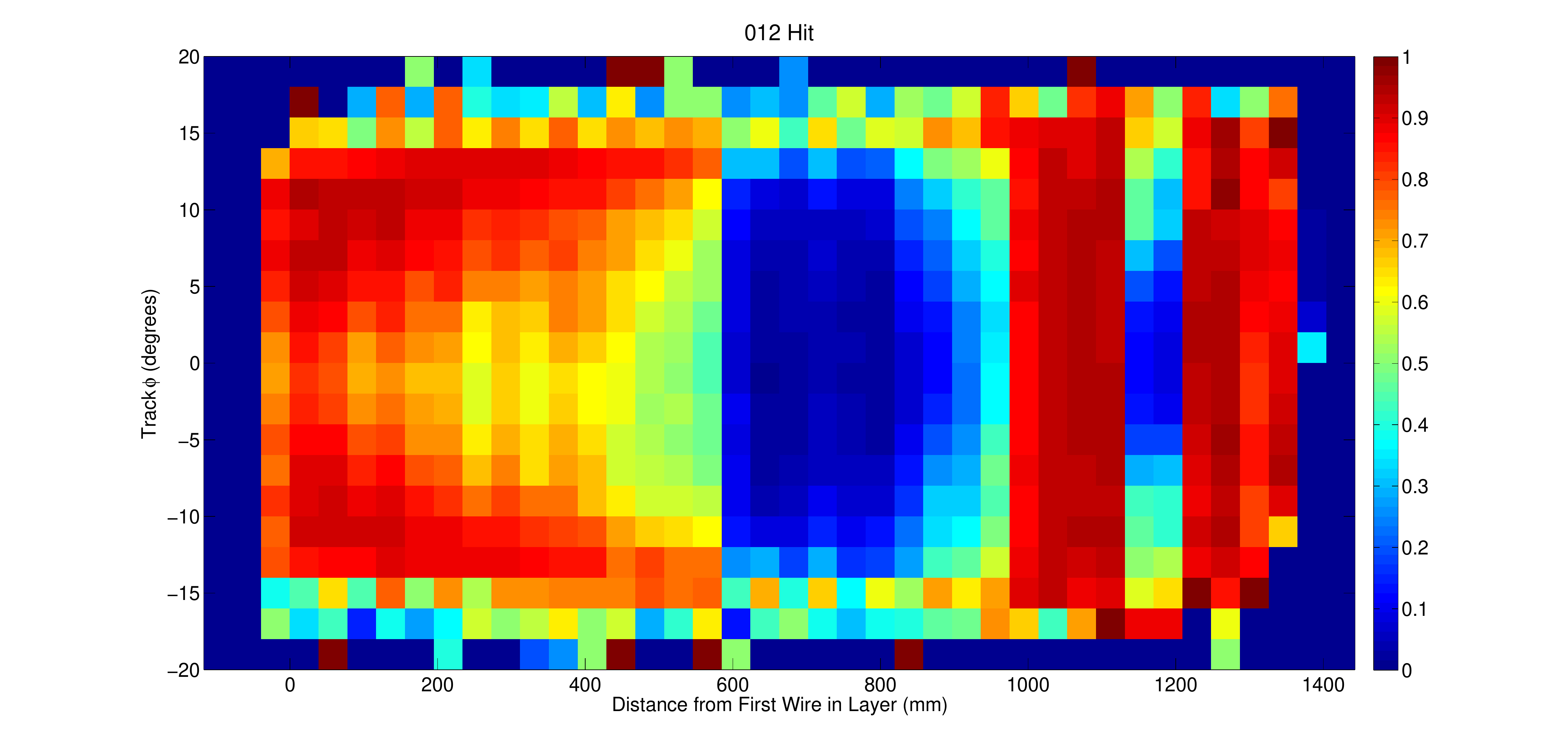}}
    \caption[Efficiency for all three wires to fire in the left innermost drift chamber cell layer]{Probability for all three wires in the innermost layer of left drift chambers cells
    to fire for a track in \ep running as a function of position in the drift chamber plane.  In addition to a reduction in efficiency due to high rates (visible at distance of $<600$ mm from the
    first wire), this layer contained a defective high voltage supply card that significantly reduced the efficiency of a 390 mm block in a correlated way.  Additionally a single
    defective wire channel at distance 1170 mm reduced the all-three hit probability in that region. }
    \label{fig:badwc}
    \end{figure}

    If the drift chambers had been highly efficient for track detection throughout their volumes, they would have affected the acceptance of the detector
    very minimally.  This, however, was not the case for the conditions of the OLYMPUS experiment.  Furthermore, because the beam environment differed
    between \ep and \pp experiment modes (predominantly through the backgrounds caused by M{\o}ller and Bhabha scattering of beam leptons from the atomic
    electrons in the target gas which caused hits in the innermost drift chamber layers), it was critical to precisely measure the efficiencies in each running
    mode to avoid an artificial shift in \ratio due to an asymmetry in drift chamber efficiencies in the two modes.  In particular, the efficiency of the drift
    chambers was lowest in the innermost layers where the wires were exposed to higher rates from particles in the target which had relatively low energy, but were not
    so low energy as to be completely contained away from the chambers by the magnetic field.  An example of the inner chamber efficiency is show in Figure \ref{fig:badwc}, which was
    additionally reduced by a defective high voltage supply card in the central region.  In the outer layers, the efficiency was typically much higher.
    
    An additional factor of consideration was that blocks of five adjacent wire chamber cells shared a single high voltage distribution card and discriminator card.
    Thus, issues with either the high voltage applied to the wires or the low voltage that powered the a discriminator card could simultaneously affect up to 15 wires,
    introducing the possibility of correlated inefficiencies between wires.   While studies indicated that correlation did not occur between wires that did not share a card,
    significant efficiency correlations were observed for wires connected to the same card.  Since a track in the drift chambers can be constructed without hits in each layer,
    it was critical to model such correlation effects so as to avoid over-predicting the number of reconstructible events in simulation.
    
\begin{figure}[thb!]
    \centerline{\includegraphics[width=1.15\textwidth]{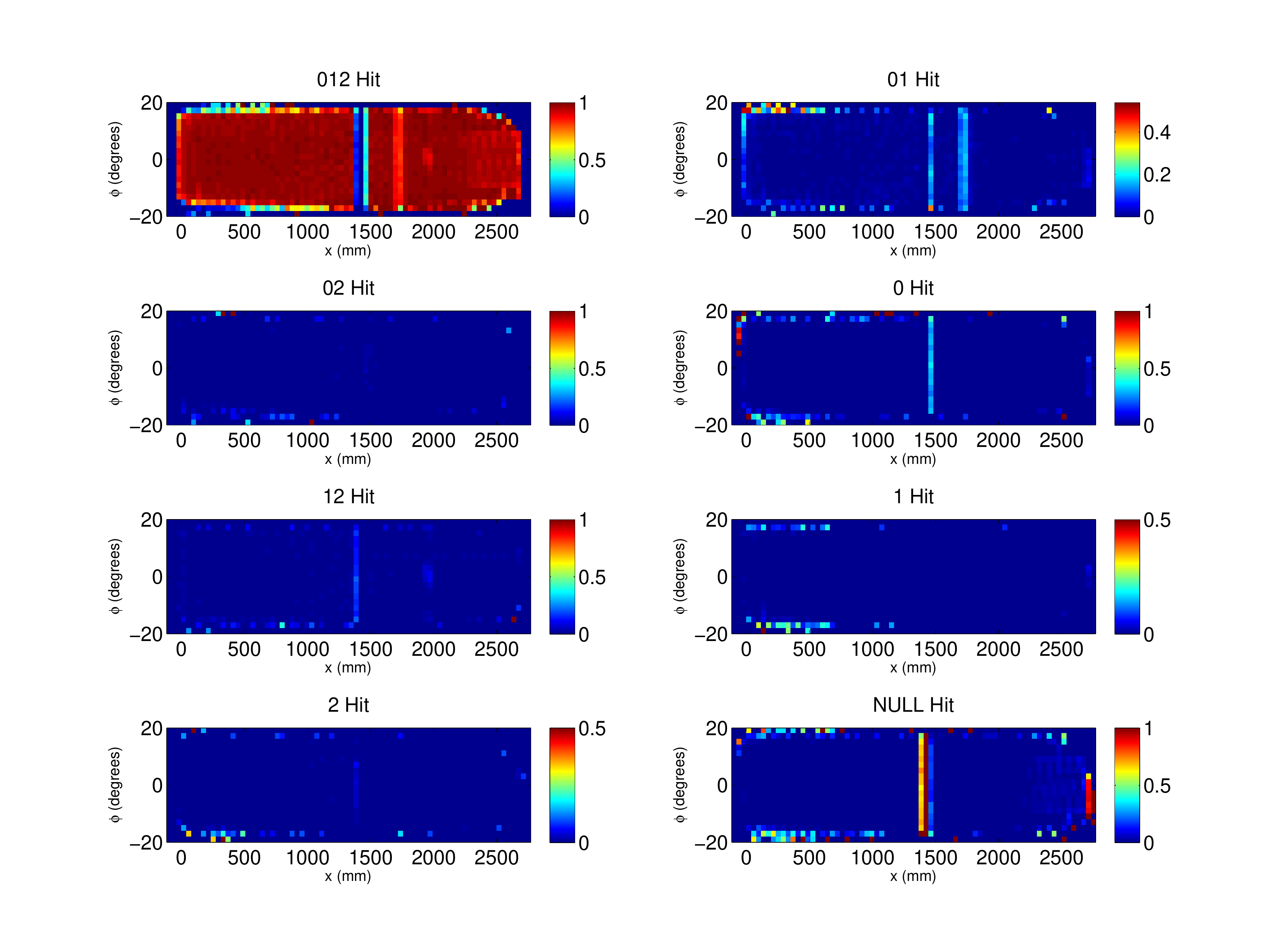}}
    \caption[Set of correlated efficiency maps for the outermost cell layer in the left chamber]{Probability maps for the eight possible combinations  of hits in the outermost left cell layer
    during \pp running, where $x$ is the distance from the first wire and $\phi$ is the track azimuthal angle.  The three layers are labeled 0, 1, and 2, and the plot titles indicate which
    wires are hit in a given map.  Note that the dominant probability is for all three wires to fire, outside of
    the region of a known disconnected cell at $x\approx1400$ mm.  Additionally, lower single wire efficiencies occur at two other locations but do not cause a significant correlated probability of
    missing all three hits in the cell layer.}
    \label{fig:wcsl5}
    \end{figure}

    To account for such effects, the drift chambers were not modeled with simple efficiency maps for each layer in the same fashion as the planes of the 12\dg telescopes
    (Section \ref{sec:12eff}).  For each cell layer in the wire chambers (three single-wire layers), eight maps were calculated corresponding to the $2^3$ possible combinations
    of hit/no hit for the three wire layers.  To determine the efficiency for a given cell layer, the data was tracked with the layer completely excluded from consideration (both
    in terms of contributing hits to possible tracks and in being required to match a track pattern (Section \ref{sec:track})).  The data used for the efficiencies mapping were a sample
    of $\sim$100 data runs of each lepton species beam sampled evening across the entirety of the dataset (approximately 10\% of the full dataset used for the analysis).  Each data run was
    tracked six times, with a different cell layer masked for each reconstruction (removing the corresponding layers on both the left and right sides).  A rough elastic selection
    was applied to the tracks to assure reasonable track quality, and the masked layer was checked for hits corresponding to the selected track.  Hits were required to be within
    several times the resolutions described in the next section to be counted as associated with a track and thus mark a wire efficient.  This cutoff was found to be very stable once outside
    of the resolution peak.  As described, for each layer the probability for each of the eight wire hit combinations was calculated as a function of position in the cell layer plane and maps
    of these probabilities were generated for each layer.  In the simulation, tracks were tested at the cell layer level using a random draw against these maps to determine which hits
    (if any from the track) would be passed from that cell layer to the reconstruction of simulation.  Separate sets of maps were constructed for position and electron beam operation since
    the noise conditions in the innermost layers differed significantly enough to induce an efficiency difference.  An example set of maps for an outer layer, where the efficiency was
    quite uniform and high, is shown in Figure \ref{fig:wcsl5}.
    
    Wires that were known to be disconnected or to have malfunctioning readout equipment were excluded from the analysis of both data and simulation entirely, rather
    than implementing 0\% efficiency in the maps.  This approach is superior due to the fact that while a track may pass predominantly through the region of a deactivated
    wire, it may also pass through the active region of an adjacent wire and produce a hit that reconstructs to the position in the deactivated cell.  This was especially
    common in the forward portions of the drift chambers where tracks passed wires with large angles relative to the normal vector of the wire planes.  Thus, the ``dead'' regions
    due to inactive wires had soft efficiency edges rather than hard cutoffs in acceptance, which was modeled in the maps as described.  This complete model of the drift
    chamber efficiencies permitted OLYMPUS to achieve a high degree of data/simulation agreement in final yields (up to the effect of TPE), providing evidence for the use of
    the method encapsulated in Equation \ref{eq:rat}.
    
    \subsubsection{Time Resolution}
    
    The inherent time resolution of the drift chambers was dominated by the physics of ionization drift rather than any time scales
    inherent to the capabilities of the TDCs used to measure times for each wire.  Dispersion caused by random interactions of the drifting ions
    with the drift gas worsened the resolution roughly linearly as a function of the distance between the point of the initial ionization and
    the wire.  The resolution was found to vary from approximately 20 ns in the vicinity of the wire to slightly more than 30 ns near the edges
    of the cell.  The resolution was measured by examining the width of the distribution of the difference between the distance reconstructed from
    the globally reconstructed track involving the drift chamber time in question and the distance predicted by the time-to-distance (TTD) function for a hit,
    which was then converted to a time width via inversion of the TTD function (see Reference \cite{schmidt}).  In this way, the method captures at least a portion of the uncertainty in
    the TTD function for a given cell in addition to resolution widening caused by the drift gas.
    
    To apply these resolutions to simulation, simulated hits were first produced in the form of distances from drift chamber wires and then converted to times
    using the inverse TTD function.  This time was then smeared according to a Gaussian of the appropriate width according to its time (distance from the wire).
    These ``smeared'' time hits were then saved as the experimental data analogs of the experimentally measured drift times and passed to the track reconstruction
    algorithm in an identical fashion to data.
    
  \subsection{Time-of-Flight System}
    
    Due to the importance of the ToF system in the analysis (as the main component of the trigger), it was critical to the results of the experiment to properly
    calibrate and simulate the system. Timing offsets, ADC calibrations, etc. for each bar were determined using a detailed data-driven approach,
    which is discussed in detail in Reference \cite{russell}.  The efficiency of the ToF bars was not simulated using a mapping as in other systems, but rather was
    modeled by measuring the response of each bar and the attenuation length of scintillation light as a function of position in the bar in data to create a model
    of the scintillator response and efficiency for implementation in the simulation.  Such a model was necessary due to the fact that while special triggers were included
    at a prescaled rate in the dataset to allow for data-driven direct ToF efficiency measurements, it was found that these triggers were generally swamped by forward event
    noise and provided very little useful data for the majority of ToF bars.
    
\begin{figure}[thb!]
    \centerline{\includegraphics[width=1.15\textwidth]{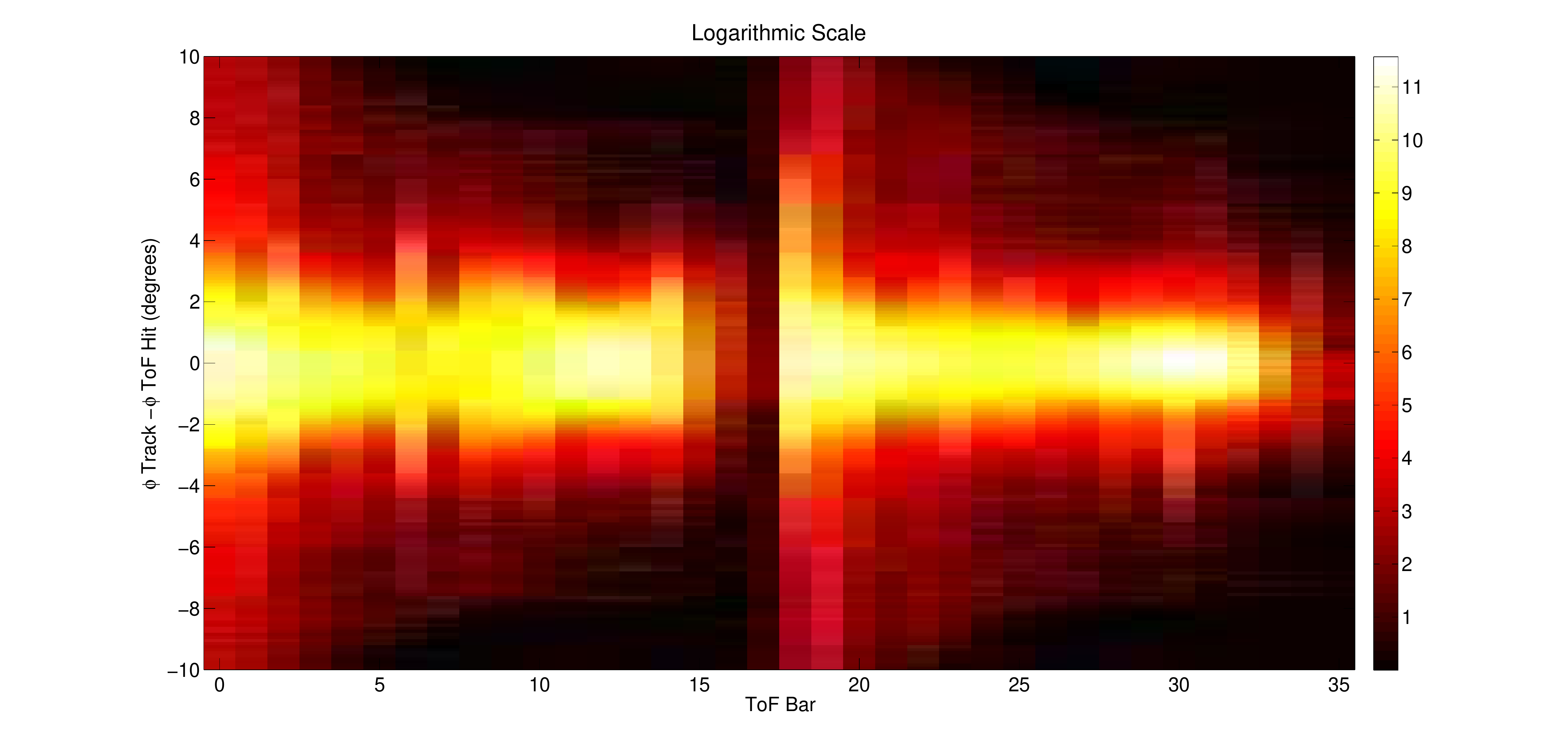}}
    \caption[Difference between ToF hit $\phi$ and reconstructed track $\phi$ by bar]{Difference between ToF hit $\phi$ and reconstructed track $\phi$ by bar for \ep data, after event
    pair selection but prior to background subtraction.  Note that the logarithm of the counts is shown by the color scale of the histogram so as to make the mid-angle bars with fewer
    counts visible in the figure.  In general, the left ToF bars (indices 0-17) had poorer resolutions than the right bars (indices 18-35), but all bars were properly calibrated to match
    tracks on average.  Such resolution differences were implemented in the simulation so as to make simulated events match data events in reconstruction and analysis.}
    \label{fig:tofphi}
    \end{figure}

    Quantities such as timing offsets were determined using an iterative approach
    of matching ToF hit positions reconstructed from the PMT timing difference and positions projected from reconstructed trajectories in the drift chambers (excluding
    the ToF hit information from the trajectory fit).  An example of the success of this methodology is shown in Figure \ref{fig:tofphi}, which shows the excellent agreement
    between the reconstructed track $\phi$ and the associated ToF hit $\phi$ position.
    
  \subsection{Reconstruction}
  \label{sec:reconrev}
  
  In general, the reconstruction algorithm based on the elastic-arms algorithm (EAA) (References \cite{OHLSSON1,OHLSSON2}) used for OLYMPUS (Section \ref{sec:track}), performed extremely well.  This was achieved
  by a careful tuning of the EAA parameters to best model the OLYMPUS tracking environment.  This process is described in detail in Reference \cite{russell}.  The extensions to EAA implemented to 
  improve its performance with regards to correctly resolving the wire side ambiguities of the drift chamber hits is described in Reference \cite{schmidt}. To test the efficiency of the
  tracker, and to provide a basis for the tuning of the EAA parameters, the capability of the tracker was tested using simulation tracks for which the true trajectories are known, but that have been
  converted to the data format with time smearing and the introduction of the wire side ambiguity of the drift chamber hits.  Through the tuning of the parameters, tracking efficiency for such tracks
  (proper identification of the correct wire side for all drift chamber hits and reconstruction parameters within acceptable bounds of the known true values) was in excess of 99\% for the tracking
  of all three relevant particle types throughout the entirety of the acceptance.  The difference in efficiencies between the different particle types was within the statistical uncertainty of
  the tests \cite{russell2}.  While there was no effective method for testing the tracking of data events in a similar fashion, the high efficiency for simulation tracking and the similarity
  of efficiency for all particle types provided high confidence in the robustness of the tracker.  This was further buoyed by the comparison of data to simulation event selection, which indicated
  very good agreement (Section \ref{sec:datasim}).
  
  Resolutions with respect to various reconstruction parameters are discussed in Section \ref{sec:pairsel} in the context of the various kinematic parameters used for the
  elastic event selection, and the figures of Appendix \ref{chap:kincuts} illustrate the resolutions on a number of kinematic parameters over the full detector acceptance
  for both \ep and \pp data.

\section{Method of the Analysis}
\label{sec:mainana}

This section describes the methodology of the elastic \pmp event selection analysis conducted by the author and used
to produce the majority of the results presented in this work.  The analysis presented here is one of several $R_{2\gamma}$
analyses conducted for OLYMPUS, each of which used different choices of particle identification, kinematic cuts, background
subtraction models.  The variation in these analyses provides a useful measure of certain elements of the systematic uncertainty
in the final result and is discussed later in this chapter.  Details on the alternate analyses may be found in any of the other
OLYMPUS PhD theses (References \cite{schmidt,russell,oconnor}), although other analyses were conducted that have not yet been published.

As previously noted, this analysis method was applied to both the experimental data and the events generated from simulation, which were
constructed in such a way so as to precisely mimic the format of the experimental data.  This allows completely equal treatment of data
and simulation throughout the entire analysis (track reconstruction, particle identification, kinematic cuts, etc.).  From this point forward,
any use of the simulated events makes use of the default event weight (see Section \ref{sec:gen}), which was the exponentiated radiative corrections
model based on the prescription of Maximon and Tjon \cite{MaximonPhysRevC.62.054320}, unless explicitly noted otherwise.

\subsection{Particle Identification and ToF Hit Association}
\label{sec:partid}

In this analysis, the first step was conduct particle identification, i.e., the proper association of particle trajectories reconstructed
by the tracking algorithm (Section \ref{sec:track}) with the true particle type associated with the track.  The reconstruction algorithms
used for OLYMPUS attempted to fit a lepton or proton to any matched pattern in the data, and thus the reconstructed data contained tracks
of different particle types associated with the same drift chamber hits.  First, the list of all pairs of leptons corresponding to the beam species
and protons in opposite sectors as identified by the tracking algorithm was constructed.  This process eliminated all leptons of the opposite
charge.

For each such pair, the validity of the tracker's particle type assignment was assessed using information from the ToF scintillators.  For each
track, the trajectory was projected to the ToF panels and the expected bar to be hit in association with the track was calculated.  Then, any actual
registered ToF hits within a tolerance of the two bars surrounding the projected hit bar were considered to be a hit that could be associated with the
track.

\begin{figure}[thb!]
    \centerline{\includegraphics[width=1.15\textwidth]{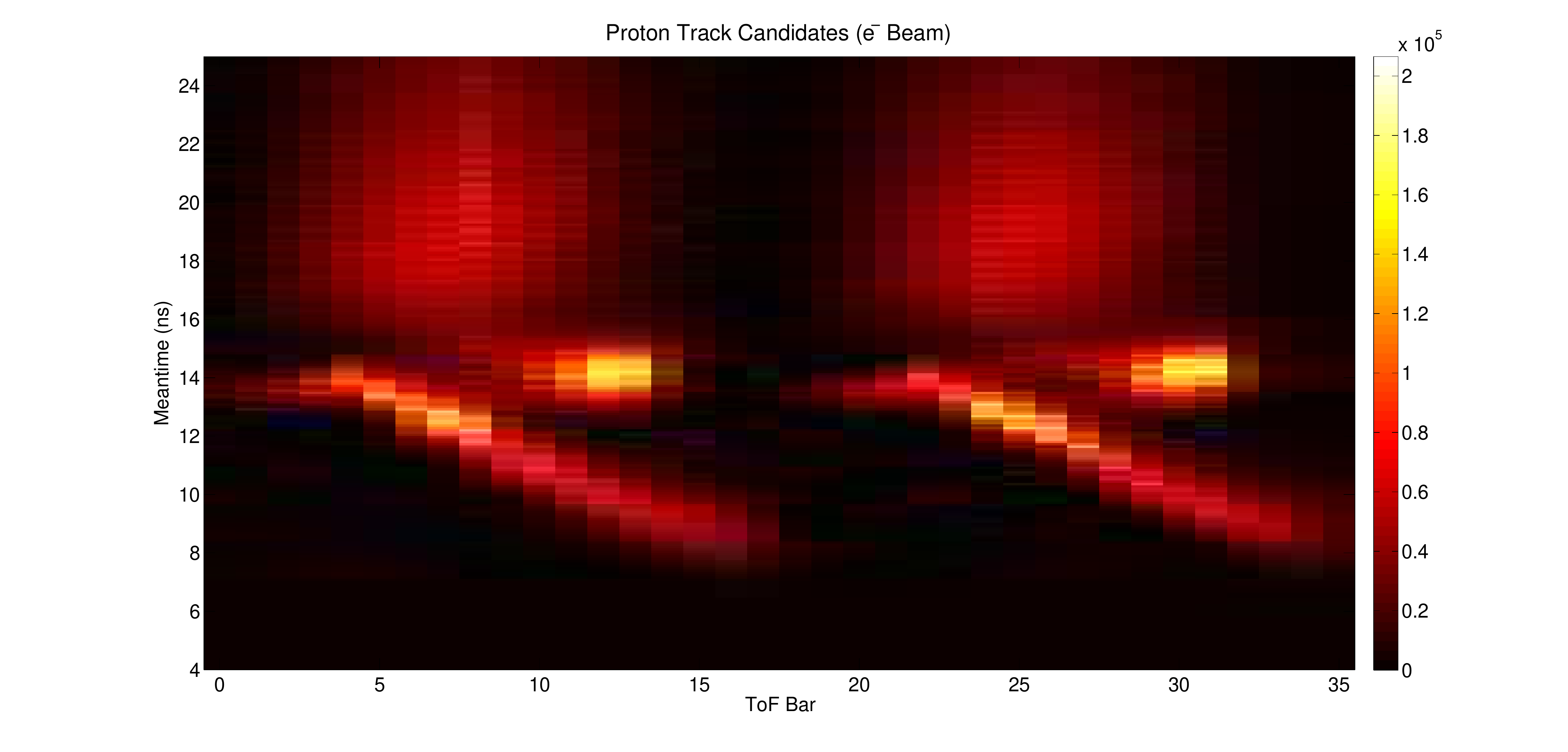}}
    \caption[Meantime distribution by ToF bar for proton candidates in $e^-$ beam data]{Distribution of ToF meantimes by bar for tracks identified
    by the reconstruction algorithm as protons in $e^-$ beam data.}
    \label{fig:pemt}
    \end{figure}

\begin{figure}[thb!]
    \centerline{\includegraphics[width=1.15\textwidth]{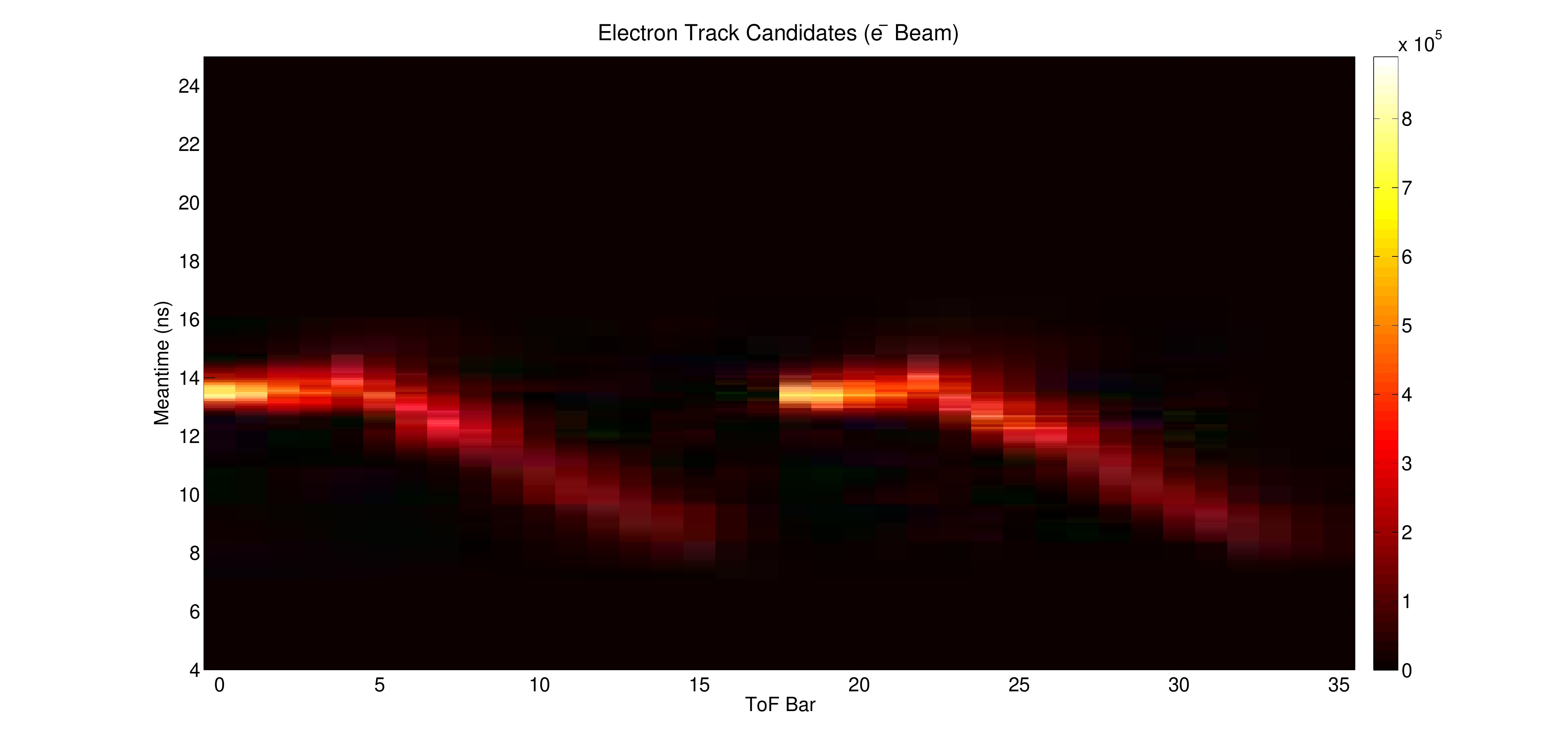}}
    \caption[Meantime distribution by ToF bar for electron candidates in $e^-$ beam data]{Distribution of ToF meantimes by bar for tracks identified
    by the reconstruction algorithm as electrons in $e^-$ beam data.}
    \label{fig:emt}
    \end{figure}

\begin{figure}[thb!]
    \centerline{\includegraphics[width=1.15\textwidth]{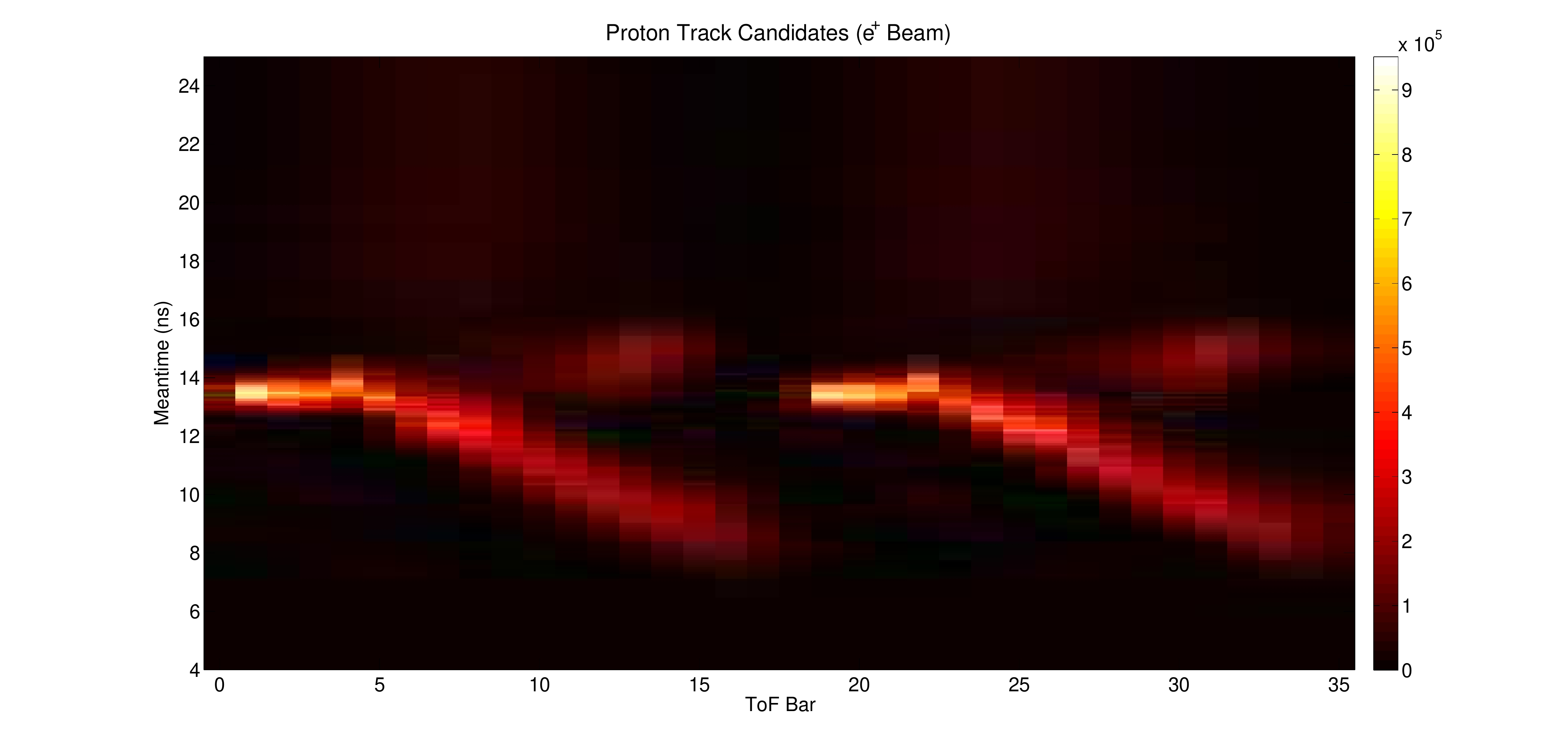}}
    \caption[Meantime distribution by ToF bar for proton candidates in $e^+$ beam data]{Distribution of ToF meantimes by bar for tracks identified
    by the reconstruction algorithm as protons in $e^+$ beam data.}
    \label{fig:ppmt}
    \end{figure}

\begin{figure}[thb!]
    \centerline{\includegraphics[width=1.15\textwidth]{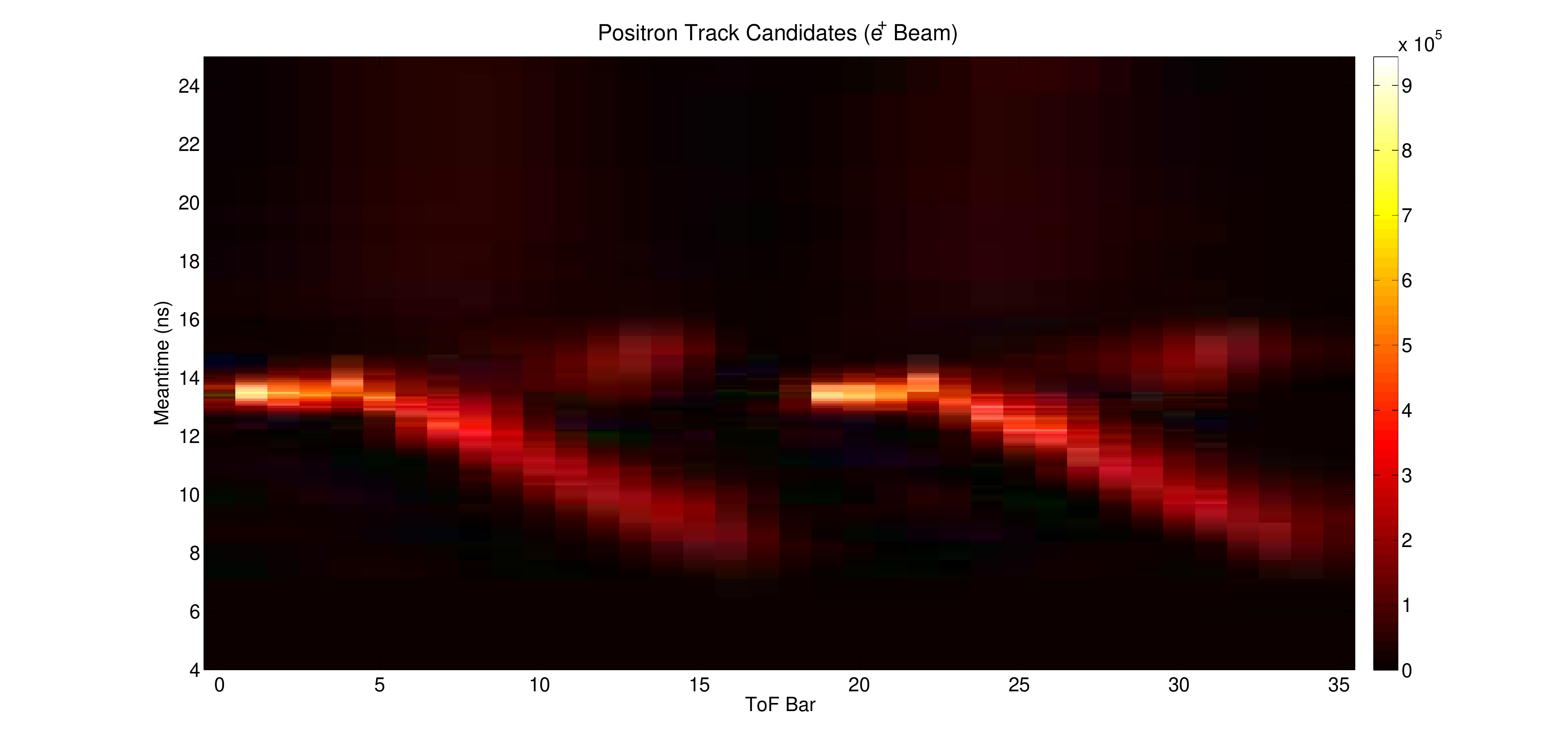}}
    \caption[Meantime distribution by ToF bar for positron candidates in $e^+$ beam data]{Distribution of ToF meantimes by bar for tracks identified
    by the reconstruction algorithm as positrons in $e^+$ beam data.}
    \label{fig:pmt}
    \end{figure}

For each ToF hit, the likelihood that the hit corresponded to the identified particle of the track in question was assessed using the meantime of hit,
i.e., the time recorded for the hit in the ToF bar as the mean of the times recorded for the upper and lower PMT hits corresponding to the time 
elapsed between the beam bunch crossing the target and the scattered particle striking the ToF (up to the correction for travel time of the beam between
the crossing time location and the scattering vertex).  The use of ToF ADC information was considered, but was not ultimately used for several reasons:
\begin{enumerate}
 \item protons could deposit a range of energies in the scintillator bars (including small amounts of energies similar to lepton depositions),
 \item it was difficult to assess the total energy deposition of particles that possibly passed through the edges of two bars, and
 \item generally the calibration of the ToF ADCs was not as well constrained as that of the TDCs for timing information.
\end{enumerate}
The distributions of the meantimes for tracks associated with each particle type for events with each beam species are shown in Figures
\ref{fig:pemt}, \ref{fig:emt}, \ref{fig:ppmt}, and \ref{fig:pmt}.  The electron candidate plot (Figure \ref{fig:emt}) most clearly shows the band
of meantimes corresponding to elastic events (since actual protons tracks tracked as leptons would be tracked with positive curvature resulting
in a positron identification).  Each of the other three plots shows the ambiguity introduced by tracking all events with both lepton and proton
assumptions that must be deconvolved by the particle identification portion of the analysis. As suggested by the clean separation of the elastic
electron meantime band (and the visible separation of bands corresponding to leptons and protons in the other plots),
a simple bar-by-bar cut on meantime would achieve much of this goal.  This approach worked very well in the backward regions of the detector
where the meantime separation is large (as in the 12\dg \pmp analysis (Section \ref{sec:12ana})), but struggled in the intermediate and
forward regions where the meantime bands blend together.

\begin{figure}[thb!]
    \centerline{\includegraphics[width=1.15\textwidth]{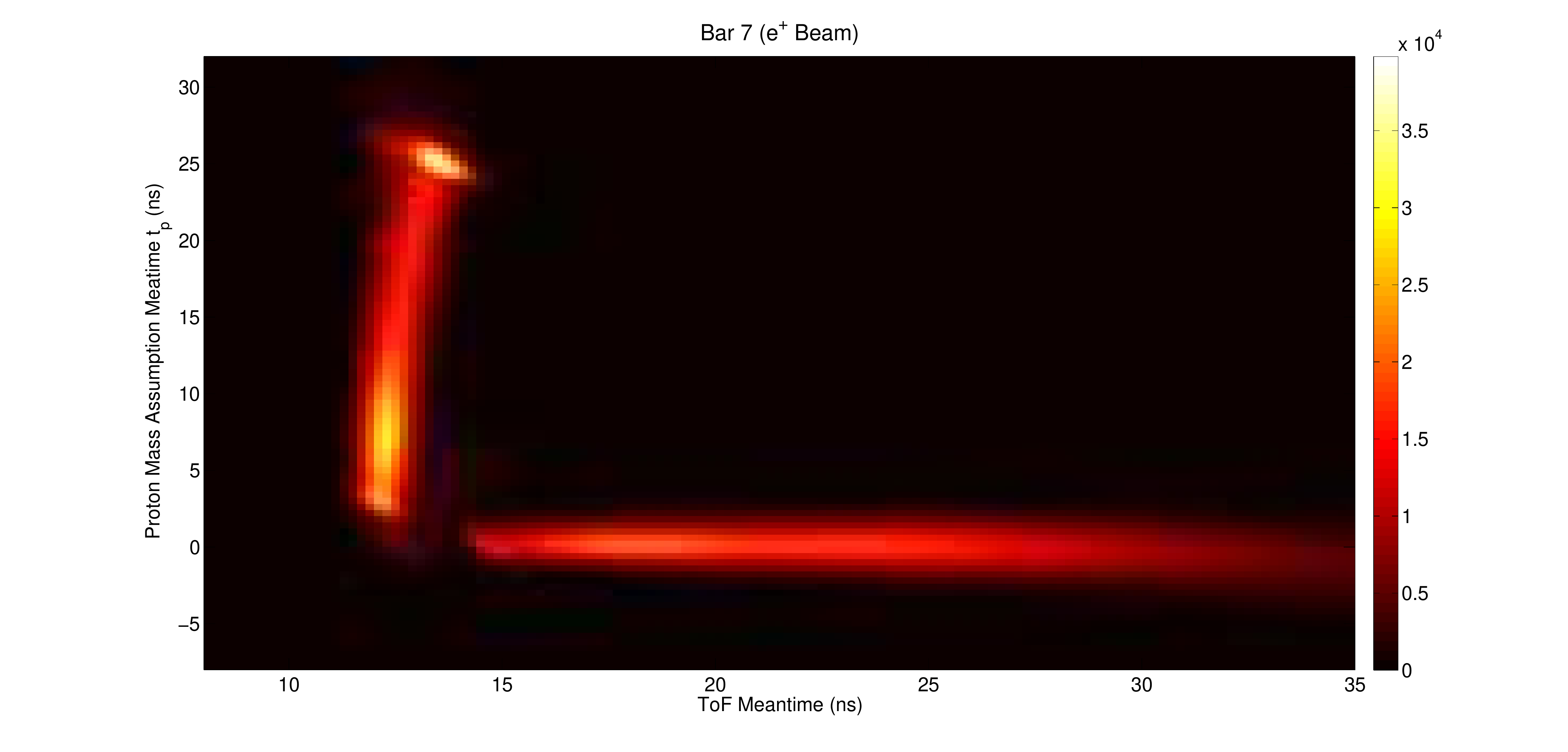}}
    \caption[ToF meantime vs. meantime assuming proton mass for Bar 7 in $e^+$ beam data]{Histogram of track pairs by the associated ToF meantime and the meantime assuming the particle
    to have proton mass $\overline{t}_p$ (Equation \ref{eq:mttp}) for Bar 7 (a mid-to-forward bar on the left) in $e^+$ beam data.
    Note that the band at $\overline{t}_p=0$ corresponding to true protons is well-separated 
    from the wrongly identified particles in the roughly vertical band. The curvature of the vertical band bends towards the ToF meantime gap, illustrating why a simple meantime is difficult since
    small meantimes for real protons were obfuscated by wrongly identified tracks in the upper portion of the vertical band.}
    \label{fig:mt2d}
    \end{figure}

To achieve better particle identification in the forward regions, a two-dimensional cut was developed using the reconstructed momentum of the track
in question to predict the meantime according to the speed of the particle's travel from the vertex to the ToF bar \footnote{The author wishes to acknowledge R. Russell \cite{russell} for the 
essential method of this particle identification cut.}.  Since essentially all leptons in the data sample
had $\beta = \frac{v}{c} \approx 1$ a simple maximum meantime cut, tuned bar-by-bar, was sufficient for lepton identification.
More interesting were the protons, whose large mass causes significant variation of $\beta$ as a function of their momentum.
For each track, the path length of the particle's trajectory from the scattering vertex to the ToF panels $l$ was calculated by the tracking algorithm.  Then, the value of $\beta$ for
each track was calculated using the momentum as reconstructed by the tracking algorithm $\left|\mathbf{p}\right|$, but assuming the particle had proton mass.  That is:
\begin{equation}
 \beta_p = \frac{\left|\mathbf{p}\right|}{\left|\mathbf{p}\right|^2 + m_p^2}.
\end{equation}
This was used to predict a meantime under the assumption of proton mass for the particle:
\begin{equation}
 \overline{t}_p = \frac{l}{\beta_p c}.
 \label{eq:mttp}
\end{equation}
Track pairs were then histogrammed by the measured ToF meantime and the quantity $\overline{t}_p$.  An example of such a histogram for a relatively forward ToF bar in $e^+$ beam data (where
the positron-proton disambiguation is most difficult) is shown in Figure \ref{fig:mt2d}.  As can be seen this histogram produces a clear separation between correctly and wrongly identified proton
candidates.  For each ToF bar, a linear cut between these two bands was optimized to as to produce clear separation between the bands, and thus properly identify tracks that corresponded
to real protons.

All lepton-proton pairs were analyzed in this fashion, attempting to find a ToF hit that could be properly associated with the particle type as identified by the tracking algorithm. 
Any \pmp pair without a valid ToF hit combination was rejected.  In the event that a single pair was found to possibly validly correspond to multiple ToF hit combinations
(an effect that occurred $<1\%$ of the time and was typically due to a particle striking near the edge of a bar and depositing energy in the adjacent bar as well), the best
combination for ToF assignment was determined as the pair of ToF hits with the best correlation of vertex times computed from the ToF meantimes and track path lengths
(i.e., predicting the most similar scatter time for the two particles).

\subsection{Elastic \pmp Pair Selection}
\label{sec:pairsel}

All track pairs with valid ToF hits were then tested by a set of kinematic and fiducial cuts designed to identify pairs with approximately elastic kinematics.  The general
philosophy in placing these cuts was to keep them wide (so as to avoid effects from varying resolutions) and to apply fiducial cuts only to the proton (since the acceptance of
the detector for protons was identical for \ep and \pp events, while the lepton acceptance was different due to the magnetic field).  For the same reason, the reconstructed
parameters of the proton were used to derive quantities such as $Q^2$ and $\epsilon$ so that comparisons of \ep and \pp data in these parameters was on the most equal footing
possible.

First, fiducial cuts were applied at $\left|\phi\right|<\pm11.5^\circ$ from the horizontal on the proton in each sector, which avoided the acceptance edges due to the drift chamber frames throughout
the acceptance, and the reconstructed vertex $z$ of the proton at $\left|z\right|=\pm 350$ mm from the center of the target.  While other analyses consider fiducial cuts that vary as a function of
the scattering angle, which allowed for the recovery of more events in regions excluded by these cuts, this approach was chosen for this analysis to limit the sensitivity
to errors in simulated acceptance of the target and to provide a cross-check on such methods.

Then each pair was tested against the following elastic kinematic cuts:
\begin{enumerate}
 \item Vertex time from ToF meantime and track path length correlation (corrected for vertex position): $\left|\Delta t\right| = \left|t_p-t_{e^\pm}\right| < 5$ ns (Figures \ref{fig:cut1e} and \ref{fig:cut1p})
 
 \item Vertex $z$ correlation: $\left|\Delta z\right| = \left|z_p-z_{e^\pm}\right| < 175$ mm (Figures \ref{fig:cut2e} and \ref{fig:cut2p})
 
 \item Electron-proton elastic angle correlation: $ \left|\theta_p - \theta_{p,\text{elas}}(\theta_{e^\pm})\right| < 7^\circ$ (Figures \ref{fig:cut3e} and \ref{fig:cut3p})
 
 \item Beam energy reconstructed from the track momenta: $\left| E_{\text{beam},p}-E_\text{beam}\right| < 1000$ MeV (Figures \ref{fig:cut4e} and \ref{fig:cut4p})
 
 \item Beam energy reconstructed from the $\theta$ of both tracks assuming elastic kinematics: $ \left|E_{\text{beam},\theta}-E_\text{beam}\right| < 350$ MeV (Figures \ref{fig:cut5e} and \ref{fig:cut5p})
 
 \item Single-arm missing energy of the lepton over the square of the energy as computed by the expected
          elastic energy from the reconstructed $\theta_{e^\pm}$: $\Delta E'_\theta/E'^2 < 0.0048$ MeV$^{-1}$ (Figures \ref{fig:cut6e} and \ref{fig:cut6p})
          
 \item Longitudinal (beam direction) momentum balance: $p_{z,p} + p_{z,e^\pm} - p_\text{beam} > -500 $ MeV (Figures \ref{fig:cut7e} and \ref{fig:cut7p})
 
 \item Coplanarity $\left|\Delta\phi-180^\circ\right| = \left|\phi_\text{right}-\phi_\text{left}-180^\circ\right|<7.5^\circ$ (Figures \ref{fig:cut8e} and \ref{fig:cut8p})
\end{enumerate}
Appendix \ref{chap:kincuts} presents 2D histograms of the data events by each of these cut parameters and the lepton scattering angle $\theta_{e^\pm}$ after application of all selection cuts (but
before background subtraction) for each beam species, as noted for each listed cut.  As can be seen in these figures, the cuts were chosen to be quite permissive in any individual cut allowing the collection
of the multiple exclusive kinematic cuts to produce a relatively clean elastic sample.  The remaining background was predominantly at high lepton $\theta$, low total momentum, and very similar between
the two modes of operation making it straightforward to separate and subtract in the final steps of the analysis.
Note that some of these cuts are heavily correlated (such as the elastic angle correlation and the beam energy fro angles), but that both cuts were applied so as to
make an effective cut that was not axis-parallel in either.  Each of these cut values (with the exception of coplanarity) was chosen as approximately five times RMS width of the distribution
of the parameter with the other (non-correlated) cuts applied
at its widest across the acceptance so as to reduce sensitivity to the detector resolutions by erring on the side of including more background.
The elastic \pmp candidate sample passed to the background subtraction methods, however, still only
contained less than 30\% background throughout the whole acceptance due to the strength of the multiple cuts that were applied using the full exclusive reconstruction of the events.
Other analyses chose deliberately tighter cuts fit to the distributions \cite{schmidt,russell,oconnor}.

If no pairs were found with good ToF hit assignments passing all cuts, remaining pairs were assessed for inclusion in the sample if they passed at least five of the seven cuts, with
strict limits still enforced for the lower bounds of the reconstructed beam energy from both angles and momenta and the longitudinal momentum balance.  If multiple pairs met these criteria,
the one with the most cuts passed and lowest weighted sum of cut parameters was preferred.  Most commonly, this allowed an event to be selected with one or both tracks at higher than expected
momentum, caused by the difficulty of resolving momentum from the bending in the magnetic field as trajectories become straighter at higher momenta.
If after this reduction in cut requirements no pair remained, the event was discarded as a non-elastic \pmp event.

In approximately two-thirds of events with a pair passing these conditions, a single pair was found.  Since the likelihood for simultaneous
 multiple elastic events in the detector was much less than 1\%, in the remaining third
the pair with the minimum sum of the cut parameters weighted by the width of each cut was chosen as the elastic pair.  The selected pair was referred to as an ``initial pair''
and was considered as part of the elastic candidate sample for background subtraction.

\subsection{Background Subtraction}
\label{sec:backsub}

\begin{figure}[thb!]
    \centerline{\includegraphics[width=1.15\textwidth]{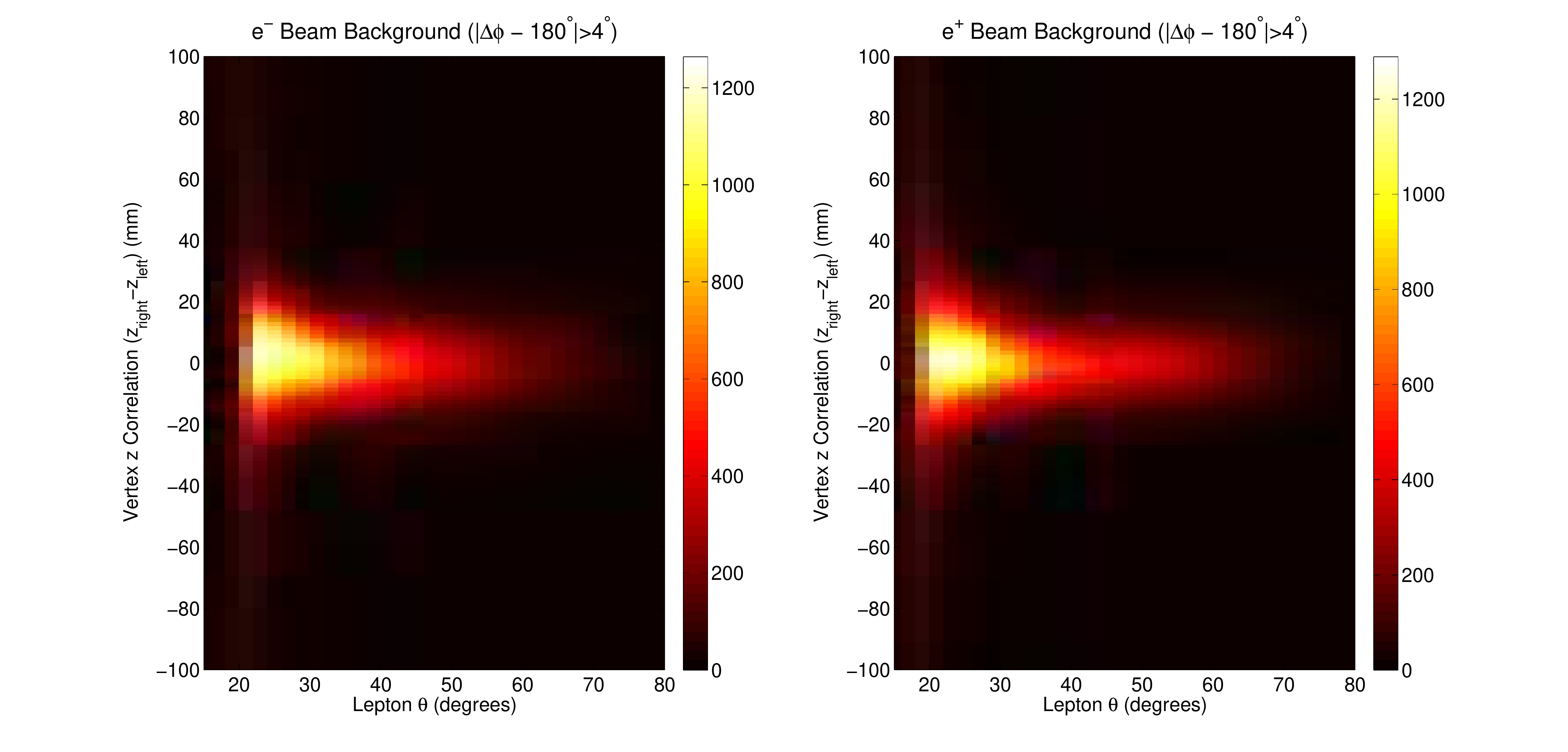}}
    \caption[Vertex $z$ correlation of the background sample]{Correlation of the reconstructed $z$ of track pairs (that of the right track minus that of the left track) for track
    pairs selected from the sidebands of the coplanarity distribution ($\left|\Delta\phi - 180^\circ\right|>4^\circ$) for \ep event selection (left) and \pp event selection (right).
    Note that the background had strong vertex correlation throughout the acceptance, indicating that the background was dominated by various scattering events from the target region
    and that this distribution was not a useful quantity on which to perform background subtraction.}
    \label{fig:zbackcor}
    \end{figure}

As noted, the final kinematic cut of the analysis (coplanarity of the lepton and positron tracks) was deliberately left open to allow the application
of a background subtraction.  Several other background models were considered, including applying background models to the $z$ vertex correlation of the track pairs and
to the elastic $\theta$ correlation of the pairs.  The former approach was discarded due to the fact that the shape of a random background (the convolution of the target
distribution for each track) was an irregular shape to fit and the fact that it was found that the background (as determined from the sidebands of the coplanarity distributions)
was found to have strong $z$ vertex correlation, as shown in Figure \ref{fig:zbackcor}.  Regarding the latter model, the asymmetry of the distribution complicated the model, which
introduced uncertainties not present in modeling the coplanarity background.  Various combinations of 2D models involving pairs of the three considered distributions were also considered,
although none were found to be as robust as the coplanarity method.  Notably, the different OLYMPUS analyses included several different approaches to both the method of background subtraction
and to the strictness of cuts prior to background subtraction (i.e., the amount of background in the sample before background subtraction and the final cut).

\begin{figure}[thb!]
    \centerline{\includegraphics[width=1.15\textwidth]{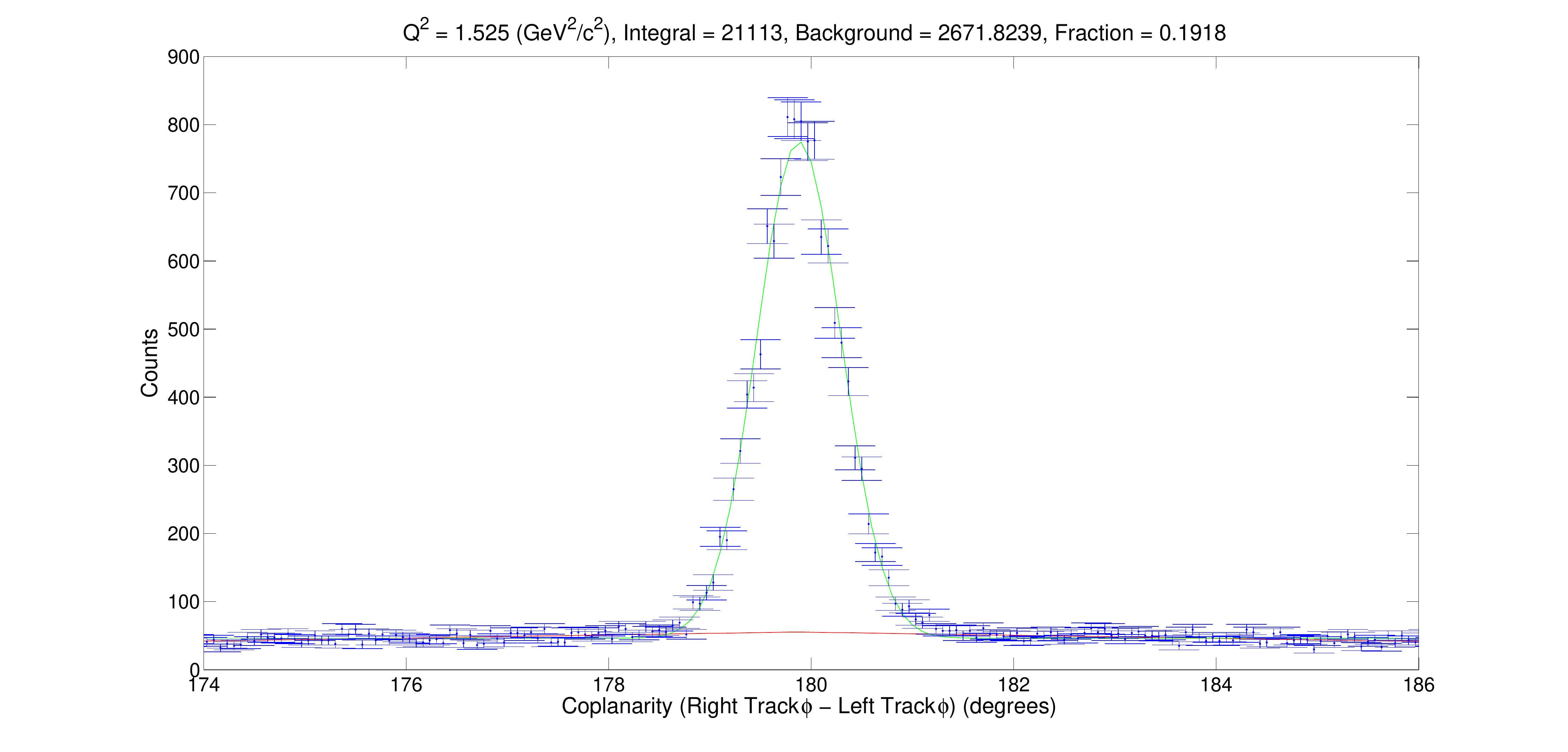}}
    \caption[Example of the background subtraction method for a single $Q^2$ bin]{Example of the background model fit for left-going electrons for the bin centered at $Q^2=1.525$ GeV$^2$ for the full
    \ep dataset used for the analysis.  The blue points represent the data (with statistical uncertainties), the green line the initial Gaussian+constant fit to the central region
    ($\left|\Delta\phi - 180^\circ\right|<6^\circ$), and the red line the triangular background fit to the regions beyond $4\sigma$ of the Gaussian peak.  For all bins within the acceptance, the model
    fit the data extremely well, producing positively sloped triangles as expected without explicitly requiring this condition.}
    \label{fig:ebacks}
    \end{figure}

\begin{figure}[thb!]
    \centerline{\includegraphics[width=1.15\textwidth]{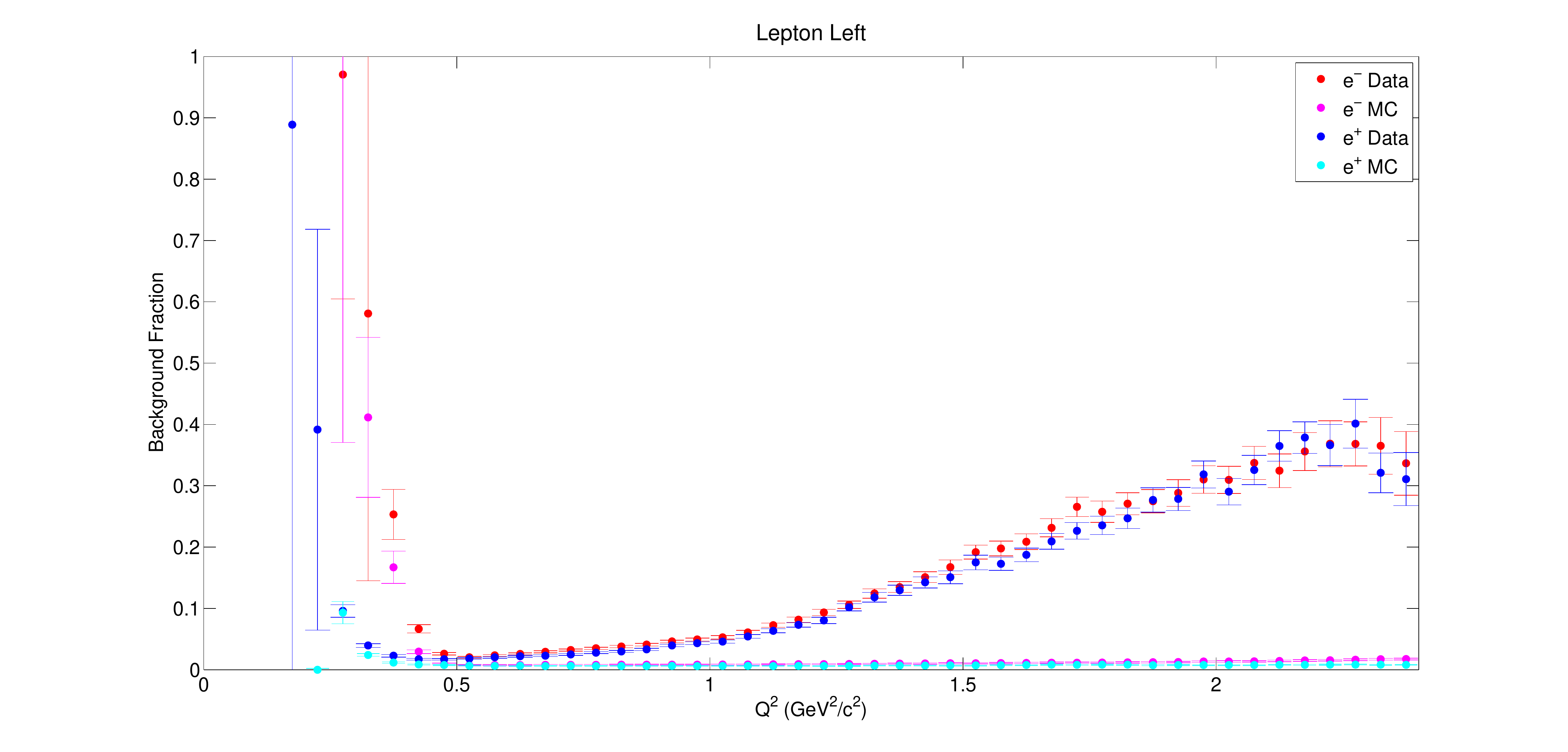}}
    \caption[Background fraction after initial event selection as a function of $Q^2$ for leptons going left]{Background fraction remaining in the final elastic event selection for leptons detected in the left sector for \ep
    and \pp events in both data and simulation.  The error bars represent the 95\% confidence bounds on the integrals of the background distributions within the final coplanarity cut.}
    \label{fig:bfleft}
    \end{figure}

\begin{figure}[thb!]
    \centerline{\includegraphics[width=1.15\textwidth]{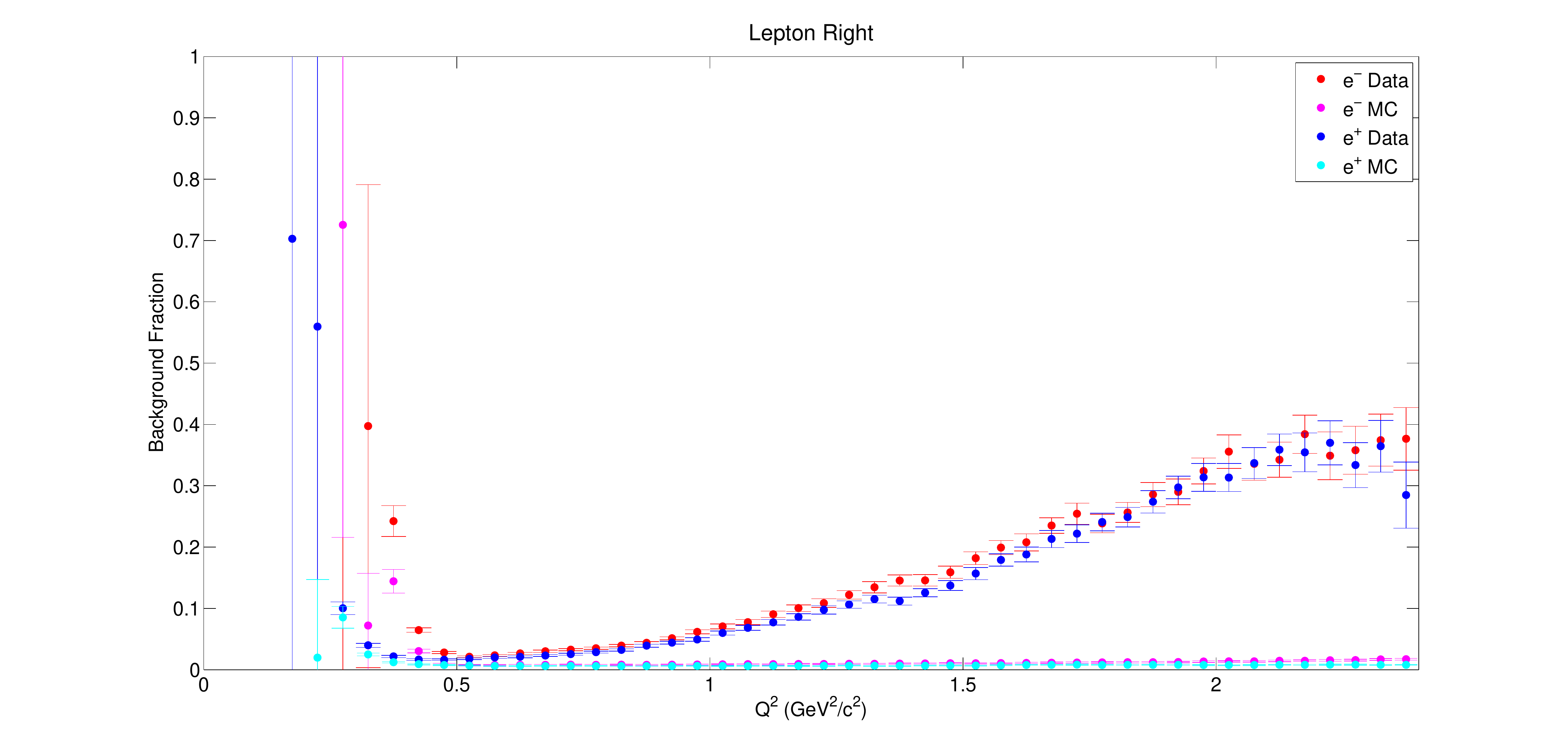}}
    \caption[Background fraction after initial event selection as a function of $Q^2$ for leptons going right]{Background fraction remaining in the final elastic event selection for leptons detected in the right sector for \ep
    and \pp events in both data and simulation.  The error bars represent the 95\% confidence bounds on the integrals of the background distributions within the final coplanarity cut.}
    \label{fig:bfright}
    \end{figure}

Since cuts involving the reconstructed momentum and elastic event $\theta$ correlation of the particles heavily
suppressed backgrounds that would be expected to also prefer $\Delta\phi\approx 180^\circ$ (such as pion electro-production \cite{PhysRev.129.1834} and $e^+e^-$ scattering),
it was expected that the background in $\Delta\phi$ would be dominated by random track pairs.  For a given small bin in $Q^2$ (or $\theta$), the corresponding $\phi$ distribution
was essentially uniform, constrained by the frames of the drift chambers.  Thus, the expected random pair background was the convolution of two uniform distributions, i.e., a triangular
distribution.  The initial \pmp pair selection was binned in a 2D histogram of $\Delta\phi$ and $Q^2$ (as determined by the proton $\theta$ assuming the beam energy and elastic kinematics), 
with bin width of 0.05 GeV$^2$ in $Q^2$ (corresponding to approximately 35 bins across the full acceptance).  The background fraction was computed independently for each $Q^2$ bin, lepton left and right,
event type (\ep or \pp), and for both data and simulation according to the following procedure:
\begin{enumerate}
 \item A single $Q^2$ bin was projected as a 1D histogram of $\Delta\phi$, as shown in Figure \ref{fig:ebacks}.
 \item A Gaussian plus constant model was fit to the central region of the histogram ($\left|\Delta\phi - 180^\circ\right|<6^\circ$) to find the peak of the coplanarity distribution.  For all bins
       within the acceptance, this peak was found to be within 0.15\dg of 180\dg, indicating excellent $\phi$ reconstruction.
 \item The sidebands of the distribution (outside of $4\sigma$ of the Gaussian fit of the previous step) were fit to a triangular distribution model ($a-b\left|x-\mu\right|$, where $\mu$ is the mean
       of the Gaussian fit in the previous step).
 \item This background distribution was passed, along with the data/simulation coplanarity distribution, to the final step of the analysis, which applied the final cut on coplanarity and determined
       the fraction of counts within that cut corresponding to background that should be subtracted.  Additionally, the 95\% confidence bounds of the background model fit were retained to estimate the 
       uncertainty in the background model.
\end{enumerate}
The resulting background fractions as a function $Q^2$ (for the final selection described in the next section), are shown in Figure \ref{fig:bfleft} and \ref{fig:bfright} for leptons going left and right
respectively for both leptons species in both data and simulation.  In general, it was found that \ep data exhibited slightly higher background levels than positron data, consistent with the beam conditions
for each species.  The very small ($\mathcal{O}(1\%)$) background levels in simulation may be attributed to radiative events in which a radiated photon causes a large deviation in $\phi$ for one of the tracks
and occasional mis-reconstructed events.  As expected, however, the background fraction in simulation was minimal.  In the experimental data, the initial pair selection resulted in a maximum background fraction
of $\sim$30\% within the acceptance of the detector ($Q^2\lesssim2.2$ GeV$^2$).  The different OLYMPUS analyses used different levels of strictness in their initial elastic selections, resulting in higher
or lower background fractions than this analysis. This difference in modeling provides a valuable test of the robust of the background subtraction models via comparison of the results produced
by each analysis.

\subsection{Final Processing}
\label{sec:fproc}

\begin{figure}[thb!]
    \centerline{\includegraphics[width=1.15\textwidth]{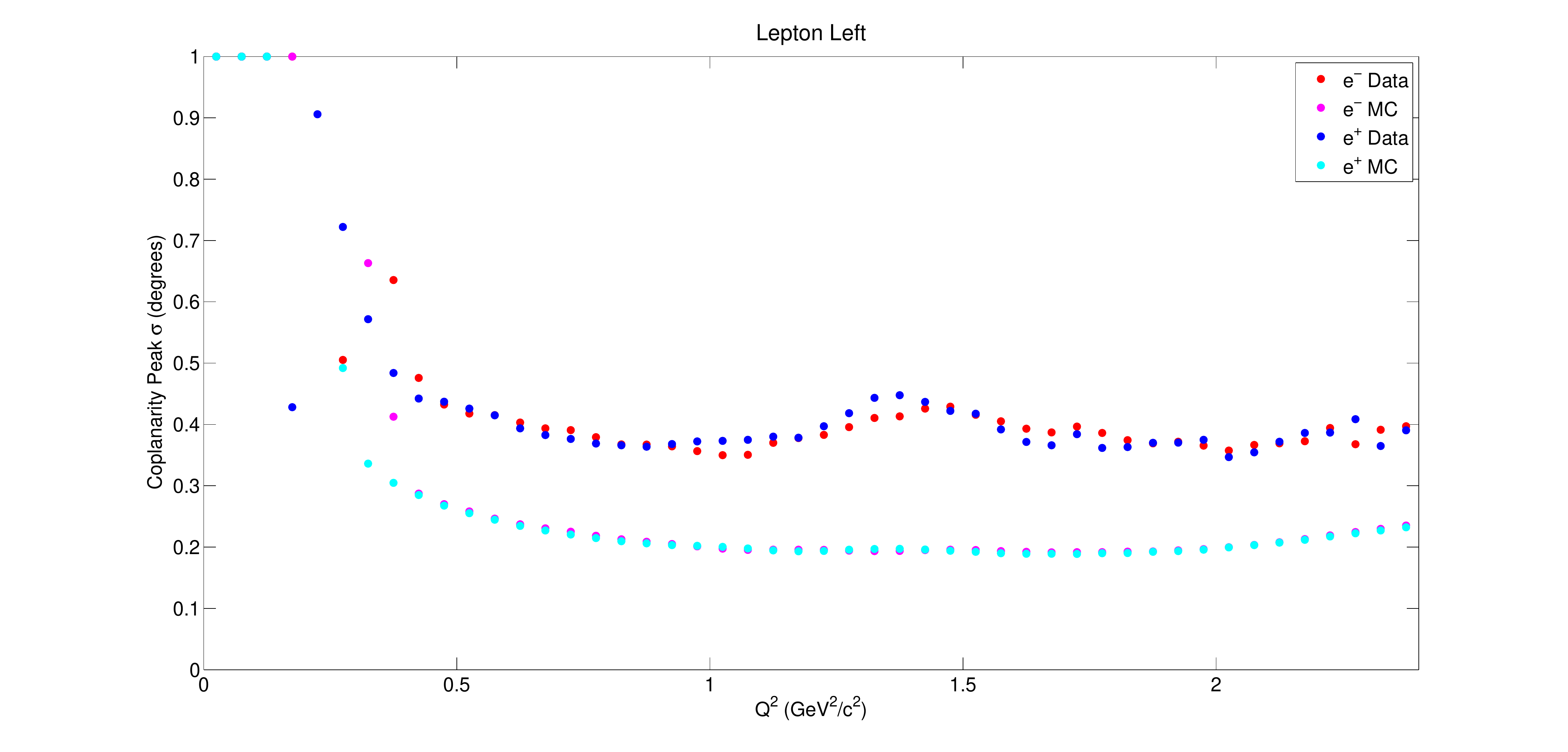}}
    \caption[Width of the $\Delta\phi$ distribution as a function of $Q^2$ for leptons going left]{The $1\sigma$ width of the Gaussian fit to the coplanarity distribution as a function
    of $Q^2$ for leptons in the left sector.  The loss of resolution in data near $Q^2 = 1.4$ GeV$^2$ corresponds to the lepton passing through the inefficient region of the left-inner drift
    chamber (see Figure \ref{fig:badwc}), which caused the lepton tracks to be comprised of fewer hits.  The purity of the simulation data sample resulted in a maintenance of resolution in this region
    despite the loss of hits due to the efficiency map implementation.}
    \label{fig:copwleft}
    \end{figure}

\begin{figure}[thb!]
    \centerline{\includegraphics[width=1.15\textwidth]{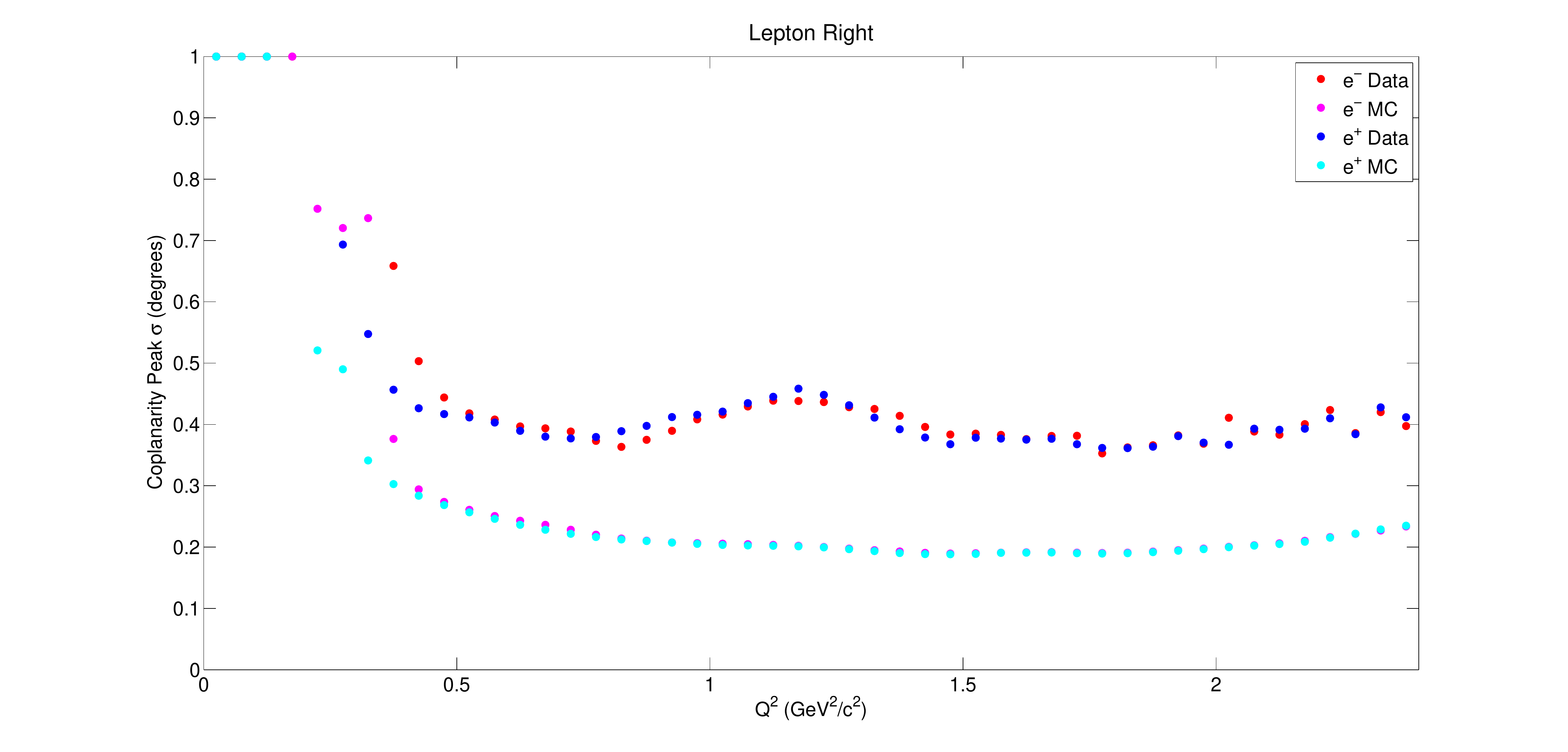}}
    \caption[Width of the $\Delta\phi$ distribution as a function of $Q^2$ for leptons going right]{The $1\sigma$ width of the Gaussian fit to the coplanarity distribution as a function
    of $Q^2$ for leptons in the left sector.  The loss of resolution in data near $Q^2 = 1.2$ GeV$^2$ corresponds to the left-going proton passing through the inefficient region of the left-inner drift
    chamber (see Figure \ref{fig:badwc}), which caused the proton tracks to be comprised of fewer hits.  The purity of the simulation data sample resulted in a maintenance of resolution in this region
    despite the loss of hits due to the efficiency map implementation.}
    \label{fig:copwright}
    \end{figure}

With the background model in place, the final event selection was made by placing the final available cut on $\Delta\phi$ and integrating the number of counts remaining to produce
the data and simulation total elastic event yields.  This $\Delta\phi$ cut was placed at the $3\sigma$ width of the Gaussian distribution fitted to each $Q^2$ bin in the background modeling
process described in the previous section.  These distribution widths are shown as functions of $Q^2$ for leptons going left and right in Figures \ref{fig:copwleft} and \ref{fig:copwright} respectively.
Note, that since the simulation resolution in $\Delta\phi$ was slightly better than the data resolution, the fit cut width of the data sample
for each lepton species was applied to both the data and simulation of that species so as to maintain the consistent treatment of data and simulation as in the initial cuts.  For each $Q^2$ bin
of width 0.05 GeV$^2$, region within the $3\sigma$ data width was integrated for both the coplanarity histogram and the background model.  The integral of the background model was subtracted from
the integral of the coplanarity histogram to produce the final yields of elastic \pmp events that were used to generate the final results of the experiment.

\section{Comparison of Data and Simulation}
\label{sec:datasim}

\begin{figure}[thb!]
  \centerline{\includegraphics[width=1.15\textwidth]{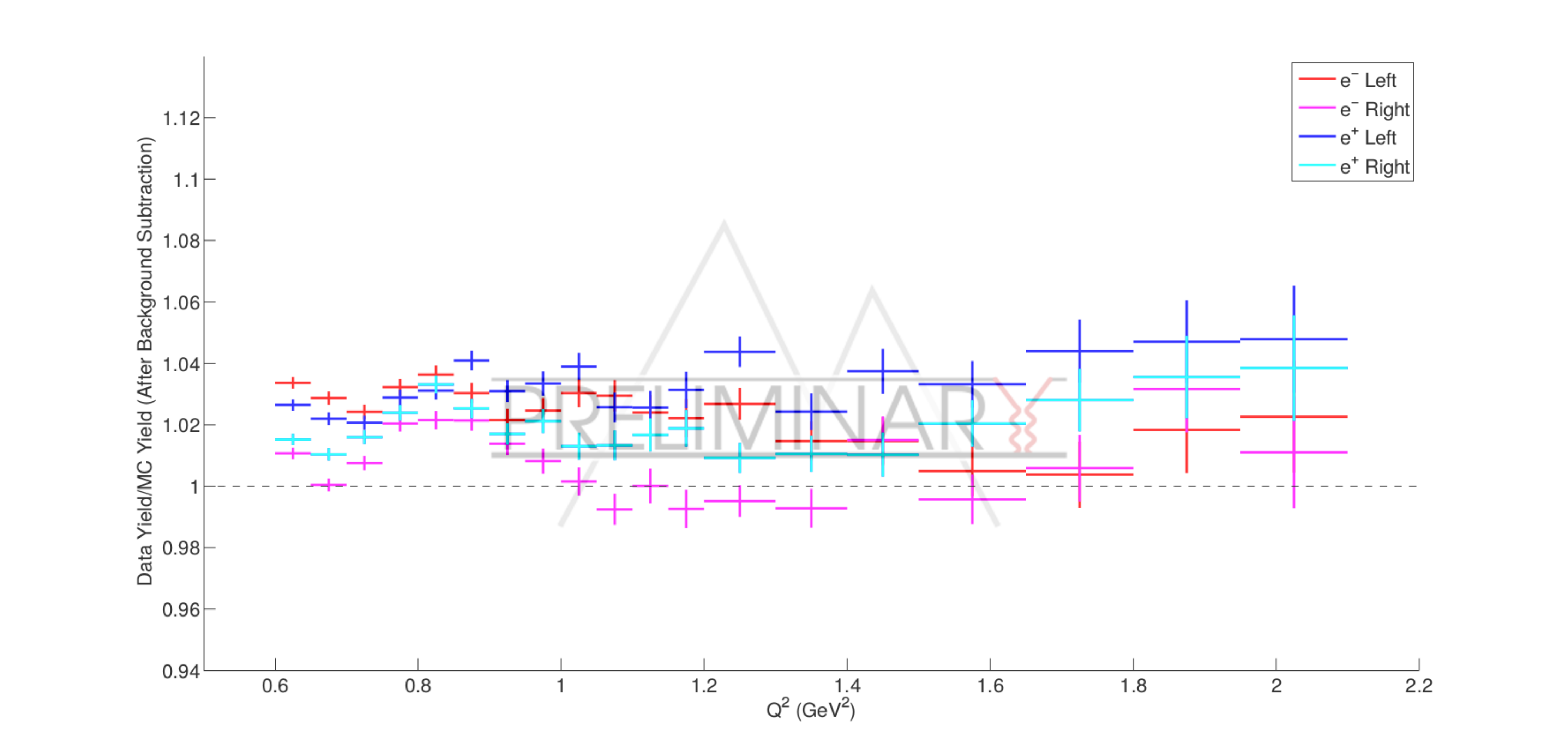}}
  \caption[Ratio of data yields over simulation yields by lepton species and detector side with the Kelly form factor model]{The ratio of data yields over simulation
  yields (background-subtracted) by lepton species and detector side with the Kelly form factor model \cite{PhysRevC.66.065203}.  Given that uncertainties in the determination of the elastic
  proton form factors are of the order of at least several percent, especially at the high end of the $Q^2$ range, the overall agreement between the data and simulation yields
  is strong.  Note that these results
  are normalized to the slow control luminosity, and are thus subject to a shared systematic uncertainty in the absolute value of all points of several percent.}
  \label{fig:datamckelly}
  \end{figure}

With the full analysis method applied, comparisons between the data and simulation elastic event yields after background subtraction
provide insight into the quality of the data and analysis as well as a preliminary measure of the absolute elastic \ep and \pp cross-sections.
Figure \ref{fig:datamckelly} presents the ratio of the background-subtracted data and simulation yields for \ep and \ep events, separated by
whether the lepton was detected in the left or right sector of the spectrometer.  For this figure, the simulation was conducted using the Kelly
parametrization of the elastic proton form factor \cite{PhysRevC.66.065203} and the Maximon and Tjon radiative corrections prescription \cite{MaximonPhysRevC.62.054320}.
Encouragingly, the data and simulation show agreement on the level of a few percent across the entire acceptance, which is consistent with the expectations in deviations
due to uncertainties in the knowledge of the form factors.  Differences in the ratios computed for leptons going left and right are indicative of the approximate systematic
uncertainty in the determination of the absolute elastic \pmp cross sections that are possible with the OLYMPUS data.  Additionally, since the simulation events used for
Figure \ref{fig:datamckelly} were generated using the slow control luminosity, the fact that the data/simulation ratio is several percent above unity across the entire
acceptance is very consistent with the measurement of the absolute luminosity from the 12\dg system ($\mathcal{L}_{e^\pm}/\mathcal{L}_\text{SC}\approx 1.046\pm0.024$, see 
Section \ref{sec:12res}).  Thus, careful analyses of the OLYMPUS data may be able to provide information on the absolute \ep and \pmp cross sections, and also information
regarding the elastic form factors, with uncertainties on the order of $\sim$5\%.

\begin{figure}[thb!]
  \centerline{\includegraphics[width=1.15\textwidth]{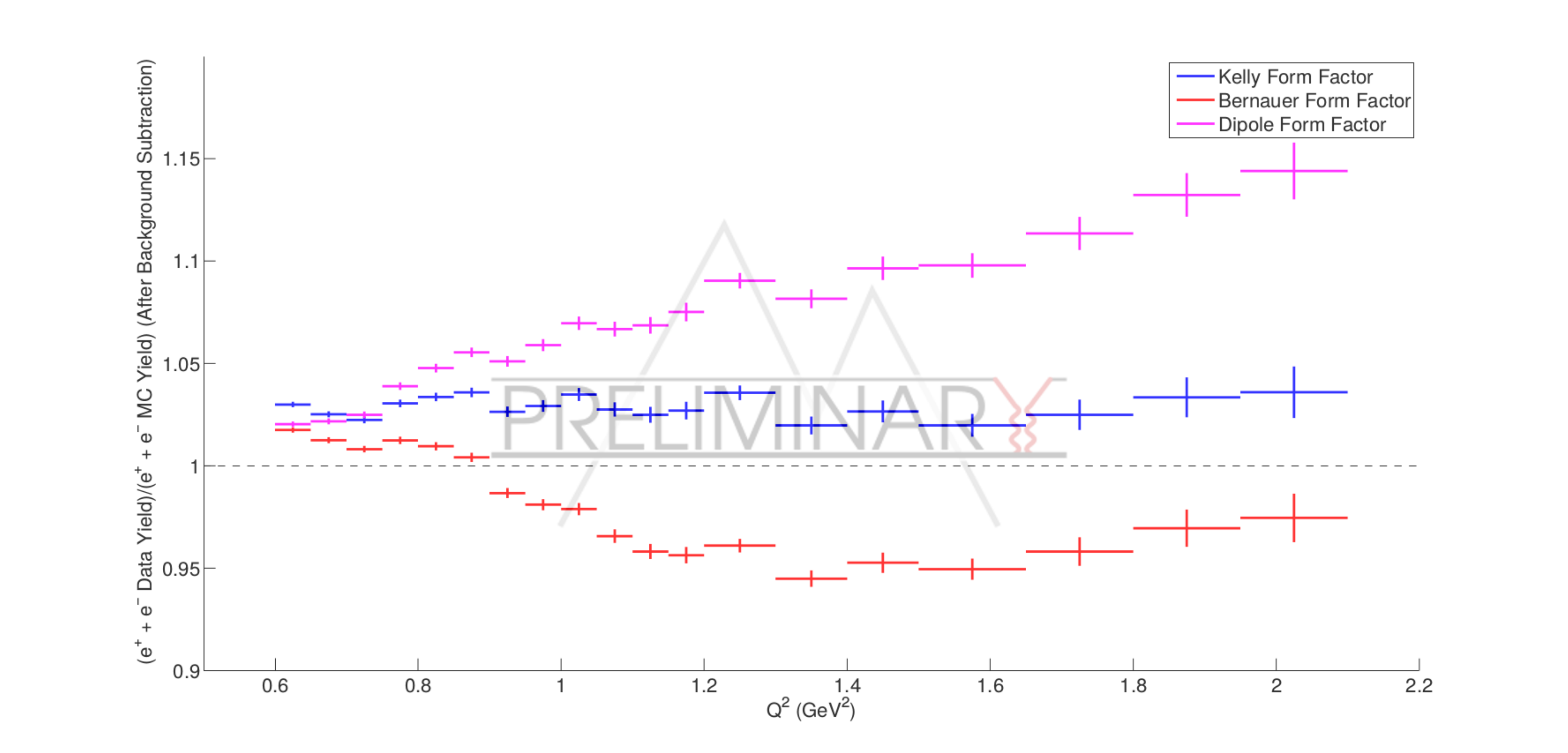}}
  \caption[Background-subtracted lepton-summed data yields over simulation for several form factor models]{Ratio of the summed elastic \ep and \pp yields of each species in
  data and simulation for several form factor models \cite{PhysRevC.66.065203,BerFFPhysRevC.90.015206}. In the summed yield of \ep and \pp, effects from TPE and $\mathcal{O}(\alpha^3)$ 
  radiative corrections from lepton vertices cancel.  Given the power of the OLYMPUS data to discriminate between models, there exists the potential in the OLYMPUS data to inform future
  models of the proton elastic form factors.  Note that these results are normalized to the slow control luminosity, and are thus subject to a shared systematic uncertainty
  in the absolute value of all points of several percent.}
  \label{fig:datamcsum}
  \end{figure}

Figure \ref{fig:datamcsum} shows the ratio of the total \ep and \pp elastic event yields in data and simulation for three different form factor parametrizations:
the Kelly model \cite{PhysRevC.66.065203}, the Bernauer model \cite{BerFFPhysRevC.90.015206}, and the dipole form factor (Equation \ref{eq:dipff}).  The sum of the \ep and \pp yields
is a particularly useful quantity for examination of the behavior of the different form factor models, since any effects from TPE or lepton charge-odd radiative corrections that
are present in the data but not the simulation are canceled in this sum, providing a means of determining the absolute cross section in the absence of such effects.  As expected, the models
agree at low $Q^2$ where existing data constrains the models well, but considerable deviations are apparent as $Q^2$ increases.  Since the luminosity normalization is shared between
the models, which effectively removes uncertainty contributions from the absolute normalization when comparing the relative values of the ratios in different models, the large deviations
between the form factor models at higher $Q^2$ indicate that the OLYMPUS data can provide constraints on future form factor models.  The possibility of these additional physics results
will be considered in future OLYMPUS publications.
  
\section{Systematic Uncertainties}
\label{sec:mainsys}

While the final analyses of the systematic uncertainties associated with the OLYMPUS $R_{2\gamma}$ result were still underway at the time of writing,
this section provides a preliminary assessment of the systematic uncertainties associated with the results presented in Chapter \ref{Chap7}, with a particular
focus on identifying the remaining dominant uncertainties for consideration in the later stages of the analysis.  Note that the systematic uncertainties from
the MIE luminosity determination (Section \ref{sec:miesys}) and the 12\dg system \pmp measurement, used either as a luminosity determination or as an additional kinematic point
(Section \ref{ss:12sys}), are independent from the effects discussed here.  In particular, any uncertainty from the luminosity determination applies as a constant shift shared by
all points in the determination of $R_{2\gamma}$, rather than as a point-to-point uncertainty.

The estimates presented here should be considered preliminary, and a final analysis (including additional detail) will accompany the published OLYMPUS results.

\subsection{Detector Acceptance and Efficiency}

The complexities of the wire chamber acceptance, efficiency, and its interaction with the magnetic field (both in terms of particle trajectories and the time-to-distance
calibration) make a simulation-based approach to estimating the systematic uncertainties associated with such effects infeasible. Due to issues with the drift chamber
efficiency, uncertainties in the geometric survey of the detector, and the difficulty of the drift chamber time-to-distance parametrization, these sources of systematic uncertainty
are the dominant contributions to the overall uncertainty, and thus it is critical to properly assess them.  The redundancy of the left and right detector
systems in OLYMPUS provides a means of estimating the systematic uncertainties from these effects based on examination of the data.

\begin{figure}[thb!]
  \centerline{\includegraphics[width=1.15\textwidth]{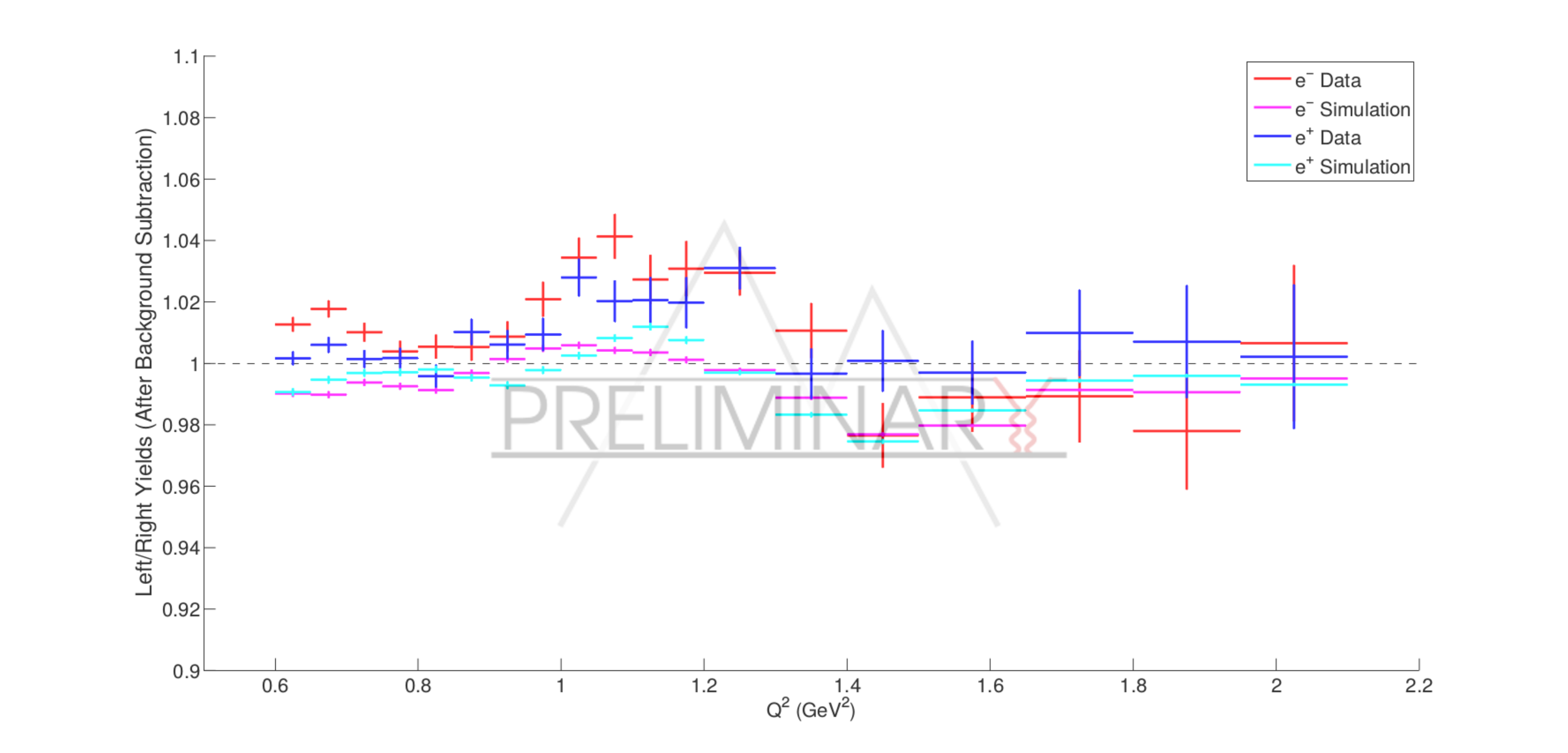}}
  \caption[Ratio of lepton left/right background-subtracted yields in data and simulation]{Ratio of the lepton left/right background-subtracted elastic \pmp yields in data and simulation,
  separated by lepton species.  Structures in this ratio that differ between data and simulation are indicative of effects such as differences in acceptance, detector efficiency, and magnetic
  field between the left and right sectors that are not fully accounted for in the simulation.  These effects contribute to the total systematic uncertainty estimate.  The dominant
   ``peak-valley'' structure between $Q^2=1.1$ GeV$^2$ and $Q^2=1.6$ GeV$^2$ is known to be related to the highly inefficient region of the left
  drift chamber (see Figure \ref{fig:badwc}), which reduces the tracking resolution significantly and affects event selection.}
  \label{fig:leftright}
  \end{figure}

\begin{figure}[thb!]
  \centerline{\includegraphics[width=1.15\textwidth]{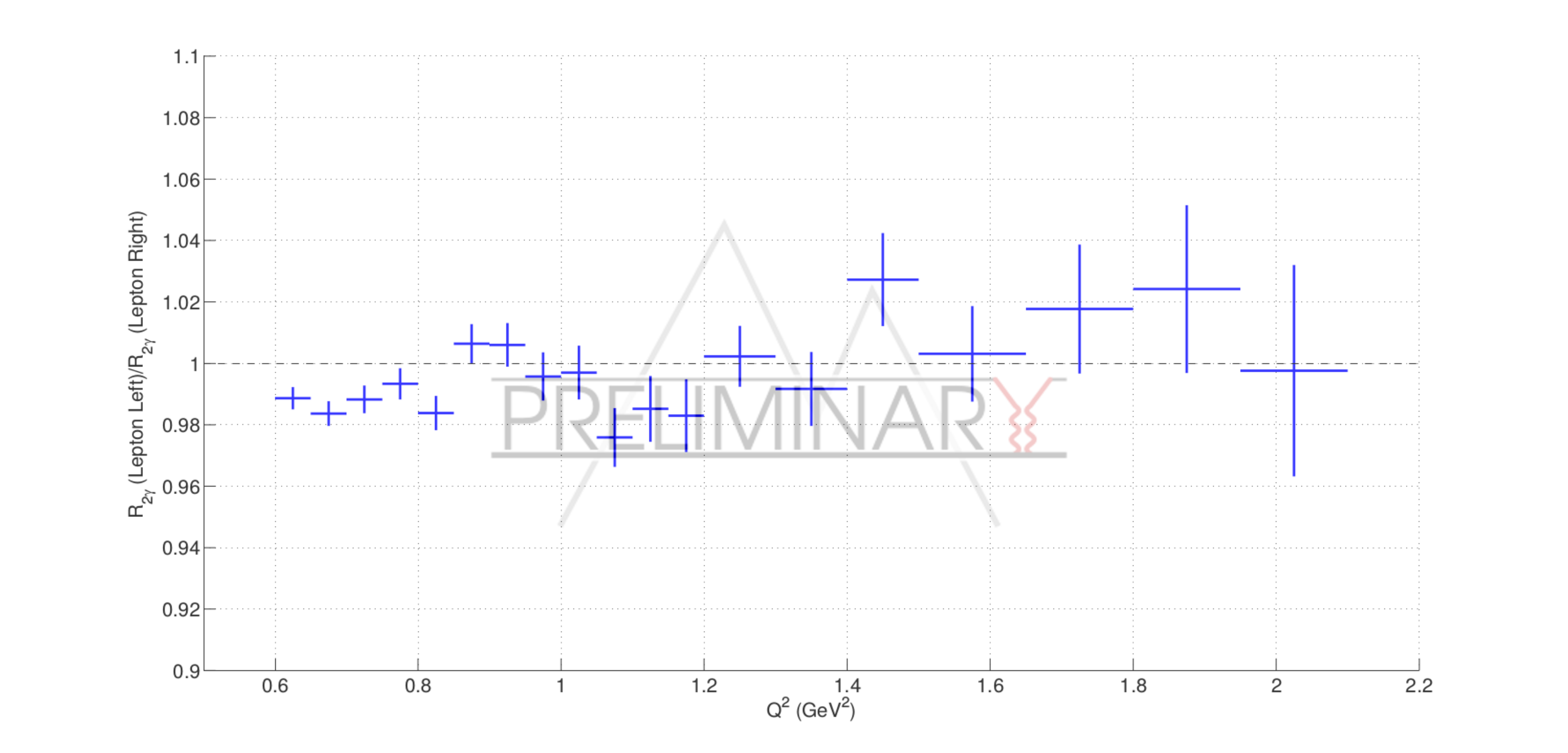}}
  \caption[Ratio of the $R_{2\gamma}$ results for leptons going left and right in the detector]{Ratio of the $R_{2\gamma}$ results for events in which the lepton was detected in the left and right
  sectors of the detector.  The error bars are statistical, and it should be noted that the lepton-left and lepton-right datasets are statistically independent.
  Deviations from unity of this ratio that cannot be attributed to statistics provide a means of estimating systematic uncertainties due to effects such as acceptance, magnetic field,
  and detector efficiency.}
  \label{fig:ratleftright}
  \end{figure}

Ideally, the elastic event yields separated by
the lepton going left or right in the detector would be identical, or at least any deviations between the left and right yields would be completely accounted for
in simulation.  Any unexplained left/right deviations provide a first-order estimate of systematic uncertainties due to detector acceptance, efficiency, and magnetic field.  Figure
\ref{fig:leftright} shows the lepton-left/right ratio of background-subtracted yields for \ep and \pp events in data and simulation, and Figure \ref{fig:ratleftright} shows the ratio
of the \rtg results for the separate lepton-left and lepton-right samples.  While the simulation captures some of the structure in the ratio of left/right yields in data, the structure
is not fully accounted for by the inefficiencies implemented in simulation.  In particular, the ``peak-valley'' structure between $Q^2=1.1$ GeV$^2$ and $Q^2=1.6$ GeV$^2$ is known to 
be related to the highly inefficient region of the left drift chamber (see Figure \ref{fig:badwc}), which reduces the tracking resolution significantly and affects event selection.  Most
likely, the applied resolutions in the simulation did not degrade the hit quality sufficiently in simulation to precisely match the conditions of data tracking, leading to less difference
in the left/right comparison in simulation than data.

  Deviations between the data and simulation in Figure \ref{fig:leftright} provide indications of the systematic uncertainty on a possible absolute cross section extraction, while deviations
  from unity in Figure \ref{fig:ratleftright} provide an indication of how such uncertainties affect the \ep and \pp yields differently and thus provide an estimate of the uncertainty of the 
  \rtg result. While the lepton-left and lepton-right data sample are statistically independent, there remain deviations in the value of \rtg of $\sim$1\% that cannot be attributed to statistical
  variation between the samples, and thus this is taken as an estimate of the systematic uncertainty due to detector acceptance and efficiency effects (which are additionally heavily convolved
  with the effects of the magnetic field uncertainty).  At this stage, this the dominant systematic uncertainty for the \rtg result, and work is ongoing to reduce it if possible.
  
\subsection{Elastic and Fiducial Cuts}

To control uncertainties due to the choices of the elastic and fiducial event cuts used in the \rtg analysis, these cuts were varied and the results recomputed to examine their
effect on the value of \rtg.  Additionally, a potentially stronger analysis of the effects of choices in the \pmp event selection is provided by the comparison of the independent
analyses of the OLYMPUS data \cite{schmidt,russell} discussed in Section \ref{sec:indana}.  It remains useful, however, to consider the effects of the cut choices within a single
analysis.

\begin{figure}[thb!]
  \centerline{\includegraphics[width=1.15\textwidth]{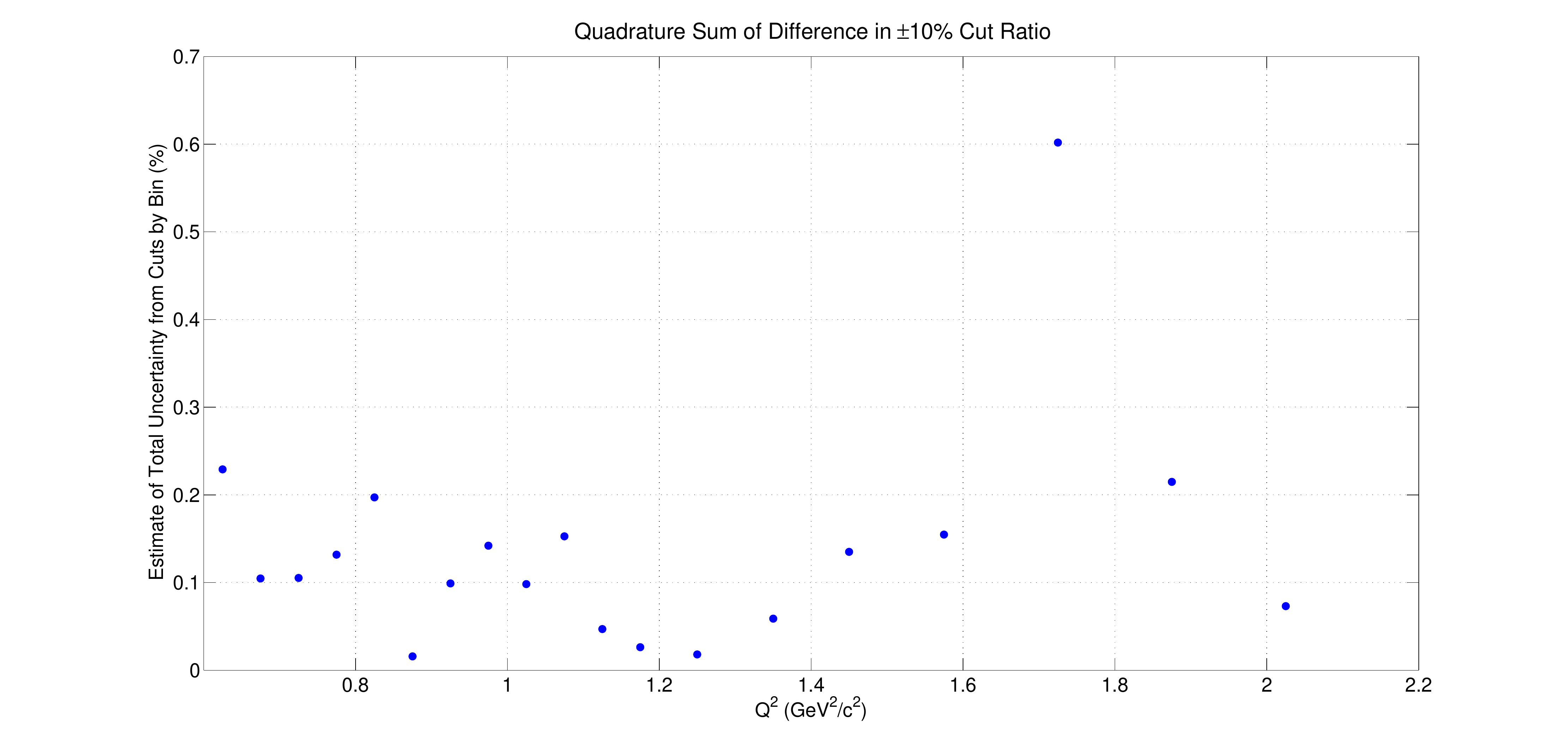}}
  \caption[Estimate of the total systematic uncertainty due to event selection cuts]{Estimate of the total systematic uncertainty due to event selection cuts, computed as the quadrature sum
  of the variations in \rtg caused by varying the boundaries of each of the cuts listed in Sections \ref{sec:pairsel} and \ref{sec:fproc} by $\pm10\%$.  For most $Q^2$ bins, this uncertainty
  is on the order of 0.2\% or smaller, with the exception of the bin at $Q^2=1.725$ GeV$^2$.}
  \label{fig:cutsys}
  \end{figure}

Each of the eight fiducial, pair selection, and elastic cuts applied in the analysis (listed in Section \ref{sec:mainana}) was varied individually by $\pm10\%$ in range, and
the resulting change in the value of \rtg between these variations was computed bin-by-bin.  The quadrature sum of the variations between the values of \rtg at $\pm10\%$ cut ranges for
the eight cuts is shown in Figure \ref{fig:cutsys}.  For most $Q^2$ bins, the sum of the variations is less than 0.2\%, with the notable exception of the bin at $Q^2=1.725$ GeV$^2$.  To an
extent, this lack of effect is expected due to the strategy used for this analysis, which placed loose cuts and allowed the exclusively reconstructed kinematics to determine event selection
rather than placing tight elastic cuts that would be more sensitive to detector resolutions.  The cause of the larger uncertainty in the $Q^2=1.725$ GeV$^2$ bin is under investigation, but
is dominated by the uncertainty associated with the fiducial cut on the reconstructed $z$ vertex of event pairs.

Additionally, effects from the choices made in particle identification (i.e., the cuts made in the measured vs. expected ToF meantime discussed in Section \ref{sec:partid}) were also examined
by a similar procedure of varying cut boundaries.  In most ToF bars, the separation between leptons and protons was extremely clear and thus variations in the cut had essentially no effect.
In the regions where pair particle identification is most ambiguous (the ToF bars in the central region of each detector side), differences between the independent analyses (Section \ref{sec:indana}),
which each treated particle identification in a different manner, are likely a more robust measure of the systematic uncertainties due to such analysis choices.
  
\subsection{Background Subtraction}

Due to the qualitative nature of the various choices made in choosing a background subtraction model (the kinematic quantities on which subtraction is performed,
the models used for the elastic peak and background in those quantities, etc.), the final assessment of the systematic uncertainty due to background subtraction will involve
a comparison of the independent \rtg analyses using the OLYMPUS data, each of which applied different approaches to this problem.  It is useful, however, to examine the uncertainties
inherent to the background subtraction method used in the analysis presented in this work (Section \ref{sec:backsub}).

As shown in Figures \ref{fig:bfleft} and \ref{fig:bfright}, an uncertainty in the background subtraction arises from the uncertainty in the fit of the background model to the
data.  This uncertainty increases for data as a function of $Q^2$ as the statistics of the data decrease, but remains relatively constant and comparably small for simulation.  In the 
highest $Q^2$ bins used for the analysis, the 95\% confidence interval for the background fit to the absolute \ep and \pp yields in data reaches approximately 2\%.  Much of this variation
is statistical, however, and thus captured somewhat in the final statistical uncertainties associated with those bins.  An additional cross check on the background subtraction was conducted
by varying the range in the coplanarity histogram away from the peak to which the background model was fit between outside of $3\sigma$ to outside of $5\sigma$ of the peak.  These alterations
to the procedure did not affect the background fraction more than the confidence intervals associated with the background fit.  Thus, conservatively, a systematic uncertainty of $\sim$0.5\% may
be ascribed to the background subtraction in the highest $Q^2$ bins, with much smaller contributions at lower $Q^2$.

\subsection{Radiative Corrections}

  While a concerted effort was made to properly account for radiative corrections in the analysis of the OLYMPUS data via the full-simulation scheme
  described in Section \ref{sec:radgen}, the choice of radiative corrections prescription that is used affects the final \rtg result due to the different
  approximations and assumptions made by different models.  The OLYMPUS radiative generator was designed to permit the OLYMPUS results to be presented
  using a variety of different radiative corrections schemes so as to facilitate comparison with other data.  This uncertainty may be eliminated by matching radiative corrections schemes when
  comparing different experiment results or experimental data to theoretical or phenomenological models, but it is worthwhile
  to assess the overall effect of different models on the \rtg results.
  
\begin{figure}[thb!]
  \centerline{\includegraphics[width=1.15\textwidth]{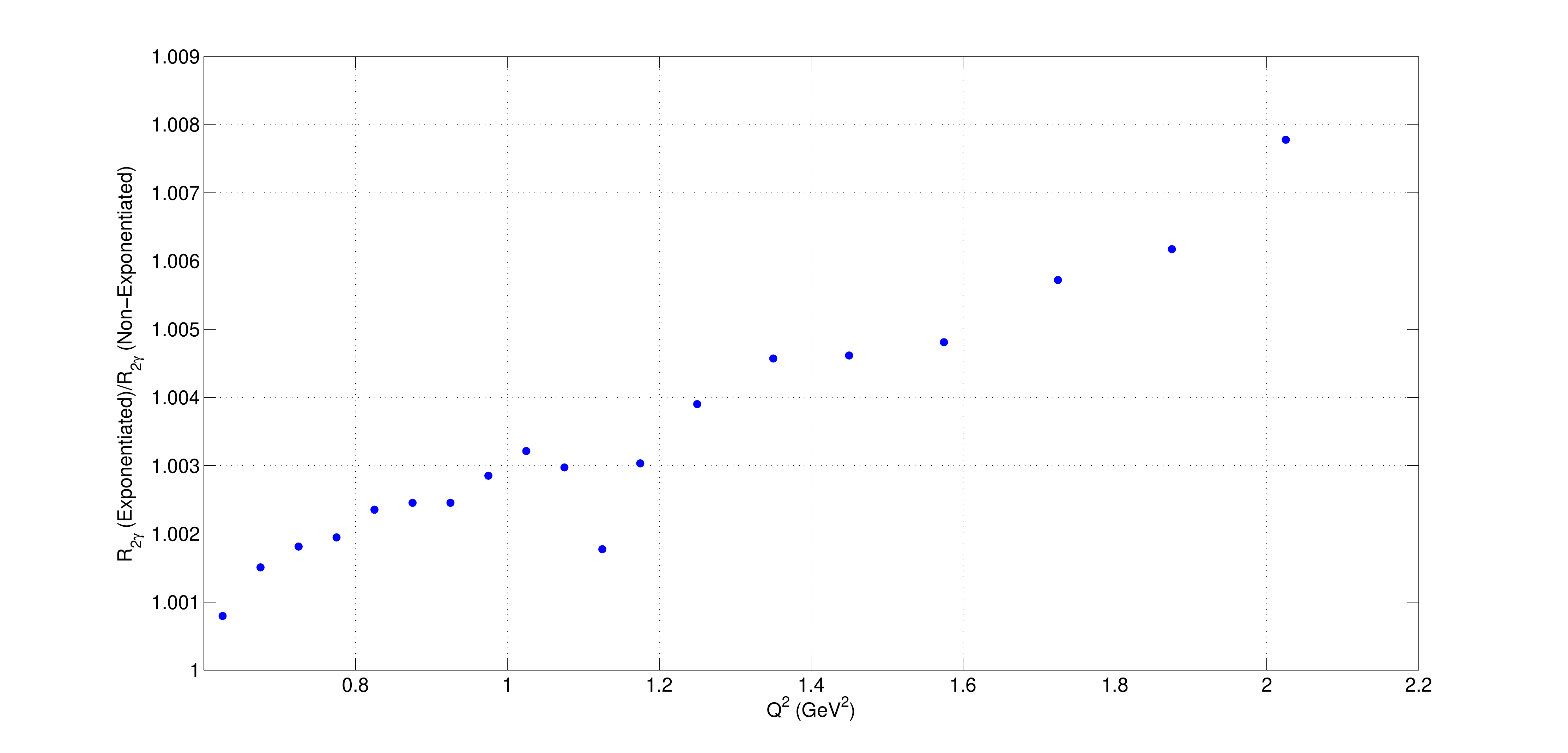}}
  \caption[Ratio of the $R_{2\gamma}$ results for simulation using exponentiated and non-exponentiated radiative corrections]{Ratio of the $R_{2\gamma}$ results for simulation using
  exponentiated and non-exponentiated radiative corrections under the prescription of Maximon and Tjon \cite{MaximonPhysRevC.62.054320}.
  The statistical uncertainties for each data point are suppressed since they are completely correlated in each radiative corrections model.}
  \label{fig:ratexpo}
  \end{figure}

\begin{figure}[thb!]
  \centerline{\includegraphics[width=1.15\textwidth]{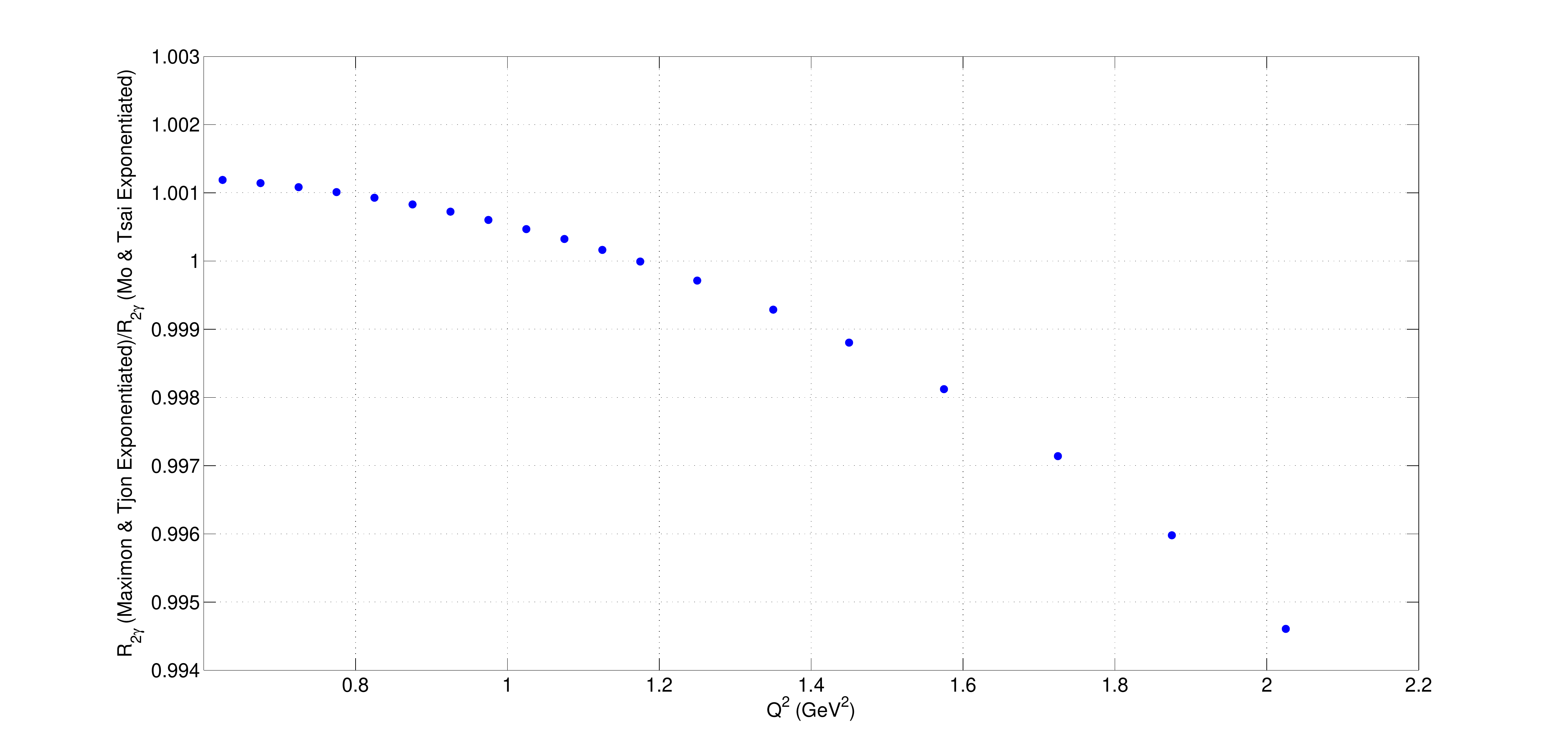}}
  \caption[Ratio of the $R_{2\gamma}$ results for simulation using the Maximon/Tjon and Mo/Tsai radiative corrections prescriptions]{Ratio of the $R_{2\gamma}$ results for simulation using
  the Maximon/Tjon \cite{MaximonPhysRevC.62.054320} and Mo/Tsai \cite{MoRevModPhys.41.205} radiative corrections prescriptions.  For each simulation, the radiative corrections were
  applied with exponentiation. The statistical uncertainties for each data point are suppressed since they
  are completely correlated in each radiative corrections model.}
  \label{fig:ratmaxmo}
  \end{figure}

  Figures \ref{fig:ratexpo} and \ref{fig:ratmaxmo} present the ratios of the \rtg results that arise from making key choices in the application of radiative corrections.
  The former shows the ratio of the result when the Maximon and Tjon radiative corrections scheme \cite{MaximonPhysRevC.62.054320} is applied using exponentiated and
  non-exponentiated methods, while the latter shows the ratio of the results using the Maximon and Tjon scheme and the Mo and Tsai scheme \cite{MoRevModPhys.41.205} (the two
  most commonly applied radiative \pmp prescriptions).  Each of these ratios shows deviation from unity that increases with $Q^2$, reaching 0.5\%-1.0\% at the upper end of the
  OLYMPUS acceptance.  This illustrates the importance of matching radiative corrections schemes when comparing datasets and predictions and the importance of understanding
  the exact methods used to apply radiative corrections in different datasets.

\subsection{Form Factors}

\begin{figure}[thb!]
  \centerline{\includegraphics[width=1.12\textwidth]{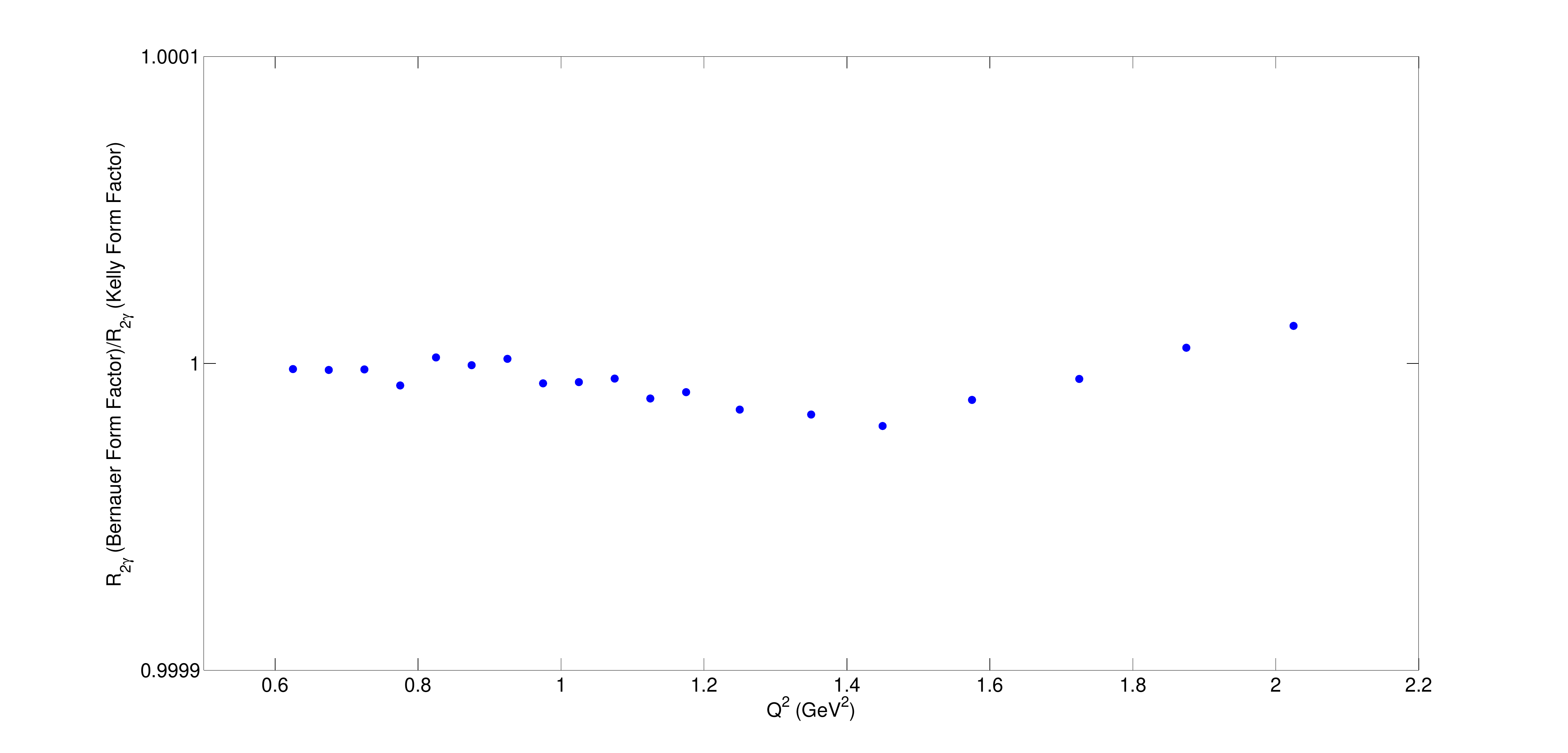}}
  \caption[Ratio of the $R_{2\gamma}$ results for simulation with the Kelly and Bernauer form factor models]{Ratio of the $R_{2\gamma}$ results for simulations conducted
  with the Kelly \cite{PhysRevC.66.065203} and Bernauer \cite{BerFFPhysRevC.90.015206} form factor models.  As expected, the total effect from the choice of form factor model is extremely
  small (much less than 0.01\% across the full acceptance).  The statistical uncertainties for each data point are suppressed since they are completely correlated in each form factor model.}
  \label{fig:ratkb}
  \end{figure}

While the choice of proton elastic form factor model significantly affects the extracted absolute \ep and \pp cross sections (Section \ref{sec:datasim}), it is expected in an analysis
of \ratio that effects from the form factor model used should be extremely small in the absence of any errors in the analysis that introduce a bias in the way the kinematics (in particular the value
of $Q^2$) for \ep and \pp events are computed.  In Figure \ref{fig:ratkb} presents the ratio of the \rtg results computed using the Kelly \cite{PhysRevC.66.065203} and Bernauer \cite{BerFFPhysRevC.90.015206}
form factor models.  Across the full acceptance, the deviation between the two is much less than 0.01\%, indicating that this is a minimal effect in the \rtg result and providing a sanity check
on the basic calculations of the reconstructed kinematics.  For reference, a similar ratio between the \rtg results using these form factors, the dipole form factor, or even treating the proton
as a point particle gives extremely similar results, providing confidence in the implementation of the form factors in the radiative event generator.

\subsection{Discussion of Current Total Systematic Uncertainty Estimate}

Due to the dominant effects of the detector acceptance and efficiency that contribute to the current systematic uncertainty estimate for this analysis, analogous effects that were
shown to be small in the 12\dg systematic uncertainty analysis (Section \ref{ss:12sys}), such as beam energy and position,  are not considered here, but will be considered as part of the final OLYMPUS results.
References \cite{schmidt} and \cite{russell} include discussion of additional systematic effects such as those from the ToF detector and track reconstruction efficiencies, but these effects were found
to be small relative to the aforementioned dominant effects.  For the preliminary results shown in Chapter \ref{Chap7}, a conservative 1.5\% bin-to-bin total systematic uncertainty
is quoted for the \rtg results in light of the fact that final analysis of the \rtg results and the systematic uncertainties was ongoing at the time of writing.
This, however, is likely a considerable overestimate in most bins (especially at low $Q^2$) and thus should only be considered in the preliminary context of these
results.  Systematic uncertainties due to the luminosity normalization and for the independent bin constructed using the results from the 12\dg system are separate from
this estimate and are discussed in detail in Sections \ref{sec:miesys} and \ref{ss:12sys}, respectively.

%% file: chap7.tex
% Chapter 7
%
% Discussion of Results
%

\chapter{Results and Discussion}
\label{Chap7}

With the methodology of the analysis in place, and in combination with the relative luminosity analyses presented in Chapter \ref{Chap5}, the OLYMPUS
result for $R_{2\gamma}$ (Equation \ref{eq:rat}) may be constructed.  This chapter presents the results of the analysis described in Chapter \ref{Chap6},
the result for the value of \ratio as measured in the 12\dg system using the MIE luminosity normalization, as well as a preliminary comparison of the different
$R_{2\gamma}$ analyses conducted using the OLYMPUS data.  As mentioned in Section \ref{sec:datasim}, further physics results from the OLYMPUS data pertaining
to the absolute elastic \pmp cross section and proton form factors may be part of future publications.
While such analyses are not part of this work, it is expected that subsequent publications will include results of this nature.

The following results represent $\sim$3.1 fb$^{-1}$ of data, approximately evenly split between the two leptons species.  This corresponded to a total yield (after background subtraction)
of approximately $4\cdot 10^6$ accepted elastic \pmp events for each lepton species.  The simulation dataset, generated according
to the principles described in Section \ref{sec:sim}, included approximately $2\cdot 10^9$ radiative elastic events, which ensured that the statistical uncertainty of
the simulated dataset was negligible in comparison to both the statistical and systematic uncertainties on the data in all regions of the OLYMPUS acceptance.  The strong statistical
power of the OLYMPUS dataset allows examination of the ratio in continuous fashion across the entire OLYMPUS acceptance.

The results that follow should be considered preliminary and subject to further investigations of the systematic uncertainties in the analysis.  It is expected, however, that these results
will accurately represent the effective trends in $R_{2\gamma}$ that will be presented in subsequent OLYMPUS publications, as indicated by the constraints on the systematic uncertainties discussed
in Section \ref{sec:mainsys} and the consistency of the independent analyses of the OLYMPUS data presented in Section \ref{sec:indana}.

  \section{Note on the Choice of the Luminosity Normalization}
  
  As discussed in Section \ref{sec:alllumi}, the 12\dg (Table \ref{tab:12results}) and MIE (Equation \ref{eq:mie}) species-relative luminosity measurements were found to be extremely consistent,
  both as a function of run number (matching the systematic variations relative to slow control, indicated by the lack of time-varying structure in Figure \ref{fig:12mie}) and in value (since the
  two determinations are well within the estimated uncertainties of each other).  Since the result for $R_{2\gamma}$ requires only a relative luminosity measurement, rather than an absolute
  integrated luminosity for each species, the result of the MIE method was chosen as the species-relative luminosity normalization for the results shown, which permits the use of the 12\dg system
  result as an additional kinematic point for the measurement of $R_{2\gamma}$.  Since the 12\dg elastic \pmp measurement exhibits systematic uncertainties that are largely independent from the 
  reconstruction of elastic events in the main spectrometer, this point provides a valuable cross check on the measurement $R_{2\gamma}$ in the rest of the acceptance as well as a precise and valuable
  check on the expectation that TPE effects go to zero at high-$\epsilon$.  The value of $R_{2\gamma}$ at $\theta\approx 12^\circ$ is discussed in Section \ref{sec:12TPE}.
  
  For possible future analyses of the elastic proton form factors and \pmp cross section, it will likely be necessary to use the measurement of the absolute luminosities for each species from the 12\dg
  system, since the MIE absolute luminosity analysis is subject to large uncertainties due to event selection cuts.  The strong consistency in the relative measurement between the two methods, as well
  as the agreement between the 12\dg absolute estimate and the simulation at forward angles (Section \ref{sec:datasim}), provides a high level of confidence in the absolute estimate for future analyses
  of the OLYMPUS data.
  
  \section{$R_{2\gamma}$ Results as a Function of $\epsilon$ and $Q^2$}
  \label{sec:thegoddamnresults}

\begin{figure}[thb!]
    \centerline{\includegraphics[width=1.15\textwidth]{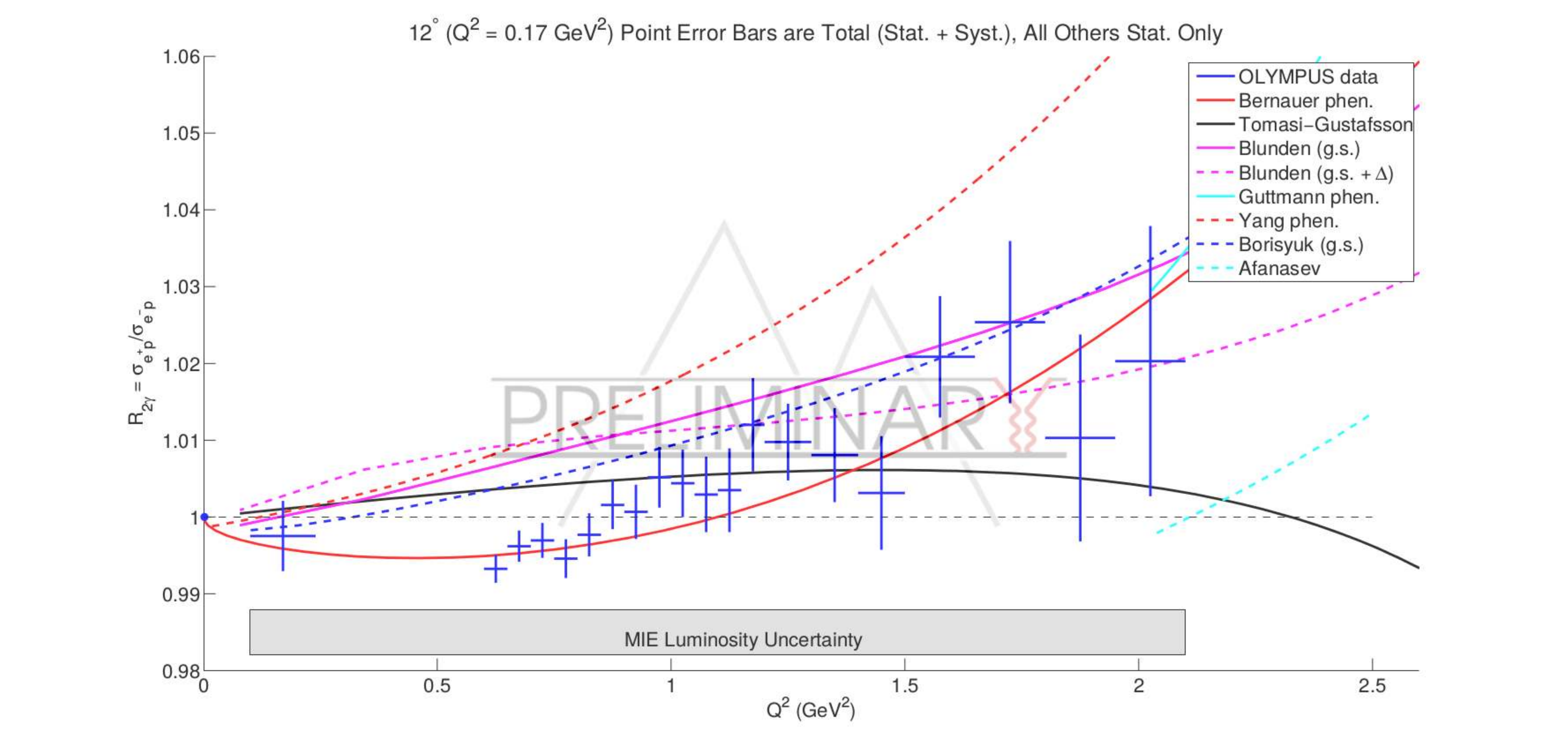}}
    \caption[Result for $R_{2\gamma}$ as a function of $Q^2$ (fine bins)]{Preliminary results for \rtg from the analysis presented in Chapter \ref{Chap6}, binned finely
    as a function of $Q^2$.  The error bars on the points represent the statistical uncertainty of the analysis, with the exception of the point at $Q^2=0.165$ GeV$^2$
    where the error bar represents the total statistical plus systematic uncertainty.  The gray box represents the total (statistical plus systematic) uncertainty
    from the MIE luminosity analysis, which applies as a constant normalization shift to all data points.  The data points in this plot are summarized in Table \ref{tab:finebins}.
    (Phenomenological models: \cite{BerFFPhysRevC.90.015206,Chen:2007ac, Guttmann:2010au}, theoretical models:
    \cite{Blunden:2003sp,Chen:2004tw,Afanasev:2005mp,Blunden:2005ew, Kondratyuk:2005kk, Borisyuk:2006fh,TomasiGustafsson:2009pw})}
    \label{fig:ratq2}
    \end{figure}

\begin{figure}[thb!]
    \centerline{\includegraphics[width=1.15\textwidth]{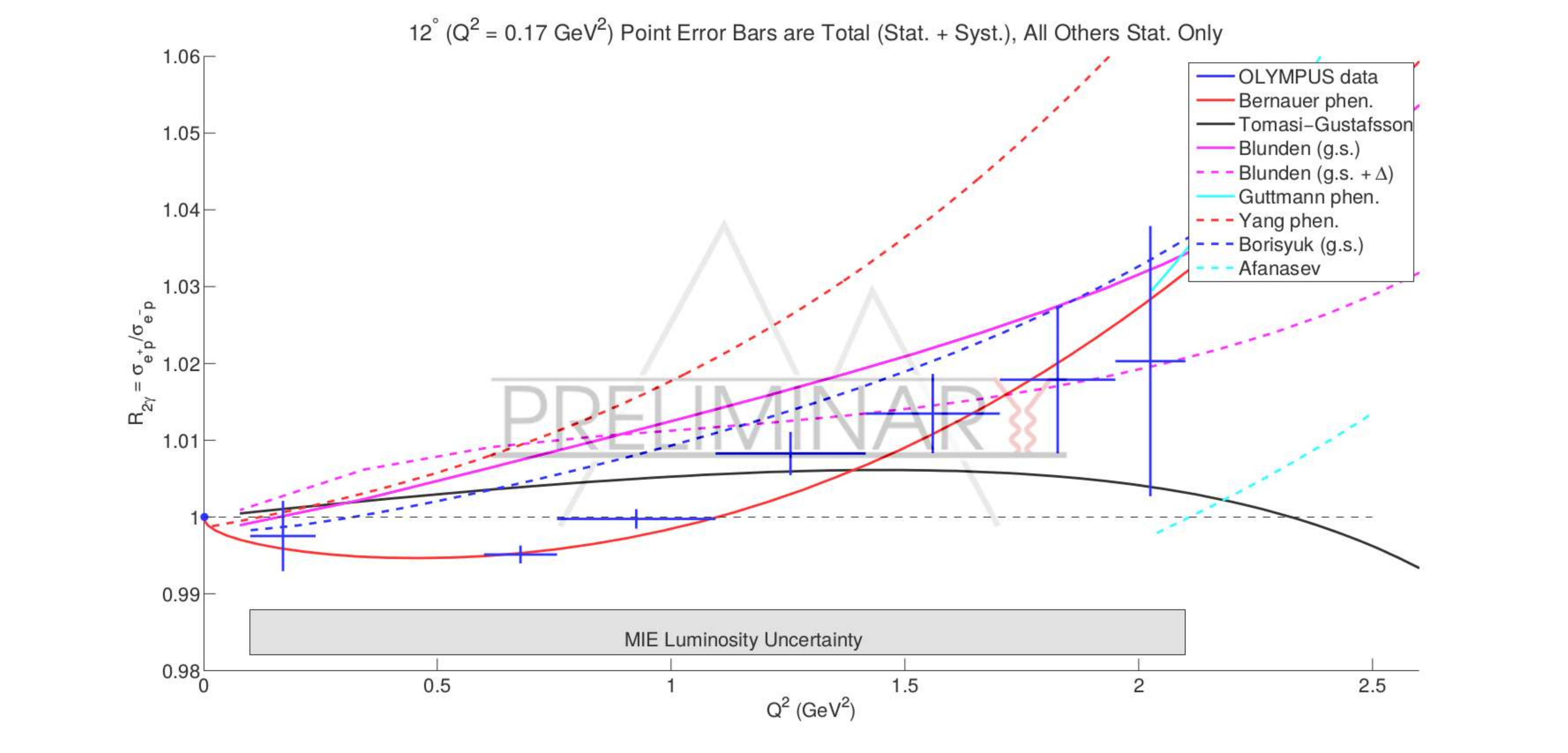}}
    \caption[Result for $R_{2\gamma}$ as a function of $Q^2$ (wide bins)]{Preliminary results for \rtg from the analysis presented in Chapter \ref{Chap6}, binned coarsely
    as a function of $Q^2$, approximately matching the bins represented by the data points in the projections of Figure \ref{fig:projections}.
    The error bars on the points represent the statistical uncertainty of the analysis, with the exception of the point at $Q^2=0.165$ GeV$^2$
    where the error bar represents the total statistical plus systematic uncertainty.  The gray box represents the total (statistical plus systematic) uncertainty
    from the MIE luminosity analysis, which applies as a constant normalization shift to all data points.  The data points in this plot are summarized in Table \ref{tab:widebins}.
    (Phenomenological models: \cite{BerFFPhysRevC.90.015206,Chen:2007ac, Guttmann:2010au}, theoretical models:
    \cite{Blunden:2003sp,Chen:2004tw,Afanasev:2005mp,Blunden:2005ew, Kondratyuk:2005kk, Borisyuk:2006fh,TomasiGustafsson:2009pw})}
    \label{fig:ratq2bb}
    \end{figure}

\begin{figure}[thb!]
    \centerline{\includegraphics[width=1.15\textwidth]{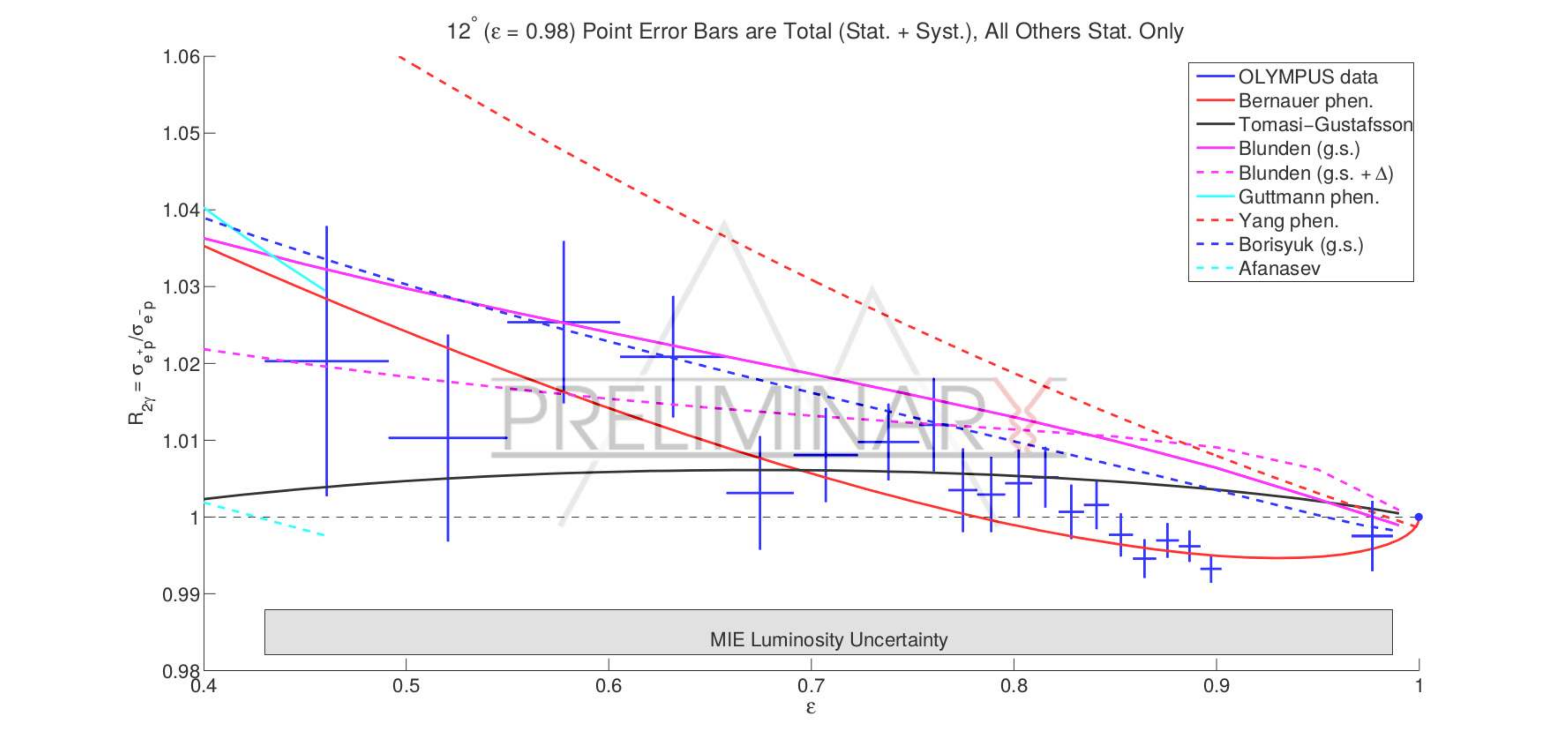}}
    \caption[Result for $R_{2\gamma}$ as a function of $\epsilon$ (fine bins)]{Preliminary results for \rtg from the analysis presented in Chapter \ref{Chap6}, binned finely
    as a function of $\epsilon$.  The error bars on the points represent the statistical uncertainty of the analysis, with the exception of the point at $\epsilon=0.98$
    where the error bar represents the total statistical plus systematic uncertainty.  The gray box represents the total (statistical plus systematic) uncertainty
    from the MIE luminosity analysis, which applies as a constant normalization shift to all data points.  The data points in this plot are summarized in Table \ref{tab:finebins}.
    (Phenomenological models: \cite{BerFFPhysRevC.90.015206,Chen:2007ac, Guttmann:2010au}, theoretical models:
    \cite{Blunden:2003sp,Chen:2004tw,Afanasev:2005mp,Blunden:2005ew, Kondratyuk:2005kk, Borisyuk:2006fh,TomasiGustafsson:2009pw})}
    \label{fig:rateps}
    \end{figure}

\begin{figure}[thb!]
    \centerline{\includegraphics[width=1.15\textwidth]{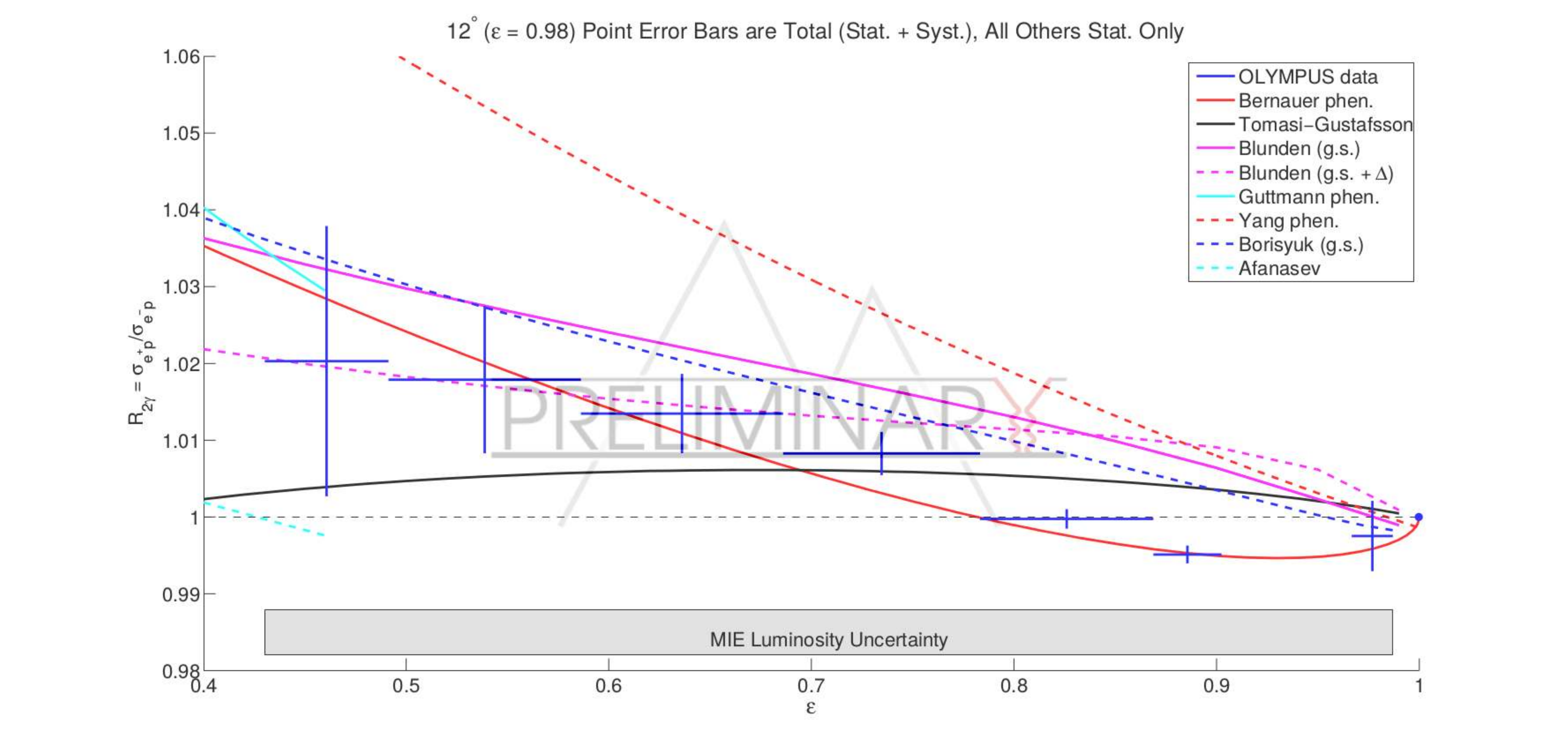}}
    \caption[Result for $R_{2\gamma}$ as a function of $\epsilon$ (wide bins)]{Preliminary results for \rtg from the analysis presented in Chapter \ref{Chap6}, binned coarsely
    as a function of $\epsilon$, approximately matching the bins represented by the data points in the projections of Figure \ref{fig:projections}.
    The error bars on the points represent the statistical uncertainty of the analysis, with the exception of the point at $\epsilon=0.98$
    where the error bar represents the total statistical plus systematic uncertainty.  The gray box represents the total (statistical plus systematic) uncertainty
    from the MIE luminosity analysis, which applies as a constant normalization shift to all data points.  The data points in this plot are summarized in Table \ref{tab:widebins}.
    (Phenomenological models: \cite{BerFFPhysRevC.90.015206,Chen:2007ac, Guttmann:2010au}, theoretical models:
    \cite{Blunden:2003sp,Chen:2004tw,Afanasev:2005mp,Blunden:2005ew, Kondratyuk:2005kk, Borisyuk:2006fh,TomasiGustafsson:2009pw})}
    \label{fig:ratepsbb}
    \end{figure}

    Figures \ref{fig:ratq2}, \ref{fig:ratq2bb}, \ref{fig:rateps}, and \ref{fig:ratepsbb} present the results for \rtg as found by the analysis
    described in Chapter 6 as a function of $Q^2$ and $\epsilon$ in two different binnings for each.  In each figure, the point at $\theta=12^\circ$
    ($Q^2=0.165$ GeV$^2$, $\epsilon=0.98$) is shown with its full systematic plus statistical uncertainty estimate (Section \ref{sec:12TPE}) while the
    error bars on the other points are statistical only.  As discussed in Section \ref{sec:mainsys}, the total systematic uncertainty for each point for these
    results may be considered on the order of 1.5\%, although for many bins this is likely an overestimate.  The magnitude of the total uncertainty from the MIE
    luminosity determination, which would apply as a common shift to all data points simultaneously, is represented by the gray box above the horizontal axis of each figure.
    Additionally, each figure presents the theoretical \cite{BerFFPhysRevC.90.015206,Chen:2007ac, Guttmann:2010au} and phenomenological
    \cite{Blunden:2003sp,Chen:2004tw,Afanasev:2005mp,Blunden:2005ew, Kondratyuk:2005kk, Borisyuk:2006fh,TomasiGustafsson:2009pw}
    models initially presented with the OLYMPUS projections in Figure \ref{fig:projections}.  The results and models were computed using the Maximon and Tjon prescription for 
    radiative corrections \cite{MaximonPhysRevC.62.054320} and the methodology described in Section \ref{sec:radgen}. Tables \ref{tab:finebins} and \ref{tab:widebins} list
    the values presented by the plots in each binning, as well as the average $Q^2$ and $\epsilon$ for each bin.
  
    As previously noted, these results are preliminary and are subject to further analysis, particularly relating to the final estimate of the systematic
    uncertainty in each bin.  As discussed in Section \ref{sec:indana}, it is highly likely that the final OLYMPUS results will strongly resemble the results
    presented here, but at the time of writing the finalization of the OLYMPUS results for publication was still underway.
  
    \begin{table}[thb!]
    \begin{center}
    \begin{tabular}{ccccc}
    \hline
    $\left<Q^2\right>$ (GeV$^2$) & $\left<\epsilon\right>$ & \rtg & Stat. Uncertainty & Syst. Uncertainty \\
    \hline\hline 
    0.165 & 0.978 & 0.9975 & $\pm0.0003$ & $\pm0.0046$ \\
    \hline 
    0.624 & 0.898 & 0.9932 & $\pm0.0018$ & $\sim\pm0.0150$ \\
    \hline
    0.674 & 0.887 & 0.9962 & $\pm0.0020$ & $\sim\pm0.0150$ \\
    \hline
    0.724 & 0.876 & 0.9970 & $\pm0.0023$ & $\sim\pm0.0150$ \\
    \hline
    0.774 & 0.865 & 0.9946 & $\pm0.0025$ & $\sim\pm0.0150$ \\
    \hline
    0.824 & 0.853 & 0.9977 & $\pm0.0028$ & $\sim\pm0.0150$ \\
    \hline
    0.874 & 0.841 & 1.0016 & $\pm0.0032$ & $\sim\pm0.0150$ \\
    \hline
    0.924 & 0.829 & 1.0007 & $\pm0.0036$ & $\sim\pm0.0150$ \\
    \hline
    0.974 & 0.816 & 1.0052 & $\pm0.0040$ & $\sim\pm0.0150$ \\
    \hline
    1.024 & 0.803 & 1.0044 & $\pm0.0044$ & $\sim\pm0.0150$ \\
    \hline
    1.074 & 0.789 & 1.0029 & $\pm0.0049$ & $\sim\pm0.0150$ \\
    \hline
    1.124 & 0.775 & 1.0035 & $\pm0.0055$ & $\sim\pm0.0150$ \\
    \hline
    1.174 & 0.761 & 1.0120 & $\pm0.0061$ & $\sim\pm0.0150$ \\
    \hline
    1.246 & 0.739 & 1.0098 & $\pm0.0050$ & $\sim\pm0.0150$ \\
    \hline
    1.347 & 0.708 & 1.0081 & $\pm0.0062$ & $\sim\pm0.0150$ \\
    \hline
    1.447 & 0.676 & 1.0031 & $\pm0.0074$ & $\sim\pm0.0150$ \\
    \hline
    1.568 & 0.635 & 1.0209 & $\pm0.0079$ & $\sim\pm0.0150$ \\
    \hline
    1.718 & 0.581 & 1.0254 & $\pm0.0106$ & $\sim\pm0.0150$ \\
    \hline
    1.868 & 0.524 & 1.0103 & $\pm0.0135$ & $\sim\pm0.0150$ \\
    \hline
    2.038 & 0.456 & 1.0203 & $\pm0.0177$ & $\sim\pm0.0150$ \\
    \hline
    \end{tabular}

    \end{center}
    \caption[\rtg results in the finer binning (Figures \ref{fig:ratq2} and \ref{fig:rateps})]{Values of the preliminary \rtg results for the finer binning
    presented in Figures \ref{fig:ratq2} and \ref{fig:rateps}, including the mean $Q^2$ and $\epsilon$ of all events in each bin.}
    \label{tab:finebins}
    \end{table}
    
    \begin{table}[thb!]
    \begin{center}
    \begin{tabular}{ccccc}
    \hline
    $\left<Q^2\right>$ (GeV$^2$) & $\left<\epsilon\right>$ & \rtg & Stat. Uncertainty & Syst. Uncertainty \\
    \hline\hline 
    0.165 & 0.978 & 0.9975 & $\pm0.0003$ & $\pm0.0046$ \\
    \hline
    0.666 & 0.889 & 0.9951 & $\pm0.0012$ & $\sim\pm0.0150$ \\
    \hline
    0.879 & 0.840 & 0.9997 & $\pm0.0013$ & $\sim\pm0.0150$ \\
    \hline
    1.220 & 0.747 & 1.0083 & $\pm0.0028$ & $\sim\pm0.0150$ \\
    \hline
    1.534 & 0.647 & 1.0135 & $\pm0.0052$ & $\sim\pm0.0150$ \\
    \hline
    1.809 & 0.547 & 1.0179 & $\pm0.0097$ & $\sim\pm0.0150$ \\
    \hline
    2.039 & 0.456 & 1.0203 & $\pm0.0177$ & $\sim\pm0.0150$ \\
    \hline
    2.238 & 0.372 & 0.9813 & $\pm0.0184$ & $\sim\pm0.0150$ \\
    \hline
    \end{tabular}

    \end{center}
    \caption[\rtg results in the wider binning (Figures \ref{fig:ratq2bb} and \ref{fig:ratepsbb})]{Values of the preliminary \rtg results for the wider binning
    presented in Figures \ref{fig:ratq2bb} and \ref{fig:ratepsbb}, including the mean $Q^2$ and $\epsilon$ of all events in each bin.}
    \label{tab:widebins}
    \end{table}
    
    \subsection{The \ratio Ratio at $\epsilon\approx 0.98$ ($\theta\approx 12^\circ$)}
    \label{sec:12TPE}
    
    While the 12\dg measurement of \ratio may be taken as a normalization point for the relative luminosity of electron and positron
    data collected by the experiment or used in conjunction with the MIE measurement to provide a combined high-precision estimate,
    it may also be taken as an additional measurement of $R_{2\gamma}$ at $\epsilon \approx 0.98$, $Q^2 \approx 0.165$ GeV$^2$ if the MIE
    is used as an independent measure of the relative luminosity.  This is especially valuable due to the fact that the MIE measurement, while
    depending in part on elastic \pmp scattering, was at a much more forward angle ($\epsilon \approx 0.99975$, $Q^2 \approx 0.002$ GeV$^2$)
    where the expectation of $R_{2\gamma}=1$ is extremely strong (physics considerations demand that $R_{2\gamma}=1$ at $\epsilon=1$).  Taking this
    approach and utilizing the results of Sections \ref{sec:12res} and \ref{sec:mieres}, the result is:
    \begin{equation}
    R_{2\gamma}\left(\epsilon = 0.98,Q^2 = 0.165\:\text{GeV}^2 \right) = 0.9975 \pm 0.0010\:(\text{stat.}) \pm 0.0053\:(\text{syst.}),
    \end{equation}
    where the uncertainties are the result of adding the uncertainties of the individual measurements in quadrature, since the uncertainties of
    the two measurements are essentially completely independent.  The measurement is quite valuable for several reasons:
    \begin{enumerate}
    \item it provides a consistency check on the overall scale of $R_{2\gamma}$ as measured in the main spectrometer by providing
	  a measurement from an independent system with largely independent systematic uncertainties,
    \item it constrains the behavior of $R_{2\gamma}$ at high $\epsilon$,
    \item it provides some discriminating power between different models for $R_{2\gamma}$ in the high-$\epsilon$ region, and
    \item it offers insight into the offset that should be applied to the luminosity normalization points used for the VEPP-3 \rtg results so
	  as to allow direct comparison of those data to models and other experiments.
    \end{enumerate}
    This measurement constitutes a new benchmark for \ratio in the forward scattering ($\theta\approx 10^\circ$) region, providing a previously
    unachieved level of total uncertainty of $\sim$0.54\%.
  
  \section{Comparison of Independent Analyses}
  \label{sec:indana}
  
  At the time of writing, three independent $R_{2\gamma}$ analyses of the OLYMPUS data were available: the two analyses described in References \cite{schmidt} and 
  \cite{russell} and the analysis described in this work.  Each analysis made use of the same
  sets of tracked and simulated data, but applied different approaches to the analysis.  Differences in the analyses included the approaches to particle identification,
  application of cuts in different kinematic variables, varying stringency in shared kinematic variable cuts, different background subtraction methods and models, as well
  as different orderings of analysis components.
  
\begin{figure}[thb!]
  \centerline{\includegraphics[width=1.15\textwidth]{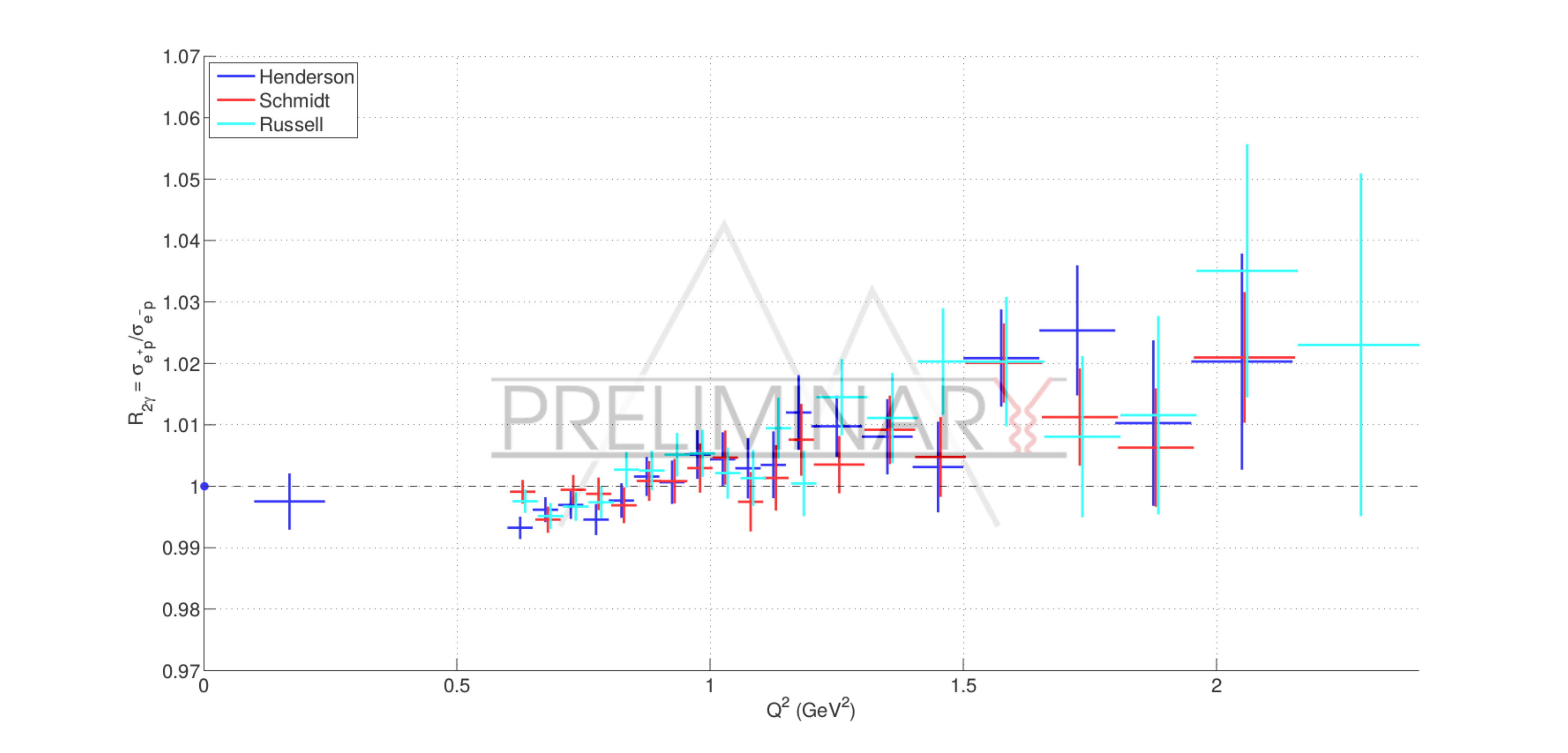}}
  \caption[Comparison of three $R_{2\gamma}$ analyses (fine bins)]{Results for $R_{2\gamma}$ as a function of $Q^2$ in the binning presented in Figure \ref{fig:ratq2} as found by three different
  analyses of the OLYMPUS dataset: Schmidt \cite{schmidt}, Russell \cite{russell}, and Henderson (this work).  Each analysis used the same data and simulation but
  different approaches to particle identification, elastic event selection, background subtraction, etc.  In general, the statistical uncertainties shown for the separate analyses are highly, but
  not entirely, correlated due to differences in the total number of events accepted by each analysis.}
  \label{fig:comp}
  \end{figure}

\begin{figure}[thb!]
  \centerline{\includegraphics[width=1.15\textwidth]{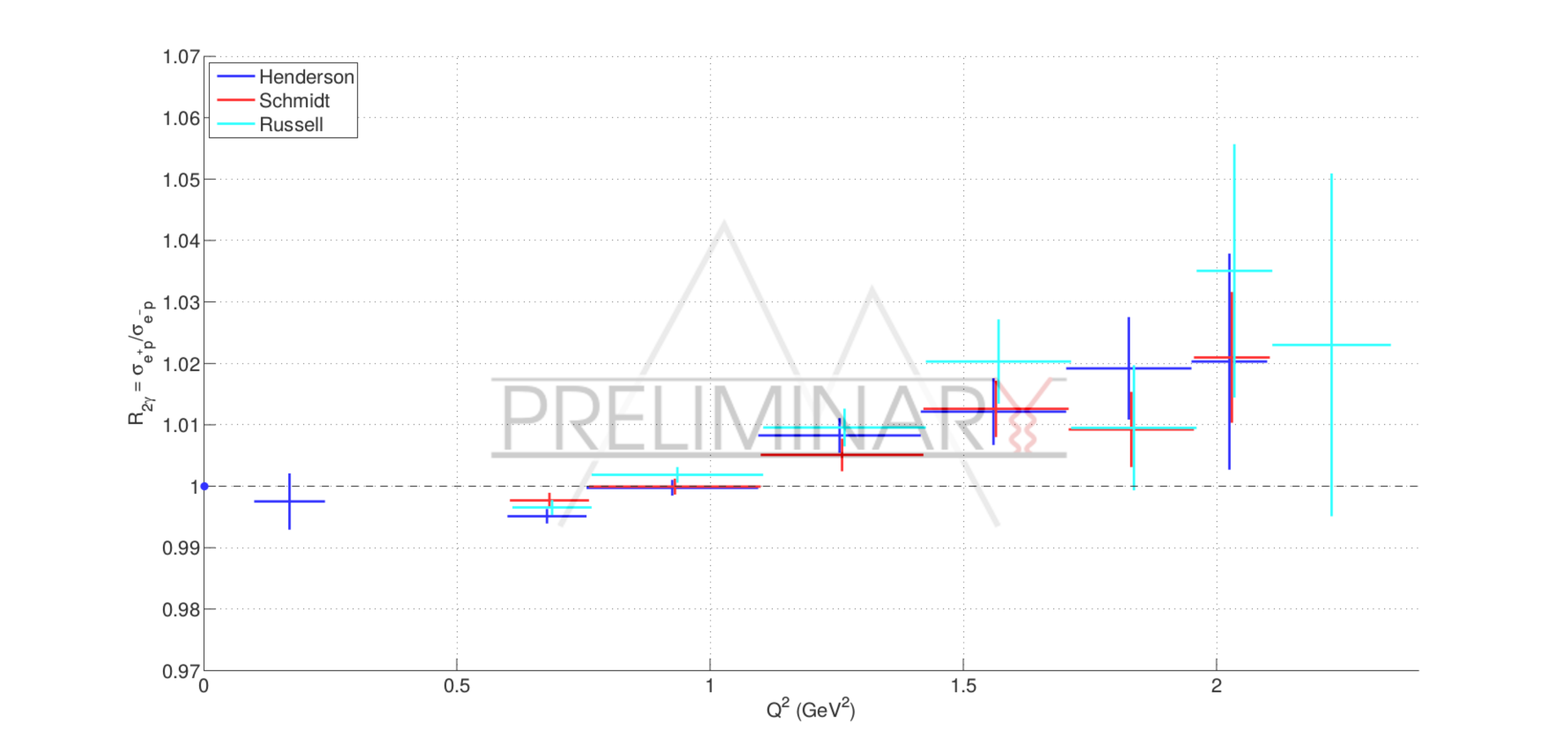}}
  \caption[Comparison of three $R_{2\gamma}$ analyses (wide bins)]{Results for $R_{2\gamma}$ as a function of $Q^2$ in the binning presented in Figure \ref{fig:ratq2bb} as found by three different
  analyses of the OLYMPUS dataset: Schmidt \cite{schmidt}, Russell \cite{russell}, and Henderson (this work).  Each analysis used the same data and simulation but
  different approaches to particle identification, elastic event selection, background subtraction, etc.  In general, the statistical uncertainties shown for the separate analyses are highly, but
  not entirely, correlated due to differences in the total number of events accepted by each analysis.}
  \label{fig:compbb}
  \end{figure}

  Figures \ref{fig:comp} and \ref{fig:compbb} show the results for \rtg from the three analyses plotted together as a function of $Q^2$ in the same binning schemes as Figures
  \ref{fig:ratq2} and \ref{fig:ratq2bb} respectively.  The three analyses agree extremely well in terms of the general trend as a function of $Q^2$, in that it rises monotonically from
  a few tenths of a percent below unity at $Q^2=0.6$ GeV$^2$ to 2-3\% above one at the upper end of the $Q^2$ acceptance.  The point-to-point variations are typically on the order of a
  few tenths of a percent, which is consistent with the systematic uncertainty associated with choices of kinematic cuts and background subtraction models discussed in Section \ref{sec:mainsys}.
  Certain deviations between the analyses that are larger (such as the difference between the Henderson analysis and the other two in the $Q^2=1.725$ GeV$^2$ bin) are likely associated with
  localized increases in systematic uncertainties arising from analysis choices, such as those shown for the analysis of this work in Figure \ref{fig:cutsys}.  At the time of writing, several
  additional analyses were underway to provide further cross-checks of the final results prior to publication, but in general the analyses show strong consistency and provide a reasonable
  level of confidence in the general features of the OLYMPUS \rtg result.

\section{Implications for Two-Photon Exchange and \\ the Proton Form Factors}
  
The results of the three TPE experiments (OLYMPUS, VEPP-3 \cite{vepp3PhysRevLett.114.062005}, and CLAS \cite{PhysRevLett.114.062003,ass}) consistently indicate the presence of a contribution
from TPE to elastic \pmp scattering, as each shows a generally increasing trend in the value of \ratio with increasing $Q^2$/decreasing $\epsilon$.  While the experiments differ in kinematic coverage,
each provides results that are consistent with an effect of $\lesssim$3\% at $\epsilon\approx 0.4$.  The general magnitude of this effect is consistent with a number of the theoretical
and phenomenological models (as shown in the OLYMPUS results figures), while models predicting larger effects (such as the Yang phenomenological model \cite{Chen:2007ac}) are well excluded.  The addition of the 
high-statistics OLYMPUS data begins to place stronger constraints on the more similar models, as it provides the first indications of the shape of the evolution of \rtg as a function
of $Q^2$ or $\epsilon$.  In particular, the OLYMPUS data strongly suggests that \rtg indeed passes below unity at $\epsilon\gtrsim 0.85$ and $Q^2\lesssim 0.6$ GeV$^2$ as initially suggested by the CLAS
results (see Figure \ref{fig:crap}).

While the OLYMPUS data appear to agree well with the Bernauer phenomenological model \cite{BerFFPhysRevC.90.015206}, which would suggest that the form factor discrepancy may be explained by
TPE contributions to elastic \pmp scattering, the assumptions underlying the model as well as the relative consistency of the combined experimental data with the general trends and magnitudes
of other models remain a question that must be examined carefully by the hadronic physics community prior to drawing any definitive conclusions.  Since most theoretical calculations do not predict
$R_{2\gamma}<1$ at high $\epsilon$, as indicated by the experimental data, this behavior in particular will need to be considered as models and calculations for the proton form factors are
assessed.  As noted, however, the three experiments
provide a clear indication of the presence of TPE in \pmp scattering, and the strength of the OLYMPUS data should provide strong guidance in discriminating between proton form factor and TPE models
going forward.

\section{Outlook}

OLYMPUS has measured the \ratio ratio to high statistical precision over the kinematic range $(0.4 \leq \epsilon \leq 0.9)$, $(0.6 \leq Q^2 \leq 2.2)$ GeV$^2/c^2$, with systematic
uncertainties on the order of 1-2\%. The OLYMPUS data indicates that \ratio is below unity at high-$\epsilon$ and then rises monotonically by several percent across the range of the data.
The systematic uncertainties are currently dominated by knowledge of the drift chamber acceptance and efficiency, and will likely be reduced as further studies are completed prior
to the publication of the final \rtg results.  Additionally, the OLYMPUS 12\dg telescope tracking system, in combination with the multi-interaction event luminosity determination, has provided
an extremely precise measurement (0.56\% total uncertainty) of \rtg at high $\epsilon$/small $\theta$.  This result provides a valuable normalization point for the VEPP-3 TPE
experiment \cite{vepp3PhysRevLett.114.062005}, providing guidance for the absolute scale of their luminosity normalization point at similar kinematics.

While the three TPE experiments indicate that there is a significant contribution from TPE to elastic \pmp scattering, whether the effect is large enough to fully account for the 
$\mu_pG_E/G_M$ ratio discrepancy will likely be a topic of considerable debate in the hadronic physics community.  In particular, the validity of different models in the kinematic
ranges relevant to the experimental data will need to be carefully considered before drawing definitive conclusions.  While the measurement of a very large TPE effect ($\sim$6\% at
$\epsilon=0.4$) would have likely provided a complete and convincing resolution of the discrepancy, there may be value in an additional measurement of \rtg at higher $Q^2$ (where the form factor
ratio discrepancy is more significant) to provide final confirmation of the trends suggested by the TPE experiments below $Q^2=2$ GeV$^2$.  Such an experiment, however, would be extremely challenging
due to the reduced elastic cross section at higher $Q^2$ in combination with increased cross sections for background processes (which differ between \ep and \pp scattering).

Future OLYMPUS publications will include the \rtg result (after final analyses of the systematic uncertainties are conducted) and likely also limited results on the absolute cross section and
proton elastic form factors, as discussed in Section \ref{sec:datasim}.  The consistency between the analyses presented in Reference \cite{schmidt}, Reference \cite{russell}, and this work
suggest that the final result from OLYMPUS will strongly resemble the preliminary results shown, but further analyses will be conducted to provide greater confidence in the OLYMPUS results.
The OLYMPUS results provide the most statistically significant measurement of \rtg over a wide kinematic range, including extending measurements to higher $Q^2$ values where the form factor
discrepancy is more significant.  The data will be valuable in distinguishing between different phenomenological and theoretical TPE models, providing insight into the structure of the proton
and the possible causes of the form factor discrepancy.

%% file: appa.tex
% Appendix A
%
% Description of the Event Display, intended to highlight the fact
% that is was a major project, a useful tool for the analysis, and
% an adaptable program for future experiments
%

\chapter{The OLYMPUS 3D Event Display}
\label{chap:ed}

As a tool for the development of the OLYMPUS analysis, a three-dimensional event display was created for the OLYMPUS analysis framework that permitted the visualization of
the detector geometry, the results of hit and track reconstruction, and the results of simulated events.  The program functions as a standalone visualization tool to allow
adaptation for other purposes, but is useful when integrated with an analysis framework (such as the one developed by J.C. Bernauer for OLYMPUS) to permit the simultaneous analysis of the events
being visualized.  The event display was extremely useful in diagnosing issues both with data reconstruction and simulation by allowing direct inspection of events, especially those
events for which a particular procedure was not functioning.

\begin{figure}[thb!]
\centerline{\includegraphics[width=1.0\textwidth]{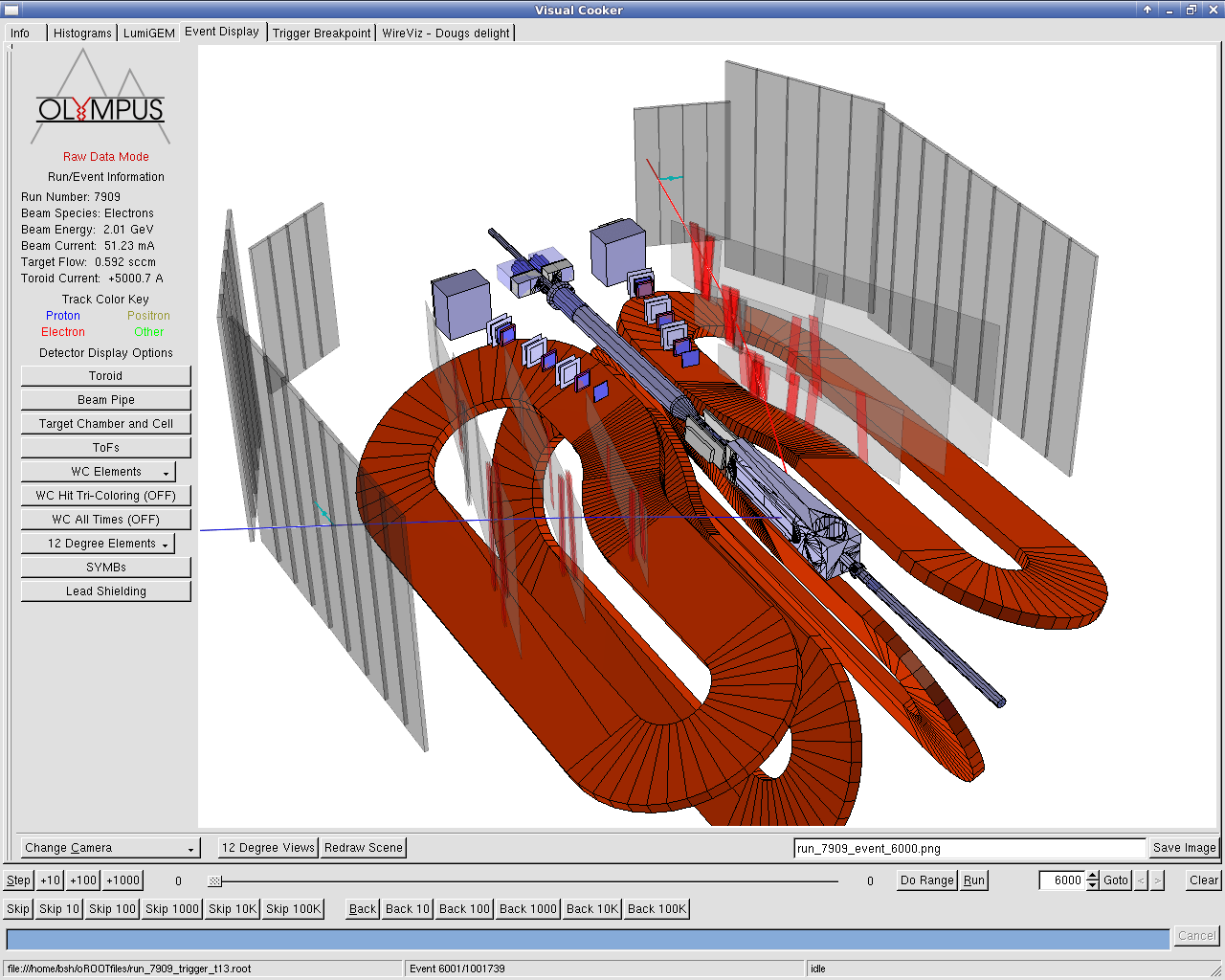}}
\caption[User interface of the OLYMPUS 3D Event Display]{The user interface of the OLYMPUS 3D event display, running in conjunction with the OLYMPUS
analysis framework for reconstruction of events from data.  The main display area can be used to visualize the geometry of the detector, hits in the various detector elements,
and the results of track reconstruction. The controls surrounding it provide options for changing visualization options, stepping through events, and the ability to save the display
window as an image.}
\label{fig:ed}
\end{figure}

The event display utilizes elements from the ROOT Event Visualization Environment originally developed for the ALICE experiment \cite{Brun1997,alice} and OpenGL to produce the detector
visualization \cite{Shreiner:2009:OPG:1696492}.  The detector geometry is imported via the GDML format discussed in Section \ref{sec:detmod}\cite{gdml}, and thus the display is capable of
visualizing any geometry in that format.
The display provides functionality to control which elements of the detector are displayed, how and which elements of the data and reconstruction
are displayed, completely control the camera view of the detector and to switch between orthographic and perspective viewing modes, and to save images of the display.  The display also provides
information relevant to the events displayed and inherits functionality from the OLYMPUS analysis framework that allows direct access to specific events in a file and viewing of analysis
histograms as they are filled by the events being displayed.  Figure \ref{fig:ed} shows an example of the interface, displaying the detector and a reconstructed elastic \ep event.

The source code for the OLYMPUS Event Display (written in C++ using ROOT libraries) is available from the author for academic purposes, and is easily adaptable for other
particle physics experiments.  Notably, Geant4 includes the functionality to export geometries written in the Geant4 geometry framework to GDML, providing immediate compatibility
with the OLYMPUS event display \cite{Agostinelli:2002hh}.

%% file: appb.tex
\chapter{Histograms of Data Events by the Quantities used for the Elastic Event Selection}
\label{chap:kincuts}

This appendix provides 2D histograms of events after initial pair selection for each of the kinematic cut parameters used in the
elastic pair selection described in Section \ref{sec:pairsel} and the reconstructed lepton angle $\theta_{e^\pm}$.
Two histograms are presented for each parameter: one corresponding to elastic \ep event selection and the other to \pp selection.
Each histogram has been normalized by the Rosenbluth cross section (using dipole form factors for simplicity) as a function of $\theta_{e^\pm}$ to make the scale of
the counts across the entire histogram more uniform and thus make the resolution of the reconstructed parameters and shape of the background at back angles
visible at all $\theta_{e^\pm}$.  In general, the corresponding simulation histograms were effectively identical up to the removal of the background and slightly
better resolutions in some parameters.

%  \item Vertex time from ToF meantime and track path length correlation (corrected for vertex position): $\left|\Delta t\right| = \left|t_p-t_{e^\pm}\right| < 5$ ns (Figures \ref{fig:cut1e} and \ref{fig:cut1p})
%  
%  \item Vertex $z$ correlation: $\left|\Delta z\right| = \left|z_p-z_{e^\pm}\right| < 175$ mm (Figures \ref{fig:cut2e} and \ref{fig:cut2p})
%  
%  \item Electron-proton elastic angle correlation: $ \left|\theta_p - \theta_{p,\text{elas}}(\theta_{e^\pm})\right| < 7^\circ$ (Figures \ref{fig:cut3e} and \ref{fig:cut3p})
%  
%  \item Beam energy reconstructed from the track momenta: $\left| E_{\text{beam},p}-E_\text{beam}\right| < 1000$ MeV (Figures \ref{fig:cut4e} and \ref{fig:cut4p})
%  
%  \item Beam energy reconstructed from the $\theta$ of both tracks assuming elastic kinematics: $ \left|E_{\text{beam},\theta}-E_\text{beam}\right| < 350$ MeV (Figures \ref{fig:cut5e} and \ref{fig:cut5p})
%  
%  \item Single-arm missing energy of the lepton over the square of the energy as computed by the expected
%           elastic energy from the reconstructed $\theta_{e^\pm}$: $\Delta E'_\theta/E'^2 < 0.0048$ MeV$^{-1}$ (Figures \ref{fig:cut6e} and \ref{fig:cut6p})
%           
%  \item Longitudinal (beam direction) momentum balance: $p_{z,p} + p_{z,e^\pm} - p_\text{beam} > -500 $ MeV (Figures \ref{fig:cut7e} and \ref{fig:cut7p})
%  
%  \item Coplanarity $\left|\Delta\phi-180^\circ\right| = \left|\phi_\text{right}-\phi_\text{left}-180^\circ\right|<7.5^\circ$ (Figures \ref{fig:cut8e} and \ref{fig:cut8p})

    \begin{figure}[thb!]
    \centerline{\includegraphics[width=1.15\textwidth]{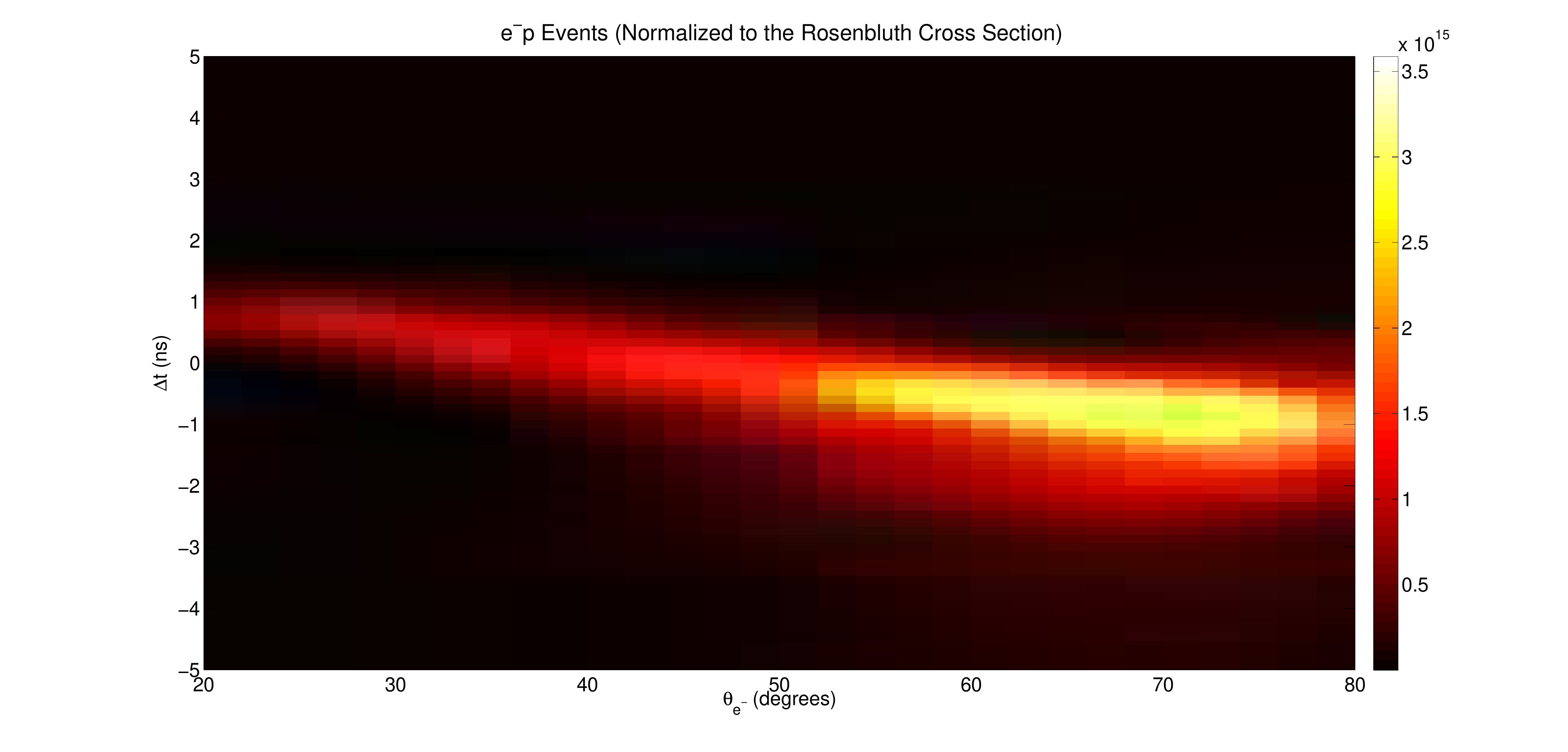}}
    \caption[Histogram of the $\Delta t$ cut parameter for the \ep initial pair selection]{Histogram of the $\Delta t$ cut parameter (uncorrected for vertex position) for the \ep initial pair selection (before background subtraction)
    as a function of the electron
    scattering angle, normalized to the Rosenbluth cross section.}
    \label{fig:cut1e}
    \end{figure}
    
    \begin{figure}[thb!]
    \centerline{\includegraphics[width=1.15\textwidth]{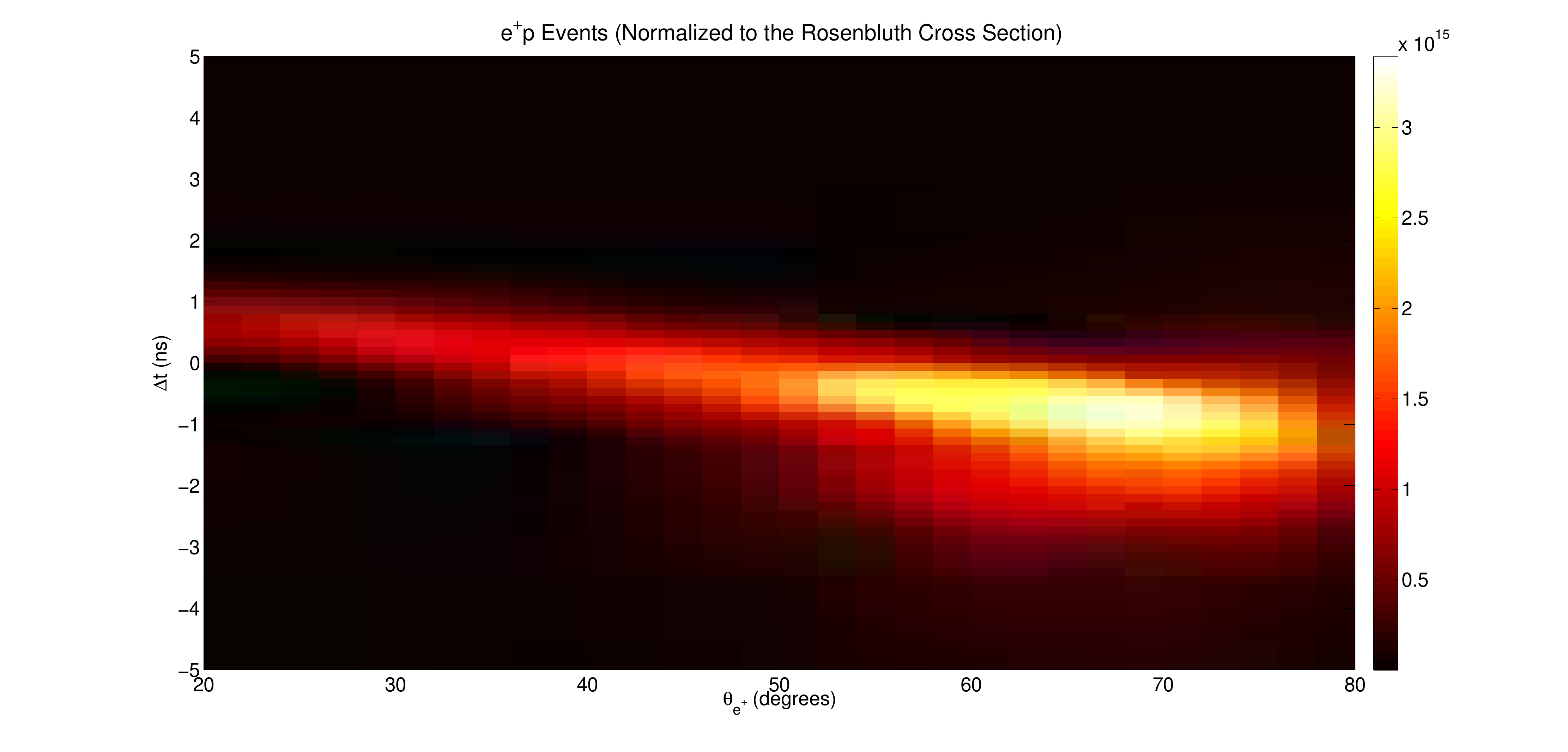}}
    \caption[Histogram of the $\Delta t$ cut parameter for the \pp initial pair selection]{Histogram of the $\Delta t$ cut parameter (uncorrected for vertex position) for the \pp initial pair selection (before background subtraction)
    as a function of the positron
    scattering angle, normalized to the Rosenbluth cross section.}
    \label{fig:cut1p}
    \end{figure}
    
    \begin{figure}[thb!]
    \centerline{\includegraphics[width=1.15\textwidth]{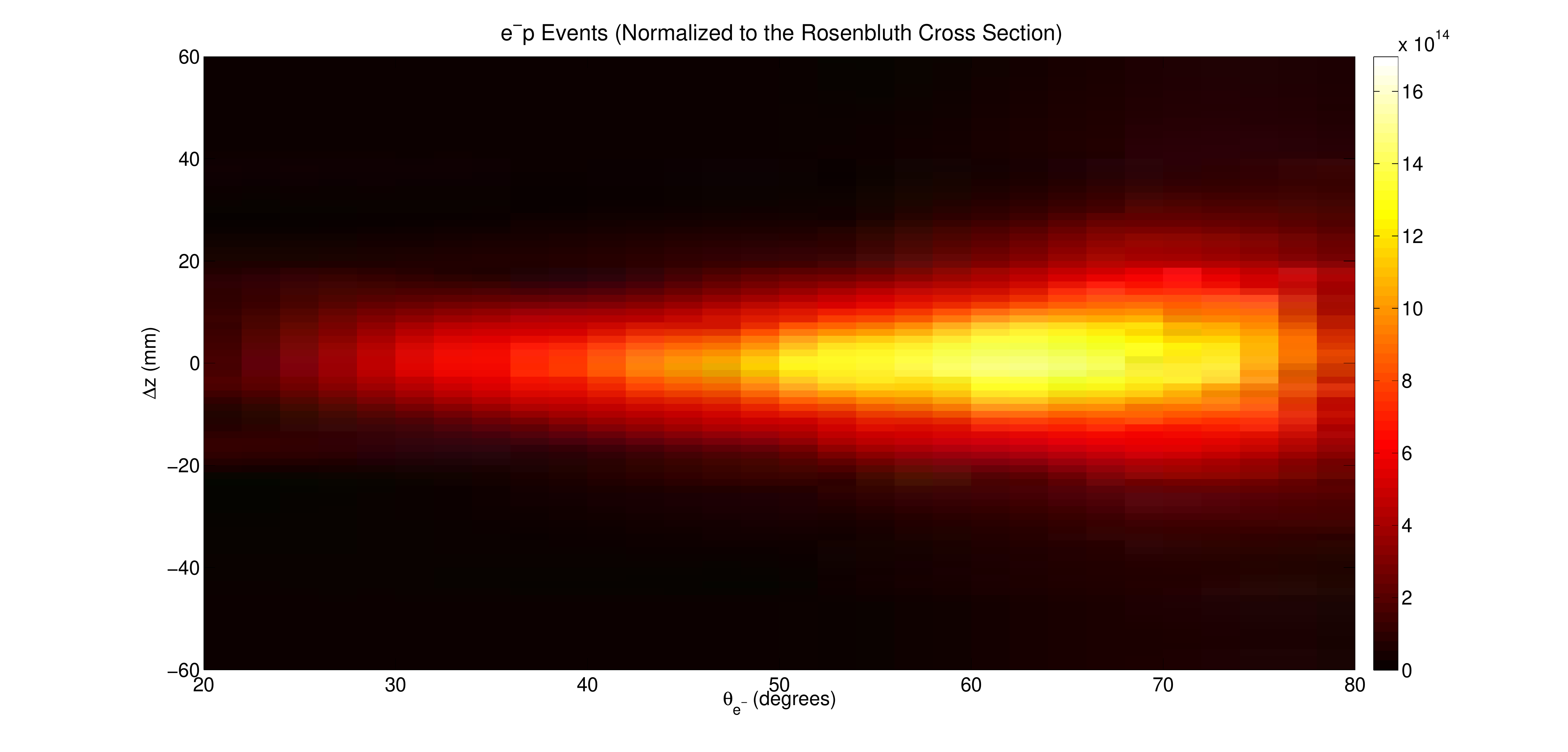}}
    \caption[Histogram of the $\Delta z$ cut parameter for the \ep initial pair selection]{Histogram of the $\Delta z$ cut parameter for the \ep initial pair selection (before background subtraction)
    as a function of the electron
    scattering angle, normalized to the Rosenbluth cross section.}
    \label{fig:cut2e}
    \end{figure}
    
    \begin{figure}[thb!]
    \centerline{\includegraphics[width=1.15\textwidth]{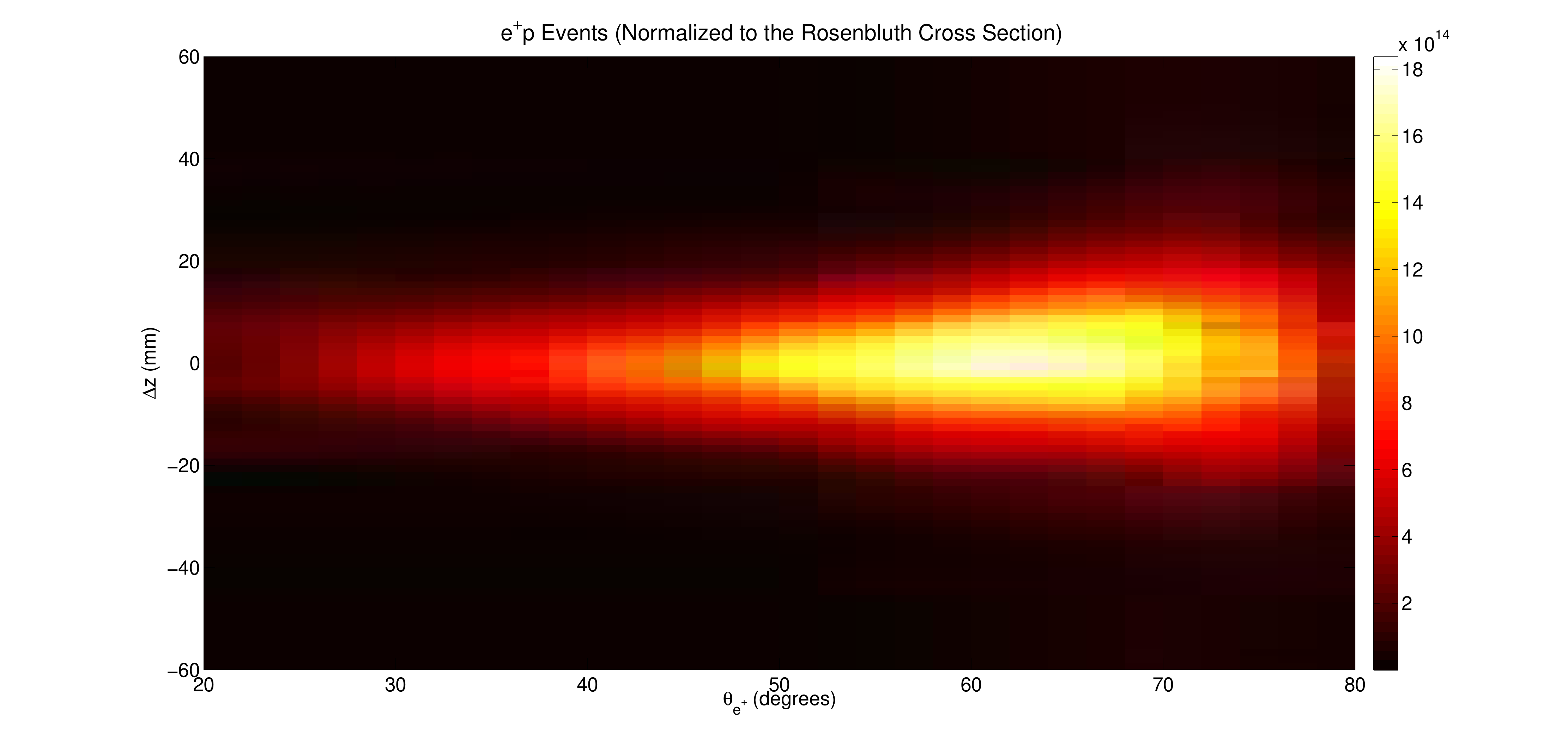}}
    \caption[Histogram of the $\Delta z$ cut parameter for the \pp initial pair selection]{Histogram of the $\Delta z$ cut parameter for the \pp initial pair selection (before background subtraction)
    as a function of the positron
    scattering angle, normalized to the Rosenbluth cross section.}
    \label{fig:cut2p}
    \end{figure}
    
    \begin{figure}[thb!]
    \centerline{\includegraphics[width=1.15\textwidth]{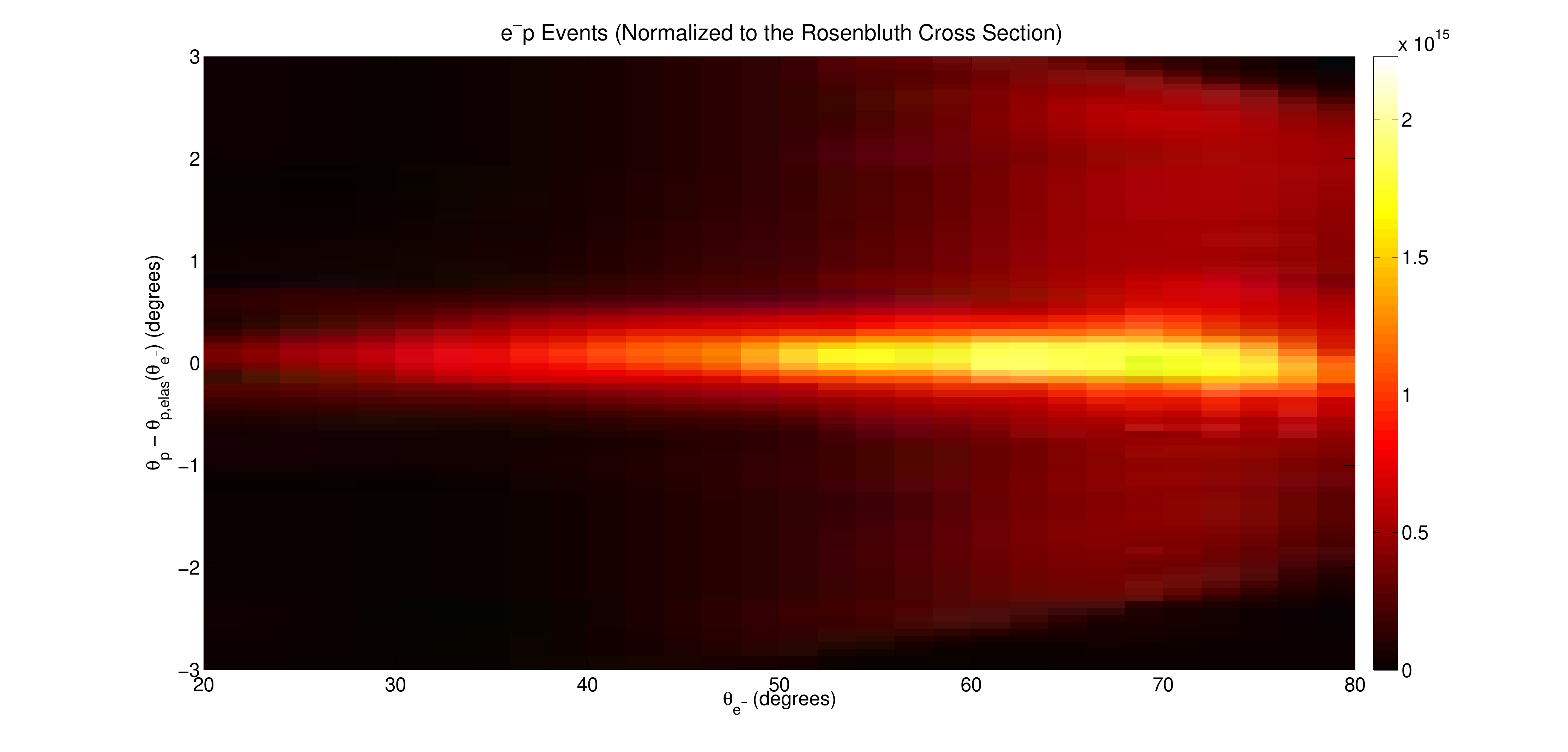}}
    \caption[Histogram of the elastic angle correlation cut parameter for the \ep initial pair selection]{Histogram of the elastic angle correlation cut parameter for the \ep initial pair selection 
    (before background subtraction) as a function of the electron
    scattering angle, normalized to the Rosenbluth cross section.}
    \label{fig:cut3e}
    \end{figure}
    
    \begin{figure}[thb!]
    \centerline{\includegraphics[width=1.15\textwidth]{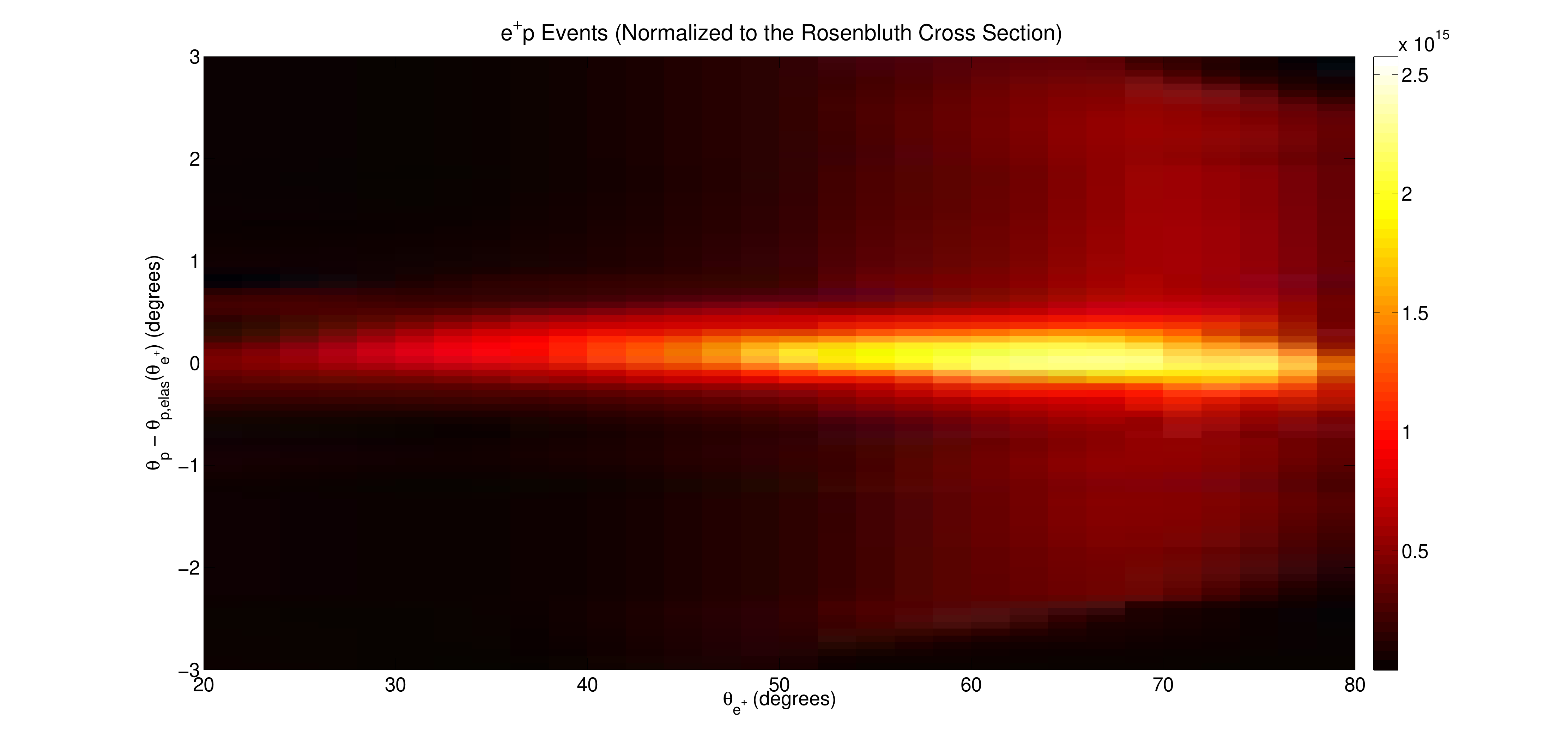}}
    \caption[Histogram of the elastic angle correlation cut parameter for the \pp initial pair selection]{Histogram of the elastic angle correlation cut parameter for the \pp initial pair selection
    (before background subtraction) as a function of the positron
    scattering angle, normalized to the Rosenbluth cross section.}
    \label{fig:cut3p}
    \end{figure}
    
    \begin{figure}[thb!]
    \centerline{\includegraphics[width=1.15\textwidth]{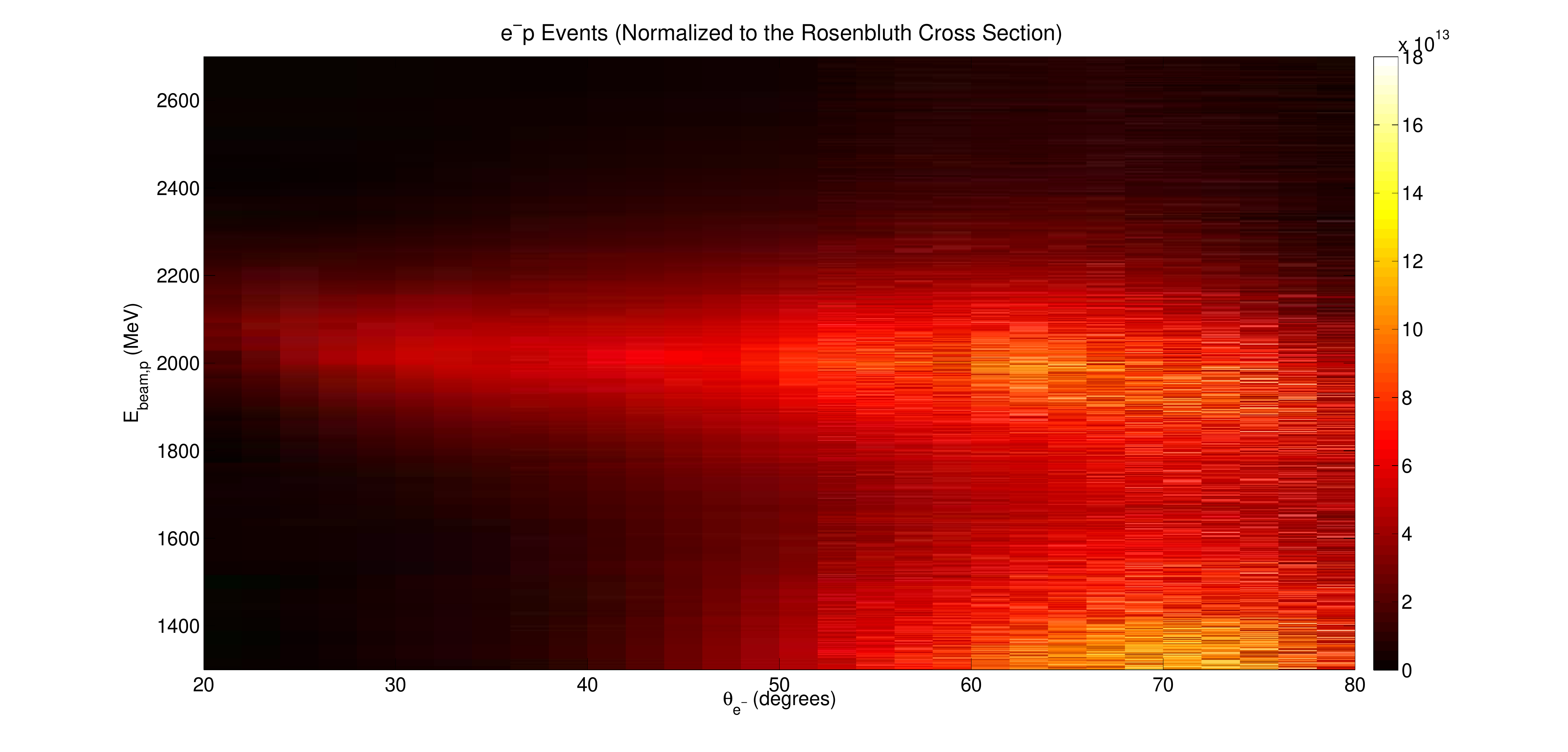}}
    \caption[Histogram of the $E_{\text{beam},p}$ cut parameter for the \ep initial pair selection]{Histogram of the $E_{\text{beam},p}$ cut parameter for the \ep initial pair selection (before background subtraction) as a function of the electron
    scattering angle, normalized to the Rosenbluth cross section.}
    \label{fig:cut4e}
    \end{figure}
    
    \begin{figure}[thb!]
    \centerline{\includegraphics[width=1.15\textwidth]{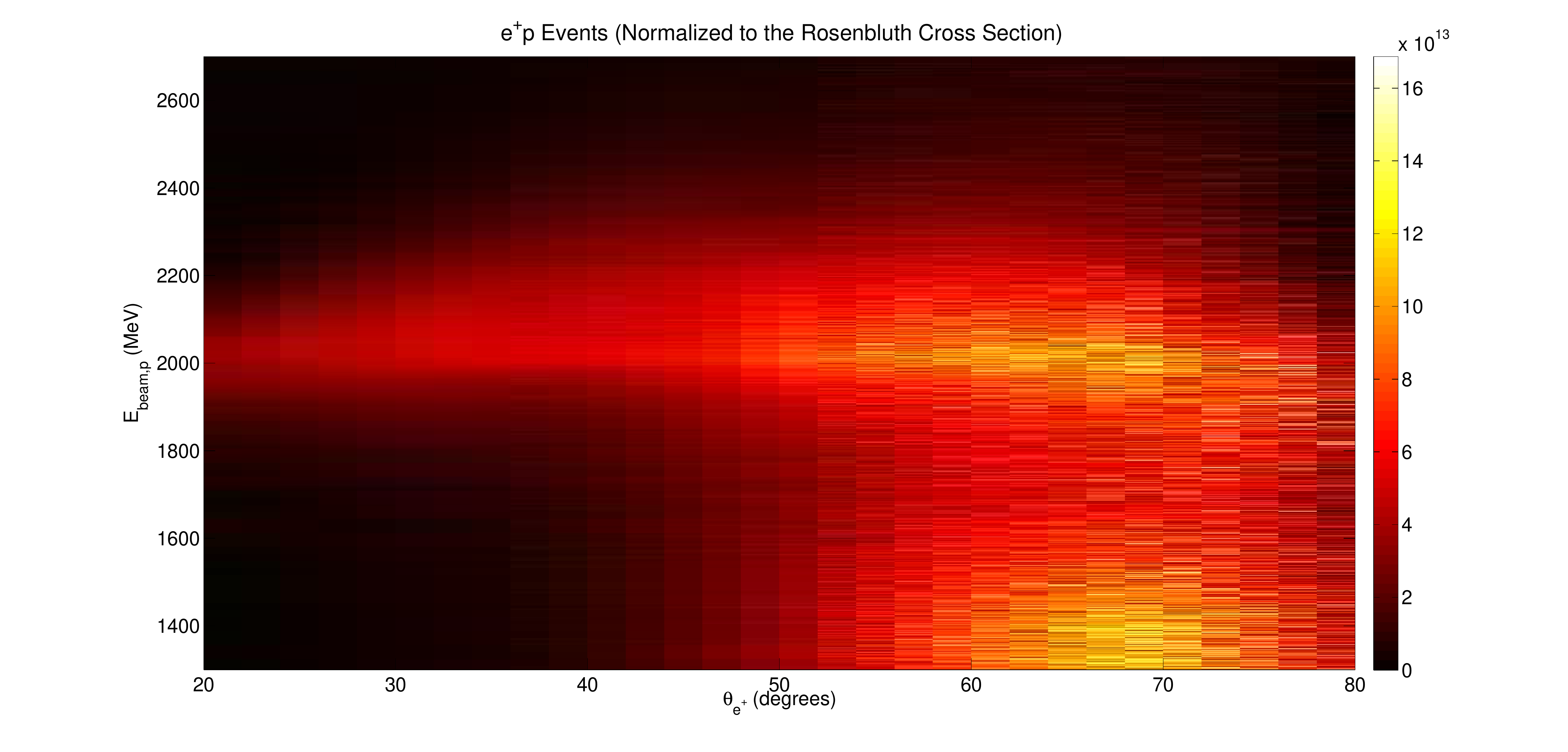}}
    \caption[Histogram of the $E_{\text{beam},p}$ cut parameter for the \pp initial pair selection]{Histogram of the $E_{\text{beam},p}$ cut parameter for the \pp initial pair selection (before background subtraction) as a function of the positron
    scattering angle, normalized to the Rosenbluth cross section.}
    \label{fig:cut4p}
    \end{figure}
    
    \begin{figure}[thb!]
    \centerline{\includegraphics[width=1.15\textwidth]{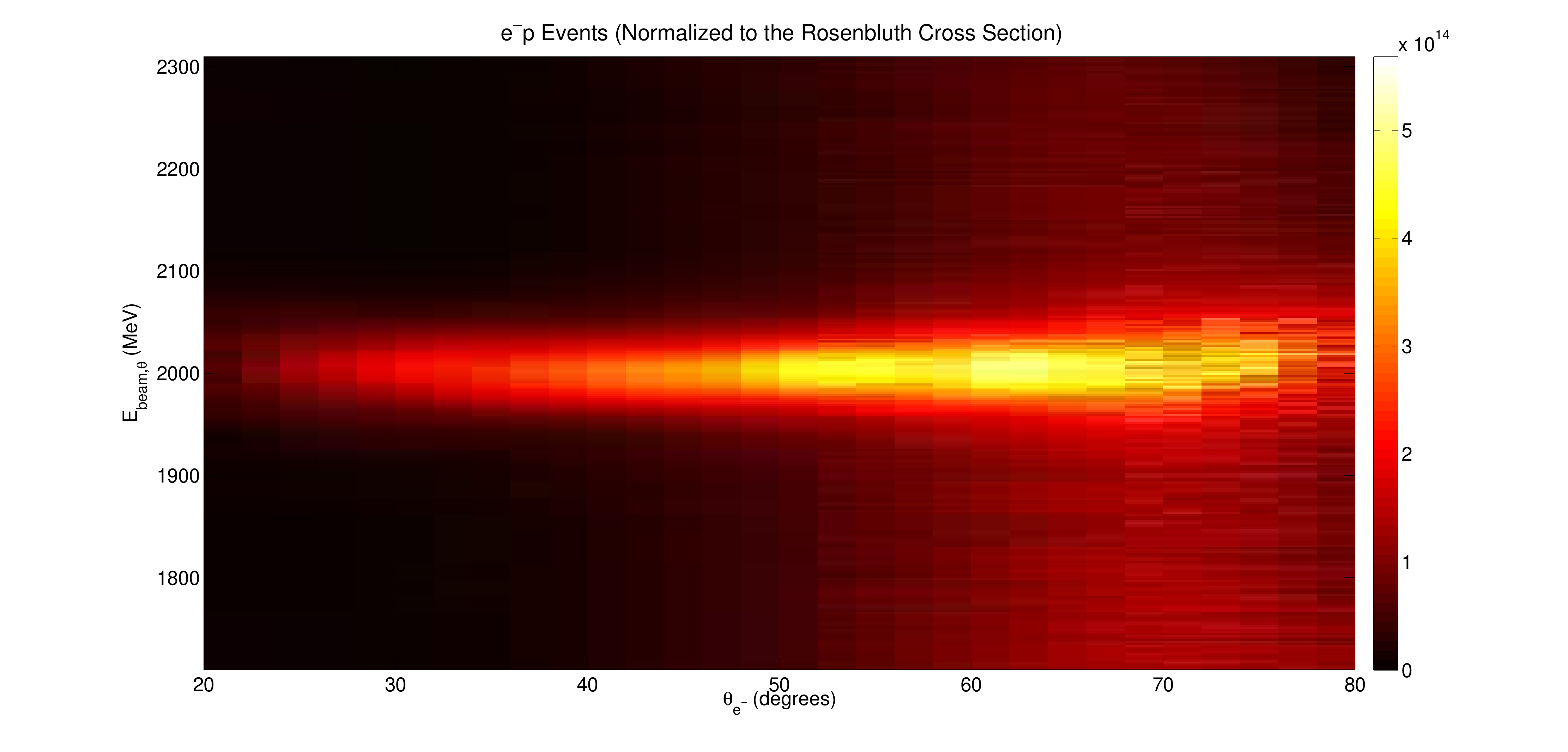}}
    \caption[Histogram of the $E_{\text{beam},\theta}$ cut parameter for the \ep initial pair selection]{Histogram of the $E_{\text{beam},\theta}$ cut parameter for the \ep initial pair selection (before background subtraction) as a function of the electron
    scattering angle, normalized to the Rosenbluth cross section.}
    \label{fig:cut5e}
    \end{figure}
    
    \begin{figure}[thb!]
    \centerline{\includegraphics[width=1.15\textwidth]{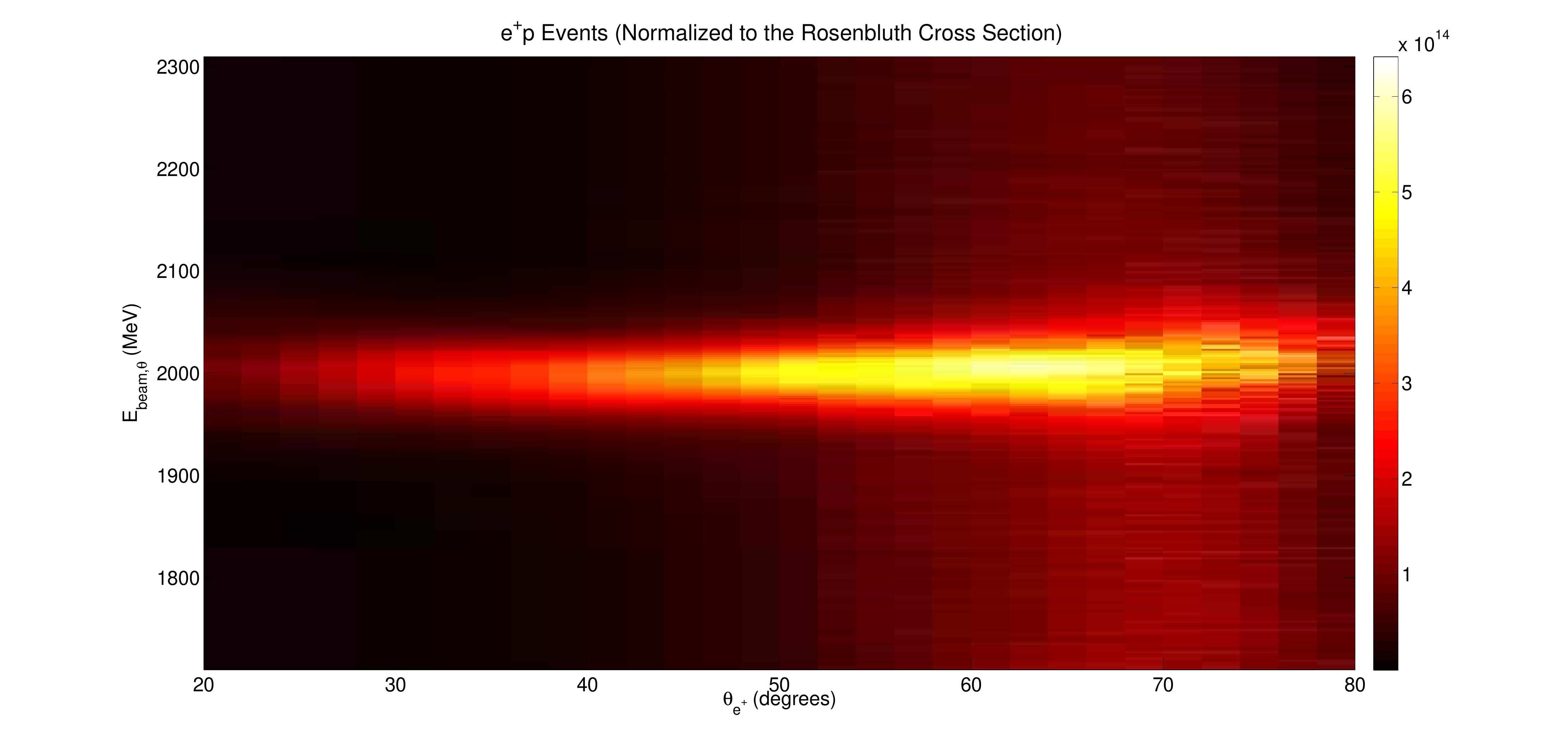}}
    \caption[Histogram of the $E_{\text{beam},\theta}$ cut parameter for the \pp initial pair selection]{Histogram of the $E_{\text{beam},\theta}$ cut parameter for the \pp initial pair selection 
    (before background subtraction) as a function of the positron
    scattering angle, normalized to the Rosenbluth cross section.}
    \label{fig:cut5p}
    \end{figure}
    
    \begin{figure}[thb!]
    \centerline{\includegraphics[width=1.15\textwidth]{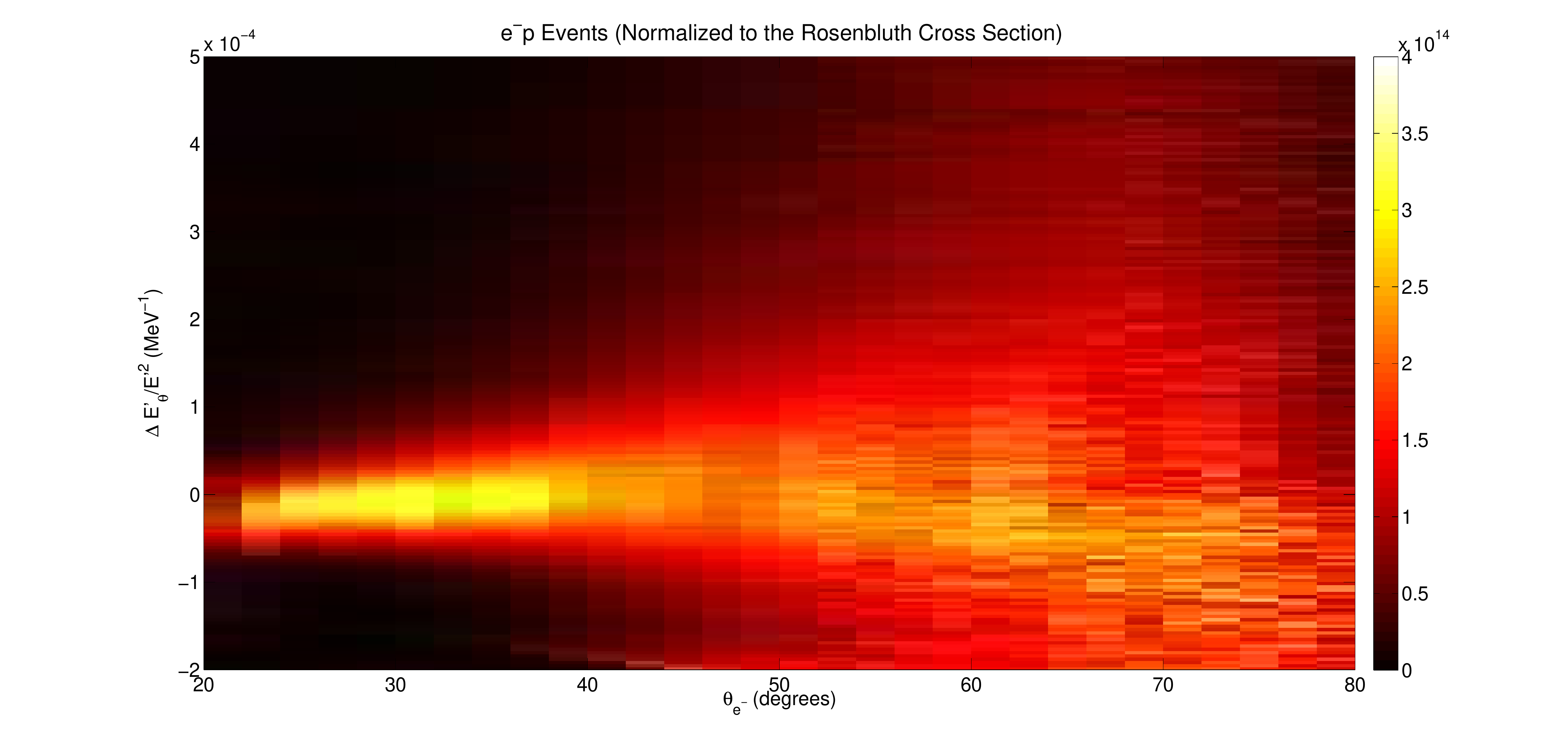}}
    \caption[Histogram of the $\Delta E'_\theta/E'^2$ cut parameter for the \ep initial pair selection]{Histogram of the $\Delta E'_\theta/E'^2$ cut parameter for the \ep initial pair selection (before background subtraction) as a function of the electron
    scattering angle, normalized to the Rosenbluth cross section.}
    \label{fig:cut6e}
    \end{figure}
    
    \begin{figure}[thb!]
    \centerline{\includegraphics[width=1.15\textwidth]{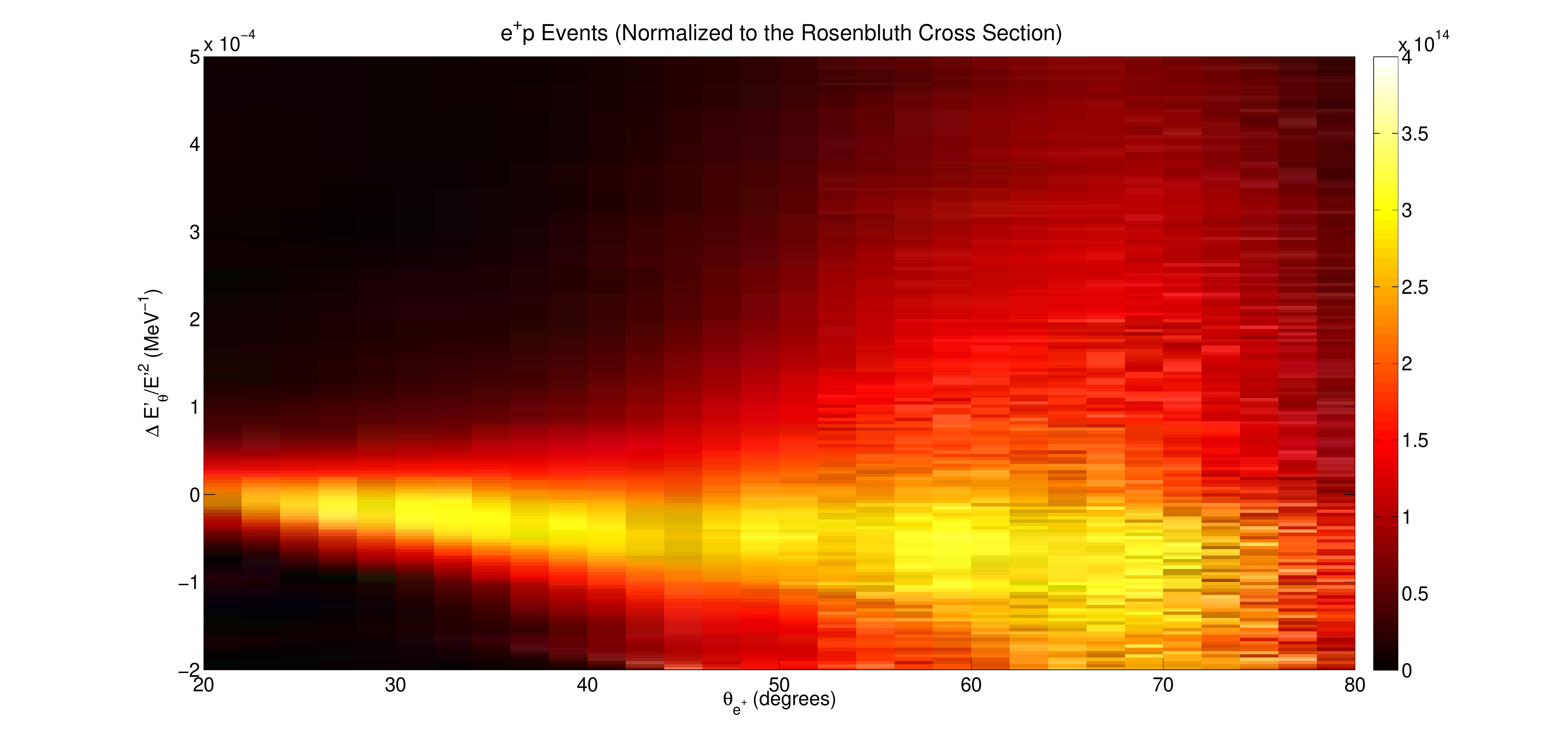}}
    \caption[Histogram of the $\Delta E'_\theta/E'^2$ cut parameter for the \pp initial pair selection]{Histogram of the $\Delta E'_\theta/E'^2$ cut parameter for the \pp initial pair selection 
    (before background subtraction) as a function of the positron
    scattering angle, normalized to the Rosenbluth cross section.}
    \label{fig:cut6p}
    \end{figure}
    
    \begin{figure}[thb!]
    \centerline{\includegraphics[width=1.15\textwidth]{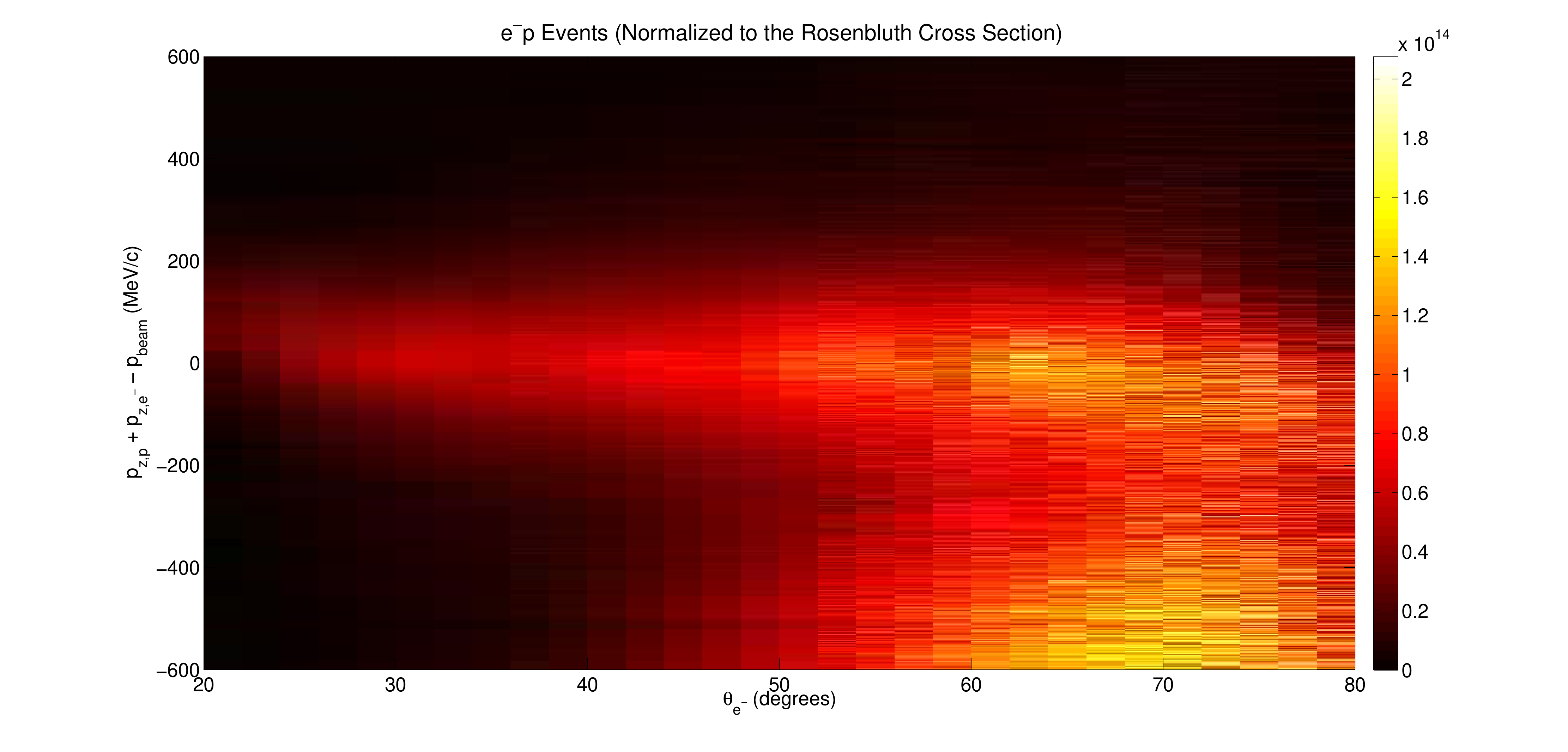}}
    \caption[Histogram of the $p_z$ balance cut parameter for the \ep initial pair selection]{Histogram of the $p_z$ balance cut parameter for the \ep initial pair selection (before background subtraction) as a function of the electron
    scattering angle, normalized to the Rosenbluth cross section.}
    \label{fig:cut7e}
    \end{figure}
    
    \begin{figure}[thb!]
    \centerline{\includegraphics[width=1.15\textwidth]{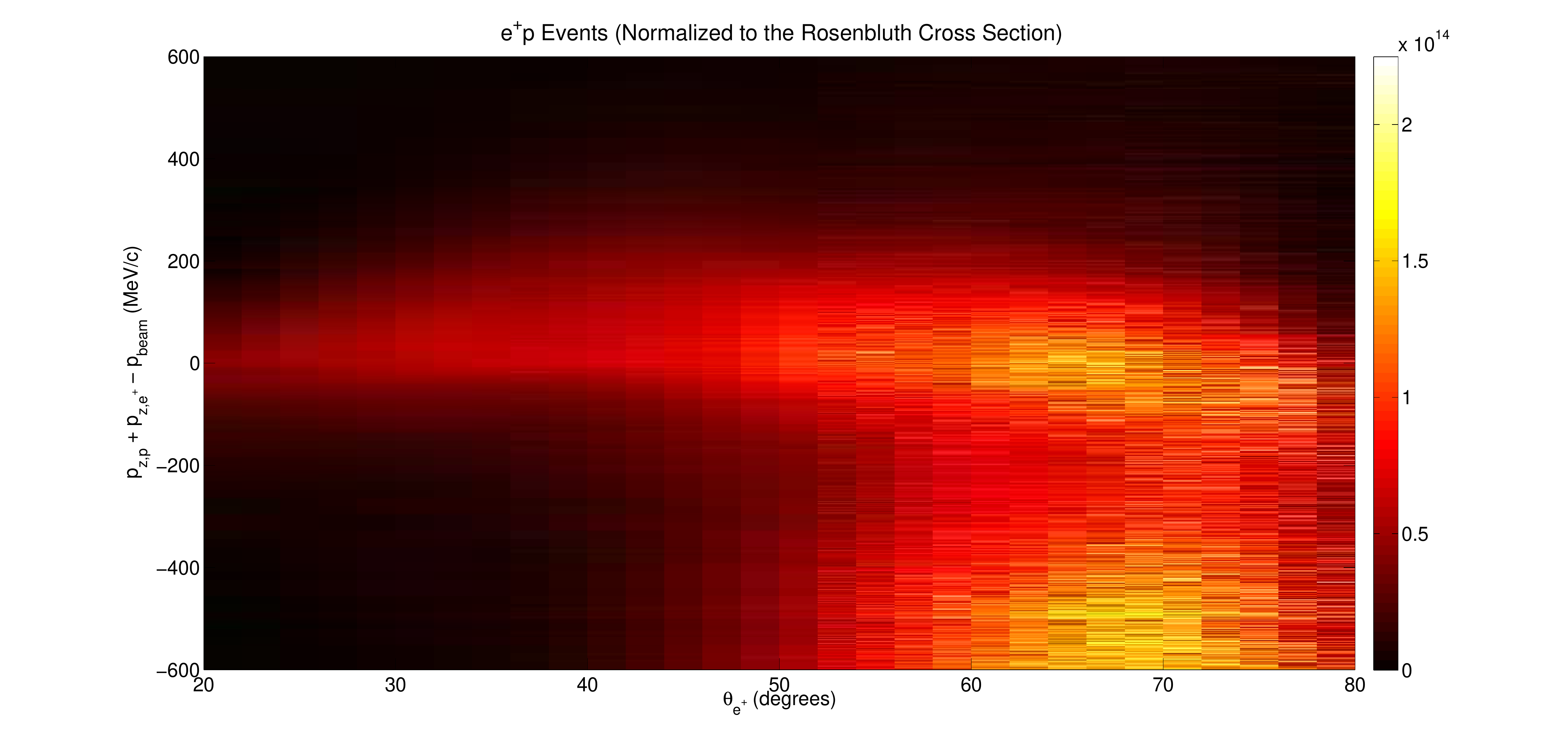}}
    \caption[Histogram of the $p_z$ balance cut parameter for the \pp initial pair selection]{Histogram of the $p_z$ balance cut parameter for the \pp initial pair selection 
    (before background subtraction) as a function of the positron
    scattering angle, normalized to the Rosenbluth cross section.}
    \label{fig:cut7p}
    \end{figure}
    
    \begin{figure}[thb!]
    \centerline{\includegraphics[width=1.15\textwidth]{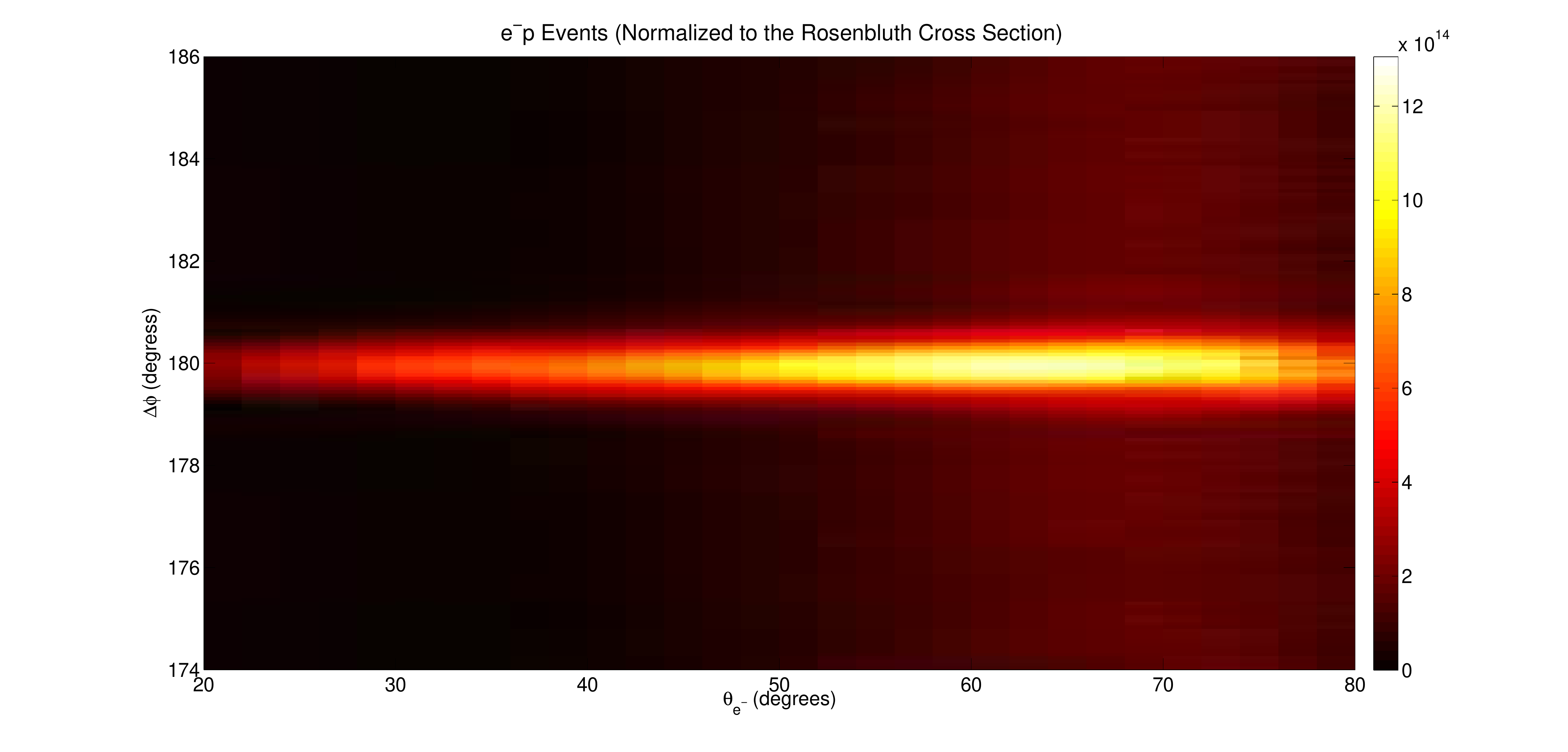}}
    \caption[Histogram of the $\Delta \phi$ cut parameter for the \ep initial pair selection]{Histogram of the $\Delta\phi$ cut parameter for the \ep initial pair selection (before background subtraction)
    as a function of the electron
    scattering angle, normalized to the Rosenbluth cross section.}
    \label{fig:cut8e}
    \end{figure}
    
    \begin{figure}[thb!]
    \centerline{\includegraphics[width=1.15\textwidth]{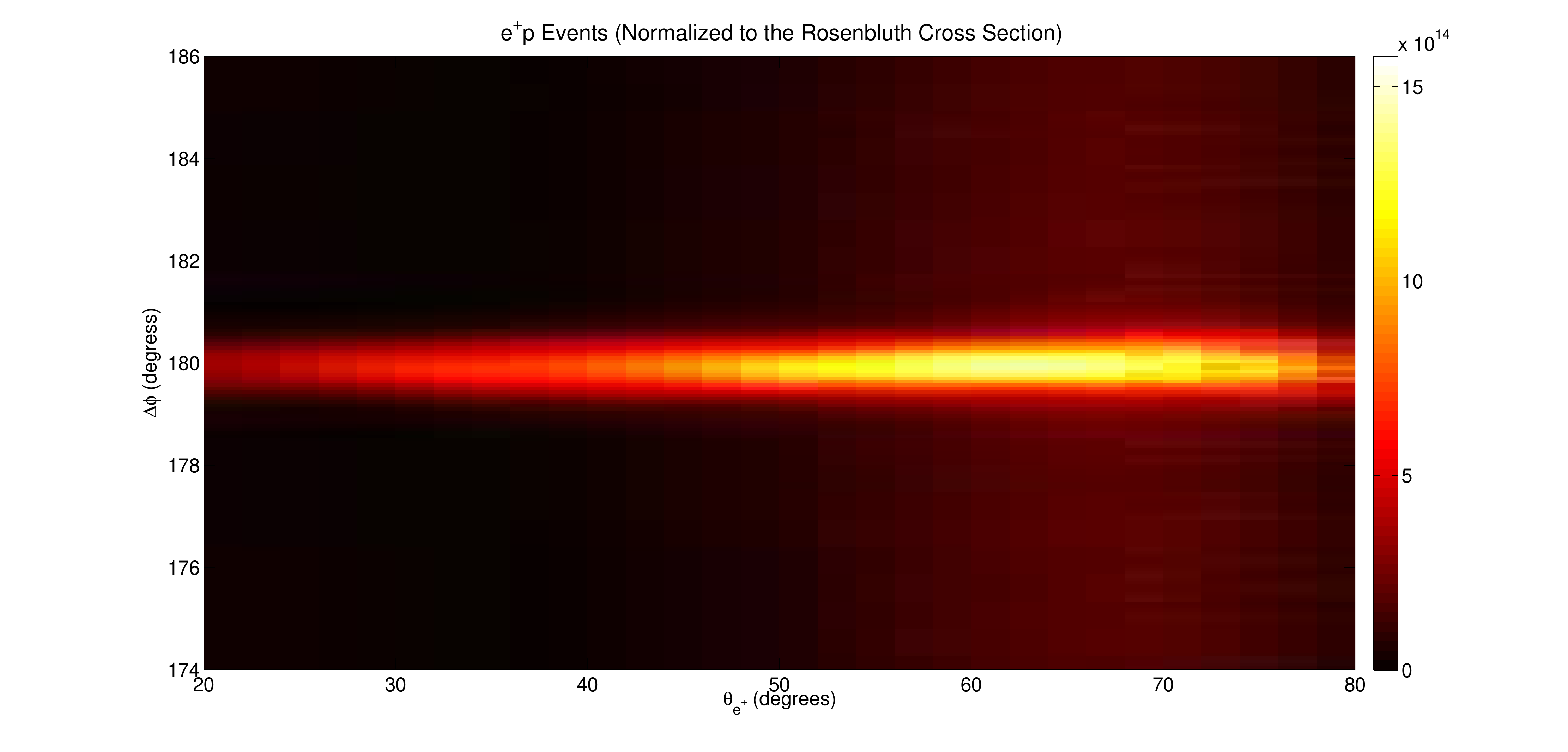}}
    \caption[Histogram of the $\Delta \phi$ cut parameter for the \pp initial pair selection]{Histogram of the $\Delta\phi$ cut parameter for the \pp initial pair selection (before background subtraction)
    as a function of the positron
    scattering angle, normalized to the Rosenbluth cross section.}
    \label{fig:cut8p}
    \end{figure}

%% file: biblio.tex
%% This defines the bibliography file (main.bib) and the bibliography style.
%% If you want to create a bibliography file by hand, change the contents of
%% this file to a `thebibliography' environment.  For more information 
%% see section 4.3 of the LaTeX manual.
\begin{singlespace}

% %\nocite{*}  % Temporary while working on bibliograhy (REMOVE WHEN FINISHED)
\bibliographystyle{ieeetr}
\bibliography{references}
\end{singlespace}

%% file: Henderson_Thesis.bbl
\begin{thebibliography}{100}

\bibitem{doi:10.1080/14786431003659230}
E.~Rutherford, ``{Collision of $\alpha$ particles with light atoms. IV. An
  anomalous effect in nitrogen},'' {\em Philosophical Magazine}, vol.~90,
  no.~sup1, pp.~31--37, 1919 (2010).

\bibitem{Rutherford374}
E.~Rutherford, ``{Bakerian Lecture. Nuclear Constitution of Atoms},'' {\em
  Proceedings of the Royal Society of London A: Mathematical, Physical and
  Engineering Sciences}, vol.~97, no.~686, pp.~374--400, 1920.

\bibitem{Chadwick0}
J.~Chadwick, ``Possible existence of a neutron,'' {\em Nature}, vol.~129,
  no.~3252, p.~312, 1932.

\bibitem{Chadwick1}
J.~Chadwick, ``{Bakerian Lecture. The Neutron},'' {\em Proceedings of the Royal
  Society of London A: Mathematical, Physical and Engineering Sciences},
  vol.~142, no.~846, pp.~1--25, 1933.

\bibitem{stern1}
{Frisch, R. and Stern, O.}, ``{\"{U}ber die magnetische Ablenkung von
  Wasserstoffmolekülen und das magnetische Moment des Protons. I},'' {\em Z.
  Phys.}, vol.~85, pp.~4--16, 1933.

\bibitem{stern2}
{Frisch, R. and Stern, O.}, ``{\"{U}ber die magnetische Ablenkung von
  Wasserstoffmolekülen und das magnetische Moment des Protons. II},'' {\em Z.
  Phys.}, vol.~85, pp.~17--24, 1933.

\bibitem{PhysRev.98.217}
R.~Hofstadter and R.~W. McAllister, ``Electron scattering from the proton,''
  {\em Phys. Rev.}, vol.~98, pp.~217--218, Apr 1955.

\bibitem{PhysRev.102.851}
R.~W. McAllister and R.~Hofstadter, ``{Elastic Scattering of 188-MeV Electrons
  from the Proton and the Alpha Particle},'' {\em Phys. Rev.}, vol.~102,
  pp.~851--856, May 1956.

\bibitem{RevModPhys.28.214}
R.~Hofstadter, ``Electron scattering and nuclear structure,'' {\em Rev. Mod.
  Phys.}, vol.~28, pp.~214--254, Jul 1956.

\bibitem{GELLMANN1964214}
M.~Gell-Mann, ``A schematic model of baryons and mesons,'' {\em Physics
  Letters}, vol.~8, no.~3, pp.~214 -- 215, 1964.

\bibitem{Zweig:570209}
G.~Zweig, ``{An SU$_3$ model for strong interaction symmetry and its breaking;
  Version 2},'' p.~80 p, Feb 1964.

\bibitem{PhysRevLett.23.930}
E.~D. Bloom {\em et~al.}, ``{High-Energy Inelastic $e-p$ Scattering at
  6$^\circ$ and 10$^\circ$},'' {\em Phys. Rev. Lett.}, vol.~23, pp.~930--934,
  Oct 1969.

\bibitem{PhysRevLett.23.935}
M.~Breidenbach {\em et~al.}, ``Observed behavior of highly inelastic
  electron-proton scattering,'' {\em Phys. Rev. Lett.}, vol.~23, pp.~935--939,
  Oct 1969.

\bibitem{doi:10.1142/9789814525220_0008}
P.~Skands, {\em Introduction to QCD}, ch.~8, pp.~341--420.

\bibitem{Agashe:2014kda}
K.~A. Olive {\em et~al.}, ``{Review of Particle Physics},'' {\em Chin. Phys.},
  vol.~C38, p.~090001, 2014.

\bibitem{pdperk}
D.~H. Perkins, ``Proton decay experiments,'' {\em Ann. Rev. Nucl. Sci.},
  vol.~34, pp.~1--52, 1984.

\bibitem{pdproc}
G.~Senjanovi\'{c}, ``Proton decay and grand unification,'' {\em AIP Conference
  Proceedings}, vol.~1200, no.~1, 2010.

\bibitem{jps}
R.~L. Jaffe, ``Where does the proton really get its spin?,'' {\em Physics
  Today}, pp.~24--30, September 1995.

\bibitem{doi:10.1142}
H.~Kolster {\em et~al.}, {\em The BLAST Polarized H/D Target}, ch.~5,
  pp.~37--41.

\bibitem{Carlson201559}
C.~E. Carlson, ``The proton radius puzzle,'' {\em Progress in Particle and
  Nuclear Physics}, vol.~82, pp.~59 -- 77, 2015.

\bibitem{Perdrisat2007694}
C.~Perdrisat, V.~Punjabi, and M.~Vanderhaeghen, ``Nucleon electromagnetic form
  factors,'' {\em Progress in Particle and Nuclear Physics}, vol.~59, no.~2,
  pp.~694 -- 764, 2007.

\bibitem{PhysRevC.69.022201}
J.~Arrington, ``Implications of the discrepancy between proton form factor
  measurements,'' {\em Phys. Rev. C}, vol.~69, p.~022201, Feb 2004.

\bibitem{PhysRev.79.615}
M.~N. Rosenbluth, ``High energy elastic scattering of electrons on protons,''
  {\em Phys. Rev.}, vol.~79, pp.~615--619, Aug 1950.

\bibitem{ff3}
J.~Litt {\em et~al.}, ``{Measurement of the ratio of the proton form factors,
  $G_E/G_M$, at high momentum transfers and the question of scaling},'' {\em
  Physics Letters B}, vol.~31, no.~1, pp.~40 -- 44, 1970.

\bibitem{ff4}
W.~Bartel {\em et~al.}, ``{Measurement of proton and neutron electromagnetic
  form factors at squared four-momentum transfers up to 3 (GeV/c)$^2$},'' {\em
  Nuclear Physics B}, vol.~58, no.~2, pp.~429 -- 475, 1973.

\bibitem{ff8}
L.~Andivahis {\em et~al.}, ``{Measurements of the electric and magnetic form
  factors of the proton from $Q^2$= 1.75 to 8.83~(GeV/c)$^2$},'' {\em Phys.
  Rev.}, vol.~D 50, pp.~5491--5517, 1994.

\bibitem{ff9}
R.~C. Walker {\em et~al.}, ``{Measurements of the proton elastic form factors
  for $1\le Q^2\le 3$ (GeV/c)$^2$ at SLAC},'' {\em Phys. Rev.}, vol.~D 49,
  pp.~5671--5689, June 1994.

\bibitem{pol11}
C.~B. Crawford {\em et~al.}, ``Measurement of the proton's electric to magnetic
  form factor ratio from
  $^{1}\stackrel{\ensuremath{\rightarrow}}{\mathrm{h}}(\stackrel{\ensuremath{\rightarrow}}{e},{e}^{\ensuremath{'}}p)$,''
  {\em Phys. Rev. Lett.}, vol.~98, p.~052301, Jan 2007.

\bibitem{pol3}
V.~Punjabi {\em et~al.}, ``{Proton elastic form factor ratios to
  $Q^2=3.5$~GeV$^2$ by polarization transfer},'' {\em Phys. Rev.}, vol.~C 71,
  p.~055202, 2005.

\bibitem{pol12}
M.~K. Jones {\em et~al.}, ``Proton ${G}_{E}/{G}_{M}$ from beam-target
  asymmetry,'' {\em Phys. Rev. C}, vol.~74, p.~035201, Sep 2006.

\bibitem{pol7}
A.~J.~R. Puckett {\em et~al.}, ``Recoil polarization measurements of the proton
  electromagnetic form factor ratio to ${Q}^{2}=8.5\text{ }\text{
  }{\mathrm{gev}}^{2}$,'' {\em Phys. Rev. Lett.}, vol.~104, p.~242301, Jun
  2010.

\bibitem{pol9}
A.~J.~R. Puckett {\em et~al.}, ``Final analysis of proton form factor ratio
  data at ${Q}^{2}=4.0$, 4.8, and 5.6 gev${}^{2}$,'' {\em Phys. Rev. C},
  vol.~85, p.~045203, Apr 2012.

\bibitem{pol13}
M.~Paolone {\em et~al.}, ``Polarization transfer in the
  $^{4}\mathrm{He}(\stackrel{\ensuremath{\rightarrow}}{e},{e}^{\ensuremath{'}}\stackrel{\ensuremath{\rightarrow}}{p})^{3}\mathbf{H}$
  reaction at ${Q}^{2}=0.8$ and $1.3\text{ }\text{ }(\mathrm{GeV}/c{)}^{2}$,''
  {\em Phys. Rev. Lett.}, vol.~105, p.~072001, Aug 2010.

\bibitem{PhysRevC.68.034325}
J.~Arrington, ``How well do we know the electromagnetic form factors of the
  proton?,'' {\em Phys. Rev. C}, vol.~68, p.~034325, Sep 2003.

\bibitem{ff10}
M.~E. Christy {\em et~al.}, ``{Measurements of electron-proton elastic cross
  sections for $0.4<Q^2<5.5$~(GeV/c)$^2$},'' {\em Phys. Rev.}, vol.~C 70,
  p.~015206, 2004.

\bibitem{ff11}
I.~A. Qattan {\em et~al.}, ``{Precision Rosenbluth measurement of the proton
  elastic form factors},'' {\em Phys. Rev. Lett.}, vol.~94, p.~142301, 2005.

\bibitem{BerFFPhysRevC.90.015206}
J.~C. Bernauer {\em et~al.}, ``Electric and magnetic form factors of the
  proton,'' {\em Phys. Rev. C}, vol.~90, p.~015206, Jul 2014.

\bibitem{Milner:2014}
R.~Milner, D.~Hasell, M.~Kohl, U.~Schneekloth, {\em et~al.}, ``{The OLYMPUS
  Experiment},'' {\em Nucl. Instrum. Meth.}, vol.~A 741, no.~0, pp.~1--17,
  2014.

\bibitem{Blunden:2003sp}
P.~G. Blunden, W.~Melnitchouk, and J.~A. Tjon, ``{Two photon exchange and
  elastic electron proton scattering},'' {\em Phys. Rev. Lett.}, vol.~91,
  p.~142304, 2003.

\bibitem{Chen:2004tw}
Y.~C. Chen {\em et~al.}, ``Partonic calculation of the two-photon exchange
  contribution to elastic electron-proton scattering at large momentum
  transfer,'' {\em Phys. Rev. Lett.}, vol.~93, p.~122301, Sep 2004.

\bibitem{Afanasev:2005mp}
A.~V. Afanasev {\em et~al.}, ``Two-photon exchange contribution to elastic
  electron-nucleon scattering at large momentum transfer,'' {\em Phys. Rev.},
  vol.~D72, p.~013008, Jul 2005.

\bibitem{Blunden:2005ew}
P.~G. Blunden, W.~Melnitchouk, and J.~A. Tjon, ``Two-photon exchange in elastic
  electron-nucleon scattering,'' {\em Phys. Rev.}, vol.~C72, p.~034612, Sep
  2005.

\bibitem{Kondratyuk:2005kk}
S.~Kondratyuk, P.~G. Blunden, W.~Melnitchouk, and J.~A. Tjon, ``{$\Delta{}$}
  resonance contribution to two-photon exchange in electron-proton
  scattering,'' {\em Phys. Rev. Lett.}, vol.~95, p.~172503, Oct 2005.

\bibitem{Borisyuk:2006fh}
D.~Borisyuk and A.~Kobushkin, ``{Box diagram in the elastic electron-proton
  scattering},'' {\em Phys. Rev.}, vol.~C74, p.~065203, 2006.

\bibitem{TomasiGustafsson:2009pw}
E.~Tomasi-Gustafsson {\em et~al.}, ``{Compilation and analysis of charge
  asymmetry measurements from electron and positron scattering on nucleon and
  nuclei},'' {\em Phys. Atom. Nucl.}, vol.~76, pp.~937--946, 2013.

\bibitem{Chen:2007ac}
Y.-C. Chen, C.-W. Kao, and S.-N. Yang, ``{Is there model-independent evidence
  of the two-photon-exchange effect in the electron-proton elastic scattering
  cross-section?},'' {\em Phys. Lett.}, vol.~B652, pp.~269--274, 2007.

\bibitem{Guttmann:2010au}
J.~Guttmann {\em et~al.}, ``Determination of two-photon exchange amplitudes
  from elastic electron-proton scattering data,'' {\em The European Physical
  Journal A-Hadrons and Nuclei}, vol.~47, pp.~1--5, 2011.
\newblock 10.1140/epja/i2011-11077-4.

\bibitem{Yount:1962aa}
D.~Yount and J.~Pine, ``Scattering of high-energy positrons from protons,''
  {\em Phys. Rev.}, vol.~128, pp.~1842--1849, Nov 1962.

\bibitem{Browman:1965zz}
A.~Browman, F.~Liu, and C.~Schaerf, ``{Positron-proton scattering},'' {\em
  Phys.Rev.}, vol.~139, pp.~B1079--B1085, 1965.

\bibitem{Bouquet:1968aa}
B.~Bouquet {\em et~al.}, ``Backward scattering of positrons and electrons on
  protons,'' {\em Physics Letters B}, vol.~26, no.~3, pp.~178 -- 180, 1968.

\bibitem{Mar:1968qd}
J.~Mar {\em et~al.}, ``{A comparison of electron-proton and positron-proton
  elastic scattering at four-momenta transfers up to 5.0 (GeV/$c$)$^2$},'' {\em
  Phys.Rev.Lett.}, vol.~21, pp.~482--484, 1968.

\bibitem{Brun1997}
R.~Brun and F.~Rademakers, ``{ROOT - An Object Oriented Data Analysis
  Framework, Proceedings AIHENP'96 Workshop, Lausanne, Sep. 1996},'' {\em Nucl.
  Inst. \& Meth. in Phys. Res.}, vol.~A 389, pp.~81--86, 1997.

\bibitem{schmidt}
A.~Schmidt, {\em {Measuring the lepton sign asymmetry in elastic
  electron-proton scattering with OLYMPUS}}.
\newblock PhD thesis, Massachusetts Institute of Technology, Cambridge,
  Massachusetts, 2016.

\bibitem{russell}
R.~Russell, {\em {A Measurement of the Two-Photon Exchange Effect in Elastic
  Electron-Proton Scattering with OLYMPUS}}.
\newblock PhD thesis, Massachusetts Institute of Technology, Cambridge,
  Massachusetts, 2016.

\bibitem{oconnor}
C.~O'Connor, {\em The Contribution of Two Photon Exchange in Elastic
  Lepton-Proton Scattering}.
\newblock PhD thesis, Massachusetts Institute of Technology, Cambridge,
  Massachusetts, 2017.

\bibitem{griffiths}
D.~Griffiths, {\em Introduction to Elementary Particles}.
\newblock Weinheim, Germany: Wiley-VCH, 2~ed., 2008.

\bibitem{grupen}
C.~Grupen and B.~Shwartz, {\em Particle Detectors}.
\newblock Cambridge, United Kingdom: Cambridge University Press, 2~ed., 2008.

\bibitem{peskin}
M.~E. Peskin and D.~V. Schroeder, {\em An Introduction to Quantum Field
  Theory}.
\newblock Boulder, Colorado: Westview Press, 1995.

\bibitem{PhysRevD.67.034503}
S.~Aoki {\em et~al.}, ``{Light hadron spectrum and quark masses from quenched
  lattice QCD},'' {\em Phys. Rev. D}, vol.~67, p.~034503, Feb 2003.

\bibitem{PhysRevLett.92.022001}
C.~T.~H. Davies {\em et~al.}, ``High-precision lattice qcd confronts
  experiment,'' {\em Phys. Rev. Lett.}, vol.~92, p.~022001, Jan 2004.

\bibitem{Durr1224}
S.~D{\"u}rr {\em et~al.}, ``Ab initio determination of light hadron masses,''
  {\em Science}, vol.~322, no.~5905, pp.~1224--1227, 2008.

\bibitem{PhysRev.73.279}
M.~E. Rose, ``The charge distribution in nuclei and the scattering of high
  energy electrons,'' {\em Phys. Rev.}, vol.~73, pp.~279--284, Feb 1948.

\bibitem{0954-3899-34-7-S03}
J.~Arrington, C.~D. Roberts, and J.~M. Zanotti, ``Nucleon electromagnetic form
  factors,'' {\em Journal of Physics G: Nuclear and Particle Physics}, vol.~34,
  no.~7, p.~S23, 2007.

\bibitem{Arrington:2011dn}
J.~Arrington, P.~Blunden, and W.~Melnitchouk, ``{Review of two-photon exchange
  in electron scattering},'' {\em Prog. Part. Nucl. Phys.}, vol.~66,
  pp.~782--833, 2011.

\bibitem{Carlson:2007sp}
C.~E. Carlson and M.~Vanderhaeghen, ``{Two-Photon Physics in Hadronic
  Processes},'' {\em Ann. Rev. Nucl. Part. Sci.}, vol.~57, pp.~171--204, 2007.

\bibitem{Pohl2010}
R.~Pohl {\em et~al.}, ``The size of the proton,'' {\em Nature}, vol.~466,
  pp.~213--216, Jul 2010.

\bibitem{RevModPhys.80.633}
P.~J. Mohr, B.~N. Taylor, and D.~B. Newell, ``{CODATA recommended values of the
  fundamental physical constants: 2006},'' {\em Rev. Mod. Phys.}, vol.~80,
  pp.~633--730, Jun 2008.

\bibitem{PhysRevLett.84.5496}
M.~Niering {\em et~al.}, ``Measurement of the hydrogen
  $1\mathit{S}$-$2\mathit{S}$ transition frequency by phase coherent comparison
  with a microwave cesium fountain clock,'' {\em Phys. Rev. Lett.}, vol.~84,
  pp.~5496--5499, Jun 2000.

\bibitem{Sick200362}
I.~Sick, ``On the rms-radius of the proton,'' {\em Physics Letters B},
  vol.~576, no.~1–2, pp.~62 -- 67, 2003.

\bibitem{Born1926}
M.~Born, ``Quantenmechanik der sto{\ss}vorg{\"a}nge,'' {\em Zeitschrift f{\"u}r
  Physik}, vol.~38, no.~11, pp.~803--827, 1926.

\bibitem{wtf}
S.~Boffi, C.~Giusti, F.~D. Pacati, and M.~Radici, {\em Electromagnetic Response
  of Atomic Nuclei}.
\newblock New York: Oxford University Press, 1996.

\bibitem{Mott425}
N.~F. Mott, ``The scattering of fast electrons by atomic nuclei,'' {\em
  Proceedings of the Royal Society of London A: Mathematical, Physical and
  Engineering Sciences}, vol.~124, no.~794, pp.~425--442, 1929.

\bibitem{Mott658}
N.~F. Mott, ``The scattering of electrons by atoms,'' {\em Proceedings of the
  Royal Society of London A: Mathematical, Physical and Engineering Sciences},
  vol.~127, no.~806, pp.~658--665, 1930.

\bibitem{PhysRev.87.688}
L.~L. Foldy, ``{The electromagnetic properties of Dirac particles},'' {\em
  Phys. Rev.}, vol.~87, pp.~688--693, Sep 1952.

\bibitem{ff5}
L.~N. Hand, D.~G. Miller, and R.~Wilson, ``Electric and magnetic form factors
  of the nucleon,'' {\em Rev. Mod. Phys.}, vol.~35, pp.~335--349, Apr 1963.

\bibitem{PhysRev.126.2256}
R.~G. Sachs, ``High-energy behavior of nucleon electromagnetic form factors,''
  {\em Phys. Rev.}, vol.~126, pp.~2256--2260, Jun 1962.

\bibitem{PhysRevC.66.065203}
J.~J. Kelly, ``{Nucleon charge and magnetization densities from Sachs form
  factors},'' {\em Phys. Rev. C}, vol.~66, p.~065203, Dec 2002.

\bibitem{ff6}
T.~Janssens {\em et~al.}, ``{Proton form factors from elastic electron-proton
  scattering},'' {\em Phys. Rev.}, vol.~142, pp.~922--931, 1966.

\bibitem{ff14}
D.~H. Coward {\em et~al.}, ``Electron-proton elastic scattering at high
  momentum transfers,'' {\em Phys. Rev. Lett.}, vol.~20, pp.~292--295, Feb
  1968.

\bibitem{ff1}
L.~E. Price {\em et~al.}, ``Backward-angle electron-proton elastic scattering
  and proton electromagnetic form factors,'' {\em Phys. Rev. D}, vol.~4,
  pp.~45--53, Jul 1971.

\bibitem{ff2}
C.~Berger {\em et~al.}, ``{Electromagnetic form factors of the proton at
  squared four-momentum transfers between 10 and 50 fm$^{−2}$ },'' {\em
  Physics Letters B}, vol.~35, no.~1, pp.~87 -- 89, 1971.

\bibitem{ff12}
K.~M. Hanson and other, ``Large-angle quasielastic electron-deuteron
  scattering,'' {\em Phys. Rev. D}, vol.~8, pp.~753--778, Aug 1973.

\bibitem{ff7}
F.~Borkowski {\em et~al.}, ``{Electromagnetic Form-Factors of the Proton at Low
  Four-Momentum Transfer},'' {\em Nucl. Phys.}, vol.~B 93, p.~461, 1975.

\bibitem{ff15}
A.~F. Sill {\em et~al.}, ``Measurements of elastic electron-proton scattering
  at large momentum transfer,'' {\em Phys. Rev. D}, vol.~48, pp.~29--55, Jul
  1993.

\bibitem{pol1}
O.~Gayou {\em et~al.}, ``{Measurements of the elastic electromagnetic
  form-factor ratio $\mu_p G_{E_p}/G_{M_p}$ via polarization transfer},'' {\em
  Phys. Rev.}, vol.~C 64, p.~038202, 2001.

\bibitem{pol2}
O.~Gayou {\em et~al.}, ``{Measurement of $G_{E_p}/G_{M_p}$ in $\vec
  ep\rightarrow e\vec p$ to $Q^2 = 5.6$~GeV$^2$},'' {\em Phys. Rev. Lett.},
  vol.~88, p.~092301, Mar 2002.

\bibitem{pol4}
G.~MacLachlan {\em et~al.}, ``{The ratio of proton electromagnetic form factors
  via recoil polarimetry at $Q^2 = 1.13$~(GeV/c)$^2$},'' {\em Nucl. Phys.},
  vol.~A764, pp.~261--273, 2006.

\bibitem{pol5}
M.~K. Jones {\em et~al.}, ``{$G_{E_p}/G_{M_p}$ Ratio by Polarization Transfer
  in $\vec e p\rightarrow e\vec p$},'' {\em Phys. Rev. Lett.}, vol.~84,
  pp.~1398--1402, Feb 2000.

\bibitem{pol6}
G.~Ron {\em et~al.}, ``Measurements of the proton elastic-form-factor ratio
  ${\ensuremath{\mu}}_{p}{G}_{E}^{p}/{G}_{M}^{p}$ at low momentum transfer,''
  {\em Phys. Rev. Lett.}, vol.~99, p.~202002, Nov 2007.

\bibitem{pol8}
X.~Zhan {\em et~al.}, ``{High-precision measurement of the proton elastic form
  factor ratio at low $Q^2$},'' {\em Physics Letters B}, vol.~705, no.~1–2,
  pp.~59 -- 64, 2011.

\bibitem{pol10}
G.~Ron {\em et~al.}, ``Low-${Q}^{2}$ measurements of the proton form factor
  ratio ${\ensuremath{\mu}}_{p}{G}_{E}/{G}_{M}$,'' {\em Phys. Rev. C}, vol.~84,
  p.~055204, Nov 2011.

\bibitem{fernow}
R.~C. Fernow, {\em Introduction to Experimental Particle Physics}.
\newblock Cambridge, United Kingdom: Cambridge University Press, 1986.

\bibitem{ak1}
A.~I. Akhiezer and R.~M. P., ``Polarization phenomena in electron scattering by
  protons in the high energy region,'' {\em Sov. Phys. Dokl.}, vol.~13, p.~572,
  Dec 1968.

\bibitem{ak2}
A.~I. Akhiezer and R.~M. P., ``Polarization effects in the scattering of
  leptons by hadrons,'' {\em Sov. J. Part. Nucl.}, vol.~4, p.~277, 1974.

\bibitem{PhysRevC.23.363}
R.~G. Arnold, C.~E. Carlson, and F.~Gross, ``Polarization transfer in elastic
  electron scattering from nucleons and deuterons,'' {\em Phys. Rev. C},
  vol.~23, pp.~363--374, Jan 1981.

\bibitem{RevModPhys.41.236}
N.~Dombey, ``Scattering of polarized leptons at high energy,'' {\em Rev. Mod.
  Phys.}, vol.~41, pp.~236--246, Jan 1969.

\bibitem{ff13}
G.~Simon {\em et~al.}, ``Absolute electron-proton cross sections at low
  momentum transfer measured with a high pressure gas target system,'' {\em
  Nuclear Physics A}, vol.~333, no.~3, pp.~381 -- 391, 1980.

\bibitem{PhysRev.122.1898}
Y.-S. Tsai, ``Radiative corrections to electron-proton scattering,'' {\em Phys.
  Rev.}, vol.~122, pp.~1898--1907, Jun 1961.

\bibitem{MoRevModPhys.41.205}
L.~W. Mo and Y.~S. Tsai, ``Radiative corrections to elastic and inelastic
  $\mathrm{ep}$ and $\mathrm{up}$ scattering,'' {\em Rev. Mod. Phys.}, vol.~41,
  pp.~205--235, Jan 1969.

\bibitem{MaximonPhysRevC.62.054320}
L.~C. Maximon and J.~A. Tjon, ``Radiative corrections to electron-proton
  scattering,'' {\em Phys. Rev. C}, vol.~62, p.~054320, Oct 2000.

\bibitem{PhysRev.74.1759}
W.~A. McKinley and H.~Feshbach, ``{The Coulomb scattering of relativistic
  electrons by nuclei},'' {\em Phys. Rev.}, vol.~74, pp.~1759--1763, Dec 1948.

\bibitem{PhysRev.102.537}
R.~R. Lewis, ``{Potential scattering of high-energy electrons in second Born
  approximation},'' {\em Phys. Rev.}, vol.~102, pp.~537--543, Apr 1956.

\bibitem{PhysRev.106.561}
S.~D. Drell and M.~A. Ruderman, ``Proton polarizability correction to
  electron-proton scattering,'' {\em Phys. Rev.}, vol.~106, pp.~561--563, May
  1957.

\bibitem{PhysRev.113.741}
S.~D. Drell and S.~Fubini, ``Higher electromagnetic corrections to
  electron-proton scattering,'' {\em Phys. Rev.}, vol.~113, pp.~741--744, Jan
  1959.

\bibitem{1969190}
J.~A. Campbell, ``Algebraic computation of radiative corrections for
  electron-proton scattering,'' {\em Nuclear Physics B}, vol.~1, no.~5, pp.~283
  -- 300, 1967.

\bibitem{PhysRev.180.1541}
J.~A. Campbell, ``Accuracy of radiative corrections to electromagnetic
  scattering from protons,'' {\em Phys. Rev.}, vol.~180, pp.~1541--1546, Apr
  1969.

\bibitem{PhysRev.184.1860}
G.~K. Greenhut, ``Two-photon exchange in electron-proton scattering,'' {\em
  Phys. Rev.}, vol.~184, pp.~1860--1867, Aug 1969.

\bibitem{Guichon:2003qm}
P.~A.~M. Guichon and M.~Vanderhaeghen, ``{How to reconcile the Rosenbluth and
  the polarization transfer method in the measurement of the proton
  form-factors},'' {\em Phys. Rev. Lett.}, vol.~91, p.~142303, 2003.

\bibitem{REKALO2004322}
M.~Rekalo and E.~Tomasi-Gustafsson, ``Polarization phenomena in e∓n elastic
  scattering, for axial parametrization of two-photon exchange,'' {\em Nuclear
  Physics A}, vol.~742, no.~3, pp.~322 -- 334, 2004.

\bibitem{Anderson:1966zzf}
R.~Anderson {\em et~al.}, ``{Scattering of positrons and electrons from
  protons},'' {\em Phys.Rev.Lett.}, vol.~17, pp.~407--409, 1966.

\bibitem{Cassiday:1967aa}
G.~Cassiday {\em et~al.}, ``Comparison of elastic positron-proton and
  electron-proton scattering cross sections,'' {\em Phys. Rev. Lett.}, vol.~19,
  pp.~1191--1192, Nov 1967.

\bibitem{Bartel:1967aa}
W.~Bartel {\em et~al.}, ``Scattering of positrons and electrons from protons,''
  {\em Physics Letters B}, vol.~25, no.~3, pp.~242 -- 245, 1967.

\bibitem{PhysRevLett.114.062003}
D.~Adikaram {\em et~al.}, ``Towards a resolution of the proton form factor
  problem: New electron and positron scattering data,'' {\em Phys. Rev. Lett.},
  vol.~114, p.~062003, Feb 2015.

\bibitem{ass}
D.~Rimal {\em et~al.}, ``Measurement of two-photon exchange effect by comparing
  elastic $e^\pm p$ cross sections.'' Pre-print: arXiv:1603.00315, 2016.

\bibitem{vepp3PhysRevLett.114.062005}
I.~A. Rachek {\em et~al.}, ``{Measurement of the two-photon exchange
  contribution to the elastic ${e}^{\ifmmode\pm\else\textpm\fi{}}p$ scattering
  cross sections at the VEPP-3 storage ring},'' {\em Phys. Rev. Lett.},
  vol.~114, p.~062005, Feb 2015.

\bibitem{Alarcon20001111c}
R.~Alarcon and the BLAST~Collaboration, ``{BLAST: A new tool for intermediate
  energy nuclear physics},'' {\em Nuclear Physics A}, vol.~663–664, no.~0,
  pp.~1111c--1114c, 2000.

\bibitem{Hasell:2009zza}
D.~Hasell, T.~Akdogan, R.~Alarcon, W.~Bertozzi, E.~Booth, {\em et~al.}, ``{The
  BLAST experiment},'' {\em Nucl. Instrum. Meth.}, vol.~A 603, pp.~247--262,
  2009.

\bibitem{DORIStab}
{DESY Accelerator Division (M-Division)}, ``{DORIS}.''
  \url{http://doris.desy.de/index_eng.html}, 2015.

\bibitem{DORISrep}
G.~A. Voss, ``{Report on DORIS},'' {\em IEEE Trans. Nucl. Sci.}, vol.~NS-22,
  no.~3, pp.~1363--1365, 1975.

\bibitem{tdr}
``{Technical Design Report for the OLYMPUS Experiment},'' tech. rep., The
  OLYMPUS Collaboration, 2010.

\bibitem{Hillert2006}
W.~Hillert, ``{The Bonn Electron Stretcher Accelerator ELSA: Past and
  future},'' {\em The European Physical Journal A - Hadrons and Nuclei},
  vol.~28, no.~1, pp.~139--148, 2006.

\bibitem{ALBRECHT1987245}
H.~Albrecht {\em et~al.}, ``Observation of $ {B}^0-\overline{B^0}$ mixing,''
  {\em Physics Letters B}, vol.~192, no.~1, pp.~245 -- 252, 1987.

\bibitem{DORIShist}
``{DORIS}.''
  \url{http://www.desy.de/research/facilities__projects/doris/index_eng.html},
  2016.

\bibitem{Bernauer201420}
J.~Bernauer {\em et~al.}, ``{The OLYMPUS internal hydrogen target},'' {\em
  Nucl. Instrum. Meth.}, vol.~A 755, no.~0, pp.~20--27, 2014.

\bibitem{uwe1}
U.~Schneekloth. Personal communication.

\bibitem{Cheever:2006xt}
D.~Cheever {\em et~al.}, ``{A highly polarized hydrogen/deuterium internal gas
  target embedded in a toroidal magnetic spectrometer},'' {\em Nucl. Instrum.
  Meth.}, vol.~A556, pp.~410--420, 2006.

\bibitem{Airapetian:2004yf}
A.~Airapetian {\em et~al.}, ``{The HERMES polarized hydrogen and deuterium gas
  target in the HERA electron storage ring},'' {\em Nucl. Instrum. Meth.},
  vol.~A540, pp.~68--101, 2005.

\bibitem{Bernauer20169}
J.~Bernauer {\em et~al.}, ``{Measurement and tricubic interpolation of the
  magnetic field for the OLYMPUS experiment},'' {\em Nucl. Instrum. Meth.},
  vol.~A 823, pp.~9 -- 14, 2016.

\bibitem{Dow2009146}
K.~A. Dow {\em et~al.}, ``{Magnetic field measurements of the BLAST
  spectrometer},'' {\em Nucl. Instrum. Meth.}, vol.~A 599, no.~2–3, pp.~146
  -- 151, 2009.

\bibitem{bicron}
Bicron, ``Organic scintillation materials,'' 2015.

\bibitem{blum}
W.~Blum, W.~Riegler, and L.~Rolandi, {\em Particle Detection with Drift
  Chambers}.
\newblock Berlin, Germany: Springer-Verlag, 2~ed., 2008.

\bibitem{dont}
B.~Foster, ``{Whisker growth in test cells},'' {Proceedings of the Workshop on
  Radiation Damage to Wire Chambers, Berkeley, CA}, 1986.

\bibitem{refId0}
{Kohl, Michael}, ``{The Muon Scattering Experiment (MUSE) at PSI and the proton
  radius puzzle},'' {\em EPJ Web of Conferences}, vol.~81, p.~02008, 2014.

\bibitem{Balewski:2014pxa}
J.~Balewski {\em et~al.}, ``{The DarkLight Experiment: A Precision Search for
  New Physics at Low Energies},'' 2014.

\bibitem{dief1}
J.~Diefenbach. Personal communication.

\bibitem{Sauli20162}
F.~Sauli, ``{The gas electron multiplier (GEM): Operating principles and
  applications},'' {\em Nucl. Instrum. Meth.}, vol.~A 805, pp.~2 -- 24, 2016.

\bibitem{kohl1}
M.~Kohl. Personal communication.

\bibitem{French:2001xb}
M.~French {\em et~al.}, ``{Design and results from the APV25, a deep sub-micron
  CMOS front-end chip for the CMS tracker},'' {\em Nucl. Instrum. Meth.},
  vol.~A466, pp.~359--365, 2001.

\bibitem{Musico:2011lia}
P.~Musico {\em et~al.}, ``{Hybrid silicon mustrip and GEM tracker for JLab
  Hall-A high luminosity experiments},'' {\em IEEE Nucl. Sci. Symp. Conf.
  Rec.}, vol.~2011, pp.~1306--1308, 2011.

\bibitem{Uvarov:cros3}
N.~Bondar {\em et~al.}, ``{Third Generation Coordinate ReadOut System --
  CROS-3},'' {\em PNPI High Energy Physics Division Main Scientific Activities
  2002-2006}, p.~334, 2007.

\bibitem{Veenhof:1998tt}
R.~Veenhof, ``{GARFIELD, recent developments},'' {\em Nucl. Instrum. Meth.},
  vol.~A419, pp.~726--730, 1998.

\bibitem{Andreev:2001kr}
A.~Andreev {\em et~al.}, ``{Multiwire proportional chambers in the HERMES
  experiment},'' {\em Nucl. Instrum. Meth.}, vol.~A 465, pp.~482--497, May
  2001.

\bibitem{moller}
{M{\o}ller, C.} {\em Ann. Phys.}, vol.~14, p.~531, 1932.

\bibitem{bhabha}
{Bhabha, H.J.} {\em Proc. Roy. Soc. (London)}, vol.~A 154, p.~195, 1935.

\bibitem{PerezBenito20166}
R.~P. Benito {\em et~al.}, ``Design and performance of a lead fluoride detector
  as a luminosity monitor,'' {\em Nucl. Instrum. Meth.}, vol.~A 826, pp.~6 --
  14, 2016.

\bibitem{KOBIS1998625}
S.~K\"{o}bis {\em et~al.}, ``{Proceedings of the Fifth International Conference
  on Advanced Technology and Particle Physics: The analog trigger processor for
  the new parity violation experiment at MAMI},'' {\em Nuclear Physics B -
  Proceedings Supplements}, vol.~61, no.~3, pp.~625 -- 629, 1998.

\bibitem{A4M}
R.~Kothe, {\em Aufbau und Betrieb einer schnellen Kalorimeterelektronik f\"{u}r
  ein Experiment zur Messung der Parit\"{a}tsverletzung in der elastischen
  Elektronenstreuung}.
\newblock PhD thesis, Johannes Gutenberg-Universit\"{a}t, Mainz, Germany, 2008.

\bibitem{Baunack:2011pb}
S.~Baunack {\em et~al.}, ``{Realtime calibration of the A4 electromagnetic lead
  fluoride calorimeter},'' {\em Nucl. Instrum. Meth.}, vol.~A640, pp.~58--68,
  2011.

\bibitem{ANDERSON1990385}
D.~Anderson {\em et~al.}, ``Lead fluoride: An ultra-compact cherenkov radiator
  for em calorimetry,'' {\em Nucl. Instrum. Meth.}, vol.~A 290, no.~2, pp.~385
  -- 389, 1990.

\bibitem{epics}
S.~A. Lewis, ``{Overview of the Experimental Physics and Industrial Control
  System: EPICS},'' tech. rep., Lawrence Berkeley National Laboratory.

\bibitem{bonndaq}
C.~Schmidt, {\em Development of a new data acquisition system for the CB-ELSA
  experiment}.
\newblock PhD thesis, Universit\"{a}t Bonn, Bonn, Germany, 2004.

\bibitem{zebra}
J.~Zoll, ``{Zebra Reference Manual book FZ},'' tech. rep., CERN Program
  Library.

\bibitem{mea2}
``{MEA2 - Vermessung}.'' \url{http://geo.desy.de/index_ger.html}.
\newblock Accessed: 2016-07-14.

\bibitem{gdml}
R.~Chytracek {\em et~al.}, ``Geometry description markup language for physics
  simulation and analysis applications,'' {\em IEEE Transactions on Nuclear
  Science}, vol.~53, pp.~2892--2896, Oct 2006.

\bibitem{Agostinelli:2002hh}
S.~Agostinelli {\em et~al.}, ``{GEANT4: A Simulation toolkit},'' {\em Nucl.
  Instrum. Meth.}, vol.~A 506, pp.~250--303, 2003.

\bibitem{bernauer3}
J.~Bernauer. Personal communication.

\bibitem{bpm}
U.~Schneekloth {\em et~al.}, ``{Calibration of OLYMPUS/DORIS beam position
  monitors},'' {Proceedings of IBIC2014, Monterey, CA}, 2014.

\bibitem{NME:NME1296}
F.~Lekien and J.~Marsden, ``Tricubic interpolation in three dimensions,'' {\em
  International Journal for Numerical Methods in Engineering}, vol.~63, no.~3,
  pp.~455--471, 2005.

\bibitem{crawford}
C.~B. Crawford, {\em {Precision Measurement of the Proton Electric to Magnetic
  Form Factor Ratio with BLAST}}.
\newblock PhD thesis, Massachusetts Institute of Technology, Cambridge,
  Massachusetts, 2005.

\bibitem{Biagi1999234}
S.~Biagi, ``{Monte Carlo simulation of electron drift and diffusion in counting
  gases under the influence of electric and magnetic fields},'' {\em Nucl.
  Instrum. Meth.}, vol.~A 421, no.~1–2, pp.~234 -- 240, 1999.

\bibitem{DELLORSO1990436}
M.~Dell'orso and L.~Ristori, ``A highly parallel algorithm for track finding,''
  {\em Nucl. Instrum. Meth.}, vol.~A 287, no.~3, pp.~436 -- 438, 1990.

\bibitem{oconnor1}
C.~O'Connor. Personal communication.

\bibitem{OHLSSON1}
M.~Ohlsson, C.~Peterson, and A.~L. Yuille, ``Track finding with deformable
  templates — the elastic arms approach,'' {\em Computer Physics
  Communications}, vol.~71, no.~1, pp.~77 -- 98, 1992.

\bibitem{OHLSSON2}
M.~Ohlsson, ``Extensions and explorations of the elastic arms algorithm,'' {\em
  Computer Physics Communications}, vol.~77, no.~1, pp.~19 -- 32, 1993.

\bibitem{MeisterPhysRev.130.1210}
N.~Meister and D.~R. Yennie, ``Radiative corrections to high-energy scattering
  processes,'' {\em Phys. Rev.}, vol.~130, pp.~1210--1229, May 1963.

\bibitem{PhysRevC.64.054610}
R.~Ent {\em et~al.}, ``{Radiative corrections for ${(e,e}^{\ensuremath{'}}p)$
  reactions at GeV energies},'' {\em Phys. Rev. C}, vol.~64, p.~054610, Oct
  2001.

\bibitem{esepp}
A.~V. Gramolin {\em et~al.}, ``A new event generator for the elastic scattering
  of charged leptons on protons,'' {\em Journal of Physics G: Nuclear and
  Particle Physics}, vol.~41, no.~11, p.~115001, 2014.

\bibitem{spuds}
C.~S. Epstein and R.~G. Milner, ``{QED Radiative Corrections to Low-Energy
  M{\o}ller and Bhabha Scattering}.'' Pre-print: arXiv:1602.07609, 2016.

\bibitem{YENNIE1961379}
D.~Yennie {\em et~al.}, ``The infrared divergence phenomena and high-energy
  processes,'' {\em Annals of Physics}, vol.~13, no.~3, pp.~379 -- 452, 1961.

\bibitem{vacpolweb}
F.~Ignatov, ``Vacuum polarization.'' \url{http://cmd.inp.nsk.su/~ignatov/vpl/}.
\newblock Accessed: 2016-05-23.

\bibitem{vacpolpres}
F.~Ignatov, ``Calculation of the vacuum polarization,'' The 4th Meeting of the
  Working Group on Rad. Corrections and MC Generators for Low Energies, IHEP,
  Beijing, 2008.

\bibitem{bettini}
A.~Bettini, {\em Introduction to Elementary Particle Physics}.
\newblock Cambridge, United Kingdom: Cambridge University Press, 2008.

\bibitem{maxwell1}
J.~C. Maxwell, ``{Illustrations of the dynamical theory of gases.---Part I. On
  the Motions and Collisions of Perfection Elastic Spheres},'' {\em Phil.
  Mag.}, vol.~19, pp.~19--32, 1860.

\bibitem{maxwell2}
J.~C. Maxwell, ``{Illustrations of the dynamical theory of gases.---Part II. On
  the process of diffusion of two or more kinds of moving particles among one
  another},'' {\em Phil. Mag.}, vol.~20, pp.~21--37, 1860.

\bibitem{rothvac}
A.~Roth, {\em Vacuum Technology}.
\newblock Amsterdam: North-Holland Publishing Company, 2~ed., 1982.

\bibitem{Knudsen}
M.~Knudsen, ``{Thermischer Molekulardruck der Gase in R\"{o}hren},'' {\em Ann.
  Physik.}, vol.~33, p.~1435, 1910.

\bibitem{0034-4885-49-10-001}
W.~Steckelmacher, ``Knudsen flow 75 years on: the current state of the art for
  flow of rarefied gases in tubes and systems,'' {\em Reports on Progress in
  Physics}, vol.~49, no.~10, p.~1083, 1986.

\bibitem{sutherland}
W.~Sutherland, ``{The viscosity of gases and molecular force},'' {\em Phil.
  Mag.}, vol.~36, pp.~507--531, 1893.

\bibitem{weast}
R.~Weast, ed., {\em CRC Handbook of Chemistry and Physics}.
\newblock Boca Raton, Florida: CRC Press, 62~ed., 1981.

\bibitem{kcos}
M.~Knudsen, ed., {\em The kinetic theory of gases: some modern aspects}.
\newblock Methuen's monographs on physical subjects, London: Methuen, 1952.

\bibitem{0022-3727-11-4-011}
W.~Steckelmacher, ``Molecular flow conductance of long tubes with uniform
  elliptical cross-section and the effect of different cross-sectional
  shapes,'' {\em Journal of Physics D: Applied Physics}, vol.~11, no.~4,
  p.~473, 1978.

\bibitem{comsol}
{COMSOL, inc.}, ``{COMSOL Multiphysics Software}.''
  \url{https://www.comsol.com/products}.

\bibitem{ZHANG2012513}
F.~Pan {\em et~al.}, ``The 18th international vacuum congress (ivc-18): The
  positional and angular distribution of molecules flowing through cylindrical
  tube in free molecular flow,'' {\em Physics Procedia}, vol.~32, pp.~513 --
  524, 2012.

\bibitem{Ackerstaff1998230}
K.~Ackerstaff {\em et~al.}, ``{The HERMES Spectrometer},'' {\em Nucl. Instrum.
  Meth.}, vol.~A 417, no.~2–3, pp.~230 -- 265, 1998.

\bibitem{1748-0221-7-03-C03042}
Y.~Takeuchi {\em et~al.}, ``{Signal shape and charge sharing between electrodes
  of GEM in dimethyl ether},'' {\em Journal of Instrumentation}, vol.~7,
  no.~03, p.~C03042, 2012.

\bibitem{kgem}
S.~H. Han {\em et~al.}, ``{Charge-sharing and electron-transfer characteristics
  of a gas electron multiplier (GEM)},'' {\em Journal of the Korean Physical
  Society}, vol.~40, no.~5, pp.~820--825, 2002.

\bibitem{geant4e}
P.~Arce, ``{GEANT4E: Error propagation for track reconstruction inside the
  GEANT4 framework},'' CHEP 2006, Mumbai, February 13--17, 2006.

\bibitem{levmar}
P.~E. Gill and W.~Murray, ``{Algorithms for the Solution of the Nonlinear
  Least-Squares Problem},'' {\em SIAM J. Numer. Anal.}, vol.~15, pp.~977--992,
  1997.

\bibitem{minpack}
J.~J. Mor\'{e}, ``{The MINPACK Project},'' in {\em Sources and Development of
  Mathematical Software} (W.~J. Cowell, ed.), pp.~88--111, Upper Saddle River,
  NJ: Prentice-Hall, 1984.

\bibitem{cminpack}
F.~Devernay, ``{C/C++ Minpack},'' 2014.

\bibitem{russell1}
R.~Russell. Personal communication.

\bibitem{bernauer1}
J.~Bernauer. Personal communication.

\bibitem{brinker1}
F.~Brinker. Personal communication.

\bibitem{Albrecht:1996gr}
H.~Albrecht {\em et~al.}, ``{Physics with ARGUS},'' {\em Phys. Rept.},
  vol.~276, pp.~223--405, 1996.

\bibitem{bernauer2}
J.~Bernauer. Personal communication.

\bibitem{Actis2010}
S.~Actis {\em et~al.}, ``Quest for precision in hadronic cross sections at low
  energy: Monte carlo tools vs. experimental data,'' {\em The European Physical
  Journal C}, vol.~66, no.~3, pp.~585--686, 2010.

\bibitem{KellyPhysRevC.70.068202}
J.~J. Kelly, ``Simple parametrization of nucleon form factors,'' {\em Phys.
  Rev. C}, vol.~70, p.~068202, Dec 2004.

\bibitem{rimal}
D.~Rimal, {\em Proton Form Factor Puzzle and the CEBAF Large Acceptance
  Spectrometer (CLAS) Two-Photon Exchange Experiment}.
\newblock PhD thesis, Florida International University, Miami, Florida, 2014.

\bibitem{Benisch2001314}
T.~Benisch {\em et~al.}, ``{The luminosity monitor of the HERMES experiment at
  DESY},'' {\em Nucl. Instrum. Meth.}, vol.~A 471, no.~3, pp.~314 -- 324, 2001.

\bibitem{compsat}
A.~Schmidt and J.~C. Bernauer, ``The comparator saturation hypothesis,'' tech.
  rep., The OLYMPUS Collaboration, 2016.

\bibitem{russell2}
R.~Russell. Personal communication.

\bibitem{PhysRev.129.1834}
L.~N. Hand, ``Experimental investigation of pion electroproduction,'' {\em
  Phys. Rev.}, vol.~129, pp.~1834--1846, Feb 1963.

\bibitem{alice}
M.~Tadel, ``{Raw-data Display and Visual Reconstruction Validation in ALICE},''
  {Proceedings of CHEP 07, Victoria, Canada}, 2007.

\bibitem{Shreiner:2009:OPG:1696492}
D.~Shreiner, {\em OpenGL Programming Guide: The Official Guide to Learning
  OpenGL, Versions 3.0 and 3.1}.
\newblock Addison-Wesley Professional, 7th~ed., 2009.

\end{thebibliography}
